\RequirePackage{silence} 
\documentclass[ twoside,openright,titlepage,numbers=noenddot,
                headinclude,footinclude,cleardoublepage=empty,abstract=on,
                BCOR=5mm,paper=a4,fontsize=11pt
                ]{scrreprt}

\PassOptionsToPackage{utf8}{inputenc}
  \usepackage{inputenc}

\PassOptionsToPackage{LGR,T2A,T1}{fontenc} 
  \usepackage{duerer} 
  \usepackage{fontenc}

\PassOptionsToPackage{
  drafting=false,    
  tocaligned=false, 
  dottedtoc=true,  
  eulerchapternumbers=true, 
  linedheaders=false,       
  floatperchapter=true,     
  eulermath=false,  
  beramono=true,    
  palatino=true,    
  style=classicthesis 
}{classicthesis}

\newcommand{\myTitle}{The geometry of dissipation\xspace}

\newcommand{\myName}{Asier López-Gordón\xspace}

\newcommand{\myFaculty}{Put data here\xspace}

\newcommand{\myUni}{Universidad Autónoma de Madrid\xspace}

\newcommand{\myVersion}{\classicthesis}

\providecommand{\mLyX}{L\kern-.1667em\lower.25em\hbox{Y}\kern-.125emX\@}

\PassOptionsToPackage{russian,ngerman,spanish,english}{babel} 
    \usepackage{babel}

\usepackage{textgreek}

\usepackage{csquotes}

\usepackage[style=ext-numeric,citestyle=numeric-comp,sorting=nyt,sortcites, backend=biber, giveninits=true, maxnames=99]{biblatex}
\DeclareFieldFormat[article,periodical]{volume}{\mkbibbold{#1}}
\DeclareFieldFormat[article,periodical]{number}{\mkbibparens{#1}}
\renewbibmacro{in:}{%
  \ifentrytype{article}
    {}
    {\printtext{\bibstring{in}\intitlepunct}}}

\DeclareFieldFormat{issuedate}{#1}
\renewcommand{\jourvoldelim}{\addcomma\space}

\renewbibmacro*{journal+issuetitle}{%
  \usebibmacro{journal}%
  \setunit*{\jourvoldelim}%
  \iffieldundef{series}
    {}
    {\setunit*{\jourserdelim}%
     \printfield{series}%
     \setunit{\servoldelim}}%
  \usebibmacro{volume+number+eid}%
  \setunit{\addcolon\space}%
  \usebibmacro{issue}%
  \setunit{\bibpagespunct}%
  \printfield{pages}%
  \setunit{\space}%
  \usebibmacro{issue+date}%
  \newunit}

\renewbibmacro*{note+pages}{%
  \printfield{note}%
  \newunit}

\DeclareFieldFormat[article,periodical]{date}{\mkbibparens{#1}}

\renewbibmacro*{publisher+location+date}{%
  \printlist{publisher}%
  \iflistundef{location}
    {\setunit*{\addcomma\space}}
    {\setunit*{\addcolon\space}}%
  \printlist{location}%
  \setunit*{\addcomma\space}%
  \usebibmacro{date}%
  \newunit}

\AtEveryBibitem{\clearfield{month}}
\AtEveryBibitem{\clearfield{day}}

\AtEveryBibitem{
  \ifentrytype{online}{}{
    \ifentrytype{thesis}{}{
      \ifentrytype{unpublished}{}{ 
        \clearfield{url}
      }
    }
  }
  \clearfield{language}
}

\DeclareSourcemap{
  \maps[datatype=bibtex]{
    \map[overwrite=true]{
      \step[fieldset=language, null]
      \step[fieldset=address, null]
     \step[fieldset=pagetotal, null]
    }
  }
}

\DeclareSourcemap{
    \maps[datatype=bibtex]{
      \map{
        \step[fieldsource=eprint,final]
        \step[fieldset=url,null]
        \step[fieldset=doi,null]
      }  
    }
  }

\AtEveryBibitem{\clearfield{pubstate}} 
\AtEveryBibitem{\clearfield{eprintclass}} 
\AtEveryBibitem{\clearfield{version}} 

\PassOptionsToPackage{fleqn}{amsmath}       
  \usepackage{amsmath}

\usepackage{graphicx} %
\usepackage{scrhack} 
\usepackage{xspace} 
\PassOptionsToPackage{printonlyused,smaller}{acronym}
  \usepackage{acronym} 

\usepackage{tabularx} 
\usepackage{subcaption}

\usepackage{listings}
\usepackage{classicthesis}

\hypersetup{%
  colorlinks=true, linktocpage=true, pdfstartpage=3, pdfstartview=FitV,%
  breaklinks=true, pageanchor=true,%
  pdfpagemode=UseNone, %
  plainpages=false, bookmarksnumbered, bookmarksopen=true, bookmarksopenlevel=1,%
  hypertexnames=true, pdfhighlight=/O,
  urlcolor=CTurl, linkcolor=CTlink, citecolor=CTcitation, 
  pdftitle={\myTitle},%
  pdfauthor={\textcopyright\ \myName, \myUni, \myFaculty},%
  pdfsubject={},%
  pdfkeywords={},%
  pdfcreator={pdfLaTeX},%
  pdfproducer={LaTeX with hyperref and classicthesis}%
}

\makeatletter
\@ifpackageloaded{babel}%
  {%
    \addto\extrasamerican{%
    }%
    \addto\extrasngerman{%
    }%
    }{\relax}
\makeatother

\newcommand{\insquote}[2]{
  \begin{flushleft}
  {\slshape
  #1} \\ \medskip
  --- #2
\end{flushleft}
}

\usepackage[hyphenbreaks]{breakurl}
\usepackage{xurl}
\hypersetup{breaklinks=true}

\usepackage{tikz}
\usetikzlibrary{cd,babel,shapes.multipart,patterns,patterns.meta}

\usepackage{mathtools}
\usepackage{mathrsfs}
\usepackage{amsfonts}
\usepackage{amsthm}
\usepackage{bm}
\usepackage{amssymb}

\newcommand\restr[2]{{
  \left.\kern-\nulldelimiterspace 
  #1 
  \right|_{#2} 
}}

\newcommand*{\NN}{\mathbb{N}}
\newcommand*{\FF}{\mathbb{F}}
\newcommand*{\ZZ}{\mathbb{Z}}

\newcommand*{\RR}{\mathbb{R}}
\let\R\RR

\newcommand*{\TT}{\mathbb{T}}
\newcommand*{\BB}{\mathbb{B}}
\newcommand*{\Sp}{\mathbb{S}}

\newcommand*{\action}{\mathcal{A}} 
\newcommand*{\zaction}{\mathcal{Z}} 

\newcommand{\dd}{\mathrm{d}}
\let\d\dd

\newcommand{\DD}{\mathrm{D}}

\newcommand{\Legp}{\FF^{f+} L_{d}}
\newcommand{\Legm}{\FF^{f-} L_{d}}

\newcommand{\Cinfty}{\mathscr{C}^\infty}
\newcommand{\Ctwo}{\mathscr{C}^2}

\newcommand{\Czero}{\mathscr{C}^0}
\newcommand{\T}{{\mathsf T}}
\newcommand{\cT}{\T^\ast}
\newcommand{\V}{\mathrm{V}} 
\newcommand{\Hor}{\mathrm{H}} 
\newcommand{\Com}{\mathrm{C}} 

\newcommand{\Lie}{\mathscr{L}}

\newcommand{\X}{\mathfrak{X}}

\renewcommand{\L}{L}

\newcommand{\Reeb}{R}
\newcommand{\Rayl}{\mathcal{R}} 
\newcommand{\evol}{\mathcal{E}}

\newcommand{\Rt}{\Reeb_{t}}
\newcommand{\Rz}{\Reeb_{z}}
\let\Rs\Rz

\newcommand{\parder}[2]{\frac{\partial #1}{\partial #2}}

\newcommand{\tparder}[2]{\partial #1/\partial #2}
\newcommand{\parderr}[3]{\frac{\partial^2 #1}{\partial #2\partial #3}}

\newcommand{\tparderr}[3]{\partial^2 #1/\partial #2\partial #3}

\DeclareMathOperator{\hor}{hor} 

\DeclareMathOperator{\Ima}{Im}
\DeclareMathOperator{\Diff}{Diff}

\let\rank\rk
\let\corank\crk

\DeclareMathAlphabet{\mathpzc}{OT1}{pzc}{m}{it}
\def\d{\mathrm{d}}
\DeclareMathOperator{\id}{id}
\let\Id\id
\DeclareMathOperator{\pro}{pr}
\DeclareMathOperator{\Ad}{Ad}
\DeclareMathOperator{\vol}{vol}
\DeclareMathOperator{\sn}{sn}
\DeclareMathOperator{\cn}{cn}

\DeclareMathOperator{\grad}{grad}

\DeclarePairedDelimiter{\sqbracks}{[}{]}
\newcommand*{\lieBr}[1]{\sqbracks{#1}}
\DeclarePairedDelimiter{\jacBr}{\lbrace}{\rbrace}
\DeclarePairedDelimiter{\set}{\{}{\}}
\DeclarePairedDelimiter{\spn}{\langle}{\rangle}
\DeclarePairedDelimiter{\ray}{\langle}{\rangle_+} 
\DeclarePairedDelimiter{\gen}{\langle}{\rangle}

\newcommand{\cd}{\mathscr{H}} 
\let\cdist\cd
\newcommand{\Ldist}{\mathfrak{L}} 

\newcommand{\pr}[1]{\pro_{#1}}

\newcommand{\liedv}[1]{\Lie_{#1}}
\newcommand*{\contr}[1]{\iota_{#1}}

\newcommand{\lsharp}{\sharp_{\Lambda}}
\newcommand{\tensors}{\mathcal{T}}

\newcommand{\Sendo}{\mathcal{S}} 
\newcommand{\Sendoadj}{\mathcal{S}^\ast} 
\newcommand{\sode}{\Gamma} 
\newcommand{\incl}{i}

\newcommand{\tauomega}{_{(\tau, \, \omega)}} 
\newcommand{\dtomega}{_{(\dd t, \, \omega)}} 
\newcommand{\taueta}{_{(\tau, \, \eta)}} 
\newcommand{\Halpha}{_{H, \, \alpha}} 
\newcommand{\Lbeta}{_{L, \, \beta}} 
\newcommand{\LRayl}{_{L, \, \Rayl}} 

\newcommand{\norm}[1]{\left\lvert\left\lvert #1 \right\rvert\right\rvert}

\newcommand{\mommap}{\mathbf{J}}

\newcommand*{\flowm}{\mathcal{F}^-}

\newcommand{\hybrid}{\mathscr{H}}
\newcommand{\qtilde}{\tilde{\mathcal {Q}}}
\newcommand{\ptilde}{\tilde{\mathcal {P}}}

\usepackage{enumitem}
\setlist[enumerate,1]{label={\roman*)}}

\usepackage{hyphenat}

\usepackage{etoolbox}
\pretocmd{\oldpart}{\cleardoublepage}{}{}

\makeatletter
\def\ttl@tocpart{
  \def\ttl@a{\protect\numberline{\thepart}\@gobble{}}}

\let\classic@l@part\l@part
\renewcommand\l@part[2]{%
  \begingroup
  \renewcommand{\numberline}[1]{\textsc{##1} }%
  \classic@l@part{#1}{#2}%
  \endgroup
}
\makeatother

\usepackage[symbols,nogroupskip,sort=none,toc=false]{glossaries-extra}

\glsxtrnewsymbol[description={set of all sections of $\pi\colon E\to M$}]{sections}{\ensuremath{\Gamma(E)}}
\glsxtrnewsymbol[description={set of all vector fields on $M$}]{vector fields}{\ensuremath{\X(M)}}
\glsxtrnewsymbol[description={set of all diffeomorphisms from $M$ to $M$}]{diffeomorphisms}{\ensuremath\Diff(M)}
\glsxtrnewsymbol[description={exterior algebra on $V$}]{exterior algebra}{\ensuremath{\bigwedge^\bullet(V)}}
\glsxtrnewsymbol[description={set of all $p$-forms on $M$}]{p-forms}{\ensuremath{\Omega^p(M)}}
\glsxtrnewsymbol[description={set of all differential forms on $M$}]{differential forms}{\ensuremath{\ensuremath{\Omega^\bullet(M)}}}
\glsxtrnewsymbol[description={bundle of tensors of type $(k, l)$ on $M$}]{tensors, bundle of}{\ensuremath{\tensors^k_l(\T M)}}
\glsxtrnewsymbol[description={tangent bundle of $Q$}]{tangent bundle}{\ensuremath{\tau_Q\colon \T Q \to Q}}
\glsxtrnewsymbol[description={cotangent bundle of $Q$}]{cotangent bundle}{\ensuremath{\pi_Q\colon \cT Q \to Q}}
\glsxtrnewsymbol[description={canonical one-form of $\cT Q$}]{canonical one-form}{\ensuremath{\theta_Q}}
\glsxtrnewsymbol[description={canonical symplectic form of $\cT Q$}]{canonical symplectic form}{\ensuremath{\omega_Q=-\dd \theta_Q}}
\glsxtrnewsymbol[description={exterior derivative of $\alpha$}]{exterior derivative}{\ensuremath{\dd \alpha}}
\glsxtrnewsymbol[description={inner product of $\alpha$ with $X$}]{inner product}{\ensuremath{\contr{X}\alpha}}
\glsxtrnewsymbol[description={Lie derivative of $A$ with respect to $X$}]{Lie derivative}{\ensuremath{\liedv{X}A}}
\glsxtrnewsymbol[description={span of $\Lambda$}]{span of Lambda}{\ensuremath{\spn{\Lambda}}}
\glsxtrnewsymbol[description={ray generated by $\Lambda$}]{ray generated by Lambda}{\ensuremath{\ray{\Lambda}}}
\glsxtrnewsymbol[description={span of $\{v_1, \ldots, v_k\}$}]{span of v, w}{\ensuremath{\spn{\{ v_1, \ldots, v_k\}}}}
\glsxtrnewsymbol[description={canonical pairing of $\alpha$ with $v$}]{pairing}{\ensuremath{\langle \alpha, v\rangle}}
\glsxtrnewsymbol[description={orbit of $x$ by the action of $G$}]{orbit}{\ensuremath{\mathcal{O}(x)=G\cdot x}}
\glsxtrnewsymbol[description={isotropy subgroup of $x$}]{isotropy subgroup}{\ensuremath{G_x}}
\glsxtrnewsymbol[description={tangent lift of the action $\Phi$}]{tangent lift}{\ensuremath{\Phi^{\T}}}
\glsxtrnewsymbol[description={cotangent lift of the action $\Phi$}]{cotangent lift}{\ensuremath{\Phi^{\cT}}}
\glsxtrnewsymbol[description={infinitesimal generator of $\xi$ on $M$}]{infinitesimal generator}{\ensuremath{\xi_M}}
\glsxtrnewsymbol[description={vertical lift of $X$}]{vertical lift}{\ensuremath{X^\V}}
\glsxtrnewsymbol[description={complete lift of $X$}]{complete lift}{\ensuremath{X^\Com}}
\glsxtrnewsymbol[description={horizontal lift of $X$}]{horizontal lift}{\ensuremath{X^\Hor}}
\glsxtrnewsymbol[description={vertical endomorphism}]{vertical endomorphism}{\ensuremath{\Sendo}}
\glsxtrnewsymbol[description={adjoint operator of $\Sendo$}]{adjoint of the vertical endomorphism}{\ensuremath{\Sendoadj}}
\glsxtrnewsymbol[description={\textsc{sode}}]{SODE}{\ensuremath{\sode}}
\glsxtrnewsymbol[description={Poincaré--Cartan one-form}]{Poincaré--Cartan one-form}{\ensuremath{\theta_L}}
\glsxtrnewsymbol[description={Poincaré--Cartan 2-form}]{Poincaré--Cartan 2-form}{\ensuremath{\omega_L=-\dd \theta_L}}
\glsxtrnewsymbol[description={Jacobi manifold}]{Jacobi manifold}{\ensuremath{(M, \Lambda, E)}}
\glsxtrnewsymbol[description={Poisson manifold}]{Poisson manifold}{\ensuremath{(M, \Lambda)}}
\glsxtrnewsymbol[description={symplectic manifold}]{Symplectic manifold}{\ensuremath{(M, \omega)}}
\glsxtrnewsymbol[description={cosymplectic manifold}]{Cosymplectic manifold}{\ensuremath{(M, \tau, \omega)}}
\glsxtrnewsymbol[description={co-oriented contact manifold}]{Contact manifold}{\ensuremath{(M, \eta)}}
\glsxtrnewsymbol[description={cocontact manifold}]{Cocontact manifold}{\ensuremath{(M, \tau, \eta)}}
\glsxtrnewsymbol[description={flat isomorphism defined by $\omega$}]{flat, symplectic}{\ensuremath{\flat_\omega}}
\glsxtrnewsymbol[description={flat isomorphism defined by $(\tau, \omega)$}]{flat, cosymplectic}{\ensuremath{\flat\tauomega}}
\glsxtrnewsymbol[description={flat isomorphism defined by $\eta$}]{flat, contact}{\ensuremath{\flat_\eta}}
\glsxtrnewsymbol[description={flat isomorphism defined by $(\tau, \eta)$}]{flat, cocontact}{\ensuremath{\flat\taueta}}
\glsxtrnewsymbol[description={annihilator of $D$}]{annihilator}{\ensuremath{D^{\circ}}}
\glsxtrnewsymbol[description={Jacobi orthogonal complement of $D$}]{Jacobi orthogonal complement}{\ensuremath{D^{\perp_\Lambda}}}
\glsxtrnewsymbol[description={symplectic orthogonal complement of $D$}]{symplectic orthogonal complement}{\ensuremath{D^{\perp_\omega}}}
\glsxtrnewsymbol[description={Reeb vector field}]{Reeb vector field}{\ensuremath{\Reeb}}
\glsxtrnewsymbol[description={time Reeb vector field}]{time Reeb vector field}{\ensuremath{\Rt}}
\glsxtrnewsymbol[description={contact Reeb vector field}]{contact Reeb vector field}{\ensuremath{\Rz}}
\glsxtrnewsymbol[description={Hamiltonian vector field of $f$}]{Hamiltonian vector field}{\ensuremath{X_f}}
\glsxtrnewsymbol[description={Evolution vector field of $f$}]{Evolution vector field}{\ensuremath{\evol_f}}
\glsxtrnewsymbol[description={(Herglotz--)Euler--Lagrange vector field}]{Euler--Lagrange vector field}{\ensuremath{\sode_L}}
\glsxtrnewsymbol[description={fiber derivative of $f$}]{fiber derivative}{\ensuremath{\FF f}}
\glsxtrnewsymbol[description={Lagrangian energy}]{energy}{\ensuremath{E_L}}
\glsxtrnewsymbol[description={action functional}]{action}{\ensuremath{\mathcal{A}}}
\glsxtrnewsymbol[description={path space from $q_1$ to $q_2$}]{path space}{\ensuremath{\Omega(q_1, q_2, [a, b])}}
\glsxtrnewsymbol[description={bundle coordinates of $\T Q$}]{Bundle coordinates of the tangent bundle}{\ensuremath{(q^i, v^i)}}
\glsxtrnewsymbol[description={symplectic Darboux coordinates}]{Darboux coordinates of a symplectic manifold}{\ensuremath{(q^i, p_i)}}
\glsxtrnewsymbol[description={cosymplectic Darboux coordinates}]{Darboux coordinates of a cosymplectic manifold}{\ensuremath{(t, q^i, p_i)}}
\glsxtrnewsymbol[description={contact Darboux coordinates}]{Darboux coordinates of a contact manifold}{\ensuremath{(q^i, p_i, z)}}
\glsxtrnewsymbol[description={cocontact Darboux coordinates}]{Darboux coordinates of a cocontact manifold}{\ensuremath{(t, q^i, p_i, z)}}
\glsxtrnewsymbol[description=$n$-dimensional torus]{torus}{\ensuremath{\TT^n}}
\glsxtrnewsymbol[description=$n$-dimensional sphere]{sphere}{\ensuremath{\Sp^n}}
\glsxtrnewsymbol[description=Rayleigh potential]{Rayleigh}{\ensuremath{\Rayl}}
\glsxtrnewsymbol[description=momentum map]{momentum map}{\ensuremath{\mommap}}

\usepackage{cleveref}
\crefname{equation}{equation}{equations}
\crefname{figure}{figure}{figures}
\crefname{subsection}{subsection}{subsections}
\crefname{proposition}{proposition}{propositions}
\Crefname{proposition}{Proposition}{Propositions}
\crefname{lemma}{lemma}{lemmas}
\crefname{definition}{definition}{definitions}
\crefname{corollary}{corollary}{corollaries}
\crefname{example}{example}{examples}
\crefformat{part}{Part #2\textsc{#1}#3}
\crefrangeformat{part}{Parts #3\textsc{#1}#4 to #5\textsc{#2}#6}
\crefmultiformat{part}{Parts #2\textsc{#1}#3}%
{ and~#2\textsc{#1}#3}{, #2\textsc{#1}#3}{ and~#2\textsc{#1}#3}

\theoremstyle{plain}
\newtheorem{theorem}{Theorem}[chapter]
\newtheorem{lemma}[theorem]{Lemma}
\newtheorem{proposition}[theorem]{Proposition}
\newtheorem{corollary}[theorem]{Corollary}  
\theoremstyle{definition}
\newtheorem{definition}{Definition}[chapter]
\newtheorem{example}{Example}[chapter]
\newtheorem{counterexample}[example]{Counterexample}
\newtheorem{remark}{Remark}[chapter]

\usepackage{ifthen}

\renewcommand*{\graffito}[1]{}

\usepackage[dvipsnames]{xcolor}
\usepackage{pagecolor}
\definecolor{UBCblue}{HTML}{f7ac00}
\definecolor{UBCgrey}{HTML}{ffe4b2}
\definecolor{complementaryorange}{rgb}{1.0, 0.5, 0.0} 
\definecolor{softwhite}{rgb}{1.0, 0.98, 0.94}    
\definecolor{gold}{rgb}{1.0, 0.84, 0.0}          
\definecolor{coral}{rgb}{1.0, 0.5, 0.31}         
\definecolor{mintgreen}{rgb}{0.6, 1.0, 0.6}      
\definecolor{silvergrey}{rgb}{0.75, 0.75, 0.75}  

\begin{document}
\frenchspacing
\raggedbottom
\pagenumbering{roman}
\pagestyle{plain}
\newpagecolor{RoyalBlue}
\begin{titlepage}
    \begin{addmargin}[-1cm]{-3cm}
    {\color{silvergrey}
    \begin{center}
        \large

        \hfill

        \vspace*{4cm}

        \begingroup
            \color{mintgreen}
            \Huge\textdubf{THE GEOMETRY OF DISSIPATION} \\ \bigskip

        \endgroup

        \vfill
    \end{center}

    \begin{flushright}
        \LARGE{Asier López-Gordón}

        \vfill


    \end{flushright}

    }
  \end{addmargin}
\end{titlepage}
\newpagecolor{white}












\cleardoublepage
\thispagestyle{empty}
\phantomsection
\pdfbookmark[1]{Dedication}{Dedication}

\vspace*{3cm}




\begin{center}
    \usefont{\encodingdefault}{pzc}{m}{n}
    \LARGE
    Dedicated to the loving memory of my grandparents. 
\end{center}

\cleardoublepage\include{FrontBackmatter/Foreword} 
\cleardoublepage
\phantomsection
\pdfbookmark[1]{Abstract}{Abstract}
\addcontentsline{toc}{chapter}{\tocEntry{Abstract}}
\begingroup
\let\clearpage\relax
\let\cleardoublepage\relax
\let\cleardoublepage\relax

\chapter*{Abstract}

Dissipative phenomena manifest in multiple mechanical systems. In this dissertation, different geometric frameworks for modelling non-conservative dynamics are considered. The objective is to generalize several results from conservative systems to dissipative systems, specially those concerning the symmetries and integrability of these systems. More specifically, three classes of geometric frameworks modelling dissipative systems are considered: systems with external forces, contact systems and systems with impacts. The first two allow modelling a continuous dissipation of energy over time, while the latter also permits considering abrupt changes of energy in the instants of the impacts. 

\bigskip

\noindent \textbf{Keywords:} forced system, contact Hamiltonian system, hybrid system, symmetries, reduction, Liouville--Arnold theorem, Hamilton--Jacobi equation, impulsive constraints

\bigskip

\noindent \textbf{MSC2020 codes:} 37J35, 37J39, 37J55, 53D10, 53E50, 53Z05, 65P10, 70F35, 70F40, 70H06, 70H20, 70H30, 70H33, 93C30

\vfill

\begin{otherlanguage}{spanish}
\phantomsection
\pdfbookmark[1]{Resumen}{Resumen}
\addcontentsline{toc}{chapter}{\tocEntry{Resumen}}
\chapter*{Resumen}
Los fenómenos disipativos se manifiestan en múltiples sistemas mecánicos. En esta disertación se consideran diferentes marcos geométricos para modelar dinámicas no conservativas. El objetivo es generalizar diversos resultados de sistemas conservativos para sistemas disipativos, particularmente aquellos que conciernen a las simetrías y la integrabilidad de dichos sistemas. Más concretamente, se consideran tres clases de marcos geométricos que modelan sistemas disipativos: sistemas con fuerzas externas, sistemas de contacto y sistemas con impactos. Los dos primeros permiten modelar una disipación de la energía continua a lo largo del tiempo, mientras que el último también permite considerar cambios abruptos en la energía producidos en los instantes de los impactos.
\end{otherlanguage}

\endgroup

\vfill

\cleardoublepage
\phantomsection
\pdfbookmark[1]{Agradecimientos/Acknowledgments}{acknowledgments}
\addcontentsline{toc}{chapter}{\tocEntry{Agradecimientos/Acknowledgments}}
\chapter*{Agradecimientos/Acknowledgments}

\insquote{Entre los pecados mayores que los hombres cometen, aunque algunos dicen que es la soberbia, yo digo que es el desagradecimiento, ateniéndome a lo que suele decirse: que de los desagradecidos está lleno el infierno. Este pecado, en cuanto me ha sido posible, he procurado yo huir desde el instante que tuve uso de razón, y si no puedo pagar las buenas obras que me hacen con otras obras, pongo en su lugar los deseos de hacerlas, y cuando éstos no bastan, las publico, porque quien dice y publica las buenas obras que recibe, también las recompensara con otras, si pudiera.}
{Don Quijote de la Mancha}

Quiero comenzar por agradecer sinceramente a Manuel de León, mi director de tesis, por su apoyo, sus conocimientos y su cercanía. Desde que comencé a trabajar con él, mientras cursaba el máster, siempre ha mostrado unas ganas de trabajar y de ayudar dignas de admiración.

Mi gratitud se extiende a todos mis colaboradores. 

A Manuel Lainz, mi hermano mayor académico, por sus buenas ideas al abordar un problema y por la multitud de dudas que me ha resuelto estos años.

A Leonardo Colombo por depositar su confianza en mí para múltiples proyectos y por una fructífera colaboración que ha dado lugar a más de un tercio de esta tesis. 

A Javier de Lucas por haberme brindado la oportunidad de realizar una estancia con él, por su sabiduría, su amabilidad y sus inacabables ganas de trabajar.

A Xavier Rivas, con quien charlar, sea frente a una pizarra o en el \textit{Kompania Piwna}, siempre es un placer. 

A María Emma Eyrea Irazú por presentarme los sistemas híbridos.

A Jordi Gaset por su talento construyendo ejemplos.

A Juan Carlos Marrero por sus grandes contribuciones fruto de la curiosidad espontánea.

To Bartosz M.~Zawora for his kindliness and his knowledge.

A Alexandre Anahory Simoes por introducirme a la teoría de control.


Les estoy asimismo agradecido a todos los miembros de la Red de Mecánica, Dinámica y Teoría de Campos. Especialmente a Edith Padrón por responsabilizarles de tantas actividades de la red y asegurar que los jóvenes podamos participar en las mismas. Quiero además mencionar al tristemente fallecido Miguel C.~Muñoz Lecanda, quien se interesó por mi investigación en diversas charlas y aportó comentarios constructivos de gran utilidad. 

I am grateful to Jair Koiller and Tom Mestdag for carefully reviewing this dissertation and providing constructive feedback.

A lo largo de la tesis he tenido la oportunidad de colaborar en la organización de diversos eventos, y la suerte de haberlo hecho junto con equipos fantásticos. He tenido el placer de organizar el Coloquio Junior durante el curso 2023-2024 junto con Alba García, Sergio García y Jorge Pérez; y una sesión paralela en la edición de 2023 del Congreso Bienal de la RSME junto con Miguel Ángel Berbel y Xavier Rivas. Agradezco además a todos mis compañeros del Comité Científico del BYMAT 2023, así como a los del Comité Organizador, por su estupenda labor.

Doy las gracias a Ágata Timón García-Longoria por invitarme a publicar mi primer artículo de divulgación y por su ayuda con la redacción del mismo.

I am thankful for all the wonderful people I have met in conferences and workshops these years. In particular, I want to acknowledge the remarkable hospitality from O\u{g}ul Esen, Begüm Ateşli, Ayten Gezici and the rest of the people from Gebze Technical University during the Workshop on Nonlinear Systems in 2023. 

He tenido la suerte de contar con unos compañeros de tesis estupendos en el ICMAT: Alba, Andrés, Bilson, <<Cherco>>, Diego, Enrique, Javi, Jesús, José Antonio, Miguel, Pablo, <<Quino>> y Sergio. En particular, ha sido un placer compartir despacho con Arnau, Guillermo y Rodrigo.  

Debo agradecer además al personal administrativo del ICMAT por ayudarme a enfrentarme a la leviatánica burocracia del CSIC. En particular, agradezco su paciencia a Esther Ruiz y Teresa Ruiz. 

Quiero agradecer de todo corazón a mis padres por haberme brindado siempre su amor y su apoyo incondicional, y por haberme alentado a ser una persona curiosa y trabajadora. 

I am thankful to Natasha, my partner in crime, for her love and support.

Agradezco a todos mis amigos por haber creído en mí y les pido disculpas por mis periodos cuasiermitaños por estudios o trabajo.

I acknowledge financial support from Spanish Ministry of Science and Innovation (MCIN/AEI/ 10.13039/501100011033) via the predoctoral contract PRE2020-093814.

Finally, I wish to express my sincere gratitude to Alexandra Elbakyan and Aaron Swartz, as well as all the collaborators of Sci-Hub, Z-Library, Anna's Archive, Library Genesis and similar platforms for their remarkable role in making scientific knowledge free and unrestricted to the whole humankind. 

¡Muchas gracias! Thank you very much! Dziękuję bardzo! 
\begin{otherlanguage*}{russian}
    Большое спасибо!
\end{otherlanguage*}

\cleardoublepage
\phantomsection
\pdfbookmark[1]{Publications}{publications}
\addcontentsline{toc}{chapter}{\tocEntry{Publications}}
\chapter*{Publications}


The majority of the results exposed in the present dissertation have been developed and published in the following list of articles.

\section*{Journal articles}
\begin{refsection}[journals]
    \small
    \nocite{*} 
    \printbibliography[heading=none]
\end{refsection}

\section*{Conference papers}
\begin{refsection}[conferences]
    \small
    \nocite{*} 
    \printbibliography[heading=none]
\end{refsection}

\section*{Preprints}
\begin{refsection}[preprints]
    \small
    \nocite{*} 
    \printbibliography[heading=none]
\end{refsection}


\cleardoublepage
\pagestyle{scrheadings}
\pdfbookmark[1]{\contentsname}{tableofcontents}
\setcounter{tocdepth}{2} 
\setcounter{secnumdepth}{3} 
\manualmark
\markboth{\spacedlowsmallcaps{\contentsname}}{\spacedlowsmallcaps{\contentsname}}
\tableofcontents
\automark[section]{chapter}
\renewcommand{\chaptermark}[1]{\markboth{\spacedlowsmallcaps{#1}}{\spacedlowsmallcaps{#1}}}
\renewcommand{\sectionmark}[1]{\markright{\textsc{\thesection}\enspace\spacedlowsmallcaps{#1}}}
\clearpage
\begingroup
    \let\clearpage\relax
    \let\cleardoublepage\relax
    \phantomsection
    \pdfbookmark[1]{List of figures}{List of figures}
    \addcontentsline{toc}{chapter}{\tocEntry{List of figures}}
    \listoffigures

    \vspace{8ex}


    \newpage




    
    \phantomsection
    \pdfbookmark[1]{Symbols}{Symbols}
    \addcontentsline{toc}{chapter}{\tocEntry{Symbols}}
    \printunsrtglossary[type=symbols,style=long]

\endgroup

\cleardoublepage
\pagestyle{scrheadings}
\pagenumbering{arabic}
\cleardoublepage


\chapter{Introduction}\label{ch:introduction}


\insquote{La filosofia è scritta in questo grandissimo libro che continuamente ci sta aperto innanzi agli occhi (io dico l’universo), ma non si può intendere, se prima non s’impara a intender la lingua, e conoscer i caratteri ne’quali è scritto. Egli è scritto in lingua matematica, e i caratteri son triangoli, cerchi ed altre figure geometriche, senza i quali mezzi è impossibile a intenderne umanamente parola; senza questi è un aggirarsi vanamente per un oscuro laberinto.
    \\
Philosophy is written in this grand book, the universe, which stands continually open to our gaze. But the book cannot be understood unless one first learns to comprehend the language and read the letters in which it is composed. It is written in the language of mathematics, and its characters are triangles, circles, and other geometric figures without which it is humanly impossible to understand a single word of it; without these, one wanders about in a dark labyrinth.
}{Galileo Galilei, \emph{Il Saggiatore} (1623)}

Einstein's equivalence principle states the laws of physics are the same for all observers. Consequently, the natural language for formulating the laws of physics is a coordinate-independent one, that is, the language of differential geometry. It is well-known among theoretical physicists that the role of differential geometry in general relativity is crucial. As a matter of fact, some authors refer to general relativity as \emph{geometrodynamics} \cite{M.T.W2017}. Albeit unfamiliar for most physicists, there are also geometric structures underlying the classical Newtonian, Lagrangian and Hamiltonian formalisms of mechanics. In particular, symplectic manifolds are the natural framework for Hamiltonian mechanics. Classical concepts such as trajectories, conserved quantities, symmetries or Poisson brackets can be given a geometrical interpretation. Providing a geometrical background to dynamical systems is not merely mathematically interesting or aesthetically pleasing. Understanding the geometry underlying a dynamical system provides a plethora of methods for studying the system, either quantitatively or qualitatively, analytically or numerically. For instance, if a Lie group of symmetries preserves a Hamiltonian system, then it is possible to get rid of the redundant degrees of freedom by projecting the dynamics onto a phase space with fewer dimensions.

Autonomous (that is, time-independent) Hamiltonian and Lagrangian systems are \emph{conservative}, namely, the total energy is preserved along the curves satisfying the Euler--Lagrange or Hamilton equations. Moreover, Hamiltonian flows preserve the symplectic structure. From the viewpoint of statistical mechanics, this implies that the volume of phase space occupied by an ensemble of states remains constant with time. Nevertheless, several dynamical systems appearing in physics and engineering are \emph{dissipative}, that is, systems whose total energy is not conserved along their evolution. In addition, in many dissipative dynamical systems, the volume of phase space occupied by an ensemble of states decreases with time. 
The dissipative character of a system may manifest as a continuous decrease in its energy over time or as an abrupt change in energy, for instance, at the instant when a collision occurs.

The objective of this doctoral dissertation is to study geometric frameworks for modelling dissipative phenomena, generalizing results regarding conservative systems. In particular, the symmetries and integrability of dissipative systems are explored. More specifically, three types of dissipative systems are considered in this dissertation: systems with external forces, contact systems and systems with impacts. The dissipation in both systems with external forces and contact systems appears as a decrease of energy over time. On the other hand, systems with impacts can also experience brusque changes of energy when the impacts occur.

It is worth mentioning that there are several other geometric structures for modelling dissipative phenomena that will not be considered in this dissertation. For instance, $b$-symplectic manifolds can be utilized for modelling fluids with dissipation \cite{C.M.M2023}. In order to model a dissipation due to entropy production, the metriplectic \cite{B.B.P+2007,C.M2020} or the \textsc{generic} \cite{E.G.P2022,G.O1997,Grmela2014} frameworks can be employed.
Homogeneous Hamiltonian control systems have also been used to model irreversible thermodynamical processes \cite{M.v2018,v.M2018}. A variational formulation of nonequilibrium thermodynamics can be found in \cite{G.Y2018}.

\section*{Structure of the dissertation}

This thesis is structured in four parts. For the sake of completeness, several fundamental concepts are reviewed in \Cref{part:preliminaries}. \Cref{ch:geometric_foundations} exposes the topics of differential geometry that will be employed throughout the rest of the dissertation. \Cref{ch:review_mechanics} explains the main results of geometric mechanics concerning conservative Hamiltonian and Lagrangian systems, and briefly presents nonholonomic and discrete mechanical systems. 

\Cref{part:forces} is devoted to systems with external forces. Forced Hamiltonian and Lagrangian systems are introduced in \Cref{ch:forced}. In \Cref{ch:forced_symmetries}, the symmetries of forced Hamiltonian and Lagrangian systems are studied and a reduction method \textit{à la} Marsden--Weinstein for forced Lagrangian systems is presented. A Hamilton--Jacobi equation for forced Hamiltonian systems is derived in \Cref{ch:forced_HJ}. In addition, the reduction and reconstruction of the Hamilton--Jacobi problem in presence of a group of symmetries is presented. In \Cref{ch:forced_discrete_HJ} a Hamilton--Jacobi theory for discrete systems with external forces is developed.

\Cref{part:contact} is devoted to (co)contact Hamiltonian and Lagrangian systems. Autonomous and non-autonomous contact Hamiltonian and Lagrangian systems are presented in \Cref{ch:contact_Hamiltonian_systems}. A classification of the symmetries of time-dependent contact Hamiltonian and Lagrangian systems, the relations between them and with dissipated quantities is carried out in \Cref{ch:contact_symmetries}. In \Cref{ch:contact_HJ}, two Hamilton--Jacobi equations for time-dependent contact Hamiltonian systems are derived. A Liouville--Arnol'd theorem for contact Hamiltonian systems is proven in \Cref{ch:contact_integrability}. In order to do so, a Liouville--Arnol'd theorem for homogeneous functions on exact symplectic manifolds is also proven. \Cref{ch:contact_stability} exposes some preliminary results concerning the stability of contact Hamiltonian systems by means of the Lyapunov method. 

\Cref{part:impacts} is devoted to systems with impacts. \Cref{ch:hybrid_systems} presents hybrid systems, a formalism that will be utilized to model certain systems with collisions. The reduction of hybrid forced systems with symmetries is studied in \Cref{ch:hybrid_reduction}. A notion of integrability of hybrid systems is proposed in \Cref{ch:integrability_hybrid}. In \Cref{ch:hybrid_HJ}, a Hamilton--Jacobi theory for forced and nonholonomic 
hybrid systems is developed. \Cref{ch:contact_impulsive} introduces contact Lagrangian systems with impulsive constraints. In order to model dissipative systems with impacts, the Herglotz principle is extended for nonsmooth curves in \Cref{ch:nonsmooth_Herglotz}. Finally, \Cref{ch:further_work} presents some possible lines of future work steming from the results of this dissertation.

\chapter{Introducción}\label{ch:introduccion}

El principio de equivalencia de Einstein establece que las leyes de la física son las mismas para todos los observadores. Por consiguiente, el lenguaje natural para formular las leyes de la física es uno independiente de coordenadas, esto es, el lenguaje de la geometría diferencial. Es bien sabido entre los físicos teóricos que la geometría diferencial es crucial en relatividad general. Más aún, algunos autores se refieren a dicha teoría como \emph{geometrodinámica} \cite{M.T.W2017}. Pese a no ser conocidas para la mayoría de los físicos, también existen estructuras geométricas que subyacen a los formalismos newtoniano, lagrangiano y hamiltoniano de la mecánica clásica. En particular, las variedades simplécticas son el marco natural para la mecánica hamiltoniana. Conceptos clásicos como el de trayectoria, cantidad conservada, simetría o corchete de Poisson pueden interpretarse geométricamente. Otorgar de un marco geométrico a un sistema dinámico no es meramente interesante desde el punto de vista matemático o estético, sino que proporciona una plétora de métodos para estudiar al sistema, tanto cualitativa como cuantitativamente, tanto analítica como numéricamente. Por ejemplo, si la acción de un grupo de Lie de simetrías preserva un sistema hamiltoniano, es posible deshacerse de los grados de libertad redundantes proyectando la dinámica a un espacio de fases de menor dimensión.

Los sistemas hamiltonianos y lagrangianos autónomos (esto es, independientes del tiempo) son \emph{conservativos}, a saber, la energía total se preserva a lo largo de las curvas que satisfacen las ecuaciones de Euler--Lagrange o las de Hamilton. Desde el punto de vista de la física estadística, esto implica que 
el volumen del espacio de fases ocupado por una colección de partículas permanece constante con el transcurso del tiempo. No obstante, existen múltiples sistemas dinámicos en física e ingeniería que son \emph{disipativos}, es decir, sistemas cuya energía total no se conserva a lo largo de su evolución.
Asimismo, en muchos sistemas dinámicos disipativos el volumen del espacio de fases ocupado por una colección de partículas disminuye con el tiempo. El caracter disipativo de un sistema puede manifestarse como una disminución continua de su energía con el paso del tiempo o como un cambio abrupto en la energía, por ejemplo, en el instante en el que se produce una colisión.

El objetivo de esta disertación doctoral es el estudio de marcos geométricos para modelar fenómenos disipativos, y generalizar resultados que se conocían para sistemas conservativos, haciendo un especial énfasis en las simetrías e integrabilidad de los sistemas disipativos. Más concretamente, en esta disertación se consideran tres marcos geométricos: sistemas con fuerzas externas, sistemas de contacto y sistemas con impactos. La disipación en los dos primeros aparece como una disminución de la energía de forma continua en el tiempo, mientras que en el tercero también pueden producirse cambios bruscos en la energía cuando los impactos tienen lugar.

Cabe mencionar que hay otras muchas estructuras geométricas con las que es posible modelar fenómenos disipativos que no se consideraran en esta disertación. Entre ellas, podemos mencionar las variedades $b$-simplécticas, que han sido empleadas para estudiar fluidos con disipación \cite{C.M.M2023}. Con el fin de modelar una disipación debida a la producción de entropía se pueden emplear los formalismos metripléctico \cite{B.B.P+2007,C.M2020} o \textsc{generic} \cite{E.G.P2022,G.O1997,Grmela2014}. Los sistemas de control hamiltonianos homogéneos también se han utilizado para representar procesos termodinámicos irreversibles \cite{M.v2018,v.M2018}. Para una formulación variacional de la termodinámica del no equilibrio véase \cite{G.Y2018}.

\section*{Estructura de la disertación}

Esta tesis está estructurada en cuatro partes. Por completitud, múltiples conceptos básicos se revisan en la Parte~\textsc{\ref{part:preliminaries}}. El Capítulo~\ref{ch:geometric_foundations} expone las nociones de geometría diferencial que se emplearán en el resto de la disertación.  El Capítulo~\ref{ch:review_mechanics} explica los principales resultados de mecánica geométrica concernientes a los sistemas hamiltonianos y lagrangianos conservativos, y presenta brevemente los sistemas no holónomos así como los sistemas mecánicos discretos.

La Parte~\textsc{\ref{part:forces}} está dedicada a los sistemas con fuerzas externas. Los sistemas hamiltonianos y lagrangianos forzados se introducen en el Capítulo~\ref{ch:forced}. En el Capítulo~\ref{ch:forced_symmetries} se estudian las simetrías de los sistemas  hamiltonianos y lagrangianos forzados, y se presenta un método de reducción a la Marsden--Weinstein para sistemas lagrangianos forzados. En el Capítulo~\ref{ch:forced_HJ} se deriva la ecuación de Hamilton--Jacobi para sistemas hamiltonianos forzados, y además se estudia la reducción y reconstrucción del problema de Hamilton--Jacobi en presencia de un grupo de simetrías. En el Capítulo~\ref{ch:forced_discrete_HJ} se desarrolla una teoría de Hamilton--Jacobi para sistemas discretos con fuerzas externas.

La Parte~\textsc{\ref{part:contact}} versa sobre los sistemas hamiltonianos y lagrangianos de (co)contacto. Los sistemas hamiltonianos y lagrangianos de contacto, tanto autónomos como no autónomos, se presentan en el Capítulo~\ref{ch:contact_Hamiltonian_systems}. En el Capítulo~\ref{ch:contact_symmetries} se lleva a cabo una clasificación de las simetrías de los sistemas hamiltonianos y lagrangianos de contacto dependientes del tiempo, y se estudian las relaciones entre ellas y con las cantidades disipadas. En el Capítulo~\ref{ch:contact_HJ} se derivan dos ecuaciones de Hamilton--Jacobi para sistemas hamiltonianos de contacto dependientes del tiempo. Un teorema de Liouville--Arnol'd para sistemas hamiltonianos de contacto se demuestra en el Capítulo~\ref{ch:contact_integrability}. Para ello, asimismo se demuestra un teorema de Liouville--Arnol'd para funciones homogéneas en variedades simplécticas exactas. En el Capítulo~\ref{ch:contact_stability} se exponen algunos resultados de un proyecto en desarrollo acerca del estudio de la estabilidad de los sistemas hamiltonianos de contacto por medio del método de Lyapunov.

La Parte~\textsc{\ref{part:impacts}} versa sobre sistemas con impactos. En el Capítulo~\ref{ch:hybrid_systems} se presentan los sistemas híbridos, un formalismo que se empleará para modelar ciertos sistemas con colisiones. La reducción de los sistemas híbridos forzados con simetrías se estudia en el 
Capítulo~\ref{ch:hybrid_reduction}. Una noción de integrabilidad en sistemas híbridos se propone en el Capítulo~\ref{ch:integrability_hybrid}. En el Capítulo~\ref{ch:hybrid_HJ}, se desarrolla una teoría de Hamilton--Jacobi theory para sistemas híbridos con fuerzas externas o con ligaduras no holónomas. En el Capítulo~\ref{ch:contact_impulsive} se introducen los sistemas lagrangianos de contacto con ligaduras impulsivas. Con el fin de modelar sistemas disipativos con impactos, el principio variacional de Herglotz se generaliza para curvas no diferenciables en el Capítulo~\ref{ch:nonsmooth_Herglotz}. Por último, en el Capítulo~\ref{ch:further_work} se presentan algunas posibles líneas de investigación que se siguen de los resultados de la presente disertación.

\part{Preliminaries}\label{part:preliminaries}
\chapter{Geometric foundations}\label{ch:geometric_foundations}

\insquote{I do not hesitate to say that mathematics deserve to be cultivated for their own sake, and the theories inapplicable to physics as well as the others. Even if the physical aim and the esthetic aim were not united, we ought not to sacrifice either.
    But more: these two aims are inseparable and the best means of attaining one is to aim at the other, or at least never to lose sight of it.} {Henri Poincaré, \emph{The value of science}, Tr.~George Bruce Halsted (1907)}

In this chapter, the fundamental concepts and results from differential geometry that will be employed are reviewed. The reader is assumed to have a basic knowledge of the geometry and topology of smooth manifolds (see, for instance, \cite{Lee2011, Lee2012, D.F.N1985, C.C.L1999, Warner1983}).
Hereinafter, unless otherwise stated, all the mappings will be assumed to be $\Cinfty$-smooth. Manifolds will be assumed to be smooth, Hausdorff and second-countable. Neighbourhoods will be understood to be open subsets.

\section{Fiber bundles, tensor fields and differential forms}

For the sake of completeness, some concepts about fiber bundles, tensors and calculus on manifolds are reviewed in this section. For further information, see \cite{A.M2008, A.M.R1988, Lee2012, d.R1989,C.C.L1999,K.S.M1993,Warner1983}.

Let $M$ and $F$ be manifolds. A \emph{fiber bundle $\pi\colon E \to M$ with model fiber $F$} is a manifold $E$ together with a surjective map $\pi\colon E \to M$ with the property that, for each $x\in M$, there exists a neighbourhood $U$ of $x$ in $M$ and a diffeomorphism $\Phi\colon \pi^{-1}(U)\to U\times F$ such that $\pi_1\circ \Phi= \pi$, where $\pi_1\colon U\times F \to U$ denotes the canonical projection. The manifolds $E$ and $M$ are called \emph{total space} and \emph{base space}, respectively. The map $\pi$ is called the \emph{projection}, and the preimage $\pi^{-1}(x)$ of $x\in M$ is called the \emph{fiber over $x$}. The pair $(U, \Phi)$ is called a \emph{local trivialization}. In particular, a \emph{trivial fiber bundle} is one that admits a local trivialization over the entire base space (a \emph{global trivialization}).

    It is noteworthy that there are more general definitions of fiber bundle which do not require the existence of local trivializations. If $E$ and $M$ are manifolds and $\pi\colon E\to M$ is a surjective submersion, it is said that \emph{$E$ is a fibered manifold over $M$}. 

Given a fiber bundle $\pi\colon E \to M$, a chart $(U; x^1, \ldots x^n, y^1, \ldots, y^p)$ on $E$ is called a \emph{fiber chart} if there exists a chart $(\pi(U); x^1\circ \pi, \ldots, x^n\circ \pi )$ of $M$. The coordinates $(x^1, \ldots, x^1, y^1, \ldots, y^p)$ are called \emph{bundle coordinates}. Hereinafter, with a slight abuse of notation, the coordinates $x^i\circ \pi$ in $M$ will be simply denoted by $x^i$. In bundle coordinates, the projection is simply given by 
\begin{equation}
    \pi\colon \left(x^1, \ldots x^n, y^1, \ldots, y^p\right) \mapsto  \left(x^1, \ldots x^n \right)\, .
\end{equation}

Let $\pi_X\colon X\to M$ and $\pi_Y\colon Y \to N$ be fiber bundles. A smooth map $F\colon X \to Y$ is called a \emph{bundle morphism} (or \emph{bundle homomorphism}) if there exists a smooth map $f\colon M \to N$ such that the following diagram commutes:
\begin{equation}\label{eq:bundle_homomorphism}
\begin{tikzcd}
    X \arrow[r, "F"] \arrow[d, "\pi_X"'] & Y \arrow[d, "\pi_Y"] \\
    M \arrow[r, "f"]                     & N                   
\end{tikzcd}
\end{equation}
In particular, if $\pi_X\colon X\to M$ and $\pi_Y\colon Y \to M$ are fiber bundles over $M$, then a map $F\colon X \to Y$ is a bundle morphism if and only if $\pi_Y\circ F = \pi_X$, that is, the diagram
\begin{equation}
\begin{tikzcd}
    X \arrow[rd, "\pi_X"'] \arrow[rr, "F"] &   & Y \arrow[ld, "\pi_Y"] \\
                                           & M &                      
\end{tikzcd}
\end{equation}
commutes. If a bundle morphism has an inverse which is also a bundle morphism, then it is called a {bundle isomorphism}.

A \emph{(real) vector bundle of rank $k$ over $M$} is a fiber bundle $\pi\colon E \to M$ such that:
\begin{enumerate}
    \item For each point $x\in M$, the fiber $E_x=\pi^{-1}(x)$ is endowed with the structure of a $k$-dimensional real vector space.
    \item Given a local trivialization $(U, \Phi)$, for each point $x\in U$, the restriction of $\Phi$ to $E_x$ is a vector space isomorphism from $E_x$ to $\{x\}\times \R^k \cong \R^k$.   
\end{enumerate}
A vector bundle of rank 1 is called a \emph{line bundle}.

Given a vector bundle $\pi\colon E \to M$, a \emph{section of $E$} is a map $\sigma\colon M \to E$ satisfying $\pi\circ \sigma = \Id_M$. Given an open subset $U\subseteq M$, a \emph{local section of $E$} is a section of $U$. The set of all sections of $E$ is denoted by $\Gamma(E)$. This set is a real vector space and a $\Cinfty(M)$-module.

Let $\pi_X\colon X\to M$ and $\pi_Y\colon Y \to N$ be real vector bundles. A smooth map $F\colon X \to Y$ is a \emph{vector bundle morphism} (or \emph{vector bundle homomorphism}) if it is a bundle morphism and, for each $x\in X$, the map $F_x = \restr{F}{\pi_X^{-1}(x)} \colon \pi_X^{-1}(x) \to \pi_Y^{-1}(f(x))$ is linear, where $f\colon M \to N$ is a map making the diagram~\eqref{eq:bundle_homomorphism} commutative. A \emph{vector bundle isomorphism} is a vector bundle homomorphism whose inverse is also a vector bundle homomorphism.

Let $\pi_1\colon E^1 \to M$ and $\pi_2\colon E^2 \to M$ be vector bundles over the same base space $M$. The \emph{Whitney sum of $E^1$ and $E^2$} is the vector bundle $\pi\colon E^1 \oplus E^2\to M$ whose fiber over each $x\in M$ is the direct sum $E^1_x \oplus E^2_x$, where $E^1_x=\pi_1^{-1}(x)$ and $E^2_x = \pi_2^{-1}(x)$.
 The total space is defined as the disjoint union
\begin{equation}
    E^1 \oplus E^2 = \bigsqcup_{x\in M} E^1_x \oplus E^2_x \, ,
\end{equation}
and the projection is given by
\begin{equation}
    \pi(v_1, v_2) = \pi_1(v_1) = \pi_2(v_2)\, .
\end{equation}

The two main vector bundles that will be subsequently employed are the tangent and the cotangent bundles.

Let $M$ be a manifold and $x\in M$ a point. A \emph{tangent vector at $x$} is a linear map $v\colon \Cinfty(M)\to \RR$ such that 
\begin{equation}
    v(fg) = f(x) vg + g(x) vf\, ,
\end{equation}
for every pair of functions $f, g\in \Cinfty(M)$. Equivalently, a tangent vector can be defined as the equivalence class of curves on $M$ which are tangent at $x$. The set of all tangent vectors at $x$ is called the \emph{tangent space to $M$ at $x$} and denoted by $\T_x M$.

The \emph{tangent bundle of $M$} is the vector bundle $\tau_M \colon \T M\to M$ whose fiber over a point $x\in M$ is the tangent space at that point, that is, $\tau_M^{-1}(x)=\T_x M$. The total space is given by the disjoint union of the tangent spaces at all points of $M$, namely,
\begin{equation}
    \T M = \bigsqcup_{x\in M} \T_x M\, . 
\end{equation}
A section of $\T M$ is called a \emph{vector field on $M$}. The set of all vector fields on $M$ is denoted by $\X(M)$.

Similarly, the \emph{cotangent bundle} of $M$ is the vector bundle $\pi_M \colon \cT M\to M$ whose fiber over a point $x\in M$ is the cotangent space at that point (that is, the dual vector space of $\T_x M$), namely, $\pi_M^{-1}(x)=\cT_x M = (\T_x M)^\ast$. A section of $\cT M$ is called a \emph{one-form} on $M$. The set of all one-forms on $M$ is denoted by $\Omega^{1}(M)$.

If $(U; x^1, \ldots, x^n)$ is a chart on $M$ then, for each $p\in U$, $\{\restr{\tparder{}{x^i}}{p}\}_{1\leq i\leq n}$ and $\{\restr{\dd x^i}{p}\}_{1\leq i\leq n}$ are bases of $\T_p M$ and $\cT_p M$, respectively. The induced bundle coordinates on $\T M$ and $\cT M$ are denoted by $(x^1, \ldots, x^n,$ $v^1, \ldots, v^n)$ and $(x^1, \ldots, x^n, p_1, \ldots, p_n)$, respectively.

If $X\in\X(M)$ is a vector field on $M$, an \emph{integral curve of $X$} is a differentiable curve $c\colon I\subseteq \RR\to M$ whose velocity at each point is equal to the value of $X$ at that point, namely,
\begin{equation}
    \dot{c}(t) = X_{c(t)}\, ,
\end{equation}
for all $t\in I$. A \emph{maximal integral curve} is one that cannot be extended to an integral curve on any larger open interval. Locally, the integral curves $c(t)=(x^i(t))$ of a vector field 
\begin{equation}
    X = X^i \frac{\partial}{\partial x^i}
\end{equation}
are given by
\begin{equation}
    \frac{\dd x^i}{\dd t} = X^i\big(x^1(t), \ldots, x^n(t)\big)\, .
\end{equation}

A \emph{flow domain} is an open subset $\mathcal{D} \subseteq \RR\times M$ such that, for all $x\in M$, the set $\mathcal{D}^{(x)} = \{t\in \RR\mid (t, x) \in U\}$ is an open interval containing $0$. Given a flow domain $\mathcal{D}$, a \emph{flow} is a map $\phi\colon \mathcal{D} \to M$ that satisfies the following group laws:
\begin{enumerate}
    \item $\phi(0, x) = x$ for all $x\in M$,
    \item $\phi\big(t, \phi(s, x)\big) = \phi(t+s, x)$ for all $s\in \mathcal{D}^{(x)}$ and $t\in \mathcal{D}^{(\phi(s,x))}$ such that $s+t\in \mathcal{D}^{(x)}$.
\end{enumerate}
The notation $\phi_t = \phi(t, \cdot)$ will often be employed for flows. A \emph{global flow} (also called a \emph{one-parameter group action}) is a flow whose flow domain is $\mathcal{D} = \RR \times M$.

Given a vector field $X\in \X(M)$, its flow is the map $\phi^X\colon \mathcal{D} \to M$ such that, for each $x\in M$, $c = \phi^X(\cdot, x)\colon \mathcal{D}^{(x)} \to M$ is the unique maximal integral curve with initial point $c(0)=x$. A vector field is called \emph{complete} if it generates a global flow or, equivalently, if each of its maximal integral curves is defined for all $t\in \RR$.


Let $V$ and $W$ be real vector spaces. The \emph{tensor product of $V$ and $W$} is a vector space $V\otimes W$, together with a bilinear map 
\begin{equation}
\begin{aligned}
    & \varphi \colon V\times W \to V\otimes W \\
    & \varphi(v, w)= v\otimes w
\end{aligned}
\end{equation}
such that, for any vector space $Z$ and any bilinear map $A\colon V\times W \to Z$, there exists a unique bilinear map $\bar A\colon V\otimes W \to Z$ which makes the following diagram commutative:
\begin{equation}
\begin{tikzcd}
    V\times W \arrow[r, "\varphi"] \arrow[rd, "A"'] & V\otimes W \arrow[d, "\bar A", dashed] \\
                                                & Z                                       
\end{tikzcd}
\end{equation}
If $V^\ast$ and $W^\ast$ denote the dual vector spaces of $V$ and $W$, then
\begin{equation}
    \alpha \otimes \beta\, (v\otimes w) = \alpha(v)\, \beta(w)\, ,
\end{equation}
for any $v\in V, w\in W, \alpha\in V^\ast$ and $\beta \in W^\ast$.

Given a non-negative integer $k$, a \emph{contravariant $k$-tensor on $V$} (respectively, a \emph{covariant $k$-tensor on $V$}) is an element of the $k$-fold tensor product $V\otimes \cdots\otimes V$ (respectively, $V^\ast \otimes \cdots\otimes V^\ast$). By convention, a $0$-tensor is a real number. 
A \emph{tensor of type $(k, l)$ on $V$} is an element of the tensor product
\begin{equation}
    \tensors^k_l(V)= \underbrace{V\otimes \cdots \otimes V}_{k \text{ copies}} \otimes \underbrace{V^\ast\otimes \cdots \otimes V^\ast}_{l \text{ copies}}  \, .
\end{equation}
Tensors of type $(k,l)$ will often be identified with multilinear forms on 
\begin{equation}
    \underbrace{V^\ast \times \cdots \times V^\ast}_{k \text{ copies}} \otimes \underbrace{V \otimes \cdots \otimes V}_{l \text{ copies}} \, .
\end{equation}
By setting $\tensors^k_0 (V) = \{0\}$ for $k<0$, it follows that 
\begin{equation}
    \tensors^\bullet V 
    = \bigoplus_{k\in \ZZ} \tensors^k_0 (V)
\end{equation}
is a $\ZZ$-graded algebra, the so-called \emph{tensor algebra of $V$}

The \emph{exterior algebra} (also called the \emph{alternating algebra} or the \emph{Grassmann algebra}) \emph{on $V$} is the associative algebra $\bigwedge^\bullet V$ whose product is the \emph{exterior product} (or \emph{wedge product}) $\wedge$, an associative binary operation defined by the identity $v\wedge w = -w\wedge v$ for all $v, w\in V$. By setting $\bigwedge^0 V = \RR$ and $\bigwedge^k V = \{0\}$ for $k<0$, it follows that $\bigwedge^\bullet V$ is a $\ZZ$-graded algebra, namely,
\begin{equation}
    \bigwedge\nolimits ^\bullet V = \bigoplus_{k\in \ZZ}  \bigwedge \nolimits ^k V\, ,
\end{equation}
where $\bigwedge^k V$ for $k>0$ is spanned by elements of the form $v_1 \wedge \cdots \wedge v_k$, with $v_1, \ldots, v_k\in V$.   
Moreover, if $\alpha \in \bigwedge^p V$ and $\beta \in \bigwedge^q V$, then
\begin{equation}
    \alpha\wedge \beta = (-1)^{pq} \beta \wedge \alpha\, .
\end{equation}
The exterior algebra $\bigwedge^\bullet V^\ast$ on $V^\ast$ is defined analogously.
Alternatively, the exterior algebra on $V$ can be formally defined as the quotient of the tensor algebra by the ideal generated by $\{v\otimes v\}_{v\in V}$, namely,
\begin{equation}
    \bigwedge\nolimits ^\bullet\, V = \frac{\tensors^\bullet\, V}{\mathrm{gen}\left(\{v\otimes v\}_{v\in V}\right)}\, .
\end{equation}

Let $M$ be a manifold. The \emph{bundle of tensors of type $(k, l)$ on $M$} is the vector bundle whose total space is the disjoint union of the spaces of tensors of type $(k, l)$ on $\T_x M$, namely,
\begin{equation}
    \tensors^k_l (\T M) = \bigsqcup_{x\in M}  \tensors^k_l (\T_x M)\, ,
\end{equation}
and whose projection is the canonical projection on $M$. In particular, one can identify $\tensors^1_0 (\T M)$ and $\tensors^0_1 (\T M)$ with $\T M$ and $\cT M$, respectively. A section of $\tensors^k_l (\T M) $ is called a \emph{tensor field of type $(k, l)$}. A section of $\tensors^k_0 (\T M) $ (respectively, $\tensors^0_l (\T M) $) is called a \emph{contravariant $k$-tensor field} (respectively, a \emph{covariant $l$-tensor field}).

A \emph{differential form} on $M$ is a section of the exterior algebra $\bigwedge^\bullet \cT M$ of the cotangent bundle. A \emph{$p$-form} on $M$ is a section of $\bigwedge^p \cT M$. The set of all $p$-forms on $M$ is denoted by $\Omega^p(M)$. In particular, smooth functions may be identified as $0$-forms, namely, $\Omega^0(M)=\Cinfty(M)$. 
The set of all differential forms on $M$ is denoted by $\Omega^\bullet(M)$. If $\dim M = n$, a nowhere-vanishing $n$-form on $M$ is called a \emph{volume form on $M$}. Given a $p$-form $\alpha\in \Omega^p (M)$, the $k$-th power with respect to the wedge product will be denoted by $\alpha^k$, that is,
\begin{equation}
    \alpha^k = \underbrace{\alpha\wedge \cdots \wedge \alpha}_{k\text{ times}}\, .
\end{equation}

There exists a unique family of maps $\dd\colon \Omega^p(M) \to \Omega^{p+1} (M)$ for all $p\in \NN$, called the \emph{exterior derivative}, satisfying the following properties:
\begin{enumerate}
    \item $\dd$ is $\RR$-linear,
    \item $\dd$ is an antiderivation with respect to the exterior product, that is,
    \begin{equation}
        \dd (\alpha \wedge \beta) = \dd \alpha \wedge \beta + (-1)^p \alpha \wedge \dd \beta\, ,
    \end{equation} 
    for any $\alpha \in \Omega^p(M)$ and $\beta\in \Omega^q(M)$;
    \item $\dd \circ \dd \equiv 0$,
    \item For any $f\in \Cinfty(M)$, $\dd f$ is the differential of $f$, namely, $\dd f(X) = Xf$ for any $X\in \X(M)$. 
\end{enumerate} 

A $p$-form $\alpha\in \Omega^{p}(M)$ is called \emph{closed} if $\dd \alpha = 0$, and \emph{exact} if there exists a $(p-1)$-form $\beta\in \Omega^{p-1}(M)$ such that $\alpha = \dd \beta$. Obviously, every exact form is closed. Moreover, if $\alpha$ is closed, around each point there exists a neighbourhood $U\subseteq M$ such that the restriction of $\alpha$ to $U$ is exact. The latter result is called the \emph{Poincaré lemma}.

Let $X\in\X(M)$ be a vector field and $\alpha\in \Omega^p(M)$ a $p$-form on $M$.
The \emph{inner product of $X$ and $\alpha$} is the $(p-1)$-form $\contr{X} \alpha \in \Omega^{p-1}(M)$ defined by 
\begin{equation}
    \contr{X} \alpha (X_1, \ldots, X_{p-1}) = \alpha(X, X_1, \ldots, X_{p-1})\, ,
\end{equation}
for any vector fields $X_1, \ldots, X_{p-1}\in \X(M)$. For $p=0$ (that is, $\alpha\in \Cinfty(M)$) the inner product is $\contr{X}\alpha=0$.

Let $M$ and $N$ be manifolds, let $x\in M$ be a point, and let $F\colon M\to N$ be a map. The \emph{tangent map of $F$ at $x$} is the map $\T F_x \colon \T_x M \to \T_{F(x)} N$ such that 
\begin{equation}
    \T F_x \left(\dot{c}(0)\right) = \restr{\frac{\dd}{\dd t}}{t=0}\left(F \circ c\right)(t)\, ,
\end{equation}
for any curve $c \colon I \subseteq \RR \to M$ such that $\dot{c}(0)\in \T_x M$. 
The \emph{(global) tangent map of $F$} is the map $\T F\colon \T M \to \T N$ whose restriction to $\T_x M \subset \T M$ is $\T F_x$. The \emph{cotangent map of $F$ at $x$} is the dual $\cT F_x\colon \cT_{F(x)} N \to \cT_x M$ of the tangent map of $F$ at $x$, namely,
\begin{equation}
    \left\langle \cT F_x \alpha_{F(X)}, v_x\right\rangle = \left\langle  \alpha_{F(X)}, \T F_x v_x\right\rangle\, ,
\end{equation}
for $v_x\in \T_x M$ and $\alpha_{F(x)}\in \cT_{F(x)} N$, where $\langle \cdot, \cdot \rangle$ denotes the natural pairing of vectors and covectors. If $F\colon M\to N$ and $G\colon N\to P$ are smooth maps between manifolds, then
\begin{equation}
    \T (F\circ G) = \T F \circ \T G\, , \quad \cT (F\circ G) = \cT G \circ \cT F\, .
\end{equation}

Given
a covariant $k$-tensor field $A\in \tensors^0_l (\T N) $ and a map $F\colon M \to N$,  the \emph{pullback of $A$ by $F$} is the covariant $k$-tensor field $A\in \tensors^0_l (\T M) $
such that
\begin{equation}
    \left(F^\ast A\right)_x (v_1, \ldots, v_p) = A\left(\T F_x (v_1), \ldots, \T F_x (v_p) \right) \, ,  
\end{equation}
for any $x\in M$ and any $v_1,\ldots, v_p \in \T_x M$.

Given a vector field $X\in \X(M)$ and a diffeomorphism $F\colon M \to N$, the \emph{pushforward of $X$ by $F$} is the vector field $F_\ast X\in \X(N)$ given by
\begin{equation}
    \left(F_\ast X\right)_x = \T F_{F^{-1}(x)} \left(X_{F^{-1}(x)}\right)\, ,
\end{equation}
for every $x\in M$.

Let $X\in \X(M)$ be a vector field on $M$ with flow $\phi_t$. Consider the map $\Phi_t\colon \tensors^k_l (\T_{\phi_t(x)} M) \to \tensors^k_l( \T_x M)$ given by
\begin{equation}
\begin{aligned}
    &\Phi_t (v_1\otimes \cdots \otimes v_k \otimes \alpha^1\otimes \cdots \otimes \alpha^l) 
    \\ \quad & = (\phi_t^{-1})_\ast v_1 \otimes \cdots \otimes (\phi_t^{-1})_\ast v_k 
    \otimes \phi_t^\ast \alpha^1 \otimes \cdots \otimes \phi_t^\ast \alpha^l\, , 
\end{aligned}
\end{equation}
for $v_1, \ldots, v_k \in \T_x M$ and $\alpha^1, \ldots, \alpha^l\in \cT_x M$.
Let $A\in \tensors^k_l (\T M) $ be a tensor field of type $(k, l)$. The \emph{Lie derivative of $A$ with respect to $X$} is the covariant $k$-tensor $\liedv{X} A \in \tensors^0_k (\T M)$ given by
\begin{equation}
    \left(\liedv{X} A\right)_x = \lim_{t\to 0} \frac{\Phi_t (A_x) - A_x}{t}\, .
\end{equation}  
In particular, if $A\in \tensors^0_l (\T M) $ is a covariant $l$-tensor field, then
\begin{equation}
    \liedv{X} A = \lim_{t\to 0} \frac{\phi_t^\ast A - A}{t}\, .
\end{equation}  
For a function $f\in \Cinfty(M)$, 
\begin{equation}
    \liedv{X} f = X(f)\, ,
\end{equation}
and for a vector field $Y\in \X(M)$,
\begin{equation}
    \liedv{X} Y = [X, Y]\, .
\end{equation}

Several useful relations for the so-called Cartan calculus on differential forms can be found in \Cref{table:Cartan_calculus}.

\begin{table}[t]
\begin{align*}
    &\liedv{X} \alpha = \contr{X} (\dd \alpha)+ \dd (\contr{X}\alpha)\, ,\\
	&X(f)=\contr{X} \dd f\, , \\
    &\contr{[X,Y]}\alpha =\liedv{X}\contr{Y}\alpha -\contr{Y}\liedv{X}\alpha\, , \\
    &\liedv{[X, Y]} \alpha = \liedv{X} \liedv{Y} \alpha - \liedv{Y} \liedv{X} \alpha\, , \\
    &\contr{X}(\alpha \wedge \beta)=\contr{X} \alpha \wedge \beta+(-1)^{p} \alpha \wedge \contr{X} \beta\, , \\
	& \dd \alpha(X_0,\ldots, X_p) 
    = \sum_{i=0}^p (-1)^i X_i \left( \alpha(X_0,\ldots, \widehat{X}_{i},\ldots, X_p)  \right)\\
	&\quad+\sum_{i<j} (-1)^{i+j} \alpha \left([X_i,X_j],X_0,\ldots, \widehat{X}_{i},\ldots, 
    \widehat{X}_{j},\ldots,X_p  \right)\, ,\\
	&\varphi^\ast (\contr{X}\alpha)=\contr{(\varphi^{-1})_\ast X} (\varphi^\ast \alpha)\, , \\
	&F^\ast (\dd \alpha)=\dd (F^\ast \alpha)\, , \\
	&\liedv{fX}\alpha =f\liedv{X}\alpha +\dd f \wedge \contr{X}\alpha\, , \\
	&\liedv{X} \dd \alpha =\dd \liedv{X}\alpha\, , \\
    &\liedv{X} (\alpha\wedge \beta) = (\liedv{X} \alpha) \wedge \beta + \alpha \wedge (\liedv{X} \beta)\, .
\end{align*}
\caption[Some relevant identities of Cartan calculus]{Some relevant identities of the exterior derivative, the inner product, the pullback, the pushforward and the Lie derivative of vector fields and differential forms. The first of these properties is called the \emph{Cartan (magic) formula}. Here $f\in \Cinfty(M)$ is a function, $\alpha\in \Omega^p(M)$ and $\beta\in \Omega^q(M)$ are differential forms, $X, X_0,\ldots, X_p\in \X(M)$ are vector fields, $F\colon N \to M$ is a map and $\varphi\colon N\to M$ is a diffeomorphism.
The omission of $X_k$ is denoted by~$\widehat{X}_k$.}
\label{table:Cartan_calculus}
\end{table}

\section{Lie groups and Lie algebras}

This section exposes some concepts about Lie groups, Lie algebras and their actions on manifolds. Additional details may be found in \cite{Lee2012, Warner1983, O.R2004}.

A \emph{Lie group} is a smooth manifold $G$ whose underlying set of elements is equipped with the structure of a group such that the group multiplication and inverse-assigning maps are smooth. If $G$ and $H$ are Lie groups, $H$ is a \emph{Lie subgroup of $G$} if it is a subgroup and an immersed submanifold of $G$. In particular, for every Lie group, its connected component containing the identity is a Lie subgroup.  

Any element $g\in G$ of a Lie group $G$ defines a map $L_g\colon G \to G$, called \emph{left translation}, given by $L_g h = g\cdot h$, where $\cdot$ denotes the group multiplication. This map is a diffeomorphism on $G$, and $L_{g^{-1}}$ is its inverse.

Let $M$ be a manifold and $G$ a Lie group. A \emph{(left) Lie group action of $G$ on $M$} is a smooth map $\Phi\colon G\times M \to M$ such that, for every point $x\in M$,
\begin{enumerate}
    \item $\Phi\big(\Phi(x, g_2), g_1\big) = \Phi(x, g_1 \cdot g_2)$, for all elements $g_1, g_2\in G$,
    \item $\Phi(e, x) = x$,
\end{enumerate}
where $e$ denotes the identity in $G$. Similarly, a \emph{right Lie group action of $G$ on $M$} is a smooth map $\Phi\colon M \times G \to M$ such that, for every point $x\in M$,
\begin{enumerate}
    \item $\Phi\big(g_2, \Phi(g_1, x)\big) = \Phi(g_1 \cdot g_2, x)$, for all elements $g_1, g_2\in G$,
    \item $\Phi(x, e) = x$,
\end{enumerate}
For instance, a global flow is an Abelian Lie group action of $\RR$ on $M$. 
Given a point $x\in M$, the \emph{orbit of $x$} is the set of all images of $x$ under the action by elements of $G$, namely,
\begin{equation}
   \mathcal{O}(x) = G\cdot x = \{y\in M\mid y = \Phi(g, x),\ g\in G\}\, , 
\end{equation}
and analogously for a right action.
The \emph{isotropy group of $x$} (also called the \emph{stabilizer} of $x$) is the subgroup of $G$ whose elements fix $x$, that is,
\begin{equation}
    G_x = \{g\in G \mid \Phi(g, x) = x\}\, .
\end{equation}
A Lie group action is called 
\begin{enumerate}
    \item \emph{transitive} if the orbit of any point is all of $M$,
    \item \emph{free} if every isotropy group is trivial,
    \item \emph{proper} if the map $G\times M \ni (g, x) \mapsto \big(\Phi(g,x), x \big) \in M\times M$ is proper. 
\end{enumerate}

Orbits are equivalence classes in $M$, where the equivalence relation is $x\sim y$ if $y=\Phi(g,x)$ for some $g\in G$. The set of orbits with the quotient topology is called the \emph{orbit space} and denoted by $M/G$. If certain assumptions are made on the action, the orbit space is guaranteed to be a smooth manifold. More specifically, if the Lie group action of $G$ on $M$ is smooth, free and proper, then the orbit space $M/G$ is a topological manifold of dimension equal to $\dim M - \dim G$, and has a unique smooth structure making the quotient map $\pi \colon M \to M/G$ a smooth submersion.

Let $\pi\colon E \to M$ be a fiber bundle and $G$ a Lie group. Let $\Phi\colon E \times G \to E$ be a right Lie group action of $G$ on $E$. If the action $\Phi$ is free and its orbits are exactly the fibers 
(that is, $\pi^{-1}(x)=\mathcal{O}(y)$, for every $x\in M$ and every $y\in \pi^{-1}(x)$),
then $\pi\colon E \to M$ is called a \emph{principal bundle with group $G$}.

A \emph{Lie algebra} (over $\RR$) is a real vector space $\mathfrak{g}$ endowed with a bilinear skew-symmetric map $\mathfrak{g}\times \mathfrak{g}\to \mathfrak{g}$, denoted by $(X,Y) \mapsto [X, Y]$ and called the \emph{Lie bracket of $\mathfrak{g}$}, which satisfies the \emph{Jacobi identity}, that is, 
\begin{equation}
    [X, [Y, Z]] + [Y, [Z, X]] + [Z, [X,Y]] = 0\, ,
\end{equation}
for all $X, Y \in \mathfrak{g}$. If the Lie bracket $[\cdot, \cdot]$ wants to be emphasized, the Lie algebra $\mathfrak{g}$ will be denoted by the pair $(\mathfrak{g}, [\cdot, \cdot])$.

The space of all smooth vector fields $\X(M)$ on a manifold $M$ is a Lie algebra. Given a pair of vector fields $X, Y\in\X(M)$, the Lie bracket of $X$ and $Y$ is defined by
\begin{equation}
    [X, Y] f = XY f - YX f\, ,
\end{equation}
for every function $f\in \Cinfty(M)$. If $F\colon M \to N$ is a diffeomorphism between two manifolds $M$ and $N$, 
\begin{equation} \label{eq:pushforward_Lie_bracket}
    F_\ast [X, Y] = [F_\ast X, F_\ast Y]\, ,
\end{equation}
for every pair of vector fields $X, Y\in \X(M)$.


Another important example of Lie algebra that will be considered along this dissertation is the space of smooth functions $\Cinfty(M)$ on a manifold $M$ endowed with a Jacobi bracket (see \Cref{sec:Jacobi_manifolds}). In particular, Poisson, symplectic, cosymplectic, contact and cocontact manifolds are endowed with Jacobi brackets. Furthermore, any Lie group has an associated Lie algebra.

A \emph{Lie subalgebra} $\mathfrak{h}\subseteq\mathfrak{g}$ of a Lie algebra $\mathfrak{g}$ is a vector subspace of $\mathfrak{g}$ which is closed under the Lie bracket. Therefore, $\mathfrak{h}$ is also a Lie algebra with the restriction of the Lie bracket of $\mathfrak{g}$.

If $\mathfrak{g}$ and $\mathfrak{h}$ are Lie algebras, a linear map $A\colon \mathfrak{g} \to \mathfrak{h}$ is called a \emph{Lie algebra homomorphism} if it preserves the Lie brackets, that is,
\begin{equation}
    A[X, Y]_{\mathfrak{g}} = [AX, AY]_{\mathfrak{h}}\, ,
\end{equation} 
for all $X, Y \in \mathfrak{g}$, where $[\cdot, \cdot]_{\mathfrak{g}}$ and $[\cdot, \cdot]_{\mathfrak{h}}$ denote the Lie brackets on $\mathfrak{g}$ and $\mathfrak{h}$, respectively. Similarly, a \emph{Lie algebra anti-homomorphism} is a linear map  $A\colon \mathfrak{g} \to \mathfrak{h}$ such that
\begin{equation}
    A[X, Y]_{\mathfrak{g}} = - [AX, AY]_{\mathfrak{h}}\, ,
\end{equation} 
for all $X, Y \in \mathfrak{g}$. A \emph{Lie algebra isomorphism} (respectively, \emph{anti-isomorphism}) is an invertible homomorphism (respectively, anti-homomorphism). 

If $G$ is a Lie group, a vector field $X\in \X(G)$ is called \emph{left-invariant} if $(L_g)_\ast X = X$ for all $g\in G$. 
If $X, Y\in \X(G)$ are left-invariant vector fields, equation~\eqref{eq:pushforward_Lie_bracket} implies that $[X, Y]$ is also left-invariant. Hence, left-invariant vector fields close a Lie subalgebra of $\X(G)$, which is called the \emph{Lie algebra of $G$} and denoted by $\mathrm{Lie}(G)$.

In what follows, let $\mathfrak{g}=\mathrm{Lie}(G)$ denote the Lie algebra of $G$. One can identify $\mathfrak{g}$ with the tangent space to $G$ at the identity. Indeed, for any $X\in \mathfrak{g}$, the evaluation $X_e\in \T_e G$ is a tangent vector at the identity. Conversely, given a tangent vector at the identity $v\in \T_e G$, one can extend it to a left-invariant vector field $v^L\in \mathfrak{g}$ by left translations, namely,
\begin{equation}
    \restr{v^L}{g} = \big(\T L_g\big)_e v\, .
\end{equation}
This correspondence is one-to-one in both directions, and thus bijective. The Lie bracket of $v$ and $w$ in $\T_e G$ can be computed by extending them to left-invariant vector fields, taking the bracket of the vector fields, and then evaluating the result at the identity, that is,
\begin{equation}
    [v, w] = \restr{\left[v^L, w^L\right]}{e} \, .
\end{equation} 
Hence, the map $X\mapsto \restr{X}{e}$ for $X\in \mathfrak{g}$ is an isomorphism of Lie algebras. This isomorphism implies that $\mathrm{Lie}(G)$ has the same dimension as $G$.

The \emph{exponential map} $\exp \colon \mathfrak{g} \to G$ is given by
\begin{equation}
    \exp X = \gamma(1)\, ,
\end{equation}
where $\gamma$ is the integral curve of $X$ starting at the identity.

Given a Lie group action $\Phi\colon G\times M \to M$, the \emph{infinitesimal generator associated to $\xi\in\mathfrak{g}$} is the vector field $\xi_M\in \X(M)$ defined by
\begin{equation}\label{eq:infinitesimal_generator}
    \xi_M(x) = \restr{\frac{\dd}{\dd t}}{t=0} \Phi_{\exp t\xi(x)} \, .
\end{equation} 
Every infinitesimal generator is a complete vector field. Moreover, the map $\mathfrak{g}\ni \xi \mapsto \xi_M\in \X(M)$ is a Lie algebra anti-homomorphism.

The \emph{adjoint action} $\Ad\colon G\times \mathfrak{g}\to \mathfrak{g}$ of a Lie group $G$ on its Lie algebra $\mathrm{g}$ is given by 
\begin{equation}
    \Ad_g \xi = \restr{\frac{\dd }{\dd t}}{t=0} g \exp (t \xi) g^{-1}\, ,
\end{equation}
for each $g\in G$ and $\xi \in \mathfrak{g}$. Let $\mathfrak{g}^\ast$ denote the dual of $\mathfrak{g}$. For each $g\in G$, let $\Ad_g^\ast \colon \mathfrak{g}^\ast\to \mathfrak{g}^\ast$ be the dual of the linear map $\Ad_g\colon \mathfrak{g}\to \mathfrak{g}$, namely,
\begin{equation}
    \big\langle \Ad_g^\ast \alpha, \xi \big\rangle =  \big\langle \alpha, \Ad_g \xi \big\rangle \, ,
\end{equation} 
for each $\xi \in \mathfrak{g}$ and $\alpha\in \mathfrak{g}^\ast$, where $\langle \cdot, \cdot \rangle\colon \mathfrak{g}^\ast \times \mathfrak{g}\to \RR$ denotes the natural pairing. The \emph{coadjoint action} $\Phi\colon G \times \mathfrak{g}^\ast$of $G$ on $\mathfrak{g}^\ast$ is given by 
\begin{equation}
    \Phi(g, \alpha) = \Ad^\ast_{g^{-1}} \alpha\, .
\end{equation}

A Lie group action $\Phi\colon G\times M\to M$ of $G$ on $M$ induces a Lie group action $\Phi^{\T M}\colon G \times \T M \to \T M$ of $G$ on the tangent bundle of $M$, called the \emph{tangent lift of the action $\Phi$}, given by
\begin{equation}
    \Phi^{\T}_g (v_x) = \big(\T \Phi_g \big)_{x}\,  v_x\, , 
\end{equation}
for $g\in G$ and $v_x\in \T_x M$. Similarly, it induces a Lie group action $\Phi^{\cT M}\colon G \times \cT M \to \cT M$ of $G$ on the cotangent bundle of $M$, called the \emph{cotangent lift of the action $\Phi$}, given by
\begin{equation}
    \Phi^{\cT}_g (\alpha_x) = \big(\cT \Phi_{g^{-1}} \big)_{x}\,  \alpha_x\, , 
\end{equation}
for $g\in G$ and $\alpha_x\in \cT_x M$.

\section{Distributions}

In this section, the notions of distribution and codistribution are briefly reviewed. To explore this topic further, the reader may consult \cite{Lee2012}.

Given a vector bundle $\pi\colon E\to M$, a \emph{subbundle of $E$} is a vector bundle $\pi_D\colon D \to M$ such that
\begin{enumerate}
    \item $D$ is an embedded submanifold in $E$,
    \item $\pi_D$ is the restriction of $\pi$ to $D$,
    \item for each $x\in M$, the subset $D_x= E_x\cap D$ is a vector subspace of $E_x$, and the vector space structure on $D_x$ is the one inherited from $E_x$.
\end{enumerate}
If the vector bundle $\pi\colon E\to M$ has rank $r$, and the subbundle $\pi_D\colon D \to M$ has rank $s$, then $\pi_D\colon D \to M$ is said to have \emph{corank} $r-s$.

A \emph{distribution of (co)rank $k$ on $M$} is a (co)rank-$k$ subbundle of $\T M$. Analogously, a \emph{codistribution of (co)rank $k$ on $M$} is a (co)rank-$k$ subbundle of $\cT M$.


A distribution $D$ of rank $k$ on an $n$-dimensional manifold $M$ is called:
\begin{enumerate}
    \item \emph{Integrable} if, for every $x\in M$, $D_x=\T_x N$ for some nonempty immersed submanifold $N\subseteq M$. These submanifolds are called \emph{integral manifolds of $D$}.
    \item \emph{Involutive} if, for every pair of local sections $X$ and $Y$ of $D$, their Lie bracket $[X, Y]$ is also a local section of $D$.
    \item \emph{Completely integrable} if, for each point $x\in M$, there exists a chart $(U, \varphi)$ centered at $x$ such that $\varphi(U)$ is a cube in $\RR^n$ and $D$ is spanned by $\tparder{}{x^1}, \ldots, \tparder{}{x^k}$, where $\varphi=(x^1, \ldots, x^n)$. In such chart, each slice of the form $x^{k+1}=c^{k+1}, \ldots, x^n = c^n$ for some constants $c^{k+1}, \ldots, c^n$ is an integral manifold of $D$.
\end{enumerate}
Obviously, every completely integrable distribution is also integrable. A remarkable result, known as \emph{Frobenius theorem}, is that a distribution is completely integrable if and only if it is involutive.

A distribution $D$ on $M$ is called \emph{bracket-generating} (or \emph{completely nonholonomic}) if $D^r = \T M$ for some $r\geq 2$, where $D^1 = D$ and
\begin{equation}
    D^{k+1}= D^k + [D, D^k]\, ,
\end{equation}
for $k\geq 1$.

Given a fiber bundle $\pi\colon E\to M$, the kernel $\V E=\ker \T \pi$ of the tangent map $\T \pi\colon \T E \to \T M$ is called the \emph{vertical bundle}. It is a distribution over $E$. As a matter of fact, it is an integrable distribution. A vector field $X\in \X(E)$ is called \emph{vertical} if it is a section of $\V E$. 
If $(x^i, y^a)$ are bundle coordinates, then a vertical vector field $Z\in \Gamma(\V E)$ is of the form
\begin{equation}
    Z = Z^a(x, y) \frac{\partial}{\partial y^a}\, .
\end{equation}

\subsection{Generalized distributions} \label{subsec:generalized_distributions}

The concept of distribution can be generalized by not requiring its rank to be constant (see references~\cite{C.d.M+2001,C.d.M+2001a,Vaisman1994}). This will be useful to characterize geometrically mechanical systems subject to constraints which are ``degenerate'' at certain points (see \Cref{ch:contact_impulsive}).

A \emph{generalized distribution} on $M$ is a family of vector subspaces $D =\left\{D _{x}\right\}$ of the tangent spaces $\T_{x} M$. Such a distribution is called \emph{differentiable} if for every $x \in \mathrm{dom}D $, there is a finite number of differentiable vector fields $X_1,\ldots, X_s\in D $ such that $D _x$ is spanned by $\left\{\restr{X_1}{x},\ldots, \restr{X_s}{x}  \right\}$.

The \emph{rank} of $D $ at $x\in M$ is the function $\rho(x)=\dim D _x$. Observe that, if $D$ is differentiable, then $\rho(x)$ cannot decrease in a neighbourhood of $x$, and hence $\rho$ is a lower semi-continuous function. Clearly, $D $ is a distribution in the usual sense if and only if $\rho(x)$ is a constant function. In the general case, $x\in M$ will be called a \emph{regular point} if $\rho(x)$ takes a constant value on an open neighbourhood of $x$ (in other words, $x$ is a local maximum of $\rho(x)$), and a \emph{singular point} otherwise. Obviously, the set $\mathscr{R}$ of the regular points of $D $ is open. Moreover, it is dense.
Indeed, if $x_0\in M\setminus \mathscr{R}$ and $U$ is a neighbourhood of $x_0$, then $\restr{\rho}{U}$ must have a maximum (since it is integer-valued and bounded), and hence $U$ contains regular points, that is, $x_0\in \overline{\mathscr{R}}$. However, $\mathscr R$ is not connected in general.

Similarly, a \emph{generalized codistribution} on $M$ is a family of vector subspaces $S =\left\{S _{x}\right\}$ of the cotangent spaces $\cT_{x} M$. Such a codistribution is called \emph{differentiable} if for every $x \in \mathrm{dom}S $, there is a finite number of differentiable 1-forms $\omega_{1}, \ldots, \omega_{s}\in S$ such that $S_{x}$ is spanned by $\left\{\restr{\omega_{1}}{x}, \ldots, \restr{\omega_{s}}{x}\right\}$. The regular and singular points are defined analogously to the ones in generalized distributions. Similarly, the set of regular points of $S$ is open, dense and, generally, non-connected.

\begin{example}
Let $M=\RR^2$ and $D _{(x,y)}=\left\langle \partial_x, \varphi(y) \partial_y  \right\rangle$, where $\varphi(y)$ is a $\Cinfty$-function with $\varphi(y)=0$ for $y\leq 0$ and $\varphi(y)>0$ for $y>0$. For instance,
\begin{equation}
  \varphi(y)=
  \left\{
    \begin{array}{ll}
    0, &y\leq 0,\\
     e^{-1/y^2}, & y> 0.
    \end{array}
    \right.
\end{equation}
Then, the singular points are those of the $x$-axis, while the connected components are the half-{plane}s $y<0$ (where $\rho = 1$) and $y>0$ (where $\rho =2$).
\end{example}

Given a generalized distribution $D$, its annihilator is the generalized codistribution given by
\begin{equation}
\begin{aligned}
  D ^{\circ}: \operatorname{dom} D  \subseteq M & \rightarrow \cT M \\
  x & \mapsto D _{x}^{\circ}=\left(D  _{x}\right)^{\circ}.
\end{aligned}
\end{equation}

The notions of \emph{integrable} and \emph{completely integrable generalized distribution} are defined just as in the regular case. Moreover, a differentiable generalized distribution $D$ on $M$ is said to be \emph{invariant} if there is a set $\X_0(M)\subseteq \X(M)$ of vector fields on $M$ such that, for every point $x\in M$:
\begin{enumerate}
    \item $D_x$ is spanned by the values of the vector fields of $\X_0(M)$ at $x$.
    \item For every vector field $X\in \X_0(M)$ with flow $\Phi^X$ and every $t\in \RR$ where $\phi^X_t$ is defined, one has
    \begin{equation}
        (\phi^X_t)_\ast (D_x) = D_{\phi^X_t(x)}\, .
    \end{equation} 
\end{enumerate}  
There are results due to Stefan \cite{Stefan1974}, Sussmann \cite{Sussmann1973} and Vifljancev \cite{Vifljancev1980} that extend Frobenius theorem to generalized distributions. The reader may also refer to \cite{Vaisman1994} for their proofs and additional details.

\begin{theorem}[Sussmann--Stefan--Frobenius]
    A differentiable generalized distribution is completely integrable if and only if it is invariant.
\end{theorem}

\begin{theorem}
    [Vifljancev--Frobenius]
    A differentiable generalized distribution $D$ on $M$ is completely integrable if and only if there exists a Lie subalgebra $\X_0(M)$ of $\X(M)$ such that
    \begin{enumerate}
        \item $D_x$ is spanned by the values of the vector fields of $\X_0(M)$ at $x$.
        \item For every vector field $X\in \X_0(M)$ with flow $\Phi^X$ and every $t\in \RR$ where $\phi^X_t$ is defined, 
        \begin{equation}
            \dim \big((\phi^X_t)_\ast (D_x) \big) = \dim D_x\, .
        \end{equation}
    \end{enumerate}
\end{theorem}

\section{Riemannian geometry}\label{sec:Riemannian}

Although not much of the machinery from Riemannian geometry will be employed in this dissertation, for the sake of completeness, and in order to establish the notation, some basic notions are reviewed in this section. 
Further information may be found in \cite{Lee2012,Lee2018}.

Let $M$ be a manifold. A \emph{Riemannian metric $g$ on $M$} is a smooth symmetric covariant 2-tensor field $g\in\tensors^0_2(\T M)$ that is positive definite at each point. In other words, a Riemannian metric $g$ on $M$ is a collection of (positive) inner products $g_x\colon \T_x M \times \T_x M \to \RR$ on each of the tangent spaces $\T_x M$ such that, if $X$ and $Y$ are (smooth) vector fields on $M$, the function $\langle X,Y\rangle_g \colon M\to \RR$ given by $\langle X , Y \rangle_g (x) = g_x (\restr{X}{x},\restr{Y}{x})$ is smooth. The pair $(M, g)$ is called a \emph{Riemannian manifold}. 

It is worth remarking that, as a consequence of the existence of partitions of unity, every smooth manifold admits a Riemannian metric. However, the choice of such metric is not canonical. 

The concept of Riemannian metric can be generalized by dropping the condition of positive definiteness. A \emph{pseudo-Riemannian} metric $g$ on $M$ is a smooth symmetric covariant 2-tensor field $g\in\tensors^0_2(\T M)$ whose value is nondegenerate at each point, and with the same signature everywhere on $M$. The pair $(M, g)$ is called a \emph{pseudo-Riemannian manifold}. Clearly, every Riemannian metric is also a pseudo-Riemannian metric. 
A pseudo-Riemannian metric with signature $(n-1, 1)$ (or $(1, n-1)$, depending on the sign convention employed) is called a \emph{Lorentzian metric}. This type of metrics play a crucial role in modern physics, where they are used to model gravitation in general theory of relativity. 

A pseudo-Riemannian metric $g$ on $M$ defines an isomorphism of vector bundles $\flat_g\colon \T M \to \cT M$ given by $\flat_g(v_x) = g_x(v_x, \cdot)$ for each $v_x\in\T_x M$; and an isomorphism of $\Cinfty(M)$-modules $\flat_g\colon \X(M) \to \Omega^1(M)$ given by $\flat_g(X)=g(X, \cdot)$ for each $X\in \X(M)$. The inverses of both isomorphism are denoted by $\sharp_g$. In the physics literature, $\flat_g(X)$ and $\sharp_g(\alpha)$ are frequently called ``lowering an index of $X$'' and ``raising an index of $\alpha$'', respectively.
 Moreover, 
the metric $g$ induces an inner product $\langle \langle\cdot , \cdot \rangle \rangle_{g_x}$ on each cotangent space $\cT_x M$ given by
\begin{equation}
   \langle\langle \alpha, \beta \rangle \rangle_{g_x} = \contr{\sharp_{g_x}(\alpha)} \beta\, .
\end{equation}

Let $(M, g)$ be a pseudo-Riemannian metric. Let $X$ and $Y$ be vector fields on $M$.
In local coordinates $(x^i)$ of $M$, if 
\begin{equation}
    g = g_{ij} \dd x^i \otimes \dd x^j\, ,
\end{equation} 
and 
\begin{equation}
    X = X^i \frac{\partial}{\partial x^i}\, , \quad Y = Y^i \frac{\partial}{\partial x^i}\, ,
\end{equation}
then 
\begin{equation}
    g(X, Y) = g_{ij} X^i Y^j\, ,
\end{equation}
and 
\begin{equation}
    \flat_g (X) = g_{ij} X^i\, .
\end{equation}
If $\alpha$ and $\beta$ are two one-forms on $M$, locally given by
\begin{equation}
    \alpha = \alpha_i \dd x^i\, , \quad \beta = \beta_i \dd x^i\, ,
\end{equation}
then 
\begin{equation}
    \sharp_g (\alpha) = g^{ij} \alpha_i\, ,
\end{equation}
and 
\begin{equation}
    \langle \langle \alpha, \beta \rangle \rangle  = \contr{\sharp_g(\alpha)} \beta = g^{ij} \alpha^i \beta^j\, .
 \end{equation} 
where $(g^{ij})$ denotes the inverse matrix of $(g_{ij})$. 

Given a (pseudo-)Riemannian manifold $(M, g)$ and a function $f\in \Cinfty(M)$, the \emph{(pseudo-)Riemannian gradient} is the vector field $\grad f \in \X(M)$ defined by
\begin{equation}
    \grad f = \sharp_g \dd f\, .
\end{equation}
Locally,
\begin{equation}
    \grad f = g^{ij} \parder{f}{q^i}
\end{equation}

\section{Structures of the tangent bundle}\label{sec:structures_TQ}

The tangent bundle of a manifold is endowed with certain geometric structures, which will be essential to develop a geometric formulation of Lagrangian mechanics (see \Cref{sec:Lagrangian_mechanics}). For more details, see \cite{Godbillon1969, Crampin1983, d.R1989,M.R2007,Y.I1967,Y.I1973}

Along this section, let $Q$ denote an $n$-dimensional manifold with local coordinates $(q^i)$. Let $\T Q$ denote its tangent bundle, with canonical coordinates $(q^i, v^i)$ and projection $\tau_Q\colon \T Q \to Q$. A vector field $X\in \X(\T Q)$ will be called \emph{vertical} if it is a section of the vertical bundle $\V \T Q$.

From vector fields on $Q$, one can construct vertical vector fields on $\T Q$. 
Given a pair of tangent vectors $v, w\in \T_q Q$, the \emph{vertical lift of $w$ with respect to $v$} is the tangent vector $w^{\V}_v$ at $t=0$ to the curve $c(t) = v + t w$, that is,
\begin{equation}
    w^{\V}_v = \dot{c}(0) = \restr{\frac{\dd }{\dd t}}{t=0} (v+tw) \in \T_v (\T Q)\, .
\end{equation}
If $X\in \X(Q)$ is a vector field, its \emph{vertical lift} is the vector field $X^{\V}\in \X(\T Q)$ such that its value at a point $v_q\in \T Q$ with $\tau_Q(v_q)=q$ is given by the vertical lift of $X(q)$ with respect to $v_q$, namely, 
\begin{equation}
    X^{\V}(v_q) = X(q)_{v_q}^{\V} = \restr{\frac{\dd }{\dd t}}{t=0} \big(v_q+tX(q)\big)\, .
\end{equation}
In canonical coordinates, if
\begin{equation}\label{eq:vector_field_coords}
    X = X^i \frac{\partial }{\partial q^i}\, , 
\end{equation}
then
\begin{equation}
    X^{\V} = X^i \frac{\partial }{\partial v^i}\, . 
\end{equation}

An \emph{almost tangent structure} on a manifold $M$ is a tensor field $\Sendo\in \Gamma(\tensors^1_1\, \T M)$ such that, for each $x\in M$, the kernel of the endomorphism $\Sendo_x\colon \T_x M \to \T_x M$ coincides with its image.
The \emph{canonical almost tangent structure of the tangent bundle} $\Sendo\in \Gamma(\tensors^1_1\, \T \T Q)$, also known as the \emph{vertical endomorphism}, is given by
\begin{equation}
    \Sendo_{v_q} (X) = \big( (\tau_Q)_\ast X\big)^{\V}\, ,
\end{equation}
for each $v_q\in \T Q$ and each $X\in \T_{v_q} \T Q$.
In canonical coordinates,
\begin{equation}
    \Sendo \left(\frac{\partial}{\partial q^i}\right) = \frac{\partial}{\partial v^i}\, , \quad  \Sendo \left(\frac{\partial}{\partial v^i}\right) = 0\, ,
\end{equation}
that is
\begin{equation}
    \Sendo =  \frac{\partial}{\partial v^i}\otimes \dd q^i \, .
\end{equation}
Observe that $\ker \Sendo = \Ima \Sendo = \V \T Q$.
The \emph{adjoint operator} $\Sendoadj$ of the vertical endomorphism $\Sendo$ is defined by $S^\ast (f) = f$ for $f\in \Cinfty(\T Q)$ and
\begin{equation}
    (\Sendoadj \alpha) (X_1, \ldots, X_p) = \alpha(\Sendo X_1, \ldots, \Sendo X_p)\, ,
\end{equation}
for $\alpha\in \Omega^p(\T Q)$ and $X_1,\ldots, X_p\in \X(\T Q)$. In canonical coordinates,
\begin{equation}
    \Sendoadj (\dd q^i) = 0\, , \quad \Sendoadj (\dd v^i) = \dd q^i\, .
\end{equation}

Consider the action $\Phi\colon \R\times \T Q\to \T Q$ of $\R$ on $\T Q$ by homothety on the fibers, that is, 
\begin{equation}
    \Phi(t, v_q) = e^t v_q\, ,
\end{equation}
for $t\in\RR$ and $v_q\in \T_q Q$. The infinitesimal generator of $\Phi$ is called the \emph{Liouville vector field} and denoted by $\Delta$. As a matter of fact, this construction can be done in any vector bundle. Equivalently, the Liouville vector field at $v_q\in \T_q Q$ is given by the vertical lift of $v_q$ with respect to $v_q$, namely,
\begin{equation}
    \Delta(v_q) = (v_q)^{\V}_{v_q} = \restr{\frac{\dd }{\dd t}}{t=0} (v_q+tv_q)\, .
\end{equation}
In canonical coordinates, 
\begin{equation}
    \Delta = v^i \frac{\partial}{\partial v^i}\, .
\end{equation}

Let $X\in \X(Q)$ be a vector field on $Q$ and $\Phi^X\colon \mathcal{D} \to Q$ its flow. The \emph{complete lift of $X$} is the vector field $X^{\Com}\in \X(\T Q)$ whose flow $\Phi^{X^{\Com}}\colon \T \mathcal{D}\to \T Q$ is the tangent map of the flow of $X$, that is, $\Phi^{X^{\Com}}_t (v_q) = (\T \Phi^X_t)_q\, v_q$ for $t\in \mathcal{D}^{(q)}$ and $v_q\in \T_q Q$.
In canonical coordinates, if $X$ is of the form~\eqref{eq:vector_field_coords}, then
\begin{equation}
    X^{\Com} = X^i \frac{\partial}{\partial q^i} + v^j \frac{\partial X^i}{\partial q^j} \frac{\partial}{\partial v^i}\, . 
\end{equation}

If $X, Y\in \X(Q)$ are vector fields on $Q$, their vertical and complete lifts satisfy the following identities:
\begin{equation}\label{eq:identities_lifts}
    [X^{\V}, Y^{\V}] = 0\, , \quad 
    [X^{\V}, Y^{\Com}] = [X, Y]^{\V}\, , \quad
    [X^{\Com}, Y^{\Com}] = [X, Y]^{\Com}\, .
\end{equation} 
The latter implies that the map $X\in \X(Q)\mapsto X^{\Com} \in \X(\T Q)$ is a Lie algebra homomorphism. Moreover,
\begin{equation}
    \Sendo(X^{\V}) = 0\, , \quad \Sendo(X^{\Com}) = X^{\V}\, .
\end{equation}

Given a curve $c\colon I\subseteq \RR \to Q$ on $Q$, its \emph{canonical lift to $\T Q$} is the curve $\tilde{c} \colon I\to \T Q$ given by 
\begin{equation}
    \tilde{c}(t) = \big( c(t), \dot{c}(t) \big)\, ,
\end{equation}
where $\dot{c}(t)$ denotes the tangent vector to the curve at the point $c(t)$. A curve $\sigma\colon I\subseteq \RR\to \T Q$ on $\T Q$ is called \emph{holonomic} if there exists a curve $c\colon I \to Q$ on $Q$ such that $\sigma = \tilde{c}$. 
A vector field $\sode\in \X(\T Q)$ is called a \emph{second order differential equation} (henceforth abbreviated as \emph{\textsc{sode}}, and also known as \emph{semispray}) \emph{on $Q$} if its integral curves are holonomic. Locally, a \textsc{sode} is of the form
\begin{equation}\label{eq:local_SODE}
    \sode = v^i \frac{\partial}{\partial q^i} + \sode^i \frac{\partial}{\partial v^i}\, .
\end{equation} 
Furthermore, the following statements are equivalent:
\begin{enumerate}
    \item $\sode$ is a \textsc{sode} on $Q$,
    \item $\sode$ is a section of the vector bundles $\tau_{\T Q}\colon \T \T Q \to \T Q$ and \newline $\T \tau_Q\colon \T \T Q\to \T Q$,
    \item $\Sendo(\sode) = \Delta$.   
\end{enumerate}
If $\sigma= \tilde{c}$ is an integral curve of a \textsc{sode} $\sode$, then the curve $c=\tau_Q\circ \sigma$ is called a \emph{solution of $\sode$}. A solution $c(t)=(q^i(t))$ of the \textsc{sode}~\eqref{eq:local_SODE} is locally given by the following system of $n$ second order differential equations:
\begin{equation}
    \frac{\dd^2 q^i}{\dd t^2} = \sode^i\left(q^1(t), \ldots, q^n(t), \frac{\dd q^1}{\dd t}, \ldots, \frac{\dd q^n}{\dd t}\right)\, .
\end{equation}

\section{Semibasic one-forms and bundle morphisms}\label{sec:semibasic}

In this section the notion of semibasic one-forms, and their associated bundle morphisms, are recalled. Additional details may be found in \cite{A.M2008,d.R1989,Godbillon1969}.

Given a fiber bundle $\pi\colon E\to M$, a one-form $\alpha\in \Omega^{1}(E)$ on $E$ is called \emph{semibasic} if $\alpha(Z)$ for every vertical vector field $Z\in \Gamma(\V E)$. If $(x^i, y^a)$ are bundle coordinates, then 
\begin{equation}
    \alpha = \alpha_i (x, y)\, \dd x^i\, .
\end{equation} 

There is a bijective correspondence between semibasic one-forms on $\T Q$ and bundle morphisms from $\T Q$ to $\cT Q$.
Indeed, given a bundle morphism $F\colon \T Q\to \cT Q$,
one can define an associated semibasic one-form $\alpha_F$ on $\T Q$ by 
\begin{equation}
    \alpha_F(v_q)(u_{v_q})=\left\langle F(v_q),T\tau_Q(u_{v_q})\right\rangle\, ,
\end{equation}
where $v_q\in T_qQ$ and $u_{v_q}\in T_{v_q}(\T Q)$. Conversely, given a semibasic one-form $\alpha$ on $\T Q$, one can define the following morphism of fiber bundles:
\begin{equation}
\begin{aligned}
    &F_\alpha: TQ\to \cT Q\, ,\\
    &\left\langle F_\alpha(v_q),w_q\right\rangle=\alpha(v_q)(u_{w_q})\, ,
\end{aligned}
\end{equation}
for every $v_q,w_q\in T_qQ,\ u_{w_q}\in T_{w_q}(TQ)$, with $T\tau_Q(u_{w_q})=w_q$.

Locally, if a bundle morphism $F\colon \T Q \to \cT Q$ is given by
\begin{equation}
    F(q^i,v^i)=(q^i,F_i(q,v))\, ,
\end{equation}
then the associated semibasic one-form is
\begin{equation}
    \alpha_F=F_i(q,v)\, \dd q^i\, .
\end{equation}
If a semibasic one-form is locally
\begin{equation}
\alpha=\alpha_i(q,v)\, \dd q^i\, ,
\end{equation}
then its associated bundle morphism is
\begin{equation}
    F_\alpha(q^i,v^i)=\left(q^i,\alpha_i(q^i,\dot{q}^i)\right)\, .
\end{equation}

Furthermore, it is worth noting that a one-form $\alpha\in \Omega^1(\T Q)$ is semibasic if and only if it is in the kernel of adjoint operator of the vertical endomorphism, namely, $\Sendoadj (\alpha) = 0$.

\section{Connections}
In this section some essential concepts about connections are succinctly exposed. Further details may be consulted, for instance, in \cite{K.S.M1993,d.R1989, C.C.L1999}.

Let $\pi\colon E \to M$ be a vector bundle. The space of smooth sections of the vector bundle $\bigwedge^k\cT M \otimes E$ is denoted by $\Omega^k(M; E)$, and its elements are called \emph{$E$-valued $k$-forms}. A \emph{connection on $\pi\colon E \to M$} is a $\V E$-valued 1-form $\Phi\in \Omega^1(E; \V E)$ such that 
\begin{equation}
    \Phi \circ \Phi = \Phi\, , \quad \Ima \Phi = \V E\, ,
\end{equation}
which means that $\Phi$ is a projection from $\T E$ to $\V E$. The subbundle $\Hor E=\ker \Phi$ of $\T E$ is called the \emph{horizontal bundle}. The tangent bundle can be decomposed as follows:
\begin{equation}
    \T E = \V E \oplus \Hor E \, .
\end{equation}
The \emph{curvature of $\Phi$} is the $\V E$-valued 2-form $\mathfrak{R}\in \Omega^2(E; \V E)$ such that
\begin{equation}
    \mathfrak{R}(X, Y) = \Phi \big([X - \Phi (X), Y - \Phi (Y)]\big)\, ,
\end{equation}
for any vector fields $X, Y\in \X (E)$.


Let $\pi\colon E\to M$ be a principal bundle with group $G$. Denote the right action $R\colon E\times G\to E$ of $G$ on $E$ by $R_g(x) = x\cdot g$ for each $g\in G$ and $x\in E$. A connection $\Phi$ on $\pi\colon E \to M$ is called a \emph{principal connection} if it is $G$-equivariant for the principal right action, that is,
\begin{equation}
    (\T R_g) \circ \Phi = \Phi \circ (\T R_g)
\end{equation}
or, equivalently,
\begin{equation}
    (\T_x R_g) \Hor_x E = \Hor_{x\cdot g}E\, ,
\end{equation}
for each $x\in E$ and $g\in G$, where $\Hor E = \ker \Phi$ is the horizontal bundle.

Consider a connection on the tangent bundle $\tau_Q \colon \T Q \to Q$ such that the horizontal bundle is $\Hor \T Q$. Given a vector field $X\in \X(Q)$, the \emph{horizontal lift of $X$} is the vector field $X^\Hor \in \X(\T Q)$ such that
\begin{equation}
    X^\Hor_{v_q} \in \Hor \T Q_{v_q}\, , \quad \restr{\T \tau_Q} {v_q} (X^\Hor_{v_q}) = X_q\, ,
\end{equation}
for each $v_q\in \T Q$, where $q= \tau_Q (v_q)$. In canonical bundle coordinates, if
\begin{equation}
    X = X^i \parder{}{q^i} \, ,
\end{equation}
then
\begin{equation}
    X^\Hor = X^i \parder{}{q^i} - X^j \Gamma^i_j \parder{}{v^j}\, ,
\end{equation}
where $\Gamma^i_j \in \Cinfty(\T Q)$ are called the \emph{Christoffel components of $\Gamma$}. Moreover, 
\begin{equation}
  \mathfrak{R}\left(\frac{\partial}{\partial q^{a}}, \frac{\partial}{\partial q^{b}}\right)=\mathfrak{R}_{a b}^{i} \frac{\partial}{\partial q^{i}}\, , 
\end{equation}
where
\begin{equation}
  \mathfrak{R}_{a b}^{i}=\frac{\partial \Gamma_{a}^{i}}{\partial q^{b}}-\frac{\partial \Gamma_{b}^{i}}{\partial q^{a}}+\Gamma_{a}^{j} \frac{\partial \Gamma_{b}^{i}}{\partial q^{j}}-\Gamma_{b}^{j} \frac{\partial \Gamma_{a}^{i}}{\partial q^{j}}\, .
\end{equation}

\section{Jacobi manifolds} \label{sec:Jacobi_manifolds}

This section exposes the principal aspects of Jacobi structures and Jacobi brackets on manifolds. Particular cases of Jacobi structures ((co)symplectic and (co)contact structures) will be considered in the subsequent sections of this chapter.

Let $M$ be a manifold. A \emph{multivector field} on $M$ is a section of the exterior algebra $\bigwedge^\bullet \T M$ of the tangent bundle. A \emph{bivector field} is a section of $\bigwedge^2 \T M$. The Lie bracket $[\cdot, \cdot]$ on the space of vector fields $\X(M)$ can be extended to a bilinear map $[\cdot, \cdot]_{\mathrm{SN}}\colon \Gamma(\bigwedge^\bullet \X(M))\times  \Gamma(\bigwedge^\bullet \X(M))\to  \Gamma(\bigwedge^\bullet \X(M))$ on the exterior algebra of multivector fields given by
\begin{equation}
\begin{aligned}
    & {\left[X_1 \wedge \cdots \wedge X_r, Y_1 \wedge \cdots \wedge Y_n\right]_{\mathrm{SN}}} \\
    & \quad=\sum_{k,\,l} (-1)^{k+l}\left[X_k, Y_l\right] \wedge \cdots \wedge \widehat{X}_k \wedge \cdots \wedge X_r \wedge Y_1 \wedge \cdots \wedge \widehat{Y}_l \wedge \cdots \wedge Y_n\, ,
\end{aligned}
\end{equation}
where $X_1, \ldots, X_r, Y_1, \ldots, Y_n\in \X(M)$ and $\widehat{X}_k$ stands for the omission of $X_k$. The bracket $[\cdot, \cdot]_{\mathrm{SN}}$ is called the \emph{Schouten--Nijenhuis bracket}. The properties of this bracket may be found in references~\cite{Grabowski2013, Michor1987, Nijenhuis1955, Nijenhuis1955a, Schouten1940}.

Consider a map $\{\cdot, \cdot\}\colon \Cinfty(M)\times \Cinfty(M)\to \Cinfty(M)$ given by
\begin{equation}\label{eq:Jacobi_bracket}
    \{f, g\} = \Lambda(\dd f\wedge \dd g) + f E(g) - g E(f)\, ,
\end{equation}
where $\Lambda\in\Gamma(\bigwedge^2 \X(M))$ is a bivector field and $E\in \X(M)$ a vector field. Lichnerowicz \cite{Lichnerowicz1977, Lichnerowicz1977a, Lichnerowicz1978} proved that $\{\cdot, \cdot\}$ is a Lie bracket if and only if 
\begin{equation}\label{eq:Jacobi_structure}
    [\Lambda, \Lambda]_{\mathrm{SN}} = 2 E \wedge \Lambda\, , \quad [E, \Lambda]_{\mathrm{SN}} = 0 \, .
\end{equation}
If these conditions hold, then $\{\cdot, \cdot\}$ is called a \emph{Jacobi bracket on $M$}, the pair $(\Lambda, E)$ is called a \emph{Jacobi structure on $M$}, and the triple $(M, \Lambda, E)$ is called a \emph{Jacobi manifold}. By construction, the Jacobi bracket is a Lie bracket, that is, it is bilinear, skew-symmetric and satisfies the Jacobi identity. Moreover, it satisfies the so-called \emph{weak Leibniz rule}, namely, 
\begin{equation}\label{eq:weak_Leibniz_rule}
    \{f, gh\} = \{f, g\} h + \{f, h\} g + g h E(f)\, ,
\end{equation} 
for any $f, g, h\in \Cinfty(M)$, where $gh$ denotes the pointwise multiplication.

It is possible to consider an equivalent, more algebraic, definition Jacobi structures on manifolds.
Since the Jacobi bracket $\{\cdot, \cdot\}$ is a Lie bracket and satisfies the weak Leibniz rule, it defines a structure of local Lie algebra in the sense of Kirillov on $\Cinfty(M)$ (see \cite{Kirillov1976,D.L.M1991,G.L1984}). Conversely, given a local Lie algebra on $\Cinfty(M)$, there exists a Jacobi structure on $M$ such that the Jacobi bracket coincides with the algebra bracket.

A Jacobi bracket $\{\cdot, \cdot\}$ is called a \emph{Poisson bracket} if and only if it is a derivation, that is, it satisfies the \emph{Leibniz rule}:
\begin{equation}
    \{f, gh\} = \{f, g\} h + \{f, h\} g\, ,
\end{equation} 
for any $f, g, h\in \Cinfty(M)$. Clearly, a Jacobi structure $(\Lambda, E)$ defines a Poisson bracket if and only if $E=0$. In that case, $\Lambda$ is called a \emph{Poisson structure on $M$}, and the pair $(M, \Lambda)$ is called a \emph{Poisson manifold}.

A collection of functions $f_1, \ldots, f_k\in \Cinfty(M)$ are said to be \emph{in involution (with respect to $\{\cdot, \cdot\}$)} if $\{f_i, f_j\} = 0$ for all $i,j\in \{1, \ldots, k\}$.

\begin{remark}\label{remark:Jacobi_orthogonal}
Let $(M, \Lambda, E)$ be a Jacobi manifold with Jacobi bracket $\{\cdot, \cdot\}$. Then (see references~\cite{D.L.M1991,I.L.M+1997})
\begin{enumerate}
    \item The bivector $\Lambda$ induces a $\Cinfty(M)$-module morphism $\lsharp\colon \Omega^1(M)\to \X(M)$ given by $\lsharp(\alpha) = \Lambda(\alpha, \cdot)$. 
    \item Given a function $f\in \Cinfty(M)$, its \emph{Hamiltonian vector field with respect to $(\Lambda, E)$} is the vector field $X_f\in \X(M)$ given by 
    \begin{equation}\label{eq:Hamiltonian_vf_Jacobi}
        X_f = \lsharp(\dd f) + f E 
        \, .
    \end{equation}
    In particular, $E$ is the Hamiltonian vector field of $f\equiv 1$.
    \item The \emph{characteristic distribution of $(M, \Lambda, E)$} is given by 
    \begin{equation}
        C= \lsharp(\cT M) + \langle E \rangle\, .
    \end{equation}
    \item If $D$ is a distribution on $M$, then its \emph{Jacobi orthogonal complement} is the distribution $D^{\perp_\Lambda} = \lsharp(D^\circ)$, where $D^\circ$ denotes the annihilator of $D$.
    \item A submanifold $N\hookrightarrow M$ is called 
    \begin{enumerate}
        \item \emph{coisotropic} if $\T N^{\perp_\Lambda} \subseteq \T N$,
        \item \emph{Legendre--Lagrangian} if $\T N^{\perp_\Lambda} = \T N \cap C$.
    \end{enumerate}
\end{enumerate}
\end{remark}

\begin{remark}\label{remark:Jacobi_equivalent}
    Let $a\in \Cinfty(M)$ be a nowhere-vanishing function. Define the bracket $\{\cdot, \cdot\}^a \colon \Cinfty(M) \times \Cinfty(M) \to \Cinfty(M)$ by
    \begin{equation}
        \{f, g\}^a = \frac{1}{a} \{af, ag\}\, ,
    \end{equation}
    for any $f, g\in \Cinfty(M)$. Then, this new bracket $\{\cdot, \cdot\}^a$ is also a Jacobi bracket on $M$ (see references~\cite{D.L.M1991,Lichnerowicz1978}). Its associated Jacobi structure is $(\Lambda^a, E^a)$, where
    \begin{equation}
        \Lambda^a = a \Lambda\, ,\quad E^a = \lsharp(\dd a) + a E\, .
    \end{equation}
    For each $f\in \Cinfty(M)$, let $X_f$ and $X_f^a$ denote its Hamiltonian vector fields with respect to $(\Lambda, E)$ and $(\Lambda^a, E^a)$, respectively. Then,
    \begin{equation}
        X_f^a = X_{af}\, ,
    \end{equation}
    and, in particular,
    \begin{equation}
        E^a = X_a\, .
    \end{equation}
    It is said that \emph{$(\Lambda^a, E^a)$ is $a$-conformal to $(\Lambda, E)$}, or that \emph{$(\Lambda, E)$ and $(\Lambda^a, E^a)$ are conformally equivalent}. An equivalence class formed by all the Jacobi structures on $M$ conformally equivalent to a given Jacobi structure is called a \emph{conformal Jacobi structure on $M$}.
    
    It is noteworthy that the Jacobi orthogonal complement depends only on the conformal Jacobi structure, that is, $D^{\perp_\Lambda} = D^{\perp_{\Lambda^a}}$ for any distribution $D$ on $M$ and for any conformal factor $a$. Consequently, if a submanifold $N\hookrightarrow M$ is coisotropic (respectively, Legendre--Lagrangian) with respect to a Jacobi structure $(\Lambda, E)$ on $M$, then it is also coisotropic (respectively, Legendre--Lagrangian) with respect to any Jacobi structure conformally equivalent to $(\Lambda, E)$.
\end{remark}

\section{Symplectic manifolds}\label{sec:symplectic}

This section is devoted to the main aspects of symplectic geometry that will be employed throughout this dissertation. References \cite{Lee2012,A.M2008,d.R1989,M.S2017,L.M1987,Godbillon1969} may be consulted for additional details. 

Let $M$ be a manifold. A \emph{symplectic form} $\omega\in \Omega^2(M)$ is a closed 2-form on $M$ which is non-degenerate, that is, the map $\T_x M\ni v\mapsto \contr{v} \omega(x)\in \cT_x M$ is an isomorphism of vector spaces. The pair $(M, \omega)$ is called a \emph{symplectic manifold}.

A 2-form $\omega\in \Omega^2(M)$ is non-degenerate if and only $\vol_{\omega} = \omega^n$ is a volume form.
Moreover, the non-degeneracy condition implies that $M$ is even-dimensional. Along the rest of the section, let $\dim M=2n$. 

Around each point $x\in M$ of a symplectic manifold $(M, \omega)$, there exist coordinates $(q^1, \ldots, q^n, p_1, \ldots, p_n)$ centered at $x$ in which the symplectic form reads
\begin{equation}
    \omega = \dd q^i \wedge \dd p_i\, .
\end{equation}
These coordinates are called \emph{Darboux coordinates} or \emph{canonical coordinates}. 

A symplectic form $\omega$ on $M$ defines an isomorphism of $\Cinfty(M)$-modules $\flat_\omega\colon \X(M)\to \Omega^1(M)$ given by $\flat_\omega(X) = \contr{X} \omega$. Its inverse is denoted by $\sharp_\omega$, and both maps are called \emph{musical isomorphisms}. 

Let $(M_1, \omega_1)$ and $(M_2, \omega_2)$ be symplectic manifolds. A map $F\colon M_1 \to M_2$ is called a \emph{symplectic transformation} if $F^\ast \omega_2 = \omega_1$. If a symplectic transformation is a diffeomorphism, it is called a \emph{symplectomorphism}, and $(M_1, \omega_1)$ and $(M_2, \omega_2)$ are said to be \emph{symplectomorphic}. A vector field $X\in \X(M)$ on a symplectic manifold $(M, \omega)$ is called an \emph{infinitesimal symplectomorphism} if its flow is made of symplectomorphisms, or, equivalently, $ \liedv{X} \omega = 0$.
Obviously, every symplectomorphism $\Phi\colon M \to M$ or infinitesimal symplectomorphism $X\in \X(M)$ preserve the volume form $\vol_\omega$, namely,
\begin{equation}
    \Phi^\ast \vol_\omega = \vol_\omega\, , \quad \liedv{X}\vol_\omega = 0\, .
\end{equation}

It is worth noting that, as a consequence of the existence of Darboux coordinates, all symplectic manifolds of the same dimension are locally symplectomorphic. More specifically, given a pair of $2n$-dimensional symplectic manifolds $(M_1, \omega_1)$ and $(M_2, \omega_2)$, with respective charts of Darboux coordinates $(U_1, \varphi_1)$ and $(U_2, \varphi_2)$, then $\varphi_2^{-1}\circ \varphi_1\colon U_1 \to U_2$ is a symplectomorphism from $(U_1, \restr{\omega_1}{U_1})$ to $(U_2, \restr{\omega_2}{U_2})$.

Let $(M, \omega)$ be a $2n$-dimensional symplectic manifold. Given a function $f\in \Cinfty(M)$, the \emph{Hamiltonian vector field of $f$ with respect to $\omega$} is the vector field $X_f\in \X(M)$ given by 
\begin{equation}\label{eq:Hamiltonian_vf_symplectic}
    X_f = \sharp_\omega (\dd f)\, .
\end{equation}
With a straightforward computation, one can verify that every Hamiltonian vector field is an infinitesimal symplectomorphism. Conversely, from the Poincaré lemma it follows that any infinitesimal symplectomorphism is locally a Hamiltonian vector field. In Darboux coordinates,
\begin{equation}
   X_f =  \frac{\partial f}{\partial p_i} \frac{\partial }{\partial q^i} - \frac{\partial f}{\partial q^i}\frac{\partial }{\partial p_i}\, .
\end{equation}

The bilinear operation $\{\cdot, \cdot\}_\omega \colon \Cinfty(M) \times \Cinfty(M) \to \Cinfty(M)$ given by 
\begin{equation}\label{eq:Poisson_bracket_symplectic}
    \{f, g\}_\omega = \omega(X_f, X_g)
\end{equation}
is a Poisson bracket. In Darboux coordinates,
\begin{equation}
    \{f, g\}_\omega = \frac{\partial f}{\partial q^i} \frac{\partial g}{\partial p_i}-\frac{\partial f}{\partial p_i} \frac{\partial g}{\partial q^i}\, .
\end{equation}
The Poisson bracket defined by $\omega$ satisfies the identity
\begin{equation}\label{eq:Lie_algebra_antihomomorphism_symp}
    X_{\{f, g\}_\omega} = -[X_f, X_g]\, ,
\end{equation}
and hence the map $f\mapsto X_f$ is a Lie algebra anti-homomorphism. 
The Poisson structure $\Lambda_\omega$ defined by $\omega$ is given by
\begin{equation}
    \Lambda_\omega (\alpha, \beta) = \omega \big(\sharp_\omega (\alpha), \sharp_\omega (\beta)\big) \, ,
\end{equation}
for $\alpha, \beta \in \Omega^1(M)$. Observe that the Hamiltonian vector field defined by the symplectic form $\omega$ coincides with the one defined by the induced Poisson structure $\Lambda_\omega$ 
(\emph{cf.~}equations~\eqref{eq:Hamiltonian_vf_Jacobi} and \eqref{eq:Hamiltonian_vf_symplectic}).

If $D$ is a distribution on $M$, then its \emph{symplectic orthogonal complement} is the distribution $D^{\perp_\omega}$ given by
\begin{equation}
    D^{\perp_\omega}_x = \sharp_\omega(D^\circ_x) = \left\{v\in \T_x M\mid \omega(v, w) = 0 \text{ for all } w\in D_x \right\}\, .
\end{equation}
A submanifold $N\hookrightarrow M$ of a symplectic manifold is called:
\begin{enumerate}
    \item \emph{symplectic} if $\T N\cap \T N^{\perp_\omega} = \{0\}$,
    \item \emph{isotropic} if $\T N\subseteq \T N^{\perp_\omega}$,
    \item \emph{coisotropic} if $\T N^{\perp_\omega}\subseteq \T N$,
    \item \emph{Lagrangian} if $\T N= \T N^{\perp_\omega}$.
\end{enumerate}
Observe that a submanifold is coisotropic with respect to the symplectic orthogonal complement defined by $\omega$ if and only if it is coisotropic with respect to the Jacobi orthogonal complement defined by $\Lambda_\omega$. Since $\sharp_\omega$ is an isomorphism, the characteristic distribution is the whole tangent bundle, that is, $C= \T M$. Hence, a submanifold is Lagrangian with respect to $\omega$ if and only if it is Legendre--Lagrangian with respect to $\Lambda_\omega$.
Moreover,
\begin{enumerate}
    \item $N$ is symplectic if and only if $\restr{\omega}{N}$ is non-degenerate,
    \item $N$ is isotropic if and only if $\restr{\omega}{N}=0$,
    \item $N$ is Lagrangian if and only if $\restr{\omega}{N}=0$ and $\dim N =  n$,
    \item $N$ is Lagrangian if and only if it is isotropic and coisotropic.
\end{enumerate}

The ``symplectic creed'' by Alan Weinstein \cite{Weinstein1981} states that ``Everything is a Lagrangian submanifold''. After that, Weinstein says that ``In practice, the symplectic creed means that one should try to express
objects and constructions in symplectic geometry in terms of Lagrangian
submanifolds''. For instance, closed one-forms are characterized by Lagrangian submanifolds (see the following paragraph). Furthermore, the notion of Lagrangian submanifold allows to give a geometric description of integrable systems and solutions to the Hamilton--Jacobi problem (see \Cref{sec:integrable_systems,sec:Hamilton-Jacobi}, respectively). 

The archetypal example of a symplectic manifold is the cotangent bundle $\cT Q$ of a manifold $Q$. The \emph{tautological one-form} $\theta_Q\in \Omega^1(Q)$ is defined by 
\begin{equation}
    \restr{\theta_Q}{\alpha_q} = (\cT \pi_{Q})_{\alpha_q}\, \alpha_q\, , 
\end{equation}
for $\alpha_q\in \cT_q Q$. It is also called the \emph{canonical one-form}, the \emph{Liouville one-form} or the \emph{Poincaré one-form}. Its is the unique one-form $\theta_Q\in \Omega^1(\cT Q)$ on $\cT Q$ such that $\alpha^\ast \theta_Q = \theta_Q$ for any one-form $\alpha\in \Omega^1(Q)$ on $Q$. If $(q^1, \ldots, q^n, p_1, \ldots, p_n)$ are bundle coordinates, then $\theta_Q = p_i\, \dd q^i$. Hence, $\omega_Q = - \dd \theta_Q$ is a symplectic form and the bundle coordinates are Darboux coordinates. The 2-form $\omega_Q$ is called the \emph{canonical symplectic form of the cotangent bundle}.
%
Furthermore, a one-form $\alpha\in \Omega^1(Q)$ on $Q$ is closed if and only if $\Ima \alpha$ is a Lagrangian submanifold of $(\cT Q, \omega_Q)$. 

\subsection{Exact symplectic manifolds}\label{subsec:exact_symplectic}

Let $M$ be a manifold. A \emph{symplectic potential} is a one-form $\theta\in \Omega^1(M)$ such that $\omega = -\dd \theta$ is a symplectic form on $M$. The pair $(M, \theta)$ is called an \emph{exact symplectic manifold}. 

A symplectic manifold $(M, \omega)$ is an exact symplectic manifold if and only if there exists a vector field $X\in \X(M)$ such that
\begin{equation}
    \liedv{X} \omega = \omega\, .
\end{equation}
In that case, $\omega = -\dd \theta$, where
\begin{equation}
    \theta = -\contr{X} \omega + \alpha\, ,
\end{equation}
for some closed one-form $\alpha\in \Omega^1(M)$. The \emph{Liouville vector field $\Delta$ of $(M, \theta)$} is given by
\begin{equation}
    \contr{\Delta} \omega = - \theta\, .
\end{equation}
A tensor field $A\in \tensors^k_l (\T M) $ is called \emph{homogeneous of degree $k$} if
\begin{equation}
    \liedv{\Delta} A = k A\, . 
\end{equation}

Let $(M_1, \theta_1)$ and $(M_2, \theta_2)$ be exact symplectic manifolds. A map $F \colon M_1 \to M_2$ an \emph{exact symplectic transformation} if $F^\ast \theta_2 = \theta_1$. An \emph{exact symplectomorphism} (or \emph{homogeneous symplectomorphim}) is an exact symplectic transformation which is also a diffeomorphism. Obviously, every exact symplectic transformation (respectively, exact symplectomorphism) is a symplectic transformation (respectively, symplectomorphism). A vector field $X\in \X(M)$ on an exact symplectic manifold $(M, \theta)$ is called an \emph{infinitesimal exact symplectomorphism} (or \emph{infinitesimal homogeneous symplectomorphim}) is its flow is made of exact symplectomorphisms, or, equivalently,
\begin{equation}
    \liedv{X} \theta = 0\, .
\end{equation}
Clearly, every infinitesimal exact symplectomorphism is an infinitesimal symplectomorphism.

\begin{proposition}\label{proposition:homogeneous_Hamiltonian}
    Let $(M, \theta)$ be an exact symplectic manifold.
    Given a vector field $Y\in \X(M)$, the following statements are equivalent:
    \begin{enumerate}
        \item \label{item:homogeneous_symp} $Y$ is an infinitesimal exact symplectomorphism,
        \item \label{item:commute_Delta} $Y$ is an infinitesimal symplectomorphism and commutes with the Liouville vector field $\Delta$,
        \item \label{item:Hamiltonian} $Y$ is the Hamiltonian vector field of $f= \theta(Y)$ and $f$ is a homogeneous function of degree $1$.
    \end{enumerate}
\end{proposition}

\begin{proof}
    The first step is to show the equivalence between statements \ref{item:homogeneous_symp} and \ref{item:commute_Delta}.
    Suppose that $Y\in \X(M)$ is an infinitesimal symplectomorphism. Then,
    \begin{equation}
        \contr{[Y, \Delta]} \omega = \liedv{Y} \contr{\Delta} \omega - \contr{\Delta} \liedv{Y} \omega = - \liedv{Y} \theta\, ,
    \end{equation}
    which implies that $Y$ is an infinitesimal exact symplectomorphism if and only if it commutes with $\Delta$.

    The next step is to show that \ref{item:homogeneous_symp} implies \ref{item:Hamiltonian}.
    Suppose that $Y\in \X(M)$ is an infinitesimal exact symplectomorphism. Then, 
    \begin{equation}
        0 = \liedv{Y} \theta = \contr{Y} \dd \theta + \dd \contr{Y} \theta\, ,
    \end{equation} 
    which implies that 
    \begin{equation}
        \contr{Y} \omega = \dd f\, ,
    \end{equation}
    for $f = \contr{Y} \theta$.

    
    The last step is to show that \ref{item:Hamiltonian} implies \ref{item:commute_Delta}. Suppose that $f\in \Cinfty(M)$ is a homogeneous function of degree $1$ and $X_f$ is its Hamiltonian vector field. Then, 
    \begin{equation}
        \contr{[\Delta, X_f]} \omega = \liedv{\Delta} \contr{X_f} \omega - \contr{X_f} \liedv{\Delta} \omega = \liedv{\Delta} \dd f - \dd f = 0\, . 
    \end{equation}
\end{proof}

For instance, for any manifold $Q$, its cotangent bundle with the canonical one-form, $(\cT Q, \theta_Q)$ is an exact symplectic manifold. Moreover, a diffeomorphism $\varphi \colon \cT Q \to \cT Q$ is an exact symplectomorphism if and only if it is the cotangent map $\varphi=\cT F$ of diffeomorphism $F\colon Q \to Q$ on $Q$. This implies that 
\begin{equation}
    \pi_Q \circ \cT F = F \circ \pi_Q\, .
\end{equation}



An interesting topological fact is that every exact symplectic manifold is non-compact. This can be proven by means of the Stokes theorem for differential forms. Consequently, compact manifolds with a trivial even cohomology group $H^{2k}(M),\, k=0, \ldots, n$, such as spheres $\Sp^{2n}$ with $n>1$, can never be symplectic.

\subsection{Momentum maps and symplectic reduction}

Consider a Lie group $G$ with Lie algebra $\mathfrak{g}$, and let $\mathfrak{g}^\ast$ be the dual of $\mathfrak{g}$. Let $(M,\omega)$ be a connected symplectic manifold. A Lie group action $\Phi:G\times M\to M$ of $G$ on $M$ is called a \emph{symplectic action} if $\Phi_g\colon M\to M$ is a symplectomorphism for each $g\in G$. A map $\mommap\colon M \to \mathfrak{g}^\ast$ is called a \emph{momentum map for the action $\Phi$} if
\begin{equation}
    \dd \mommap^\xi = \contr{\xi_M} \omega\,
\end{equation} 
where $\xi_M\in \X(M)$ denotes the infinitesimal generator associated to $\xi \in \mathfrak{g}$ (see equation~\eqref{eq:infinitesimal_generator}), and $\mommap^\xi\colon M \to \RR$ is the function given by
\begin{equation}
    \mommap^\xi(x) = \left\langle \mommap(x) , \xi \right\rangle\, .
\end{equation} 
In other words, $\mommap$ is a momentum map provided that, for every $\xi \in \mathfrak{g}$, the Hamiltonian vector field of $\mommap^\xi$ the infinitesimal generator of $\xi$, namely, $X_{\mommap^\xi} = \xi_M$.

The \emph{affine action} of $G$ on $\mathfrak{g}^\ast$ associated with the momentum map $\mommap:M\to \mathfrak{g}^\ast$ is given by
\begin{equation}
\begin{aligned}
G\times \mathfrak{g}^\ast & \to \mathfrak{g}^\ast\\
(g,\mu) & \mapsto \Ad_{g^{-1}}^\ast \mu 
		+ \mommap  \left( \Phi_g(x)  \right)
		- \Ad_{g^{-1}}^\ast \left( \mommap(x)  \right).
\end{aligned}
\end{equation}
This definition is independent of the choice of $x\in M$. The isotropy group of $\mu\in \mathfrak{g}$ under the affine action is denoted by $G_\mu$.


A momentum map $\mommap$ is called \emph{$\Ad^\ast$-equivariant} if, for each $g\in G$, the diagram
\begin{equation}
\begin{tikzcd}
M \arrow[rr, "\Phi_g"] \arrow[d, "\mommap"']                     &  & M \arrow[d, "\mommap"] \\
\mathfrak{g}^\ast \arrow[rr, "\Ad^\ast_{g^{-1}}"'] &  & \mathfrak{g}^\ast  
\end{tikzcd}
\end{equation}
commutes. In other words, 
\begin{equation}
	\mommap \left( \Phi_g(x)  \right) = \Ad_{g^{-1}}^\ast \mommap(x)\, ,
\end{equation}
for each $g\in G$ and each $x\in M$. 
If $\mommap$ is an $\Ad^\ast$-equivariant momentum map, then 
\begin{equation}
	\left\{\mommap^\xi, \mommap^\eta  \right\}_\omega = \mommap^{[\xi,\eta]}
\end{equation}
for each $\xi,\eta\in \mathfrak{g}$, where $\left\{\cdot,\cdot  \right\}_{\omega}$ is the Poisson bracket defined by $\omega$. Hence, the map $\xi \mapsto \mommap^\xi$ is an homomorphism of Lie algebras.

Let $\Phi\colon G\times M \to M$ be a Lie group action of $G$ on $M$. 
A differential form $\alpha \in \Omega^\bullet(M)$ is called \emph{$\Phi$-invariant} if $\Phi^\ast_g\, \alpha = \alpha$ for all $g\in G$. If the action of $G$ on $M$ is understood, a $\Phi$-invariant differential form is also called \emph{$G$-invariant}. 

Let $(M, \theta)$ be a exact symplectic manifold. A Lie group action $\Phi \colon G \times M \to M$ of $G$ on $M$ is called an \emph{exact symplectic action} if $\Phi_g\colon M\to M$ is an exact symplectomorphism for each $g\in G$. 

If $\Phi$ is an exact symplectic action, then the map $\mommap\colon M \to \mathfrak{g}^\ast$ given by
\begin{equation}
	\left\langle \mommap(x), \xi \right \rangle = \left( \contr{\xi_M} \theta  \right)(x)\, ,
\end{equation}
is an $\Ad^\ast$-equivariant momentum map, called the \emph{natural momentum map}.

Consider a Lie group action $\Phi\colon G\times Q\to Q$ of $G$ on $Q$ and its cotangent lift $\Phi^{\cT}\colon G\times \cT Q \to \cT Q$. Since $\pi_Q \circ \Phi^{\cT}_g = \Phi_g \circ \pi_Q$ for each $g\in G$, the infinitesimal generators of the actions of $G$ on $Q$ and $\cT Q$ are related by 
\begin{equation}
    \T \pi_Q \circ \xi_{\cT Q} = \xi_Q \circ \pi_Q\, .
\end{equation}
Since the cotangent lift is an exact symplectic action, it has a natural momentum map $\mommap\colon M \to \mathfrak{g}$, given by
\begin{equation}
	\left\langle \mommap(\alpha_q), \xi \right \rangle = \left( \contr{\xi_{\cT Q}} \theta_Q  \right)(\alpha_q) = \alpha_q \big(\xi_Q (q)\big) \, .
\end{equation}


The following theorem, due to Marsden and Weinstein, is the basis of symplectic reduction (see \cite{M.W1974, O.R2004, A.M2008} for the proof and further information).

\begin{theorem}[Symplectic point reduction] \label{thm:symplectic_point_reduction}
    Let $(M, \omega)$ be a connected symplectic manifold, and let $G$ be a Lie group. Let $\Phi\colon G\times M \to M$ be a symplectic action of $G$ on $M$.
    Then the following statements hold:
    \begin{enumerate}
        \item The quotient space $M_\mu\coloneqq \mommap^{-1}(\mu)/G_\mu$ has a symplectic form $\omega_\mu$ uniquely characterized by the relation
        \begin{equation}
            \pi_\mu^\ast \omega_\mu = \incl_\mu^\ast \omega\, .
        \end{equation}
        Here the maps $\incl_\mu: \mommap^{-1}(\mu)\hookrightarrow M$ and $\pi_\mu:\mommap^{-1}(\mu)\to \mommap^{-1}(\mu)/G_\mu$ denote the inclusion and the projection, respectively. The pair $(M_\mu,\omega_\mu)$ is called the \emph{symplectic point reduced space}.
        \item Let $f\colon M \to \RR$ be a $\Phi$-invariant function on $M$, and $X_f\in \X(M)$ its Hamiltonian vector field. The flow $F_t$ of $X_f$ induces a flow $F_t^\mu$ on $M_\mu$ given by
        \begin{equation}
            \pi_{\mu} \circ F_{t} \circ \incl_{\mu}=F_{t}^{\mu} \circ \pi_{\mu}\, .
        \end{equation}
        \item The vector field $X_{f_{\mu}}$  generated by the flow $F_t^\mu$ on $(M_\mu,\omega_\mu)$ is the Hamiltonian vector field of the function $f_\mu\colon  M_\mu \to \RR$ defined by
        \begin{equation}
            f_\mu \circ \pi_\mu = f \circ \incl_\mu.
        \end{equation}
        The vector fields $X_f$ and $X_{f_\mu}$ are $\pi_\mu$-related, that is,
        \begin{equation}
            \T \pi_{\mu}\circ X_{f} \circ \incl_\mu =  X_{f_\mu} \circ \pi_\mu\, .
        \end{equation}
        \item If $f,h\colon M \to \RR$ are $\Phi$-invariant functions, then $\left\{f,h  \right\}$ is also $G$-invariant and its associated reduced function is given by 
        \begin{equation}
           \big(\left\{f,h  \right\}\big)_\mu = \left\{f_\mu,h_\mu  \right\}_{\omega_\mu}\, ,
        \end{equation}
        where $\left\{\cdot,\cdot  \right\}_{\omega_\mu}$ denotes the Poisson bracket defined by $\omega_\mu$ on $M_\mu$.
    \end{enumerate}

\end{theorem}


\section{Cosymplectic manifolds}\label{sec:cosymplectic}

In this section, some aspects of cosymplectic geometry are briefly reviewed. For more details, refer to \cite{C.D.Y2013, Albert1989, C.L.L1992, d.T1996, d.S1993}.

A \emph{cosymplectic manifold} is a triple $(M, \tau, \omega)$ consisting of a $(2n+1)$-dimensional manifold $M$, a closed one-form $\tau\in \Omega^1(M)$, and a closed 2-form $\omega\in \Omega^2(M)$, such that
\begin{equation}
    \vol\tauomega = \tau \wedge \omega^n
\end{equation}
is a volume form. The pair $(\tau, \omega)$ is called a \emph{cosymplectic structure on $M$}.

The standard example of a cosymplectic manifold is the \emph{extended cotangent bundle} $\RR\times \cT Q$ of a manifold $Q$, with a cosymplectic structure induced by the canonical symplectic form $\omega_Q$ on $\cT Q$. More precisely, let $t$ denote the canonical coordinate of $\RR$, and let $\pi_2\colon \RR \times \cT Q\to \cT Q$ be the canonical projection. Then, $(\RR\times \cT Q, \dd t, \pi_2^\ast \omega_Q)$ is a cosymplectic manifold.

Let $(M, \tau, \omega)$ be a cosymplectic manifold.
Around each point $x\in M$, there exist coordinates $(t, q^1, \ldots, q^n, p_1, \ldots, p_n)$ centered at $x$ such that
\begin{equation}
   \tau = \dd t\, ,\quad  \omega = \dd q^i \wedge \dd p_i\, .
\end{equation}
These coordinates are called \emph{Darboux coordinates} or \emph{canonical coordinates}. 
The map $\flat\tauomega\colon \T M \to \cT M$ given by
\begin{equation}
    \flat\tauomega\colon v \mapsto \contr{v} \omega + (\contr{v}\tau) \tau
\end{equation}
is a vector bundle isomorphism. It can be extended to a $\Cinfty(M)$-module isomorphism $\flat\tauomega\colon\X (M) \to \Omega^1(M)$. Its inverse is denoted by $\sharp\tauomega$.
There is a distinguished vector field $\Reeb\in \X(M)$ such that $\flat\tauomega(\Reeb) = \tau$, that is,
\begin{equation}
    \contr{\Reeb} \tau = \tau\,, \quad \contr{\Reeb} \omega = 0\, .
\end{equation}
Given a function $f\in \Cinfty(M)$, its \emph{Hamiltonian vector field with respect to $(\tau, \omega)$} is the vector field $X_f\in \X(M)$ such that
\begin{equation}\label{eq:Ham_vf_cosymp}
    \flat\tauomega (X_f) = \dd f - \Reeb(f) \tau\, ,
\end{equation}
and its \emph{evolution vector field} is $\evol_f = X_f + \Reeb$. In Darboux coordinates,
\begin{equation}
    \Reeb = \parder{}{t}\, ,\quad X_f =  \frac{\partial f}{\partial p_i} \frac{\partial }{\partial q^i} - \frac{\partial f}{\partial q^i}\frac{\partial }{\partial p_i}\, .
 \end{equation}

Equation~\eqref{eq:Ham_vf_cosymp} is equivalent to 
\begin{equation}
    \contr{X_f} \tau = 0\, ,\quad \contr{X_f} \omega = \dd f - \Reeb(f) \tau\, ,
\end{equation}
and thus
\begin{equation}
    \contr{\evol_f} \tau = 1\, ,\quad \contr{\evol_f} \omega = \dd f - \Reeb(f) \tau\, .
\end{equation}
Hence, both Hamiltonian and evolution vector fields preserve the volume form, that is,
\begin{equation}
    \liedv{X_f} \vol\tauomega = \liedv{\evol_f} \vol\tauomega = 0\, .
\end{equation}

Furthermore, the cosymplectic structure $(\tau, \omega)$ defines a Poisson bracket in $\Cinfty(M)$ by 
\begin{equation}
    \{f, g\}\tauomega = \omega(X_f, X_g)\, ,
\end{equation}
for each pair of functions $f, g\in \Cinfty(M)$. The Poisson structure $\Lambda\tauomega$ defined by $(\tau, \omega)$ is given by
\begin{equation}
    \Lambda\tauomega (\alpha, \beta) = \omega \big(\sharp\tauomega (\alpha), \sharp\tauomega (\beta)\big) \, ,
\end{equation}
for $\alpha, \beta \in \Omega^1(M)$. Observe that the Hamiltonian vector field with respect to $(\tau, \omega)$ coincides with the Hamiltonian vector field with respect to $\Lambda\tauomega$.

\section{Contact manifolds}\label{sec:contact}

In this section, the main aspects of contact manifolds are introduced. For more details, the reader may consult \cite{Arnold1978,d.L2019a,d.L2019b,Geiges2008, L.M1987}. Roughly speaking, a contact distribution is a distribution that is as far as possible from being integrable. The origins of contact geometry can be attributed to Sophus Lie, who introduced the concept of \emph{contact transformations} (Berührungstransformation) in order to study systems of differential equations. The history of contact geometry and topology with an extended list of references may be found in \cite{Geiges2001}.

Let $M$ be a manifold. 
A corank-1 distribution $D$ on $M$ can be locally written as the kernel of a one-form. If it is possible to write globally $D = \ker \alpha$ for some one-form $\alpha\in \Omega^1(M)$, then $D$ is said to be \emph{co-orientable}\footnote{As the name suggests, co-orientability is, in a certain sense, the dual notion of orientability. By means of an auxiliar Riemannian metric $g$ on $M$ (which always exists by partitions of unity), one can construct an orthogonal complement $D^{\perp_g}$ of $D$. Since $D^{\perp_g}$ is a line bundle, it is orientable if and only if there exists a global section $X\in \Gamma(D^{\perp_g})$. If $D$ is globally $\ker \alpha$, then one can define a global section $X = \sharp_g (\alpha)$ of $D^{\perp_g}$. Conversely, if a global section $X$ of $D^{\perp_g}$ exists, one can define a global one-form $\alpha = \flat_g(X)\in \Omega^1(M)$ such that $D=\ker \alpha$. Hence, $D$ is co-orientable if and only if $D^{\perp_g}$ is orientable.}. If the one-form $\alpha$ is fixed, then $D$ is said to be \emph{co-oriented}.

Frobenius theorem implies that a corank-1 distribution $D$ on $M$ locally given by $D=\ker \alpha$ is integrable if and only if 
\begin{equation}
    \alpha \wedge \dd \alpha \equiv 0\, .
\end{equation}

Let $M$ be a $(2n+1)$-dimensional manifold. A \emph{contact structure} (or \emph{contact distribution}) $\cd$ on $M$ is a maximally non-integrable corank-1 distribution $\cd$ on $M$, that is, if $\cd$ is locally given by $\cd = \ker \alpha$, then
\begin{equation}
    \alpha \wedge (\dd \alpha)^n \neq 0\, .
\end{equation}
The pair $(M, \cd)$ is called a \emph{contact manifold}.
If $\cd$ is a co-orientable contact structure, then a one-form $\eta\in \Omega^1(M)$ such that $\cd= \ker \eta$ is called a \emph{contact form}. The pair $(M, \eta)$ is called a \emph{co-oriented contact manifold}. 

Let $(M, \eta)$ be a co-oriented contact manifold. The contact form $\eta$ on $M$ defines a volume form
\begin{equation}
    \vol_\eta = \eta \wedge (\dd \eta)^n \, .
\end{equation}
It also defines a vector bundle isomorphism $\flat_\eta \colon \T M \to \cT M$ by
\begin{equation}
    \flat_\eta (v) = \eta(v) \eta + \contr{v} \dd \eta \, ,
\end{equation}
which induces an isomorphism of $\Cinfty(M)$-modules $\flat_\eta \colon \X(M) \to \Omega^1(M)$. The inverses of both isomorphisms are denoted by $\sharp_\eta$. There exists an unique vector field $\Reeb\in \X(M)$, called the \emph{Reeb vector field}, such that $\flat_\eta(\Reeb) = \eta$, that is,
\begin{equation}
    \contr{\Reeb} \dd \eta = 0\, , \quad \contr{\Reeb} \eta = 1\, .
\end{equation}
Given a function $f\in \Cinfty(M)$, the \emph{Hamiltonian vector field of $f$ with respect to $\eta$} is the vector field $X_f\in \X(M)$ such that
\begin{equation}
    \flat_\eta (X_f) = \dd f - \left(\Reeb f + f\right) \eta\, ,
\end{equation}
or, equivalently,
\begin{equation}\label{eqs:contact_Hamiltonian_vf_liedv}
    \eta(X_f) = -f \, , \quad \liedv{X_f} \eta = - \Reeb (f) \eta\, .
\end{equation}
Observe that, in particular, the Reeb vector field is the Hamiltonian vector field of the function $f\equiv -1$. 

Since $\flat_\eta\colon \T M \to \cT M$ is an isomorphism, the tangent bundle of $M$ can be written as the following Whitney sum decomposition:
\begin{equation}
    \T M = \ker \eta \oplus \ker \dd \eta\, .
\end{equation}

Around each point $x\in M$ of a co-oriented contact manifold $(M, \eta)$, there exist coordinates $(q^1, \ldots, q^n, p_1, \ldots, p_n, z)$ centered at $x$ in which the contact form reads
\begin{equation}
    \eta = \dd z - p_i \dd q^i \, .
\end{equation}
These coordinates are called \emph{Darboux coordinates} or \emph{canonical coordinates}. In these coordinates,
the Reeb vector field is written as
\begin{equation}
    \Reeb = \frac{\partial}{\partial z}\, ,
\end{equation}
and the Hamiltonian vector field of $f$ is
\begin{equation}
    X_f = \frac{\partial f}{\partial p_i} \frac{\partial}{\partial q^i} - \left( \frac{\partial f}{\partial q^i} + p_i \frac{\partial f}{\partial z}\right) \frac{\partial}{\partial p_i} + \left(  p_i \frac{\partial f}{\partial p_i} - f\right)\frac{\partial}{\partial z}\, . 
\end{equation}

    Let $(M_1, \eta_1)$ and $(M_2, \eta_2)$ be co-oriented contact manifolds. A diffeomorphism $\Phi\colon M_1 \to M_2$ is called\footnote{In the contact topology community (see, for instance, \cite{Geiges2008}), they usually call contactomorphism what in this dissertation is referred to as a conformal contactomorphism.} a \emph{$f$-conformal contactomorphism} if $\Phi^\ast\eta_2 = f \eta_1$ for some nowhere-vanishing function $f$ on $M_1$ called the \emph{conformal factor}. A \emph{(strict) contactomorphism} is a conformal contactomorphism with conformal factor $f\equiv 1$. 

   Let $(M, \eta)$ be a co-oriented contact manifold. An \emph{infinitesimal conformal (respectively, strict) contactomorphism} is a vector field $X\in \mathfrak{X}(M)$ whose flow is a one-parameter group of conformal (respectively, strict) contactomorphisms.
In light of the right-hand side equation in \eqref{eqs:contact_Hamiltonian_vf_liedv}, every Hamiltonian vector field is an infinitesimal conformal contactomorphism. Conversely, if $X\in \X(M)$ is a conformal contactomorphism, then it is the Hamiltonian vector field of $f=-\eta(X)$. Furthermore, $X_f$ is an infinitesimal strict contactomorphism if and only if $\Reeb(f) = 0$.

Observe that, unlike in symplectic and cosymplectic manifolds, Hamiltonian vector fields do not preserve the volume form. Indeed,
\begin{equation}
    \liedv{X_f} \vol_\eta = -(n+1) \Reeb (f) \vol_\eta\, ,
\end{equation}
for any $f\in \Cinfty(M)$.

Every co-oriented contact manifold $(M, \eta)$ has a Jacobi structure $(\Lambda, E)$, where
\begin{equation}
    \Lambda(\alpha, \beta) = - \dd \eta \big( \sharp_\eta (\alpha), \sharp_\eta (\beta) \big)\, , \quad E = - \Reeb\, ,
\end{equation}
for any pair of one-forms $\alpha, \beta \in \Omega^1(M)$.
Hence, the associated Jacobi bracket $\left\{\cdot, \cdot  \right\}$ is given by 
 \begin{equation}
   \left\{f, g  \right\} = -\dd \eta \big(\sharp_\eta (\dd f), \sharp_\eta (\dd g)\big) - f \Reeb(g) +g\Reeb(f)\, ,
 \end{equation}
 for any pair of functions $f, g \in \Cinfty(M)$. 
Moreover, this Jacobi structure induces the objects summarized in the following remark (\emph{cf.}~\Cref{remark:Jacobi_orthogonal} for the analogous results on an arbitrary Jacobi manifold). 

 \begin{remark}\label{remark:Jacobi_orthogonal_contact}
    Let $(M, \eta)$ be a co-oriented contact manifold with associated Jacobi structure $(\Lambda, -\Reeb)$. Then:
    \begin{enumerate}
        \item The bivector $\Lambda$ induces a $\Cinfty(M)$-module morphism $\lsharp\colon \Omega^1(M)\to \X(M)$ given by 
        \begin{equation}
            \lsharp(\alpha) = \Lambda(\alpha, \cdot) = \sharp_\eta(\alpha) - \alpha(\Reeb) \Reeb \, .    
        \end{equation}
        \item The morphism $\lsharp$ is not an isomorphism. As a matter of fact,
        \begin{equation}
            \ker \lsharp = \langle \eta \rangle\, , \quad \Ima \lsharp = \cd = \ker \eta\, .
        \end{equation}
        \item Given a function $f\in \Cinfty(M)$, its Hamiltonian vector field with respect to $\eta$ coincides with its Hamiltonian vector field with respect to $(\Lambda, -\Reeb)$. Indeed,
        \begin{equation}\label{eq:Hamiltonian_vf_Jacobi_contact}
            X_f = \lsharp(\dd f) - f \Reeb = \sharp_\eta(\dd f) - (\Reeb f + f) \Reeb \, .
        \end{equation}
        \item The characteristic distribution of $(M, \Lambda, E)$ is the complete tangent bundle, namely, $C = \T M$.
        \item If $D$ is a distribution on $M$, then its Jacobi orthogonal complement is the distribution $D^{\perp_\Lambda} = \lsharp(D^\circ)$, where $D^\circ$ denotes the annihilator of $D$.
        \item A submanifold $N\hookrightarrow M$ is called 
        \begin{enumerate}
            \item \emph{isotropic} if $\T N \subseteq \T N^{\perp_\Lambda}$,
            \item \emph{coisotropic} if $\T N^{\perp_\Lambda} \subseteq \T N$,
            \item \emph{Legendrian} if $\T N^{\perp_\Lambda} = \T N$.
        \end{enumerate}
    \end{enumerate}
\end{remark}

Equivalently, $N$ is:
\begin{enumerate}
    \item isotropic if and only if $\restr{\eta}{\T N} = 0$,
    \item Legendrian if and only if it is isotropic and $n$-dimensional.
\end{enumerate}
\begin{remark}\label{remark:Jacobi_bracket_contact}
The Jacobi bracket defined by $\eta$ can also be written as
\begin{equation} \label{eq:Jacobi_bracket_contact_vfs}
    \{f,g\}  
        = X_f(g) + g\Reeb(f) \, ,
\end{equation}
or 
\begin{equation}
    \{f,g\} = - \eta\left([X_f, X_g]\right)\, ,
\end{equation}
for any $f, g\in \Cinfty(M)$.
The latter equation implies that the map $f \mapsto X_f$ is a Lie algebra anti-isomorphism between $\Cinfty(M)$ with the Jacobi bracket and the set of infinitesimal conformal contactomorphisms with the Lie bracket. Its inverse is given by $X \mapsto - \eta(X)$. The restriction of these map is also an anti-isomorphism between the set of infinitesimal strict contactomorphisms and the set of smooth functions preserved by the Reeb flow, that is, $\{f\in \Cinfty(M) \mid \Reeb (f) = 0\}$.
\end{remark}

\begin{remark}\label{remark:conformal_equivalence_contact}
    Given a co-orientable contact manifold $(M, \cdist)$, the set of contact forms $\eta$ on $M$ such that $\ker \eta =\cdist$ form an equivalence class called the \emph{conformal class of $\eta$}. It is clear that $\eta$ and $\tilde{\eta}$ belong to the same conformal class if and only if $\tilde{\eta} = a \eta$ for some nowhere-vanishing function $a\in \Cinfty(M)$. Thus, (infinitesimal) conformal contactomorphisms are diffeomorphisms (vector fields) preserving the conformal class. Furthermore, if two contact forms belong to the same conformal class then their Jacobi structures are conformally equivalent. More specifically, if $\tilde{\eta} = a \eta$, then
    \begin{enumerate}
        \item The Reeb vector field $\Reeb_{\tilde{\eta}}$ of $\tilde{\eta}$ is the Hamiltonian vector field of $-1/a$ with respect to $\eta$, namely,  $\Reeb_{\tilde{\eta}} =X_{-1/a}$.
        \item The Jacobi structure defined by $\tilde{\eta}$ is $(\Lambda_{\tilde{\eta}}, - \Reeb_{\tilde{\eta}})$, where $\Lambda_{\tilde{\eta}} = 1/a\, \Lambda_{\eta}$.
        \item The Jacobi brackets $\{\cdot, \cdot\}_{\eta}$ and $\{\cdot, \cdot\}_{\tilde{\eta}}$ defined by $\eta$ and $\tilde{\eta}$ are related by $\jacBr{f,g}_{\bar{\eta}} = a \jacBr{f/a, g/a}_{\eta}$.
    \end{enumerate}
\end{remark}

The paradigmatic example of a co-oriented contact manifold is the extended cotangent bundle $\cT Q \times \RR$ of a manifold $Q$, with a contact form induced by the tautological one-form $\theta_Q$ of $\cT Q$. More precisely, if $z$ denotes the canonical coordinate of $\RR$ and $\pi_1 \colon \cT Q \times \RR \to \cT Q$ the natural projection, then $\eta_Q = \dd z - \pi_1^\ast \theta_Q$ is a contact form on $\cT Q \times \RR$, called the \emph{canonical contact form}. If $(q^i)$ are coordinates on $Q$, then the induced bundle coordinates $(q^i, p_i, z)$ on $\cT Q\times \RR$ are Darboux coordinates for $\eta_Q$, namely, 
\begin{equation}
    \eta_Q = \dd z - p_i \dd q^i\, , \quad \Reeb = \parder{}{z}\, .
\end{equation}

\subsection{Symplectization}\label{subsec:symplectization}

In this subsection, some results concerning the symplectization of contact manifolds are concisely reviewed. For more details, see \cite{Lainz2022, G.G2022a, L.M1987}.

\begin{definition}
    Given two exact symplectic manifolds $(M_1, \theta_1)$ and $(M_2, \theta_2)$, a \emph{homogeneous symplectomorphism} is a diffeomorphism $\Phi\colon M_1 \to M_2$ such that $\Phi^\ast \theta_2 = \theta_1$. 
    
    Given a symplectic manifold $(M, \theta)$, an \emph{infinitesimal homogeneous symplectomorphism} is a vector field $X\in \X(M)$ whose flow is a one-parameter group of homogenous symplectomorphisms or, equivalently, $\liedv{X}\theta = 0$.
\end{definition}

Clearly, every (infinitesimal) homogeneous symplectomorphism is an (infinitesimal) symplectomorphism. Moreover, if $\Delta_1$ and $\Delta_2$ are the Liouville vector fields of $(M_1, \theta_1)$ and $(M_2, \theta_2)$, and $\Phi\colon M_1 \to M_2$ is a homogeneous symplectomorphism, then $\Phi_\ast \Delta_1 = \Delta_2$.

\begin{definition}
    Let $\cdist$ be a contact distribution on a manifold $M$. Consider an exact symplectic manifold $(M^\Sigma, \theta)$ and a (locally trivial) fiber bundle $\Sigma\colon M^\Sigma \to M$. This fiber bundle is called a \emph{symplectization} if
    \begin{equation}\label{eq:def_symplectization}
        \Sigma_\ast \Ldist = \cdist\, ,
    \end{equation}
    outside the singular points, where $\mathfrak L = \ker \theta$ is called the \emph{Liouville distribution}.
\end{definition}


\begin{proposition}
    Let $(M, \eta)$ and $(M^\Sigma, \theta)$ be a contact manifold and an exact symplectic manifold, respectively.
    A (locally trivial) fiber bundle $\Sigma\colon M^\Sigma \to M$ is a symplectization if, and only if, there exists a nowhere-vanishing function $\sigma\colon M^\Sigma \to \R$ such that
    \begin{equation}
        \sigma \left(\Sigma^\ast \eta\right) = \theta\, .
    \end{equation}
    The function $\sigma$ is called the \emph{conformal factor} of $\Sigma$.
\end{proposition}

\begin{proof}
    Since $\cdist$ and $\Sigma_\ast \Ldist$ are both distributions of corank $1$ on $M$, in order to ensure that $\Sigma$ is a symplectization it is enough to verify that $\Sigma_\ast \Ldist \subseteq \cdist$.
    By taking duals, this condition is equivalent to 
    \begin{equation}
        \Sigma^\ast (\cdist^\circ) \subseteq \Ldist^\circ\, ,
    \end{equation}
    but $\cdist^\circ= \langle\eta\rangle$ and $\Ldist^\circ = \langle\theta\rangle$. Thus,
    \begin{equation}\label{eq_symp}
        \sigma \left(\Sigma^\ast \eta\right) = \theta\, , 
    \end{equation}
    for some nowhere-vanishing function $\sigma$. 
    

\end{proof}




\begin{proposition}
    Let $(M, \eta)$ be a co-oriented contact manifold, $(M^\Sigma, \theta)$ an exact symplectic manifold, and $\Sigma \colon M^\Sigma \to M$ a symplectization with conformal factor $\sigma$. Then, $\ker \T \Sigma = \langle \Delta\rangle$, where $\Delta$ denotes the Liouville vector field of $(M^\Sigma, \theta)$. Moreover, the conformal factor is a homogeneous function of degree $1$, namely, $\Delta(\sigma) = \sigma$. 
\end{proposition}

\begin{proof}
    The fact that $\cdist = \ker \eta \subseteq \Ima T\Sigma$ and $\Ima \T \Sigma$ being an integrable distribution imply that $\Ima \T \Sigma \supseteq [\cdist, \cdist] = \T M$. Taking into account that the rank of $\Ima \T \Sigma$ is $2n+1$, the rank of $\ker \T \Sigma$ must be $1$. Therefore, the proof is completed by showing that $\Delta$ belongs to the distribution $\ker \T \Sigma$.

    Observe that $\contr{\Delta} \theta = 0$, and thus $\contr{\Delta} (\Sigma^\ast \eta) = 0$. On the other hand,
    \begin{equation}\label{eq:eta_contr_Delta}
        \sigma \left(\Sigma^\ast \eta\right) = \contr{\Delta} \dd \big(\sigma \left(\Sigma^\ast \eta\right)\big) 
        = \Delta(\sigma) (\Sigma^\ast \eta) + \sigma \contr{\Delta} (\Sigma^\ast \dd \eta)
    \end{equation}
    Let $\Reeb$ be the Reeb vector field of $(M, \eta)$. Consider a vector field $\tilde{\Reeb}\in \X(M^\Sigma)$ such that $\Sigma_\ast \tilde{\Reeb} = \Reeb$. Then, 
    \begin{equation}
        \contr{\tilde{\Reeb}} (\Sigma^\ast \eta) = \Sigma^\ast (\contr{\Reeb} \eta) = 1  \, , \quad
       \contr{\tilde{\Reeb}} (\Sigma^\ast \dd \eta) = \Sigma^\ast (\contr{\Reeb} \dd \eta) = 0 \, . 
    \end{equation}
    Contracting both sides of equation \eqref{eq:eta_contr_Delta} with $\tilde{\Reeb}$ yields $\Delta(\sigma) = \sigma$, which in turn implies that 
    \begin{equation}
        0 = \contr{\Delta} (\Sigma^\ast \dd \eta) = \Sigma^\ast \left(\contr{\Sigma_\ast \Delta} \dd \eta\right)\, .
    \end{equation}
    Since $\Sigma$ is a submersion, $\contr{\Sigma_\ast \Delta} \dd \eta = 0$. Therefore, $\flat_\eta (\Sigma_\ast \Delta) = 0$, and hence $\Sigma_\ast \Delta = 0$.
\end{proof}

\begin{theorem}[Simplectization of contactomorphisms]\label{thm:symplectization_contactomorphism}

    Let $(M_1, \eta_1)$ and $(M_2, \eta_2)$ be co-oriented contact manifolds. 
    Consider the symplectizations $\Sigma_1 \colon M_1^\Sigma \to M_1$ and $\Sigma_2\colon M_2^\Sigma \to M_2$ with symplectic potentials $\theta_1 \in \Omega^1(M_1^\Sigma)$ and $\theta_2 \in \Omega^1(M_2^\Sigma)$, and conformal factors $\sigma_1\in \Cinfty(M_1^\Sigma)$ and $\sigma_2\in \Cinfty(M_2^\Sigma)$, namely, $\sigma_1 \Sigma_1^\ast \eta_1 = \theta_1$ and $\sigma_2 \Sigma_2^\ast \eta_2 = \theta_2$.
    
    If $F^\Sigma \colon M_1^\Sigma \to M_2^\Sigma$ is a homogeneous symplectomorphism, then it can be projected onto a map $F\colon M_1 \to M_2$ making the diagram
    \begin{equation}\label{eq:symplectization_morphisms_diagram}
        \begin{tikzcd}
            M_1^\Sigma \arrow[r, "F^\Sigma"] \arrow[d, "\Sigma_1"] & M_2^\Sigma \arrow[d, "\Sigma_2"] \\
            M_1 \arrow[r, "F"]                                   & M_2                           
            \end{tikzcd}
    \end{equation}
    commutative. The map $F\colon M_1 \to M_2$ is a contactomorphism. 
    
    Conversely, if $F\colon M_1 \to M_2$ is a contactomorphism with conformal factor $f\in \Cinfty (M_1)$ (that is, $F^\ast \eta_2 = f \eta_1$), then $F^\Sigma \colon M_1^\Sigma \to M_2^\Sigma$ is a homogeneous symplectomorphism if and only if 
    \begin{equation}
        {(F^\Sigma)}^\ast(\sigma_2) = \frac{\sigma_1}{\Sigma_1^\ast(f)}\, .
    \end{equation}
    In that case, the map $F^\Sigma$ is called the \emph{symplectization} of $F$ and the diagram~\eqref{eq:symplectization_morphisms_diagram} commutes.
    

    Moreover, the symplectization is functorial. More specifically, if $(M_3, \eta_3)$ is a co-oriented contact manifold, $(M_3^\Sigma, \theta_3)$ an exact symplectic manifold, and $G\colon M_2 \to M_3$ a contactomorphism, then the symplectization $G^\Sigma\colon M_2^\Sigma \to M_3^\Sigma$ of $G$ verifies ${(G \circ F)}^\Sigma = G^\Sigma \circ F^\Sigma$. Hence, the following diagram is commutative:
    \begin{equation}\label{eq:diagram_functioral_symplectization}
        \begin{tikzcd}
            M^\Sigma_1 \arrow[r, "F^\Sigma"'] \arrow[d, "\Sigma_1"] \arrow[rr, "(G \circ F)^\Sigma", bend left] & M^\Sigma_2 \arrow[d, "\Sigma_2"] \arrow[r, "G^\Sigma"'] & M^\Sigma_3 \arrow[d, "\Sigma_3"] \\
            M_1 \arrow[r, "F"] \arrow[rr, "G \circ F"', bend right]                                           & M_2 \arrow[r, "G"]                                    & M_3                 
            \end{tikzcd}
    \end{equation}
\end{theorem}

\begin{proof}

Let $\Delta_1$ and $\Delta_2$ denote the Liouville vector fields of $(M_1^\Sigma, \theta_1)$ and $(M_2^\Sigma, \theta_2)$, respectively. Let $\cdist_1 = \ker \eta_1$ and $\cdist_2 = \ker \eta_2$ denote the contact distributions, and $\Ldist_1 = \ker \theta_1$ and $\Ldist_2 = \ker \theta_2$ the Liouville distributions.

Suppose that $F^\Sigma\colon M_1^\Sigma\to M_2^\Sigma$ is a homogeneous symplectomorphism. Then, ${(F^\Sigma)}_\ast\Delta_1 = \Delta_2$, $\ker {(\Sigma_1)}_\ast = \gen{\Delta_1}$ and $\ker {(\Sigma_1)}_\ast = \gen{\Delta_1}$. This implies that the map $F^\Sigma$ projects onto a map $F$ such that the diagram \eqref{eq:symplectization_morphisms_diagram} commutes. Using the commutative diagram, one obtains
\begin{equation}
\begin{aligned}
     F_\ast \cdist_1 & = {(\Sigma_2)_\ast \circ (F^\Sigma)_\ast \circ ((\Sigma_1)_\ast)^{-1}} (\cdist_1) \\
     & = (\Sigma_2)_\ast \circ (F^\Sigma)_\ast\, \Ldist_1 = 
    (\Sigma_2)_\ast\, \Ldist_2 = \cdist_2\, .
\end{aligned}
\end{equation}
Thus, $F$ is a contactomorphism.

Conversely, suppose that $F\colon M_1 \to M_2$ is a contactomorphism with conformal factor $f$. 
Then,
\begin{equation}
    \begin{aligned}
        {(F^\Sigma)}^\ast \theta_2 &= {(F^\Sigma)}^\ast (\sigma_2 {(\Sigma_2)}^\ast \eta_2) 
        ={(F^\Sigma)}^\ast \left[\sigma_2 {\big(F\circ \Sigma_1 \circ (F^\Sigma)^{-1}\big)}^\ast \eta_2\right] 
        \\ & 
        = {(F^\Sigma)}^\ast(\sigma_2) (\Sigma_1^\ast F^\ast \eta_2) 
        =
        {(F^\Sigma)}^\ast(\sigma_2) (\Sigma_1^\ast (f \eta_1)) \\ &=
        {(F^\Sigma)}^\ast(\sigma_2) \Sigma_1^\ast(f) \frac{1}{\sigma_1} \theta_1\, ,
    \end{aligned}
\end{equation}
and hence ${(F^\Sigma)}^\ast \theta_2 = \theta_1$ if and only if
\begin{equation}
    {(F^\Sigma)}^\ast(\sigma_2) = \frac{\sigma_1}{\Sigma_1^\ast(f)}\, .
\end{equation}
This implies that each contactomorphism has a unique symplectization.



The commutativity of the diagram \eqref{eq:diagram_functioral_symplectization} follows from the fact that the composition of contactomorphisms (respectively, exact symplectomorphisms) is a contactomorphism (respectively, exact symplectomorphism), and the uniqueness of the symplectization of each contactomorphism.
\end{proof}



The fact that the symplectization preserves the composition of morphisms permits to symplectize the vector fields. 
Let $X\in \X(M)$ be an infinitesimal contactomorphism, and let $\phi_t$ be its flow. Notice that
\begin{equation}
 {\phi}_{t+r}^\Sigma = {(\phi_r \circ \phi_t)}^\Sigma =  {\phi}^\Sigma_t \circ {\phi}^\Sigma_r\, ,
\end{equation}
and thus ${\phi}_t^\Sigma$ is also a flow. Since it is the symplectization of contactomorphisms, $\phi_t^\Sigma$ is a flow of homogeneous symplectomorphisms. Hence, its infinitesimal generator, ${X}^\Sigma$ is an infinitesimal homogeneous symplectomorphism. 
\begin{theorem}[Symplectization of vector fields]\label{thm:symp_fields}
    Let $\cdist$ be a contact distribution on the manifold $M$. Let $(M^\Sigma, \theta)$ be an exact symplectic manifold. A symplectization $\Sigma\colon M^\Sigma\to M$ provides a bijection between infinitesimal contactomorphisms $X$ on $M$ and infinitesimal homogeneous symplectomorphisms $X^\Sigma$ on $M^\Sigma$ such that $X$ and $X^\Sigma$ are related by the bijection if and only if
    \begin{equation}
        \Sigma_\ast X^\Sigma = X\, .
    \end{equation}
    If $(M, \eta)$ is a co-oriented contact manifold with $\ker \eta = \cdist$, the vector field $X^\Sigma$ can also be characterized as the one that projects onto $X$ and satisfies
    \begin{equation}
        X^\Sigma(\sigma) = -a_X \sigma\, ,
    \end{equation}
    where $a_X$ is the conformal factor of $X$, namely, $\liedv{X}\eta = a_X\eta$.
    It can also be written as
    \begin{equation}
        X^\Sigma = X^\sigma - a_X  \Delta\, ,
    \end{equation}
    where $X^\sigma$ is the vector field that projects onto $X$ and satisfies $X^\sigma(\sigma) = 0$.
    In particular, for the Reeb vector field $\Reeb \in \X(M)$,
    \begin{equation}
        {\Reeb}^\Sigma = \Reeb^\sigma\, .
    \end{equation}
    Moreover, the symplectization is a Lie algebra isomorphism, that is,
    \begin{equation}\label{eq:symplectization_fields}
        \lieBr{X,Y}^\Sigma = \lieBr{X^\Sigma, Y^\Sigma}\, , 
    \end{equation} 
    for any pair of infinitesimal contactomorphisms $X,Y\in \X(M)$.
\end{theorem}

In order to proof this Theorem, as well as other subsequent results, the following Lemma is required (see \cite[Lemma~6.19.2]{K.S.M1993}).
\begin{lemma} \label{lemma:Kolar}
    Consider the category of all $n$-dimensional smooth manifolds with local diffeomorphisms between them.
    Let $F$ be a vector bundle functor, that is, a functor which associates a vector bundle $\pi_M\colon F(M)\to M$ to each $n$-dimensional manifold $M$ and a vector bundle homomorphism to each local diffeomorphism $f\colon M \to N$. Given a vector field $X \in \X(M)$ with flow $\phi_t$, there is a vector field $X^F\in \X(F(M))$, given by
    \begin{equation}
        X^F = \restr{\frac{\dd}{\dd t}}{t=0} F(\phi_t)\, .
    \end{equation}
    Then, 
    \begin{equation}
        \lieBr{X, Y}^F = \lieBr{X^F, Y^F}\, , 
    \end{equation}
    for any $X, Y \in \X(M)$.
\end{lemma}

\begin{proof}[Proof of \Cref{thm:symp_fields}]
   By \Cref{thm:symplectization_contactomorphism}, given the flow $\phi$ of an infinitesimal contactomorphism $X$ there exists a unique flow $\phi^\Sigma$ which is made of exact symplectomorphisms and projects onto $\phi$. Hence, there exists a unique infinitesimal homogeneous symplectomorphisms $X^\Sigma$, which is the infinitesimal generator of $\phi^\Sigma$. 
    Similarly, one can see that every infinitesimal homogeneous symplectomorphism projects onto an infinitesimal contactomorphism.

    Consider the contact form $\eta$ on $M$ with conformal factor $\sigma$ (that is, $\sigma \left(\Sigma^\ast \eta\right) = \theta$) and let $X$ be an infinitesimal contactomorphism with conformal factor $a_X$ (that is, $\liedv{X} \eta= a_X \eta$).
    Then, since $X^\Sigma$ is an infinitesimal homogeneous symplectomorphism, one has
    \begin{equation}
        0 = \liedv{X^\Sigma} \theta = \liedv{X^\Sigma}\big(\sigma \left(\Sigma^\ast\eta\right)\big) = 
        ({X^\Sigma}(\sigma)  + \sigma a_X) \eta\, ,
    \end{equation}
   which implies that ${X^\Sigma}(\sigma)  = - \sigma a_X$.

    Since $\ker \T \Sigma = \gen{\Delta}$, the vector field $X^\Sigma$ is of the form $X^\sigma + b\Delta$ for some function $b\in \Cinfty(M^\Sigma)$. Thus,
    \begin{equation}
        -a_X \sigma  = X^\Sigma(\sigma) = b \Delta (\sigma) = b \sigma\, ,
    \end{equation}
    since $\sigma$ is homogeneous of degree $1$. Hence, $b = -a_X  b$.
     
    Finally, it is clear that the symplectization is a vector bundle functor. Hence, by \Cref{lemma:Kolar}, equation~\eqref{eq:symplectization_fields} holds.
\end{proof}

Since the contact Hamiltonian vector fields and the functions on $(M, \eta)$ on one side, and the homogeneous Hamiltonian vector fields and the homogeneous functions on $(M^\Sigma, \theta)$ on the other side, are in one-to-one correspondence, there is a bijection between both sets of functions.

\begin{theorem}\label{thm:symp_functions}
    Let $(M, \eta)$ be a co-oriented contact manifold and $(M^\Sigma, \theta)$ an exact symplectic manifold.
    For each $f\in \Cinfty(M)$, let $X_f\in \X(M)$ denote the Hamiltonian vector field with respect to $\eta$. 
    For each $f^\Sigma\in \Cinfty(M^\Sigma)$, let $X_{f^\Sigma}\in \X(M^\Sigma)$ denote the Hamiltonian vector field with respect to $\theta$. 
    Given a symplectization $\Sigma\colon M^\Sigma \to M$ with conformal factor $\sigma$, there is a bijection between functions $f$ on $M$ and homogeneous functions of degree 1 $f^\Sigma$ on $M^\Sigma$ such that
    \begin{equation}
        \Sigma_\ast \left(X_{f^\Sigma}\right) = X_f\, .
    \end{equation}
    This bijection is given by
    \begin{equation}
        f^\Sigma = -\sigma \left(\Sigma^\ast f\right)\, .
    \end{equation}
    Moreover, one has
    \begin{equation}
        \left\{f^\Sigma, g^\Sigma \right\}_\theta = \left\{f, g \right\}^\Sigma_\eta\, ,
    \end{equation}
    where $\{\cdot, \cdot\}_\theta$ and $\{\cdot, \cdot\}_\eta$ denote the Poisson bracket of $(M^\Sigma, \theta)$ and the Jacobi bracket of $(M, \eta)$, respectively. Note that this implies that $\Sigma$ is a conformal Jacobi morphism, that is, $\left\{\sigma \left(\Sigma^\ast f\right), \sigma \left(\Sigma^\ast g\right)\right\}_\theta = \sigma \left(\Sigma^\ast \{f, g\}_\eta\right)$.
    
\end{theorem}

\begin{proof}
    Indeed,
    \begin{equation}
    \begin{aligned}
        f^\Sigma & = \theta(X_{f^\Sigma}) = \theta({(X_f)^\Sigma })
        = \sigma \left(\Sigma^\ast(\eta)((X_f)^\Sigma )\right) \\
        & = \sigma \left(\Sigma^\ast(\eta(X_{f}))\right) = -\sigma \left(\Sigma^\ast f\right)\, .
    \end{aligned}
    \end{equation}
    The other claims follow from \Cref{remark:Jacobi_bracket_contact}, together with equation~\eqref{eq:Lie_algebra_antihomomorphism_symp}, and the fact that the symplectization preserves the Lie brackets (equation~\eqref{eq:symplectization_fields}).
\end{proof}


\begin{remark}
    Note that a function $\hat{f}: M^\Sigma \to \R$ is homogeneous of degree $0$ if and only if one can write $\hat{f} = \Sigma^\ast f$ for some $f:M \to \R$. This follows from \Cref{thm:symp_functions} and the fact that a function $\hat{f}$ is homogeneous of degree $0$ if and only if the function $-\sigma \hat{f}$ is homogeneous of degree $1$.
\end{remark}

\section{Cocontact manifolds}\label{sec:cocontact}

Cocontact geometry was defined by de León, Gaset, Gràcia, Muñoz-Lecanda and Rivas in \cite{d.G.G+2022}. It is a geometric structure which combines, in a certain manner, contact and cosymplectic structures. In fact, a natural way to generate a cocontact manifold is as the product of a contact manifold or a cosymplectic manifold with a real line (see \Cref{example:cocontact_contact,example:cocontact_cosymplectic}).

    A \emph{cocontact manifold} is a triple $(M,\tau,\eta)$, where $M$ is a $(2n+2)$-manifold, and $\tau$ and $\eta$ are one-forms on $M$ such that $\dd\tau = 0$ and 
    \begin{equation}
        \vol\taueta = \tau\wedge\eta\wedge(\dd\eta)^n
    \end{equation}
    is a volume form.
    The pair $(\tau, \eta)$ is called a \emph{cocontact structure} on $M$.

Given a cocontact manifold $(M,\tau,\eta)$, the distribution $\cd = \ker\eta$ is called the \emph{horizontal} or \emph{contact distribution}. It is worth noting that this distribution has corank one and is maximally non-integrable.

\begin{example}\label{example:cocontact_contact}
	Let $(P,\eta_0)$ be a co-oriented contact manifold, and consider the product manifold $M = \RR\times P$ with the canonical projection $\pi_2 \colon \RR \times P \to P$. Denoting by $t$ the canonical coordinate in $\RR$ and by $\eta=\pi_2^\ast \eta_0$ the pullback of $\eta_0$ to $M$, the pair $(\dd t, \eta)$ is a cocontact structure on $M$.
\end{example}

\begin{example}\label{example:cocontact_cosymplectic}
	Let $(P,\tau,\omega)$ be a cosymplectic manifold such that $\omega = - \dd \theta$ for some $\theta\in \Omega^1(P)$, and consider the product manifold $M = P\times\RR$ with the canonical projection $\pi_1 \colon P \times \RR \to P$. Denoting by $z$ the canonical coordinate in $\RR$, define the one-form $\eta = \dd z - \pi_1^\ast \theta$. Then, $(\tau, \eta)$ is a cocontact structure on $M = P\times\RR$.
\end{example}

    The quintessential cocontact manifold is $\RR \times \cT Q \times \RR$, for some manifold $Q$, with a cocontact structure induced by the tautological one-form $\theta_Q$ of $\cT Q$.
    Consider the following product manifolds and canonical projections:
	\begin{equation}\label{eq:digram_projections_cocontact}
		\begin{tikzcd}
			& \RR\times\cT Q\times\RR \arrow[dl, swap, "\rho_1"] \arrow[dr, "\rho_2"] \arrow[dd, "\pi"] & \\
			\RR\times\cT Q \arrow[dr, swap, "\pi_2"] & & \cT Q\times\RR \arrow[dl, "\pi_1"] \\
			& \cT Q &
		\end{tikzcd}
	\end{equation}
    Let $\theta_1 = \pi_1^\ast\theta_Q$ and $\theta_2 = \pi_2^\ast\theta_Q$.
    As it was mentioned in previous sections, $(\dd t, \theta_2)$ is a cosymplectic structure on $\RR\times\cT Q$, and $\eta_1 = \dd z - \theta_1$ is a contact form on $\cT Q\times\RR$. Additionally, consider the one-forms $\theta = \rho_1^\ast\theta_2 = \rho_2^\ast\theta_1 = \pi^\ast\theta_Q$ and $\eta = \dd z - \theta$ on $\RR\times\cT Q\times\RR$. Then, $(\dd t, \eta)$ is a cocontact structure on $\RR\times\cT Q\times\RR$. 
    
    Let $(q^i)$ be local coordinates in $Q$, and $(t, q^i, p_i, z)$ the induced bundle coordinates in $\RR \times \cT Q\times \RR$, namely,
    \begin{equation}
    \begin{aligned}
        & \rho_1 (t, q^i, p_i, z) = (t, q^i, p_i)\, , \\
        & \rho_2 (t, q^i, p_i, z) = (q^i, p_i, z)\, , \\
        & \pi (t, q^i, p_i, z) = \pi_2(t, q^i, p_i) = \pi_1(q^i, p_i, z) = (q^i, p_i)\, .
    \end{aligned}
    \end{equation}
    The local expression of the one-form $\eta$ is then
    \begin{equation}
        \eta = \dd z - p_i\dd q^i\,.
    \end{equation}

In fact, any cocontact structure is locally of this form. Indeed, around every point $p\in M$ of a cocontact manifold $(M,\tau,\eta)$, there exists a local chart $(U; t, q^i, p_i, z)$ of \emph{canonical} or \emph{Darboux coordinates} such that
\begin{equation}
    \restr{\tau}{U} = \dd t\,, \quad \restr{\eta}{U} = \dd z - p_i\dd q^i\,.
\end{equation}

A cocontact structure $(\tau, \eta)$ on a manifold $M$ defines an isomorphism of vector bundles $\flat\taueta\colon \T M \to \cT M$ given by
\begin{equation}
    \flat\taueta\colon v\mapsto (\contr{v}\tau)\tau + \contr{v}\dd\eta + (\contr{v}\eta)\eta\, . 
\end{equation}
This map is naturally extended to an isomorphism of $\Cinfty(M)$-modules $\flat\taueta\colon \X(M) \to \Omega^1(M)$. Its inverse is denoted by $\sharp\taueta$. 
There exist two distinguished vector fields $\Rt$, $\Rz\in \X(M)$ such that 
\begin{equation}
    \tau = \flat\taueta(\Rt)\,, \quad \eta = \flat\taueta (\Rz)\, ,
\end{equation}
or, equivalently,
\begin{equation}
\begin{aligned}
    \contr{\Rt}\tau = 1\,,\qquad \contr{\Rt}\eta = 0\,,\qquad \contr{\Rt}\dd\eta = 0\,,\\
    \contr{\Rz}\tau = 0\,,\qquad \contr{\Rz}\eta = 1\,,\qquad \contr{\Rz}\dd\eta = 0\,.
\end{aligned}
\end{equation}
The vector fields $\Rt$ and $\Rz$ are called \emph{time Reeb vector field} and \emph{contact Reeb vector field}, respectively.

Let $f\in \Cinfty(M)$ be a function on a cocontact manifold $(M, \tau, \eta)$. The \emph{(cocontact) Hamiltonian vector field of $f$ with respect to $(\tau,\eta)$} is 
the vector field $X_f\in \X(M)$ defined by
\begin{equation}\label{eq:cocontact_Hamiltonian_vf}
    \flat\taueta(X_f)=\dd f-\left(\Rz f + f\right)\eta+\left(1-\Rt f\right)\tau\, .
\end{equation}
This equation is equivalent to the following set of equations:
\begin{equation}\label{eq:cocontact_Hamiltonian_vf_contr}
        \liedv{X_f}\eta = -\Rz(f)\eta-\Rt(f)\tau\,,\qquad
    	\contr{X_f}\eta = -f\,,\qquad
    	\contr{X_f}\tau = 1\,.
\end{equation}
In Darboux coordinates $(t, q^i, p_i, z)$, the Reeb vector fields are written
\begin{equation}
    \Rt = \parder{}{t}\,,\quad \Rz = \parder{}{z}\,.
\end{equation}
and the Hamiltonian vector field of $f\in \Cinfty(M)$ reads
\begin{equation}
   X_f = \parder{}{t} + \parder{f}{p_i}\parder{}{q^i} - \left(\parder{f}{q^i} + p_i\parder{f}{z}\right)\parder{}{p_i} + \left(p_i\parder{f}{p_i} - f\right)\parder{}{z}\,.
\end{equation}

Let $(M_1, \tau_1, \eta_1)$ and $(M_2, \tau_2, \eta_2)$ be cocontact manifolds. A diffeomorphism $\Phi\colon M_1 \to M_2$ is called a \emph{$f$-conformal cocontactomorphism} if $\Phi^\ast \tau_2 = \tau_1$ and $\Phi^\ast\eta_2 = f \eta_1$ for some nowhere-vanishing function $f$ on $M_1$ called the \emph{conformal factor}. A \emph{(strict) cocontactomorphism} is a conformal cocontactomorphism with conformal factor $f\equiv 1$. If the conformal factor $f$ is understood, an $f$-conformal contactomorphism will be called simply a conformal contactomorphism.

Let $(M, \tau, \eta)$ be a cocontact manifold. An \emph{infinitesimal conformal (respectively, strict) cocontactomorphism} is a vector field $X\in \mathfrak{X}(M)$ whose flow is a one-parameter group of conformal (respectively, strict) cocontactomorphisms. In other words, $X$ is an infinitesimal conformal cocontactomorphism if $\liedv{X} \tau = 0$ and $\liedv{X} \eta = g \eta$ for some function $g\in \Cinfty(M)$. This function $g$ is called the \emph{(infinitesimal) conformal factor}, and $X$ is called an \emph{infinitesimal $g$-conformal cocontactomorphism}.

In light of the left-hand side equation in \eqref{eq:cocontact_Hamiltonian_vf_contr}, not every Hamiltonian vector field is an infinitesimal conformal cocontactomorphism. As a matter of fact, $X_f$ is an infinitesimal conformal cocontactomorphism if and only if $\Rt(f) = 0$; it is an infinitesimal strict cocontactomorphism if and only if 
$\Rt(f) = 0$ and $\Rz(f) = 0$. Furthermore, the volume form $\vol\taueta$ is not preserved by the flow of $X_f$, namely,
\begin{equation}
    \liedv{X_f} \vol\taueta = -(n+1) \Rz(f) \vol\taueta\, .
\end{equation}


\begin{proposition}\label{prop:cocontactomorphism_Reebs}
    If $\Phi:M\to M$ is a cocontactomorphism, then it preserves the Reeb vector fields, that is, 
    \begin{equation}
        \Phi_\ast \Rt= \Rt\, , \quad \Phi_\ast \Rz= \Rz\, .
    \end{equation}
\end{proposition}
\begin{proof}
   Suppose that $\Phi$ is a cocontactomorphism. Then, 
   \begin{align*}
		& \contr{\Phi_\ast^{-1}\Rt}(\Phi^\ast\d\eta) = \Phi^\ast(\contr{\Rt}\d\eta) = 0\,,\\
		& \contr{\Phi_\ast^{-1}\Rt}(\Phi^\ast\tau) = \Phi^\ast(\contr{\Rt}\tau) = 1\,,\\
		& \contr{\Phi_\ast^{-1}\Rt}(\Phi^\ast\eta) = \Phi^\ast(\contr{\Rt}\eta) = 0\,.
	\end{align*}
	Since $\Phi^\ast\eta = \eta$ and $\Phi^\ast\tau = \tau$, the uniqueness of the time Reeb vector field implies that $\Phi_\ast \Rt = \Rt$. Analogously, one can see that the contact Reeb vector field is also preserved. 
\end{proof}

\begin{corollary}\label{corollary:cocontactomorphism_Reebs}
    If a vector field $Y\in \mathfrak{X}(M)$ is an infinitesimal cocontactomorphism, then it commutes with the Reeb vector fields, namely,
    \begin{equation}
        [Y, \Rt]=[Y, \Rz]=0\, .
    \end{equation}
\end{corollary}

It is worth noting that the converse is false. 

\begin{example}
    Consider the cocontact manifold $(\RR^4, \dd t, \d z-p\d q)$, where $(t, q, p, z)$ are the canonical coordinates. Clearly, the vector field $Y=\tparder{}{p}$ preserves the Reeb vector fields $\Rt = \tparder{}{t}$ and $\Rz = \tparder{}{z}$. However, it is not an infinitesimal cocontactomorphism. Indeed,
    \begin{equation}
        \liedv{Y} \eta = \contr{Y} \d \eta = - \dd q \neq 0.
    \end{equation}
    Similarly, one can verify that the map $\Phi\colon \RR^4 \to \RR^4,\  (t, q, p, z)\mapsto (t, q, 2p, z)$ is a diffeomorphism preserving the Reeb vector field, but it is not a cocontactomorphism.
\end{example}


A cocontact manifold $(M, \tau, \eta)$ has a Jacobi structure $(\Lambda, E)$, where
\begin{equation}
    \Lambda(\alpha,\beta) = -\dd\eta\big(\sharp\taueta(\alpha),\sharp\taueta(\beta)\big)\,, \quad E = -\Rz\, ,
\end{equation}
for any pair of one-forms $\alpha, \beta \in \Omega^1(M)$.
The Jacobi bracket is thus 
\begin{equation}\label{eq:Jacobi_bracket_cocontact}
    \left\{f,g\right\} 
    = -\dd \eta \left(\sharp\taueta \dd f, \sharp\taueta \dd g\right) - f \Rz(g) + g \Rz(f)\,.
\end{equation}
for any pair of functions $f, g\in \Cinfty(M)$.

\begin{proposition}\label{prop:Jacobi_bracket_cocontact}
    Let $(M, \tau, \eta)$ be a cocontact manifold. Then, the Jacobi bracket defined by $(\tau, \eta)$ can be written as 
    \begin{equation}
        \{f, g\} = -X_g(f) - \Rz (g) f + \Rt(f)\, ,
    \end{equation}
    for any pair of functions $f, g \in \Cinfty(M)$.
\end{proposition}

\begin{proof}
    From the definition of Hamiltonian vector field \eqref{eq:cocontact_Hamiltonian_vf},
    \begin{equation}
        \sharp\taueta \dd f = X_f + \left(\Rz(f) + f\right) \Rz - \left(1 - \Rt(f) \right) \Rt\, ,
    \end{equation}
    which combined with equations~\eqref{eq:cocontact_Hamiltonian_vf_contr} implies that
    \begin{equation}
        \contr{\sharp\taueta (\dd f)}\, \dd \eta = \contr{X_f} \dd \eta
        = \dd f - \Rz(f) \eta -\Rt(f) \tau\, ,
    \end{equation}
    and thus
    \begin{equation}
        \dd \eta \big(\sharp\taueta (\dd f), \sharp\taueta (\dd g)\big) 
        = X_g(f)  + \Rs(f) g - \Rt(f)
        \,.
    \end{equation}
    By replacing this expression in the right-hand side of equation~\eqref{eq:Jacobi_bracket_cocontact}, the result follows.
\end{proof}

The bivector $\Lambda$ induces a $\Cinfty(M)$-module morphism $\lsharp\colon \Omega^1(M)\to \X(M)$ given by 
\begin{equation}
    \lsharp(\alpha) = \Lambda(\alpha, \cdot) = \sharp_{(\tau,\eta)}(\alpha) - \alpha(\Rt) \Rt - \alpha(\Rz) \Rz \, .    
\end{equation}

\begin{proposition}
    The morphism $\lsharp$ defined above is not an isomorphism. As a matter of fact,
    \begin{equation}
        \ker \lsharp = \langle \{\tau, \eta \}\rangle\, , \quad \Ima \lsharp = \ker \tau\cap \ker \eta
        \, .
    \end{equation}
\end{proposition}
\begin{proof}
    Clearly, $\tau, \eta \in \ker \lsharp$. Conversely, each $\alpha \in \ker \lsharp$ satisfies $\sharp\taueta\alpha =  \alpha(\Rt) \Rt + \alpha(\Rz) \Rz$, and therefore $\alpha =  \alpha(\Rt) \eta + \alpha(\Rz) \tau$. Hence, $\ker\lsharp= \langle \{\tau, \eta \}\rangle$.

    On the other hand,
    \begin{equation}
        \eta (\lsharp \alpha) = \eta (\sharp\taueta \alpha) - \alpha (\Rz)
        \, ,
    \end{equation}
    but
    \begin{equation}
        \left\langle\eta , \sharp\taueta (\alpha)\right\rangle 
        = \left\langle \flat\taueta (\Rz) , \sharp\taueta (\alpha)\right\rangle 
        = \left\langle \Rz, \alpha \right\rangle \, ,
    \end{equation}
    and thus $\lsharp\alpha \in \ker \eta$. By an analogous argument, one may proof that $\lsharp\alpha \in \ker \tau$.
\end{proof}

It is worth remarking that the so-called cocontact Hamiltonian vector field, as defined in equation~\eqref{eq:cocontact_Hamiltonian_vf}, is not a Hamiltonian vector field with respect to the Jacobi structure. Indeed, given a cocontact manifold $(M, \tau, \eta)$ with associated Jacobi structure $(\Lambda, - \Rz)$, the cocontact Hamiltonian vector field of $f\in \Cinfty(M)$ with respect to $(\tau, \eta)$ is given by
\begin{equation}
    X_f 
    = \lsharp(\dd f) - f \Rz + \Rt \, ,
\end{equation}
and therefore the Hamiltonian vector field with respect to $(\Lambda, -\Rz)$ is $X_f-\Rt$.

As in a co-oriented contact manifold, the characteristic distribution of a cocontact manifold is the complete tangent bundle, namely, $C = \T M$. Moreover, the definitions of isotropic, coisotropic and Legendrian submanifolds are analogous (see \Cref{remark:Jacobi_orthogonal_contact}). In addition, given a $(2n+2)$-dimensional cocontact manifold $(M, \tau, \eta)$, a submanifold $\incl\colon N \hookrightarrow M$ is:
\begin{enumerate}
    \item isotropic if and only if $\incl^\ast \tau = 0$ and $\incl^\ast \eta = 0$,
    \item Legendrian if and only if it is coisotropic and $n$-dimensional.
\end{enumerate}

\chapter{A review on geometric mechanics}\label{ch:review_mechanics}

\insquote{The miracle of the appropriateness of the language of mathematics for the formulation of the laws of physics is a wonderful gift which we neither understand nor deserve. We should be grateful for it and hope that it will remain valid in future research and that it will extend, for better or for worse, to our pleasure, even though perhaps also to our bafflement, to wide branches of learning.} {Eugene Wigner, \emph{Comm.~Pure~Appl.~Math.}, 13:1-14 (1960)}

There is a plethora of dynamical systems that can be described by the flow of a vector field on a manifold. In particular, a solution of a system of ordinary differential equations 
\begin{equation}
    \frac{\dd x^i}{\dd t} (t) = X^i\big(x^1(t), \ldots, x^n(t)\big)\, , \quad i=1,\ldots,n 
\end{equation}
can be understood as an integral curve of the vector field
\begin{equation}
    X = X^i\parder{}{x^i}\in \X(\RR^n) \, ,
\end{equation}
where $(x^1, \ldots, x^n)$ are the canonical coordinates of $\RR^n$.
Furthermore, if this vector field has some additional properties, such as being a Hamiltonian vector field with respect to a Jacobi structure on the manifold, these properties can be exploited in order to study the dynamics of the system. 

Consider a dynamical system described by the flow of a vector field $X$ on a manifold $M$. A function $f\in \Cinfty(M)$ is called a \emph{conserved quantity} (also known as \emph{constant of the motion} or \emph{first integral}) if, for any integral curve $c\colon I\subseteq \RR \to M$ of $X$, the composition $f\circ c$ is a constant function. Equivalently, $f$ is a conserved quantity if and only if it is a solution of the partial differential equation
\begin{equation}
    X(f) = 0\, .
\end{equation}

In this chapter, the Hamiltonian and Lagrangian frameworks for classical mechanical systems are presented in the language of differential geometry. Moreover, some results concerning the symmetries, constants of the motion and the Hamilton--Jacobi theory for Hamiltonian and Lagrangian systems are reviewed. Additionally, integrable systems and nonholonomic systems are introduced. Henceforth, the reader will be assumed to be familiarized with symplectic geometry (see \Cref{sec:symplectic} and references therein). For an in-depth introduction to geometric mechanics, refer to \cite{A.M2008,d.R1989,Godbillon1969,M.R1999,L.M1987, C.I.M+2015j,Souriau2008}.

\section{Hamiltonian formalism} \label{sec:Hamiltonian_mechanics}

An (\emph{autonomous}, or \emph{time-independent}) \emph{Hamiltonian system} is a triple $(M, \omega, H)$, formed by a symplectic manifold $(M, \omega)$ and a \emph{Hamiltonian function} $H\in \Cinfty(M)$. The dynamics of $(M, \omega, H)$, that is, the trajectories of the physical system it describes, are the integral curves of the Hamiltonian vector field $X_H=\sharp_\omega \dd H$ of $H$. In Darboux coordinates $(q^i, p_i)$, this vector field reads
\begin{equation}
    X_H = \parder{H}{p_i} \parder{}{q^i} - \parder{H}{q^i} \parder{}{p_i}\, .
\end{equation}
Hence, a curve $c\colon I\subseteq \RR \to M,\, c(t)=(q^i(t), p_i(t))$ is an integral curve of $X_H$ if and only if it satisfies the \emph{Hamilton equations}:
\begin{equation}\label{eq:Hamilton_equations}
    \frac{\dd q^i(t)}{\dd t} = \parder{H}{p_i} \big(c(t)\big)\, , \quad 
    \frac{\dd p_i(t)}{\dd t} = -\parder{H}{q^i} \big(c(t)\big)\, .
\end{equation}
In numerous physical systems modeled by a Hamiltonian system, the Hamiltonian function can be regarded as the energy of the system. 

Let $(M, \omega, H)$ be a Hamiltonian system. 
A function $f\in \Cinfty(M)$ is
a conserved quantity if and only if
\begin{equation}
    \{f, H\}_\omega = 0\, ,
\end{equation}
where $\{\cdot, \cdot\}_\omega$ denotes the Poisson bracket defined by $\omega$. 
In particular, the Hamiltonian function $H$ is a conserved quantity. Moreover, the volume form $\vol_\omega$ is also preserved by the dynamics, that is,
\begin{equation}
    \liedv{X_H} \vol_\omega = 0\, .
\end{equation}
Hence, this kind of systems are useful to model mechanical systems with a conservative behaviour. However, in order to describe non-conservative phenomena such as damping, friction, systems with constraint forces, or certain thermodynamical processes, alternative mathematical depictions are required.

\subsection{Non-autonomous Hamiltonian systems}

A \emph{non-autonomous} (or \emph{time-dependent}) \emph{Hamiltonian system} is a tuple $(M, \tau, \omega, H)$, formed by a cosymplectic manifold $(M, \tau, \omega)$ and a \emph{Hamiltonian function} $H\in \Cinfty(M)$ (see \Cref{sec:cosymplectic} for a brief review on cosymplectic geometry). Its dynamics are given by the evolution vector field $\evol_H$ of $H$. In Darboux coordinates $(t, q^i, p_i)$, this vector field is written
\begin{equation}
    \evol_H = \parder{}{t} + \parder{H}{p_i} \parder{}{q^i} - \parder{H}{q^i} \parder{}{p_i}\, .
\end{equation}
A function $f\in \Cinfty(M)$ is a \emph{conserved quantity} if and only if $\evol_H (f)=0$, or, equivalently,
\begin{equation}
    \{f, H\}\tauomega + \Reeb(f) = 0\, .
\end{equation}
Notice that, unlike in autonomous Hamiltonian systems, the Hamiltonian function is, in general, no longer a conserved quantity. As a matter of fact,
\begin{equation}\label{eq:Hamiltonian_evolution_cosymp}
    \liedv{\evol_H} H = \Reeb(H)\, ,
\end{equation}
and thus $H$ is a conserved quantity if and only if it is time-independent. Nevertheless, the volume form is preserved along the evolution, that is,
\begin{equation}
    \liedv{\evol_H} \vol\tauomega = 0\, .
\end{equation}

\section{Lagrangian formalism} \label{sec:Lagrangian_mechanics}

Along this section, let $Q$ denote an $n$-dimensional manifold with local coordinates $(q^i)$. Denote by $(q^i, v^i)$ the induced bundle coordinates on the tangent bundle $\tau_Q\colon \T Q\to Q$. Physically, $Q$ will represent the space of positions, or configuration space, of a mechanical system, and $\T Q$ its space of positions and velocities. Refer to \Cref{sec:structures_TQ} and references therein for the structures on $\T Q$ that will be employed.

Fixing two points $q_1$ and $q_2$ in $Q$, and an interval $[a, b]$, the \emph{path space from $q_1$ to $q_2$} is the following set of curves:
\begin{equation}\label{eq:path_space}
    \Omega(q_1, q_2, [a,b]) = \left\{c\in \Ctwo \big([a, b]\to Q\big) \mid  c(a) = q_1,\, c(b) = q_2 \right\}\, .
\end{equation}
It can be proven that $\Omega(q_1, q_2, [a,b])$ is an infinite-dimensional smooth manifold. 
Its tangent space $\T_c  \Omega(q_1, q_2, [a,b])$ at $c$ is the set of maps $v\in \Ctwo \big([a, b]\to \T Q\big)$ such that $\tau_q\circ v = c$ and $v(a) = v(b) = 0$. A tangent vector $v\in \T_c  \Omega(q_1, q_2, [a,b])$ is called an \emph{infinitesimal variation of the curve $c$ subject to fixed endpoints}.

A \emph{Lagrangian system} is a pair $(Q, L)$, consisting of a manifold $Q$ and a function $L\in \Cinfty(\T Q)$, called the \emph{Lagrangian function}. The \emph{Lagrangian energy} is the function 
\begin{equation}
    E_L = \Delta(L) - L\, ,
\end{equation}
where $\Delta$ is the Liouville vector field.

Let $(Q, L)$ be a Lagrangian system, and let $q_1$ and $q_2$ be two fixed points in $Q$. The \emph{action (functional)} is the map $\action \colon \Omega(q_1, q_2, [a,b]) \to \RR$ given by
\begin{equation}\label{eq:action_functional}
    \action (c) = \int_a^b L \big(c(t), \dot{c}(t)\big)\, \dd t\, .
\end{equation}
\emph{Hamilton's principle} states that, between two specified configurations $q_1=c(a),\, q_2=c(b)\in Q$ at two specified times $a, b\in \RR$, the dynamics of $(Q, L)$ is a critical point $c\in \Omega(q_1, q_2, [a,b])$ of the action $\action$, that is, $\dd \action (c) = 0$. A curve $c\in \Omega(q_1, q_2, [a,b])$ is a critical point of $\action$ if and only if it satisfies the \emph{Euler--Lagrange equations}:
\begin{equation}\label{eq:Euler-Lagrange}
    \frac{\dd }{\dd t} \parder{L}{v^i} \big(c(t), \dot{c}(t)\big) -  \parder{L}{q^i} \big(c(t), \dot{c}(t)\big) = 0\, .
\end{equation}

Given a function $f\in \Cinfty(\T Q)$, its \emph{fiber derivative} is the map $\FF f\colon \T Q \to \cT Q$ such that
\begin{equation}
    \FF f(v) \cdot w = \restr{\frac{\dd }{\dd t}}{t=0} f (L + tw)\, ,
\end{equation}
where $v, w\in \T_q Q$. In other words, $\FF f(v) \cdot w$ is the derivative of $f$ at $v$ along the fiber $\T_q Q$ in the direction $w$. It bears mentioning that the fiber derivative is a bundle morphism from $\T Q$ to $\cT Q$. In bundle coordinates, 
\begin{equation}
    \FF f\colon \left(q^i, v^i \right) \mapsto \left(q^i, \parder{f}{v^i} \right)\, .
\end{equation}

Given a Lagrangian system $(Q, L)$, the \emph{Legendre transform} (or \emph{Legendre transformation}) is the fiber derivative $\FF L$ of the Lagrangian function $L$. By means of the Legendre transform, one can construct the differential forms 
\begin{equation}
    \theta_L = (\FF L)^\ast \theta_Q\, , \quad \omega_L = (\FF L)^\ast \omega_Q = - \dd \theta_L\, ,
\end{equation}
which are called the \emph{Poincaré--Cartan one-form} and \emph{Poincaré--Cartan two-form}, respectively. 
Here $\theta_Q$ denotes the tautological form and $\omega_Q = - \dd \theta_Q$ the canonical symplectic form on the cotangent bundle $\cT Q$. 
The Poincaré--Cartan one-form is also given by
\begin{equation}
    \theta_L = \Sendoadj (\dd L)\, ,
\end{equation}
where $\Sendoadj\colon \cT Q \to \cT Q$ denotes the adjoint operator of the vertical endomorphism.
In canonical coordinates, 
\begin{equation}\label{eq:Poincare-Cartan_1_form_coords}
    \theta_L = \parder{L}{v^i} \dd q^i\, .
\end{equation}
Equivalently, the Legendre transform can be defined as the bundle morphism $\FF L\colon \T Q \to \cT Q$ associated to the semibasic one-form $\theta_L\in \Omega^1(\T Q)$ and \emph{vice versa} (see \Cref{sec:semibasic}). 

Furthermore, the following assertions are equivalent:
\begin{enumerate}
    \item the Poincaré--Cartan two-form $\omega_L$ is a symplectic form,
    \item the Legendre transform $\FF L$ is a local diffeomorphism,
    \item the Hessian matrix
    \begin{equation}\label{eq:Hessian_Lagrangian}
        (W_{ij}) = \left(\parderr{L}{v^i}{v^j}\right)
    \end{equation}
    is regular.
\end{enumerate}
If these equivalent conditions are verified, the Lagrangian function $L$ and the Lagrangian system $(Q, L)$ are said to be \emph{regular}. In addition, if $\FF L$ is a (global) diffeomorphism, then $L$ and $(Q, L)$ are said to be \emph{hyper-regular}.

Let $(Q, L)$ be a regular Lagrangian system. By construction, the triple $(\T Q, \omega_L, E_L)$ is a Hamiltonian system. The Hamiltonian vector field of $E_L$ with respect to $\omega_L$ is a \textsc{sode}, called the \emph{Euler--Lagrange vector field of $(Q, L)$} and denoted by $\sode_L$. In bundle coordinates,
\begin{equation}
    \sode_L = v^i\parder{}{q^i} + W^{ji}\left( \parder{L}{q^j} - v^k\parderr{L}{q^k}{v^j}   \right)\parder{}{v^i}\, ,
\end{equation}
where $(W^{ij})$ denotes the inverse of the Hessian matrix $(W_{ij})$. Furthermore, the solutions of $\sode_L$ satisfy the Euler--Lagrange equations~\eqref{eq:Euler-Lagrange}.

If $(Q, L)$ is hyper-regular, the Hamiltonian system $(\cT Q, \omega_Q, H)$ with
\begin{equation}
    H=E_L \circ (\FF L)^{-1}
\end{equation}
is equivalent to $(\T Q, \omega_L, E_L)$, that is, $\FF L$ is a symplectomorphism and
\begin{equation}
    \sode_L = (\FF L)_\ast X_H\, ,
\end{equation}
where $X_H$ is the Hamiltonian vector field of $H$ with respect to $\omega_Q$.  

A Lagrangian function $L\in \Cinfty(\T Q)$ is called \emph{mechanical} if it is of the form
\begin{equation}\label{eq:mechanical_Lagrangian}
    L(v) = \frac{1}{2} g(v, v) - V\circ \tau_Q (v)\, , 
\end{equation}
where $g$ is a Riemannian metric on $Q$, and $V\in\Cinfty(Q)$ is a function called the \emph{potential (function)}. 
Every mechanical Lagrangian is hyper-regular. Indeed, if $L$ is of the form~\eqref{eq:mechanical_Lagrangian}, then
\begin{equation}
    \FF L = \flat_g\, ,
\end{equation}
which is a bundle isomorphism between $\T Q$ and $\cT Q$ (see \Cref{sec:Riemannian}). In bundle coordinates,
\begin{equation}
    \FF L \left(q^i, v^i\right) = \left(q^i, g_{ij}\, v^j\right)\, ,
\end{equation}
where $g=g_{ij}\dd q^i \otimes \dd q^j$.

\section{Symmetries and conserved quantities} \label{sec:symmetries}

The connection between symmetries and conserved quantities has captivated the interest of mathematicians and physicists ever since Emmy Noether's groundbreaking work \cite{Noether1971} (see also \cite{Neeman1999, Kosmann-Schwarzbach2011, Sardanashvily2016}). Furthermore, reduction procedures can be used in order to simplify the description of a dynamical system whose group of symmetries is known. This section is devoted to the types of symmetries of Hamiltonian and Lagrangian systems, as well as their relation with conserved quantities. Further details can be found in \cite{Roman-Roy2020,d.R1989} (see also \cite{Crampin1983,Prince1983,Prince1985,C.L.M1989,Lunev1990,Sarlet1983,S.C1981, L.M.R1999, M.M1986,d.M1994,d.M1996,C.M1989,C.F1994,F.P1990,M.K.M2007}).

Let $(M, \omega, H)$ be a Hamiltonian system. A \emph{dynamical symmetry} is a diffeomorphism $\Phi\colon M \to M$ such that, for every integral curve $c$ of $X_H$, the curve $c \circ \Phi$ is also an integral curve $c$ of $X_H$; or, equivalently, $\Phi_\ast X_H = X_H$. An \emph{infinitesimal dynamical symmetry} is a vector field $Y\in \X(M)$ whose flow is made of dynamical symmetries, that is, $[Y, X_H]=0$. 

Let $f\in \Cinfty(M)$ be a conserved quantity for $(M, \omega, H)$. Given a dynamical symmetry $\Phi\colon M \to M$  and an infinitesimal dynamical symmetry $Y\in \X(M)$, the functions $f\circ \Phi$ and $Y(f)$ are also conserved quantities.

A diffeomorphism $\Phi\colon M \to M$ is called a \emph{Hamiltonian symmetry} if it preserves the Hamiltonian function, namely, $\Phi^\ast H = H$. A vector field $Y\in \X(M)$ is called an \emph{infinitesimal Hamiltonian symmetry} if its flow is made of Hamiltonian symmetries, or, equivalently, $X(H)=0$. It is worth noting that being an (infinitesimal) Hamiltonian symmetry does not imply being an (infinitesimal) dynamical symmetry. Nevertheless, if an (infinitesimal) symplectomorphism is an (infinitesimal) Hamiltonian symmetry, then it is an (infinitesimal) dynamical symmetry. 

An (infinitesimal) Hamiltonian symmetry which is also an (infinitesimal) symplectomorphism is called an \emph{(infinitesimal) Cartan symmetry}\footnote{In some references (for instance, \cite{d.R1989}), for a vector field $Y\in \X(M)$ on an exact symplectic manifold $(M, \theta)$ to be considered an infinitesimal Cartan symmetry it is required that $\liedv{Y}\theta$ is exact. Nevertheless, with the definition of Cartan symmetry employed in this dissertation, this will occur locally but, in general, not globally.}.

Suppose that $Y\in \X(M)$ is an infinitesimal Cartan symmetry. Then, $\contr{Y} \omega$ is closed, and thus, for each $x\in M$, there exists a neighbourhood $U\ni x$ and a function $f\in \Cinfty(U)$, unique up to an additive constant, such that 
\begin{equation}
    \restr{\contr{Y}\omega}{U} = \dd f\, .
\end{equation}
This function $f$ is a constant of the motion for $(U, \restr{\omega}{U}, \restr{H}{U})$. Conversely, if $f\in \Cinfty(M)$ is a conserved quantity for $(M, \omega, H)$, then its Hamiltonian vector field $X_f$ is an infinitesimal Cartan symmetry.

In the remainder of this section, let $(Q, L)$ be a regular Lagrangian system. A Cartan symmetry (respectively, a dynamical symmetry) $\Phi\colon \T Q \to \T Q$ is called a \emph{Noether symmetry} (respectively, a \emph{Lie symmetry}) if $\Phi = \T \varphi$ for some diffeomorphism $\varphi\colon Q \to Q$. An infinitesimal Cartan symmetry  (respectively, an infinitesimal dynamical symmetry) $Y\in \X (\T Q)$ is called an \emph{infinitesimal Noether symmetry} (respectively, an \emph{infinitesimal Lie symmetry}) if $Y=X^\Com$ for some $X\in \X(Q)$. 

If $Y\in \X(\T Q)$ is an infinitesimal symplectomorphism for $\omega_L$, then, for each point $v\in \T Q$, there exists a neighbourhood $U\ni v$ and a function $f\in\Cinfty(U)$ such that $\liedv{Y} \theta_L = \dd f$. Furthermore, $Y$ is an infinitesimal Cartan symmetry for the Hamiltonian system $(\T Q, \omega_L, E_L)$ if and only if $f-\restr{(\Sendo Y)(L)}{U}\colon U \to \RR$ is a conserved quantity. In particular, $Y=X^\Com$ is an infinitesimal Noether symmetry if and only if $f-\restr{X^\V(L)}{U}\colon U \to \RR$ is a conserved quantity.

A diffeomorphism $\Phi\colon \T Q\to \T Q$ is called a \emph{symmetry of the Lagrangian $L$} if $\Phi^\ast L = L$. In addition, if $\Phi = \T \varphi$ for some diffeomorphism $\varphi\colon Q \to Q$, then it is called a \emph{natural symmetry of the Lagrangian $L$}. A vector field $Y\in \X(\T Q)$ is called an \emph{infinitesimal (natural) symmetry of the Lagrangian $L$} if its flow is made of (natural) symmetries of the Lagrangian $L$. Equivalently, $Y$ is an infinitesimal symmetry of the Lagrangian if $Y(L)=0$; and an infinitesimal natural symmetry of the Lagrangian if, additionally, $Y=X^\Com$ for some $X\in \X(Q)$. Moreover, a vector field $Y=X^C$ is an infinitesimal natural symmetry of the Lagrangian if and only if $X^\V (L)$ is a conserved quantity. It is also worth remarking that an (infinitesimal) natural symmetry of the Lagrangian $L$ is also an (infinitesimal) Noether symmetry for $(\T Q, \omega_L, E_L)$.

\section{Integrable systems} \label{sec:integrable_systems}

In this section, the notion of integrability in the sense of Liouville is presented, and the Liouville--Arnol'd theorem is introduced. Further details may be found in \cite{Arnold1978,Audin2004,B.F2004,Sardanashvily2015,Wiggins2003}.


A \emph{completely integrable} (or \emph{Liouville integrable}) \emph{Hamiltonian system} is a triple $(M, \omega, F)$, where $(M, \omega)$ is a $2n$-dimensional symplectic manifold, and $F\colon M \to \RR^n$ is a map given by
\begin{equation}
    F(x) = \big(f_1(x), \ldots, f_n(x)\big)\, ,
\end{equation}
where $f_1, \ldots, f_n\in \Cinfty(M)$ are functions such that:
\begin{enumerate}
    \item\label{item:IS_1} they are functionally independent, that is, their differentials are linearly independent almost everywhere,
    \item\label{item:IS_2} they are in \emph{involution}, that is, $\{f_i, f_j\}_\omega=0$ for any $i,j=1,\ldots, n$,
    \item\label{item:IS_3} their Hamiltonian vector fields $X_{f_i}$ are complete.
\end{enumerate}
The functions $f_1, \ldots, f_n$ are called \emph{integrals}. In physical applications, one of the integrals is usually regarded as the Hamiltonian function. For instance, $(M, \omega, f_1)$ is a Hamiltonian system and $f_1, \ldots, f_n$ are conserved quantities\footnotemark. 

\footnotetext{
In some references (for instance, \cite{B.F2004}), a Hamiltonian system $(M, \omega, H)$ is called \emph{Liouville integrable} if there exists a set of $n=\tfrac{1}{2}\dim M$ constants of the motion $f_1, \ldots, f_n$ with the properties \ref{item:IS_1}, \ref{item:IS_2} and \ref{item:IS_3} above. 
It is easy to see that, on a $2n$-dimensional symplectic manifold, the maximum number of functions that can be simultaneously in involution and functionally independent is $n$. Hence, if a Hamiltonian system $(M, \omega, H)$ is Liouville integrable, then its Hamiltonian function $H$ is functionally dependent of its integrals $f_1, \ldots, f_n$. Therefore, this second definition of Liouville integrable system does not provide additional information with respect to the one above.}

The following result is the basis of the theory for integrable systems.

\begin{theorem}[Liouville--Arnol'd]\label{theorem:Liouville-Arnold}
    Let $(M, \omega, F)$ be a $2n$-dimensional completely integrable Hamiltonian system. Let $M_\Lambda=F^{-1}(\Lambda)$ be a regular level set of $F$ for some $\Lambda\in \RR^n$, that is, $\rank \T F_x=n$ for every point $x\in M_\Lambda$. Then,
    \begin{enumerate}
        \item $M_\Lambda$ is a Lagrangian submanifold of $(M, \omega)$. Moreover, it is invariant with respect to the flow of $X_{f_1}, \ldots, X_{f_n}$.
        \item Every compact and connected component of $M_\Lambda$ is diffeomorphic to the $n$-dimensional torus $\TT^n$, called the \emph{Liouville torus}.
        \item Each Liouville torus has a neighbourhood in $M$ given by the direct product of the torus and a ball, namely, $U = \TT^n\times \BB^n$.
        \item In this neighbourhood $U$, there are coordinates $(\varphi^i, s_i)$, called \emph{action-angle coordinates}, such that:
        \begin{enumerate}
            \item they are Darboux coordinates, namely, $\omega= \dd \varphi^i \wedge \dd s_i$,
            \item the \emph{action coordinates} $s_i$ are functions of the integrals $f_1, \ldots, f_n$,
            \item considering one of the integrals $f_i$ as a Hamiltonian function, its Hamilton equations are given by
            \begin{equation}
                \frac{\dd \varphi^i}{\dd t} = \Omega^i (s_1, \ldots, s_n)\, , \quad  \frac{\dd s_i}{\dd t} = 0\, ,
            \end{equation}
            for some functions $\Omega^i\in \Cinfty(U)$. 
        \end{enumerate}
    \end{enumerate} 
\end{theorem}

\section{Hamilton--Jacobi theory} \label{sec:Hamilton-Jacobi}

The \emph{Hamilton--Jacobi equation} is the first-order nonlinear partial differential equation
\begin{equation}\label{eq:HJ_classical}
    H\left(q^i, \parder{S}{q^i}\right) = E\, ,
\end{equation}
where $H\colon \RR^{2n}\to \RR$ is the Hamiltonian function and $E$ is a constant representing the energy of the system. In the case of non-autonomous Hamiltonians, the Hamilton--Jacobi equation takes the form
\begin{equation}\label{eq:HJ_classical_time}
    H\left(t, q^i, \parder{S}{q^i}\right) = \parder{S}{t}\, .
\end{equation}
In this section, the Hamilton--Jacobi equation~\eqref{eq:HJ_classical} is derived in two ways: first, through the classical, more analytical, approach of generating functions for canonical transformations of coordinates; and then through a geometric approach, based on projecting the Hamiltonian vector field and obtaining a section of the cotangent bundle to reconstruct the original vector field from the projected vector field. Moreover, an intrinsic (coordinate free) form of the Hamilton--Jacobi equation for Hamiltonian systems on cotangent bundles is presented. Refer to \cite{A.M2008,Arnold1978,C.G.M+2006,E.d.L+2022,d.M.V2014} for additional details.


Let $(M, \omega)$ be a $2n$-dimensional symplectic manifold. If $(q^i, p_i)$ and $(Q^i, P_i)$ are two sets of Darboux coordinates centered at $x\in M$, then there exists a neighbourhood $U$ of $x$ such that $U$ is diffeomorphic to $\RR^{2n}$ and 
\begin{equation}\label{eq:generating_function}
    p_i \dd q^i - P_i \dd Q^i = \dd F\, , 
\end{equation}
for some $F\in \Cinfty(U)$. Assume that, in some neighbourhood $V\subseteq U$, the Jacobian matrix 
\begin{equation}
    \parder{(Q^i, q^j)}{(p_k, q^l)}
\end{equation}
is not singular. Then, it is possible to express $F(p, q) = S(q, Q)$ for some function $S\colon \RR^n\times \RR^n\to \RR$, which is called the \emph{generating function}. From equation~\eqref{eq:generating_function}, it follows that 
\begin{equation}
    p_i = \parder{S}{q^i}\, , \quad P_i = - \parder{S}{Q^i}\, .
\end{equation}
Let $H$ be a Hamiltonian function of $(M, \omega)$. Suppose that the coordinates $(Q^i, P_i)$ are such that the Hamiltonian function only depends on the coordinates $P_i$, namely,
\begin{equation}\label{eq:HJ_K}
    H\left(q^i, \parder{S}{q^i}\right) = K(P_i)\, ,
\end{equation}
for some function $K\colon \RR^n \to \RR$. Thus, Hamilton equations are given by
\begin{equation}
    \frac{\dd Q^i}{\dd t} = \parder{K}{P_i}\, , \quad \frac{\dd P_i}{\dd t} = 0\, .
\end{equation}
The right-hand side equations imply that $P_i(t)$, and hence $K(P_i(t))$, are constant functions. Therefore, after fixing the initial conditions, equation~\eqref{eq:HJ_K} leads to the Hamilton--Jacobi equation~\eqref{eq:HJ_classical}.

In the rest of the section, the geometric approach to the Hamilton--Jacobi theory will be considered. Let $\pi_Q\colon \cT Q \to Q$ be the cotangent bundle of $Q$, with canonical symplectic form $\omega_Q$. Consider the Hamiltonian system $(\cT Q, \omega_Q, H)$, with Hamiltonian vector field $X_H$. Given a one-form $\gamma\in \Omega^1(Q)$ (that is, a section of $\pi_Q\colon \cT Q \to Q$), the vector field 
\begin{equation}
    X_H^\gamma = \T \pi_Q \circ X_H \circ \gamma 
\end{equation}
is a projection of $X_H$ along $\Ima \gamma$. The objective is to find a one-form $\gamma$ which maps integral curves of $X_H^\gamma$ into integral curves of $X_H$. This occurs if and only if the vector fields $X_H^\gamma$ and $X_H$ are $\gamma$-related, that is,
\begin{equation}\label{eq:HJ_gamma_related}
    X_H \circ \gamma = \T\gamma \circ X_H^\gamma\, .
\end{equation}
In other words, the diagram
\begin{equation}
\begin{tikzcd}
        \cT Q \arrow[rr, "X_{H}"] \arrow[d, "\pi_{Q}"] &  & \T \cT Q \arrow[d, "\T\pi_{Q}"']    \\
        Q \arrow[rr, "X_H^{\gamma}"'] \arrow[u, "\gamma", bend left, shift left]    &  & \T Q \arrow[u, "\T\gamma"', bend right, shift right]
        \end{tikzcd}
\end{equation}
is commutative. Notice that $X_H^\gamma$ and $X_H$ are $\gamma$-related if and only if $X_H$ is tangent to $\Ima \gamma$. In addition, equation~\eqref{eq:HJ_gamma_related} is equivalent to the following equations in bundle coordinates $(q^i, p_i)$:
\begin{equation}
    - \frac{\partial H} {\partial q^i} \circ \gamma(x) = \frac{\partial H} {\partial p_j} \circ \gamma(x) \frac{\partial \gamma_i} {\partial q^j}(x)\, .
\end{equation}
Assuming that $\gamma$ is closed, these equations are equivalent to 
\begin{equation}
    - \frac{\partial H} {\partial q^i}\circ \gamma(x) = \frac{\partial H} {\partial p_j}\circ \gamma(x) \frac{\partial \gamma_j} {\partial q^i}(x)\, 
\end{equation}
for any point $x\in Q$.
Therefore, if $\gamma$ is closed, equation~\eqref{eq:HJ_gamma_related} holds if and only if 
\begin{equation}\label{eq:HJ_geometric}
    \dd (H \circ \gamma) = 0\, .
\end{equation}
By the Poincaré lemma, around each point $q\in Q$, there exists a neighbourhood $U$ such that $\restr{\gamma}{U} = \dd S$ for some function $S\in \Cinfty(U)$, and
\begin{equation}\label{eq:HJ_geometric_local}
    H\circ \dd S = E_U\, 
\end{equation}
for some constant $E_U$. In local coordinates $(q^i)$ of $Q$, equation~\eqref{eq:HJ_geometric_local} is the Hamilton--Jacobi equation \eqref{eq:HJ_classical}. Equation~\eqref{eq:HJ_geometric} also receives the name of \emph{Hamilton--Jacobi equation for $(Q, H)$}, and a one-form $\gamma$ that satisfies it is called a \emph{solution of the Hamilton--Jacobi problem for $(Q, H)$}.

The geometric Hamilton--Jacobi theory is summarized as follows.
Given a closed one-form $\gamma\in \Omega^1(Q)$, the following assertions are equivalent:
\begin{enumerate}
    \item $\gamma$ is a solution of the Hamilton--Jacobi problem for $(Q, H)$,
    \item if $c\colon I \subseteq \RR \to Q$ is an integral curve of $X_H^\gamma$, then $\gamma\circ c$ is an integral curve of $X_H$;
    \item $X_H$ is tangent to $\Ima \gamma$. 
\end{enumerate}

A \emph{complete solution of the Hamilton--Jacobi problem for $(Q, H)$} is a local diffeomorphism $\Phi\colon Q \times \RR^n \to \cT Q$ such that, for each $\Lambda\in\RR^n$, the map $\Phi_\Lambda = \Phi(\cdot, \Lambda)$ is solution of the Hamilton--Jacobi problem for $(Q, H)$.

If there exists a complete solution of the Hamilton--Jacobi problem for a Hamiltonian system, then it is (locally) completely integrable (see \Cref{sec:integrable_systems}). 
Let $\Phi\colon Q \times \RR^n \to \cT Q$ be a complete solution of the Hamilton--Jacobi problem for $(Q, H)$. Around each point $q\in Q$, there exists a neighbourhood $U$ such that $\restr{\Phi}{U\times \RR^n}\colon U\times \RR^n \to V=\Phi(U\times \RR^n)$ is a diffeomorphism. Let $\pi_a\colon Q\times \RR^n\to \RR$ denote the projection on the $a$-th component of $\RR^n$, and define the $n$ functions 
\begin{equation}
    f_a = \pi_a \circ \restr{\Phi}{U\times \RR^n}^{-1}\colon V \to \RR\, .
\end{equation}
By construction, these functions are functionally independent, that is, their differentials are linearly independent. Moreover, observe that 
\begin{equation}
  \Phi_\Lambda(U)= \bigcap_{a=1}^n f_a^{-1}(\Lambda_a)\, ,
\end{equation}
for $\Lambda = (\Lambda_1, \ldots, \Lambda_n)\in \RR^n$. 
Since $X_H$ is tangent to $\Phi_\Lambda(U)$, the functions $f_a$ are conserved quantities.
In addition,
\begin{equation}
  \restr{\left\{f_a,f_b \right\}_{\omega_V}}{\Phi_\Lambda(U)}
  = \restr{\omega_V(X_{f_a}, X_{f_b})}{\Phi_\Lambda(U)} = 0\, ,
\end{equation}
where $\omega_V$ denotes the restriction of $\omega_Q$ to $V$. Hence, $(V, \omega_V, F)$ is a completely integrable system, with 
\begin{equation}
    F = (f_1, \ldots, f_n) = \pi_2 \circ \restr{\Phi}{U\times\RR^n}^{-1}\colon V \to \RR^n\, ,
\end{equation}
where $\pi_2 \colon Q \times \RR^n\to \RR^n$ denotes the canonical projection. 

Furthermore, a complete solution of the Hamilton--Jacobi problem can be employed to compute action-angle coordinates. There exists a neighbourhood $W$ around each point of $Q$ such that, for each $\Lambda\in \RR^n$ there is a function $S_\Lambda\in \Cinfty(W)$ such that $\restr{\Phi_\Lambda}{W} = \dd S_\Lambda$. Hence, the function $S\in \Cinfty(W\times \RR^n)$ defined by $S(\cdot, \Lambda)=S_\Lambda$ is a generating function. From this generating function, it is possible to construct action-angle coordinates (see \cite{A.M2008,Arnold1978,Goldstein1980}).

It is worth remarking that, since all symplectic manifolds of the same dimension are locally symplectomorphic, this Hamilton--Jacobi theory may be applied locally to any symplectic manifold by means of Darboux coordinates. Additionally, the Hamilton--Jacobi theory has been extended for Hamiltonian systems $(M, \omega, H)$ on arbitrary fiber bundles $\pi\colon M\to B$ (see \cite{G.M.P2021,G.P2016}).

\section{Routhian reduction}\label{sec:Routh}

This section reviews the Routhian reduction for Lagrangian systems with symmetries. 
After presenting the classical theory, the modern interpretation as a particular case of symplectic reduction is introduced.
Refer to \cite{L.C.V2010, Marsden1992} for additional details; see also \cite{G.L.C2014,G.U2019,G.M.Y2016,Mestdag2008,Leone2018,M.R.S2000,Capriotti2017,L.A.C2012,L.M.V2011,L.L2010,C.M2008}.

As in \Cref{sec:Lagrangian_mechanics}, along this section, $Q$ will denote an $n$-dimensional manifold with local coordinates $(q^i)$, and $(q^i, v^i)$ will denote induced bundle coordinates on the tangent bundle $\tau_Q\colon \T Q\to Q$. 
 
Consider a regular Lagrangian system $(Q, L)$. The easiest type of symmetry the Lagrangian function can exhibit is a \emph{cyclic coordinate}, that is, $\tparder{L}{q^i} = 0$ for one of the coordinates $q^i$. In that case, the corresponding Euler--Lagrange equation simplifies to
\begin{equation}
    \frac{\dd }{\dd t} \parder{L}{v^i} \big(c(t), \dot{c}(t)\big) = 0\, ,
\end{equation}
which implies that the associated canonical momenta $p_i=\tparder{L}{v^i}$ is a conserved quantity. The idea of Routhian reduction is to employ this conserved quantity to consider a new Lagrangian function, the Routhian, which leads to the same Euler--Lagrange equations as the original one.

Suppose that $Q=\RR^n$ and $(q^i)$ are the canonical coordinates.
Assuming that $q^1$ is a cyclic coordinate and $\tparderr{L}{v^1}{v^1}\neq 0$, there exists a function $\psi$ such that $\tparder{L}{v^1} = \mu$ is equivalent to $v^1 = \psi(q^2, \ldots, q^n, v^2, \ldots, v^n)$. 
The \emph{Routhian} is the function $R^\mu \colon \RR^{2(n-1)}\to \RR$ defined by $R^\mu = L - v^1 \mu$, where all instances of $v^1$ are replaced by $\psi$. Then, any solution $c(t)=(q^1(t), \ldots, q^n(t))$ of the Euler--Lagrange equations
\begin{equation}
    \frac{\dd }{\dd t} \parder{L}{v^i} \big(c(t), \dot{c}(t)\big) -  \parder{L}{q^i} \big(c(t), \dot{c}(t)\big) = 0\, , \quad i=1\ldots, n
\end{equation}
with momentum $p_1 \coloneqq \tparder{L}{v^1} = \mu$ projects onto a solution $\hat{c}=(q^2(t), \ldots, q^n(t))$ of the Euler--Lagrange equations 
\begin{equation}
    \frac{\dd }{\dd t} \parder{R^\mu}{v^k} \big(\hat{c}(t), \dot{\hat{c}}(t)\big) -  \parder{R^\mu}{q^k} \big(c(t), \dot{c}(t)\big) = 0\, , \quad k=2\ldots, n\, .
\end{equation}
Conversely, any solution of the Euler–-Lagrange equations for $R^\mu$ can be lifted to a solution of the Euler-–Lagrange equations for $L$ with momentum $p_1=\mu$.

The presence of the cyclic coordinate $q^1$ can be interpreted geometrically as the invariance of the Lagrangian function $L$ under the Abelian Lie group action of $\RR$ by translations along the $q^1$-direction, and the conserved quantity $p_1$ is the associated momentum map. Analogously, one could consider an $\Sp^1$ action by rotations, leading to the conservation of a component of angular momentum. This idea can be extended to not necessarily Abelian Lie group actions as follows.

Let $Q$ be an $n$-dimensional manifold and $G$ a Lie group with Lie algebra $\mathfrak{g}$ and dual $\mathfrak{g}^\ast$. Consider the Lie group action $\Phi\colon G\times Q \to Q$ and its tangent lift $\Phi^\T \colon G \times \T Q \to \T Q$. Suppose that $(Q, L)$ is a regular Lagrangian system such that $L$ is $\Phi^\T$-invariant. Let $A$ be a principal connection on the principal bundle $\pi\colon Q \to Q/G$, and let $A_\mu\in \Omega^{1}(Q)$ be the one-form given by its pairing with $\mu \in \mathfrak{g}^\ast$. Let $G_\mu <G$ denote the isotropy subgroup in $\mu$ under the coadjoint action. Let $\mathfrak{g}_\mu$ denote the Lie algebra of $G_\mu$ and $\mathfrak{g}_\mu^\ast$ its dual.

The momentum map $\mommap_L \colon \T Q \to \mathfrak{g}^\ast$ is the natural momentum map defined by the Poincaré--Cartan one-form $\theta_L$, namely,
\begin{equation}
     \left\langle \mommap_L(x), \xi \right \rangle = \left( \contr{\xi_{\T Q}} \theta_L  \right)(x) 
    \, ,
\end{equation}
for each $\xi \in \mathfrak{g}$. For a regular value $\mu$ of $\mommap_L$, the function $R^\mu\in \Cinfty(\T Q)$ is defined by $R^\mu = L - A_\mu$. By construction, this function is $G_\mu$-invariant. Moreover, the restriction 
of $R^\mu$ to $\mommap_L^{-1}(\mu)$
can be projected into a function $[R^\mu]$ on the quotient $\mommap_L^{-1}(\mu)/G_\mu$.

Furthermore, if the Lagrangian function $L$ satisfies an additional regularity condition, then the quotient manifold $\mommap_L^{-1}(\mu)/G_\mu$ is diffeomorphic to the fiber product $\T(Q/G) \times_{Q/G} Q/G_\mu$. More precisely, $L$ is called \emph{$G_\mu$-regular} if the mapping $\xi \in \mathfrak{g}_\mu \mapsto \mommap_L(v_q + \xi_{\T Q}(q))\in \mathfrak{g}_\mu^\ast$ is a diffeomorphism. If $L$ is $G_\mu$-regular, then there exists a diffeomorphism $\psi_\mu\colon \T(Q/G) \times_{Q/G} Q/G_\mu \to \mommap_L^{-1}(\mu)/G_\mu$. The \emph{Routhian} is the function $\mathscr{R}^\mu = \psi_\mu^\ast [R^\mu]$. The one-form $A_\mu$ is, by construction, $\Phi^{\cT}$-invariant, and satisfies $\mommap_L\circ A_\mu = \mu$. Thus, $\contr{\xi_Q}A_\mu\in \Cinfty(Q)$ is a constant function and
\begin{equation}
    \contr{\xi_Q} \dd A_\mu = \liedv{\xi_Q} A_\mu - \dd \contr{\xi_Q}A_\mu = 0\, .
\end{equation}
Consequently, there exists a unique 2-form $\rho_\mu \in \Omega^1(Q/G)$ such that $\pi^\ast\rho_\mu = \dd A_\mu$. Define the 2-form $B_\mu = \pi^\ast_{Q/G}\, \rho_\mu\in \Omega^1(\cT(Q/G))$, where $\pi_{Q/G}\colon \cT(Q/G) \to Q/G$ denotes the canonical projection.

The symplectic manifold $(\T Q, \omega_L)$ reduces to $\T(Q / G) \times_{Q/G} Q / G_\mu$ endowed with the symplectic form
\begin{equation}
    \omega_\mu = \left(\FF \mathscr{R}^\mu\right)^*\left(\pi_1^* \omega_{Q / G}+\pi_2^* B_\mu\right)\, ,
\end{equation}
where
\begin{equation}
    \pi_1: \cT(Q / G) \times_{Q/G} Q / G_\mu \to \cT(Q / G)\, ,
\end{equation}
and
\begin{equation}
    \pi_2: \cT(Q / G) \times_{Q/G} Q / G_\mu \to Q / G_\mu
\end{equation}
are the canonical projections.
The \emph{Routhian energy} $E_{\mathscr{R}^\mu}$ is defined by
\begin{equation}
    E_{\mathscr{R}^\mu}(v_x, y) = \left\langle \FF \mathscr{R}^\mu (v_x, y), (v_x, y) \right\rangle - \mathscr{R}^\mu (v_x, y)\, .
\end{equation}
The Euler--Lagrange vector field $\sode_L$ of $L$ (that is, the Hamiltonian vector field of $E_L$ with respect to $\omega_L$) projects onto the Hamiltonian vector field of $E_{\mathscr{R}^\mu}$ with respect to $\omega_\mu$.

\section{Nonholonomic systems}\label{sec:nonholonomic}

In the previous sections of this chapter, the dynamics considered were unconstrained, that is, the trajectories could in principle take any position or velocity. Nonholonomic systems are, roughly speaking, mechanical systems with constraints on their velocities that cannot be derived from constraints on their positions. Usually, the constraints considered are linear or affine in the velocities. Lagrangian and Hamiltonian systems subject to linear nonholonomic constraints are briefly reviewed in this section. Further information and examples can be found in \cite{d.M1996a, Bloch2015, CortesMonforte2002, L.M1995, v.M1994, B.S1993, S.C.S1995, S.S.C1996, B.M.Z2005, G.d.M+2009, G.M2020}.

As in previous sections, let $Q$ denote an $n$-dimensional manifold with local coordinates $(q^i)$, and let $(q^i, v^i)$ be the induced bundle coordinates on the tangent bundle $\tau_Q\colon \T Q\to Q$. 

A \emph{nonholonomic Lagrangian system} is a triple $(Q, L, D)$, where $(Q, L)$ is a regular Lagrangian system and $D$ is a distribution on $Q$. 
Given two fixed points $q_1$ and $q_2$ in $Q$ and an interval $[a, b]$, the \emph{admissible path space from $q_1$ to $q_2$} is the subset of the path space from $q_1$ to $q_2$ given by
\begin{equation}
    \Omega_D(q_1, q_2, [a,b]) = \left\{c\in \Omega(q_1, q_2, [a,b]) \mid  \dot{c}(t)\in D_{c(t)} \right\}\, .
\end{equation}
Its elements are called \emph{admissible paths} or \emph{admissible curves}. 
The \emph{space of admissible variations} is the subspace $\mathcal{V}_c$ of the tangent space to $\Omega(q_1, q_2, [a,b])$ at $c$ given by
\begin{equation}
    \mathcal{V}_c = \left\{v \in \T_c \Omega(q_1, q_2, [a,b])\mid v(t)\in D_{c(t)}  \right\}\, .
\end{equation}
\emph{Lagrange--d'Alembert--Hölder principle}\footnote{In the literature, this principle can be found named as Lagrange--d'Alembert principle or as Hölder principle.} states that, between two specified configurations $q_1=c(a),\, q_2=c(b)\in Q$ at two specified times $a, b\in \RR$, the dynamics of $(Q, L, D)$ is an admissible path $c\in \Omega_D(q_1, q_2, [a,b])$ such that
\begin{equation}
    \left\langle \dd \action (c), v\right\rangle = 0\, ,
\end{equation}
for all admissible variations $v\in \mathcal{V}_c$.

It is worth mentioning that, unlike Hamilton's principle, Lagrange--d'Alembert--Hölder principle is not \textit{stricto sensu} a variational principle, since the motion of the nonholonomic system is not a critical point of any functional in the sense of the calculus of variations. This other approach, that is, extremizing the action functional among all the admissible curves, leads to the so-called \emph{vakonomic dynamics} (see, for instance, \cite{Arnold1988,C.d.M+2002, d.M.M2000}). The resulting equations are not equivalent to the nonholonomic equations of motion, except in certain cases. The empirical evidence suggests that the dynamical behaviour of systems subject to non-sliding type constraints is well-modeled by nonholonomic systems rather than vakonomic systems (see~\cite{L.M1995, Kai2013}).

Henceforth, assume that the so-called \emph{admissibility condition} holds, that is, for all $x\in D$, 
\begin{equation}
    \dim (\T_x D)^\circ =  \dim \Sendoadj \big((\T_x D)^\circ\big)\, ,
\end{equation}
where $\Sendoadj$ denotes the adjoint operator of the vertical endomorphism, and the annihilator of $\T_x D$ is taken in $\cT_x \T Q$, namely,
\begin{equation}
    (\T_x D)^\circ = \left\{ \alpha \in \cT_x \T Q\mid \alpha(v) = 0\, \forall\, v\in \T_x D\right\}\, .
\end{equation}
Consider the distribution $F$ on $\T Q$ whose annihilator $F^\circ$ is given by
\begin{equation}
    F^\circ_x = \Sendoadj \big((\T_x D)^\circ\big)
\end{equation}
The dynamics of $(Q, L, D)$ are then given by
\begin{equation}\label{eq:nonholonomic_SODE}
\begin{aligned}
    & \restr{\left(\contr{X}\omega_L - \dd E_L\right)}{D} \in F^\circ\, ,\\
    & \restr{X}{D} \in \T D\, .
\end{aligned}
\end{equation} 
If the so-called \emph{compatibility condition}
\begin{equation}
    F^{\perp_{\omega_L}}\cap \T M = \{0\}
\end{equation}
is satisfied, then \cref{eq:nonholonomic_SODE} have a unique solution, denoted by $\sode_{(L, D)}$ and called the \emph{nonholonomic Euler--Lagrange vector field}. If the Hessian matrix $(W_{ij})$ of $L$ with respect to the velocities \eqref{eq:Hessian_Lagrangian} is definite, then this condition is automatically satisfied. In particular, this is the case of mechanical Lagrangians. Hereinafter, the compatibility condition will be assumed to be satisfied.

Suppose that the annihilator $D^\circ$ of $D$ is locally spanned by the one-forms $\mu^a=\mu^a_i \dd q^i\in \Omega^1(Q), \, a=1,\ldots, m$, where $m=\corank D$. Then, the admissibility condition means that this one-forms are linearly independent at any point. Moreover, the first of the equations~\eqref{eq:nonholonomic_SODE} can be written as
\begin{equation}
    \restr{\left(\contr{X}\omega_L - \dd E_L\right)}{D} = \lambda_a\, \tau_Q^\ast \mu^a\, .
\end{equation}
where $\lambda_a$ are Lagrange multipliers.
The vector fields
\begin{equation}
    Z^a = \sharp_{\omega_L} (\tau_Q^\ast \mu^a) 
\end{equation}
span $F^{\perp_{\omega_L}}=\sharp_{\omega_L} (F^\circ)$. In bundle coordinates,
\begin{equation}
    Z^a = \mu^a_i W^{ij} \parder{}{v^j}\, ,
\end{equation}
where $(W^{ij})$ denotes the inverse of the Hessian matrix $(W_{ij})$. Thus, a solution $\sode_{(L, D)}$ to the equations~\eqref{eq:nonholonomic_SODE} is of the form
\begin{equation}\label{eq:nh_vf}
    \sode_{(L, D)} = \sode_{L} + \lambda_a Z^a \, ,
\end{equation}
where $\sode_L$ is the Euler--Lagrange vector field of $(Q, L)$. 
The Lagrangian multipliers can be determined by imposing the tangency condition
\begin{equation}
    \restr{\sode_{(L, D)}}{D}\in\T D\, .
\end{equation}
From the expression~\eqref{eq:nh_vf}, it is clear that $\sode_{(L, D)}$ is a \textsc{sode}.

Given a curve $c\in \Omega(q_1, q_2, [a,b])$, the following statements are equivalent:
\begin{enumerate}
    \item $c$ is the trajectory between $q_1$ and $q_2$ according to Lagrange--d'Alembert--Hölder principle,
    \item $c$ is a solution of the \textsc{sode} $\sode_{(L, D)}$,
    \item $c$ satisfies the \emph{nonholonomic Euler--Lagrange equations}:
    \begin{equation}\label{eq:Euler-Lagrange-nh}
    \begin{aligned}
        & \frac{\dd }{\dd t} \parder{L}{v^i} \big(c(t), \dot{c}(t)\big) -  \parder{L}{q^i} \big(c(t), \dot{c}(t)\big) = \lambda_a\big(c(t)\big)\, \mu^a_i\big(c(t)\big) \, \\
        & \mu^a\big(c(t), \dot{c}(t)\big) = 0\, .
    \end{aligned}
    \end{equation}
\end{enumerate}

\section{Discrete mechanics}\label{sec:discrete_mechanics}

In this section, some aspects of discrete Lagrangian mechanics and variational integrators are briefly reviewed. The quintessential reference for these and related topics is the monographic by Marsden and West \cite{M.W2001}.

Frequently, equations of motion cannot be integrated analytically, becoming necessary the study them by means of other techniques such as numerical methods.
A straightforward way in which one could study numerically a Lagrangian (respectively, Hamiltonian) system would be to compute the Euler--Lagrange (respectively, Hamilton) equations, and then discretize this system of ordinary differential equations, for instance, by a Runge--Kutta method. The usual numerical methods are designed to minimize the error of the approximated solution with respect to the exact solution of the differential equations, but they do not take into account other properties of the original dynamical system, such as rotational symmetry or conservation of the symplectic structure.
This lead to the introduction of the so-called geometric integrators, numerical integrators constructed to take into account the geometry underlying the dynamical system being discretized. In particular, if the continuous dynamics can be derived from a variational principle, a variational integrator is a discrete counterpart of the variational principle leading to the discrete dynamics.

Roughly speaking, the idea of discrete Lagrangian mechanics consists on replacing the tangent space $\T Q$ by $Q\times Q$, so that nearby points become the discrete analogue of a tangent vector. The action functional, given by the integral of the Lagrangian function along the curves solutions to the Euler--Lagrange equations, is replaced by a discrete action functional, given by the sum of the discrete Lagrangian evaluated along the discrete curves solving the discrete Euler--Lagrange equations. 

It is noteworthy that the discrete Lagrangian formalism may also be used to study discrete dynamical systems that do not necessarily have a continuous counterpart.
Although they will not be employed in this dissertation, it is worth mentioning that Lie groupoids are a natural geometric structure to study discrete mechanics (see \cite{M.D.M2006a, C.d.M+2006}). In this formalism, the role of the continuous counterpart is played by the associated Lie algebroid.

Let $Q$ be an $n$-dimensional manifold. A \emph{discrete Lagrangian (function)} is a function $L_d\in \Cinfty(Q\times Q)$. The pair $(Q, L_d)$ will be called a \emph{discrete Lagrangian system}. Consider an increasing sequence $\{t_k = kh\mid k=0, \ldots N\}\subset \RR$, which represents the sequence of times, with $h$ being regarded as the time step. Let $q_0$ and $q_N$ be two fixed points in $Q$. The \emph{discrete path space from $q_0$ to $q_N$} is the set 
\begin{equation}
    \Omega \left(q_0, q_N, \{t_k\}_{k=0}^N \right) 
    = \left\{q_d \colon\{t_k\}_{k=0}^N\to Q\mid q_d(0)=q_0\, ,\, q_d(N) = q_N \right\}\, .
\end{equation}
Each element $q_d$ of the discrete path space is called a \emph{discrete trajectory}, and it will be identified with its image $\{q_k\}_{k=0}^N$. 
The \emph{discrete action (map)} $\action_d \colon  \Omega \left(\{t_k\}_{k=0}^N\, , Q \right) \to \RR$ is defined by
\begin{equation}\label{eq:discrete_action}
    \action_d (q_d) = \sum_{k=0}^{N-1} L_d(q_k, q_{k+1})\, .
\end{equation}
The discrete path space is isomorphic to $Q^{N-1}$, and thus it can be endowed with a smooth product manifold structure. The smoothness of the discrete Lagrangian makes the discrete action map smooth as well. 

The tangent space $\T_{q_d}  \Omega \left(q_0, q_N, \{t_k\}_{k=0}^N \right)$ to $ \Omega \left(q_0, q_N, \{t_k\}_{k=0}^N \right)$ at $q_d$ is the set of maps $v_{q_d}\colon \{t_k\}_{k=0}^N \to \T Q$ such that $\pi_Q \circ v_{q_d} = q_d$ and $v_{q_d}(0) = v_{q_d}(t_N) = 0$. Each of these maps is called a \emph{discrete variation}, and it will be identified with its image, namely, $v_{q_d} = \{(q_k, v_k)\}_{k=0}^N$.

A discrete curve $q_d\in \Omega \left(q_0, q_N, \{t_k\}_{k=0}^N \right)$ satisfies the \emph{discrete Hamilton's principle} if it is a critical point of the discrete action, that is, $\dd \action_d(q_d)=0$. This occurs if and only if the discrete curve satisfies the \emph{discrete Euler--Lagrange equations}:
\begin{equation}
    \DD_2 L_d (q_k, q_{k-1}) + \DD_1 L_d (q_k, q_{k+1}) = 0\, ,
\end{equation}
where $\DD_1$ and $\DD_2$ denote the derivatives with respect to the first and second argument, respectively.

Let $(Q, L)$ be a regular Lagrangian system. Consider a sufficiently small time step $h\in \RR$ and a sufficiently small open subset $U\subseteq Q$ such that such that the Euler--Lagrange equations for $L$ with boundary conditions $c_{0, \, 1}(0) = q_0\in U$ and $c_{0, \, 1}(h) = q_{1}\in U$ have a unique solution $c_{0, \, 1}\colon [0, h]\to Q$. The \emph{exact discrete Lagrangian} $L_d^{\mathrm{ex}}\in \Cinfty(U\times U)$ is given by
\begin{equation}
    L_d^{\mathrm{ex}}(q_0, q_{1}) = \int_0^h L\big(c_{0, \, 1}(t), \dot{c}_{0, \, 1}(t)\big)\, \dd t\, .
\end{equation}
The exact discrete Lagrangian receives that name because the solutions of its discrete Euler--Lagrange equations are points in the solutions of the Euler--Lagrange equations for $L$. More precisely, solutions $c\colon [0, t_N]\to Q$ of the Euler--Lagrange equations for $L$ and solutions $\{q_k\}_{k=0}^N$ of the discrete Euler--Lagrange equations for $ L_d^{\mathrm{ex}}$ are related by
\begin{equation}
\begin{array}{ll}
    q_k = c(t_k)\, , \quad &  k=0, \ldots, N\, ,\\
    c(t) = c_{k,\, k+1} (t)\, , & t_k \leq t \leq t_{k+1}\, ,
\end{array}
\end{equation}
where $c_{k,\, k+1}\colon[t_k, t_{k+1}]\to Q$ is the unique solution of the Euler--Lagrange equations for $L$ satisfying $c_{k,\, k+1}(kh) = q_k$ and $c_{k,\, k+1}((k+1)h) = q_{k+1}$.

Of course, in practice is not useful to use the exact discrete Lagrangian, since its computation requires to know the solutions of the Euler--Lagrange equations, which are the equations that one wants to integrate numerically in the first place. However, there is a remarkable result stating that, compared with the solutions of the continuous Euler--Lagrange equations, the order of the error of the solutions discrete Euler--Lagrange equations coincides with the order of the approximation of a discrete Lagrangian with respect to the exact discrete Lagrangian, namely,
\begin{equation}
    L_d(q_k, q_{k+1}) = L_d^{\mathrm{ex}}(q_k, q_{k+1}) + \mathcal{O}(h^r) \Longrightarrow q_k = c(t_k) + \mathcal{O}(h^r) \, ,
\end{equation}
where $\{q_k\}$ is the solution of the discrete Euler--Lagrange equations for $L_d$ and $c$ the solution of the Euler--Lagrange equations for $L$. 

\part{Systems with external forces}\label{part:forces}
\chapter{Systems with external forces}\label{ch:forced}


In the present chapter, autonomous and non-autonomous forced Hamiltonian and Lagrangian systems with external forces are reviewed. Moreover, the Rayleigh potential is introduced. More details can be found in \cite{Godbillon1969,d.R1989,Minguzzi2015,Goldstein1980,Gantmakher1970,Lurie2002}.

\section{Forced Hamiltonian systems}


Along this section, let $Q$ be an $n$-dimensional manifold with local coordinates $(q^i)$. The bundle coordinates of the cotangent bundle $\pi_Q\colon\cT Q\to Q$ are denoted by $(q^i, p_i)$.
Let $\omega_Q\in \Omega^1(\cT Q)$ denote the canonical one-form, and let $\omega_Q = - \dd \theta_Q$ denote the canonical symplectic form. The Hamiltonian vector field of $f\in \Cinfty(\cT Q)$ with respect to $\omega_Q$ is denoted by $X_f$. 

A \emph{forced Hamiltonian system} is a triple $(Q, H, \alpha)$, formed by a manifold $Q$, a function $H\in \Cinfty(\cT Q)$, and a semibasic one-form $\alpha\in \Omega^1(\cT Q)$.
The function $H$ is called the \emph{Hamiltonian function}, and the semibasic one-form $\alpha$ is called the \emph{external force}. Let $Z_\alpha = \sharp_{\omega_Q} \alpha \in \X(\cT Q)$, that is, 
\begin{equation}
    \contr{Z_\alpha} \omega_Q = \alpha \, .
\end{equation}
The dynamics of $(Q, H, \alpha)$ is given by the vector field $X\Halpha = X_H + Z_\alpha$, that is,
\begin{equation}
    \contr{X\Halpha} \omega = \dd H + \alpha\, .
\end{equation}
This vector field is called the \emph{forced Hamiltonian vector field of $(H, \alpha)$}. 
In bundle coordinates, an external force is of the form
\begin{equation}
    \alpha = \alpha_i\, \dd q^i\, .
\end{equation} 
Then, 
\begin{equation}
    Z_\alpha = -\alpha_i \parder{}{p_i}\, ,
\end{equation}
and
\begin{equation}
    X\Halpha = \parder{H}{p_i} \parder{}{q^i} - \left(\parder{H}{q^i} + \alpha_i \right) \parder{}{p_i}\, .
\end{equation}
Observe that the vector field $Z_\alpha$ is vertical, namely, $\T \pi_Q (Z_\alpha) = 0$. A curve $c\colon I \subseteq \RR \to M, \, c(t) = (q^i(t), p_i(t))$ is an integral curve of $X\Halpha$ if an only if it satisfies the \emph{forced Hamilton equations}:
\begin{equation}\label{eq:forced_Hamilton_equations}
        \frac{\dd q^i(t)}{\dd t} = \parder{H}{p_i} \big(c(t)\big)\, , \quad 
        \frac{\dd p_i(t)}{\dd t} = -\parder{H}{q^i} \big(c(t)\big) - \alpha_i \big(c(t)\big) \, .
\end{equation} 

Let $(Q, H, \alpha)$ be a forced Hamiltonian system. 
A function $f\in \Cinfty(\cT Q)$ 
is a conserved quantity if and only if it is a solution of the partial differential equation
\begin{equation}
    X\Halpha(f) = 0\, .
\end{equation}
Unlike in Hamiltonian systems, the Hamiltonian function is not a conserved quantity in general. In fact,
\begin{equation}
    X\Halpha (H) = Z_\alpha (H)\, , 
\end{equation}
or, in bundle coordinates,
\begin{equation}
    X\Halpha (H) = -\alpha_i \parder{H}{p_i}\, .
\end{equation}
Moreover,
\begin{equation}
    \liedv{X\Halpha} \vol_{\omega_Q} = n \liedv{Z_\alpha} (\omega_Q) \wedge \omega_Q^{n-1} = n\dd \alpha \wedge \omega_Q^{n-1}\, ,
\end{equation}
and thus the flow of $X\Halpha$ preserves the volume form defined by the symplectic form $\omega_Q$ if and only if $\alpha$ is closed.

If $\alpha$ is exact, then there exists a function $f\in \Cinfty(\cT Q)$ such that $\alpha = \dd f$. Since $\alpha$ is semibasic, this function is of the form $f = \tilde{f} \circ \pi_Q$ for some function $\tilde{f} \in \Cinfty(Q)$. Then, the forced Hamiltonian vector field of $(H, \alpha)$ is equal to the Hamiltonian vector field of $H+f$, namely, $X\Halpha = X_{H+f} = X_H + X_f$. Therefore, the forced Hamiltonian system $(Q, H, \omega)$ is equivalent to the Hamiltonian system $(\cT Q, \omega_Q, H+f)$. By the Poincaré lemma, if $\alpha$ is closed then it is locally equivalent to a Hamiltonian system. The case of interest is the one in which $\alpha$ is not closed, in other words, it is non-conservative external force.

\section{Forced Lagrangian systems}

Along this section, the notations from \Cref{sec:Lagrangian_mechanics} will be employed.
A \emph{forced Lagrangian system} is a triple $(Q, L, \beta)$ such that the pair $(Q, L)$ is a Lagrangian system and $\beta\in \Omega^1(\T Q)$ is a semibasic one-form. This one-form is called the \emph{external force}. In bundle coordinates,
\begin{equation}
    \beta = \beta_i\, \dd q^i\, .
\end{equation}
Equivalently, an external force can be regarded as a bundle morphsim from $\T Q$ to $\cT Q$ (see \Cref{sec:semibasic}).

\emph{Lagrange--d'Alembert principle} states that, between two specified configurations $q_1=c(a),\, q_2=c(b)\in Q$ at two specified times $a, b\in \RR$, the dynamics of $(Q, L)$ is a curve $c\in \Omega(q_1, q_2, [a,b])$ such that
\begin{equation}
    \contr{v} \dd \action (c) = \int_a^b  \contr{v}\beta \circ c (t)\, \dd t\, ,
\end{equation} 
for any variation $v\in \T_c \Omega(q_1, q_2, [a,b])$, where $\action$ denotes the action functional \eqref{eq:action_functional}.  A curve $c\in \Omega(q_1, q_2, [a,b])$ satisfies the Lagrange--d’Alembert principle if and only if it verifies the \emph{forced Euler--Lagrange equations}:
\begin{equation}\label{eq:Euler-Lagrange_forced}
    \frac{\dd }{\dd t} \parder{L}{v^i} \big(c(t), \dot{c}(t)\big) -  \parder{L}{q^i} \big(c(t), \dot{c}(t)\big) = -\beta_i\big(c(t), \dot{c}(t)\big)\, .
\end{equation}

A forced Hamiltonian system $(Q, L, \beta)$ is called \emph{regular} (respectively, \emph{hyper-regular}) if $L$ is regular (respectively, hyper-regular). Hereinafter, every forced Lagrangian system will be assumed to be regular. 

Given a forced Hamiltonian system $(Q, L, \beta)$, consider the symplectic manifold $(\T Q, \omega_L)$, where $\omega_L$ is the Poincaré--Cartan 2-form defined by $L$. Consider the vector field $Z_\beta = \sharp_{\omega_L} \beta \in \X (\T Q)$, that is,
\begin{equation}
    \contr{Z_\beta} \omega_L = \beta\, .
\end{equation}
The dynamics of $(Q, L, \beta)$ is given by the vector field $\sode\Lbeta = \sode_L + Z_\beta$, where $\sode_L$ denotes the Euler--Lagrange vector field, namely,
\begin{equation}
    \contr{\sode\Lbeta} \omega_L = \dd E_L + \beta\, .
\end{equation}
The vector field $\sode\Lbeta$ is called the \emph{forced Euler--Lagrange vector field}. In bundle coordinates,
\begin{equation}
    Z_\beta = -\beta_i \parder{}{v^i}\, ,
\end{equation}
and
\begin{equation}
    \sode\Lbeta =  v^i\parder{}{q^i} + W^{ji}\left( \parder{L}{q^j} - v^k\parderr{L}{q^k}{v^j}-\beta_i   \right)\parder{}{v^i}\, ,
\end{equation}
where $(W^{ij})$ denotes the inverse of the Hessian matrix
\begin{equation}
    (W_{ij})=\left(\parderr{L}{v^i}{v^j}\right)\, .
\end{equation}
It is worth remarking that $Z_\beta$ is a vertical vector field and $\sode\Lbeta$ is a \textsc{sode}. Moreover, the solutions of $\sode\Lbeta$ satisfy the forced Euler--Lagrange equations~\eqref{eq:Euler-Lagrange_forced}. A function $f\in \Cinfty(\T Q)$ is a \emph{conserved quantity} for $(Q, L, \beta)$ if and only if $\sode\Lbeta (f) =0$.

If $(Q, L, \beta)$ is hyper-regular, then it is equivalent to the forced Hamiltonian system $(Q, H, \alpha)$ with
\begin{equation}
    H = \big((\FF L) ^{-1} \big)^\ast E_L\, , \quad \alpha = \big((\FF L) ^{-1} \big)^\ast \beta\, .
\end{equation}
In other words, $\FF L$ is a symplectomorphism between $(\T Q, \omega_L)$ and $(\cT Q, \omega_Q)$, and
\begin{equation}
    \sode\Lbeta = (\FF L)_\ast X\Halpha\, .
\end{equation}


\begin{definition}
    A forced Lagrangian system $(Q, L, \beta)$ is said to be \emph{Rayleighable} if there exists a function $\Rayl\in \Cinfty(\T Q)$ such that $\alpha = \Sendoadj\circ \dd \Rayl$, where $\Sendoadj$ denotes the adjoint operator of the vertical endomorphism. The function $\Rayl$ is called a \emph{Rayleigh potential} for $\beta$. The triple $(Q, L, \Rayl)$ is called a \emph{Rayleigh system}.
\end{definition}

Let $(Q, L, \beta)$ be a forced Lagrangian system and $\Rayl$ a Rayleigh potential for $\beta$. Then, the forced Euler--Lagrange vector field $\sode\Lbeta$ will be denoted by $\sode\LRayl$. In bundle coordinates,
\begin{equation}
    \beta = \parder{\Rayl}{v^i} \dd q^i\, ,
\end{equation}
and
\begin{equation}
    \sode\LRayl = v^i\parder{}{q^i} + W^{ji}\left( \parder{L}{q^j} - v^k\parderr{L}{q^k}{v^j}- \parder{\Rayl}{v^i}   \right)\parder{}{v^i}\, .
\end{equation}

\begin{remark}\label{remark:Rayleigh_gauge}
    If $\Rayl$ is a Rayleigh potential for $\beta$, then, for any $f\in \Cinfty(Q)$, the function $\Rayl + f\circ \tau_Q$ is also a Rayleigh potential for $\beta$.
\end{remark}

J.~W.~Strutt, 3rd Baron Rayleigh, introduced the so-called \emph{dissipation function} in order to study external forces which depend linearly on the velocities \cite{Strutt1871,Minguzzi2015,Goldstein1980,Gantmakher1970,Lurie2002}. This dissipation function is a Rayleigh potential given by a quadratic form on $\T Q$, namely, 
\begin{equation}
    \Rayl(v_q) = \frac{1}{2} \overline{\Rayl} (v_q, v_q)\, ,
\end{equation}
where $\overline{\Rayl} \in \tensors^0_2(\T Q)$ is a symmetric tensor field. In bundle coordinates,
\begin{equation}
    \overline{\Rayl} = \Rayl_{ij}\, \dd q^i \otimes \dd q^j\, ,
\end{equation}
and
\begin{equation}\label{eq:Rayleigh_quadratic}
    \Rayl = \frac{1}{2} \Rayl_{ij}\, v^i v^j\, .
\end{equation}
Therefore,
\begin{equation}
    \beta = \Sendoadj\circ \dd \Rayl = \Rayl_{ij}\, v^i \dd q^j\, .
\end{equation}

Given a Rayleigh system $(Q, L, \Rayl)$ with forced Euler--Lagrange vector field $\sode\LRayl$, the evolution of energy is given by
\begin{equation}
    \sode\LRayl (E_L) = \contr{\sode\LRayl} \contr{\sode_L} \omega_L 
    = - \contr{\sode_L} (\dd E_L + \Sendoadj \circ \dd \Rayl) 
    = -\Delta (\Rayl)\, . 
\end{equation}
In particular, if $\Rayl$ is of the form \eqref{eq:Rayleigh_quadratic}, then
\begin{equation}
    \sode\LRayl (E_L) = -2\Rayl\, .
\end{equation}

In order to study the conserved quantities of Rayleigh systems, the following bracket was introduced in \cite{deLeon2022b}.

\begin{definition}
    The \emph{dissipative bracket} is a bilinear map $[\cdot,\cdot]_L\colon \Cinfty (\T Q)\times \Cinfty (\T Q)\to \Cinfty (\T Q)$ given by
    \begin{equation}
      [f,g]_L = \Sendo \circ X_f(g)\, , \label{dissipative_bracket}
    \end{equation}
    where $\Sendo$ is the vertical endomorphism, and $X_f\in \X(\T Q)$ denotes the Hamiltonian vector field of $f$ with respect to $\omega_L$.
\end{definition}

\begin{proposition}
    The dissipative bracket $[\cdot,\cdot]$ on $(\T Q,\omega_L)$ verifies the following properties:
    \begin{enumerate}
    \item $[f,g]_L = [g,f]_L$ (it is symmetric),
    \item $[f,gh]_L = [f,h]_Lg+[f,g]_L h$ (it verifies the Leibniz rule),
    \end{enumerate}
    for all functions $f,g,h\in \Cinfty(\T Q)$.
\end{proposition}
\begin{proof}
    In local coordinates,
    \begin{equation}
    [f,g]_L = W^{ij} \frac{\partial f} {\partial v^j} \frac{\partial g} {\partial v^i}\, ,
    \end{equation}
    where $(W^{ij})$ is the inverse matrix of the Hessian matrix $(W_{ij})$ of the Lagrangian function $L$. From this expression both assertions can be easily proven.
\end{proof}

Since the dissipative bracket is bilinear and verifies the Leibniz rule, it is a derivation, or a so-called Leibniz bracket \cite{O.P2004}.

\begin{proposition}
  Given a Rayleigh system $(Q, L, \Rayl)$, a function $f\in \Cinfty(\T Q)$ is a constant of the motion if and only if
  \begin{equation}
    \left\{f,E_L  \right\}_{\omega_L} - [f,\Rayl]_L = 0\, ,
    \label{double_bracket_Rayleigh}
  \end{equation}
  where $\left\{\cdot,\cdot  \right\}_{\omega_L}$ denotes the Poisson bracket defined by $\omega_L$.
\end{proposition}

\begin{proof}
    As a matter of fact, if $\beta = \Sendoadj \circ \dd \Rayl$ is the external force defined by $\Rayl$, then
    \begin{equation}
    \begin{aligned} 
        [f,\Rayl]_L
        &= \Sendo \circ X_f (\Rayl)
        = \contr{\Sendo \circ X_f} \dd \Rayl
        = \contr{X_f} (\Sendoadj \circ \dd \Rayl )
        = \contr{X_f} \beta\\
        &= \contr{X_f} \contr{Z_{\beta}} \omega_L
        = -\contr{Z_{\beta}} \contr{X_f} \omega_L
        = -\contr{Z_{\beta}} \dd f
        = -Z_{\beta}(f)\, ,
    \end{aligned}
    \end{equation}
    and
    \begin{equation}
    \left\{f,E_L  \right\}
    = \omega_L(X_f,\sode_{L})
    = \contr{\sode_{L}}\contr{X_f}\omega_L
    = \sode_{L}(f)\, ,
    \end{equation}
    Thus
    \begin{equation}\begin{aligned}
    \left\{f,E_L  \right\} - [f,\Rayl]
    = \sode_{L}(f) + Z_{\beta}(f)
    = \sode_{L, \beta} (f)\, . \label{eq:proof_bracket_conserved}
    \end{aligned}\end{equation}
    In particular, the right-hand side vanishes if and only if $f$ is a constant of the motion.
\end{proof}

    Other types of dissipative systems, particularly thermodynamical systems, exhibit a ``double bracket'' dissipation, that is, their dynamics are described in terms of two brackets. As a matter of fact, the dissipative bracket defined above has certain similarities with other types of brackets.

    The dissipative bracket $[\cdot,\cdot]_L$ resembles the dissipative bracket $(\cdot,\cdot)$ appearing in the metriplectic framework \cite{C.M2020,B.B.P+2007}. Both brackets are symmetric and bilinear. However, the latter requires the additional assumption that $(E_L,f)$ vanishes identically for every function $f$ on $TQ$. Clearly, this requirement does not hold for $[\cdot,\cdot]_L$.

    On the other hand, the dissipative bracket $[\cdot,\cdot]_L$ can also be related with the so-called Ginzburg-Landau bracket $[\cdot,\cdot]_{\mathrm{GL}}$ \cite{G.O1997}. This bracket, together with symmetry and bilinearity, satisfies the positivity condition, that is, $[f,f]_{\mathrm{GL}}$ is a non-negative function for all $f\in \Cinfty(\T Q)$. 
    See also \cite{B.K.M+1996a} for various types of systems with double bracket dissipation.


\section{Non-autonomous forced Hamiltonian systems}

Let $\pi_2\colon \RR \times \cT Q \to \cT Q$ be the canonical projection, and let $(t, q^i, p_i)$ be the canonical bundle coordinates on $\RR \times \cT Q$. Consider the cosymplectic manifold $(\RR \times \cT Q, \dd t, \omega)$, where $\omega = \pi_2^\ast \omega_Q$. The Reeb vector field of $(\dd t, \omega)$ is $\Reeb = \tparder{}{t}$.
The Hamiltonian and evolution vector fields of $f\in \Cinfty(\RR \times \cT Q)$ with respect to $(\dd t, \omega)$ are denoted by $X_f$ and $\evol_f$, respectively.

A \emph{non-autonomous} (or \emph{time-depedent}) \emph{forced Hamiltonian system} is a triple $(Q, H, \alpha)$, where $Q$ is an $n$-dimensional manifold, $H\in \Cinfty(\RR \times\cT Q)$ is a function and $\alpha \in \Omega^1(\RR\times \cT Q)$ is a one-form such that $\alpha$ is semibasic with respect to the vector bundle $\RR \times \cT Q \to Q$.
The function $H$ and the one-form $\alpha$ are called the \emph{Hamiltonian function} and the \emph{external force}, respectively. Define the vector field $Z_\alpha = \sharp\dtomega \alpha\in \X(\RR \times \cT Q)$. Equivalently,
\begin{equation}
    \contr{Z_\alpha} \dd t = 0\, , \quad
    \contr{Z_\alpha} \omega = \alpha\, ,
\end{equation}
since $\alpha$ is semibasic. The dynamics of $(Q, H, \alpha)$ is given by the vector field $\evol\Halpha = \evol_H + Z_\alpha$, or, equivalently
\begin{equation}
    \contr{\evol\Halpha} \dd t = 1\, , \quad
    \contr{\evol\Halpha} \omega = \dd H + \alpha - \Reeb(H) \dd t\, .
\end{equation}
In bundle coordinates, a (time-dependent) external force is of the form
\begin{equation}
    \alpha = \alpha_i\, \dd q^i\, .
\end{equation}
Therefore,
\begin{equation}
    Z_\alpha = -\alpha_i \parder{}{p_i}\, ,
\end{equation}
and
\begin{equation}
    \evol\Halpha = \parder{}{t} + \parder{H}{p_i} \parder{}{q^i} - \left(\parder{H}{q^i} + \alpha_i \right) \parder{}{p_i}\, .
\end{equation}

\section{Non-autonomous forced Lagrangian systems}

Consider the cosymplectic manifold $(\RR \times \T Q, \dd t, \omega_L)$, where $\omega_L = \dd (\Sendoadj\circ \dd L)$, with $\Sendoadj$ the adjoint operator of the natural extension of the vertical endomorphism to $\RR \times \T Q$. The Reeb vector field of $(\dd t, \omega_L)$ is denoted by $\Reeb_L$. The Hamiltonian and evolution vector fields of $f\in \Cinfty(\RR \times \T Q)$ with respect to $(\dd t, \omega_L)$ are denoted by $X_f$ and $\evol_f$, respectively.

A \emph{non-autonomous} (or \emph{time-depedent}) \emph{forced Lagrangian system} is a triple $(Q, L, \beta)$, formed by a manifold $Q$, a function $L\in \Cinfty(\RR \times \T Q)$ and a one-form $\beta \in \Omega^1(\RR\times \cT Q)$ such that $\beta$ is semibasic with respect to the vector bundle $\RR \times \T Q \to Q$. The function $L$ and the one-form $\beta$ are called the \emph{Lagrangian function} and the \emph{external force}, respectively.
 Define the vector field $Z_\beta = \sharp\dtomega \beta\in \X(\RR \times \cT Q)$. Equivalently,
\begin{equation}
    \contr{Z_\beta} \dd t = 0\, , \quad
    \contr{Z_\beta} \omega_L = \beta\, ,
\end{equation}
since $\beta$ is semibasic. The dynamics of $(Q, L, \beta)$ is given by the vector field $\sode\Lbeta = \sode_L + Z_\beta$, or, equivalently
\begin{equation}
    \contr{\evol\Halpha} \dd t = 1\, , \quad
    \contr{\evol\Halpha} \omega = \dd H + \beta - \Reeb(H) \dd t\, .
\end{equation}
In bundle coordinates, a (time-dependent) external force is of the form
\begin{equation}
    \beta = \beta_i\, \dd q^i\, .
\end{equation}
Therefore,
\begin{equation}
    Z_\beta = -\beta_i \parder{}{p_i}\, ,
\end{equation}
and
\begin{equation}
    \evol\Halpha = \parder{}{t} + \parder{H}{p_i} \parder{}{q^i} - \left(\parder{H}{q^i} + \beta_i \right) \parder{}{p_i}\, .
\end{equation}

\chapter{Symmetries and reduction of forced mechanical systems}\label{ch:forced_symmetries}


\insquote{Symmetry is a vast subject, significant in art and nature. Mathematics lies at its root, and it would be hard to find a better one on which to demonstrate the working of the mathematical intellect.}{Hermann Weyl, \emph{Symmetry} (1952)}

This chapter is devoted to forced Lagrangian and Hamiltonian systems with symmetries. Different types of symmetries are introduced, considering the relations between them and the associated constants of the motion. Furthermore, a theory for the reduction of forced Lagrangian systems which are invariant under the action of a Lie group is presented. The results are particularized for Rayleigh systems. 

Noether's theorems for forced Lagrangian systems had been previously studied by Bahar and Kwatny \cite{B.K1987} and 
Djukic and Vujanovic \cite{D.V1975}, albeit from a variational point of view without considering the geometry. Cantrijn \cite{Cantrijn1982} considered time-dependent Lagrangian systems and defined a 2-form on $\T Q \times \RR$ that depends on the Poincaré--Cartan 2-form defined by the Lagrangian function and on the external force. Moreover, van der Schaft \cite{vanderSchaft1981,vanderSchaft1983} studied the symmetries and conservation laws for forced Hamiltonian systems making use of a framework from system theory, in which an ``observation'' manifold, together with the state space $Q$, is considered. The advantage of the formalism presented here is that, compared with the study of symmetries for unforced systems, no additional structure or objects have to be introduced besides the proper external force.

The results from this chapter were previously published in the article \cite{deLeon2021a}. They were also included in the master's thesis from the author \cite{Lopez-Gordon2021}. Nevertheless, for the sake of completeness, they have been incorporated to the present dissertation.

\section{Forced Hamiltonian systems with symmetries}

Throughout this section, consider the forced Hamiltonian system $(Q, H, \alpha)$ with forced Hamiltonian vector field $X_{H, \alpha}$. For each function $f\in \Cinfty(\cT Q)$, let $X_f\in \X(\cT Q)$ denote its Hamiltonian vector field with respect to the canonical symplectic form $\omega_Q$.

Let $\{\cdot, \cdot\}$ denote the Poisson bracket defined by $\omega_Q$. A function $f\in \Cinfty(\cT Q)$ is a constant of the motion with respect to $(Q, H, \alpha)$ if and only if
\begin{equation}
    \{f, H\} = \alpha(X_f)\, .
\end{equation}

\begin{definition}
    A \emph{dynamical symmetry} is a vector field $Y\in \X(\cT Q)$ which commutes with the forced Hamiltonian vector field, namely,
    \begin{equation}
        \liedv{Y} X_{H, \alpha} = [Y, X_{H, \alpha}] = 0\, .
    \end{equation}
\end{definition}

\begin{proposition}\label{prop:Cartan_symmetry_symplectomorphism}
    An infinitesimal symplectomorphism $Y\in \X(\cT Q)$ is a dynamical symmetry if an only if
    \begin{equation}
        \liedv{Y} (\dd H + \alpha) = 0 \, .
    \end{equation}
\end{proposition}

\begin{proof}
    Indeed,
    \begin{equation}
    \begin{aligned}
        \flat_{\omega_Q} \big( [Y, X_{H, \alpha}]\big)& = \contr{[Y, X_{H, \alpha}]} \omega_Q
         = \liedv{Y} \contr{X_{H, \alpha}} \omega_Q - \contr{X_{H, \alpha}} \liedv{Y} \omega_Q\\
         & = \liedv{Y} \left(\dd H + \alpha \right)
         \, .
    \end{aligned}
    \end{equation}
    Since $\flat_{\omega_Q}\colon \X(\cT Q) \to \Omega^1(\cT Q)$ is an isomorphism, $Y$ is a dynamical symmetry if and only if the right-hand side term vanishes. 
\end{proof}

\begin{definition}\label{def:Cartan_symmetry_forced}
    A \emph{Cartan symmetry} is a vector field $Y\in \X(\cT Q)$ such that
    \begin{enumerate}
        \item $Y$ is an infinitesimal symplectomorphism, that is, $\liedv{Y} \omega_Q = 0$,
        \item $Y(H) + \alpha (Y) = 0$.
    \end{enumerate}
\end{definition}

In the case of unforced Hamiltonian systems, every Cartan symmetry is also a dynamical symmetry. However, in the case of forced Hamiltonian systems, this is not always the case. As a matter of fact, there is the following relation between both types of symmetries.

\begin{proposition}
    A Cartan symmetry $Y\in \X(\cT Q)$ is a dynamical symmetry if and only if
    \begin{equation}
        \contr{Y} \dd \alpha = 0\, .
    \end{equation}
\end{proposition}
The proof follows directly from \Cref{prop:Cartan_symmetry_symplectomorphism} and \Cref{def:Cartan_symmetry_forced}.

\begin{proposition}\label{prop:Cartan_conserved_forced}
    Let $Y\in \X(\cT Q)$ be a vector field such that
    \begin{equation}\label{eq:prop_Cartan_conserved_forced}
        \liedv{Y} \theta_Q = \dd f\, ,
    \end{equation}
    for a function $f\in \Cinfty(\cT Q)$. Then, $Y$ is a Cartan symmetry if and only if $f-\theta_Q(Y)$ is a conserved quantity.
\end{proposition}

\begin{proof}
    By taking the exterior derivative on both sides of equation~\eqref{eq:prop_Cartan_conserved_forced}, it is clear that $Y$ is an infinitesimal symplectomorphism. Moreover, one can write
    \begin{align}
        \dd f =  \liedv{Y} \theta_Q = \contr{Y} \dd \theta_Q + \dd \contr{Y} \theta_Q\, ,
    \end{align}
    that is,
    \begin{equation}
        \dd \big(f - \theta_Q(Y)\big) = -\contr{Y} \omega_Q\, .
    \end{equation}
    Contracting both sides with $X_{H, \alpha}$ yields
    \begin{equation}
        X_{H, \alpha} \big(f - \theta_Q(Y)\big) = \contr{Y} \contr{X_{H, \alpha}} \omega_Q
        = \contr{Y} (\dd H + \alpha) = Y(H) + \alpha(Y)\, .
    \end{equation}
    In particular, the left-hand side vanishes if and only if the right-hand side does, and the result follows.
\end{proof}

By Poincaré lemma, every infinitesimal symplectomorphism satisfies equation~\eqref{eq:prop_Cartan_conserved_forced}, at least, locally. Consequently, every Cartan symmetry has an associated local conserved quantity.

\section{Forced Lagrangian systems with symmetries}

Throughout this section, the forced Lagrangian system $(Q, L, \beta)$, with forced Euler--Lagrange vector field $\sode_{L, \beta}$, will be considered. For each function $f\in \Cinfty(\T Q)$, let $X_f\in \X(\T Q)$ denote its Hamiltonian vector field with respect to the Poinceré--Cartan 2-form $\omega_L$.

The definitions and results from the previous section, \textit{mutatis mutandis}, can also be applied to forced Lagrangian systems. 
 Additionally, it is possible to consider natural symmetries, that is, those which are the complete lift from a vector field on $Q$, as well as the so-called symmetries of the forced Lagrangian.

\begin{definition}
    A vector field $Y\in \X(\T Q)$ is called a:
    \begin{enumerate}
        \item \emph{natural dynamical symmetry} (or a \emph{Lie symmetry}) if it is a dynamical symmetry and $Y=X^\Com$ for a vector field $X\in \X(Q)$,
        \item \emph{natural Cartan symmetry} (or a \emph{Noether symmetry}) if it is a Cartan symmetry and $Y=X^\Com$ for a vector field $X\in \X(Q)$,
        \item \emph{symmetry of the forced Lagrangian} if $Y(L)=\beta(Y)$,
        \item \emph{natural symmetry of the forced Lagrangian} if it is a symmetry of the forced Lagrangian and $Y=X^\Com$ for a vector field $X\in \X(Q)$.
    \end{enumerate}
\end{definition}

It is easy to see that a (natural) Cartan symmetry $Y\in \X(\T Q)$ is a (natural) symmetry of the forced Lagrangian if and only if $Y(\Delta (L)) = 0$. Conversely, a (natural) symmetry of the forced Lagrangian is a (natural) Cartan symmetry if and only if it is an infinitesimal symplectomorphism.

\begin{theorem}[Noether's theorem for forced Lagrangian systems]\label{theorem:Noether_forced}
    Let $X\in \X(Q)$ be a vector field on $Q$, with complete lift $X^\Com$ and vertical lift $X^\V$. Then, $Y=X^\Com$ is a natural symmetry of the forced Lagrangian if and only if $X^\V(L)$ is a conserved quantity.
\end{theorem}

\begin{proof}
    For any vector field $Y\in \X(\T Q)$, one can write
    \begin{equation}
    \begin{aligned}
        Y(E_L) + \beta(Y) 
        & = \contr{Y} \contr{\sode_{L, \beta}} \omega_L 
        = - \contr{Y} \liedv{\sode_{L, \beta}} \theta_L + \contr{Y} \dd \contr{\sode_{L, \beta}} \theta_L \\
        & = \contr{[\sode_{L, \beta}, Y]} - \liedv{\sode_{L, \beta}} \contr{Y} \theta_L + \contr{Y} \dd \contr{\sode_{L, \beta}} \theta_L \\
        & = \Sendo \circ [\sode_{L, \beta}, Y] (L) - \sode_{L, \beta} \big(\Sendo \circ Y (L)\big) \\
        & \quad + Y \big(\Sendo \circ \sode_{L, \beta} (L)\big) \\
        & = \Sendo \circ [\sode_{L, \beta}, Y] (L) - \sode_{L, \beta} \big(\Sendo \circ Y (L)\big) 
        + Y \big(\Delta (L)\big) \, ,
    \end{aligned}
    \end{equation}
    where on the last step it has been used that $\sode_{L, \beta}$ is a \textsc{sode}. Therefore,
    \begin{equation}
        \sode_{L, \beta} \big(\Sendo \circ Y (L)\big) = Y(L) - \beta(Y) + \Sendo \circ [\sode_{L, \beta}, Y] (L)\, .
    \end{equation}
    If $Y=X^\Com$ for a vector field $X\in \X(Q)$, then, $[\sode, Y]$ is a vertical vector field for any \textsc{sode} $\sode\in \X(\T Q)$. Therefore, $\Sendo \circ [\sode_{L, \beta}, X^\Com]$ vanishes. In addition, $\Sendo \circ X^\Com = X^\V$ and the result follows.
\end{proof}

\begin{example}[cyclic coordinate]\label{example:cyclic_coord_forced}
    Let $(\RR^n, L, \beta)$ be a forced Lagrangian system, and denote by $(q^i, v^i)$ the canonical bundle coordinates in $\T (\RR^n)$. Suppose that $\tparder{L}{q^1} = 0$ and $\beta$ is of the form 
    \begin{equation}
        \beta = \sum_{k=2}^n \beta_k \dd q^k\, .
    \end{equation}
    Then, $p_1\coloneqq \tparder{L}{v^1}$ is a constant of the motion. This can be verified directly from the forced Euler--Lagrange equations \eqref{eq:Euler-Lagrange_forced}. Equivalently, the complete lift $Y = X^\Com  = \tparder{}{q^1}$ of the vector field $X = \tparder{}{q^1}\in \X(\RR^n)$, is a natural symmetry of the forced Lagrangian and $p_1 = X^\V (L)$ is its associated conserved quantity.
\end{example}


\subsection{Rayleigh systems with symmetries}

Along this subsection, consider a Rayleigh system $(Q, L, \Rayl)$, and let $\alpha = \Sendoadj \circ \dd \Rayl$ be the external force. 

For any vector field $Y\in \X(\T Q)$, one can write $\alpha(Y) = \Sendo\circ Y (\Rayl)$. In particular, if $Y=X^\Com$ for $X\in \X(Q)$, then $\alpha(Y) = X^\V (\Rayl)$. \Cref{theorem:Noether_forced} implies the following.

\begin{proposition}
	Let $X\in \X(Q)$ be a vector field on $Q$. Then, the following assertions are equivalent:
	\begin{enumerate}
		\item $X^\Com$ is a natural symmetry of the forced Lagrangian $(Q, L, \alpha)$,
		\item $X^\Com(L)=X^\V(\Rayl)$,
		\item $X^\V(L)$ is a conserved quantity.
	\end{enumerate}
\end{proposition}

\begin{example}[Drag force]
    Consider a body of mass $m$ moving through a fluid that fully encloses it. For the sake of simplicity, suppose that the motion takes place along one dimension. Let $(q, v)$ be the canonical coordinates of $\T \R$.
    Then, the drag force of the fluid is given by
    \begin{equation}
      \alpha
      = \frac{1}{2} \rho CA v^2 \dd q\, ,
    \end{equation}
    where $C$ is a dimensionless constant depending on the body shape, $\rho$ is the mass density of the fluid, and $A$ is the area of the projection of the object on a plane perpendicular to the direction of motion (see \cite{Batchelor2000,Falkovich2011}). For further simplification, suppose that the density is uniform, and then $k=CA\rho/2$ is constant.
    In that case, $\alpha$ is Rayleighable and 
    \begin{equation}
      \Rayl=\frac{k}{3}v^3
    \end{equation}
    is a Rayleigh potential for it.
    If the body is not subject to forces besides the drag, its Lagrangian function is $L=m v^2/2$. The vector field
    \begin{equation}
        X=e^{kq/m } \parder{}{q}
    \end{equation}
    verifies that $X^\Com(L)=X^\V(\Rayl)$, and thus $X^\V(L)=me^{kq/m}v$ is a conserved quantity. In particular, if $k=0$, the vector field $X$ is the generator of translations and the conservation of linear momentum is recovered.
\end{example}

\begin{proposition}
    Let $Y\in \X(\T Q)$ be a vector field such that
    \begin{equation}
        \liedv{Y} \theta_L = \dd f\, ,
    \end{equation}
    for a function $f\in \Cinfty(\T Q)$. Then, the following statements are equivalent:
     \begin{enumerate}
     \item $Y$ is a Cartan symmetry,
     \item $Y(E_L)+\Sendo\circ Y (\Rayl)=0$,
     \item $f- \Sendo \circ Y(L)$ is a conserved quantity.
     \end{enumerate}
\end{proposition}
    
The following examples of Rayleigh systems are due to Minguzzi~\cite{Minguzzi2015}. 
\begin{example}[A rotating disk]
    Let $(\varphi, v)$ be bundle coordinates in $\T \Sp^1$. Consider the Rayleigh system $(\Sp^1, L, \Rayl)$, with
    \begin{equation}
        L = \frac{mr^{2}}{4} v^{2}\, , \quad \Rayl = \frac{\mu mgr}{2} v\, ,
    \end{equation}
    where $m, r, g$ and $\mu$ are constants. The Poincaré--Cartan one-form is $\theta_L=mr^2 v/2\  \dd \varphi$, and the external force is $\alpha=\mu mgr/2\ \dd \varphi$.

    Consider the vector field
    \begin{equation}
        Y = -rv\parder{}{\varphi} + \mu g \parder{}{v}\, .
    \end{equation}
    Clearly, $Y(E_L)=Y(L)=-\Sendo \circ Y(\Rayl)$. In addition,
    \begin{equation}
    \liedv{Y}\theta_L=
    \frac{\mu mgr^2}{2} \dd \varphi-
    \frac{mr^3}{2}v \dd v=\dd f,
    \end{equation}
    where
    \begin{equation}
    f=\frac{\mu mgr^2}{2}\varphi -\frac{mr^3}{4}v^2
    \end{equation}
    up to an additive constant. Moreover, $\Sendo \circ Y(L)=-mr^{3} v^{2}/2$,
    and thus
    \begin{equation}
    f-\Sendo \circ Y (L)=\frac{\mu mgr^2}{2}\varphi +\frac{mr^3}{4}v^2
    \end{equation}
    is a constant of the motion. Since $\alpha$ is closed, $\contr{Y}\dd \alpha=0$ is trivially satisfied. Hence, $Y$ is a dynamical symmetry as well as a Cartan symmetry. 

    However, since $\alpha$ is closed, it is a conservative force. In fact, the Euler--Lagrange equations for the Lagrangian function
    \begin{equation}
    \tilde{L}=L+\frac{\mu mgr}{2} \varphi
    \end{equation}
    coincide with the forced Euler--Lagrange equations for $(L, \alpha)$.
\end{example}

\begin{example}[The rotating stone polisher]
    Let $(x, y, \varphi, \psi, v_x, v_y, v_\varphi, v_\psi)$ be bundle coordinates in $\T(\RR^2 \times \TT^2)$. Consider the Rayleigh system $(\RR^2 \times \TT^2, L, \Rayl)$, with
    \begin{equation}
        L = \frac{m}{2}(v_x^{2}+v_y^{2} +r^{2}v_\varphi^{2}+r^{2}v_\psi^{2})\, ,
    \end{equation}
    and
    \begin{equation}
        \Rayl=2\mu mg rv_\psi +\frac{\mu mg}{2rv_\psi}(v_x^{2}+v_y^{2})\, , 
    \end{equation}
    where $m, r, g$ and $\mu$ are constants.
    The Poincaré--Cartan one-form is $\theta_L=2m(v_x\dd  x+ v_y \dd  y +r^2 v_\varphi \dd  \varphi)$, and the external force is
    \begin{equation}
        \alpha=\frac{\mu mg}{rv_\psi}(v_x\dd x + v_y \dd y).
    \end{equation}

    Consider the vector fields
    \begin{equation}
        Y_1=-2rv_\psi \parder{}{x} +\mu g \parder{}{v_x}\, , \quad
        Y_2=-2rv_\psi \parder{}{y} +\mu g  \parder{}{v_y}\, .
    \end{equation}
    One can check that $Y_a(E_L)=Y_a(L)=-S\circ Y_a(\Rayl)$ (for $a=1,2$). Moreover, $\liedv{Y_a}\theta_L=\dd f_a$,where $f_1=2\mu mgx$ and $f_2=2\mu mgy$. Additionally, $S\circ Y_1(L)=-4mr v_x$ and $S\circ Y_2(L)=-4mr v_y$. Thus, $2mrv_\psi v_x+\mu mgx$ and $2mrv_\psi v_y+\mu mgy$ are constants of the motion.
\end{example}

\section{Reduction of forced Lagrangian systems}


Let $G$ be a Lie group with Lie algebra $\mathfrak{g}$, and denote by $\mathfrak{g}^\ast$ the dual of $\mathfrak{g}$. 
Lete $(Q, L, \beta)$ be a forced Lagrangian system. Consider the Lie group action $\Phi\colon G \times Q \to Q$ of $G$ on $Q$, and its tangent lift $\Phi^\T\colon G \times \T Q \to \T Q$. Assume that the tangent lift is a free and proper action of $G$ on $\T Q$. 

Suppose that the Lagrangian function $L$ is invariant under the action $\Phi^\T$, namely, $L\circ \Phi^\T_g = L$ for all $g\in G$. Since $\Phi^\T$ is a tangent lift, it also preserves the vertical endomorphism $\Sendo$. Hence, $\Phi^T$ is an exact symplectic action for the Poincaré--Cartan one-form $\theta_L$. Therefore, the natural momentum map $\mommap\colon M \to \mathfrak{g}^\ast$, given by
\begin{equation}
	\mommap^\xi(x) = \left\langle \mommap(x), \xi \right \rangle = \left( \contr{\xi_{\T Q}} \theta_L  \right)(x)\, ,
\end{equation}
is an $\Ad^\ast$-equivariant momentum map. For each $\xi \in \mathfrak{g}$, the infinitesimal generator of its action on $\T Q$ is the complete lift of its infinitesimal generator on $Q$, namely, $\xi_{\T Q}= \xi_Q^\Com$. Therefore, 
$\theta_L(\xi_{\T Q}) = \Sendo \circ \xi_Q^\Com (L) = \xi_Q^\V (L)$, and then
\begin{equation}
	\mommap^\xi = \xi_Q^\V (L)\, ,
\end{equation}
for each $\xi \in \mathfrak{g}$.

\begin{proposition}\label{prop:Lie_subalgebra_forced_Lagrangian}
    For each $\xi \in \mathfrak{g}$, the function $\mommap^\xi$ is a conserved quantity if and only if
    \begin{equation}
        \beta(\xi_Q^\Com) = 0\, .
    \end{equation}
    If this equation holds, a necessary and sufficient condition for the infinitesimal action generated by $\xi$ to leave $\beta$ invariant is
    \begin{equation}
        \contr{\xi_Q^\Com} \dd \beta = 0\, .
    \end{equation}
    Moreover, the subset 
    \begin{equation}
        \mathfrak{g}_\beta = \left\{\xi \in \mathfrak{g} \mid \beta(\xi_Q^\Com) = 0, \, \contr{\xi_Q^\Com} \dd \beta = 0\right\}
    \end{equation}
    is a Lie subalgebra of $\mathfrak{g}$.
\end{proposition}

\begin{proof}
    By \Cref{theorem:Noether_forced}, $\mommap^\xi = \xi_Q^\V (L)$ is a conserved quantity if and only if $\xi_Q^\Com$ is a natural symmetry of the forced Lagrangian, that is, $\xi_Q^\Com(L) = \beta(\xi_Q^\Com)$. It is assumed that the action $\Phi^\T$ preserves the Lagrangian function. This implies that $\xi_Q^\Com (L) = 0$ for any $\xi \in \mathfrak{g}$. Therefore, $\mommap^\xi$ is a conserved quantity if and only if $\beta(\xi_Q^\Com)$ vanishes.

    If $\beta(\xi_Q^\Com) = 0$, then $\beta$ is $\xi$-invariant (that is, $\liedv{\xi_Q^\Com} \beta = 0$) if and only if $\contr{\xi_Q^\Com} \dd \beta = 0$ vanishes.

    For each $\xi, \chi \in\mathfrak{g}_\beta$, 
    \begin{equation}
    \begin{aligned}
        \contr{[\xi_Q^\Com, \chi_Q^\Com]} \beta 
        = \liedv{\xi_Q^\Com} \contr{\chi_Q^\Com} \beta - \contr{\chi_Q^\Com} \liedv{\xi_Q^\Com} \beta 
        = - \contr{\chi_Q^\Com} \dd \contr{\xi_Q^\Com} \beta - \contr{\chi_Q^\Com} \contr{\xi_Q^\Com} \dd \beta
        = 0\, , 
    \end{aligned}
    \end{equation}
    and
    \begin{equation}
        \begin{aligned}
            \contr{[\xi_Q^\Com, \chi_Q^\Com]} \dd\beta 
            = \liedv{\xi_Q^\Com} \contr{\chi_Q^\Com} \dd\beta - \contr{\chi_Q^\Com} \liedv{\xi_Q^\Com} \dd\beta 
            = - \contr{\chi_Q^\Com} \dd \contr{\xi_Q^\Com} \dd \beta 
            = 0\, . 
    \end{aligned}
    \end{equation}
    Since $\xi\in \mathfrak{g} \mapsto \xi_Q\in \X(Q)$ is a Lie algebra anti-homomorphism and $X \in\X (Q)\mapsto  \X^\Com \in \X(\T Q)$ is a Lie algebra homomorphism, one can write
    \begin{equation}
        [\xi_Q^\Com, \chi_Q^\Com] = \big([\xi, \chi]\big)_Q^\Com\, .
    \end{equation}
    Thus, $[\xi, \chi]\in \mathfrak{g}_\beta$ for any $\xi, \chi \in\mathfrak{g}_\beta$; in other words, $g_\beta$ is a Lie subalgebra.
\end{proof}

Observe that the infinitesimal generator $\xi_Q^\Com$ of each $\xi\in \mathfrak{g}_\beta$ is a natural symmetry of the forced Lagrangian and a natural Cartan symmetry. 

Since $\mathfrak{g}_\beta$ is a Lie subalgebra of $\mathfrak{g}$, there exists a unique connected Lie subgroup $G_\beta$ of $G$ whose Lie algebra is $\mathfrak{g}_\beta$. 
The action $\Phi^\T\colon G \times \T Q \to \T Q$ is assumed to be free, which implies that its restriction $\restr{\Phi^\T}{G_\beta\times \T Q}$ is a free action of $G_\beta$ on $\T Q$. Moreover, if $G_\beta$ is closed Lie subgroup of $G$, then the action $\restr{\Phi^\T}{G_\beta\times \T Q}$ is proper.

Let $\mommap_\beta\colon \T Q \to \mathfrak{g}_\beta^\ast$ denote the momentum map given by the restriction of the natural momentum map $\mommap$. Let $\mu$ be a regular value of $\mommap_\beta$ (that is, $(\T \mommap_\beta)_v\colon \T_v \T Q\to \T \mathfrak{g}^\ast$ is surjective for every $v\in \mommap_\beta^{-1}(\mu)$). Then, $\mommap^{-1}(\mu)$ is a submanifold. The fact that, for each $\xi\in \mathfrak{g}_\beta$, the function $\mommap^\xi$ is a conserved quantity implies that the forced Euler--Lagrange vector field is tangent to $\mommap^{-1}(\mu)$.

It is clear that $\beta$ is preserved by $\restr{\Phi^\T}{G_\beta\times \T Q}$, that is, $\restr{\Phi^\T}{G_\beta\times \T Q}^\ast \beta = \beta$. In addition, since the action is a tangent lift it preserves the Liouville vector field $\Delta \in \X(\T Q)$, namely, $\restr{\Phi^\T}{G_\beta\times \T Q}_\ast \Delta = \Delta$. This observations, together with \Cref{thm:symplectic_point_reduction} allow proving the following.

\begin{theorem}\label{theorem:reduction_forced}
    Let $G_\beta$ be the unique connected Lie subgroup of $G$ whose Lie algebra is $\mathfrak{g}_\beta$. Assume that it is a closed Lie subgroup of $G$. Let $\mu$ be a regular value of the natural momentum map $\mommap_\beta\colon \T Q \to \mathfrak{g}_\beta$. Denote by $G_{\beta, \mu}\subseteq G_{\beta}$ the isotropy subgroup of $\mu$ with respect to the coadjoint action.
    Then,
    \begin{enumerate}
        \item $\mommap_\beta^{-1}(\mu)$ is a submanifold of $\T Q$ and the forced Euler--Lagrange vector field $\sode_{L,\beta}$ is tangent to it.
        \item The quotient space $M_\mu\coloneqq \mommap_{\beta}^{-1}(\mu)/G_{\beta, \mu}$ has a symplectic form $\omega_\mu$ uniquely characterized by the relation
        \begin{equation}
            \pi_\mu^\ast \omega_\mu = \incl_{\mu}^\ast \omega_L\, ,
        \end{equation}
        where the maps $\incl_{\mu}\colon \mommap_{\beta}^{-1}(\mu)\hookrightarrow \T Q$ and $\pi_\mu\colon \mommap_{\beta}^{-1}(\mu)\to \mommap_{\beta}^{-1}(\mu)/G_{\beta, \mu}$ denote the inclusion and the projection, respectively.
        \item There is a vector field $\Delta_\mu\in \X(M_\mu)$ which is $\pi_\mu$-related to the Liouville vector field $\Delta\in \X(\T Q)$, namely,
        \begin{equation}
            \T \pi_\mu \circ \Delta \circ \incl_{\mu} = \Delta_\mu \circ \pi_\mu\, .
        \end{equation}
        \item The Lagrangian function induces a function $L_{\mu} \in \Cinfty(M_\mu)$ such that
        \begin{equation}
            L_{\mu} \circ \pi_\mu = L \circ \incl_{\mu}\, . 
        \end{equation}
        \item Similarly, the Lagrangian energy induces a function $E_{L_{\mu}} \in \Cinfty(M_\mu)$ such that
        \begin{equation}
            E_{L_{\mu}} \circ \pi_\mu = E_{L_{\mu}} \circ \incl_{\mu}\, . 
        \end{equation}
        Moreover, this function is the Lagrangian energy of $L_{\mu}$, namely, $E_{L_{\mu}} = \Delta_\mu (L_{\mu}) - L_{\mu}$.
        \item The external force induces a one-form $\beta_\mu \in \Omega^1(M_\mu)$ such that
        \begin{equation}
            \pi_\mu^\ast\, \beta_\mu = \incl_{\mu}^\ast\, \beta\, .
        \end{equation}
        \item There is a vector field $\sode_{L_\mu, \beta_\mu}\in \X(M_\mu)$ which is $\pi_\mu$-related to the forced Euler--Lagrange vector field $\sode_{L,\beta}\in \X(\T Q)$, namely,
        \begin{equation}
            \T \pi_\mu \circ \sode_{L,\beta} \circ \incl_{\mu} = \sode_{L_\mu, \beta_\mu} \circ \pi_\mu\, .
        \end{equation}
        This vector field is given by
        \begin{equation}
           \sode_{L_\mu, \beta_\mu} = \sharp_{\omega_\mu}  \left(\dd E_{L_\mu} + \beta_\mu\right)\, .
        \end{equation}
    \end{enumerate} 
\end{theorem}

\subsection{Reduction of Rayleigh systems}

In this subsection, suppose that $(Q, L, \alpha)$ is Rayleighable and $\Rayl$ is a Rayleigh potential.

\begin{proposition}
    The set
    \begin{equation}
        \mathfrak{g}_{\Rayl} = \left\{ \xi \in \mathfrak{g} \mid \xi_Q^\V (\Rayl) = \xi_Q^\Com (\Rayl) = 0\right\}\, ,
    \end{equation}
    is a Lie subalgebra of $\mathfrak{g}_\beta$. 
\end{proposition}

\begin{proof}
    First, one has to show that $\mathfrak{g}_{\Rayl}$ is contained in $\mathfrak{g}_{\beta}$. In other words, that, for every $\xi \in \mathfrak{g}_{\Rayl}$, 
    \begin{equation}
        \beta(\xi_Q^\Com) = 0\, , \quad \contr{\xi_Q^\Com} \dd \beta = 0\, .
    \end{equation}
    Indeed,
    \begin{equation}
        \beta(\xi_Q^\Com) = \Sendoadj \circ \dd \Rayl (\xi_Q^\Com) = \xi_Q^\V (\Rayl) = 0\, .
    \end{equation}
    By a direct computation in local coordinates, it is possible to show that 
    \begin{equation}
        \liedv{X^\Com} \beta = \Sendoadj \circ \dd \big(X^\Com(\Rayl)\big)\, ,
    \end{equation}
    for any $X\in \X(Q)$. Thus, $\xi_Q^\Com (\Rayl)=0$ implies that $\liedv{\xi_Q^\Com} \beta=0$, and hence $\xi \in \mathfrak{g}_\beta$.
    
    The proof is completed by showing that $\mathfrak{g}_\Rayl$ closes a Lie algebra. Taking into account that the map $\xi\in \mathfrak{g}_\Rayl \mapsto \xi_Q\in \X(Q)$ is a Lie algebra anti-homomorphism, together with the properties of the vertical and complete lifts, one can write
    \begin{equation}
        \big([\xi, \chi]\big)_Q^\Com = - \big([\xi_Q, \chi_Q]\big)^\Com = - [\xi_Q^\Com, \chi_Q^\Com]\, ,
    \end{equation}
    and 
    \begin{equation}
        \big([\xi, \chi]\big)_Q^\V = - \big([\xi_Q, \chi_Q]\big)^\V = - [\xi_Q^\Com, \chi_Q^\V] \, ,
    \end{equation}
    for any, $\xi, \chi\in \mathfrak{g}_\Rayl$. Thus,
    \begin{equation}
        \liedv{\big([\xi, \chi]\big)_Q^\Com} \Rayl = \left[\liedv{\xi_Q^\Com}, \liedv{\chi_Q^\Com}\right] \Rayl = 0\, ,
    \end{equation}
    and
    \begin{equation}
        \liedv{\big([\xi, \chi]\big)_Q^\V} \Rayl = \left[\liedv{\xi_Q^\Com}, \liedv{\chi_Q^\V}\right] \Rayl = 0\, .
    \end{equation}
    Therefore, $[\xi, \chi]\in \mathfrak{g}_{\Rayl}$ and $\mathfrak{g}_\Rayl$ closes a Lie algebra. 
\end{proof}

\begin{theorem}
    Let $G_{\Rayl}$ be the unique connected Lie subgroup of $G$ with Lie algebra $\mathfrak{g}_{\Rayl}$. Assume that it is a closed Lie subgroup of $G$. Let $\mu$ be a regular value of the natural momentum map $\mommap_{\Rayl}\colon \T Q\to \mathfrak{g}_{\Rayl}$. Denote by $\incl_{\mu}\colon \mommap_{\beta}^{-1}(\mu)\hookrightarrow \T Q$ and $\pi_\mu\colon \mommap_{\beta}^{-1}(\mu)\to \mommap_{\beta}^{-1}(\mu)/G_{\beta, \mu}\eqqcolon M_\mu$ the inclusion and the projection respectively. 
    Then, $\Rayl$ induces a function $\Rayl_\mu \in \Cinfty(M_\mu)$ such that
    \begin{equation}
        \Rayl_\mu \circ \pi_\mu = \Rayl \circ \incl_\mu\, .
    \end{equation}
    In addition, the results from \Cref{theorem:reduction_forced} can be applied by replacing $G_\beta$ with $G_{\Rayl}$.
\end{theorem}


\begin{example}
    Consider the Rayleigh system $(\RR^n\setminus \{0\}, L,\mathcal{R})$. Each point $v_q\in \T(\RR^n\setminus \{0\})$ can be identified with a pair $(q, v)\in \RR^n \times \RR^n \setminus \{0\}$. Suppose that the Lagrangian function $L$ is spherically symmetric, that is, $L(q,v)= \ell(\lVert q \rVert,\lVert{v}\rVert)$ for an $\ell \in \Cinfty(\RR^2)$, where $\lVert \cdot \rVert$ denotes the Euclidean norm in $\RR^n$. Consider the Lie group $G=\mathrm{SO}(n)$ acting by rotations on $Q$. Recall that
    \begin{equation}
        \mommap(q,v) = q \times v.
    \end{equation}
    The objective is now to find Rayleigh potentials $\mathcal{R}$ such that $\mathfrak{g}_\mathcal{R} = \mathfrak{g}$. The condition that $\xi_Q^v (\mathcal{R})=0$ implies that $\mathcal{R}$ must be spherically symmetric on the velocities. Then, the condition that $\xi_Q^c(\mathcal{R})$ is basic means that the terms which are not spherically symmetric on the positions cannot involve the velocities, that is,
    \begin{equation}
        \mathcal{R} = A(q) + B(\lVert{q}\rVert, \lVert v \rVert).
    \end{equation}
    Without loss of generality (see \Cref{remark:Rayleigh_gauge}), the Rayleigh potential can be taken as
    \begin{equation}
        \mathcal{R} =  B(\lVert{q}\rVert, \lVert v \rVert). \label{Rayleigh_rotation_invariant}
    \end{equation}
    In particular, if $\mathcal{R}$ is a quadratic form, say
    \begin{equation}
        \mathcal{R} = \frac{1}{2} R_{ij} (q)v^i v^j,
    \end{equation}
    then requirement \eqref{Rayleigh_rotation_invariant} leads to
    \begin{equation}
        R_{ij} = r_i\left( \lVert q \rVert  \right) \delta_{ij}.
    \end{equation}
    \end{example}

\chapter{Hamilton--Jacobi theory for systems with external forces}\label{ch:forced_HJ}
\insquote{The technique of generating functions for canonical transformations, developed by Hamilton and Jacobi, is the most powerful method available for integrating the differential equations of dynamics.}{Vladimir Arnol'd, \emph{Mathematical Methods of Classical Mechanics} (1978)}

In this chapter, a Hamilton--Jacobi theory for forced Hamiltonian systems is developed. Moreover, a method for the reduction and reconstruction of solutions of the Hamilton--Jacobi equation for forced Hamiltonian systems with symmetries is presented. 

The results from the present chapter had been previously published in the article \cite{deLeon2022b}.

\section{Hamilton--Jacobi theory for forced systems}


\begin{theorem}
    Consider a forced Hamiltonian system $(Q, H, \alpha)$ with forced Hamiltonian vector field $X\Halpha$. Let $\gamma \in \Omega^1(Q)$ be a closed one-form, and define the vector field
    \begin{equation}
        X\Halpha^\gamma = \T \pi_Q \circ X\Halpha \circ \gamma \in \X(Q)\, .
    \end{equation}
    Then, the following assertions are equivalent:
    \begin{enumerate}
        \item \label{thm:HJ_forced_curves} If $c\colon I \subseteq \RR \to Q$ is an integral curve of $X\Halpha^\gamma$, then $\gamma\circ c$ is an integral curve of  $X\Halpha$,
        \item \label{thm:HJ_forced_related} $X\Halpha^\gamma$ and $X\Halpha$ are $\gamma$-related, namely, $\T \gamma \circ X\Halpha^\gamma =  X\Halpha \circ \gamma$,
        \item \label{thm:HJ_forced_tangent} $X\Halpha$ is tangent to $\Ima \gamma$,
        \item \label{thm:HJ_forced_equation} $\gamma$ is a solution of the partial differential equation
        \begin{equation}\label{eq:HJ_forced}
            \gamma^\ast \left(\dd H + \alpha\right) = 0\, .
        \end{equation}
    \end{enumerate}
\end{theorem}

Equation~\eqref{eq:HJ_forced} is called the \emph{Hamilton--Jacobi equation for $(Q, H, \alpha)$}.

\begin{proof}
    The first step of the proof is to show that \ref{thm:HJ_forced_related} is a necessary and sufficient condition for \ref{thm:HJ_forced_curves} to hold. Indeed, 
    if $c$ is an integral curve of $X\Halpha^\gamma$, then
    \begin{equation}
        \frac{\dd}{\dd t} \gamma \circ c (t) = \T \gamma \circ \frac{\dd}{\dd t} c (t) = \T \gamma \circ X\Halpha^\gamma \circ c (t)\, ,
    \end{equation}
    which implies that $\gamma \circ c$ is an integral curve of $X\Halpha$ if and only if
    \begin{equation}
        X\Halpha \circ \gamma \circ c (t) = \T \gamma \circ X\Halpha^\gamma \circ c (t)\, .
    \end{equation}
    This equality holds for every integral curve $c$ if and only if $X\Halpha^\gamma$ and $X\Halpha$ are $\gamma$-related.

    Taking into account that $\T (\Ima \gamma) = \Ima (\T \gamma)$, it is clear that \ref{thm:HJ_forced_related} and \ref{thm:HJ_forced_tangent} are equivalent.

    The next step is proving the equivalence between \ref{thm:HJ_forced_related} and \ref{thm:HJ_forced_equation}. In bundle coordinates,
    \begin{equation}
        X\Halpha = \parder{H}{p_i} \parder{}{q^i} - \left(\parder{H}{q^i} + \alpha_i \right) \parder{}{p_i}\, .
    \end{equation}
    and
    \begin{equation}
        X\Halpha^\gamma (x) = \parder{H}{p_i} \circ \gamma(x) \restr{\parder{}{q^i}}{x} - \left(\parder{H}{q^i} + \alpha_i \right) \parder{}{p_i}\, .
    \end{equation}
    Hence,
    \begin{equation}
        X\Halpha \circ \gamma(x) = \parder{H}{p_i} \circ \gamma(x) \restr{\parder{}{q^i}}{x} - \left(\parder{H}{q^i} \circ \gamma(x) + \alpha_i \circ \gamma(x) \right) \restr{\parder{}{p_i}}{x}\, .
    \end{equation}
    and
    \begin{equation}
        \T \gamma \circ X\Halpha^\gamma (x) = \parder{H}{p_i} \circ \gamma(x) \restr{\parder{}{q^i}}{x} + \parder{H}{p_i} \circ \gamma(x) \parder{\gamma_j}{q^i}(x) \restr{\parder{}{p_j}}{x} \, .
    \end{equation}
    Therefore, $X\Halpha^\gamma$ and $X\Halpha$ are $\gamma$-related if and only if
    \begin{equation}
        \parder{H}{q^i} \circ \gamma(x) + \alpha_i \circ \gamma(x) + \parder{H}{p_j} \circ \gamma(x) \parder{\gamma_i}{q^j}(x) = 0\, .
    \end{equation}
    Since $\gamma$ is closed, the latter can also be written as
    \begin{equation}
        \parder{H}{q^i} \circ \gamma(x) + \alpha_i \circ \gamma(x) + \parder{H}{p_j} \circ \gamma(x) \parder{\gamma_j}{q^i}(x) = 0\, ,
    \end{equation}
    which is equation~\eqref{eq:HJ_forced} in bundle coordinates.
\end{proof}


A \emph{complete solution of the Hamilton--Jacobi equation for $(Q, H, \alpha)$} is a local diffeomorphism $\Phi\colon Q \times \RR^n \to \cT Q$ such that, for each $\Lambda\in\RR^n$, the map $\Phi_\Lambda = \Phi(\cdot, \Lambda)$ is solution of the Hamilton--Jacobi equation for $(Q, H, \alpha)$.

Consider a complete solution $\Phi\colon Q \times \RR^n \to \cT Q$ of the Hamilton--Jacobi problem for $(Q, H, \alpha)$. Around each point $q\in Q$, there exists a neighbourhood $U$ such that $\restr{\Phi}{U\times \RR^n}\colon U\times \RR^n \to V=\Phi(U\times \RR^n)$ is a diffeomorphism. Define the $n$ functions 
\begin{equation}
    f_a = \pi_a \circ \restr{\Phi}{U\times \RR^n}^{-1}\colon V \to \RR\, ,
\end{equation}
where $\pi_a\colon Q\times \RR^n\to \RR$ denotes the projection on the $a$-th component of $\RR^n$. By construction, these functions are functionally independent, that is, their differentials are linearly independent. One can write
\begin{equation}
  \Phi_\Lambda(U)= \bigcap_{a=1}^n f_a^{-1}(\Lambda_a)\, ,
\end{equation}
for $\Lambda = (\Lambda_1, \ldots, \Lambda_n)\in \RR^n$. The fact that $X\Halpha$ is tangent to $\Phi_\Lambda(U)$ implies that the functions $f_a$ are constants of the motion.
Moreover, these functions are in involution, namely,
\begin{equation}
  \restr{\left\{f_a,f_b \right\}_{\omega_V}}{\Phi_\Lambda(U)}
  = \restr{\omega_V(X_{f_a}, X_{f_b})}{\Phi_\Lambda(U)} = 0\, ,
\end{equation}
where $\omega_V$ denotes the restriction of $\omega_Q$ to $V$.


\begin{example}
    Consider the forced Hamiltonian system $(\RR^n, H, \alpha)$, with
    \begin{equation}
        H = \sum_{i=1}^n \frac{p_i^2}{2}\, , \quad
        \alpha = \sum_{i=1}^n \kappa_i p_i^2 \dd q^i\, .
    \end{equation}
    In bundle coordinates, the Hamilton--Jacobi equation reads
    \begin{equation}\label{eq:HJ_quadratic_force}
        \gamma_j \parder{\gamma_j}{q^i} + \kappa_i \gamma_i^2 = 0\, , \quad i\in\{1, \ldots, n\}\, .
    \end{equation}
    Assume that the solution is separable, namely,
    \begin{equation}
        \gamma(q^1, \ldots, q^n) = \gamma_1(q^1) \dd q^1 + \cdots + \gamma_n (q^n) \dd q^n\, ,
    \end{equation}
    where $\gamma_i$ is a function depending only on the coordinate $q^i$. This implies that $\gamma$ is closed. Then, equations~\eqref{eq:HJ_quadratic_force} reduce to
    \begin{equation}\label{eq:HJ_quadratic_force_separable}
        \gamma_i \frac{\dd \gamma_i}{\dd q^i} + \kappa_i \gamma_i^2 = 0\, , \quad i\in\{1, \ldots, n\}\, ,
    \end{equation}
    whose non-trivial solutions are of the form
    \begin{equation}
        \gamma_i = \lambda_i e^{-\kappa_i q^i}\, ,
    \end{equation}
    for some constants $\lambda_i$. The map $\Phi\colon \RR^n \times \RR^n \to \cT \RR^n$ given by
    \begin{equation}
        \Phi(q^1, \ldots, q^n; \lambda_1, \ldots, \lambda_n) = \sum_{i=1}^n \lambda_i e^{-\kappa_i q^i} \dd q^i
    \end{equation}
    is a diffeomorphism, and hence a complete solution of the Hamilton--Jacobi equation for $(\RR^n, H, \alpha)$. Its inverse is given by
    \begin{equation}
        \Phi^{-1} (q^1, \ldots, q^n, p_1, \ldots, p_n) = \left(q^1, \ldots, q^n, p_1 e^{\kappa_1 q^1}, \ldots, p_n e^{\kappa_n q^n}\right)\, .
    \end{equation} 
    Thus, the functions
    \begin{equation}
        f_a (q^i, p_i) = \pi_a \circ \Phi^{-1} (q^i, p_i) = p_a e^{\kappa_a q^a}\, , a\in \{1, \ldots, n\}
    \end{equation}
    are constants of the motion in involution.
\end{example}

The cotangent bundle with the canonical symplectic structure is the natural framework to study a Hamilton--Jacobi theory for (forced) Hamiltonian systems. Nevertheless, it is also possible to develop a Hamilton--Jacobi theory in the (forced) Lagrangian formalism (see Section IV in \cite{deLeon2022b}, see also \cite{C.G.M+2006} for the case without external forces).

\section{Reduction and reconstruction of the Hamilton--Jacobi problem}

Let $G$ be a Lie group with Lie algebra $\mathfrak{g}$, and let $\mathfrak{g}^\ast$ denote the dual of $\mathfrak{g}$. Consider a Lie group action $\Phi\colon G \times Q \to Q$ of $G$ on a manifold $Q$ and its cotangent lift $\Phi^{\cT} \colon G\times \cT Q \to \cT Q$. Henceforth, assume that both of these actions are free and proper. Let $\mommap\colon \cT Q \to \mathfrak{g}^\ast$ denote the natural momentum map, namely,
\begin{equation}
   \mommap^\xi(x) = \left\langle \mommap(x), \xi \right \rangle = \contr{\xi_{\cT Q}} \theta_Q (x)\, ,
\end{equation}
where $\xi_{\cT Q}\in \X (\cT Q)$ denotes the infinitesimal generator of the action defined by $\xi \in \mathfrak{g}$, and $\theta_Q\in \Omega^1(\cT Q)$ is the canonical one-form.

The next paragraphs review some results required for the reduction of the Hamilton--Jacobi problem. For the proofs and further details refer to \cite{M.M.O+2007,d.M.V2017}.

Let $A$ be a principal connection on $\pi_G\colon Q \to Q/G$. Consider the free and proper action of $G$ on $Q \times \mathfrak{g}$ given by $(g, q, \eta)\mapsto (\Phi(g,q), \Ad_{g^{-1}} \eta)$, where $\Ad^\ast$ denotes the coadjoint action of $G$ on $\mathfrak{g}^\ast$. The quotient
\begin{equation}
    \tilde{\mathfrak{g}}^\ast \coloneqq Q \times_G \mathfrak{g}^\ast\coloneqq \frac{Q\times \mathfrak{g}^\ast}{G}\, ,
\end{equation}
together with the projection $\pi_{\tilde{\mathfrak{g}}} \colon \tilde{\mathfrak{g}} \to Q/G, \, \pi_{\tilde{\mathfrak{g}}}([q, \xi]) = \pi_G(q)$, is a vector bundle called the \emph{adjoint bundle of $\pi_G\colon Q \to Q/G$}. Moreover, the connection $A$ induces a splitting
\begin{equation}
    \frac{\cT Q}{G} \equiv \cT \left(\frac{Q}{G}\right) \times_{Q/G}  \tilde{\mathfrak{g}}^\ast\, .
\end{equation}
Let $\operatorname{hor}_q\colon T_{\pi_G(q)}(Q/G)\to \Hor_q$ denote the horizontal lift defined by the connection $A$. The identification above is given by
\begin{equation}
\begin{aligned}
    \Psi\colon \frac{\cT Q}{G} & \longrightarrow  \cT \left(\frac{Q}{G}\right) \times_{Q/G}  \tilde{\mathfrak{g}}^\ast\\
    [\alpha_q] & \longmapsto \big[\big(\alpha_q \circ \operatorname{hor}_q, \mommap(\alpha_q)\big)\big]\, .
\end{aligned}
\end{equation}
For each point $\alpha_q\in \mommap^{-1}(\mu)$, one has $\Psi([\alpha_q])=[(\alpha_q\circ \hor_q, \mu)]$, which implies that $\Psi\left({\mommap^{-1}(\mu)}/{G}\right)$ can be identified with $\cT (Q/G)$.

Let $A_\mu\in \Omega^{1}(Q)$ be the one-form given by the contraction of $\mu\in \mathfrak{g}^\ast$ with $A$ (understanding the connection as a $\mathfrak{g}$-valued one-form). This one-form is, by construction, $\Phi^{\cT}$-invariant, and satisfies $\mommap\circ A_\mu = \mu$. Thus, $\contr{\xi_Q}A_\mu\in \Cinfty(Q)$ is a constant function and
\begin{equation}
    \contr{\xi_Q} \dd A_\mu = \liedv{\xi_Q} A_\mu - \dd \contr{\xi_Q}A_\mu = 0\, .
\end{equation}
Consequently, there exists a unique 2-form $\rho_\mu \in \Omega^1(Q/G)$ such that $\pi^\ast\rho_\mu = \dd A_\mu$. Define the 2-form $B_\mu = \pi^\ast_{Q/G}\, \rho_\mu\in \Omega^1(\cT(Q/G))$, where $\pi_{Q/G}\colon \cT(Q/G) \to Q/G$ denotes the canonical projection.

Let $\omega_\mu\in\Omega^2(\mommap^{-1}(\mu)/G)$ be the symplectic form induced by the canonical symplectic form $\omega_Q\in \cT Q$ (see \Cref{thm:symplectic_point_reduction}). Denote by $Q/G$ the quotient of $Q$ by the action $\Phi$, and let $\omega_{Q/G} \in \Omega^2(\cT (Q/G))$ be the canonical symplectic form of its cotangent bundle.
The symplectic manifolds $(\mommap^{-1}(\mu)/G, \omega_\mu)$ and $(\cT (Q/G), \omega_{Q/G}+B_\mu)$ are symplectomorphic, and the 2-form $B_\mu$ is called a \emph{magnetic term}. In addition, if $\mathfrak{L}\subseteq \mommap^{-1}(\mu)$ is a Lagrangian submanifold of $(\cT Q, \omega_Q)$, then $\pi(\mathfrak{L})$ is a Lagrangian submanifold of $(\mommap^{-1}(\mu)/G, \omega_{\mu})$.

The Hamiltonian counterpart of \Cref{prop:Lie_subalgebra_forced_Lagrangian} is as follows.

\begin{proposition}\label{prop:Lie_subalgebra_forced_Hamiltonian}
    Let $(Q, H, \alpha)$ be a forced Hamiltonian system such that $H$ is $\Phi^{\cT}$-invariant. For each $\xi \in \mathfrak{g}$, the function $\mommap^\xi$ is a conserved quantity if and only if
    \begin{equation}
        \alpha(\xi_{\cT Q}) = 0\, .
    \end{equation}
    If this equation holds, a necessary and sufficient condition for the infinitesimal action generated by $\xi$ to leave $\alpha$ invariant is
    \begin{equation}
        \contr{\xi_{\cT Q}} \dd \alpha = 0\, .
    \end{equation}
    Moreover, the subset 
    \begin{equation}
        \mathfrak{g}_\alpha = \left\{\xi \in \mathfrak{g} \mid \alpha(\xi_{\cT Q}) = 0, \, \contr{\xi_{\cT Q}} \dd \alpha = 0\right\}
    \end{equation}
    is a Lie subalgebra of $\mathfrak{g}$.
\end{proposition}

Consider a forced Hamiltonian system $(Q, H, \alpha)$.
Let $G_\alpha\subseteq G$ denote unique connected Lie subgroup of $G$ with Lie algebra $\mathfrak{g}_\alpha$. 
Without loss of generality, suppose hereafter that $G_\alpha = G$ (if that is not the case, it suffices to replace $G$ by $G_\alpha$ as the Lie group considered). Assume that $\mu$ is a fixed point of the coadjoint action, that is, $\Ad^\ast_g \mu = \mu$ for all $g\in G$. 

Since $H$ and $\alpha$ are $\Phi^{\cT}$-invariant, there exists a function $H_G\in \Cinfty(\cT Q/G)$ and a one-form $\alpha_G \in \Omega^1(\cT Q/G)$ such that 
\begin{equation}
    H = H_G \circ \pi\, , \quad \alpha = \pi^\ast \alpha_G\, .
\end{equation} 
Define the function $\tilde{H}\in \Cinfty(\cT \left({Q}/{G}\right) \times_{Q/G}  \tilde{\mathfrak{g}}^\ast)$ and the one-form $\alpha_G \in \Omega^1(\cT \left({Q}/{G}\right) \times_{Q/G}  \tilde{\mathfrak{g}}^\ast)$ by
\begin{equation}
    \tilde{H} = H_G \circ \Psi^{-1}\, , \quad \tilde{\alpha} = \left(\Psi^{-1}\right)^\ast \tilde{\alpha}\, .
\end{equation}
Finally, consider $\tilde{H}_\mu\in \Cinfty(\cT(Q/G))$ and $\tilde{\alpha}_\mu \in \Omega^1(\cT(Q/G))$ such that
\begin{equation}
    \tilde{H}_\mu\left(\tilde{\varsigma}_{\tilde{q}}\right) = \tilde{H}\left(\tilde{\varsigma}_{\tilde{q}},[q, \mu]\right) \, , \quad
    \tilde{\alpha}_\mu\left(\tilde{\varsigma}_{\tilde{q}}\right)=\tilde{\alpha}\left(\tilde{\varsigma}_{\tilde{q}},[q, \mu]\right)\, .
\end{equation}
for each $\left(\tilde{\varsigma}_{\tilde{q}}\right) \in \T_{\tilde{q}}^\ast(Q / G)$, where $\tilde{q}=[q] \in Q / G$. Since $\Phi^{\cT}$ is a cotangent lift action, $\tilde{\alpha}_\mu$ is a semibasic one-form on $\cT(Q/G)$. 
The triple $(Q/G, \tilde{H}_{\mu}, \tilde{\alpha}_{\mu})$ is called the \emph{reduced Hamiltonian system of $(Q, H, \alpha)$}.


\begin{proposition}[Reduction]
    Let $\gamma$ be a $\Phi^{\cT}$-invariant solution of the Hamilton-Jacobi equation for $(H,\alpha)$. Let $\mathfrak{L} = \Ima \gamma$, and $\tilde{\mathfrak{L}}=\Psi\circ \pi(L)$. Then $\gamma$ induces a one-form $\tilde \gamma_\mu\in \Omega^1(Q/G)$ such that $\Ima \tilde \gamma_\mu = \tilde {\mathfrak{L}}$ and $\tilde \gamma_\mu$ is a solution the Hamilton-Jacobi equation for $(\tilde{H}_\mu, \tilde{\alpha}_\mu)$.
\end{proposition}

The situation is summarized in the following commutative diagram:
\begin{equation}
\begin{tikzcd}
                                                                                                     & \tilde{\mathfrak{g}}^*                                                              &                                                                                           &                                                                                                                                                         \\
Q \arrow[r, "\gamma"] \arrow[d, "\pi_G"] \arrow[rrdd, "\tilde \gamma", bend right=67, shift right=3] & \cT  Q \arrow[l, "\hspace{-.25cm} \pi_Q"', bend left] \arrow[d, "\pi"] \arrow[rr, "H"] \arrow[u, "\mommap"] &                                                                                           & \mathbb{R}                                                                                                                                              \\
\frac Q G \arrow[rrd, "\tilde \gamma_\mu"]                                                           & \frac {\cT  Q}{G} \arrow[l, "p"'] \arrow[rr, "\Psi", shift left] \arrow[rru, "H_G"]   &                                                                                           & \hspace{-.1cm} \cT  \left(\frac Q G\right)  \hspace{-.1cm}   \times_{ \hspace{-.1cm}\frac{Q}{G}} \hspace{-.07cm}\tilde{\mathfrak{g}}^* \hspace{-.1cm}   \arrow[ll, "\Psi^{-1}", shift left] \arrow[u, "\tilde H"] \arrow[ld, dashed] \\
                                                                                                     &                                                                                     & \cT  \left(\frac {Q} {G}\right) \arrow[ruu, "\tilde{H}_\mu"', bend right=90, shift right=6] &                                                                                                                                                        
\end{tikzcd}
\end{equation}

\begin{proof}
    Since $\mathfrak{L}$ is $\Phi^{\cT}$-invariant and the natural momentum map is $\Ad^\ast$-equivariant, $\mommap(\mathfrak{L}) = \mu$, that is, $\mathfrak{L}\subseteq \mommap^{-1}(\mu)$. Therefore, $\pi(\mathfrak{L})$ is a Lagrangian submanifold of $(\mommap^{-1}(\mu)/G, \omega_{\mu})$. The fact that $\Psi$ is a symplectomorphism between $(\mommap^{-1}(\mu)/G, \omega_\mu)$ and $(\cT (Q/G), \omega_{Q/G}+B_\mu)$ implies that $\tilde{\mathfrak{L}}$ is a Lagrangian submanifold of $(\cT (Q/G), \omega_{Q/G}+B_\mu)$.

    Since $\gamma$ is $\Phi^{\cT}$-invariant, it induces a map $\tilde{\gamma}\colon Q \to \cT  (Q/G)$ which is also $\Phi^{\cT}$-invariant. This map in turn induces a one-form $\tilde \gamma_\mu\in \Omega^1(Q/G)$ such that $\tilde{\gamma}=\pi_G^\ast \tilde \gamma_\mu$ and $\Ima \tilde \gamma_\mu = \tilde{\mathfrak{L}}$.

    Hamilton--Jacobi equation~\eqref{eq:HJ_forced} implies that $\dd H +\alpha = \pi^\ast (\dd H_g + \alpha_G)$ vanishes along $\mathfrak{L}$. 
    Thus, $\dd H_G + \alpha_G$ vanishes along $\pi(\mathfrak{L})$.
    If $\tilde{\varsigma}_{\tilde{q}}\in \tilde{\mathfrak{L}}$, then $\Psi^{-1}(\tilde{\varsigma}_{\tilde{q}}) \in \pi (\mathfrak{L})$, and 
    \begin{equation}
    \tilde{H}_\mu (\tilde{\varsigma}_{\tilde{q}}) = H_G \circ \Psi^{-1} (\tilde{\varsigma}_{\tilde{q}}, \mu)\, , \quad \tilde{\alpha}_\mu (\tilde{\alpha}_{\tilde{q}}) = (\Psi^{-1})^* \alpha_G (\tilde{\alpha}_{\tilde{q}}, \mu)\, .
    \end{equation}
    Hence,
    \begin{equation}
    \left(\dd \tilde{H}_\mu + \tilde{\alpha}_\mu  \right) (\tilde{\alpha}_{\tilde{q}})
    = \left(\dd H_G + \alpha_G  \right) \left(\Psi^{-1}(\tilde{\alpha}_{\tilde{q}}, \mu)\right) = 0\, .
    \end{equation}
    Consequently,
    \begin{equation}
        \tilde{\gamma}_\mu^\ast \left(\dd \tilde{H}_\mu + \tilde{\alpha}_\mu  \right) = 0\, .
    \end{equation}
\end{proof}

\begin{proposition}[Reconstruction]
    Let $\tilde{\mathfrak{L}}$ be a Lagrangian submanifold of $(\cT  (Q/G), \omega_{Q/G}+B_\mu)$ for some $\mu\in \mathfrak{g}^*$ which is a fixed point of the coadjoint action. Assume that $\tilde{\mathfrak{L}} = \Ima \tilde \gamma_\mu$, where $\tilde \gamma_\mu$ is a solution of the Hamilton-Jacobi problem for $(\tilde H_\mu, \tilde \alpha_\mu)$. Let 
    \begin{equation}
    \hat{\mathfrak{L}} = \left\{ (\tilde{\varsigma}_{\tilde{q}}, [\mu]_{\tilde q}) \in \cT  (Q/G) \times_{Q/G} \tilde{\mathfrak{g}}^*\mid \tilde{\varsigma}_{\tilde{q}} \in \tilde{\mathfrak{L}}  \right\},
    \end{equation}
    and take
    \begin{equation}
    \mathfrak{L} = \pi^{-1} \circ \Psi^{-1} (\hat{\mathfrak{L}}).
    \end{equation}
    Then
    \begin{enumerate}
    \item $\mathfrak{L}$ is a $\Phi^{\cT}$-invariant Lagrangian submanifold of $\cT  Q$,
    \item $\mathfrak{L}= \Ima \gamma$, where $\gamma$ is a solution of the Hamilton-Jacobi problem for $(H, \alpha)$.
    \end{enumerate}
\end{proposition}

\begin{proof}
    The mapping $\tilde{\gamma}_\mu: Q/G\to \cT  (Q/G)$ induces a $\Phi^{\cT}$-invariant mapping
    \begin{equation}
    \tilde{\gamma}=\pi^*\tilde \gamma_\mu:Q \to \cT  (Q/G),
    \end{equation}
    which in turn induces a $\Phi^{\cT}$-invariant mapping $\gamma:Q\to \cT  Q$. Let $\tilde \alpha_{\tilde q} \in \tilde{\mathfrak{L}}$, so $(\tilde{\varsigma}_{\tilde{q}}, [\mu]_{\tilde q}) \in \cT  (Q/G) \times_{Q/G} \tilde{\mathfrak{g}}^*$, and 
    \begin{equation}
    \pi^{-1} \circ \Psi^{-1} (\tilde{\varsigma}_{\tilde{q}}, [\mu]_{\tilde q}) \in \mathfrak{L},
    \end{equation}
    and then
    \begin{equation}
    \left(\tilde{\alpha}_\mu + \dd \tilde{H}_\mu   \right) (\tilde{\alpha}_{\tilde{q}})
    = \left(\dd H_G + \alpha_G  \right) \left(\Psi^{-1}(\tilde{\alpha}_{\tilde{q}}, \mu)\right) = 0.
    \end{equation}

\end{proof}

\begin{example}[Calogero--Moser system with an external force linear in the momenta]
    Consider the forced Hamiltonian system $(\RR^2, H, \alpha)$, whose
    Hamiltonian function
    \begin{equation}
    H = \frac{1}{2} \left(p_1^2 + p_2^2 + \frac{1}{(q^1-q^2)^2}  \right)\, ,
    \end{equation}
    and its external force is
    \begin{equation}
    \alpha = (p_1 + p_2) (\dd q^1- \dd q^2)\, ,
    \end{equation}
    where $(q^1, q^2, p_1, p_2)$ are canonical coordinates in $\cT \RR^2 \simeq \RR^4$.
    Consider the action of $\RR$ on $\RR^2$ given by
    \begin{equation}
    \begin{aligned}
    \Phi : \RR \times \RR^2 &\to \RR^2\\
    \left(r,(q^1, q^2)  \right) & \mapsto (r+q^1, r+q^2).  
    \end{aligned}
    \end{equation}
    Clearly, both $H$ and $\alpha$ are invariant under the corresponding lifted action on $\cT  Q$. The momentum map is\footnote{In this case, the momentum map is naturally identified with its pairing with the generator of the Lie algebra.}
    \begin{equation}
    \begin{aligned}
    \mommap: \cT  \RR^2 &\to \RR\\
    (q^1, q^2, p_1, p_2) & \mapsto p_1 + p_2.
    \end{aligned}
    \end{equation}
    Then
    \begin{equation}
    \mommap^{-1}(\mu) = \left\{(q^1, q^2, p_1, p_2) \in \cT  \RR^2 \mid p_2 = \mu - p_1  \right\}.
    \end{equation}
    The quotient $\mommap^{-1}(\mu)/\RR$ can be identified with $\RR^2\simeq \cT \RR$, with coordinates $(q,p)$, and the natural projection
    \begin{equation}
    \begin{aligned}
        \pi : \mommap^{-1}(\mu) &\to \mommap^{-1}(\mu)/\RR\\
        (q^1, q^2, p, \mu-p) &\mapsto (q=q^1-q^2, p).
    \end{aligned}
    \end{equation}
    This allows to introduce a reduced Hamiltonian function 
    \begin{equation}
        \tilde{H}_\mu = \frac{1}{2} \left((\mu-p)^2 + p^2 + \frac{1}{q^2}\right)\, ,  
    \end{equation}
    and a reduced external force
    \begin{equation}
        \tilde{\alpha}_\mu = \mu \dd q\, .
    \end{equation}
    Let $\tilde{\gamma}\in \Omega^1(\RR)$ be a closed one-form. Then
    \begin{equation}
    \tilde{\gamma}^*(\dd \tilde{H}_\mu + \tilde{\alpha}_\mu) = \left(\mu - \frac{1}{q^3} -\mu \frac{\partial \tilde{\gamma}} {\partial q} \right) \dd q\, ,
    \end{equation}
    which implies that $\tilde{\gamma}$ is a solution of the Hamilton--Jacobi problem for $(\tilde{H}_\mu, \tilde{\alpha}_\mu)$ if and only if
    \begin{equation}
    \mu - \frac{1}{q^3} -\mu \frac{\partial \tilde{\gamma}} {\partial q} = 0\, .
    \end{equation}
    Hence,
    \begin{equation}
    \tilde{\gamma}_\lambda = \left(q + \frac{1}{2\mu q^2} + \lambda\right) \dd q
    \end{equation}
    is a complete solution depending on the parameter $\lambda \in \RR$. The associated generating function is
    \begin{equation}
    \tilde S_\lambda (q) = \frac{1}{2} q^2 - \frac{1}{2\mu q} + \lambda q ,
    \end{equation}
    where, without loss of generality, the integration constant has been taken as zero. A complete solution $\gamma_\lambda$ of the Hamilton--Jacobi for $(H, \alpha)$ is thus given by
    \begin{equation}
    \begin{aligned}
    \gamma_\lambda & =  \left(q^1 - q^2 + \frac{1}{2\mu (q^1-q^2)^2} + \lambda\right) \dd q^1\\
    & \quad +  \left(\mu - q^1 + q^2 - \frac{1}{2\mu (q^1-q^2)^2} - \lambda\right) \dd q^2\, ,
    \end{aligned}
    \end{equation}
    and its associated generating function is
    \begin{equation}
    S_\lambda (q^1, q^2) = \tilde S_\lambda(q^1-q^2) + \mu q^2\, .
    \end{equation}
\end{example}


\chapter{Discrete Hamilton--Jacobi theory for systems with external forces}\label{ch:forced_discrete_HJ}

\insquote{A mathematical problem does not cease being mathematical just because we have discretized it. The purpose of discretization is to render mathematical problems, often approximately, in a form accessible to efficient calculation by computers. [...]
We can still ask proper mathematical questions with uncompromising rigour and seek answers with the full mathematical etiquette of precise definitions, statements and proofs. The rules of the game do not change at all.}{Arieh Iserles, \emph{A First Course in the Numerical Analysis of Differential Equations} (2009)}

This chapter is devoted to discrete mechanical systems subject to external forces. The discrete counterparts of some results from previous chapters are here studied. The main new results are the notion of discrete Rayleigh potential, a Noether's theorem for forced discrete Lagrangian systems, and two Hamilton--Jacobi equations for these systems. These results had been previously published in the article \cite{deLeon2022a}. Refer to \Cref{sec:discrete_mechanics} for a review on discrete Lagrangian mechanics.

\section{Discrete forced systems}

Let $Q$ be an $n$-dimensional manifold. A \emph{discrete external force} is a one-form $f_d =(f_d^+, f_d^-) \in \Omega^1(Q\times Q)$. A \emph{forced discrete Lagrangian system} is a triple $(Q, L_d, f_d)$ formed by a discrete Lagrangian system $(Q, L_d)$ and a discrete external force $f_d$. The $j$-th \emph{discrete action sum} is defined by
\begin{equation}
  \action_d^j (c_d) = \sum_{k=0}^{j-1} L_d(q_k, q_{k+1})\, ,
\end{equation}
for $1\leq j\leq N$. In particular, the \emph{discrete action} is the $N$-th action sum, namely, $\action_d (c_d) =  \action_d^{N} (c_d)$.
The \emph{discrete Lagrange--d'Alembert principle} states that the dynamics of the forced discrete Lagrangian system $(Q, L_d, f_d)$ is given by discrete curves $c_d=\{q_k\}_{k=0}^N$ that satisfy 
\begin{equation}\label{eq:discrete_Lagrange_dAlembert}
    \dd \action_d(c_d) \cdot v_{c_d} + \sum_{k=0}^{N-1} \left[f_d^-(q_k, q_{k+1}) \cdot v_k + f_d^+(q_k, q_{k+1})\cdot v_{k+1}\right] = 0\, ,
\end{equation}
for all variations $v_{c_d} = \{v_k\}_{k=0}^N \in \T_{c_d}  \Omega \left(q_0, q_N, \{t_k\}_{k=0}^N \right)$.
A necessary and sufficient condition for a discrete curve $\{q_k\}_{k=0}^N$ to satisfy the discrete Lagrange--d'Alembert principle is that 
\begin{equation}\label{eq:forced_discrete_Euler-Lagrange}
    \DD_2 L_d (q_k, q_{k-1}) + \DD_1 L_d (q_k, q_{k+1})
    + f_d^+ (q_{k-1}, q_k) + f_d^- (q_k, q_{k+1})
    = 0\, ,
\end{equation}
where $\DD_1$ and $\DD_2$ denote the derivatives with respect to the first and second argument, respectively. These equations are called the \emph{forced discrete Euler--Lagrange equations}.

Let $(Q, L_d, f_d)$ be a forced discrete Lagrangian system, with $f_d=(f_d^+, f_d^-)$.
The \emph{right forced discrete Legendre transform} is the map given by
\begin{equation}
\begin{aligned}
     \Legp\colon Q \times Q &\to \cT Q\\
     (q_{j}, q_{j+1}) &\mapsto \big(q_{j+1},\, \DD_2 L_d(q_j, q_{j+1})+f_d^+(q_j, q_{j+1})\big) \, .
\end{aligned}
\end{equation}
Similarly, the \emph{left forced discrete Legendre transform} is defined by
\begin{equation}
\begin{aligned}
     \Legm\colon Q \times Q &\to \cT Q\\
     (q_{j}, q_{j+1}) &\mapsto \big(q_{j},\, - \DD_1 L_d(q_j, q_{j+1}) - f_d^-(q_j, q_{j+1})\big) \, .
\end{aligned}
\end{equation}
Thus, the corresponding momenta are given by
\begin{equation}
    p_{j,\, j+1}^+ = \DD_2 L_d(q_j, q_{j+1}) + f_d^+(q_j, q_{j+1})\, ,
\end{equation}
and
\begin{equation}
    p_{j,\, j+1}^- = - \DD_1 L_d(q_j, q_{j+1}) - f_d^-(q_j, q_{j+1})\, .
\end{equation}
If a discrete curve $\{q_k\}_{k=0}^N$ is a solution of the forced discrete Euler--Lagrange equations~\eqref{eq:forced_discrete_Euler-Lagrange}, then
\begin{equation}
    p_{j-1,\, j}^+ =  p_{j,\, j+1}^- \eqqcolon p_j\, .
\end{equation}
The \emph{discrete Hamiltonian flow} is defined as
\begin{equation}\label{eq:discrete_Hamiltonian_flow}
    \mathcal{F}_d^H = \Legp \circ \left(\Legm\right)^{-1} \, ,
\end{equation}
so that
\begin{equation}
    \mathcal{F}_d^H \colon (q_{j}, p_{j}) \mapsto (q_{j+1}, p_{j+1})\, ,
\end{equation}
along a solution $\{q_k\}_{k=0}^N$ of the forced discrete Euler--Lagrange equations.
Similarly, the \emph{discrete Lagrangian flow} is defined as
\begin{equation}
    \mathcal{F}_{L_d}^{f_d} = \left(\Legm\right)^{-1} \circ \Legp \, ,
\end{equation}
and it maps each pair of points in a solution $\{q_k\}_{k=0}^N$ to the next pair of points, namely, 
\begin{equation}
    \mathcal{F}_{L_d}^{f_d} \colon (q_{j-1}, q_{j}) \mapsto (q_{j}, q_{j+1})\, .
\end{equation}

\subsection{Forced discrete Hamilton equations}

Henceforth, the discrete curves $\{q_k\}_{k=0}^N$ and their corresponding momenta $\{p_k\}_{k=0}^N$ will be identified with their canonical coordinates. Equivalently, assume that $Q=\RR^n$. Assume that $\tparder{p_k}{q_k}$ is invertible for every $k\in \{0, \ldots, N\}$.

The \emph{right discrete Hamiltonian} $H_d^+\colon \cT Q \to \RR$ is defined by
\begin{equation}
    H_d^+ (q_j, p_{j+1}) = p_{j+1} \cdot q_{j+1} - L_d(q_j, q_{j+1})\, ,
\end{equation}
where $p_{j+1} \cdot q_{j+1}$ denotes the canonical dot product in $\RR^n$. 
Deriving both sides with respect to $q_j$ and with respect to $p_{j+1}$, and making use of the forced discrete Euler--Lagrange equations~\eqref{eq:forced_discrete_Euler-Lagrange} yields
\begin{subequations}
\begin{flalign}
    &\left[  q_{j+1} -\DD_{2} H_{d}^{+}\left(q_{j}, p_{j+1}\right)  \right] 
    \frac{\partial p_{j+1}} {\partial q_{j+1} }
       =       -f_d^+(q_j,q_{j+1})\, , 
       \label{eq:right_discrete_Hamilton_a} \\
    &p_{j} =\DD_{1} H_{d}^{+}\left(q_{j}, p_{j+1}\right) 
           - f_d^-\left(q_{j}, q_{j+1}\right)\, , 
           \label{eq:right_discrete_Hamilton_b}
\end{flalign} \label{eq:right_discrete_Hamilton}
\end{subequations}
These equations are called the \emph{forced right discrete Hamilton equations}.
Similarly, the \emph{left discrete Hamiltonian} $H_d^-\colon \cT Q \to \RR$ is defined by
\begin{equation}
    H_d^- (q_{j+1}, p_{j}) = - p_{j} \cdot q_{j} - L_d(q_j, q_{j+1})\, .
\end{equation}
Deriving both sides with respect to $q_{j+1}$ and with respect to $p_{j}$, and making use of the forced discrete Euler--Lagrange equations~\eqref{eq:forced_discrete_Euler-Lagrange}, one can obtain the \emph{forced left discrete Hamilton equations}:
\begin{subequations}
\begin{flalign} 
    &\left[  q_{j} +\DD_{2} H_{d}^{-}\left(q_{j+1}, p_{j}\right)  \right] 
    \frac{\partial p_{j}} {\partial q_{j} }
    =       f_d^-(q_j,q_{j+1})  \label{eq:left_discrete_Hamilton_a}\, ,\\
    &p_{j+1} =-\DD_{1} H_{d}^{-}\left(q_{j+1}, p_{j}\right) 
        + f_d^+\left(q_{j}, q_{j+1}\right)\, . 
        \label{eq:left_discrete_Hamilton_b}
\end{flalign} \label{eq:left_discrete_Hamilton}
\end{subequations}

\subsection{Discrete Rayleigh forces}

\begin{definition}
    A forced discrete Lagrangian system $(Q, L_d, f_d)$ is said to be \emph{Rayleighable} if there exists a function $\Rayl_d \in \Cinfty(Q\times Q)$ such that
    \begin{equation}
      f_d^+(q_0,q_1)= -\DD_2\Rayl_d(q_0,q_1)\, ,
    \end{equation}
    and
    \begin{equation}
      f_d^-(q_0,q_1)=\DD_1\Rayl_d(q_0,q_1)\, ,
    \end{equation}
    for each $(q_0,q_1)\in Q\times Q$.
    The function $\Rayl_d$ is called a \emph{discrete Rayleigh potential} for $f_d$. The triple $(Q, L_d, \Rayl_d)$ is called a \emph{discrete Rayleigh system}.

    The \emph{modified discrete Lagrangians} are the functions $L_d^+=L_d+\Rayl_d$ and $L_d^-=L_d-\Rayl_d$. 
\end{definition}

\begin{remark}
    A forced discrete Lagrangian system $(Q, L_d, f_d)$ is Rayleighable if and only if 
    \begin{equation}\label{eq:condition_discrete_Rayleighable}
        \DD_1 f_d^+(q_0, q_1) = - \DD_2 f_d^-(q_0, q_1)\, ,
    \end{equation}
    for all $(q_0, q_1)\in Q\times Q$.
\end{remark}

The forced discrete Euler--Lagrange equations can be written in terms of the modified discrete Lagrangians as
\begin{equation}\label{discrete_forced_EL_Rayleigh}
    \DD_{2}  L_{d}^-\left(q_{k-1}, q_{k}\right)+\DD_{1}  L_{d}^+\left(q_{k}, q_{k+1}\right)=0\, .
\end{equation}
The forced discrete Legendre transforms read
\begin{equation}
\begin{aligned}
    &\Legp: \left(q_j, q_{j+1}\right) &\mapsto \left(q_{j+1}, \DD_{2}  L_{d}^-\left(q_j, q_{j+1}\right)\right)\, ,\\
    &\Legm: \left(q_j, q_{j+1}\right) &\mapsto \left(q_{j}, -\DD_{1} L_{d}^+\left(q_j, q_{j+1}\right)\right)\, ,
\end{aligned}
\end{equation}
Therefore, their corresponding momenta are
\begin{equation}
\begin{aligned}
    &p_{j,j+1}^+=\DD_{2}  L_{d}^-\left(q_j, q_{j+1}\right)\, ,\\
    &p_{j,j+1}^- = -\DD_{1}  L_{d}^+\left(q_j, q_{j+1}\right)\, .
\end{aligned}
\end{equation}
Denoting $\DD_1^2=\DD_1\circ \DD_1$ and $\DD_2^2 = \DD_2 \circ \DD_2$, the forced right discrete Hamilton equations read
\begin{equation}
\begin{aligned}
    &\left[  q_{j+1} -\DD_{2} H_{d}^{+}\left(q_{j}, p_{j+1}\right)  \right] 
    \DD_2^2 L_d^-(q_j,q_{j+1})
    =       -\DD_2\Rayl_d(q_j,q_{j+1}) , \\
    &p_{j} =\DD_{1} H_{d}^{+}\left(q_{j}, p_{j+1}\right) 
        + \DD_1\Rayl_d(q_j,q_{j+1}), 
\end{aligned}
\end{equation}
and the forced left discrete Hamilton equations read
\begin{equation}
\begin{aligned} 
    &\left[  q_{j} +\DD_{2} H_{d}^{-}\left(q_{j+1}, p_{j}\right)  \right] 
    \DD_1^2 L_d^+ (q_j, q_{j+1})
    =  \DD_1 {R}_d (q_j, q_{j+1}) \, , \\
    &p_{j+1} =-\DD_{1} H_{d}^{-}\left(q_{j+1}, p_{j}\right) 
        + \DD_2 {R}_d \left(q_{j}, q_{j+1}\right). \label{eq:left_discrete_Hamilton_Rayleigh}
\end{aligned}
\end{equation}

\begin{proposition}[Equivalent discrete Rayleigh systems]
    Consider two discrete Rayleigh systems $(Q, L_d, \Rayl_d)$ and $(Q, \tilde{L}_d, \tilde{\Rayl}_d)$, with $\tilde{L}_d = L_d + \phi$ and $\tilde{\Rayl}_d= \Rayl_d + \chi$ for some functions $\phi,\chi\in \Cinfty(Q\times Q)$. Then, $(L_d, \Rayl_d)$ and $(\tilde{L}_d, \tilde{\Rayl}_d)$ are \emph{equivalent} (that is, they lead to the same forced discrete Euler--Lagrange equations~\eqref{discrete_forced_EL_Rayleigh}) if and only if 
    \begin{subequations}
    \begin{equation}
    \tilde{L}_d(q_0, q_1) 
    = L_d (q_0, q_1) + \psi(q_0) + \varphi_1(q_1) 
    + \varphi_0(q_0) - \psi(q_1)  \, ,
    \end{equation}
    and
    \begin{equation}
    \tilde{\Rayl}_d(q_0, q_1) 
    = \Rayl_d (q_0, q_1)  + \psi(q_0) - \varphi_1(q_1) 
    - \varphi_0(q_0) + \psi(q_1)\,  ,
    \end{equation}
    \end{subequations}
    for some functions $\psi, \varphi_0, \varphi_1$ on $Q$.

    In other words, $(L_d^+, L_d^-)$ and $(\tilde L_d^+, \tilde L_d^-)$ are equivalent if and only if
    \begin{subequations}
    \begin{equation}
    \tilde L_d^+(q_0, q_1) = L_d^+(q_0, q_1) + 2\psi (q_0) +2\varphi_1 (q_1)\, ,
    \end{equation}
    and
    \begin{equation}
    \tilde L_d ^-(q_0, q_1) = L_d^-(q_0, q_1) + 2\varphi_0(q_0) - 2\psi(q_1)\, ,
    \end{equation}
    \end{subequations}
    where $L_d^\pm = \L_d \pm \Rayl_d$ and $\tilde{L}_d^\pm = \tilde{L}_d \pm \tilde{\Rayl}_d$ denote the modified Lagrangians.
\end{proposition}

\begin{proof}
    Let $L_d^\pm = L_d \pm \Rayl_d$, and $\tilde{L}_d^\pm = L_d^\pm + \rho^\pm$ for some functions $\rho^+, \rho^-\in \Cinfty(Q\times Q)$. Clearly, $(L_d^-, L_d^+)$ and $(\tilde{L}_d^-, \tilde{L}_d^+)$ lead to the same forced discrete Euler--Lagrange equations \eqref{discrete_forced_EL_Rayleigh} if and only if
    \begin{equation}
    \DD_2\, \rho^- (q_{k-1}, q_k) + \DD_1\, \rho^+(q_k, q_{k+1}) = 0,  
    \label{condition_equivalent_Rayleigh_discrete}
    \end{equation}
    which implies that
    \begin{equation}
    \DD_1 \DD_2\, \rho^-(q_{k-1}, q_{k}) = \DD_1 \DD_2\, \rho^+ (q_k, q_{k+1}) = 0,
    \end{equation}
    and thus
    \begin{equation}
    \rho^\pm (q_0, q_{1}) = \varphi_0^\pm (q_0) + \varphi^\pm_1 (q_1) 
    \end{equation}
    for some functions $\varphi_0^\pm, \varphi_1^\pm$ on $Q$.
    Equation~\eqref{condition_equivalent_Rayleigh_discrete} implies that
    \begin{equation}
    \left(\varphi^-_1 \right)'(q_k) + \left(\varphi^+_0  \right)'(q_k) = 0\, ,
    \end{equation}
    and then
    \begin{equation}
    \varphi^+_0 (q_k) = -\varphi^-_1(q_k) + b\, ,
    \end{equation}
    for some constant $b$. Denoting $\psi=1/2\ \varphi^+_0$, $\varphi_0= 1/2\ \varphi_0^-$ and $\varphi_1=1/2\ \varphi_1^+$, one can write
    \begin{equation}
    \rho^+(q_0, q_1) = 2\psi(q_0) + 2\varphi_1(q_1)\, ,
    \end{equation}
    and
    \begin{equation}
    \rho^-(q_0, q_1) =  2\varphi_0 (q_0) - 2\psi(q_1) 
    +b\, .
    \end{equation}
    The constant $b$ can be absorbed in $\varphi_0$.
    Therefore, $(L_d^-, L_d^+)$ and $(\tilde{L}_d^-, \tilde{L}_d^+)$ lead to the same forced discrete Euler--Lagrange equations if and only if
    \begin{subequations}
    \begin{equation}
    \tilde L_d^+(q_0, q_1) = L_d^+(q_0, q_1) + 2\psi (q_0) +2\varphi_1 (q_1) ,
    \end{equation}
    and
    \begin{equation}
    \tilde L_d ^-(q_0, q_1) = L_d^-(q_0, q_1) + 2\varphi_0(q_0) - 2\psi(q_1) ,
    \end{equation}
    \end{subequations}
    for some functions $\psi, \varphi_0, \varphi_1$ on $Q$ 
    and some constant $b$.
    Obviously, $\tilde{L}_d^\pm$ are the modified Lagrangians associated with $(\tilde L_d, \Rayl_d)$ if and only if
    \begin{equation}
    \tilde L_d = \frac{1}{2} \left(\tilde L_d^+ + \tilde L_d^-  \right),
    \end{equation}
    and
    \begin{equation}
    \tilde{\Rayl}_d = \frac{1}{2} \left(\tilde L_d^+ - \tilde L_d^-  \right),
    \end{equation}
    from where the result follows.
\end{proof}

\subsection{Exact discrete Lagrangian and external force}

Let $(Q, L, \alpha)$ be a regular forced Lagrangian system. Consider a sufficiently small time step $h\in \RR$ and a sufficiently small open subset $U\subseteq Q$ such that such that the forced Euler--Lagrange equations for $L$ with boundary conditions $c_{0, \, 1}(0) = q_0\in U$ and $c_{0, \, 1}(h) = q_{1}\in U$ have a unique solution $c_{0, \, 1}\colon [0, h]\to Q$. The \emph{exact discrete Lagrangian} $L_d^{\mathrm{ex}}\in \Cinfty(U\times U)$ is given by
\begin{equation}
    L_d^{\mathrm{ex}}(q_0, q_{1}) = \int_0^h L\big(c_{0, \, 1}(t), \dot{c}_{0, \, 1}(t)\big)\, \dd t\, .
\end{equation}
The \emph{exact discrete external force} $f_d^{\mathrm{ex}} = (f_d^{\mathrm{ex}+}, f_d^{\mathrm{ex}-}) \in \Omega^1(U \times U)$ is defined by
\begin{equation}
\begin{aligned}
    &f_{d}^{\mathrm{ex}+}\left(q_{0}, q_{1}\right) = -\int_{0}^{h} \alpha\big(c_{0, \, 1}(t), \dot{c}_{0, \, 1}(t)\big) \cdot \frac{\partial q(t)}{\partial q_{1}} \dd t\, , \\
    &f_{d}^{\mathrm{ex}-}\left(q_{0}, q_{1}\right)= - \int_{0}^{h} \alpha\big(c_{0, \, 1}(t), \dot{c}_{0, \, 1}(t)\big)  \cdot \frac{\partial q(t)}{\partial q_{0}} \dd t\, .
\end{aligned}
\end{equation}
The minus signs have been included in order to be consistent with Lew, Marsden, Ortiz and West's sign criterion for discrete external forces \cite{M.W2001,L.M.O+2004}, which is different from the sign criterion employed in the previous chapters (for instance, compared the continuous and discrete versions of Lagrange--d'Alembert principle). 
It can be shown \cite{M.W2001} that the exact discrete system is equivalent to the corresponding continuous system.
More precisely, solutions $c\colon [0, t_N]\to Q$ of the forced Euler--Lagrange equations for $(L, \alpha)$ and solutions $\{q_k\}_{k=0}^N$ of the forced discrete Euler--Lagrange equations for $(L_d^{\mathrm{ex}}, f_d^{\mathrm{ex}})$ are related by
\begin{equation}
\begin{array}{ll}
    q_k = c(t_k)\, , \quad &  k=0, \ldots, N\, ,\\
    c(t) = c_{k,\, k+1} (t)\, , & t_k \leq t \leq t_{k+1}\, ,
\end{array}
\end{equation}
where $c_{k,\, k+1}\colon[t_k, t_{k+1}]\to Q$ is the unique solution of the forced Euler--Lagrange equations for $L$ satisfying $c_{k,\, k+1}(kh) = q_k$ and $c_{k,\, k+1}((k+1)h) = q_{k+1}$.

\begin{example}[Harmonic oscillator with Rayleigh dissipation]\label{example_harmonic_oscillator}
    Consider the Rayleigh system $(\RR, L, \Rayl)$, with
    \begin{equation}
      L= \frac{1}{2}m v^2 - \frac{1}{2}kq^2\, , \quad \Rayl=\frac{r}{2} v^2\, ,
    \end{equation}
    for some costants $m, k, r\in \RR$.
    Then the forced Euler--Lagrange equations yield
    \begin{equation}
      m\ddot{q}+r\dot{q}+kq=0\, . \label{harmonic_oscillator_Rayleigh_ODE}
    \end{equation}
    Suppose that $4km>r^2$, and let
    \begin{equation}
      a \coloneqq \frac{r}{2m}\, ,\quad 
      b \coloneqq \frac{\sqrt{4km-r^2}}{2m}\, .
    \end{equation}
    The solution of the ordinary differential equation \eqref{harmonic_oscillator_Rayleigh_ODE} with boundary values $q(0)=q_0$ and $q(h)=q_1$ is 
    \begin{equation}
      q(t) = e^{-a t} \left[ 
          q_0 \cos(b t)  + c \sin(b t) 
        \right]\, ,
    \end{equation}
    where
    \begin{equation}
      c = \frac{e^{a h}q_1-\cos(bh) q_0}{\sin(bh)}\, .
    \end{equation}
    The exact discrete Lagrangian is given by
    \begin{equation}
    \begin{aligned} 
      L_d
      =&
      \frac{1}{16} (\coth (b h)-1)^2
      \left[\frac{\left(m (a-b)^2-k\right) \left(q_0 -q_1  e^{h (a+b)}\right)^2}{a-b}
        \right.\\&\left.
       +\frac{\left(m (a+b)^2-k\right) \left(q_0  e^{2 b h}-q_1  e^{h (a+b)}\right)^2}{a+b}
      \right.\\&\left.
      +\frac{2 \left(m \left(b^2-a^2\right)+k\right) \left(q_0  e^{2 b h}-q_1  e^{h (a+b)}\right) \left(q_0 -q_1  e^{h (a+b)}\right)}{a}
      \right.\\&\left.
        + e^{-2 h (a+b)} \left(\frac{e^{4 b h} \left(k-m (a-b)^2\right) \left(q_0 -q_1  e^{h (a+b)}\right)^2}{a-b}
      \right.\right.\\&\left.\left.
    -\frac{2 e^{3 b h} \left(m \left(b^2-a^2\right)+k\right) \left(q_0  e^{b h}-q_1  e^{a h}\right) \left(q_0 -q_1  e^{h (a+b)}\right)}{a}
      \right.\right.\\&\left.\left.
     + \frac{\left(k-m (a+b)^2\right) \left(q_0  e^{2 b h}-q_1  e^{h (a+b)}\right)^2}{a+b}\right)
     \right]\, ,
    \end{aligned}
    \end{equation}
    and the exact discrete force is $f_d = (f_d^+, f_d^-)$, where
    \begin{equation}
    \begin{aligned}
        f_d^+ (q_0,q_1) = \frac{1}{2} r \left(\frac{b q_0 \sinh (a h) \csc (b h)}{a}-q_1\right)\, , \\
        f_d^-(q_0,q_1) = \frac{r (a q_0-b q_1 \sinh (a h) \csc (b h))}{2 a} \, .
    \end{aligned}
    \end{equation}
    One can check that condition \eqref{eq:condition_discrete_Rayleighable} holds, and thus $f_d$ is Rayleighable. As a matter of fact, 
    \begin{equation}
      \Rayl_d (q_0,q_1) = \frac{1}{4} r \left(q_0^2+q_1^2\right) -\frac{b q_0 q_1 r \sinh (a h) \text{csch}(b h)}{2 a}
    \end{equation}
    is a discrete Rayleigh potential from which $f_d$ can be derived.
\end{example}


\subsection{Midpoint rule}
Consider a regular forced Lagrangian system $(Q, L, \alpha)$. The associated \emph{midpoint rule discrete Lagrangian} $L_d\in \Cinfty(Q\times Q)$ is given by \cite{M.W2001}
\begin{equation}
  L_d^{\frac{1}{2}}\left( q_0, q_1 \right) = h L \left( \frac{q_0 + q_1}{2}, \frac{q_1-q_0}{h}  \right)\, ,
\end{equation}
where $h\in \RR$ is the time step.
Similarly, the \emph{midpoint rule discrete force} $f_d^{\frac{1}{2}}=\left(f_d^{\frac{1}{2}+}, f_d^{\frac{1}{2}-}\right)\in \Omega^1(Q\times Q)$ is given by\footnote{Once again the minus sign is included in order to compensate the different sign criteria employed in the continuous and discrete Lagrange--d'Alembert principles.}
\begin{equation}
  f_d^{\frac{1}{2}+}\left( q_0, q_1  \right) 
  = f_d^{\frac{1}{2}-}\left( q_0, q_1  \right) 
  = -\frac{h}{2} \alpha \left( \frac{q_0 + q_1}{2}, \frac{q_1-q_0}{h}  \right)\, .
\end{equation}
By equation \eqref{eq:condition_discrete_Rayleighable}, the midpoint rule discrete force is Rayleighable if and only if 
\begin{equation}
    \DD_1 \alpha(q, v) = 0\, ,
\end{equation}
that is, if $\alpha$ only depends on the velocities.

\begin{remark}
Let $(Q, L, \Rayl)$ be a Rayleigh system such that $\partial\Rayl/\partial q^i=0$ for all $i=1, \ldots, \dim Q$. Then the associated midpoint rule discrete force is Rayleighable. As a matter of fact, the function $\mathcal R_d^{\frac{1}{2}}\in \Cinfty(Q\times Q)$ given by
\begin{equation}
  \mathcal R_d^{\frac{1}{2}}(q_0, q_1) = \frac{h^2}{2} \Rayl\left( \frac{q_0+q_1}{2}, \frac{q_1-q_0}{h} \right)
\end{equation}
is a discrete Rayleigh potential for $f_d^{\frac{1}{2}}$, which will be called the \emph{midpoint rule discrete Rayleigh potential}.

\end{remark}

\begin{example} \label{example_Marsden_West}
Consider a Rayleigh system $(\RR^2, L, \Rayl)$, with
\begin{equation}
  L (q, v) = \frac{1}{2} \norm{v}^2 - \norm{q}^2 \left( \norm{q}^2 -1  \right)^2\, , 
\end{equation} 
and
\begin{equation}
  \Rayl (q, v) = \frac{1}{2} k \norm{v}^2\, ,
\end{equation}
for some constant $k$. Here $q=(q^1, q^2)$ are the Cartesian coordinates in $\RR^2$, $(q, v )= (q^1, q^2 , v^1, v^2)$ are the induced fibred coordinates in $T\RR^2$, and $\norm{\cdot}$ denotes the Euclidean norm in $\RR^2$. Hence, the external force defined by $\Rayl$ is 
\begin{equation}
    \alpha(q,v) = \Sendoadj \circ \dd \Rayl(q,v) = k v^1 \dd q^1 + k v^2 \dd q^2\, .
\end{equation}

For a time step $h$, the midpoint rule discrete Lagrangian is
\begin{equation}
    L_d^{\frac{1}{2}} (q_0, q_1) = \frac{h}{2} \norm{\frac{q_1-q_0}{h}}^2 - h\norm{\frac{q_0+q_1}{2}}^2 \left( \norm{\frac{q_0+q_1}{2}}^2 -1  \right)^2\, ,
\end{equation}
and the exact discrete force is $f_d^{\frac{1}{2}} = \left(f_d^{\frac{1}{2}+}, f_d^{\frac{1}{2}-}\right)$, with
\begin{equation}
    f_d^{\frac{1}{2}+} (q_0, q_1) = f_d^{\frac{1}{2}-} (q_0, q_1) 
    = - \frac{kh}{2} \frac{q_1^1 - q_0^1}{h} \dd q^1 - \frac{kh}{2} \frac{q_1^2 - q_0^2}{h} \dd q^2 \, ,
\end{equation}
where $(q_0^1, q_0^2)$ and  $(q_1^1, q_1^2)$ are the canonical coordinates of $q_0$ and $q_1$, respectively. Notice that $\tparder{\Rayl}{q^1} = \tparder{\Rayl}{q^2} = 0$, and thus $f_d^{\frac{1}{2}}$ is Rayleighable. Indeed, the function 
\begin{equation}
    \Rayl^{\frac{1}{2}} (q_0, q_1) = \frac{kh^2}{4} \norm{\frac{q_1-q_0}{h}}^2
\end{equation}
is its midpoint rule discrete Rayleigh potential.


Henceforth, let $k=10^{-3}$, corresponding to Example 3.2.3 from \cite{M.W2001}. In Figure~\ref{fig_energy_Marsden_West} the evolution of the energy of the system is plotted, with the initial conditions $q_0^1=0$, $\dot{q}_0=\left(1/2, 0 \right)$ and $E_L(q_0, \dot{q}_0)=11/40$. The variational midpoint rule and the standard fourth-order Runge--Kutta method (see, for instance, \cite{Newman2013,ButcherJohnC.2016}) are compared with a benchmark numerical integration of high precision. Observe that the variational midpoint rule reproduces the energy dissipation correctly, whereas the Runge--Kutta or other standard integrators do not. This effect is specially relevant when the external force is small compared to the magnitude of the conservative dynamics and the time period of integration (see \cite{H.L1999}).

\begin{figure}[t]
    \centering
    \includegraphics[width=.9\linewidth]{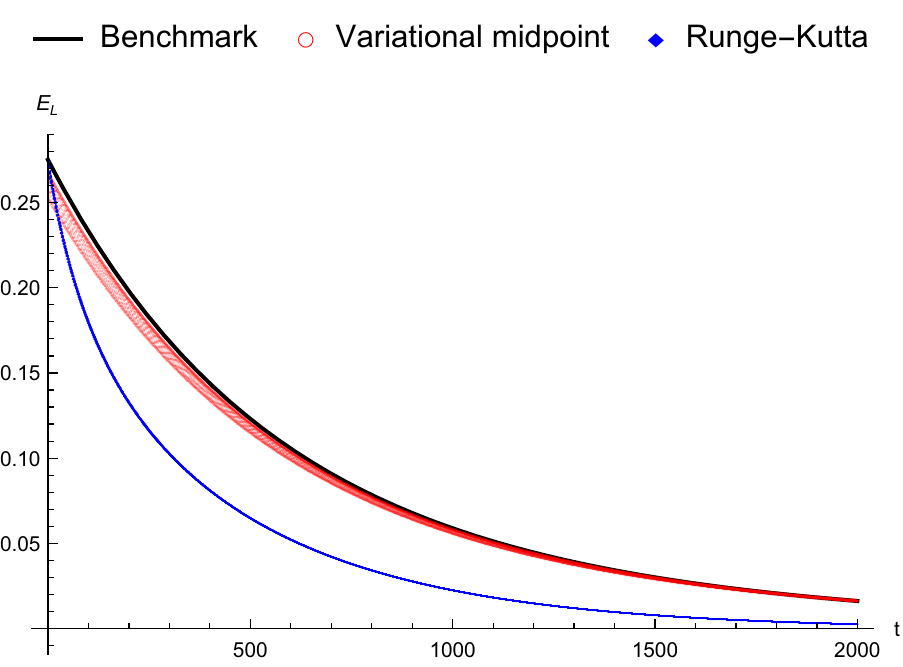}
    \caption[Energy of a Rayleigh system computed with the variational midpoint and fourth-order Runge--Kutta methods]{Energy of a Rayleigh system computed with the variational midpoint and fourth-order Runge--Kutta methods. Observe the remarkable supremacy of the former, despite being a lower order method.}
    \label{fig_energy_Marsden_West}
\end{figure}

\end{example}



\subsection{Discrete Noether's theorem}

Consider the left Lie group action $\Phi:G\times Q \to Q$ of $G$ on $Q$, and let $\xi_Q\in \X(Q)$ be the infinitesimal generator of this action. This action can be lifted to $Q\times Q$ by the product
\begin{equation}
  \Phi_g^{Q\times Q} (q_0, q_1) = \left(\Phi_g (q_0), \Phi_g (q_1)  \right).
\end{equation}
For each $\xi \in \mathfrak{g}$, the infinitesimal generator of this action is the vector field $\xi_{Q\times Q} (q_0, q_1) = (\xi_Q(q_0), \xi_Q(q_1))\in \X(Q\times Q)$.

Let $\mathfrak{g}$ denote the Lie algebra of $G$, and $\mathfrak{g}^\ast$ its dual. The \emph{discrete momentum maps} $\mommap_{L_d}^{f+},\, \mommap_{L_d}^{f-}\colon Q\times Q\to \mathfrak{g}^*$ are defined by
\begin{equation}
\begin{aligned}
  & \left\langle \mommap_{L_d}^{f+}(q_0, q_1), \xi   \right\rangle 
  = \left\langle \Legp (q_0,q_1), \xi_Q (q_1) \right\rangle\, ,   \\
  & \left\langle \mommap_{L_d}^{f-}(q_0, q_1), \xi   \right\rangle 
  = \left\langle \Legm (q_0,q_1), \xi_Q (q_0) \right\rangle\, . 
\end{aligned}
\end{equation}
If $\left\langle \mommap_{L_d}^{f+}, \xi  \right\rangle = \left\langle \mommap_{L_d}^{f-}, \xi  \right\rangle $, for some $\xi \in \mathfrak g$, one can define the function
\begin{equation}
    \begin{aligned}
      \mommap_d^\xi\colon Q\times Q &\to \RR \\
      (q_0, q_1) & \mapsto \left\langle \mommap_{L_d}^{f+}, \xi  \right\rangle  (q_0, q_1) = \left\langle \mommap_{L_d}^{f-}, \xi  \right\rangle  (q_0, q_1)\, .
    \end{aligned}
\end{equation}  



\begin{theorem}[Discrete forced Noether's theorem]\label{theorem:Noether_discrete}
    Consider a forced discrete Lagrangian system $(Q, L_d, f_d)$. Let $G$ be a Lie group acting on $Q$ and let $\mathfrak{g}$ be the Lie algebra of $G$. Then, the following statements are equivalent:
    \begin{enumerate}
        \item The function $\mommap_d^\xi$ is defined, and it is a constant of the motion, namely, $ \mommap_d^\xi  (q_{N-1}, q_N) = \mommap_d^\xi (q_0, q_1)$ along the discrete Lagrangian flow.
        \item $\xi_{Q\times Q}(L_d) + f_d(\xi_{Q\times Q}) = 0$.
    \end{enumerate} 
\end{theorem}

\begin{proof}
    One can write
    \begin{equation}
    \begin{aligned}
    \left\langle \dd L_d + f_d, \xi_{Q\times Q}  \right\rangle (q_0, q_1)
    & = \left(\DD_1 L_d+ f_d^-  \right)  (q_0, q_1) \cdot \xi_Q(q_0) 
    \\ & \quad
        + \left(\DD_2 L_d+ f_d^+  \right)  (q_0, q_1) \cdot \xi_Q(q_1) \\
    & = \Legp(q_0, q_1) \cdot \xi_q(q_1) 
    \\ & \quad 
    - \Legm (q_0, q_1) \cdot \xi_Q(q_0) \\
    & = \left(\mommap_{L_d}^{f+} - \mommap_{L_d}^{f-}  \right)(q_0,q_1) \cdot \xi\, .
    \end{aligned}
    \end{equation}
    Thus, $\langle \mommap_{L_d}^{f+}, \xi \rangle = \langle \mommap_{L_d}^{f-}, \xi \rangle \eqqcolon \mommap^\xi$ if and only if $\xi_{Q\times Q}(L_d) + f_d(\xi_{Q\times Q})$ vanishes.
    Moreover, 
    \begin{equation}
    \begin{aligned}
    &  \sum_{k=0}^{N-1} \left\langle \dd L_d + f_d, \xi_{Q\times Q}  \right\rangle (q_k, q_{k+1})\\
    & = \sum_{k=1}^{N-1}\left[\DD_{2} L_{d}\left(q_{k-1}, q_{k}\right)+\DD_{1} L_{d}\left(q_{k}, q_{k+1}\right)
    \right.\\ & \quad  \left.
    +f_{d}^{+}\left(q_{k-1}, q_{k}\right)+f_{d}^{-}\left(q_{k}, q_{k+1}\right)\right] \cdot \xi_{Q}\left(q_{k}\right)\\
    &\quad +\left[\DD_{2} L_{d}\left(q_{N-1}, q_{N}\right)+f_{d}^{+}\left(q_{N-1}, q_{N}\right)\right] \cdot \xi_{Q}\left(q_{N}\right)
    \\ & \quad 
    +\left[\DD_{1} L_{d}\left(q_{0}, q_{1}\right)+f_{d}^{-}\left(q_{0}, q_{1}\right)\right] \cdot \xi_{Q}\left(q_{0}\right)\\
    &=\left[\DD_{2} L_{d}\left(q_{N-1}, q_{N}\right)+f_{d}^{+}\left(q_{N-1}, q_{N}\right)\right] \cdot \xi_{Q}\left(q_{N}\right)
    \\ & \quad 
    +\left[\DD_{1} L_{d}\left(q_{0}, q_{1}\right)+f_{d}^{-}\left(q_{0}, q_{1}\right)\right] \cdot \xi_{Q}\left(q_{0}\right)\\
    & =\Legp \left(q_{N-1}, q_{N}\right) \cdot \xi_{Q}\left(q_{N}\right)-\Legm \left(q_{0}, q_{1}\right) \cdot \xi_{Q}\left(q_{0}\right)\\
    & =\left\langle \mommap_{L_d}^{f+} (q_{N-1}, q_N) - \mommap_{L_d}^{f-} (q_0, q_1), \xi  \right\rangle
    \, ,
    \end{aligned}
    \end{equation}
    where the forced discrete Euler--Lagrange equations \eqref{eq:forced_discrete_Euler-Lagrange} have been used. Therefore,
    \begin{equation}
        \mommap_d^\xi  (q_{N-1}, q_N) - \mommap_d^\xi (q_0, q_1)\, ,
    \end{equation}
    that is, $\mommap_d^\xi$ is a constant of the motion, if and only if $\xi_{Q\times Q}(L_d) + f_d(\xi_{Q\times Q})$ vanishes. 
\end{proof}




\begin{theorem}
Consider a forced discrete Lagrangian system $(Q, L_d, f_d)$. Let $G$ be a Lie group acting on $Q$ and let $\mathfrak{g}$ be the Lie algebra of $G$.
Suppose that $L_d$ is $G$-invariant. Then, for each $\xi\in \mathfrak{g}$,
\begin{enumerate}
\item $\mommap_d^\xi$ is a constant of the motion if and only if 
\begin{equation}
  f_d(\xi_{Q\times Q}) = 0 \, . \label{eq_constant_motion_subalgebra}
\end{equation}
\item If the equation above holds, then $\xi$ leaves $f_d$ invariant if and only if
\begin{equation}
  \contr{\xi_{Q\times Q}} \dd f_d = 0 \, .
\end{equation}
\end{enumerate}
Moreover, the vector subspace $\mathfrak{g}_{f_d}$ of $\mathfrak{g}$ given by
\begin{equation}
  \mathfrak{g}_{f_d} = \left\{\xi \in \mathfrak{g}\mid  f_d(\xi_{Q\times Q}) = 0,\ \contr{\xi_{Q\times Q}} \dd f_d=0   \right\}
\end{equation}
is a Lie subalgebra of $\mathfrak g$.
\end{theorem}

\begin{proof}
    Obviously, $L_d$ is $G$-invariant if and only if
    \begin{equation}
    \xi_{Q\times Q} (L_d) = 0\, ,
    \end{equation}
    for each $\xi \in \mathfrak{g}$. 
    Combining this with \Cref{theorem:Noether_discrete} implies that $\mommap_d^\xi$ is a constant of the motion if and only if 
    \begin{equation}
    f_d(\xi_{Q\times Q}) = 0\, .
    \end{equation}
    If this equation holds, $f_d$ is $\xi$-invariant if and only if
    \begin{equation}
    \liedv{\xi_{Q\times Q}} f_d = \contr{\xi_{Q\times Q}} \dd f_d = 0\, .
    \end{equation}
    Let $\xi, \eta\in \mathfrak{g}_{f_d} = \left\{\xi \in \mathfrak{g}\mid  f_d(\xi_{Q\times Q}) = 0,\ \contr{\xi_{Q\times Q}} \dd f_d=0   \right\}$. Then,
    \begin{equation}
    \begin{aligned}
    f_d \left([\xi_Q, \eta_Q]^c  \right)
    &= f_d \left([\xi_{Q\times Q}, \eta_{Q\times Q}]  \right)
    = \contr{[\xi_{Q\times Q}, \eta_{Q\times Q}]} f_d
    \\ &
    = \liedv{\xi_{Q\times Q}} \contr{\eta_{Q\times Q}} f_d 
        - \contr{\eta_{Q\times Q}} \liedv{\xi_{Q\times Q}} f_d\\
    &= \xi_{Q\times Q}\left(f_d(\eta_{Q\times Q})  \right)
        -\eta_{Q\times Q}\left(f_d(\xi_{Q\times Q})  \right)
        \\ &\quad
        - \contr{\eta_{Q\times Q}} \contr{\xi_{Q\times Q}} \dd f_d
    = 0 \, ,
    \end{aligned}
    \end{equation}
    and
    \begin{equation}
    \begin{aligned}
    \contr{[\xi_Q, \eta_Q]^c} \dd f_d
    &= \contr{[\xi_{Q\times Q}, \eta_{Q\times Q}] } \dd f_d
    = \liedv{\xi_{Q\times Q}} \contr{\eta_{Q\times Q}} \dd f_d 
        - \contr{\eta_{Q\times Q}} \liedv{\xi_{Q\times Q}} \dd f_d\\
    &= \liedv{\xi_{Q\times Q}} \contr{\eta_{Q\times Q}} \dd f_d 
        - \contr{\eta_{Q\times Q}} \dd \liedv{\xi_{Q\times Q}} f_d
    = 0 \, .
    \end{aligned}
    \end{equation}
    Since $\xi\in \mathfrak{g}\mapsto \xi_Q\in \X(Q)$ is a Lie algebra anti-homomorphism, this proves that $[\xi, \eta]\in \mathfrak{g}_{f_d}$.
\end{proof}

This is the discrete analogue from \Cref{prop:Lie_subalgebra_forced_Lagrangian}.
The first statement was previously found by Marsden and West \cite[Theorem 3.2.1]{M.W2001}.

Define the vector field $\widehat \xi_{Q\times Q}\in \mathfrak{X}(Q\times Q)$ given by 
\begin{equation}
  \widehat \xi_{Q\times Q}(q_0, q_1) = \left(\xi_Q(q_0), -\xi_Q(q_1)  \right)\, .
\end{equation}
In other words, $\widehat{\xi}_{Q\times Q}$ is an infinitesimal generator of the action $\widehat{\Phi}\colon G\times Q\times Q \to Q\times Q$ given by
\begin{equation}
    \widehat{\Phi}_g (q_0, q_1) = \left(\Phi_g(q_0), \Phi_{g^{-1}}(q_1)\right)\, .
\end{equation}
Let $(Q, L_d, \Rayl_d)$ be a discrete Rayleigh system, with associated discrete external force $f_d=(f_d^+, f_d^-)$. Then,
\begin{equation}
\begin{aligned}
    \widehat{\xi}_{Q\times Q} (\mathcal R_d) (q_0, q_1) 
   & = \DD_1  \mathcal R_d (q_0, q_1) \cdot \xi_Q(q_0)
    - \DD_2 \mathcal R_d (q_0, q_1) \cdot \xi_Q(q_1)\\
    &= f_d^-(q_0, q_1) \cdot \xi_Q(q_0) + f_d^+(Qq_0, q_1) \cdot \xi_Q(q_1)
    \\ &
    = f_d(q_0, q_1) \cdot \xi_{Q\times Q}(q_0, q_1)\, .
\end{aligned}
\end{equation}
Combining this with \Cref{theorem:Noether_discrete} implies the following.

\begin{proposition}
Let $(Q, L_d, \Rayl_d)$ be a discrete Rayleigh system. Given $\xi \in \mathfrak{g}$, the function $J^\xi$ is defined and it is a constant of the motion if and only if
    \begin{equation}
        \xi_{Q\times Q}(L_d) +   \widehat{\xi}_{Q\times Q} (\mathcal R_d)  = 0\, .
    \end{equation}
\end{proposition}

\begin{example} Consider a discrete Rayleigh system $(\RR^2, L_d, \mathcal R_d)$ of the form
\begin{equation}
\begin{aligned}
  &L_d(r_1, \theta_1, r_2, \theta_2) 
    = \frac{h}{2} \left(\frac{r_2-r_1}{h}  \right)^2
    +\frac{h}{2} \left(\frac{r_1+r_2}{2}  \right)^2 \left(\frac{\theta_2-\theta_1}{h}  \right)^2
    \\ & 
    - V \left(\frac{r_1+r_2}{2}  \right)
    \, ,\\
    &\mathcal R_d (r_1, \theta_1, r_2, \theta_2)  
    = F \left(r_1, r_2, {\theta_1+\theta_2}  \right)\, ,
\end{aligned}
\end{equation}
where $(r_1, \theta_1, r_2, \theta_2)$ are the coordinates in $\RR^2\times \RR^2$ induced by the polar coordinates $(r, \theta)$ in $\RR^2$. The function $V$ depends only on the combination $(r_1+r_2)/2$, while $F$ depends on $r_1,\, r_2$ and $\theta_1+\theta_2$. For instance, $(\RR^2, L_d, \mathcal R_d)$ could be the midpoint rule discretization of the Rayleigh system $(\RR^2, L, \mathcal R)$ from \Cref{example_Marsden_West}.

Consider the action of $\mathbb{S}^1$ on $\RR^2$ by rotations. The infinitesimal generator of this action is $\xi_{\RR^2}= \partial/\partial \theta$, and thus
\begin{equation}
  \xi_{\RR^2\times \RR^2}= \frac{\partial  } {\partial \theta_1} + \frac{\partial  } {\partial \theta_2}, \qquad
  \widehat \xi_{\RR^2\times \RR^2}= \frac{\partial  } {\partial \theta_1} - \frac{\partial  } {\partial \theta_2} \, .
\end{equation}
 Clearly, $\xi_{\RR^2 \times \RR^2}(L_d)=0$ and $\widehat \xi_{\RR^2 \times \RR^2}(\mathcal R_d)=0$. Hence,
 \begin{equation}
   \mommap^\xi (r_1, \theta_1, r_2, \theta_2) = \left(\frac{r_1+r_2}{2}  \right)^2 \left(\frac{\theta_2-\theta_1}{h}  \right)
   - \DD_3 F(r_1, r_2, \theta_1+ \theta_2 )
 \end{equation}
is a constant of the motion. 
In particular, if $\DD_3 F(r_1, r_2, \theta_1+ \theta_2 )=0$, then the angular momentum is conserved.
\end{example}

After studying the symmetries of a dynamical system under a symmetry group, the next natural question is how to reduce the system by the action of this group. As a matter of fact, a process for reduction and reconstruction of forced Lagrangian systems with symmetries has been recently proposed by Caruso, Fernández, Tora and Zuccalli \cite{C.F.T+2023}.


\section{Discrete Hamilton--Jacobi theory}

In terms of the right discrete Hamiltonian, the $k$-th discrete action sum evaluated along a discrete curve $c_d = \{q_j\}_{j=0}^{N}$ can be written as
\begin{equation}
    \action_d^k(q_k) = \sum_{j=0}^{k-1} \left[p_{j+1} \cdot q_{j+1} - H_d^+ (q_j, p_{j+1})  \right]\, ,
\end{equation}
which is seen as a function of the endpoint $q_k$ and the discrete end time $k$.
Therefore,
\begin{equation}\label{eq:right_subs_actions}
  \action_d^{k+1}(q_{k+1}) - \action_d^k(q_k) = p_{k+1} \cdot q_{k+1} - H_d^+ (q_k, p_{k+1}) \, ,
\end{equation}
where $p_{k+1}$ is considered to be a function of $q_k$ and $q_{k+1}$.
Deriving both sides of this equation with respect to $q_{k+1}$ yields
\begin{equation}
  \DD  \action_d^{k+1}(q_{k+1}) = p_{k+1} + \parder{p_{k+1}}{q_{k+1}} \cdot \left[ q_{k+1}-\DD_2 H_d^+(q_k, p_{k+1})\right]\, ,
\end{equation}
The discrete curve being a solution of the right discrete Hamilton equations~\eqref{eq:right_discrete_Hamilton} implies that
\begin{equation}\label{eq:right_der_action}
  \DD  \action_d^{k+1}(q_{k+1}) = p_{k+1} - f_d^+(q_k, q_{k+1})\, .
\end{equation}
Define the maps $\gamma^+_k, \, \gamma^-_k \colon Q\times Q \to \cT Q$ by
\begin{subequations}
\begin{equation} \label{eq:definition_gamma_plus}
  \gamma^+_k \colon (q_k,q_{k+1}) \mapsto \DD \action_d^{k+1} (q_{k+1}) + f_d^+(q_k,q_{k+1}) \, ,
\end{equation}
and
\begin{equation} \label{eq:definition_gamma_minus}
  \gamma^-_k \colon (q_k,q_{k+1}) \mapsto \DD \action_d^{k} (q_{k}) - f_d^-(q_k,q_{k+1})\, . 
\end{equation}
\end{subequations}
Let $\gamma_k =(\gamma^-_k, \gamma^+_k)\colon Q \times Q \to \cT Q \times \cT Q \simeq \cT (Q\times Q)$.

The combination of equations~\eqref{eq:right_subs_actions} and \eqref{eq:right_der_action} yields
\begin{equation}\label{eq:FRDHJ}
  \action_d^{k+1}(q_{k+1}) - \action_d^k(q_k) - \gamma_k^+(q_k, q_{k+1}) \cdot q_{k+1} + H_d^+ \left(q_k,\, \gamma_k^+(q_k, q_{k+1}) \right) = 0 \, .
\end{equation}
This equation will be called the \emph{forced right discrete Hamilton--Jacobi equation}.

Similarly, in terms of the left discrete Hamiltonian, the $k$-th discrete action reads
\begin{equation}
  \action_d^k(q_k) = - \sum_{j=0}^{k-1} \left[p_{j} \cdot q_{j}  + H_d^- (q_{j+1}, p_{j})  \right]\, ,
\end{equation}
Therefore, 
\begin{equation}\label{eq:left_subs_actions}
  \action_d^{k+1}(q_{k+1}) - \action_d^{k}(q_{k}) = -p_{k} \cdot q_{k}  - H_d^- (q_{k+1}, p_{k})\, ,
\end{equation}
which derived with respect to $q_k$ yields
\begin{equation}
  -\DD  \action_d^{k}(q_{k}) = - p_k  - \parder{p_{k}}{q_{k}} \cdot \left[ q_{k}+\DD_2 H_d^-(q_{k+1}, p_{k})\right]\, ,
\end{equation}
and using equation~\eqref{eq:left_discrete_Hamilton_a} one can write
\begin{equation}\label{eq:left_der_action}
  \DD  \action_d^{k}(q_{k}) =  p_k  + f_d^-(q_{k}, q_{k+1}) \, . 
\end{equation}
The combination of equations~\eqref{eq:left_subs_actions} and \eqref{eq:left_der_action} yields
\begin{equation}\label{eq:FLDHJ}
  \action_d^{k+1}(q_{k+1}) - \action_d^{k}(q_{k}) + \gamma^-_ k(q_k, q_{k+1}) \cdot q_{k}  + H_d^- (q_{k+1}, \gamma^-_k(q_k, q_{k+1})) = 0\, ,
\end{equation}
This equation will be called the \emph{forced left discrete Hamilton--Jacobi equation}.

Define the maps
\begin{equation}
\begin{aligned}
  \left(\mathcal{F}_d^H   \right)^{\gamma^+_k}
  =\pi_Q\circ \mathcal{F}_d^H \circ \gamma^+_k\colon Q\times Q&\to Q
  ,\\
   \left(\mathcal{F}_d^H   \right)^{\gamma^-_k}
  =\pi_Q\circ \mathcal{F}_d^H \circ \gamma^-_k\colon Q\times Q&\to Q
  \, ,
\end{aligned}
\end{equation}
and
\begin{equation}
\begin{aligned} 
  \left(\mathcal{F}_d^H   \right)^{\gamma_k}
  \colon Q\times Q&\to Q\times Q\\
    (q_{j-1},q_{j})  
    &\mapsto  \left( \left(\mathcal{F}_d^H   \right)^{\gamma^-_k} (q_{j-1},q_{j}),
      \left(\mathcal{F}_d^H   \right)^{\gamma^+_k}  (q_{j-1},q_{j})  \right)
    \, ,
\end{aligned}
\end{equation}
where $\mathcal{F}_d^H$ is the discrete Hamiltonian flow \eqref{eq:discrete_Hamiltonian_flow}. 

\begin{theorem}\label{theorem:discrete_HJ}
  Let $\left\{q_k  \right\}_{k=0}^N\subset Q$ be a sequence of points such that
  \begin{equation}
    q_{k+1} = \left(\mathcal{F}_d^H  \right)^{\gamma^+_k} (q_{k-1}, q_k)\, . 
  \end{equation}
  Suppose that the maps $\{\action_d^k\}_{k=0}^{N-1}$ and $\{\gamma^-_k\}_{k=0}^{N-1}$ satisfy the forced left discrete Hamilton--Jacobi equation \eqref{eq:FLDHJ} along the sequence of points $\left\{q_k  \right\}_{k=0}^N$.
  Then, the sequence of points $\left\{(q_k,p_k)  \right\}_{k=0}^N\subset \cT Q$ with
  \begin{equation}
    (p_k, p_{k+1}) = \gamma(q_{k},q_{k+1})
  \end{equation}
  is a solution of the forced left discrete Hamilton equations \eqref{eq:left_discrete_Hamilton}.

\end{theorem}

\begin{lemma}\label{lemma:discrete_HJ_left}
  Consider the maps $\flowm_k\colon Q\times Q\to Q$ implicitly defined by
  \begin{equation}\label{eq:implicit_definition_flowm}
    \begin{aligned}
     q_k &= 
     f_d^- \left( q_k,  \flowm_k(q_{k-1},q_k) \right) 
     \left[ \DD_{q_k} \gamma^-_k(q_k, \flowm_k(q_{k-1},q_k))   \right]^{-1}\\
     &\ - \DD_2 H_d^- \left( \flowm_k(q_{k-1},q_k), \gamma^-_k(q_k, \flowm_k(q_{k-1},q_k))  \right)
    \end{aligned}
    \end{equation}
    and let $\left\{q_k  \right\}_{k=0}^N\subset Q$ be a sequence of points such that
    \begin{equation}
      q_{k+1} = \flowm_k (q_{k-1}, q_k), 
      \label{q_k_hat_flowm}
    \end{equation}
  Suppose that the maps $\{\action_d^k\}_{k=0}^{N-1}$ and $\{\gamma^-_k\}_{k=0}^{N-1}$ satisfy the forced left discrete Hamilton--Jacobi equation \eqref{eq:FLDHJ} along the sequence $\left\{q_k  \right\}_{k=0}^N$.
  Then, the sequence of points $\left\{(q_k,p_k)  \right\}_{k=0}^N\subset \cT Q$ with
  \begin{equation}
    (p_k, p_{k+1}) = \gamma_k(q_{k},q_{k+1})
  \end{equation}
  is a solution of the forced left discrete Hamilton equations \eqref{eq:left_discrete_Hamilton}.
\end{lemma}

\begin{proof}
  Along the sequence $\left\{(q_k,p_k)  \right\}_{k=0}^N$, the forced left discrete Hamilton equations \eqref{eq:left_discrete_Hamilton} reads
  \begin{equation}
  \begin{aligned}
      \action_d^{k+1} \left(\flowm_k(q_{k-1},q_k)  \right)& - \action_d^k(q_k)
      + \gamma^-_k \left(q_k, \flowm_k(q_{k-1},q_k)  \right) \cdot q_k\\
      &+ H_d^- \left(\flowm_k(q_{k-1},q_k), \gamma^-_k \left(q_k, \flowm_k(q_{k-1},q_k)  \right)  \right) = 0\, ,
  \end{aligned}
  \end{equation}
  Deriving with respect to $q_k$ yields
  \begin{equation}
  \begin{aligned}
    & \DD_2 H_d^- \left(\flowm_k(q_{k-1},q_k), \gamma^-_k \left(q_k, \flowm_k(q_{k-1},q_k)  \right)  \right) 
            \DD_{q_k} \gamma^-_k \left(q_k, \flowm_k(q_{k-1},q_k) \right)
            \\ & 
    + \DD_1 H_d^-  \left(\flowm_k(q_{k-1},q_k), \gamma^-_k \left(q_k, \flowm_k(q_{k-1},q_k)  \right)  \right) \DD_2 \flowm_k(q_{k-1},q_k)
      \\ & 
    + \DD \action_d^{k+1}  \left(\flowm_k(q_{k-1},q_k)  \right) \DD_2\flowm_k(q_{k-1},q_k)
    - \DD \action_d^k(q_k)
    \\ &
    + \DD_{q_k} \gamma^-_k \left(q_k, \flowm_k(q_{k-1},q_k)  \right) \cdot q_k
    \\ & 
    + \gamma^-_k \left(q_k, \flowm_k(q_{k-1},q_k)  \right) 
    = 0\, .
  \end{aligned}
  \end{equation}
  By equation~\eqref{eq:implicit_definition_flowm}, this equation can be written as
  \begin{equation}
    \begin{aligned}
      & \DD_1 H_d^-  \left(\flowm_k(q_{k-1},q_k), \gamma^-_k \left(q_k, \flowm_k(q_{k-1},q_k)  \right)  \right) \DD_2 \flowm_k(q_{k-1},q_k)
        \\ & 
      + \DD \action_d^{k+1}  \left(\flowm_k(q_{k-1},q_k)  \right) \DD_2\flowm_k(q_{k-1},q_k)
      - \DD \action_d^k(q_k)
      \\ &
      + f_d^- \left( q_k,  \flowm_k(q_{k-1},q_k) \right)
      \\ &
      + \gamma^-_k \left(q_k, \flowm_k(q_{k-1},q_k)  \right) 
      = 0\, ,
    \end{aligned}
    \end{equation}
 This equation together with \eqref{eq:definition_gamma_minus} imply that
  \begin{equation}
  \begin{aligned}
    \DD \action_d^{k+1} & \left(\flowm_k(q_{k-1},q_k)  \right) 
    \\ &  
    + \DD_1 H_d^- \left(\flowm_k(q_{k-1},q_k), \gamma^-_k \left(q_k, \flowm_k(q_{k-1},q_k)  \right)  \right) 
    = 0\, .
  \end{aligned}
  \end{equation}
  Using the definition of $\gamma^+_k$ \eqref{eq:definition_gamma_plus}, one can write
  \begin{equation}
  \begin{aligned}
     \gamma^+_k & \left(q_k, \flowm_k(q_{k-1},q_k)   \right)
      - f_d^+ \left(q_k, \flowm_k(q_{k-1},q_k)   \right)
      \\ &
      + \DD_1 H_d^- \left(\flowm_k(q_{k-1},q_k), \gamma^-_k \left(q_k, \flowm_k(q_{k-1},q_k)  \right) \right)
      =0\, ,
  \end{aligned}
  \end{equation}
  that is,
  \begin{equation}
      p_{k+1} - f_d^+(q_k, q_{k+1}) + \DD_1 H_d^- (q_{k+1}, p_k)=0\, ,
  \end{equation}
  which, together with the definition of $\flowm_k$ prove that the forced left discrete Hamilton equations are satisfied. 
\end{proof}

\begin{proof}[Proof of \Cref{theorem:discrete_HJ}]
  By \Cref{lemma:discrete_HJ_left}, it suffices to prove that \eqref{eq:implicit_definition_flowm} is verified by $\flowm_k = \left(\mathcal{F}_d^H  \right)^{\gamma^+_k}$. Indeed,
  \begin{equation}
    \begin{aligned}
     f_d^- &\left( q_k,   \left(\mathcal{F}_d^H  \right)^{\gamma^+_k}(q_{k-1},q_k) \right) 
     \left[ \DD_{q_k} \gamma^-_k(q_k,  \left(\mathcal{F}_d^H  \right)^{\gamma^+_k}(q_{k-1},q_k))   \right]^{-1}\\
     &\ - \DD_2 H_d^- \left(  \left(\mathcal{F}_d^H  \right)^{\gamma^+_k}(q_{k-1},q_k), \gamma^-_k(q_k,  \left(\mathcal{F}_d^H  \right)^{\gamma^+_k}(q_{k-1},q_k))  \right)\\
     & = f_d^- \left( q_k,  q_{k+1} \right) 
     \left[ \DD_{q_k} \gamma^-_k(q_k,  q_{k+1})   \right]^{-1}
      - \DD_2 H_d^- \left( q_{k+1}, \gamma^-_k(q_k, q_{k+1})  \right)
      \\
      & = f_d^- \left( q_k,  q_{k+1} \right) 
      \left(\parder{p_k}{q_k}   \right)^{-1}
       - \DD_2 H_d^- \left( q_{k+1}, p_k \right)\, ,
    \end{aligned}
    \end{equation}
    along the sequence of points $\left\{(q_k,p_k)  \right\}_{k=0}^N$ with $q_{k+1} = \left(\mathcal{F}_d^H  \right)^{\gamma^+_k} (q_{k-1}, q_k)$ and $(p_k, p_{k+1}) = \gamma(q_{k},q_{k+1})$. By construction, the left discrete Hamilton equations are satisfied along the discrete Hamiltonian flow $\mathcal{F}_d^H$. Hence, equation~\eqref{eq:left_discrete_Hamilton_a} is satisfied along $\left(\mathcal{F}_d^H  \right)^{\gamma^+_k}=\pi_Q\circ \mathcal{F}_d^H \circ \gamma^+_k$, and the result follows.
\end{proof}


Alternatively, Elnatanov and Schiff \cite{E.S1996} derived a discrete Hamilton--Jacobi equation by considering generating functions of canonical transformations (see also \cite{O.B.L2011,d.S2018}).


\subsection{Hamilton-Jacobi theory for discrete Rayleigh systems}

If the discrete force is Rayleighable and $\Rayl_d$ is a discrete Rayleigh potentials, one can write
\begin{equation}
\begin{aligned}
    &\gamma^-_k = \DD \action_d^k \circ \pi_1 - \DD_1 \Rayl_d,
    &\gamma^+_k = \DD \action_d^k \circ \pi_2 - \DD_2 \Rayl_d.
\end{aligned}
\end{equation}
In addition, defining the functions
\begin{equation}
\begin{aligned}
    \tilde{\action}_d^{k,\, l} \colon Q\times Q &\to \RR\\
    (q_k, q_l) & \mapsto \action_d^k (q_k) + \action_d^l(q_l)\, ,
\end{aligned}
\end{equation}
and $G_d^{k}= \tilde{\action}_d^{k,\, k+1} - \Rayl_d$, one can write
\begin{equation}
    \gamma_k = \DD \tilde{\action}_d^{k,\, k+1} - \DD \Rayl_d = \DD G_d^k\, ,
\end{equation}
that is, 
\begin{equation}
    \gamma^-_k = \DD_1 G_d^k\, , \quad
    \gamma^+_k = \DD_2 G_d^k\, .
\end{equation}
Thus, the forced right and left discrete Hamilton--Jacobi equations read
\begin{equation}
\begin{aligned}
  \action_d^{k+1}\left(q_{k+1}\right)&-\action_d^{k}\left(q_{k}\right)
  -\DD_2 G_d^k (q_k ,q_{k+1})\cdot q_{k+1}\\
  & +H_{d}^{+}\left(q_{k}, \DD_2 G_d^k (q_k,q_{k+1})\right)=0\, ,
\end{aligned}
\end{equation}
and
\begin{equation}\label{eq:FLDHJ_Rayleigh}
\begin{aligned}
   \action_d^{k+1}\left(q_{k+1}\right)&-\action_d^{k}\left(q_{k}\right)
  +\DD_1 G_d (q_k,q_{k+1}) q_{k}\\
  &+H_{d}^{-}\left(q_{k+1}, \DD_1 G_d (q_k,q_{k+1})\right)=0\, , 
\end{aligned}
\end{equation}
respectively.

\Cref{theorem:discrete_HJ} can be particularized for discrete Rayleigh systems as follows.

\begin{corollary}
  Let $\left\{c_k  \right\}_{k=0}^N\subset Q$ be a sequence of points such that
  \begin{equation}
    q_{k+1} = \pi_Q \circ \mathcal{F}_d^H \circ \DD_2 G_d^k (q_{k-1}, q_k)\, .
  \end{equation}
  Suppose that the maps $\action_d$ and $G_d$ satisfy equation~\eqref{eq:FLDHJ_Rayleigh} along this sequence of points.
  Then, the sequence of points $\left\{(q_k,p_k)  \right\}_{k=0}^N\subset \cT Q$ with
  \begin{equation}
    (p_k, p_{k+1}) = \DD G_d^k(q_{k}, q_{k+1})
  \end{equation}
  is a solution of the forced left discrete Hamilton equations. 
\end{corollary}

\begin{example}
  \begin{figure}[h!]
    \centering
    \begin{subfigure}[t]{.45\linewidth}
      \centering
      \includegraphics[width=\linewidth]{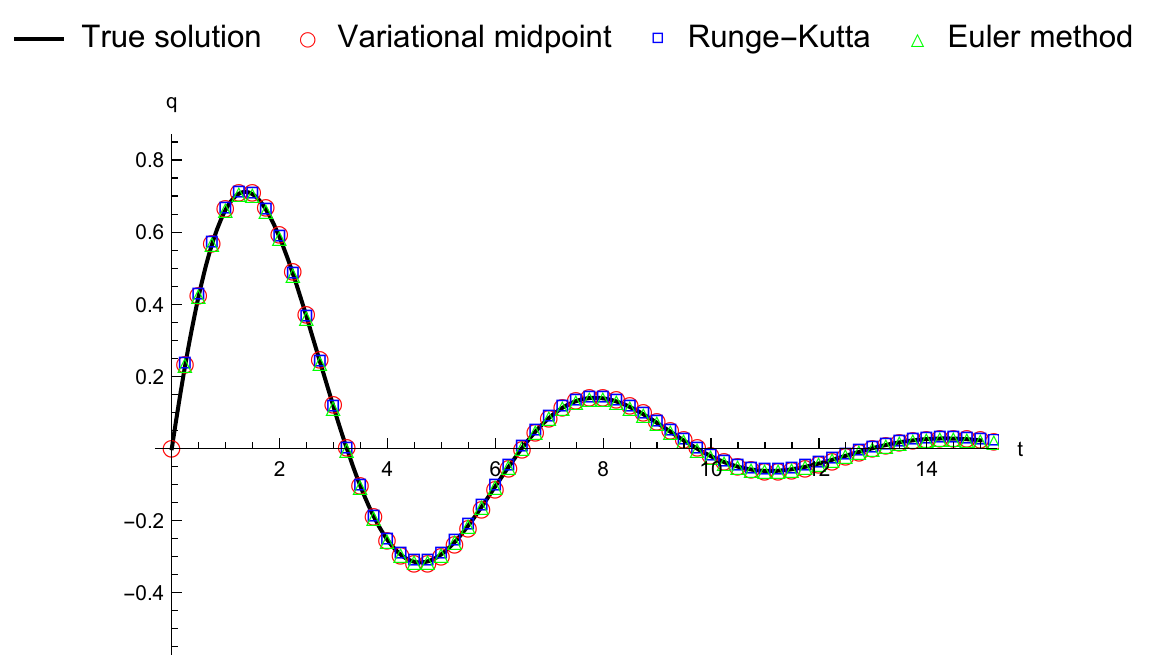}
      \caption{Trajectory}
      \label{plot_positions}
    \end{subfigure}
    \begin{subfigure}[t]{.45\linewidth}
      \centering
      \includegraphics[width=\linewidth]{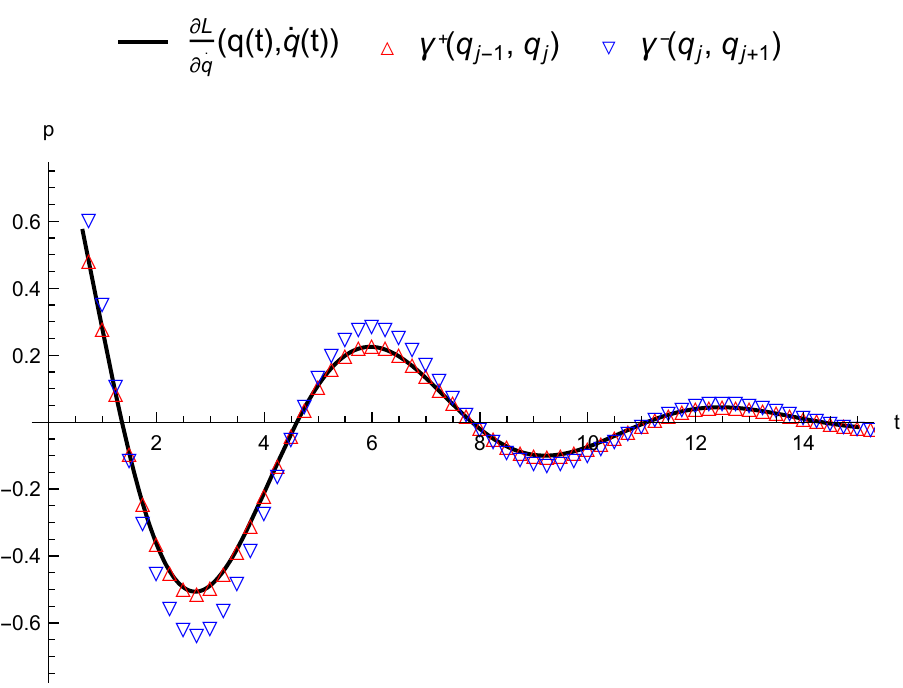}
      \caption{Momenta}
      \label{plot_momenta}
    \end{subfigure}
    \begin{subfigure}[t]{.45\linewidth}
      \centering
      \includegraphics[width=\linewidth]{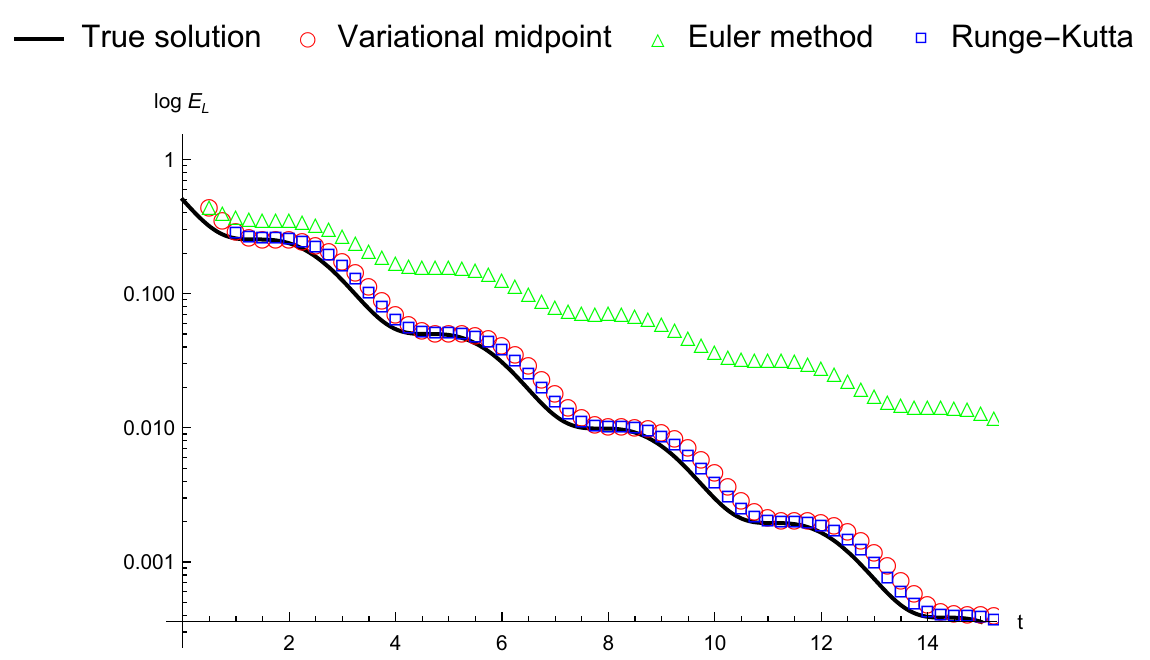}
      \caption{Energy}
      \label{plot_energy}
    \end{subfigure}
    \begin{subfigure}[t]{.45\linewidth}
      \centering
      \includegraphics[width=\linewidth]{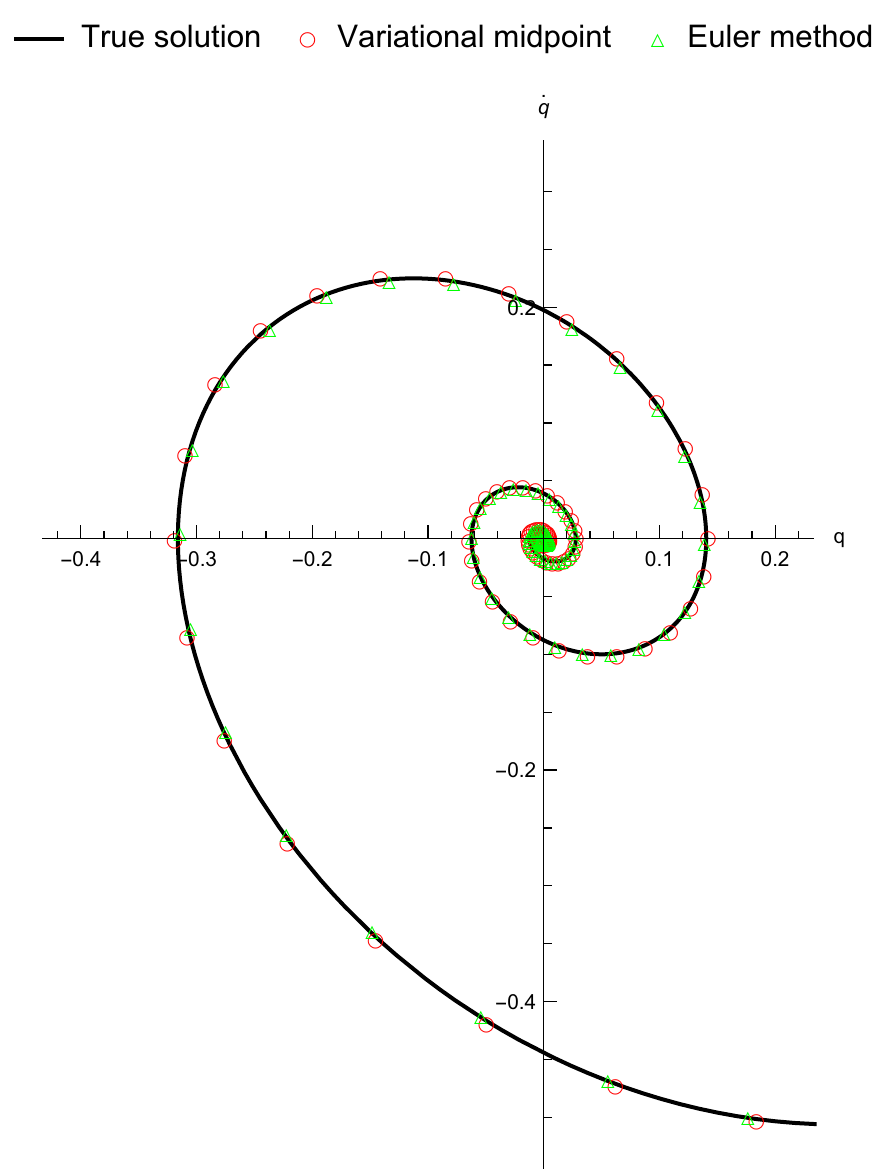}
      \caption{Space of velocities}
      \label{plot_phase_space}
    \end{subfigure}
    \caption[Exact curves and numerical approximations for a harmonic oscillator subject to a linear Rayleigh dissipation]{Harmonic oscillator subject to a linear Rayleigh dissipation.
    The true curves, given by the solution of the continuous forced Euler--Lagrange equation, are compared with the ones obtained from the variational midpoint rule and the Euler and Runge-.Kutta methods.
    }
    \label{plot_harmonic_oscillator}
  \end{figure}

  Consider the discrete Rayleigh system $(\RR, L_d, \Rayl_d)$, with
  \begin{equation}
    L_d (q_{j-1}, q_{j}) = \frac{h}{2} m \left(\frac{q_{j}-q_{j-1}}{h}  \right)^2 - \frac{h}{2}k \left(\frac{q_{j-1}+q_j}{2}  \right)^2\, ,
  \end{equation}
  and
  \begin{equation}
    \Rayl_d (q_{j-1}, q_{j}) =  r  \left(\frac{q_{j}-q_{j-1}}{2}  \right)^2\, ,
  \end{equation}
  where $m, k$ and $r$ are constants and $h$ is the time step.
  The corresponding discrete force $f_d=(f_d^+, f_d^-)$ is given by
  \begin{equation}
    f_d^+(q_{j-1}, q_{j}) = f_d^-(q_{j-1}, q_{j}) = -r \frac{q_{j}-q_{j-1}}{2}.
  \end{equation}
  This system corresponds to the harmonic oscillator with a Rayleigh force (see Example \ref{example_harmonic_oscillator}) discretized via the midpoint rule. The modified discrete Lagrangians are 
  \begin{equation}
    L_d^\pm (q_{j-1},q_{j}) = \frac{h}{2} m  \left(\frac{q_{j}-q_{j-1}}{h}  \right)^2 
    -\frac{h}{2} k  \left(\frac{q_{j}+q_{j-1}}{2}  \right)^2
    \pm  r  \left(\frac{q_{j}-q_{j-1}}{2}  \right)^2\, .
  \end{equation}
  The discrete Legendre transforms are given by
  \begin{equation}
  \begin{aligned}
    \Legp&\colon (q_{j-1}, q_{j})\\
     &\mapsto \left(q_{j}, \frac{m}{h}(q_{j}-q_{j-1}) - \frac{1}{4} kh (q_{j}+q_{j-1}) -r \frac{q_{j}-q_{j-1}}{2} \right)\, ,  \\
    \Legm&\colon (q_{j-1}, q_{j}) \\ 
    &\mapsto \left(q_{j-1}, \frac{m-r}{h}(q_{j}-q_{j-1}) + \frac{1}{4} kh (q_{j}+q_{j-1}) + r \frac{q_{j}-q_{j-1}}{2} \right)\, ,
  \end{aligned}
  \end{equation}
  Thus, the discrete Hamiltonian flow is
  \begin{equation}
  \begin{aligned}
    \mathcal{F}_{d}^{H}: (q_{j}, p_{j})  \mapsto 
    &\left(\frac{h^2 (-k) q_j+4 h p_j+2 h q_j r+4 m q_j}{h^2 k+2 h r+4 m},
    \right.\\ & \quad \left.
    -\frac{h^2 k p_j+4 h k m q_j+2 h p_j r-4 m p_j}{h^2 k+2 h r+4 m} \right)\, .
  \end{aligned}
  \end{equation}
  On the other hand, the discrete action is
  \begin{equation}
    \action_d^j (q_j) = \sum_{k=0}^{j-1} \left[\frac{h}{2} m \left(\frac{q_{j+1}-q_j}{h}  \right)^2
    - \frac{h}{2}k \left(\frac{q_{j+1}+q_{j}}{2}  \right)^2
    \right] \, .
  \end{equation}
  Therefore,
  \begin{equation}
  \begin{aligned}
    G_d^j(q_j, q_{j+1}) 
  &= \sum_{k=0}^{j-2} \left[\frac{h}{2} m \left(\frac{q_{j+1}-q_j}{h}  \right)^2
    - \frac{h}{2}k \left(\frac{q_{j+1}+q_{j}}{2}  \right)^2
    \right]\\
    &+ m h \left(\frac{q_{j}-q_{j-1}}{h}  \right)^2
    - kh \left(\frac{q_{j}+q_{j-1}}{2}  \right)^2\\
  & +\frac{h}{2} m \left(\frac{q_{j+1}-q_j}{h}  \right)^2
    -\frac{h}{2} k \left(\frac{q_{j+1}+q_j}{2}  \right)^2
    -r  \left(\frac{q_{j+1}-q_{j}}{2}  \right)^2\, .
  \end{aligned}
  \end{equation}
  Then
  \begin{equation}
  \begin{aligned}
    \gamma^-_j   (q_j, q_{j+1}) 
    & =  \left(\frac{m}{h} - \frac{3}{4} kh \right) q_j -  \left(\frac{m}{h} - \frac{1}{2} kh\right) q_{j+1} \\
    & \quad - \left(\frac{2m}{h} + \frac{kh}{2}\right) q_{j-1} +  r  \frac{q_{j+1}-q_{j}}{2} \, ,
  \end{aligned}
  \end{equation}
  and
  \begin{equation}
    \gamma^+_j (q_j, q_{j+1}) 
    = \frac{m}{h} (q_{j+1} - q_j)
    -\frac{1}{4}kh (q_j + q_{j+1})
    -r  \frac{q_{j+1}-q_{j}}{2}\, .
  \end{equation}
  Hence, the projected flow $\flowm_k = \pi_Q \circ \mathcal{F}_d^H \circ \DD_2 G_d^k$ is given by
  \begin{equation}\label{example_flowm}
  \begin{aligned}
    \flowm_j &: (q_{j-1}, q_{j})\\ & \mapsto
    \frac{h^2 (-k) (q_{j-1}+2 q_{j})+2 h q_{j-1} r-4 m (q_{j-1}-2 q_{j})}{h^2 k+2 h r+4 m}\, .
  \end{aligned}
  \end{equation}
  The left discrete Hamiltonian is
  \begin{equation}
    H_d^-(q_{j+1}, p_j) = -p_j q_j - \frac{h}{2} m \left(\frac{q_{j+1} - q_{j}}{h}  \right)^2 + \frac{h}{2}k \left(\frac{q_{j}+q_{j+1}}{2}  \right)^2.
  \end{equation}
  By construction, the maps $\{\action_d^k\}$ and $\{\gamma^-_k\}$ satisfy equation \eqref{eq:FLDHJ_Rayleigh}. 
  Then, the sequence of points $\left\{(q_j,p_k)  \right\}_{k=0}^N$, with $q_{j+1}=\flowm_k(q_{j-1}, q_j)$ and $p_k = \gamma^-(q_j, q_{j+1})$, 
  is a solution of the forced left discrete Hamilton equations for $(H_d^-, f_d)$. Observe that the points given by $q_{j+1}=\flowm(q_{j-1}, q_j)$ coincide with the solutions of the forced discrete Euler--Lagrange equations \eqref{discrete_forced_EL_Rayleigh}.

  In Figure \ref{plot_harmonic_oscillator} the position, momenta and energy of the system are plotted as a function of time, as well as the velocities as a function of the positions; comparing the solutions of the continuous forced Euler--Lagrange equations with the points given by the flow $\flowm$ (or, equivalently, the forced discrete Euler--Lagrange equations) as well as the Euler and fourth-order Runge--Kutta methods. However, this simple linear and 1-dimensional system does not show the advantages of variational methods over the standard numerical methods (see Example \ref{example_Marsden_West}).
\end{example}

\part{Autonomous and non-autonomous contact Hamiltonian systems}\label{part:contact}
\chapter{Contact Hamiltonian systems}\label{ch:contact_Hamiltonian_systems}

\insquote{He said that the geometry of the dream-place he saw was abnormal, non\hyp{}Euclidean, and loathsomely redolent of spheres and dimensions apart from ours.}{H.~P.~Lovecraft, \emph{The Call of Cthulu} (1926)}

Symplectic manifolds are the natural framework for time\hyp{}independent Hamiltonian systems (see \Cref{sec:Hamiltonian_mechanics}). Similarly, cosymplectic geometry may be employed to describe time-dependent Hamiltonian systems. Both autonomous and non-autonomous Hamiltonian systems have a Lagrangian counterpart, which can be derived from the Hamilton's variational principle. In order to model a dissipative phenomena one may consider that an external force is acting on the system.
Several results can be extended from Hamiltonian (respectively, Lagrangian) systems to forced Hamiltonian (respectively, Lagrangian) systems in a natural manner (see \Cref{part:forces}). 
However, this formalism has some disadvantages: one has to consider a semibasic one-form besides the Hamiltonian (or Lagrangian) function; and, unlike Hamilton's principle, Lagrange--D'Alembert principle is not strictly speaking a variational principle, since it does not seek for critical points of any functional. Therefore, it is interesting to consider a geometric formulation for dissipative systems such that the dynamics depends only on a Hamiltonian (or Lagrangian) function, and it can be derived from a purely variational principle. This is achieved by considering the dynamics given by the Hamiltonian vector field with respect to a co-orientable contact structure, whose flow does not preserve its Hamiltonian function, neither its volume form. Furthermore, by considering a Lagrangian function which depends on the action functional, one obtains a variational principle, the so-called \emph{Herglotz principle}, leading to dissipative dynamics, which is equivalent to the one given by the contact Hamiltonian vector field.

In addition, there are dynamical systems which exhibit both a dissipative behaviour and an explicit time dependence. These systems include mechanical systems subjected to external forces that vary with time. In particular, they encompass controlled systems in which the controls are used to compensate the damping effects. Further examples of such systems include physical systems with time-dependent mass that experience a linear damping. One may describe time-dependent dissipative Hamiltonian or Lagrangian systems by the use of cocontact geometry.

In this chapter, autonomous and non-autonomous contact Hamiltonian and Lagrangian systems are introduced, and Herglotz variational principle is presented. Refer to \Cref{sec:contact,sec:cocontact} for an introduction to contact and cocontact manifolds, respectively. See \cite{d.G.G+2022, G.G.M+2020a, Gaset2023, Rivas2021, Lainz2022, d.L2019, d.L2019a, d.L2019b, Bravetti2017, B.C.T2017, B.L.M+2020} for further details. In this dissertation, the contact Hamiltonian systems considered will be co-oriented. For an approach to contact Hamiltonian dynamics on not necessarily co-orientable contact manifolds, refer to the recent works by Grabowska and Grabowski \cite{G.G2022a, G.G2022b, G.G2023, G.G2023a} and references therein.

\section{Contact Hamiltonian systems}\label{sec:contact_Hamiltonian}

A \emph{contact Hamiltonian system} is a triple $(M, \eta, H)$, formed by a co-oriented contact manifold $(M, \eta)$ and a \emph{Hamiltonian function} $H\in\Cinfty(M)$. The dynamics of $(M, \omega, H)$, that is, the trajectories of the dynamical system it describes, are the integral curves of the Hamiltonian vector field $X_H$ of $H$ with respect to $\eta$. In Darboux coordinates $(q^i, p_i, z)$, this vector field is written
\begin{equation}
        X_H = \frac{\partial H}{\partial p_i} \frac{\partial}{\partial q^i} - \left( \frac{\partial H}{\partial q^i} + p_i \frac{\partial H}{\partial z}\right) \frac{\partial}{\partial p_i} + \left(  p_i \frac{\partial H}{\partial p_i} - H\right)\frac{\partial}{\partial z}\, . 
\end{equation}
Hence, a curve $c\colon I\subseteq \RR \to M,\, c(t)=(q^i(t), p_i(t), z(t))$ is an integral curve of $X_H$ if and only if it satisfies the \emph{contact Hamilton equations}:
\begin{equation}\label{eq:Hamilton_equations_contact}
\begin{aligned}
        &\frac{\dd q^i(t)}{\dd t} = \parder{H}{p_i} \big(c(t)\big)\, , \\
        &\frac{\dd p_i(t)}{\dd t} = -\parder{H}{q^i}\big(c(t)\big) - p_i(t) \parder{H}{z}\big(c(t)\big)\, , \\
        &\frac{\dd z(t)}{\dd t} = p_i(t) \frac{\partial H}{\partial p_i}\big(c(t)\big) - H\big(c(t)\big)\, .
\end{aligned}
\end{equation}
A function $f\in \Cinfty(M)$ is a \emph{conserved quantity} if and only if $X_H (f)=0$. Notice that the Hamiltonian function is not, in general, a conserved quantity. As a matter of fact, 
\begin{equation}
        \liedv{X_H} H = -\Reeb(H) H\, ,
\end{equation}
where $\Reeb$ denotes the Reeb vector field of $(M, \eta)$. Moreover, the volume form $\vol_\eta$ is also not preserved by the dynamics, namely,
\begin{equation}
        \liedv{X_H} \vol_\eta = -(n+1) \Reeb (H) \vol_\eta\, .
\end{equation}
Therefore, if $H$ is nowhere-vanishing and $c\colon I\subseteq \RR \to M$ is an integral curve of $X_H$, then
\begin{equation}\label{eq:dissipation_Hamiltonian_time}
        H \circ c(t) = H\circ c(0)\, \exp\left(-\int_0^t \Reeb(H)(c(s))\, \dd s \right)\, ,
\end{equation}
and
\begin{equation}
       \vol_\eta \circ\, c(t) = \vol_\eta \circ\, c(0)\, \exp\left(-(n+1)\int_0^t \Reeb(H)(c(s))\, \dd s \right)\, .
\end{equation}
In particular, if $\Reeb(H)$ is a positive constant, then $H\circ c$ and $\vol_\eta\circ\, c$ decrease exponentially with time. Regarding $H$ as the energy of the system, this models an exponential dissipation of the energy. The evolution of $H$ along the flow of $X_H$ motivates the following definition.

\begin{definition}\label{def:dissipated_quantity}
        Let $(M, \eta, H)$ be a contact Hamiltonian system. Let $\Reeb$ and $X_H$ denote the Reeb vector field and the Hamiltonian vector field of $H$ with respect to $\eta$, respectively. A \emph{dissipated quantity} is a solution $f\in \Cinfty(M)$ of the partial differential equation
        \begin{equation}
                \liedv{X_H} f = -\Reeb(H) f\, .
        \end{equation}
\end{definition}
In order to emphasize that the rate of dissipation $-\Reeb(H)$ depends on the Hamiltonian function $H$, some authors (for instance, \cite{Lainz2022}) refer to dissipated quantities as ``$H$-dissipated quantities''. Nevertheless, it is noteworthy that this rate depends not only on the Hamiltonian function but also on the contact form. Instead, if there is a risk of ambiguity, in this dissertation, it will be stated that $f$ is a \emph{dissipated quantity with respect to  $(M, \eta, H)$}.

\section{Contact Lagrangian systems}\label{sec:contact_Lagrangian}

Throughout this section, let $Q$ denote an $n$-dimensional manifold with local coordinates $(q^i)$. Denote by $(q^i, v^i)$ the induced bundle coordinates on the tangent bundle $\tau_Q\colon \T Q\to Q$. See \Cref{sec:structures_TQ} and references therein for the structures on $\T Q$ that will be employed. 

Consider the product manifold $\T Q \times \RR$ with the natural projections $\pi_1\colon \T Q \times \RR \to \T Q$ and $\pi_2\colon \T Q \times \RR \to \RR$. Let $(q^i, v^i, z)$ denote the induced bundle coordinates on $\T Q \times \RR$. A \emph{contact Lagrangian system} is a pair $(Q, L)$, where $Q$ is a manifold and $L\in \Cinfty(\T Q\times \RR)$ is an \emph{(action-dependent) Lagrangian function}. 

Let $(Q, L)$ be a contact Lagrangian system.  Given two fixed points $q_1, q_2\in Q$ and an interval $[a, b]$, let $\Omega(q_1, q_2,[a,b])$ denote the path space from $q_1$ to $q_2$ (see equation~\eqref{eq:path_space}). Consider the operator
\begin{equation}
        \zaction\colon \Omega\big(q_1, q_2,[a,b]\big) \to \Ctwo\big([a, b]\to \RR \big)
\end{equation}
that assigns to each curve $c\in \Omega(q_1, q_2,[a,b])$ the function $\zaction(c)$ that is the solution of the following Cauchy problem:
\begin{equation}
\begin{aligned}
        &\frac{\dd \zaction(c) (t)}{\dd t} = L \big(c(t), \dot{c}(t), \zaction(c)(t)\big)\, , \\ 
        &\zaction(c)(a) = z_a\, . 
\end{aligned}
\end{equation}
The quantity $Z(c)(t)$ can be interpreted as the action of the curve $c$ at time $t$. The \emph{Herglotz action functional} is the map $\action\colon \Omega(q_1, q_2,[a,b])\to \RR$ that assigns to each curve the solution of the Cauchy problem above evaluated at the endpoint, namely,
\begin{equation}
        \action\colon c \mapsto \zaction (c) (b)\, .
\end{equation}
The trajectories of the dynamical system described by $(Q, L)$ are given by the following variational principle.
A curve $c\in \Omega(q_1, q_2, [a,b])$ satisfies the \emph{Herglotz variational principle} if it is a critical point of the Herglotz action functional, that is, 
\begin{equation}
        \dd \action(c) = 0\, .
\end{equation}
A curve $c\in \Omega(q_1, q_2, [a,b])$ is a critical point of $\action$ if and only if it satisfies the \emph{Herglotz--Euler--Lagrange equations}:
\begin{equation}\label{eq:Herglotz_variational}
\begin{aligned}
        \frac{\dd }{\dd t} &\parder{L}{v^i} \big(c(t), \dot{c}(t), \zaction(c)(t) \big) 
        -  \parder{L}{q^i} \big(c(t), \dot{c}(t), \zaction(c)(t)\big)\\
        &-  \parder{L}{v^i} \big(c(t), \dot{c}(t), \zaction(c)(t)\big) \parder{L}{z} \big(c(t), \dot{c}(t), \zaction(c)(t)\big)
         = 0\, .
\end{aligned}
\end{equation}

For more details on the Herglotz variational principle refer to \cite{Herglotz1930, G.G.G1996, d.L2019, Georgieva2011}.

By replacing the path space $\Omega(q_1, q_2, [a,b])\subset \Ctwo([a,b]\to Q)$ with a space of curves which are not differentiable, it is possible to develop a variational principle describing the trajectories of mechanical systems with damping and collisions, the so-called \emph{nonsmooth Herglotz variational principle} (see \Cref{ch:nonsmooth_Herglotz}).

Since $\T (\T Q\times \RR)$ can be identified with the Whitney sum $(\T (\T Q)\times \RR)\oplus (\T Q \times \T \RR)$, every operation acting on tangent vectors on $\T Q$ can act on tangent vectors on $\T Q \times \RR$. Hence, the geometric structures on $\T Q$ can be naturally extended to $\T Q \times \RR$. The vertical endomorphism $\Sendo\in \Gamma(\tensors^1_1(\T \T Q))$ of the tangent bundle induces a vertical endomorphism on $\T Q \times \RR$ which, with a slight abuse of notation, is also denoted by $\Sendo$. Similarly, the Liouville vector field $\Delta\in \X(\T Q)$ induces a Liouville vector field on $\T Q \times \RR$, also denoted by $\Delta$. In bundle coordinates,
\begin{equation}
        \Sendo = \parder{}{v^i} \otimes \dd q^i \, , \quad \Delta = v^i  \parder{}{v^i}\, .
\end{equation}
Given a curve $c\colon I\subseteq \RR \to Q\times \RR$, where $c(t)=(c_Q(t), c_{\RR}(t))$, its \emph{canonical lift to $\T Q\times \RR$} is the curve $\tilde{c}\colon I \to \T Q \times \RR$ given by
\begin{equation}
        \tilde{c}(t) = \big(\tilde{c}_Q(t), c_{\RR}(t))\, ,
\end{equation}
where $\tilde{c}_Q$ denotes the canonical lift of $c_Q$ to $\T Q$. A curve $\sigma\colon I\subseteq \RR\to \T Q\times \RR$ on $\T Q\times \RR$ is called \emph{holonomic} if there exists a curve $c\colon I \to Q\times \RR$ on $Q\times \RR$ such that $\sigma = \tilde{c}$. In other words, $c=(c_Q, c_{\RR})$ is a holonomic curve on $\T Q\times \RR$ if and only if $c_Q$ is a holonomic curve on $\T Q$.
A vector field $\sode\in \X(\T Q\times \RR)$ is called an (\emph{action-dependent}) \emph{second order differential equation} (hereinafter abbreviated as \emph{\textsc{sode}}) \emph{on $Q \times \RR$} if its integral curves are holonomic. Equivalently, $\sode$ is an action-dependent \textsc{sode} on $Q \times \RR$ if and only if its projection $\T \pi_1 (\sode)$ is a \textsc{sode} on $Q$.
Locally, an action-dependent \textsc{sode} is of the form
\begin{equation}\label{eq:local_SODE_cocontact}
    \sode = v^i \frac{\partial}{\partial q^i} + \sode^i \frac{\partial}{\partial v^i} + f \parder{}{z}\, .
\end{equation} 
Moreover, $\sode$ is a \textsc{sode} if and only if $\Sendo (\sode) = \Delta$. If $\sigma=(\sigma_{\T Q}, c_{\RR})$ is an integral curve of a \textsc{sode} $\sode$, then the curve $c=(\tau_Q\circ \sigma_{\T Q}, c_{\RR})$ is called a \emph{solution of $\sode$}.

Let $(Q, L)$ be a contact Lagrangian system. The \emph{Legendre transform} is the map $\FF L \colon \T Q \times \RR \to \cT Q \times \RR$ given by
\begin{equation}
        \FF L (v, z) = \big( \FF \ell_z (v), z \big)\, ,
\end{equation} 
where $\ell_z = L(\cdot, z)$ is the Lagrangian function with $z$ fixed, and $\FF \ell_z\colon \T Q\to \cT Q$ is the fiber derivative as defined in \Cref{sec:Lagrangian_mechanics}. In bundle coordinates, 
\begin{equation}
        \FF L (q^i, v^i, z) = \left(q^i, \parder{L}{v^i}, z\right)\, .
\end{equation}
The \emph{Poincaré--Cartan one-form} $\theta_L \in \Omega^1(\T Q \times \RR)$ is the natural generalization from the one on $\T Q$, namely,
\begin{equation}
        \theta_L  = \Sendoadj \dd L = \FF L^\ast (\pi_1^\ast \theta_Q) \, .
\end{equation}
Then, one can define the one-form 
\begin{equation}
        \eta_L = \dd z - \theta_L \, ,
\end{equation}
or, in bundle coordinates, 
\begin{equation}
        \eta_L = \dd z - \parder{L}{v^i} \dd q^i\, .
\end{equation}
The following assertions are equivalent:
\begin{enumerate}
        \item the one-form $\eta_L$ is a contact form,
        \item the Legendre transform $\FF L$ is a local diffeomorphism,
        \item the Hessian matrix
        \begin{equation}
                (W_{ij}) = \left(\parderr{L}{v^i}{v^j}\right)
        \end{equation}
        is regular.
\end{enumerate}
If these equivalent conditions are satisfied, the action-dependent Lagrangian function $L$ and the contact Lagrangian system $(Q, L)$ are said to be \emph{regular}, and the one-form $\eta_L$ is called the \emph{Lagrangian contact form}. In addition, if $\FF L$ is a (global) diffeomorphism, then $L$ and $(Q, L)$ are said to be \emph{hyper-regular}.

If $(Q, L)$ is regular, the Reeb vector field of $(\T Q\times \RR, \eta_L)$ is given by $\Reeb_L = \sharp_{\eta_L} \eta_L$. In bundle coordinates,
\begin{equation}
        \Reeb_L = \parder{}{z} - W^{ji} \parderr{L}{z}{v^j} \parder{}{v^i}\, .
\end{equation}
The \emph{Lagrangian energy} is given by
\begin{equation}
        E_L = \Delta(L) - L = v^i \parder{L}{v^i} - L\, .
\end{equation}
Its evolution along the flow of the Reeb vector field is given by
\begin{equation}
        \liedv{\Reeb_L} E_L = - \parder{L}{z}\, .
\end{equation}
Thus, a function $f\in \Cinfty(\RR\times \T Q \times \RR)$ is a dissipated quantity with respect to $(\T Q \times \RR, \eta_L, E_L)$ if and only if 
\begin{equation}
        \liedv{\sode_L} f =  \parder{L}{z} f\, .
\end{equation}

Let $(Q, L)$ be a regular contact Lagrangian system. The triple $(\T Q\times \RR , \eta_L, E_L)$ is a contact Hamiltonian system. The Hamiltonian vector field of $E_L$ with respect to $\eta_L$ is an action-dependent \textsc{sode}, called the \emph{Herglotz--Euler--Lagrange vector field of $(Q, L)$} and denoted by $\sode_L$. In bundle coordinates,
\begin{equation}
    \sode_L = v^i\parder{}{q^i} + W^{ji}\left( \parder{L}{q^j} - v^k\parderr{L}{q^k}{v^j} - L \parderr{L}{z}{v^j} + \parder{L}{z} \parder{L}{v^j}  \right)\parder{}{v^i}
    + L \parder{}{z}\, ,
\end{equation}
where $(W^{ij})$ denotes the inverse of the Hessian matrix $(W_{ij})$. Furthermore, a solution $(c(t), z(t))$ of $\sode_L$ satisfies the Herglotz--Euler--Lagrange equations, namely,
\begin{equation}
\begin{aligned}
        &\frac{\dd }{\dd t} \parder{L}{v^i} \big(c(t), \dot{c}(t), z(t) \big) 
        -  \parder{L}{q^i} \big(c(t), \dot{c}(t), z(t)\big)\\
        &\qquad -  \parder{L}{v^i} \big(c(t), \dot{c}(t), z(t)\big) \parder{L}{z} \big(c(t), \dot{c}(t), z(t)\big)
         = 0\, , \\
        & \frac{\dd z (t)}{\dd t} = L \big(c(t), \dot{c}(t), z(t) \big) \, .
\end{aligned}
\end{equation}

\section{Time-dependent contact Hamiltonian systems} \label{sec:cocontact_Hamiltonian}

A \emph{time-dependent} contact Hamiltonian system (also called a \emph{non\hyp{}autonomous contact Hamiltonian system} or a \emph{cocontact Hamiltonian system}) is a tuple $(M, \tau, \eta, H)$, consisting of a cocontact manifold $(M, \tau, \eta)$ and a \emph{Hamiltonian function} $H\in \Cinfty(M)$. The dynamics of $(M, \tau, \eta, H)$ are the integral curves of the Hamiltonian vector field $X_H$ of $H$ with respect to $(\tau, \eta)$. In Darboux coordinates $(t, q^i, p_i, z)$, this vector field reads
\begin{equation}
        X_H = \parder{}{t} + \parder{H}{p_i}\parder{}{q^i} - \left(\parder{H}{q^i} + p_i\parder{H}{z}\right)\parder{}{p_i} + \left(p_i\parder{H}{p_i} - H\right)\parder{}{z}\,.
\end{equation}
Thus, up to a shift of the origin of the time parameter $t$, an integral curve of $X_H$ is of the form $c\colon I\subseteq \RR \to M,\, c(t)=(t, q^i(t), p_i(t), z(t))$ and satisfies the contact Hamilton equations~\eqref{eq:Hamilton_equations_contact}.
A function $f\in \Cinfty(M)$ is a \emph{conserved quantity} if and only if $X_H (f)=0$. Moreover, the notion of dissipated quantities can be naturally extended to non-autonomous contact Hamiltonian systems (\textit{cf.}~Definition~\ref{def:dissipated_quantity}) as follows.

\begin{definition}\label{def:dissipated_quantity_time}
        Let $(M, \tau, \eta, H)$ be a time-dependent contact Hamiltonian system. Let $\Rz = \sharp_{(\tau, \eta)}\eta, \, \Rt = \sharp_{(\tau, \eta)}\tau$ and $X_H$ denote the Reeb vector fields and the Hamiltonian vector field of $H$ with respect to $(\tau, \eta)$, respectively. A \emph{dissipated quantity} is a solution $f\in \Cinfty(M)$ of the partial differential equation
        \begin{equation}
                \liedv{X_H} f = -\Rz(H) f\, .
        \end{equation}
\end{definition}

If there is a risk of ambiguity, it will be said that $f$ is a \emph{dissipated quantity with respect to $(M, \tau, \eta, H)$}.

It is worth noting that the Hamiltonian function is not, in general, a dissipated quantity. Indeed,
\begin{equation}
        \liedv{X_H} H = -\Rz(H) f + \Rt(H)\, .
\end{equation}
and thus $H$ is a dissipated quantity if and only if $\Rt(H) = 0$. This resembles the case of time-dependent Hamiltonian systems on cosymplectic manifolds, where the Hamiltonian is a conserved quantity if and only if $\Reeb(H) = 0$ (\textit{cf.}~equation~\eqref{eq:Hamiltonian_evolution_cosymp}).

\begin{remark}\label{remark:extended_contact_phase_space}
        An alternative approach for describing contact Hamiltonian systems with time dependence is to consider the so-called extended contact phase space (see~\cite{B.G2021}). This space consists on the extended cotangent bundle $\cT Q \times \RR \times \RR$, endowed with a contact form $\eta = \pi^\ast \eta_Q+H\dd t$ depending on the Hamiltonian function $H\in \Cinfty(\cT Q \times \RR \times \RR)$, where $\pi_1\colon (\cT Q \times \RR) \times \RR \to (\cT Q \times \RR)$ is the canonical projection, $\eta_Q$ is the canonical contact form on $\cT Q \times \RR$ and $t$ is the canonical coordinate of the second copy of $\RR$. The cocontact formalism has more advantages, since one may consider more general manifolds, and the canonical cocontact structure on $\RR \times \cT Q \times \RR$ is independent on the Hamiltonian function.
\end{remark}

\section{Time-dependent contact Lagrangian systems} \label{sec:cocontact_Lagrangian}

Given a smooth $n$-dimensional manifold $Q$, consider the product manifold $\R\times\T Q\times \R$ equipped with bundle coordinates $(t, q^i, v^i, z)$.  A \emph{time-dependent contact Lagrangian system} (also called a \emph{non-autonomous contact Lagrangian system} or a \emph{cocontact Lagrangian system}) is a pair $(Q, L)$, where $Q$ is a manifold and $L\in \Cinfty(\RR\times \T Q\times \RR)$ is an \emph{(time and action-dependent) Lagrangian function}.

Let $(Q, L)$ be a time-dependent contact Lagrangian system. Given two fixed points $q_1, q_2\in Q$ and an interval $[a, b]$, consider the operator
\begin{equation}
        \zaction\colon \Omega\big(q_1, q_2,[a,b]\big) \to \Ctwo\big([a, b]\to \RR \big)
\end{equation}
that assigns to each curve $c\in \Omega(q_1, q_2,[a,b])$ the function $\zaction(c)$ that is the solution of the following Cauchy problem:
\begin{equation}
\begin{aligned}
        &\frac{\dd \zaction(c) (t)}{\dd t} = L \big(t, c(t), \dot{c}(t), \zaction(c)(t)\big)\, , \\ 
        &\zaction(c)(a) = z_a\, . 
\end{aligned}
\end{equation}
By defining the Herglotz action functional and the Herglotz variational principle as in the autonomous case (see \Cref{sec:contact_Lagrangian}), one obtains that the trajectories of $(Q, L)$ are given by the \emph{time-dependent Herglotz--Euler--Lagrange equations}:
\begin{equation}\label{eq:Herglotz_variational_time}
\begin{aligned}
        \frac{\dd }{\dd t} &\parder{L}{v^i} \big(t, c(t), \dot{c}(t), \zaction(c)(t) \big) 
        -  \parder{L}{q^i} \big(t, c(t), \dot{c}(t), \zaction(c)(t)\big)\\
        &-  \parder{L}{v^i} \big(t, c(t), \dot{c}(t), \zaction(c)(t)\big) \parder{L}{z} \big(t, c(t), \dot{c}(t), \zaction(c)(t)\big)
                = 0\, .
\end{aligned}
\end{equation}

Consider the canonical projections
\begin{equation}
\begin{aligned}
	\tau_1&\colon \R\times\T Q\times\R\to\R\ ,&& \tau_1(t, v_q, z) = t\,,\\
	\tau_2&\colon\R\times\T Q\times\R\to\T Q\ ,&& \tau_2(t, v_q, z) = v_q\,,\\
	\tau_3&\colon\R\times\T Q\times\R\to\R\ ,&& \tau_3(t, v_q, z) = z\,,\\
	\tau_0&\colon\R\times\T Q\times\R\to \R\times Q\times\R\ ,&& \tau_0(t, v_q, z) = (t, q, z)\,, 
\end{aligned}
\end{equation}
which are summarized in the following diagram:
\begin{equation}\label{eq:projections_cocontact_Lagrangian}
    \begin{tikzcd}
        && \R\times\T Q\times\R \arrow[ddll, swap, "\tau_1", start anchor={south west}] \arrow[ddrr, "\tau_3", start anchor={south east}] \arrow[d, "\tau_2"] \arrow[dd, bend right=40, swap, "\tau_0"] && \\
        && \T Q  \arrow[dd, swap, bend left=80, pos=0.7, swap, "\tau_Q"] && \\
        \R && \R\times Q\times\R \arrow[ll, swap, "\mathrm{pr}_1"] \arrow[d, swap, "\mathrm{pr}_2"] \arrow[rr, crossing over, "\mathrm{pr}_3"] && \R \\
        && Q 
    \end{tikzcd}
\end{equation}
The geometric structures of the tangent bundle $\T Q$ can be naturally extended to $\RR \times \T Q \times \RR$. These extensions are completely analogous to the ones to $\T Q \times \RR$ (see \Cref{sec:contact_Lagrangian}). The vertical endomorphism $\Sendo \in \Gamma (\tensors^1_1 (\T(\RR \times \T Q\times \RR )))$ and the Liouville vector field $\Delta \in \X(\RR \times \T Q\times \RR)$ are locally
\begin{equation}\label{eq:Sendo_Liouville_vf_local}
        \Sendo = \parder{}{v^i} \otimes \dd q^i \, , \quad \Delta = v^i  \parder{}{v^i}\, .
\end{equation}
Given a curve $c\colon I\subseteq \RR \to \RR\times  Q\times \RR$, where $c(t)=(c_1(t), c_Q(t), c_{2}(t))$, its \emph{canonical lift to $\RR\times \T Q\times \RR$} is the curve $\tilde{c}\colon I \to \RR\times \T Q \times \RR$ given by
\begin{equation}
        \tilde{c}(t) = \big(c_1(t), \tilde{c}_Q(t), c_{2}(t))\, ,
\end{equation}
where $\tilde{c}_Q$ denotes the canonical lift of $c_Q$ to $\T Q$. A curve $\sigma\colon I\subseteq \RR\to \RR\times \T Q\times \RR$ is called \emph{holonomic} if there exists a curve $c\colon I \to \RR\times  Q\times \RR$ such that $\sigma = \tilde{c}$. In other words, $c=(c_1, c_Q, c_2)$ is a holonomic curve on $\RR\times \T Q\times \RR$ if and only if $c_Q$ is holonomic curve on $\T Q$.
A vector field $\sode\in \X(\RR\times \T Q\times \RR)$ is called an (\emph{action and time-dependent}) \emph{second order differential equation} (hereinafter abbreviated as \emph{\textsc{sode}}) \emph{on $\RR\times Q \times \RR$} if its integral curves are holonomic. Equivalently, $\sode$ is an action and time-dependent \textsc{sode} on $\RR\times Q \times \RR$ if and only if its projection $\T \tau_2 (\sode)$ is a \textsc{sode} on $Q$.
Locally, an action and time-dependent \textsc{sode} is of the form
\begin{equation}\label{eq:local_SODE_contact}
    \sode = f \parder{}{t} + v^i \frac{\partial}{\partial q^i} + \sode^i \frac{\partial}{\partial v^i} + g \parder{}{z}\, .
\end{equation} 
Furthermore, $\sode$ is a \textsc{sode} if and only if $\Sendo (\sode) = \Delta$. If $\sigma=(c_1, \sigma_{\T Q}, c_2)$ is an integral curve of a \textsc{sode} $\sode$, then the curve $c=\tau_0\circ \sigma = (c_1, \tau_Q\circ \sigma_{\T Q}, c_2)$ is called a \emph{solution of $\sode$}.

Let $(Q, L)$ be a time-dependent contact Lagrangian system. The \emph{Legendre transform} is the map $\FF L \colon \RR \times \T Q \times \RR \to \RR \times  \cT Q \times \RR$ given by
\begin{equation}
        \FF L (v, z) = \big( \FF \ell_{(t,\, z)} (v), z \big)\, ,
\end{equation} 
where $\ell_{(t,\, z)} = L(t, \cdot, z)$ is the Lagrangian function with $t$ and $z$ fixed, and $\FF \ell_{(t,\, z)}\colon \T Q\to \cT Q$ is the fiber derivative as defined in \Cref{sec:Lagrangian_mechanics}. In bundle coordinates, 
\begin{equation}
        \FF L (t, q^i, v^i, z) = \left(t, q^i, \parder{L}{v^i}, z\right)\, .
\end{equation}
The \emph{Poincaré--Cartan one-form} $\theta_L \in \Omega^1(\RR \times\T Q \times \RR)$ is the natural generalization from the one on $\T Q$, namely,
\begin{equation}
        \theta_L  = \Sendoadj \dd L = \FF L^\ast (\tau_2^\ast \theta_Q) \, .
\end{equation}
Then, one can define the one-form 
\begin{equation}
        \eta_L = \dd z - \theta_L = \dd z - \parder{L}{v^i} \dd q^i\, .
\end{equation}
The following assertions are equivalent:
\begin{enumerate}
        \item the pair $(\dd t, \eta_L)$ is a cocontact structure on $\RR \times\T Q \times \RR$,
        \item the Legendre transform $\FF L$ is a local diffeomorphism,
        \item the Hessian matrix
        \begin{equation}
                (W_{ij}) = \left(\parderr{L}{v^i}{v^j}\right)
        \end{equation}
        is regular.
\end{enumerate}
If these equivalent conditions are satisfied, the time and action\hyphen{}dependent Lagrangian function $L$ and the cocontact Lagrangian system $(Q, L)$ are said to be \emph{regular}. In addition, if $\FF L$ is a (global) diffeomorphism, then $L$ and $(Q, L)$ are said to be \emph{hyper-regular}.

Let $(Q, L)$ be a regular time-dependent contact Lagrangian system. The tuple $(\RR \times \T Q\times \RR, \dd t,  \eta_L, E_L)$ is a time-dependent contact Hamiltonian system. The Hamiltonian vector field of $E_L$ with respect to $(\dd t, \eta_L)$ is a time and action-dependent \textsc{sode}, called the \emph{time-dependent Herglotz--Euler--Lagrange vector field of $(Q, L)$} and denoted by $\sode_L$. In bundle coordinates,
\begin{equation}\label{eq:Herglotz_vf_time}
\begin{aligned}
         \sode_L = \parder{}{t} &+ v^i\parder{}{q^i} + W^{ji}\left( \parder{L}{q^j} - v^k\parderr{L}{q^k}{v^j} 
         \right.\\ & \left.
         - L \parderr{L}{z}{v^j} + \parder{L}{z} \parder{L}{v^j}  \right)\parder{}{v^i}+ L \parder{}{z}\, ,
\end{aligned}
\end{equation}
where $(W^{ij})$ denotes the inverse of the Hessian matrix $(W_{ij})$. 
Thus, up to a shift of the origin of the time parameter $t$, a solution of $\sode_L$ is of the form $\sigma\colon I\subseteq \RR \to \RR\times Q\times \RR ,\, \sigma(t)=(t, c(t), z(t))$ and satisfies the time-dependent Herglotz--Euler--Lagrange equations, namely,
\begin{equation}
\begin{aligned}
        &\frac{\dd }{\dd t} \parder{L}{v^i} \big(t, c(t), \dot{c}(t), z(t) \big) 
        -  \parder{L}{q^i} \big(t, c(t), \dot{c}(t), z(t)\big)\\
        &\qquad -  \parder{L}{v^i} \big(t, c(t), \dot{c}(t), z(t)\big) \parder{L}{z} \big(t, c(t), \dot{c}(t), z(t)\big)
         = 0\, , \\
        & \frac{\dd z (t)}{\dd t} = L \big(t, c(t), \dot{c}(t), z(t) \big) \, .
\end{aligned}
\end{equation}

If $(Q, L)$ is regular, the Reeb vector fields of $(\RR\times\T Q\times \RR, \dd t, \eta_L)$ are given by $\Rt^L = \sharp_{(\dd t, \, \eta_L)} \dd t$ and $\Rz^L = \sharp_{(\dd t, \, \eta_L)} \eta_L$. In bundle coordinates,
\begin{equation}
        \Rt^L = \parder{}{t} - W^{ji} \parderr{L}{t}{v^j} \parder{}{v^i}\, , \quad
        \Rz^L = \parder{}{z} - W^{ji} \parderr{L}{z}{v^j} \parder{}{v^i}\, .
\end{equation}
The \emph{Lagrangian energy} is given by
\begin{equation}
        E_L = \Delta(L) - L = v^i \parder{L}{v^i} - L\, .
\end{equation}
Its evolution along the flow of the Reeb vector fields is given by
\begin{equation}
        \liedv{\Rt^L} E_L = - \parder{L}{t}\, ,\quad \liedv{\Rz^L} E_L = - \parder{L}{z}\, .
\end{equation}
Hence, a function $f\in \Cinfty(\RR\times \T Q \times \RR)$ is a dissipated quantity with respect to $(\RR\times\T Q \times \RR, \eta_L, E_L)$ if and only if 
\begin{equation}
        \liedv{\sode_L} f =  \parder{L}{z} f\, .
\end{equation}
In particular, $E_L$ is a dissipated quantity if and only if $L$ is time-independent, that is, $\tparder{L}{t} = 0$.
\chapter{Symmetries of time-dependent contact systems}\label{ch:contact_symmetries}

\insquote{Without any underlying symmetry properties, the job of proving interesting results becomes extremely unpleasant. The enjoyment of one's tools is an essential ingredient of successful work.}{Donald Knuth, \emph{The Art Of Computer Programming} (1998)}

This chapter is devoted to the symmetries of time-dependent contact Hamiltonian and Lagrangian systems and their relation with dissipated quantities. A theorem characterizing a class of symmetries which are in bijection with dissipated quantities is proven. Other classes of symmetries preserving (up to a conformal factor) additional structures, such as the contact form or the Hamiltonian function, are studied and classified. Moreover, making use of the structures of the extended tangent bundle, additional types of symmetries for time-dependent contact Lagrangian systems are considered. These results are illustrated with some examples. It is worth mentioning that the results presented are also applicable to time-independent contact systems.

The present chapter is based on the article \cite{Gaset2023}. An alternative approach for studing the symmetries and dissipated quantities of time-dependent contact systems can be found in \cite{B.G2021}. Their results are restricted to the so-called extended contact phase space (see \Cref{remark:extended_contact_phase_space}). Some notions of symmetries for time-independent contact Hamiltonian and Lagrangian systems were previously studied in \cite{d.L2020, G.G.M+2020a, Gaset2021a}. 
However, the literature lacked from a systematic classification of symmetries, considering the structures they preserve and the relations between them, as well as which sets of symmetries close Lie algebras (or Lie groups).
The types of symmetries and their relations are summarized in Figures \ref{fig:infinitesimal_symmetries}, \ref{fig:infinitesimal_symmetries_Lagrangian} and \ref{fig:infinitesimal_symmetries_Lagrangian_Hamiltonian}.

\section[Symmetries of cocontact Hamiltonian systems]{Symmetries and dissipated quantities of cocontact Hamiltonian systems}\label{sec:symmetries_cocontact_Hamiltonian}

Along this section, $X_f$ will denote the Hamiltonian vector field of $f$ with respect to the given cocontact structure.

It is well-known that the conserved quantities of a Hamiltonian system may be characterized by the Poisson bracket defined by the underlying symplectic structure. Similarly, the dissipated quantities of a (time-dependent) contact Hamiltonian system can be characterized by means of the Jacobi bracket as follows. 
\begin{proposition}\label{prop:bracket_dissipated}
        Let $(M,\tau,\eta,H)$ be a cocontact Hamiltonian system. A function $f\in \Cinfty(M)$ is a dissipated quantity if and only if 
        \begin{equation}
                \{f, H\} = \Rt(f)\, ,
        \end{equation}
        where $\{\cdot,\cdot\}$ is the Jacobi bracket associated to the cocontact structure $(\tau,\eta)$.
\end{proposition}
The proof follows from \Cref{prop:Jacobi_bracket_cocontact} and \Cref{def:dissipated_quantity_time}. Moreover, from the definitions of conserved and dissipated quantities, the following relations can be derived.

\begin{proposition}\label{prop:relations_dissipated_conserved_quantities}
        Consider the cocontact Hamiltonian system $(M,\tau,\eta,H)$. Then, for any dissipated quantities $f_1, f_2\in \Cinfty(M)$, any conserved quantities $g_1, g_2\in \Cinfty(M)$ and any constants $a_1, a_2, a_3\in \RR$: 
        \begin{enumerate} 
            \item in the restriction to $M\setminus f_2^{-1}(0)$, the function $f_1/f_2$ is a conserved quantity,
            \item $f_1 g_1$ is a dissipated quantity,
            \item $a_1 f_1 + a_2 f_2$ is a dissipated quantity,
            \item $a_1 g_1 + a_2 g_2 + a_3$ is a conserved quantity.
        \end{enumerate}
\end{proposition}

In the case of Hamiltonian systems, there is a local one-to-one correspondence (up to an additive constant) between Cartan symmetries and conserved quantities (see \Cref{sec:symmetries}). Similarly, it is possible to define a notion of symmetries on cocontact Hamiltonian systems which are in bijection with dissipated quantities. 

\begin{theorem}[Noether's theorem]\label{theorem:Noether_cocontact}
        Consider the cocontact Hamiltonian system $(M,\tau,\eta,H)$. Let $Y\in \X(M)$ be a vector field.
        If $\eta([Y,X_H]) = 0$ and $\contr{Y}\tau = 0$, then $f = -\contr{Y}\eta$ is a dissipated quantity. Conversely, given a dissipated quantity $f\in \Cinfty(M)$, the vector field $Y = X_f - \Rt$
        verifies $\eta([Y,X_H]) = 0$, $\contr{Y}\tau = 0$ and $f = -\contr{Y}\eta$.
\end{theorem}
    
\begin{proof}
        Let $f = -\contr{Y}\eta$, where $Y$ satisfies $\eta([Y,X_H]) = 0$ and $\contr{Y}\tau = 0$. Then,
        \begin{align*}
                \liedv{X_H}f &= -\liedv{X_H}\contr{Y}\eta
                = -\contr{Y}\liedv{X_H}\eta - \contr{[X_H,Y]}\eta
                \\&
                = \contr{Y}\left( \Rz(H)\eta + \Rt(H)\tau \right)
                \\&
                = \Rz(H)\contr{Y}\eta
                = -\Rz(H) f\,,
        \end{align*}
        and thus $f$ is a dissipated quantity.

        On the other hand, given a dissipated quantity $f$, let $Y = X_f - \Rt$. Then, it is clear that $f = -\contr{Y}\eta$. In addition, $\contr{Y}\tau = 0$, and
        \begin{align*}
                \contr{[X_H,Y]}\eta 
                &= \liedv{X_H} \contr{Y} \eta -  \contr{Y} \liedv{X_H} \eta
                = -\liedv{X_H} f  + \contr{Y} \left(\Rz(H) \eta +\Rt(H) \tau\right)
                \\
                &= \Rz(H) f - \Rz(H) \contr{Y} \eta = 0\,,
        \end{align*}
        where equations \eqref{eq:cocontact_Hamiltonian_vf_contr} have been used.
\end{proof}

This result motivates the following definition.

\begin{definition}
        Let $(M,\tau,\eta,H)$ be a cocontact Hamiltonian system. A \emph{generalized infinitesimal dynamical symmetry} is a vector field $Y\in\X(M)$ such that $\eta([Y,X_H]) = 0$ and $\contr{Y}\tau = 0$.
\end{definition}

If there is a risk of ambiguity, it will be said that $Y$ is a \emph{generalized infinitesimal dynamical symmetry of $(M, \tau, \eta, H)$}, and analogously for other types of symmetries defined below.

It is noteworthy that, despite the condition $\tau(Y)=0$, the dissipated quantity $f$ associated to a generalized infinitesimal dynamical symmetry $Y$ may be time-dependent. Indeed, if $Y$ is given by
\begin{equation}
        Y = Y^{q^i}\parder{}{q^i}+Y^{p_i}\parder{}{p_i}+Y^{z}\parder{}{z}
\end{equation}
in Darboux coordinates, then 
\begin{equation}
\begin{aligned}
    \liedv{\Rt} f &= - \liedv{\Rt} \contr{Y} \eta = -\contr{[\Rt, Y]} \eta - \contr{Y} \liedv{\Rt} \eta
    = -\eta\big([\Rt, Y] \big)\\
    &= - \parder{Y^z}{t} + p_i\parder{Y^{q^i}}{t} \, .
\end{aligned}
\end{equation}


It is also interesting to study other types of symmetries which preserve more properties of the system, such as the dynamical vector field or the Hamiltonian function. 

Firstly, one can consider vector fields or diffeomorphisms preserving the Hamiltonian vector field $X_H$ of $H$. 

\begin{definition}
        Let $(M,\tau,\eta,H)$ be a cocontact Hamiltonian system. A vector field $Y\in\X(M)$ is called an \emph{infinitesimal dynamical symmetry} if $\liedv{Y} X_H = [Y,X_H] = 0$ and $\contr{Y}\tau = 0$.
\end{definition}

Since $\tau$ is closed, on a neighbourhood $U$ of each point $x\in M$, it is exact, namely, $\restr{\tau}{U} = \dd \sigma$ for some $\sigma\in \Cinfty(U)$. The condition $\contr{Y} \tau=0$ implies that the flow $\phi^Y_s$ of $Y$ preserves the local function $\sigma$, that is, $(\phi^Y_s)^\ast \sigma = \sigma$. For simplicity's sake, one can make this condition global by assuming that $M$ is the product manifold of a co-oriented contact manifold with a real line, and taking $\sigma = t$, for $t$ the canonical coordinate of $\RR$. 

\begin{definition}
        Let $(N, \eta_0)$ be a co-oriented contact manifold and consider the cocontact Hamiltonian system $(\RR \times N, \dd t, \pi_2^\ast \eta_0, H)$, where $\pi_2\colon  \RR \times N\to N$ denotes the canonical projection, and $t$ the canonical coordinate of $\RR$ (see \Cref{example:cocontact_contact}). A diffeomorphism $\Phi\colon \RR \times N\to \RR \times N$ is called a \emph{dynamical symmetry} if $\Phi_\ast X_H = X_H$ and $\Phi^\ast t = t$.
\end{definition}

Notice that if a vector field $Y\in \X(\RR \times N)$ is an infinitesimal dynamical symmetry, then its flow is made of dynamical symmetries.

Generalized infinitesimal dynamical symmetries receive that name since they satisfy weaker conditions than infinitesimal dynamical symmetries. Indeed, it is clear that every infinitesimal dynamical symmetry is a generalized infinitesimal dynamical symmetry. One may also define a generalization of dynamical symmetries as follows:

\begin{definition}\label{def:gendynsym}
        Let $(N, \eta_0)$ be a co-oriented contact manifold and consider the cocontact Hamiltonian system $(\RR \times N, \dd t, \eta, H)$, where $\pi_2\colon$  $\RR \times N\to N$ denotes the canonical projection, $t$ the canonical coordinate of $\RR$ and $\eta = \pi_2^\ast \eta_0$. A \emph{generalized dynamical symmetry} is a diffeomorphism $\Phi\colon \RR \times N\to \RR \times N$ such that $\eta(\Phi_\ast X_H) = \eta(X_H)$ and $\Phi^\ast  t = t$.
\end{definition}

The following counterexample shows that, unlike other symmetries with infinitesimal counterparts, the flow of a generalized infinitesimal dynamical symmetry is not necessarily made of generalized dynamical symmetries.

\begin{counterexample}\label{counterexample:cocontact_gen_inf_dyn_symmetry}
        Consider the cocontact Hamiltonian system $(\RR^4\setminus\{0\}, \tau, \eta, H)$, with $\tau= \dd t,\ \eta = \dd z - p \dd x$ and 
        \begin{equation}
        H=\frac{p^2}{2} + z\, ,
        \end{equation}
        where $(t, x, p, z)$ are the canonical coordinates in $\RR^4$. 
        The family of diffeomorphisms
        \begin{equation}
        \begin{aligned}
        \Phi^r\colon \RR^4\setminus\{0\} &\to \RR^4\setminus\{0\}\\
        \left( t, x, p, z\right ) & \mapsto \left( t, x, p+r, z \right)
        \end{aligned}
        \end{equation}
        for $r\in\RR$, is the flow of the vector field $Y =\tparder{}{p}$. 
        The Hamiltonian vector field of $H$ is given by
        \begin{equation}
        X_H = \parder{}{t} + p \parder{}{x} - p \parder{}{p} + \left( \frac{p^2}{2} - z\right) \parder{}{z}\,,
        \end{equation}
        thus
        \begin{equation}
        \Phi^r_\ast X_H = \parder{}{t} + (p-r) \parder{}{x} - (p-r) \parder{}{p} + \left( \frac{(p-r)^2}{2} - z\right) \parder{}{z}\,,
        \end{equation}
        and $\eta(\Phi^r_\ast X_H)\neq\eta(X_H)$. 
        Hence, $Y$ is a generalized infinitesimal dynamical symmetry, but $\Phi^r$ is not a generalized dynamical symmetry for $r\neq 0$. 
\end{counterexample}

The (infinitesimal) dynamical symmetries defined above are the counterparts of (infinitesimal) dynamical symmetries in Hamiltonian systems (see \Cref{sec:symmetries}). They are interesting since they map trajectories of the system onto other trajectories. More specifically, if $c\colon I\subseteq \RR \to \RR \times N$ is an integral curve of $X_H$ and $\Phi$ is a dynamical symmetry of $(\RR \times N, \dd t, \eta, H)$, then $\Phi \circ c$ is also an integral curve of~$X_H$. In addition, infinitesimal dynamical symmetries close a Lie algebra, and dynamical symmetries close a Lie group.

\begin{proposition}
        Let $(M,\tau,\eta,H)$ be a cocontact Hamiltonian system. Then, infinitesimal dynamical symmetries close a Lie subalgebra of $(\X(M), [\cdot, \cdot])$. In other words, given two infinitesimal dynamical symmetries $Y_1,Y_2\in\X(M)$, its Lie bracket $[Y_1,Y_2]$ is also an infinitesimal dynamical symmetry.
\end{proposition}

\begin{proof}
        If $Y_1$ and $Y_2$ are infinitesimal dynamical symmetries, the Jacobi identity implies that
        \begin{equation}
         [[Y_1,Y_2],X_H] = [Y_2,[X_H,Y_1]] + [Y_1,[Y_2,X_H]] = 0\,. 
        \end{equation}
        In addition,
        \begin{equation}
         \contr{[Y_1,Y_2]}\tau = \liedv{Y_1}\contr{Y_2}\tau - \contr{Y_2}\liedv{Y_1}\tau = -\contr{Y_2}\left( \contr{Y_1}\d\tau + \dd \contr{Y_1}\tau \right) = 0\,. 
        \end{equation}
\end{proof}

\begin{proposition}
        Let $(N, \eta_0)$ be a co-oriented contact manifold and consider the cocontact Hamiltonian system $(\RR \times N, \dd t, \pi_2^\ast \eta_0, H)$, where $\pi_2\colon  \RR \times N\to N$ denotes the canonical projection, $t$ the canonical coordinate of $\RR$ and $\eta = \pi_2^\ast \eta_0$. Then, dynamical symmetries form a Lie subgroup of $\Diff(\RR \times N)$, that is, for any pair of dynamical symmetries $\Phi_1$ and $\Phi_2$, the composition $\Phi_1 \circ \Phi_2$ is also a dynamical symmetry.
\end{proposition}

\begin{proof}
        If $\Phi_1$ and $\Phi_2$ are dynamical symmetries, then 
        \begin{equation}
                (\Phi_1\circ \Phi_2)_\ast X_H = (\Phi_1)_\ast (\Phi_2)_\ast X_H 
                = (\Phi_1)_\ast X_H = X_H\, ,
        \end{equation} 
        and 
        \begin{equation}
        (\Phi_1\circ \Phi_2)^\ast t = \Phi_2^\ast \Phi_1^\ast t = \Phi_2^\ast t = t\, .
        \end{equation}
        Obviously, the identity map of $M$ is a dynamical symmetry. Finally, if $\Phi$ is a dynamical symmetry, then
        \begin{equation}
        X_H = (\Phi^{-1}\circ \Phi)_\ast X_H = \Phi^{-1}_\ast \Phi_\ast X_H = \Phi^{-1}_\ast X_H\, ,
        \end{equation}
        and similarly $(\Phi^{-1})^\ast t = t$. This proves that dynamical symmetries form a group under composition.
\end{proof}

On the contrary, generalized infinitesimal dynamical symmetries do not close a Lie algebra, as the counterexample below shows. 

\begin{counterexample}
        Consider the cocontact Hamiltonian system from \Cref{counterexample:cocontact_gen_inf_dyn_symmetry}.
        Given the vector fields
        \begin{equation}
                Y=\frac{\partial}{\partial p}\, ,\qquad Z=\frac{x}{2}\frac{\partial}{\partial x}+\frac{p}{2}\frac{\partial}{\partial p}+(z+p)\frac{\partial}{\partial z}\,,
        \end{equation}
        the reader can validate that $Y$ is a generalized infinitesimal dynamical symmetry and $Z$ is an infinitesimal dynamical symmetry. Nevertheless,
        \begin{equation}
         [Y,Z]=\frac12\frac{\partial}{\partial p}+\frac{\partial}{\partial z}
        \end{equation}
    is not a generalized infinitesimal symmetry.
\end{counterexample}

A natural type of objects that conserve the geometry of the system are (infinitesimal) conformal cocontactomorphisms (see \Cref{sec:cocontact}). Since the function $H$ is independent of the cocontact structure $(\tau, \eta)$, in general conformal cocontactomorphisms are not generalized dynamical symmetries. The necessary and sufficient condition is shown in the next result.

\begin{proposition}
        Let $(M,\tau,\eta,H)$ be a cocontact Hamiltonian system.
        \begin{enumerate}
            \item Let $\Phi\colon M\rightarrow M$ be an $f$-conformal cocontactomorphism, that is, $\Phi^\ast \eta=f\eta$ and $\Phi^\ast \tau=\tau$. Then, $\eta(\Phi_\ast X_H)=\eta(X_H)$ if and only if $\Phi^\ast H=fH$.
            \item Suppose that the cocontact Hamiltonian system is of the form $(\RR \times N, \dd t, \pi_2^\ast \eta_0, H)$, where $\pi_2\colon  \RR \times N\to N$ denotes the canonical projection on the co-oriented contact manifold $(N, \eta_0)$, with $t$ the canonical coordinate of $\RR$ and $\eta = \pi_2^\ast \eta_0$. Then, $\Phi$ is a generalized dynamical symmetry if and only if $\Phi^\ast H=fH$ and $\Phi^\ast t=t$.
            \item Let $Y\in\X(M)$ be an infinitesimal $g$-conformal cocontactomorphism, that is, $\liedv{Y}\eta=g\eta$ and $\liedv{Y}\tau=0$. Then, $\eta([Y,X_H])=0$ if and only if $\liedv{Y}H=gH$. In particular, $Y$ is a generalized infinitesimal dynamical symmetry if and only if $\liedv{Y}H=gH$ and $\contr{Y}\tau=0$.
        \end{enumerate}
\end{proposition}

\begin{proof} 
        The Hamiltonian vector field $X_H$ verifies $\contr{X_H} \eta = -H$, and thus, for any diffeomorphism $\Phi\colon M \to M$,
        \begin{equation}
                \Phi^\ast H=-\Phi^\ast (\contr{X_H}\eta)=- \contr{\Phi_\ast X_H}\Phi^\ast \eta=-f\contr{\Phi_\ast X_H}\eta \,.
        \end{equation}
        If $\Phi$ is a generalized dynamical symmetry, then $\contr{\Phi_\ast X_H}\eta=\contr{X_H}\eta$, and therefore $\Phi^\ast H=fH$. Conversely, if $\Phi^\ast H=fH$, then
        \begin{equation}
                f\contr{\Phi_\ast X_H}\eta=-\Phi^\ast H=-fH=f\contr{X_H}\eta\,.
        \end{equation}
        Since the conformal factor $f$ is nowhere-vanishing, $\contr{\Phi_\ast X_H}\eta=\contr{X_H}\eta$.
        Similarly, for any vector field $Y\in \X(M)$, 
        \begin{equation}
        \begin{aligned}
                \liedv{Y}H&=-\liedv{Y}(\contr{X_H}\eta)=- \contr{[Y,X_H]}\eta-\contr{X_H}\liedv{Y}\eta\\
                &=- \contr{[Y,X_H]}\eta-g\contr{X_H}\eta=- \contr{[Y,X_H]}\eta+gH \,.
        \end{aligned}
        \end{equation}
\end{proof}

This proposition justifies the following definitions. 

\begin{definition}
        Let $(M,\tau,\eta,H)$ be a cocontact Hamiltonian system. A vector field $Y\in \X(M)$ is called an \emph{infinitesimal $\rho$-conformal Hamiltonian symmetry} is a vector field $Y\in\X(M)$ such that
        \begin{equation}
                \contr{Y}\tau = 0\ ,\qquad \liedv{Y}\eta = \rho\eta\, ,\qquad \liedv{Y} H = \rho H\,,
        \end{equation}
        where $\rho\in \Cinfty(M)$ is called the \emph{(infinitesimal) conformal factor}.
        If $Y$ is an infinitesimal strict cocontactomorphism (that is, if $\rho \equiv 0$), then $Y$ is called an \emph{infinitesimal strict Hamiltonian symmetry}.

        If the infinitesimal conformal factor is understood or irrelevant, an infinitesimal $\rho$-conformal Hamiltonian symmetry will be simply called an \emph{infinitesimal conformal Hamiltonian symmetry}.
\end{definition}

\begin{definition}
        Let $(N, \eta_0)$ be a co-oriented contact manifold and consider the cocontact Hamiltonian system $(\RR \times N, \dd t, \eta, H)$, where $\pi_2\colon$  $\RR \times N\to N$ denotes the canonical projection, $t$ the canonical coordinate of $\RR$ and $\eta = \pi_2^\ast \eta_0$. A diffeomorphism $\Phi\colon \RR\times N \to \RR\times N$ is called an \emph{$f$-conformal Hamiltonian symmetry} if
        \begin{equation}
                \Phi^\ast t = t \ ,\qquad \Phi^\ast\eta = f\eta\ ,\qquad \Phi^\ast H 
                = fH\,,
        \end{equation}
        where $f\in\Cinfty(M)$ is a nowhere-vanishing function called the \emph{conformal factor}.
        If $\Phi$ is a strict cocontactomorphism (that is, if $f \equiv 1$), then $\Phi$ is called a \emph{strict Hamiltonian symmetry}.

        If the conformal factor is understood or irrelevant, a $f$-conformal Hamiltonian symmetry will be simply called a \emph{conformal Hamiltonian symmetry}.
\end{definition}

Clearly, a vector field $Y\in \X(\RR \times N)$ is an infinitesimal conformal (respectively, strict) Hamiltonian symmetry if and only if its flow is made of conformal (respectively, strict) Hamiltonian symmetries.

These symmetries correspond, in time-independent contact systems, to  ``contact symmetries'' (see \cite{G.G.M+2020a}). The symplectic counterparts of (infinitesimal) strict Hamiltonian symmetries are sometimes referred to as ``(infinitesimal) Noether symmetries'' (see \Cref{sec:symmetries} and references therein).

If a conserved quantity is known, (infinitesimal) dynamical symmetries may be employed to compute additional conserved quantities. Similarly, if a dissipated quantity is known, (infinitesimal) strict Hamiltonian symmetries can be used to compute new dissipated quantities.

\begin{proposition}\label{prop:new_quantities}
        Consider the cocontact Hamiltonian system $(M,\tau,\eta,H)$.
        Let $g\in\Cinfty(M)$ be a conserved quantity and $f\in\Cinfty(M)$ a dissipated quantity.
        \begin{enumerate}
                \item If $Y \in \X(M)$ is an infinitesimal strict Hamiltonian symmetry and an infinitesimal dynamical symmetry, then $\widetilde{f}=\liedv{Y} f$ is also a dissipated quantity.
                \item If $Y\in\X(M)$ is an infinitesimal dynamical symmetry, then $\widetilde{g} = \liedv{Y} g$ is also a conserved quantity.
        \end{enumerate}
\end{proposition}


\begin{proof}
        Let $f$ and $g$ be a dissipated and a conserved quantity, respectively.
        If $Y$ is an infinitesimal dynamical symmetry, then
        \begin{equation}
         \liedv{X_H}\widetilde{g} = \liedv{X_H}\liedv{Y} g = \liedv{[X_H,Y]} g + \liedv{Y}\liedv{X_H} g = 0\,.
        \end{equation}
        Similarly, if $Y \in \X(M)$ is an infinitesimal strict Hamiltonian symmetry and an infinitesimal dynamical symmetry, then
        \begin{equation}
         \liedv{X_H}\widehat{g} = \liedv{X_H}\liedv{Y} g = \liedv{[X_H,Y]} g + \liedv{Y}\liedv{X_H} g = 0\,.
        \end{equation}
        \begin{equation}
        \begin{aligned}
                \liedv{X_H} \widetilde{f}
                & = \liedv{X_H} \left( \liedv{Y} f \right)
                =\liedv{[X_H,Y]} f + \liedv{Y} \left(\liedv{X_H} f \right)\\
                & = \liedv{Y}\left( - \liedv{\Rz} (H) f\right)
                = - \liedv{\Rz} (H)  \left( \liedv{Y} f \right)\, ,
        \end{aligned}
        \end{equation}
        where in the last step it has been used that infinitesimal cocontactomorphisms preserve the Reeb vector fields (see \Cref{corollary:cocontactomorphism_Reebs}).
\end{proof}

\begin{proposition}\label{prop:new_quantities_diffeo}
        Let $(N, \eta_0)$ be a co-oriented contact manifold and consider the cocontact Hamiltonian system $(\RR \times N, \dd t, \eta, H)$, where $\pi_2\colon$  $\RR \times N\to N$ denotes the canonical projection, $t$ the canonical coordinate of $\RR$ and $\eta = \pi_2^\ast \eta_0$.
        Let $f\in\Cinfty(\RR\times N)$ be a dissipated quantity and $g\in\Cinfty(\RR\times N)$ a conserved quantity.
        \begin{enumerate}
                \item If $\Phi\colon \RR\times N\to \RR\times N$ is a strict Hamiltonian symmetry and a dynamical symmetry, then $\widehat{f}=\Phi^\ast f$ is also a dissipated quantity.
                \item If $\Phi\colon \RR\times N\to \RR\times N$ is a dynamical symmetry, then $\widehat{g}=\Phi^\ast g$ is also a conserved quantity.
        \end{enumerate}
\end{proposition}

\begin{proof}
        Let $f$ and $g$ be a dissipated and a conserved quantity, respectively.
        Suppose that $\Phi$ is an strict Hamiltonian symmetry and a dynamical symmetry. Then,
        \begin{equation}
        \begin{aligned}
                \liedv{X_H} \widehat{f}
                & = \liedv{X_H} (\Phi^\ast f)
                = \Phi^\ast \left(\liedv{\Phi_\ast X_H} f\right)
                = \Phi^\ast \left(\liedv{X_H} f\right)\\
                & = \Phi^\ast \big(-\liedv{\Rz}(H) f\big)
                = -\liedv{\Rz}(H) \Phi^\ast f\, .
        \end{aligned}
        \end{equation}
        where in the last step it has been used that cocontactomorphisms preserve the Reeb vector fields (see \Cref{prop:cocontactomorphism_Reebs}).
        Similarly, if $\Phi$ is a dynamical symmetry, then
        \begin{equation}
                \liedv{X_H} \widehat{g}
                = \liedv{X_H} (\Phi^\ast g)
                = \Phi^\ast \left(\liedv{\Phi_\ast X_H} g\right)
                = \Phi^\ast \left(\liedv{X_H} g\right)
                = 0\, .
        \end{equation}
\end{proof}

The following counterexample proves that the results from Proposition \ref{prop:new_quantities} cannot be extended to generalized infinitesimal dynamical symmetries.

\begin{counterexample}\label{counterxample:new_quantities_generalized}
    Consider the same system as in \Cref{counterexample:cocontact_gen_inf_dyn_symmetry}. Let $Y\in \X(\RR^4\setminus \{0\})$ be the vector field $Y=\tparder{}{p}$. Observe that $[Y,X_H]\neq0$, but $\eta([Y,X_H])=0$, and thus $Y$ is a generalized infinitesimal symmetry but it is not a dynamical symmetry.
    
    The function $f(t,x,p,z) = p$ is a dissipated quantity, but $\liedv{Y} f =1$ is not a dissipated quantity. Likewise, $\liedv{Y} H = p$ is not a dissipated quantity either. Finally,
    \begin{equation}
        \liedv{Y}\frac{H}{f} =\frac12-\frac{z}{p^2}\,,
    \end{equation}
    is not a conserved quantity.
\end{counterexample}

It is also worth mentioning that preserving the Hamiltonian is not a sufficient condition for a vector field to be an infinitesimal dynamical symmetry, or for a diffeomorphism to be a dynamical symmetry. It is not a sufficient condition for being a generalized (infinitesimal) dynamical symmetry either, as the following counterexample shows.

\begin{counterexample}
    Consider the cocontact Hamiltonian system $(\RR^4, \tau, \eta, H)$, with $\tau= \dd t,\ \eta = \dd z - p \dd x$ and 
    \begin{equation}
        H=\frac{p^2}{2}\, ,
        \end{equation}
    where $(t, x, p, z)$ are the canonical coordinates in $\RR^4$. Its Hamiltonian vector field is given by
    \begin{equation}
        X_H = \parder{}{t} +  p \parder{}{x} + \frac{p^2}{2} \parder{}{z}.
    \end{equation}    
    Let $Y= z\tparder{}{z}$. The reader can verify that $Y(H) = 0$, but $[Y, X_H]\neq 0$ and $\eta([Y, X_H]) \neq 0$. Similarly, $\Phi \colon \RR^4 \to \RR^4,\ (t, x, p, z)\mapsto (t, x, p, 2 z)$ is a diffeomorphism preserving the Hamiltonian function $H$ but not the Hamiltonian vector field $X_H$.  
\end{counterexample}

Furthermore, the notion of infinitesimal $\rho$-conformal Hamiltonian symmetry may be generalized as follows.

\begin{definition}
        Given a cocontact Hamiltonian system $(M,\tau,\eta, H)$, a \emph{$(\rho, g)$-Cartan symmetry} is a vector field $Y\in\X(M)$ such that
        \begin{equation}
                \liedv{Y}\eta = \rho \eta + \dd g\,,\qquad \liedv{Y} H = \rho H + g \Rz(H)\,,\qquad \contr{Y}\tau = 0\,, 
        \end{equation}
        where $\rho,g\in\Cinfty(M)$.
\end{definition}

It is clear that a $\rho$-conformal Hamiltonian symmetry is a $(\rho, 0)$-Cartan symmetry. On the other hand, $(0,g)$-Cartan symmetries are the analogous of Cartan symmetries in symplectic Hamiltonian systems (see \Cref{sec:symmetries}). 

\begin{theorem}\label{theorem:Cartan_dissipated}
    If $Y$ is a $(\rho, g)$-Cartan symmetry of a cocontact Hamiltonian system $(M,\tau,\eta,H)$, the function $f = g - \contr{Y}\eta$ is a dissipated quantity.
\end{theorem}

\begin{proof}
Indeed,
    \begin{align*}
        \liedv{X_H} f &= \liedv{X_H}(g - \contr{Y}\eta) = \contr{X_H}\dd g - \contr{Y}\liedv{X_H}\eta - \contr{[X_H,Y]}\eta\\
        &= \contr{X_H}\dd g + \contr{Y}( \Rz(H)\eta + \Rt(H)\tau ) + \contr{[Y,X_H]}\eta\\
        &= \contr{X_H}\dd g + \Rz(H)\contr{Y}\eta + \Rt(H)\contr{Y}\tau + \liedv{Y} \contr{X_H}\eta - \contr{X_H}\liedv{Y}\eta\\
        &= \contr{X_H}\dd g + \Rz(H)\contr{Y}\eta + \Rt(H)\contr{Y}\tau - \liedv{Y} H - \contr{X_H}(\rho\eta + \dd g)\\
        &= \Rz(H)\contr{Y}\eta - \rho H - g\Rz(H) - \rho \contr{Y}\eta = -(g - \contr{X_H}\eta)\Rz(H)\\
        &= -\Rz(H) f\,.
    \end{align*}
\end{proof}

Since a Cartan symmetries has an associated dissipated quantity and dissipated quantities are in bijection with 
generalized infinitesimal dynamical symmetry, every Cartan symmetry has an associated generalized infinitesimal dynamical symmetry.

\begin{proposition}
        Let  $(M,\tau,\eta, H)$ be a cocontact Hamiltonian system.
        If $Y$ is a $(\rho, g)$-Cartan symmetry, then $Z= Y - g \Rz$ is a generalized infinitesimal dynamical symmetry.
\end{proposition}
\begin{proof}   
        Suppose that $Y$ is a $(\rho, g)$-Cartan symmetry. Then, by \Cref{theorem:Cartan_dissipated}, the function $f= g - \contr{Y}\eta$ is a dissipated quantity. Hence, by \Cref{theorem:Noether_cocontact}, $Z=X_f-\Rt$ is a generalized infinitesimal dynamical symmetry. The Hamiltonian vector field of $f$ is given by
        \begin{align}
                \flat\taueta (X_f) &= \dd g - \dd (\contr{Y} \eta) 
                - \left( \liedv{\Rz} g - \liedv{\Rz} \contr{Y} \eta + g - \contr{Y} \eta \right) \eta\\
                & \quad + \left(1 - \liedv{\Rt} g + \liedv{\Rt} \contr{Y} \eta \right) \tau\,,
        \end{align}
        but
        \begin{align}
                \liedv{\Rz} \contr{Y} \eta 
                & = \contr{[\Rz, Y]} \eta + \contr{Y} \liedv{Rz}  \eta   
                = \contr{[\Rz, Y]} \eta
                = - \contr{[Y, \Rz]} \eta\\
                & = -\liedv{Y} \contr{\Rz} \eta + \contr{\Rz} \liedv{Y} \eta
                =  \contr{\Rz} \liedv{Y} \eta\\
                & = \contr{\Rz} \left(\rho \eta + \dd g\right)
                = \rho + \liedv{\Rz} g\, .
        \end{align}
        Similarly, 
        \begin{equation}
        \liedv{\Rt} \contr{Y} \eta = \liedv{\Rt} g\, .
        \end{equation}
        In addition, 
        \begin{align}
                \dd (\contr{Y} \eta)  = \liedv{Y} \eta - \contr{Y} \dd \eta
                = \rho \eta + \dd g - \contr{Y} \dd \eta\, .
        \end{align}
        Thus,
        \begin{align}
                \flat\taueta (X_f) = \contr{Y} \dd \eta
                - \left(g - \contr{Y} \eta \right) \eta
                + \tau\,.
        \end{align}
        On the other hand,
        \begin{equation}
                \flat\taueta (Y) = (\contr{Y} \eta) \eta + \contr{Y} \dd \eta\,,
        \end{equation}
        and therefore one can write
        \begin{equation}
                \flat\taueta (X_f - Y) = -g\eta + \tau\,,
        \end{equation}
        that is,
        \begin{equation}
                X_f = Y - g \Rz + \Rt\,,
        \end{equation}
        which implies that $Z= Y - g \Rz$.
\end{proof}

It is noteworthy that if $Y$ is a $(\rho, g)$-Cartan symmetry and $Z= Y-g \Rz$ is its associated generalized infinitesimal dynamical symmetry, then the dissipated quantities associated to $Y$ and to $Z$ via \Cref{theorem:Noether_cocontact,theorem:Cartan_dissipated} coincide.

Regarding the Lie algebra structures formed by these types of symmetries, there is the following result.

\begin{proposition}[Lie algebras of symmetries]
Let  $(M,\tau,\eta, H)$ be a cocontact Hamiltonian system.
\begin{enumerate}
        \item Infinitesimal conformal Hamiltonian symmetries close a Lie subalgebra of $(\X(M), [\cdot,\cdot])$. More precisely, if $Y_1$ is a $\rho_1$-conformal Hamiltonian symmetry and $Y_2$ is a $\rho_2$-conformal Hamiltonian symmetry, then $[Y, Z]$ is a $\widetilde \rho$-conformal Hamiltonian symmetry, where $\widetilde \rho = Y_1(\rho_2) - Y_2(\rho_1)$.
        \item Infinitesimal strict Hamiltonian symmetries close a Lie subalgebra from the Lie algebra of infinitesimal conformal Hamiltonian symmetries.
\end{enumerate}
\end{proposition}

\begin{proof}
        If $Y_1$ is a $(\rho_1, g_1)$-Cartan symmetry and $Y_2$ is a $(\rho_2, g_2)$-Cartan symmetry, then
        \begin{align}
                \liedv{[Y_1, Y_2]} \eta 
                & = \liedv{Y_1} \liedv{Y_2} \eta -  \liedv{Y_2} \liedv{Y_1} \eta \\
                &= \liedv{Y_1} \left( \rho_2 \eta + \d g_2  \right)
                - \liedv{Y_2} \left( \rho_1 \eta + \d g_1  \right)\\
                & = \left(Y_1 (\rho_2) - Y_2(\rho_1)  \right) \eta 
                + \d \left(Y_1(g_2) -Y_2(g_1) \right) \\
                &\quad + \rho_2 \d g_1 + \rho_1 \dd g_2\, .
        \end{align}
        Therefore, in general, $[Y_1, Y_2]$ is not a Cartan symmetry (see \Cref{counterexample:Lie_Cartan}). However, for $g_1=g_2 \equiv 0$, 
        \begin{equation}
                \liedv{[Y_1, Y_2]} \eta = \left(Y_1 (\rho_2) - Y_2(\rho_1)  \right) \eta = \widetilde {\rho} \eta\,.
        \end{equation}
        Moreover,
        \begin{equation}
        \begin{aligned}
                \liedv{[Y_1, Y_2]} H
                & = \liedv{Y_1} \liedv{Y_2} H -  \liedv{Y_2} \liedv{Y_1} H
                = \liedv{Y_1} \left( \rho_2 H \right)
                - \liedv{Y_2} \left( \rho_1 H  \right) \\
                & = \left(Y_1 (\rho_2) - Y_2(\rho_1)  \right) H\, .
        \end{aligned}   
        \end{equation}
        and hence $[Y_1, Y_2]$ is an infinitesimal $\widetilde \rho$-conformal Hamiltonian symmetry.
        
        In particular, if $Y_1$ an $Y_2$ are infinitesimal strict Hamiltonian symmetries, then $\rho_1=\rho_2\equiv 0$, which implies that $\widetilde \rho\equiv 0$, and thus $[Y_1, Y_2]$ is an infinitesimal strict Hamiltonian symmetry.
\end{proof}
    
The counterexample below shows that Cartan symmetries do not close a Lie subalgebra.
\begin{counterexample}\label{counterexample:Lie_Cartan}
Consider the cocontact Hamiltonian system $(\RR^4, \tau, \eta, H)$, with $\tau= \dd t,\ \eta = \dd z - p \dd q$ and 
\begin{equation}
        H=e^{q-z} ,
\end{equation}
where $(t, q, p, z)$ are the canonical coordinates in $\RR^4$. 
The vector field 
\begin{equation}
        Y_1=q\frac{\partial}{\partial z}
\end{equation}
is a $(0,q)$-Cartan symmetry and
\begin{equation}
        Y_2=(p-1)e^{q-z}\frac{\partial}{\partial p}-e^{q-z}\frac{\partial}{\partial z}
\end{equation}
is a $(e^{q-z},0)$-Cartan symmetry. Their commutator is $[Y_1,Y_2]=-qY_2$, and
\begin{equation}
        \liedv{[Y_1,Y_2]}\eta=-qe^{q-z}\eta+e^{q-z}\dd q\,.
\end{equation}
There is no function $f\in \Cinfty(\RR^4)$ such that $f\eta+e^{q-z}\dd q$ is exact.
Consequently, it is not possible to write $\liedv{[Y_1,Y_2]}\eta=\rho \eta + \dd g$ for any functions $\rho, g\in \Cinfty(\RR^4)$, and hence $[Y_1, Y_2]$ is not a Cartan symmetry.
\end{counterexample}



\begin{figure}[t]
        \centering  
        \begin{tikzpicture}[every text node part/.style={align=center}]
                \tikzstyle{every node}=[font=\small]
                        \draw (0,0) node[minimum height=6cm,minimum width=9cm,draw, rounded corners] (gen) {};
                        \node[below] at (gen.north) {Generalized infinitesimal dynamical symmetries\\ $\tau(Y)=0\,,\qquad \eta\big([Y, X_H]\big)=0$};
                        \draw (-1.5,0) node[minimum height=4cm, minimum width = 6cm, fill, draw, pattern=dots, pattern color=gray!70, rounded corners] (dyn) {};
                        \node[below right] at (dyn.north west) {Infinitesimal dynamical symmetries\\ $\tau(Y)=0\,,\quad [Y, X_H]=0$};
                        \draw (2.75,-1) node[minimum height=4cm, minimum width = 8.5cm, draw, pattern=north west lines, pattern color=blue!10, rounded corners](coco){};
                        \node[below left] at (coco.north east) {Infinitesimal\\ conformal\\cocontacto-\\morphisms\\$\tau(Y)=0\,,$\\$\liedv{Y}\eta=\rho \eta$};
                        \draw (1.5,-2) node[minimum height=6cm, minimum width = 6cm, draw, pattern=north east lines, pattern color=red!20, rounded corners](cartan){};
                        \node[above] at (cartan.south) {Cartan\\ symmetries\\$\tau(Y)=0\,,$\quad $\liedv{Y}\eta=\rho \eta+\d g\,,$\\ $Y(H)=\rho H +g \Rz(H)$};
                        \draw (1.5,-1) node[minimum height=4cm, minimum width = 6cm, draw, rounded corners](conham){};
                        \node[below] at (conham.north) {Infinitesimal conformal\\ Hamiltonian symmetries\\$\tau(Y)=0\,,$\quad $\liedv{Y}\eta=\rho \eta\,,$\\ $Y(H)=\rho H$};
                        \draw (1.3,-2) node[color=yellow!5, minimum height=2cm, minimum width = 4.3cm, fill, semitransparent, rounded corners](strham){};
                        \draw (1.3,-2) node[minimum height=2cm, minimum width = 4.3cm, draw, rounded corners](strham){};
                        \node[below] at (strham.north) {Infinitesimal strict\\ Hamiltonian symmetries\\$\tau(Y)=0\,,$\quad$\liedv{Y}\eta=0\,,$\\ $Y(H)=0$};
                \end{tikzpicture}

        \caption[Classification of infinitesimal symmetries and relations between them.]{Classification of infinitesimal symmetries and relations between them. Infinitesimal dynamical symmetries, infinitesimal conformal Hamiltonian symmetries and infinitesimal strict Hamiltonian symmetries close Lie algebras, whereas Cartan symmetries and generalized infinitesimal dynamical symmetries do not close Lie algebras.}
        \label{fig:infinitesimal_symmetries}
\end{figure}
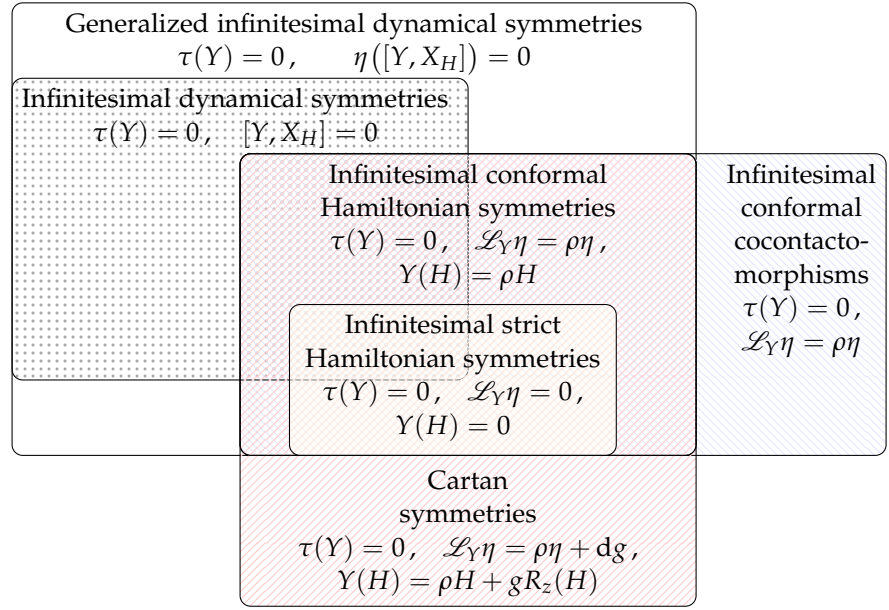

The types of symmetries and the relations between them are summarized in \Cref{fig:infinitesimal_symmetries}.


\section[Symmetries of cocontact Lagrangian systems]{Symmetries and dissipated quantities of cocontact Lagrangian systems}\label{sec:symmetries_cocontact_Lagrangian}

\begin{figure}[t]
        \centering  
        \begin{tikzpicture}[every text node part/.style={align=center}]
                \tikzstyle{every node}=[font=\small]
                \draw (2.5,-3) node[minimum height=6cm, minimum width = 6cm, fill, draw, pattern=north east lines, pattern color=red!20, rounded corners](coco){};
                \node[below] at (coco) {\\Infinitesimal symmetries\\of the Lagrangian\\$\liedv{Y} t=0\,,$\\$\liedv{Y}L=0$};
                \draw (0.3,-0.6) node[minimum height=4.7cm, minimum width = 9cm,, draw, pattern=north west lines, pattern color=blue!20, rounded corners](conham){};
                \node[below] at (conham.north) {Infinitesimal extended natural\\ symmetries of the Lagrangian\\$X=X^i(q) \frac{\partial}{\partial q^i}+ \zeta(z) \frac{\partial}{\partial z}\,, \quad Y=\bar X^\Com\,,\quad \liedv{Y}L =\zeta' L$};
                \draw (2.3,-1.6) node[color=yellow!5, minimum height=2.5cm, minimum width = 4.5cm, fill, semitransparent, rounded corners](strham){};
                \draw (2.3,-1.6) node[minimum height=2.5cm, minimum width = 4.5cm, draw, rounded corners](strham){};
                \node[below] at (strham.north) {Infinitesimal natural\\ symmetries of \\the Lagrangian\\$X=X^i(q) \frac{\partial}{\partial q^i}\,,$\\ $\quad Y=X^\Com\,,\quad \liedv{Y}L =0$};
                \draw (0,0) node[minimum height=6cm,minimum width=11cm,draw, rounded corners] (gen) {};
                \node[below] at (gen.north) {Infinitesimal extended symmetries of the Lagrangian\\ $Y=A^i(t,q, v) \parder{}{q^i} + B^i(t,q,v) \parder{}{v^i}+ \zeta(z) \parder{}{z}\,,\quad \liedv{Y} L = \zeta' L$};
                \end{tikzpicture}

        \caption[Classification of infinitesimal Lagrangian symmetries and relations between them.]{Classification of infinitesimal Lagrangian symmetries and relations between them. Infinitesimal symmetries of the Lagrangian, infinitesimal natural symmetries of the Lagrangian and infinitesimal extended natural symmetries of the Lagrangian close Lie subalgebras.}
        \label{fig:infinitesimal_symmetries_Lagrangian}
\end{figure}
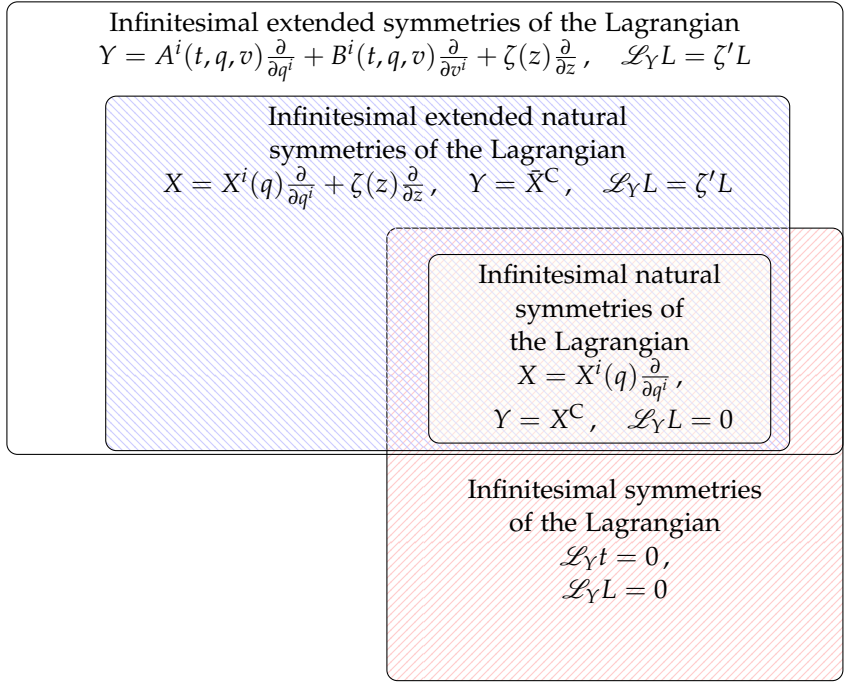

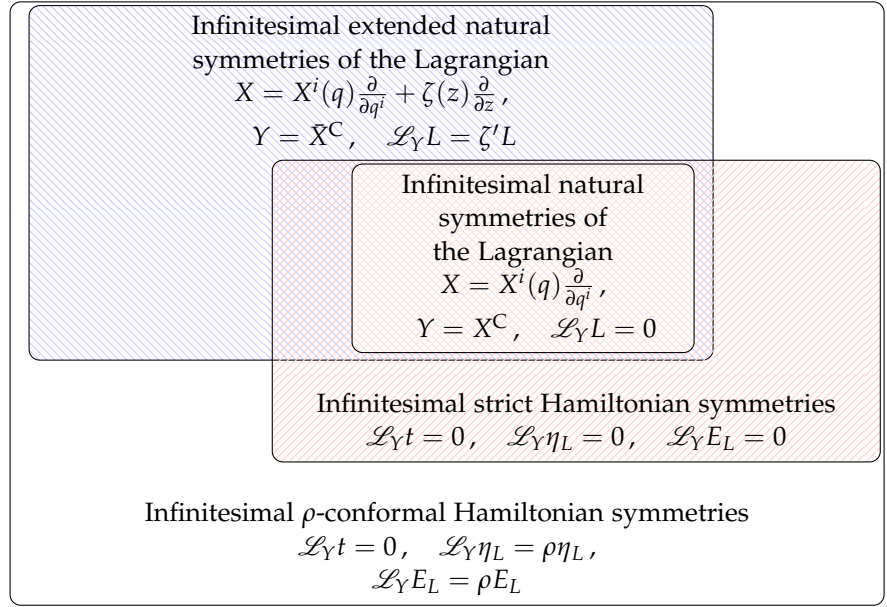
\begin{figure}[t]
        \centering  
        \begin{tikzpicture}[every text node part/.style={align=center}]
                \tikzstyle{every node}=[font=\small]
                \draw (0.3,-0.6) node[minimum height=4.7cm, minimum width = 9cm, draw, pattern=north west lines, pattern color=blue!20, rounded corners](conham){};
                \node[below] at (conham.north) {Infinitesimal extended natural\\ symmetries of the Lagrangian\\$X=X^i(q) \frac{\partial}{\partial q^i}+ \zeta(z) \frac{\partial}{\partial z}$\,,\\ $\quad Y=\bar X^\Com\,,\quad \liedv{Y}L =\zeta' L$};
                \draw (1.3,-2.2) node[minimum height=8cm, minimum width = 11.5cm, draw, rounded corners](cartan){};
                \node[above] at (cartan.south) {Infinitesimal $\rho$-conformal Hamiltonian symmetries\\$\liedv{Y} t=0\,,\quad \liedv{Y}\eta_L=\rho \eta_L\,,$\\ $\liedv{Y}E_L=\rho E_L$};
                \draw (3,-2.3) node[minimum height=4cm, minimum width = 8cm, fill, pattern=north east lines, pattern color=red!20, rounded corners](strham){};
                \draw (3,-2.3) node[minimum height=4cm, minimum width = 8cm, draw, rounded corners](strham){};
                \node[above] at (strham.south) {Infinitesimal strict Hamiltonian symmetries\\$\liedv{Y} t=0\,,\quad\liedv{Y}\eta_L=0\,,\quad \liedv{Y}E_L=0$};
                \draw (2.3,-1.6) node[color=yellow!5, minimum height=2.5cm, minimum width = 4.5cm, fill, semitransparent, rounded corners](strham){};
                \draw (2.3,-1.6) node[minimum height=2.5cm, minimum width = 4.5cm, draw, rounded corners](strham){};
                \node[below] at (strham.north) {Infinitesimal natural\\ symmetries of \\the Lagrangian\\$X=X^i(q) \frac{\partial}{\partial q^i}\,,$\\ $\quad Y=X^\Com\,,\quad \liedv{Y}L =0$};
                \end{tikzpicture}

        \caption{Relations between infinitesimal (extended) natural symmetries of the Lagrangian, conformal Hamiltonian symmetries and strict Hamiltonian symmetries.}
        \label{fig:infinitesimal_symmetries_Lagrangian_Hamiltonian}
\end{figure}

Along this section, a regular cocontact Lagrangian system $(\RR \times \T Q \times \RR, L)$ with cocontact structure $(\d t, \eta_L)$ will be considered. Since $(\RR \times \T Q \times \RR, \d t, \eta_L, E_L)$ is a cocontact Hamiltonian system, every result from the previous section can be applied to this case. Moreover, making use of the geometric structures of the tangent bundle and their natural extensions to $\RR \times \T Q \times \RR$, it is possible to consider additional types of symmetries. A summary of these symmetries and their relations can be found in \Cref{fig:infinitesimal_symmetries_Lagrangian}. The relation between (extended) natural symmetries of the Lagrangian and Hamiltonian symmetries is depicted in \Cref{fig:infinitesimal_symmetries_Lagrangian_Hamiltonian}.

Let $\pr{Q}\colon Q \times \RR \to Q$ and $\pr{\RR} \colon Q\times \RR \to \RR$ be the canonical projections, and denote other projections as in diagram \eqref{eq:projections_cocontact_Lagrangian}.
Consider a diffeomorphism
\begin{equation}
        \varphi=(\varphi_Q\circ \pr{Q}, \varphi_z\circ \pr{\RR})\colon Q \times \RR \to Q\times \RR\, ,
\end{equation}
where $\varphi_Q\colon Q \to Q$ and $\varphi_z\colon \RR \to \RR$ are diffeomorphisms.
The \emph{action-dependent lift} of $\varphi$ is the diffeomorphism
\begin{equation}
        \widetilde \varphi = (\tau_1, \T \varphi_Q \circ \tau_2, \varphi_z\circ \tau_3): \RR \times \T Q \times \RR \to \RR \times \T Q \times \RR\, , 
\end{equation}
that is,
\begin{equation}
        \widetilde{\varphi} (t, v_q, z) = \big(t, \T \varphi_Q(v_q), \varphi_z(z) \big)\, .
\end{equation}
A vector field $Y \in \X(Q\times \RR)$ is \emph{split} if it is projectable by $\pr{Q}$ and $\pr{\RR}$, that is, $\T\pr{Q}(X)\in\X(Q)$ and $\T\pr{\RR}\in \X(\RR)$. Given a split vector field $Y \in \X(Q\times \RR)$, its {\emph{action-dependent lift}} is the vector field $\bar Y^\Com \in \X(\RR \times \T Q \times R)$ whose flow is the action-dependent lift of the flow of $Y$. In bundle coordinates, a split vector field $Y$ is of the form
\begin{equation}\label{eq:local-Y}
        Y = Y^i(q) \frac{\partial}{\partial q^i}+ \zeta(z) \frac{\partial}{\partial z}\,,
\end{equation}
and its action-dependent complete lift is
\begin{equation}\label{eq:action-dependent_complete_lift}
        \bar Y^\Com=Y^i(q) \frac{\partial}{\partial q^i}
        + v^j \frac{\partial Y^i}{\partial q^j} \frac{\partial}{\partial v^i}
        + \zeta(z) \frac{\partial}{\partial z}\, .      
\end{equation}
The \emph{vertical lift} of an split $Y\in \X(Q\times \RR)$ to $\RR \times \T Q \times \RR$ is the vector field $\bar Y^\V\in \X(\RR \times \T Q \times \RR)$ given by the vertical lift of $\T \pr{Q} Y\in \X(Q)$ to $TQ$. Locally, if $Y$ has the local expression \eqref{eq:local-Y}, its vertical lift reads
\begin{equation}
        \bar Y^\V= Y^i(q) \parder{}{v^i}\,. 
\end{equation}
The following properties hold for any $X, Y \in \X(Q\times \RR)$:
\begin{equation} \label{eq:properties_lifts}
\begin{array}{lll}
        [\bar X^\Com, \Delta] = 0\,,\qquad &\Sendo (\bar X^\Com) = \bar X^\V\,,\qquad &\Sendo (\bar X^\V) = 0\,, \\ 
        \liedv{\bar X^\V}\Sendo = 0\,, &\liedv{\bar X^\Com}\Sendo = 0\,.
\end{array}
\end{equation}
where $\Sendo$ and $\Delta$ denote the vertical endomorphism and the Liouville vector field, with local expressions~\eqref{eq:Sendo_Liouville_vf_local}. 

Along the rest of the chapter, the derivative of a function $\phi\colon \RR\to \RR$ will be denoted by $\phi^{\prime}$.

\begin{definition}
        A diffeomorphism $\Phi\colon \RR \times \T Q \times \RR\to \RR \times \T Q \times \RR$ of the form
        \begin{equation}\label{eq:extended_symmetry_Lagrangian}
                \Phi\colon (t, q, v, z) \mapsto \left(t, \Phi_q(t,q,v), \Phi_v(t,q,v), \Phi_z(z) \right)\, .
        \end{equation}
        is called an \emph{extended symmetry of the Lagrangian} if $\Phi^\ast L = \Phi_z'L$. In addition, if $\Phi$ is the action-dependent lift of some $\varphi \in \Diff(Q\times \RR)$, then it is called an \emph{extended natural symmetry of the Lagrangian}.

        A vector field $Y \in \X(\RR \times \T Q \times \RR)$ of the form
        \begin{equation}
         Y=A^i(t,q, v) \parder{}{q^i} + B^i(t,q,v) \parder{}{v^i}+ \zeta(z) \parder{}{z}
        \end{equation}
        is called an \emph{infinitesimal extended symmetry of the Lagrangian} if $\liedv{Y} L = \zeta' L$. 
        In addition, if $Y$ is the action-dependent complete lift of some $X \in \X(Q\times \RR)$, then it is called an \emph{infinitesimal extended natural symmetry of the Lagrangian}.
\end{definition}

\begin{proposition}\label{prop:extLagSym}
        Let $Y\in \X(Q\times \RR)$ be a vector field such that $\bar{Y}^\Com$ is an infinitesimal extended natural symmetry of the Lagrangian and $\T \pr{\RR}(Y) = \zeta(z) \tparder{}{z}$. Then, $\bar Y^\Com$ is an infinitesimal $\zeta'$-conformal Hamiltonian symmetry of $(M,\tau,\eta_L,E_L)$.

        Let $\Phi\colon \RR \times \T Q \times \RR\to \RR \times \T Q \times \RR$ be an extended natural symmetry of the Lagrangian with $\tau_3\circ \Phi= \Phi_z\circ \tau_3$. Then, $\Phi$ is a $\Phi_z'$-conformal Hamiltonian symmetry of $(M,\tau,\eta_L,E_L)$.
\end{proposition}

\begin{proof}
        Clearly, $\contr{\bar Y^\Com}\dd t=0$. Moreover,
        \begin{equation}
        \begin{aligned}
                \liedv{\bar Y^\Com}E_L&=\liedv{\bar Y^\Com}(\Delta(L))- \liedv{\bar Y^\Com}(L)=(\Delta-1)(\liedv{\bar Y^\Com}(L))\\
                &=(\Delta-1)(\zeta'L)=\zeta'E_{L}\,,
        \end{aligned}
        \end{equation}
        since the action-dependent complete lift of a vector field commutes with the Liouville vector field (see properties \eqref{eq:properties_lifts}), and
        \begin{equation}
        \begin{aligned}
         \liedv{\bar Y^\Com}\eta_L&=\liedv{\bar Y^\Com}(\d z-\Sendoadj\d L)=\d \zeta-\Sendoadj \d \left(\liedv{\bar Y^\Com}L\right)\\
         &=\zeta'\d z-\Sendoadj \left(L\d \zeta' +\zeta'\d L\right)
         \\
         &=\zeta'\left(\d z-\Sendoadj\d L\right)=\zeta'\eta_L\, ,   
        \end{aligned}
        \end{equation}
        since $\liedv{\bar Y^\Com} \Sendo = 0$.
        Therefore, $\bar Y^\Com$ is an infinitesimal $\zeta'$-conformal Hamiltonian symmetry. 

        Similarly, taking into account that an extended natural symmetry of the Lagrangian is of the form
        \begin{equation}
                \Phi (t, v_q, z) = \big(t, \T \varphi_Q(v_q), \Phi_z(z) \big)\, ,
        \end{equation}
        then 
        \begin{equation}
        \begin{aligned}
                \Phi^\ast E_L & = \Phi^\ast \left(\liedv{\Delta} L - L\right)
                = \contr{\Phi_\ast \Delta} \Phi^\ast \dd  L - \Phi^\ast L\\
                &= \contr{\Delta} \dd (\Phi_z'  L) - \Phi_z' L
                = \Phi_z' E_L\, ,
        \end{aligned}
        \end{equation}
        and
        \begin{equation}
        \begin{aligned}
                \Phi^\ast \eta_L & = \Phi^\ast \left(\dd z -\Sendoadj \dd L\right)
                = \dd \Phi_z - \Sendoadj \Phi^\ast \dd L\\
                &= \Phi_z' \dd z - \Sendoadj \Phi_z' \dd L
                = \Phi_z' \eta_L\, ,
        \end{aligned}
        \end{equation}
        since the tangent lift of a diffeomorphism on $Q$ preserves the Liouville vector field and the vertical endomorphism on $\T Q$.
        Therefore, $\Phi$ is a $\Phi_z'$-conformal Hamiltonian symmetry.
\end{proof}

\begin{proposition}\label{prop:extsym}
        Let $Y\in \X(Q\times \RR)$ be a split vector field with $\T \pr{\RR} (Y)= \zeta(z)\tparder{}{z}$. Then, $\bar Y^\Com$ is an infinitesimal extended natural symmetry of the Lagrangian if and only if $\bar Y^\V(L)-\zeta$ is a dissipated quantity.
\end{proposition}

\begin{proof}
        By using the second of the properties~\eqref{eq:properties_lifts}, one obtains
        \begin{equation}
                \eta_L(\bar Y^\Com)=(\d z- \Sendoadj \d L) (\bar Y^\Com)=\zeta-\bar Y^\V(L)\, ,
        \end{equation}
        and therefore
        \begin{align*}
        \liedv{\sode_L} \left(\bar Y^\V(L)-\zeta\right)&+\liedv{\Rz^L}(E_L) \left(\bar Y^\V(L)-\zeta\right) \\
        &= -\liedv{\sode_L} \contr{\bar Y^\Com} \eta_L - \contr{\bar Y^\Com} \big( \Rz^L (E_L) \eta_L \big)  \\
        & = -\liedv{\sode_L} \contr{\bar Y^\Com} \eta_L + \contr{\bar Y^\Com} \big( \liedv{\sode_L} \eta_L +\Rt(E_L) \tau \big)  \\
        &=-\contr{[\sode_L, \bar Y^\Com]} \eta_L\, .
        \end{align*}
        On the other hand,
        \begin{align*}
                \contr{[\bar Y^\Com, \sode_L ]} \eta_L
                &= \liedv{\bar Y^\Com} \contr{\sode_L} \eta_L - \contr{\sode_L} \liedv{\bar Y^\Com} \eta_L\\
                &= -\liedv{\bar Y^\Com} E_L - \contr{\sode_L} \liedv{\bar Y^\Com} \left( \d z - \Sendoadj \d L \right)\\
                &= -\Delta (\liedv{\bar Y^\Com} L)+ \liedv{\bar Y^\Com} L- \contr{\sode_L} \liedv{\bar Y^\Com}  \d z+ \contr{\sode_L} \Sendoadj \d \left(\liedv{\bar Y^\Com} L\right) \\
                &= \liedv{\bar Y^\Com} L - \contr{\sode_L} \liedv{\bar Y^\Com}  \d z
                = \liedv{\bar Y^\Com} L - \contr{\sode_L} \d \zeta\\
                &= \liedv{\bar Y^\Com} L - \liedv{\sode_L} \zeta\, ,
                \end{align*}
        where it has been used that $\Sendo \sode_L = \Delta$. Thus, $\bar Y^\V(L)-\zeta$ is a dissipated quantity if and only if $\liedv{\bar Y^\Com} L - \liedv{\sode_L} \zeta$ vanishes.
\end{proof}

A particular case of extended natural symmetries are symmetries which are lifted from $Q$.

\begin{definition}
        A diffeomorphism $\Phi \colon \RR \times \T Q \times \RR\to \RR \times \T Q \times \RR$ is called a \emph{symmetry of the Lagrangian} if $\Phi^\ast L = L$ and $\Phi^\ast t=t$. In addition, if $\Phi$ is the canonical lift of a diffeomorphism $\varphi \colon Q \to Q$, then it is called a \emph{natural symmetry of the Lagrangian}.
    
        A vector field $Y \in \X(\RR \times \T Q \times \RR)$ is called an \emph{infinitesimal symmetry of the Lagrangian} if $\liedv{Y} L = 0$ and $\contr{Y} \tau=0$. In addition, if $Y$ is the complete lift of some $X \in \X(Q)$, then it is called an \emph{infinitesimal natural symmetry of the Lagrangian}.
\end{definition}

\Cref{prop:extLagSym} implies the following.
\begin{corollary}
    Every (infinitesimal) natural symmetry of the Lagrangian is an (infinitesimal) strict Hamiltonian symmetry of $(\RR \times \T Q\times \RR, \d t, \eta_{L}, E_L)$.
\end{corollary}

It is worth noting that a symmetry of the Lagrangian which is not natural is not, in general, a Hamiltonian symmetry. Moreover, in general, it is not an extended symmetry of the Lagrangian either.

\begin{counterexample}
        Let $(t,x,v,z)$ be the canonical bundle coordinates on $\RR\times \T\RR\times \RR$.
        Consider the Lagrangian function $L\in \Cinfty(\RR\times \T\RR\times \RR)$ given by
        \begin{equation}
                L(t,x,v,z)=\frac{1}{2} v^2-V(t,x,z)\, .
        \end{equation}
        Clearly, the vector field
        \begin{equation}
        Y = v \frac{\partial }{\partial x} + \frac{\partial V}{\partial x} \frac{\partial }{\partial v}
        \end{equation}
        is an infinitesimal symmetry of the Lagrangian (but it is not natural). However, $Y(E_L) \neq 0$. Moreover, $\eta_L = \d z - v \d x$, and hence
        \begin{equation}
                \liedv{Y} \eta_L 
                = - \frac{\partial V}{\partial x} \d x - v \d v \neq \rho \eta_L\, , 
        \end{equation}
        for any $\rho\in\Cinfty(\RR\times \T\RR\times \RR)$
\end{counterexample}

\Cref{prop:extsym} implies the following.
\begin{corollary}\label{corollary:natural_Lagrangian_cocontact}
    Let $Y$ be a vector field on $Q$ and assume that $L$ is regular. Then $Y^\Com$ is an infinitesimal natural symmetry of $L$ if, and only if, $Y^\V(L)$ is a dissipated quantity.
\end{corollary}

\begin{example}[Cyclic coordinate]
   Suppose that $L$ has a cyclic coordinate, namely $\tparder{L}{q^i}=0$ for some $i \in \left\{1, \ldots, n\right\}$. Then, $\bar{Y}^\Com$ is an infinitesimal natural Lagrangian symmetry, where $Y= \tparder{}{q^i}$, and its associated dissipated quantity is the corresponding momentum $\tparder{L}{v^i}$.
\end{example}

\begin{proposition}
        Infinitesimal symmetries of the Lagrangian, infinitesimal natural symmetries of the Lagrangian and infinitesimal extended natural symmetries of the Lagrangian close Lie subalgebras of $(\X(\RR \times \T Q \times \RR), [\cdot, \cdot])$.
\end{proposition}

\begin{proof}
        If $Y_1, Y_2 \in \X(\RR \times \T Q \times \RR)$ are symmetries of the Lagrangian $L$, then
        \begin{equation}
                \liedv{[Y_1, Y_2]} L = \left[\liedv{Y_1}, \liedv{Y_2} \right] L = 0\, , \quad \contr{[Y_1, Y_2]} \tau = 0\, ,
        \end{equation}
        and thus $[Y_1, Y_2]$ is a symmetry of the Lagrangian. In particular, if $Y_1=X_1^\Com$ and $Y_2=X_2^\Com$, for $X_1, X_2\in \X(Q)$, are natural symmetries of the Lagrangian, then $[Y_1, Y_2] = [X_1, X_2]^\Com$. Therefore, $[Y_1, Y_2]$ is also a natural symmetry of the Lagrangian.

        Similarly, suppose that $\bar Y_1^\Com$ and $\bar Y_2^\Com$ are extended natural symmetries of the Lagrangian with local expressions
        \begin{equation}
                Y_1 = Y_1^i(q) \frac{\partial}{\partial q^i} + \zeta_1(z) \frac{\partial}{\partial z}\, ,\quad 
                Y_2 = Y_2^i(q) \frac{\partial}{\partial q^i} + \zeta_2(z) \frac{\partial}{\partial z}\, .
        \end{equation}
        Then,
        \begin{equation}
                \liedv{[\bar Y_1^\Com, \bar Y_2^\Com]} L 
                = \left[ \liedv{\bar Y_1^\Com}, \liedv{\bar Y_2^\Com}\right] L
                = \left( \zeta_1 \zeta_2^{\prime \prime} - \zeta_2 \zeta_1^{\prime \prime}\right) L
                = \frac{\d}{\d z}\left( \zeta_1 \zeta_2^{\prime } - \zeta_2 \zeta_1^{\prime }\right) L
                \, ,
        \end{equation}
        but
        \begin{equation}
                [Y_1, Y_2]^\Com
                = \left(Y_1^i \frac{\partial Y_2^j}{\partial q^i} - Y_2^i \frac{\partial Y_1^j}{\partial q^i}\right) \frac{\partial}{\partial q^j}
                + \left( \zeta_1 \zeta_2^{\prime} - \zeta_2 \zeta_1^{\prime }\right) \frac{\partial}{\partial z},
        \end{equation}
        and therefore $[\bar Y_1^\Com, \bar Y_2^\Com]$ is an extended natural symmetry of $L$. 

\end{proof}

Finally, one can consider transformations on the ``$z$'' variable, or changes of action, which preserve the dynamics. This kind of symmetries are employed in \cite{d.G.L2022} to generate equivalent Lagrangians.
\begin{definition}
    A diffeomorphism $\Phi\colon \RR \times \T Q \times \RR\to \RR \times \T Q \times \RR$  is called a \emph{change of action} if, for any section $\gamma$ of the vector bundle $\pr{\RR \times \T Q}\colon \RR \times \T Q \times \RR \to \RR \times \T Q$, it verifies
    \begin{equation}
        \pr{\RR \times \T Q}\circ\,\Phi\circ \gamma=\id_{\RR \times \T Q}\, .
    \end{equation}
    A vector field $Z \in \mathfrak{X}(\RR\times \T Q \times \RR)$ is called an \emph{infinitesimal change of action} if $\T\pr{\RR \times \T Q} (Z)\equiv 0$.
\end{definition}

In bundle coordinates, a change of action is of the form
\begin{equation}\label{eq:change_action}
        \Phi\colon (t, q, v, z) \mapsto \left(t, v_q, \Phi_z(t, q, v, z) \right)\, .
\end{equation}
Since $\Phi$ is a diffeomorphism, $\tparder{\Phi_z}{z}$ is nowhere-vanishing.

Clearly, the flow of an infinitesimal change of action is made up of changes of action. Moreover, if $Y\in\mathfrak{X}(\RR\times Q\times\RR)$ is a \textsc{sode} and $\Phi$ is a change of action, then $\Phi_\ast Y$ is also a \textsc{sode}. 

\begin{proposition}
        A change of action $\Phi\colon \RR \times \T Q \times \RR\to \RR \times \T Q \times \RR$ of the form \eqref{eq:change_action}
        is a generalized dynamical symmetry if, and only if,  
        $\sode_L(\Phi_z)=L\circ\Phi$.

        An infinitesimal change of action $Z \in \mathfrak{X}(\RR\times \T Q \times \RR)$ with local expression
        \begin{equation}
                Z = \zeta(t,q,v, z) \parder{}{z}
        \end{equation}
        is a generalized infinitesimal dynamical symmetry if, and only if, $\zeta$ is a dissipated quantity.
\end{proposition}

\begin{proof}
        Observe that any \textsc{sode} $X\in \X(\RR \times \T Q \times \RR)$ verifies $\theta_L(X)=\Delta(L)$. 
        If $\Phi$ is a change of action, then
        \begin{equation}
                \Phi_\ast\sode_L=\sode_L^t\parder{}{t}+\sode_L^{q^i}\parder{}{q^i}+\big(\sode_L^{v^i}\circ\Phi^{-1}\big)\parder{}{v^i}+\big(\sode_L(\Phi_z)\circ\Phi^{-1}\big)\frac{\partial}{\partial z}\,.
        \end{equation}
        In addition,
        \begin{equation}
                \contr{\Phi_\ast\sode_L}\eta_L=\sode_L(\Phi_z)\circ\Phi^{-1}-\theta_L(\Gamma_L)=\sode_L(\Phi_z)\circ\Phi^{-1}-\Delta(L)
                \,.
        \end{equation}
        On the other hand, $\contr{\sode_L}\eta_L=-E_{L}=L-\Delta(L)$. Therefore, $\Phi$ is a generalized dynamical symmetry if, and only if, $\sode_L(\Phi_z)= L\circ\Phi $.


        Furthermore, if $Z$ is an infinitesimal change of action, then
        \begin{equation}
        \begin{aligned}
        \contr{[\Gamma_L, Z]} \eta_L 
        &= \liedv{\Gamma_L} \contr{Z} \eta_L - \contr{Z} \liedv{\Gamma_L} \eta_L \\
        & = \Gamma_L(\zeta) + \contr{Z} \big( \Rz^L (E_L) \eta_L +\Rt^L(E_L) \tau\big)\\
        &= \Gamma_L(\zeta) + \Rz^L (E_L) \zeta\, .
        \end{aligned}
        \end{equation}
\end{proof}
This result motivates the following definition.

\begin{definition}
    A diffeomorphism
    $\Phi\colon \RR \times \T Q \times \RR\to \RR \times \T Q \times \RR$ of the form
    \begin{equation}
        \Phi\colon (t, q, v, z) \mapsto \left(t, q, v, \Phi_z(t,q,v,z) \right)
        \end{equation}
    is called an \emph{action symmetry} if $\sode_L(\Phi_z)=L\circ\Phi $.

    A vector field $Z \in \mathfrak{X}(\T Q \times \RR)$ of the form $Z=\zeta(t, q, v, z) \tparder{}{z}$ is called an \emph{infinitesimal action symmetry} if $\zeta$ is a dissipated quantity.
\end{definition}

\section{Examples}\label{sec:symmetries_cocontact_examples}


In this section, some examples are discussed in order to illustrate some of the concepts previously presented. In particular, it is shown how symmetries and dissipated quantities can be used to study the dynamics of the $2$-body problem with time-dependent friction.

\subsection{The free particle with time-dependent mass and linear dissipation}
Consider the cocontact Hamiltonian system $(\RR\times \cT \RR \times \RR, \d t, \eta, H)$, {with natural coordinates $(t, q, p, z)$ where
$\eta=\d z -p \d q$ is the contact form and
\begin{equation}
        H = \frac{p^2}{2m(t)} + \frac{\kappa}{m(t)} z\, ,
\end{equation}
is the Hamiltonian function,} with $m$ a positive-valued function depending only on $t$, expressing the mass of the particle, and $\kappa$ a positive constant. The Hamiltonian vector field of $H$ is
\begin{equation}
         X_H =\parder{}{t} + \frac{p}{m(t)}\parder{}{q} -  p\frac{\kappa}{m(t)} \parder{}{p} + \left(\frac{p^2}{2m(t)} - \frac{\kappa}{m(t)} z\right)\parder{}{z}\,. 
\end{equation}
{
Its integral curves are given by
\begin{equation}
\begin{aligned}
        & \frac{\dd q}{\dd t} = \frac{p}{m(t)}\,,
        \\
       & \frac{\dd p}{\dd t} = -  p\frac{\kappa}{m(t)} \,,
        \\
       & \frac{\dd z}{\dd t} =  \frac{p^2}{2m(t)} - \frac{\kappa}{m(t)} z\,,		
\end{aligned}	  
\end{equation}
which yield
\begin{equation}
\begin{aligned}
        q(t) & =  \int _0^t\frac{\exp \left(\int _0^u-\frac{\kappa }{m(s)}\dd s\right) p_0}{m(u)}\dd u+q_0\, ,\\
        p(t) & = p_0 \exp \left(\int _0^t-\frac{\kappa }{m(s)}\dd s\right)\, ,\\
        z(t) & = \exp \left(\int _0^t-\frac{\kappa }{m(v)}\dd v\right) \int _0^t\frac{\exp \left( \int _0^w-\frac{\kappa }{m(s)}\dd s\right) p_0^2}{2 m(w)}\dd w \\
        & \quad +z_0 \exp \left(\int _0^t-\frac{\kappa }{m(v)}\dd v\right)\, ,
\end{aligned}
\end{equation}
where $q_0=q(0),\, p_0=p(0),\, z_0=z(0)$ are the initial conditions. The term of $H$ linear in the variable $z$ permits to model a damping phenomena. As a matter of fact, in the particular case where $m(t)$ is constant the linear momenta $p$ (and hence the velocity) of the system decreases exponentially.
}

The function
\begin{equation}
        f(t,q,p,z)=\exp \left(- \displaystyle \int _0^t\frac{\kappa}{m(s)}\d s\right)
\end{equation}
is a dissipated quantity. Hence, by \Cref{theorem:Noether_cocontact}, the vector field
\begin{equation}
Y_f = X_f - \Rt 
= -\exp \left(- \int _0^t\frac{\kappa}{m(s)} \d s\right) \parder{}{z}
= - f \Rz
\end{equation}
is a generalized infinitesimal dynamical symmetry. In addition, one can verify that $Y_f$ is an infinitesimal dynamical symmetry, that is, $Y_f$ commutes with $X_H$. Furthermore,
\begin{equation}
Y_f (H) 
= - f \Rz(H)\, ,
\end{equation}
and
\begin{equation}
\liedv{Y_f} \eta
= - f \liedv{\Rz} \eta - (\dd f) \, \contr{\Rz} \eta
= -\dd f\, ,
\end{equation}
which means that $Y_f$ is a $(0,-f)$-Cartan symmetry.

Moreover,
$f_2(t,q,p,z)=p$ is also a dissipated quantity, whose associated generalized infinitesimal dynamical symmetry is
\begin{equation}
         Y_{f_2} = \parder{}{q}\,. 
\end{equation}
It is clear that $Y_{f_2}$ is an infinitesimal dynamical symmetry. Moreover, $\liedv{Y_{f_2}} \eta = 0$
and $Y_{f_2}(H)=0$, thus $Y_{f_2}$ is an infinitesimal strict Hamiltonian symmetry.

The Lagrangian counterpart of this system is characterized by the Lagrangian function $L\colon \RR \times \T \RR \times \RR \to \RR$ given by 
\begin{equation}
L = m(t)\frac{v^2}{2} - \frac{\kappa}{m(t)} z\,.
\end{equation}
The vector field $Z\in \mathfrak{X}(\RR\times \T \RR \times \RR)$ with local expression
\begin{equation}
Z = \zeta(t,q, v,z) \parder{}{z}
= \exp \left(- \int _0^t\frac{\kappa}{m(s)} \d s\right) \parder{}{z}
\end{equation}
is an infinitesimal action symmetry, since it is an infinitesimal change of action and $\zeta$ is a dissipated quantity.

\subsection{An action-dependent central potential with time-dependent mass}
Consider a Lagrangian function $L\colon \RR \times \T \RR^2 \times \RR \to\RR$ of the form
\begin{equation}
L {(t, x, y, v_x, v_y, z)} = \frac{m(t)}{2} \left(v_x^2 + v_y^2\right) - V \left(t,(x^2+y^2), z\right)\, ,
\end{equation}
where $m(t)$ is a positive-valued function. Let $Y\in \mathfrak{X}(\RR^2)$ be infinitesimal generator of rotations on the plane, namely,
\begin{equation}
Y = -y \parder{}{x} + x \parder{}{y}\, .
\end{equation}
Its complete lift is given by
\begin{equation}
\bar{Y}^\Com = -y \parder{}{x} + x \parder{}{y} - v_y \parder{}{v_x} + v_x \parder{}{v_y} \, ,
\end{equation}
and its vertical lift is 
\begin{equation}
\bar{Y}^\V = -y \parder{}{v_x} + x \parder{}{v_y}\, .
\end{equation}
Clearly, $\bar{Y}^\Com$ is an infinitesimal natural symmetry of the Lagrangian, namely, $\bar{Y}^\Com(L)=0$. Hence, by \Cref{corollary:natural_Lagrangian_cocontact}, the funtion
\begin{equation}
        \bar{Y}^\V(L) = m(t)\left(-y v_x + x v_y\right)
\end{equation}
is a dissipated quantity. This quantity is the angular momentum for a particle with time-dependent mass.

\subsection{The two-body problem with time-dependent friction}

The two-body problem describes the dynamics of two particles under the effects of a force that depends on the distance between the particles, usually the gravitational force. To model time-dependent friction, one may add a linear term on the action in the Lagrangian function, with a time-dependent coefficient. 

Consider the cocontact Lagrangian system $(\RR^6, L)$. Denote the canonical bundle coordinates of $\RR\times \T\RR^6\times \RR$ by  $(t,\bm{q}^1,\bm{q}^2,\bm{v}^1,\bm{v}^2,z)$, where the superindex denotes each particle, and the bold notation is a shorthand for the three spatial components, namely, $\bm{q}^1=(q_1^1,q_2^1,q_3^1)$ and $\bm{q}^2=(q_1^2,q_2^2,q_3^2)$. The relative distance between the particles is $\bm{r}=\bm{q}^2-\bm{q}^1$, whose Euclidean length will be denoted $r = \vert\bm{r}\vert$.

The Lagrangian function is
\begin{equation}
        L=\frac12m_1\bm{v}^1\cdot\bm{v}^1+\frac12m_2\bm{v}^2\cdot\bm{v}^2-U(r)-\gamma(t)z\,,
\end{equation}
where $m_1,m_2\in\RR$ are the masses of the particles (assumed to be constant), $U(r)$ is the central potential and $\gamma$ is a time-dependent function. The Lagrangian energy is
\begin{equation}
        E_L=\frac12m_1\bm{v}^1\cdot\bm{v}^1+\frac12m_2\bm{v}^2\cdot\bm{v}^2+U(r)+\gamma(t)z\,,
\end{equation}
and the cocontact structure is given by the one-forms
\begin{equation}
        \eta=\d z-m_1\bm{v}^1\cdot\d\bm{q}^1-m_2\bm{v}^2\cdot\d\bm{q}^2\,, \quad \tau=\d t\,.
\end{equation}
The evolution of the system is given by the Herglotz--Euler--Lagrange vector field $\sode_L$, with local expression~\eqref{eq:Herglotz_vf_time}.
Its solutions satisfy the Herglotz--Euler--Lagrange equations:
\begin{align}
       m_1\dot{\bm{v}}^1&=\bm{F}-\gamma(t) m_1\bm{v}^1\,,\label{eq:HEL2body1}
    \\
    m_2\dot{\bm{v}}^2 &= -\bm{F}-\gamma(t) m_2\bm{v}^2\,. \label{eq:HEL2body2}
\end{align}
Here the dot notation indicates time derivative and
\begin{equation}
        \bm{F}=-\frac{\d U}{\d r}\frac{\bm{r}}{r}
\end{equation}
is the force of the potential $U$. 

Proceeding as in the classical two-body problem, consider the evolution of the center of masses
\begin{equation}
        \bm{R}=\frac{m_1\bm{q}^1+m_2\bm{q}^2}{m_1+m_2}\,.
\end{equation}
Since $\sode_L$ is a \textsc{sode},
\begin{equation}
        \dot{\bm{R}} = \sode_L(\bm{R})=\frac{m_1\bm{v}^1+m_2\bm{v}^2}{m_1+m_2}\, .
\end{equation}
Then, 
\begin{equation}
        \sode_L(\dot{\bm{R}})=-\gamma\dot{\bm{R}}\, ,
\end{equation}
that is, every component of $\dot{\bm{R}}$ is a dissipated quantity. Along a solution of $\sode_L$, it evolves as
\begin{equation}
        \dot{\bm{R}}(t)=\dot{\bm{R}}_0\, e^{-\int \gamma(t)\d t}\,.
\end{equation}
In particular, if $\gamma$ is a positive constant, as the time increases the center of mass tends to move on a line with constant speed $\dot{\bm{R}}_0$. By \Cref{theorem:Noether_cocontact}, the corresponding generalized infinitesimal dynamical symmetries are $\bm{Y_{\dot{R}}}=X_{\dot{\bm{R}}}-R^L_t$, where $R^L_t$ is subtracted to every component. A short computation shows that
\begin{equation}
        \bm{Y_{\dot{R}}}=\frac{1}{m_1+m_2}\left(\frac{\partial}{\partial \bm{q}^1}+\frac{\partial}{\partial \bm{q}^2}\right)\,.
\end{equation}
Each component of $\bm{Y_{\dot{R}}}$ is an action dependent complete lift and $\liedv{\bm{Y_{\dot{R}}}}L=0$ therefore, they are infinitesimal natural symmetries of the Lagrangian.

The fact that the center of mass is moving in a very concrete way, may indicate that one could express the system using only the relative position. Indeed, equations~\eqref{eq:HEL2body1} and \eqref{eq:HEL2body2} imply that
\begin{equation}
        \mu\ddot{\bm{r}}=-\bm{F}-\gamma\mu\dot{\bm{r}}\,,
\end{equation}
where $\mu=\dfrac{m_1m_2}{m_1+m_2}$ is the reduced mass. This equation can also be derived from the cocontact Lagrangian system $(\RR^3, L_\mu)$, with
\begin{equation}
        L_\mu=\frac{1}{2}\mu\dot{\bm{r}}\cdot\dot{\bm{r}}-U(r)-\gamma z\, .
\end{equation}
The angular momentum is given by
\begin{equation}
        \bm{L}=\mu\bm{r}\times\dot{\bm{r}}\,,
\end{equation}
where $\times$ denotes the cross product. 
Each component is a dissipated quantity:
\begin{equation}
        \Gamma_{L}(\bm{L})=-\gamma\bm{L}\,.     
\end{equation}
The angular momentum along a solution of $\sode_L$ evolves as
\begin{equation}
        \bm{L}(t)=\bm{L}_0\, e^{-\int \gamma(t)\d t}\,.
\end{equation}
Since the direction of $\bm{L}$ remains constant, the movement takes place on a plane perpendicular to $\bm{L}_0$. If $\gamma$ is a positive constant, the angular momentum tends to zero as $t$ goes to infinity. The associated generalized infinitesimal dynamical symmetries are
\begin{equation}
        \bm{Y_{L}}=X_{\bm{L}}-R^L_t=\bm{r}\times\left(\frac{1}{m_2}\frac{\partial}{\partial \bm{q}^2}-\frac{1}{m_1}\frac{\partial}{\partial \bm{q}^1}\right)-\dot{\bm{r}}\times\left(\frac{1}{m_2}\frac{\partial}{\partial \bm{v}^2}-\frac{1}{m_1}\frac{\partial}{\partial \bm{v}^1}\right)\, .
\end{equation}
Each component of $\bm{Y_{L}}$ is an action dependent complete lift and $\liedv{\bm{Y_{L}}}L=0$, therefore they are infinitesimal natural symmetries of the Lagrangian.

Finally, the Lagrangian energy $E_L$ evolves as
\begin{equation}
        \sode_L(E_L)=-R_z^L(E_L)E_L+R_t^L(E_L)=-\gamma E_L+\dot{\gamma}z\,,
\end{equation}
and it is not a dissipated quantity due to the time-dependence of $\gamma$.


\chapter{Hamilton--Jacobi theory for time-dependent contact systems}\label{ch:contact_HJ}

In this chapter, a Hamilton--Jacobi theory for time-dependent contact Hamiltonian systems is developed. 
This is done by considering the cocontact manifold $(\RR \times \cT Q\times \RR, \dd t, \eta)$, and a vector bundle with $\RR \times \cT Q \times \RR$ as total space. There are two approaches depending on the base space considered for this vector bundle. In the so-called action-independent approach, the vector bundle is $\pi_Q^t\colon \RR \times \cT Q \times \RR\to \RR \times Q$, whereas in the action-dependent approach the vector bundle is $\pi_Q^{t,z}\colon  \RR \times \cT Q \times \RR\to \RR \times Q\times \RR$. These names are due to the dependence (or independence) of sections of the bundle on the ``$z$'' variable, which can be interpreted as the action.

The theory developed here is also useful for time-independent contact Hamiltonian systems. The action-independent approach for autonomous contact systems only covers the zero level set of the Hamiltonian function. However, this can be solved by considering the time-dependent Hamilton--Jacobi equation. It is also worth mentioning that the results obtained could also be applied locally to arbitrary cocontact manifolds. As a matter of fact, the existence of Darboux coordinates implies that, for any $(2n+2)$-dimensional cocontact manifold $(M, \tau, \tilde{\eta})$, there exists a local cocontactormorphism from $(M, \tau, \tilde{\eta})$ to $(\RR \times \cT Q\times \RR, \dd t, \eta)$, where $\dim Q = n$. The results of the present chapter were previously published in the article \cite{deLeon2023}.

Along the rest of the chapter, consider the $(2n+2)$-dimensional cocontact Hamiltonian system $(\RR \times \cT Q\times \RR, \dd t, \eta, H)$, with canonical coordinates $(t, q^i, p_i, z)$ such that $\eta = \dd z - p_i \dd q^i$. Let $X_f$ denote the Hamiltonian vector field of $f\in \Cinfty(\RR \times \cT Q\times \RR)$ with respect to $(\dd t, \eta)$. Denote the projections as in diagram~\eqref{eq:digram_projections_cocontact}, namely,
\begin{equation}\label{eq:digram_projections_cocontact_chapter_10}
    \begin{tikzcd}
        & \RR\times\cT Q\times\RR \arrow[dl, swap, "\rho_1"] \arrow[dr, "\rho_2"] \arrow[dd, "\pi"] & \\
        \RR\times\cT Q \arrow[dr, swap, "\pi_2"] & & \cT Q\times\RR \arrow[dl, "\pi_1"] \\
        & \cT Q &
    \end{tikzcd}
\end{equation}

\section{The action-independent approach}\label{sec:HJ_action-independent}

Consider the vector bundle $\RR \times \bigwedge\nolimits^k\cT Q \to \RR \times Q$ with the natural projection.
Let $\pr{\R}\colon \RR \times \bigwedge\nolimits^k\cT Q \to \RR$ and $\pr{\bigwedge\nolimits^k \cT Q}\colon \RR \times \bigwedge\nolimits^k\cT Q \to \bigwedge\nolimits^k\cT Q$ be the canonical projections.
Given a section $\alpha$ of $\RR \times \bigwedge\nolimits^k\cT Q \to \RR \times Q$, let 
\begin{equation}
    \begin{aligned}
        \alpha_{(t)} \colon  Q &\longrightarrow \bigwedge\nolimits^k\cT Q\\
        x &\longmapsto \pr{\bigwedge\nolimits^k \cT Q} (\alpha(t,x))\, ,
    \end{aligned}
\end{equation}
for each $t \in \RR$.

\begin{definition}
    The \emph{exterior derivative of $\alpha$ at fixed $t$} is the section of $\RR \times \bigwedge\nolimits^{k+1}\cT Q \to \RR \times Q$ given by
   \begin{equation}
       \d_Q \alpha(t, x) = (t, \d \alpha_{(t)} (x))\,.
   \end{equation}
\end{definition}

In bundle coordinates, for a function $f\in\Cinfty(\RR\times Q)$ and a section $\alpha(t,x) = (t,\alpha_i\d_x q^i)$ of the bundle $\RR \times Q\to \RR \times \bigwedge\nolimits^k\cT Q$, their exterior derivative at fixed $t$ read
\begin{equation}
    \begin{split}
        \d_Q f &= \left(t, \frac{\partial f}{\partial  q^i} \d_x q^i\right)\,,\\
        \d_Q \alpha &= \left(t, \frac{\partial \alpha_j}{\partial q^i} \d_x q^i \wedge \d_x q^j \right)\,.
    \end{split}
\end{equation}

\begin{definition}
     Given $f\in \Cinfty(\RR\times Q)$, the \emph{$1$-jet of $f$ at fixed $t$} is the section $j_t^1f$ of $\RR \times \cT Q \times \RR \to \RR \times Q$ given by
     \begin{equation}
        j_t^1 f =(\d_Q f, f)\,.
     \end{equation}
\end{definition}


Given a section $\gamma$ of the vector bundle $\pi_Q^t\colon \RR\times \T^*Q\times \RR\to \RR\times Q$, let $X_H^\gamma\in \X(\RR \times Q)$ denote the vector field given by  
\begin{equation}
    X_H^\gamma = \T \pi_Q^t \circ X_H \circ \gamma\, .
\end{equation}
The vector fields $X_H^\gamma$ and $X_H$ are said to be \emph{$\gamma$-related} if  
\begin{equation}\label{eq:gamma_related}
    X_H \circ \gamma = \T \gamma \circ X_H^\gamma\, , 
\end{equation}
in other words, if the following diagram commutes:
\begin{equation}
    \begin{tikzcd}
        \RR \times \cT Q \times \RR \arrow[d, "\pi_Q^t"] \arrow[r, "X_H"]
        & \T(\RR \times \cT Q\times \R) \arrow[d, swap, "\T\pi_Q^t"] \\
        \RR \times Q \arrow[r, "X_H^\gamma"] \arrow[u, "\gamma", bend left]     & \T (\RR \times Q) \arrow[u, swap, "\T \gamma", bend right]
    \end{tikzcd}
\end{equation}

\begin{theorem}[Action-independent Hamilton--Jacobi theorem]\label{theorem:action_indep_HJ}
    Let $\gamma$ be the 1-jet at fixed $t$ of a function $S\in \Cinfty(\RR \times Q)$. Then, the following statements are equivalent:
    \begin{enumerate}
        \item \label{theorem:action_indep_HJ_item1} for every integral curve $c\colon I \subseteq \RR\to \RR\times Q$ of $X_H^\gamma$, the curve $\gamma\circ c$ is an integral curve of $X_H$;
        \item \label{theorem:action_indep_HJ_item2} $X_H^\gamma$ and $X_H$ are $\gamma$-related,
        \item \label{theorem:action_indep_HJ_item3} $S$ is a solution of the partial differential equation
        \begin{equation} \label{eq:HJ_Sardonian}
            H \circ j^1_t S + \frac{\partial S}{\partial t} = 0\,.
        \end{equation}
    \end{enumerate}
\end{theorem}

Equation \eqref{eq:HJ_Sardonian} is called the \emph{action-independent Hamilton--Jacobi equation for $(\RR \times \cT Q\times \RR, \dd t, \eta, H)$}, and a solution $S$ is called a \emph{generating function for $(\RR \times \cT Q\times \RR, \dd t, \eta, H)$}. If the cocontact Hamiltonian system considered is clear, equation~\eqref{eq:HJ_Sardonian} will be simply called the \emph{action-independent Hamilton--Jacobi equation}, and $S$ a \emph{generating function}.

\begin{proof}
    The equivalence between statements \ref{theorem:action_indep_HJ_item1} and \ref{theorem:action_indep_HJ_item2} is straightforward. The equivalence between statements \ref{theorem:action_indep_HJ_item2} and \ref{theorem:action_indep_HJ_item3} can be proven in bundle coordinates as follows.
    For each $x\in \RR \times Q$, let $ \gamma(x) = \left(x, \gamma_i(x), S(x)\right)$.
    Therefore,
    \begin{equation}
    \begin{aligned}
        X_H \circ \gamma(x) &= \restr{\frac{\partial}{\partial t}}{x} 
        + \frac{\partial H}{\partial p_i} \big(\gamma(x)\big)  \restr{\frac{\partial }{\partial q^i}}{x} \\ 
        & \quad - \left( \frac{\partial H }{\partial q^i} \big(\gamma(x)\big) + \gamma_i(x) \frac{\partial H}{\partial z} \big(\gamma(x)\big)\right)   \restr{\frac{\partial }{\partial p_i}}{x} \\
        & \quad + \left(\gamma_i(x) \frac{\partial H}{\partial p_i}\big(\gamma(x)\big) - H \big(\gamma(x)\big)\right) \restr{\frac{\partial }{\partial z}}{x} \,,
    \end{aligned}
    \end{equation}
    and
    \begin{equation}
    \begin{aligned}
        \T\gamma \circ X_H^\gamma (x) &= \restr{\frac{\partial}{\partial t}}{x}  
        + \frac{\partial H}{\partial p_i} \big(\gamma(x)\big) \restr{\frac{\partial }{\partial q^i}}{x} \\
        & \quad +\left(\frac{\partial \gamma_i} {\partial t}(x) + \frac{\partial H} {\partial p_j} \big(\gamma(x)\big) \frac{\partial \gamma_j} {\partial q^i}(x) \right) \restr{\frac{\partial  } {\partial p_i}}{x} \\
        & \quad + \left(\frac{\partial S}{\partial t}(x) + \frac{\partial S}{\partial q^i}(x)\frac{\partial H}{\partial p_i}\big(\gamma(x)\big) \right) \restr{\frac{\partial }{\partial z}}{x} \, .
    \end{aligned}
    \end{equation}
    Thus, $X_H^\gamma$ and $X_H$ are $\gamma$-related if and only if
    \begin{equation}\label{Sardonian_generalized_HJ_local}
        \begin{aligned} 
            - \left( \frac{\partial H }{\partial q^i}\big(\gamma(x)\big)\ + \right. & \left.  \gamma_i(x) \frac{\partial H}{\partial z}\big(\gamma(x)\big) \right)\\
            =\ & \frac{\partial \gamma_i} {\partial t}(x) + \frac{\partial H} {\partial p_j}\big(\gamma(x)\big) \frac{\partial \gamma_j} {\partial q^i}(x) \, ,\\
            \gamma_i(x) \frac{\partial H}{\partial p_i}\big(\gamma(x)\big) -\ & H\big(\gamma(x)\big)\\
            =\ & \frac{\partial S}{\partial t}(x) + \frac{\partial S}{\partial q^i}(x)\frac{\partial H}{\partial p_i}\big(\gamma(x)\big)\, ,
        \end{aligned}
    \end{equation}
    for every $x\in \RR \times Q$.
    Since $\gamma$ is the 1-jet at fixed $t$ of $S$, one can write
    \begin{equation}
        \gamma_i = \frac{\partial S}{\partial q^i}\,,
    \end{equation}
    and therefore equations~\eqref{Sardonian_generalized_HJ_local} can be written as
        \begin{equation}
        \begin{aligned}
            - \left( \frac{\partial H }{\partial q^i}\big(\gamma(x)\big)\ + \right. & \left. + \frac{\partial S}{\partial q^i}(x) \frac{\partial H}{\partial z}\big(\gamma(x)\big) \right)\\
            =\ & \frac{\partial^2 S}{\partial t \partial q^i}(x) + \frac{\partial H} {\partial p_j}\big(\gamma(x)\big) \frac{\partial S}{\partial q^i \partial q^j}(x) \, ,
            \label{Sardonian_HJ_local_a}
        \end{aligned} 
        \end{equation}
        and
        \begin{equation}    
        \begin{aligned}
            \\
                \frac{\partial S}{\partial q^i}(x) \frac{\partial H}{\partial p_i}\big(\gamma(x)\big) -\ & H\big(\gamma(x)\big)\\
            =\ & \frac{\partial S}{\partial t}(x) + \frac{\partial S}{\partial q^i}(x)\frac{\partial H}{\partial p_i}\big(\gamma(x)\big)\, .
            \label{Sardonian_HJ_local_b}
        \end{aligned}    
        \end{equation}
        \label{Sardonian_HJ_local}
    Equation~\eqref{Sardonian_HJ_local_a} implies that
    \begin{equation}
        -\d_Q(H\circ \gamma)  = \d_Q \left(\parder{S}{t} \right) \,,
        \label{eq:HJ_Sardonian_redundant}
    \end{equation}
    while equation~\eqref{Sardonian_HJ_local_b} yields
    \begin{equation}\label{eq:HJ_Sardonian_proof}
        H \circ \gamma = - \frac{\partial S}{\partial t}\, .
    \end{equation}
    Clearly, equation~\eqref{eq:HJ_Sardonian_redundant} is implied by equation~\eqref{eq:HJ_Sardonian_proof}. Using that $\gamma = j^1_t S$ completes the proof.
\end{proof}

The condition of $\gamma$ to be the 1-jet at fixed $t$ of the generating function $S$ can be expressed in a more geometric manner as follows.

\begin{proposition}\label{proposition:Legendrian_jet}
    Let $\gamma$ be a section of $\pi_Q^t\colon \RR \times \cT Q \times \RR \to \RR\times Q$. Then, for every $t \in \RR$, the following assertions are equivalent:
    \begin{enumerate}
        \item $\Ima \gamma(t, \cdot)$ is a Legendrian submanifold of $(\RR \times \cT Q\times \RR, \dd t, \eta)$,
        \item $\Ima(\rho_2\circ \gamma (t, \cdot))$ is a Legendrian submanifold of $(\cT Q\times \RR, \eta_Q)$,
        \item $\gamma$ is the 1-jet at fixed $t$ of a function $f\in \Cinfty(\RR \times Q)$, namely, 
        \begin{equation}
            \gamma(t, q) = (j_t^1 f)(q) = (\d_Q f(t, q), f(t,q)) \,,
        \end{equation}
        for each $q\in Q$.
    \end{enumerate} 
\end{proposition}
\begin{proof}
    Consider a section $\gamma(t,q) = (t, \alpha(t,q),f(t,q))$ of $\pi_Q^t\colon \RR \times \cT Q \times \RR \to \RR\times Q$. For each $t\in \RR$, let $\gamma_t = \gamma(t, \cdot) \colon Q \to \RR \times \cT Q \times \RR$.
    Then,
    \begin{equation}
    \begin{aligned}
        \gamma_{t}^\ast \eta &= \gamma_{t}^\ast (\rho_2^\ast \eta_Q ) = (\rho_2 \circ \gamma_{t})^\ast \eta_Q\\
        &= f_{(t)}^\ast \dd z - \alpha_{(t)}^\ast \theta_Q 
        = \dd f_{(t)} - \alpha_{(t)} \,,
    \end{aligned}
    \end{equation}
    and $\gamma_{t}^\ast \dd t = 0$. Using that a submanifold $N$ of a co-oriented contact manifold $(M, \eta)$ (respectively, of a cocontact manifold $(M, \tau, \eta)$) is Legendrian if an only if $\restr{\eta}{\T N}=0$ and $\dim N = n$ (respectively, $\restr{\eta}{\T N}=0, \, \restr{\tau}{\T N} = 0$ and $\dim N = n$), the results follow. 
\end{proof}

In order to study the integrability of cocontact Hamiltonian systems, it is of interest to introduce the following.

\begin{definition}\label{def:complete_sols_indep}
    A \emph{complete solution of the action-independent Hamilton--Jacobi equation} is a function $S\in \Cinfty(\RR \times Q \times \RR^{n+1})$ such that
    \begin{enumerate}
        \item $S_\lambda=S(\cdot, \lambda)\in \Cinfty(\RR\times Q)$ is a generating function for each $\lambda\in \RR^{n+1}$,
        \item the map $\Phi\colon \RR\times Q \times \RR^{n+1} \to \RR \times \cT Q \times \RR$ given by
        \begin{equation}
            \Phi(\cdot, \lambda) = j^1_t S_\lambda \, ,
        \end{equation}
        is a local diffeomorphism.
    \end{enumerate}
\end{definition}

It is worth noting that complete solutions depend on $n+1$ real parameters, one extra parameter in comparison with the (co)symplectic case (see \Cref{sec:Hamilton-Jacobi}). In order to consider complete solutions depending on just $n$ parameters, it is necessary to introduce a different approach to the Hamilton--Jacobi problem for (co)contact Hamiltonian systems (see \Cref{sec:HJ_action-dependent}).

Let $\pi_a \colon \RR\times Q\times \RR^{n+1} \to \RR$ denote the canonical projection on the $a$-th component of $\RR^{n+1}$. 

\begin{theorem}\label{thm:complete_solution_involution_indep}
   Let $S\in \Cinfty(\RR \times Q \times \RR^{n+1})$ be a complete solution of the action-independent Hamilton--Jacobi equation. Around each point $x\in \RR\times Q$, there is a neighbourhood $U\subseteq \RR\times Q$ containing $x$ such that the map
   \begin{equation}
    \begin{aligned}
        \restr{\Phi}{U\times \RR^{n+1}}\colon U \times \RR^{n+1}&\to V = \Phi(U\times \RR^{n+1})\, ,\\
        (t, q, \lambda) & \mapsto j^1_t S_\lambda(q)
    \end{aligned}
    \end{equation}
   is a diffeomorphism. 
   Then, the $n+1$ functions $f_\alpha=\pi_\alpha\circ \restr{\Phi}{U\times \RR^{n+1}}^{-1}\in \Cinfty(V), \, \alpha\in \{1, \dotsc, n+1\}$, are conserved quantities. 
   
   However, these functions are not in involution, that is, $\{f_\alpha, f_\beta \}\neq 0$ for at least one pair $\alpha, \beta\in \{1, \dotsc, n+1\}$, where $\{\cdot, \cdot\}$ denotes the Jacobi bracket defined by $(\restr{\dd t}{V}, \restr{\eta}{V})$.
\end{theorem}


\begin{proof}
One can write 
\begin{equation}
    \Phi(U, \lambda)
    = \bigcap_{\alpha=1}^{n+1} \, f_\alpha^{-1}(\lambda_i)\,,
\end{equation}
for each $\lambda = (\lambda_1, \dotsc, \lambda_{n+1}) \in \RR^{n+1}$.
Since $X_H$ is tangent to any of the submanifolds $ \Phi(U, \lambda)$, every function $f_\alpha$ is a constant of the motion, namely, $X_H (f_\alpha) = 0$.
On the other hand, the Jacobi bracket of each pair of the functions $f_\alpha, f_\beta $ is given by
\begin{equation}
    \{f_\alpha, f_\beta \} = -\dd \eta \left(\sharp_{(\dd t,\, \eta)}\dd f_\alpha, \sharp_{(\dd t,\, \eta)}\dd f_\beta \right)  -f_\alpha \Rz(f_\beta ) + f_\beta  \Rz(f_\alpha) \, .
\end{equation}
By \Cref{proposition:Legendrian_jet}, $\Phi(U, \lambda)$ is a Legendrian submanifold of $(\RR\times \cT Q \times \RR, \dd t, \eta)$. Therefore, $\dd \eta$ vanishes on $\Phi(U, \lambda)$, and thus
\begin{equation}\label{eq:Jacobi_bracket_Legendrian}
    \{f_\alpha , f_\beta \} =  -f_\alpha  \Rz(f_\beta ) + f_\beta  \Rz(f_\alpha ) \, .
\end{equation}
By contradiction, one can proof that not all these functions are in involution.
Observe that
\begin{equation}
    \Phi^{-1} (t, q^i, p_i, z) = \big(t, q^i, f_\alpha(t, q^i, p_i, z)\big)\, .
\end{equation}
Since $\Phi^{-1}$ is a local diffeomorphism, $\T \Phi^{-1}$ has rank $2n+2$, and thus $\Rz(f_\delta) = \tparder{f_\delta}{z}$ does not vanish for at least one $\delta \in \{1, \ldots, n+1\}$. Then, $f_\alpha $ and $f_\delta$ are in involution if and only if
\begin{equation}
    f_\alpha  \Rz(f_\delta) = f_\delta \Rz(f_\alpha ) \, ,
\end{equation}
which implies that $\dd_x f_\alpha $ and $\dd_x f_\delta$ are linearly dependent for $x\in\Phi(U,0)\subseteq f_\alpha ^{-1}(0)\cap f_\delta^{-1}(0)$. Therefore, $\Phi^{-1}$ is not a submersion, and hence $\Phi$ cannot be a local diffeomorphism.

\end{proof}

Complete solutions of the Hamilton--Jacobi equations may be used to integrate the dynamics of the system as follows:
\begin{enumerate}
    \item Solve the Hamilton--Jacobi equation
    \begin{equation}
        H \circ j^1_t S_\lambda + \parder{S_\lambda}{t} = 0
    \end{equation}
    for arbitrary values of $\lambda \in \RR^{n+1}$. Let $\Phi_\lambda = j_t^1 S_\lambda$.
    \item Compute the integral curves $c \colon I \subseteq \RR \to \RR \times Q,\, c(t) = (t, q^i(t))$~of $X_H^{\Phi_\lambda}$, which are given by
    \begin{equation}\label{eq:integrate_HJ_cocontact}
        \frac{\d q^i}{\d t} = \restr{\frac{\partial H}{\partial p_i}}{\Phi_\lambda(U)}\, , 
    \end{equation}
    where the restriction to $\Phi_\lambda(U)$ means that one has to write $p_i=\tparder{S_\lambda}{q^i}$ and $z=S_\lambda$.
    \item The integral curves $\tilde c$ of $X_H$ on $\Phi_\lambda(U)$ are given by $\Phi_\lambda \circ c$, namely,
    \begin{equation}
       \tilde{c}(t) =\Phi_\lambda \circ c(t)= \left(c(t), \parder{S_\lambda}{q^i} \big(c(t)\big), S_\lambda \big(c(t)\big) \right)\, .
    \end{equation}
\end{enumerate}

It is worth noting that computing the integral curves of $X_H^\gamma$ is not always straightforward. 
However, there are some relevant cases in which it is particularly easy.

\begin{example}
    Consider $Q=\RR^n$, and the Hamiltonian function $H\in \Cinfty(\RR \times \cT \RR^n\times \RR)$ is of the form
    \begin{equation}
        H = \frac{\lVert{p}\rVert^2}{2m(t)} + V(t,q,z)\, ,
    \end{equation}
    where $\lVert{\cdot}\rVert$ denotes the Euclidean norm. Suppose that the generating function is \emph{separable}, that is, $S(t, q^1, q^2, \ldots,q^n) = S_0(t) + S_1(q^1)+\dotsb + S_n(q^n)$, where $S_0$ is a function depending exclusively on $t$, and $S_i$ is a function depending exclusively on $q^i$, for $i=1, \ldots, n$. Then, equations~\eqref{eq:integrate_HJ_cocontact} simplify to
    \begin{equation}
        \frac{\d q^i}{\d t} = \frac{1}{m(t)} S_i'(q^i)\, .
    \end{equation}
\end{example}

\subsection{Example: the free particle with time-dependent mass and a linear external force}
Let $Q=\RR$ and consider the Hamiltonian function
\begin{equation}
    H = \frac{p^2}{2m(t)} - \frac{\kappa}{m(t)} z,
\end{equation}
with $m$ a positive-valued function depending only on $t$, expressing the mass of the particle, and $\kappa$ a positive constant. 
Then, the action-independent Hamilton--Jacobi equation for $H$ is given by 
\begin{equation}
    \frac{1}{2m(t)}\left( \frac{\partial S}{\partial q}\right)^2 - \frac{\kappa}{m(t)} S(t,q) + \frac{\partial S}{\partial t} = 0\, ,
\end{equation}
that is,
    \begin{equation}
    \left( \frac{\partial S}{\partial q}\right)^2 - 2 \kappa S(t,q) + 2m(t)\frac{\partial S}{\partial t} = 0\, .
    \label{HJ_example_indep}
\end{equation}
Suppose that the generating function $S$ is separable, namely, $S(t,q)= \alpha(t)+\beta(q)$, with $\alpha$ a funtion depending only on $t$ and $\beta$ a function depending only on $q$. Then, equation~\eqref{HJ_example_indep} can be written as
\begin{equation}
        \left( \frac{\d \beta}{\d q}\right)^2 - 2 \kappa \alpha(t) - 2 \kappa \beta(q) + 2m(t)\frac{\d \alpha}{\d t} = 0\, ,
\end{equation}
which implies that
\begin{align}
    &  2m(t)\frac{\d \alpha}{\d t} -2 \gamma \alpha(t) = 0\,,\\
    & \left( \frac{\d \beta}{\d q}\right)^2 - 2 \kappa \beta(q) = 0\, .
\end{align}
The solutions of this system of ordinary differential equations are
\begin{equation}
    \alpha_{\lambda_1}(t) = \lambda_1 e^{\kappa \int_0^t \frac{1}{m(s)} \d s}, \,\qquad
    \beta_{\lambda_2}(q) = \left( \sqrt{\frac{\kappa}{2}} q + \lambda_2 \right )^2\, ,
\end{equation}
for any $\lambda_1, \lambda_2\in \RR$.
Therefore, the function $S\colon \RR \times Q \times \RR^2\to \RR$ defined by
\begin{equation}
    S(t, q, \lambda_1, \lambda_2) = \lambda_1 e^{\kappa \int_0^t \frac{1}{m(s)} \d s} + \left( \sqrt{\frac{\kappa}{2}} q + \lambda_2 \right )^2,
\end{equation}
is a complete solution of the action-independent Hamilton--Jacobi equation. Its associated local diffeomorphism  $\Phi\colon \RR \times Q \times \RR^2\to \RR \times \cT Q \times \RR$ is given by
\begin{equation}
\begin{aligned}
    \Phi(t,q,\lambda) 
    &= j_t^1 S_\lambda(t,q) \\
    &= \left(t, q, \sqrt{2\kappa}\left( \sqrt{\frac{\kappa}{2}} q + \lambda_2 \right),  \lambda_1 e^{\kappa \int_0^t \frac{1}{m(s)} \d s} + \left( \sqrt{\frac{\kappa}{2}} q + \lambda_2 \right )^2\right)\, ,
\end{aligned}
\end{equation}
where $\lambda=(\lambda_1, \lambda_2)$, and its inverse is 
\begin{equation}
    \Phi^{-1} \colon (t,q, p, z) \mapsto
    \left(t, q, e^{-\kappa \int_0^t \frac{1}{m(s)} \d s} \left(z - \frac{p^2}{2\kappa}  \right),\frac{p-\kappa q}{\sqrt{2\kappa}}\right)\, .
\end{equation}
Observe that, in this case, $\Phi$ is a global diffeomorphism.
Let $\pi_a \colon \RR\times Q\times \RR^{n+1} \to \RR$ denote the canonical projection on the $a$-th component of $\RR^{2}$, namely, $\pi_1(t, q, \lambda_1, \lambda_2) = \lambda_1$ and $\pi_2(t, q, \lambda_1, \lambda_2) = \lambda_2$. 
By \Cref{thm:complete_solution_involution_indep}, the functions $f_a = \pi_a \circ \Phi^{-1} \in \Cinfty(\RR \times \cT Q \times \RR)$, for $a=1, 2$, are conserved quantities. Their expressions in bundle coordinates are
\begin{equation}
    f_1 (t, q, p,z) = e^{-\kappa \int_0^t \frac{1}{m(s)} \d s} \left(z - \frac{p^2}{2\kappa} \right)\, ,
\end{equation}
and
\begin{equation}
    f_2 (t, q, p,z) = \frac{p-\kappa q}{\sqrt{2\kappa}}\, .
\end{equation}

The Hamiltonian vector field of $H$ is given by
\begin{equation}
    X_H 
    = \parder{}{t} + \frac{p}{m(t)}\parder{}{q} + \frac{\kappa p}{m(t)} \parder{}{p} + \left(\frac{p^2}{2m(t)} + \frac{\kappa}{m(t)} z\right)\parder{}{z}\,.
\end{equation}
One can check that $X_H(f_1)=X_H(f_2)=0$. Moreover,
\begin{equation}
    X_H^{\Phi_\lambda}
    = \restr{\parder{}{t} + \frac{p}{m(t)}\parder{}{q}}{\Ima \Phi_\lambda}
    = \parder{}{t} + \frac{\sqrt{2\kappa}\left( \sqrt{\frac{\kappa}{2}} q + \lambda_2 \right)}{m(t)}\parder{}{q}
    \,,
\end{equation}
whose integral curves $\sigma(t)=(t, q(t))$ are given by
\begin{equation}
    q(t) = e^{\int _1^t\frac{\kappa }{m(s)}\d s} \left(\int _1^t\frac{\sqrt{2\kappa} e^{-\int _1^u\frac{\kappa }{m(s)}\d s }  \lambda }{m(u)}\d u+c\right)\, ,
\end{equation}
where $c$ is a constant.
Then, the integral curves of $X_H$ along $\Ima \Phi_\lambda$ are given by $\Phi_\lambda \circ \sigma(t) = (t, q(t), p(t), z(t))$, where
\begin{equation}
    p(t) = \sqrt{2\kappa}\left( \sqrt{\frac{\kappa}{2}} q(t) + \lambda_2 \right)\, ,
\end{equation}
and
\begin{equation}
    z(t) = \lambda_1 e^{\kappa \int_0^t \frac{1}{m(s)} \d s} + \left( \sqrt{\frac{\kappa}{2}} q(t) + \lambda_2 \right)^2\, .
\end{equation}

\subsection{A new approach for the Hamilton--Jacobi problem in time-independent contact Hamiltonian systems}

Along this subsection, consider the time-independent contact Hamiltonian system $(\cT Q\times \RR, \eta_Q, H)$, with $\eta_Q$ the canonical contact form. The analogous of \Cref{theorem:action_indep_HJ} for autonomous contact Hamiltonian systems was developed in \cite{d.S2017a} (see also \cite{d.L.M2021}): 

\begin{theorem}[de León--Sardón]\label{thm_action_indep_autonomous}
    Let $X_H$ be the Hamiltonian vector field of $H$ with respect to $\eta_Q$. Consider a section $\gamma$ of $\pi_Q:  \cT Q\times \RR\to  Q$ such that $\Ima \gamma$ is a Legendrian submanifold of $(\cT Q\times \RR, \eta_Q)$. Then, $X_H^\gamma=\T \pi_Q \circ X_H \circ \gamma$ and $X_H$ are $\gamma$-related if and only if 
    \begin{equation}
        H\circ \gamma = 0\,. \label{eq:HJ_action_indep_autonomous}
    \end{equation}
\end{theorem}

The problem with this approach is that it cannot be used to completely integrate the system. Indeed, equation~\eqref{eq:HJ_action_indep_autonomous} implies that every integral curve of $X_H\circ \gamma$ is contained in $H^{-1}(0)$.
This can be solved by regarding the contact Hamiltonian system $(\cT Q\times \RR, \eta, H)$ as the cocontact Hamiltonian system $(\RR \times \cT Q\times \RR, \d t, \eta, {\widehat H})$, where {$\widehat H= H \circ \rho_2$ (that is, $\widehat H (t, q, p, z) = H(q,p, z)$),}  such that $\Rt({\widehat H})=0$ and making use of \Cref{theorem:action_indep_HJ}.
Suppose that $S\in \Cinfty(\RR \times Q)$ is of the form $S = \alpha\circ \pr{Q} + \beta \circ \pr{\RR}$ for some functions $\alpha\in \Cinfty(Q)$ and $\beta\in \Cinfty(\RR)$, where 
$\pr{\RR}\colon \RR\times Q \to \RR$ and $\pr{Q} \colon \RR\times Q \to Q$ denote the canonical projections. In other words, $S$ can be written as $S(t, q^1, \dotsc, q^n) = \alpha(q^1, \dotsc q^n) + \beta(t)$, where $\alpha$ is a function depending only on the coordinates $(q^i)$ and $\beta$ is a function depending only on $t$. Then, equation~\eqref{eq:HJ_Sardonian} yields
\begin{equation}
    H \circ j^1 \alpha + \frac{\dd \beta}{\dd t} = 0\, ,
\end{equation}
that is,
\begin{equation}
    H \left(q^i, \frac{\partial \alpha}{\partial q^i}, z\right) +  \frac{\dd \beta}{\dd t} = 0\, .
    \label{HJ_contact_as_cocontact}
\end{equation}
With a suitable choice of $\alpha$ and $\beta$, one can cover energy levels distinct from $H=0$.

\begin{definition}\label{def:complete_sols_indep_autonomous}
Let $(\T^*Q\times \RR, \eta_Q, H)$ be a contact Hamiltonian system, and let $(\RR\times \cT Q\times \RR, \dd t, \eta, \widehat H = H\circ \rho_2)$ be its associated cocontact Hamiltonian system.
A \emph{complete solution of the time-dependent and action-independent Hamilton--Jacobi equation} for $(\T^*Q\times \RR, \eta_Q, H)$ is a function $S\in \Cinfty(\RR \times Q \times \RR^{n})$ such that:
\begin{enumerate}
    \item For each $\lambda\in \RR^{n}$, the function $S_\lambda=S(\cdot, \lambda)\in \Cinfty(\RR\times Q)$ is a generating function for $(\RR\times \cT Q\times \RR, \dd t, \eta, \widehat H = H\circ \rho_2)$.
    \item The map $\Phi = \rho_2 \circ \widehat{\Phi} \colon \RR\times Q \times \RR^{n} \to  \cT Q \times \RR$ is a local diffeomorphism, where $\widehat{\Phi}\colon \RR\times Q \times \RR^{n} \to \RR \times \cT Q \times \RR$ is given by
    \begin{equation}
        \widehat{\Phi}(\cdot, \lambda) = j^1_t S_\lambda \, .
    \end{equation}
\end{enumerate}

\end{definition}
Let $\pi_a \colon \RR\times Q\times \RR^{n} \to \RR$ denote the canonical projection on the $a$-th component of $\RR^{n}$. Around each point $x\in \RR\times Q$, there is a neighbourhood $U\subseteq \RR\times Q$ containing $x$ such that the map
\begin{equation}
 \begin{aligned}
     \restr{\Phi}{U\times \RR^{n}}\colon U \times \RR^{n}&\to V = \Phi(U\times \RR^{n})\subseteq\cT Q \times \RR\, ,\\
     (t, q, \lambda) & \mapsto \rho_2\circ j^1_t S_\lambda(q)
 \end{aligned}
 \end{equation}
is a diffeomorphism. 
Define the $n$ functions $f_i=\pi_i\circ \restr{\Phi}{U\times \RR^{n+1}}^{-1}\in \Cinfty(V), \, i\in \{1, \dotsc, n\}$.
Then,
\begin{equation}
    \Phi_\lambda(U\times \RR^n) = \bigcap_{i=1}^n f_i^{-1} (\lambda_i)\, ,
\end{equation}
and
\begin{equation}
    \widehat{\Phi}(U\times \RR^n) = \bigcap_{i=1}^n (f_i\circ \rho_2)^{-1} (\lambda_i)\, .
\end{equation}
This implies that the functions $f_i\circ \rho_2\in \Cinfty(\RR \times V)$ are constants of the motion for $\restr{\widehat H}{\RR \times V}$, and thus the functions $f_i$ are constants of the motion for $\restr{H}{V}$.
\begin{example}[The free particle with a linear external force]
    Consider the contact Hamiltonian system $(\cT \RR \times \RR, \d t, \eta_{\RR}, H)$, where
    \begin{equation}
        H = \frac{p^2}{2} - \kappa z\, ,
    \end{equation}
    with $\kappa$ a positive constant. Let $\widehat H = H \circ \rho_2$ be the associated time-dependent Hamiltonian.
    Then, the action-independent Hamilton--Jacobi equation for $\widehat H$ is given by 
    \begin{equation}
        \frac{1}{2}\left( \frac{\partial S}{\partial q}\right)^2 - \kappa S(t,q) + \frac{\partial S}{\partial t} = 0\, ,
    \end{equation}
    that is,
     \begin{equation}
       \left( \frac{\partial S}{\partial q}\right)^2 - 2 \kappa S(t,q) + 2 \frac{\partial S}{\partial t} = 0\, ,
        \label{HJ_example_indep_auton}
    \end{equation}
    Suppose that the generating function $S$ is separable, namely, $S(t,q)= \alpha(t)+\beta(q)$. Then, equation~\eqref{HJ_example_indep_auton} can be written as
    \begin{equation}
         \left( \frac{\d \beta}{\d q}\right)^2 - 2 \kappa \alpha(t) - 2 \kappa \beta(q) + 2\frac{\d \alpha}{\d t} = 0\, ,
    \end{equation}
    which yields
   \begin{align}
       &  2\frac{\d \alpha}{\d t} -2 \kappa \alpha(t) = 0\,,\\
       & \left( \frac{\d \beta}{\d q}\right)^2 - 2 \kappa \beta(q) = 0\, .
   \end{align}
   The solutions of these ordinary differential equations are
   \begin{equation}
       \alpha(t) = e^{\kappa t}\,, \qquad
       \beta_{\lambda}(q) = \left( \sqrt{\frac{\kappa}{2}} q + \lambda \right )^2\, .
   \end{equation}
   Therefore, 
   \begin{equation}
        S_\lambda(t,q) = S(t,q, \lambda) = e^{\kappa t} + \left( \sqrt{\frac{\kappa}{2}} q + \lambda \right )^2
   \end{equation}
   is a complete solution of the time-dependent and action-independent Hamilton--Jacobi equation for $(\cT \RR\times \RR, \eta_\RR, H)$. 
   Hence, 
   \begin{equation}
        \widehat\Phi(t,q,\lambda) = j_t^1 S_\lambda(t,q) = \left(t, q, \sqrt{2\kappa}\left( \sqrt{\frac{\kappa}{2}} q + \lambda \right),  e^{\kappa t} + \left( \sqrt{\frac{\kappa}{2}} q + \lambda \right )^2\right)\, ,
   \end{equation}
   and the map $\Phi= \rho_2 \circ \widehat{\Phi} \colon \RR \times \RR \times \RR \to \cT \RR \times \RR$ reads
   \begin{equation}
        \Phi\colon (t, q,\lambda) \mapsto \left(q, \sqrt{2\kappa}\left( \sqrt{\frac{\kappa}{2}} q + \lambda \right),  e^{\kappa t} + \left( \sqrt{\frac{\kappa}{2}} q + \lambda \right )^2\right)\, ,
   \end{equation}
   Its inverse is given by
   \begin{equation}
       \Phi^{-1} \colon (q, p, z) \mapsto
       \left(\frac{1}{\kappa}\log \left|{z-\frac{p^2}{2\kappa}}\right|, q, 
       \frac{p-\kappa q}{\sqrt{2\kappa}}
       \right)\, .
   \end{equation}
   In this case, $\Phi$ is a global diffeomorphism.
   Let $\pi\colon \RR \times \RR \times \RR\to \RR$ denote the canonical projection on the right component, namely, 
   $\pi(t, q, \lambda) = \lambda$. Then, 
   \begin{equation}
       f_1(q,p,z) = \pi\circ \Phi^{-1}(q,p,z) =  \frac{p-\kappa q}{\sqrt{2\kappa}}
   \end{equation}
   is a conserved quantity.

  The Hamiltonian vector field of $H$ with respect to $\eta_Q$ is given by
   \begin{equation}
       X_H  = p\parder{}{q} + \kappa p \parder{}{p} + \left(\frac{p^2}{2} + \kappa z\right)\parder{}{z}\,.
   \end{equation}
   One can check that $X_H(f_1)=0$. Moreover,
    \begin{equation}
       X_H^{\Phi_\lambda}
     = \restr{ p\parder{}{q}}{\Ima \Phi_\lambda}
     = \sqrt{2\kappa}\left( \sqrt{\frac{\kappa}{2}} q + \lambda \right)\parder{}{q}
     \,,
   \end{equation}
   whose integral curves $\sigma(t)=(t, q(t))$ are given by
   \begin{equation}
       q(t) = c e^{\kappa  t} -\sqrt{\frac{2}{\kappa}}\lambda\, ,
   \end{equation}
   where $c$ is a constant.
   Then, the integral curves of $X_H$ along $\Ima \Phi_\lambda$ are given by $\Phi_\lambda \circ \sigma(t) = (q(t), p(t), z(t))$, where
   \begin{equation}
       p(t) = \sqrt{2\kappa}\left( \sqrt{\frac{\kappa}{2}} q(t) + \lambda \right)
            = \kappa\, c e^{\kappa  t}\, ,
   \end{equation}
   and
   \begin{equation}
       z(t) =  e^{\kappa t} + \left( \sqrt{\frac{\kappa}{2}} q(t) + \lambda \right)^2
      =  e^{\kappa t} +  \frac{\kappa}{2} c^2 e^{2\kappa  t} \, .
   \end{equation}
\end{example}

\section{The action-dependent approach}\label{sec:HJ_action-dependent}



In the previous section, a Hamilton--Jacobi theory for time-dependent contact Hamiltonian systems has been introduced. In particular, this approach has been shown to be useful for studying time-independent contact Hamiltonian systems, where time is used as a free parameter. Nevertheless, this approach has a couple of drawbacks. First, complete solutions depend on $n+1$ parameters, instead of the $n$ parameters that are required for symplectic Hamiltonian systems (see \Cref{sec:Hamilton-Jacobi}).
Additionally, time-independent solutions only cover the zero-energy level. 

In order to solve these problems, in this section an alternative approach is proposed, considering solutions of the Hamilton--Jacobi problem depending on the action variable $z$.

Let $\pr{\bigwedge\nolimits^k \cT Q}: \RR \times \bigwedge\nolimits^k\cT Q \times \mathbb{R} \to \bigwedge\nolimits^k\cT Q$ denote the canonical projection.
Given a section $\alpha: \RR \times Q \times \mathbb{R} \to \RR \times \bigwedge\nolimits^k\cT Q \times \mathbb{R}$, let 
\begin{equation}
    \begin{aligned}
        \alpha_{(t,\, z)} :  Q &\longrightarrow \bigwedge\nolimits^k\cT Q\\
        x &\longmapsto \pr{\Lambda^k \cT Q} (\alpha(t,x,z))\, ,
    \end{aligned}
\end{equation}
for each $t, z\in \RR$.

\begin{definition}
    The \emph{exterior derivative of $\alpha$ at fixed $t$ and $z$} is the section of $\RR \times \bigwedge\nolimits^{k+1}\cT Q \times \mathbb{R} \to \RR \times Q \times \mathbb{R}$ given by
   \begin{equation}
       \d_Q \alpha(t, x, z) = (t, \d \alpha_{(t,\, z)} (x), z)\,.
   \end{equation}
\end{definition}

The coisotropic condition can be written in local coordinates as follows.

\begin{lemma}
    Assume that an $(n+2)$-dimensional submanifold $N$ of a $(2n+2)$-dimensional cocontact manifold $(M, \tau, \eta)$ is locally the zero set of the constraint functions $\left\{\phi_a\right\}_{a=1,\dotsc,n}$.
    Then, $N$ is coisotropic if and only if the following equation holds in Darboux coordinates:
     \begin{equation}\label{eq:coisotropic_coords}
        \left(\parder{\phi_a}{q^i} + p_i \parder{\phi_a}{z}\right)\parder{\phi_b}{p_i} - 
         \left(\parder{\phi_b}{q^i} + p_i\parder{\phi_b}{z}\right)\parder{\phi_a}{p_i} = 0\, .
    \end{equation}
\end{lemma}
\begin{proof}
     Assume that $(M,\tau, \eta)$ is a $(2n+2)$-dimensional cocontact manifold. Let $N\hookrightarrow M$ be a $k$-dimensional submanifold locally given as the zero set of the functions $\phi_a:U\to \RR$, with $a\in \{1, \ldots, 2n+2-k\}$. Then, the Jacobi orthogonal complement of $\T N$ is given by
    \begin{equation*}
        {\T N}^{\perp_{\Lambda}} = \left\langle\left\{Z_a\right\}_{a=1,\dotsc,2n+2-k}\right\rangle\, ,
    \end{equation*}
    where
    \begin{equation*}
        Z_a = \hat \Lambda(\d \phi_a)
        = \left(\parder{\phi_a}{q^i} + p_i \parder{\phi_a}{z}\right)\parder{}{p_i} - 
        \parder{\phi_a}{p_i} \left(\parder{}{q^i} + p_i\parder{}{z}\right)\,.
    \end{equation*}    
    Therefore, $N$ is coisotropic if and only if $Z_a(\phi_b)=0$ for all  $a,b$, which in Darboux coordinates yields 
    equation~\eqref{eq:coisotropic_coords}.
\end{proof}

\begin{proposition}\label{prop:cosiotropic_local}
    Let $\gamma$ be a section of $\RR\times \cT Q \times \mathbb{R}$ over $\RR \times Q \times \mathbb{R}$. Then $\Ima \gamma$ is a coisotropic submanifold if and only if
    \begin{equation}
        \frac{\partial \gamma_i} {\partial q^j} + \gamma_j \frac{\partial \gamma_i} {\partial z} 
        = \frac{\partial \gamma_j} {\partial q^i} + \gamma_i \frac{\partial \gamma_j} {\partial z}.
        \label{eq_coisotropic_coords}
    \end{equation}
\end{proposition}
\begin{proof}
    Equation~\eqref{eq_coisotropic_coords} is obtained by applying the previous result to the submanifold $N = \Ima \gamma$, which is locally defined by the constraints $\phi_i = p_i - \gamma_i$.
\end{proof}

Given a section $\gamma$ of the bundle $\pi_Q^{t,z}: \RR\times \T^*Q\times \RR\to \RR\times Q \times \RR$, let $X_H^\gamma \in \X(\RR \times Q \times \RR)$ denote the vector field defined by
\begin{equation}
    X_H^\gamma = \T \pi_Q^{t, z} \circ X_H \circ \gamma\, .
\end{equation}
The vector fields $X_H^\gamma$ and $X_H$ are said to be \emph{$\gamma$-related} if  
\begin{equation}
    X_H \circ \gamma = \T \gamma \circ X_H^\gamma\, , 
\end{equation}
in other words, if the following diagram commutes:
\begin{equation}
    \begin{tikzcd}
        \RR \times \cT Q \times \RR \arrow[d, "\pi_Q^{t,z}"] \arrow[r, "X_H"]                           & \T(\RR \times \cT Q\times \R) \arrow[d, swap, "\T\pi_Q^{t,z}"]                       \\
        \RR \times Q \times \RR \arrow[r, "X_H^{\gamma}"] \arrow[u, "\gamma", bend left] & \T(\RR \times Q \times \R) \arrow[u, swap, "\T \gamma", bend right]
    \end{tikzcd}
\end{equation}

Note that $\Ima \gamma $ is $(n + 2)$-dimensional, so it no longer makes sense to require it to be Legendrian. Instead, it will be required to be coisotropic.

\begin{theorem}[Action-dependent Hamilton--Jacobi Theorem]
    Let $\gamma$ be a section of $\pi_Q^{t,z}: \RR\times \T^*Q\times \RR\to \RR\times Q\times \RR$ such that $\Ima \gamma$ is a coisotropic submanifold of $(\RR\times \T^*Q\times \RR, \dd t, \eta)$. 
    Then, the following assertions are equivalent:
    \begin{enumerate}
        \item \label{theorem:action_dep_HJ_item1} for every integral curve $c\colon I \subseteq \RR\to \RR\times Q$ of $X_H^\gamma$, the curve $\gamma\circ c$ is an integral curve of $X_H$;
        \item \label{theorem:action_dep_HJ_item2} $X_H^\gamma$ and $X_H$ are $\gamma$-related,
        \item \label{theorem:action_dep_HJ_item3} $\gamma$ is a solution of the partial differential equation
        \begin{equation} \label{eq:HJ_Lainzian}
            \d_Q \left(H\circ \gamma  \right) + \frac{\partial}{\partial z} (H\circ \gamma) \gamma + \liedv{\Rt} \gamma = (H\circ \gamma) \liedv{\frac{\partial}{\partial z}} \gamma\, .
        \end{equation}
    \end{enumerate}
\end{theorem}

Equation \eqref{eq:HJ_Lainzian} is called the \emph{action-dependent Hamilton--Jacobi equation for $(\RR\times \T^*Q\times \RR, \dd t, \eta, H)$}.

\begin{proof}
    The equivalence between statements \ref{theorem:action_dep_HJ_item1} and \ref{theorem:action_dep_HJ_item2} is self-evident. By using bundle coordinates, the equivalence between statements \ref{theorem:action_dep_HJ_item2} and \ref{theorem:action_dep_HJ_item3} can be proven in the following manner. Suppose that $\gamma$ is locally written as 
    \begin{equation}
        \gamma(t, q^i, z) = \big(t, q^i, \gamma_i(t, q^i, z), z\big)\, . 
    \end{equation}
    Then, for each point $x\in \RR \times Q \times \RR$, 
    \begin{equation}
    \begin{aligned}
        X_H \circ \gamma (x) &= \restr{\frac{\partial}{\partial t}}{x}
        + \frac{\partial H}{\partial p_i} \big(\gamma(x)\big) \restr{\frac{\partial }{\partial q^i}}{x}\\
        & \quad - \left( \frac{\partial H }{\partial q^i}\big(\gamma(x)\big) + \gamma_i(x) \frac{\partial H}{\partial z}\big(\gamma(x)\big) \right) \restr{\frac{\partial }{\partial p_i}}{x} \\
        & \quad + \left(\gamma_i(x) \frac{\partial H}{\partial p_i}\big(\gamma(x)\big) -H\big(\gamma(x)\big)\right) \restr{\frac{\partial }{\partial z}}{x}\,,
    \end{aligned}
    \end{equation}
    and
    \begin{equation}
    \begin{aligned}
       \T\gamma \circ X_H^\gamma &= \restr{\frac{\partial}{\partial t}}{x} 
        + \frac{\partial H}{\partial p_i}\big(\gamma(x)\big) \restr{\frac{\partial }{\partial q^i}}{x} \\
        & \quad + \left(\frac{\partial \gamma_i} {\partial t}(x) + \frac{\partial H} {\partial p_j}\big(\gamma(x)\big) \frac{\partial \gamma_i} {\partial q^j} (x) 
        \right. \\ & \quad \left.
        + \left(\gamma_j(x) \frac{\partial H} {\partial p_j}\big(\gamma(x)\big)- H\big(\gamma(x)\big)  \right)\frac{\partial \gamma_i} {\partial z}(x)  \right) \restr{\frac{\partial  } {\partial p_i}}{x} \\
        & \quad + \left(\gamma_i(x) \frac{\partial H}{\partial p_i}\big(\gamma(x)\big) -H\big(\gamma(x)\big)\right) \restr{\frac{\partial }{\partial z}}{x}\, ,
    \end{aligned}
    \end{equation}
    Thus, $X_H^\gamma$ and $X_H$ are $\gamma$-related if and only if
    \begin{equation}\label{pre_HJ_Lainzian}
    \begin{aligned}
      - \frac{\partial H }{\partial q^i}\big(\gamma(x)\big) & - \gamma_i(x) \frac{\partial H}{\partial z}\big(\gamma(x)\big) 
      = \frac{\partial \gamma_i} {\partial t}(x) + \frac{\partial H} {\partial p_j}\big(\gamma(x)\big) \frac{\partial \gamma_i} {\partial q^j} (x) \\
      & \quad + \left(\gamma_j(x) \frac{\partial H} {\partial p_j}\big(\gamma(x)\big)- H\big(\gamma(x)\big)  \right)\frac{\partial \gamma_i} {\partial z}(x)
    \end{aligned}
    \end{equation}
    for every $x\in \RR \times Q \times \RR$.
    Taking into account that $\Ima \gamma$ is coisotropic, \Cref{prop:cosiotropic_local} implies that
    \begin{equation}
    \begin{aligned}
        \frac{\partial H }{\partial q^i}\big(\gamma(x)\big)  &+ \frac{\partial H} {\partial p_j}\big(\gamma(x)\big) \frac{\partial \gamma_j} {\partial q^i} (x)
        \\ & \quad
        + \gamma_i(x) \left(\frac{\partial H} {\partial p_j}\big(\gamma(x)\big) \frac{\partial \gamma_j} {\partial z}(x)  +\frac{\partial H}{\partial z}\big(\gamma(x)\big) \right)
        \\ & \quad
            + \frac{\partial \gamma_i} {\partial t}(x)
        = H\big(\gamma(x)\big) \frac{\partial \gamma_i} {\partial z}(x) \, .
    \end{aligned}
    \end{equation}
    The global expression for this equation is \eqref{eq:HJ_Lainzian}.
\end{proof}

\begin{definition}
Let $(\RR\times \T^*Q\times \RR, \dd t, \eta, H)$ be a cocontact Hamiltonian system.
A \emph{complete solution of the action-dependent Hamilton--Jacobi problem} for $(\RR\times \T^*Q\times \RR, \dd t, \eta, H)$ is a local diffeomorphism $\Phi\colon \RR\times Q \times \RR^n \times \RR\to \RR \times \cT Q \times \RR$ such that, for each $\lambda \in \RR^n$,
\begin{equation}
\begin{aligned}
    \Phi_\lambda \colon \RR\times Q \times \R&\longrightarrow \RR \times\cT Q \times \RR  \\
    \left(t, q^i, z\right) &\longmapsto \Phi\left(t, q^i, \lambda, z\right) 
\end{aligned}    
\end{equation}
is a solution of the action-dependent Hamilton--Jacobi problem for $(\RR\times \T^*Q\times \RR, \dd t, \eta, H)$.
\end{definition}

Let $\pi_i\colon \RR\times Q\times \mathbb{R}^n  \to \RR$ denote the canonical projection on the $i$-th component of $\RR^n$. Consider the functions $f_i=\pi_i\circ \alpha\circ \Phi^{-1}$ on $\RR \times \cT Q \times \RR$, so that the following diagram commutes:
\begin{equation}
    \begin{tikzcd}
        \RR \times Q \times \RR^n \times \RR \arrow[r, "\Phi", shift left] \arrow[d, "\alpha"] & \RR\times \cT Q \times \RR \arrow[l, "\Phi^{-1}", shift left] \arrow[d, "f_i"] \\
        \RR^n \arrow[r, "\pi_i"] & \R
    \end{tikzcd}
\end{equation}

\begin{theorem}\label{thm:complete_solution_involution}
    Let $\Phi\colon \RR\times Q \times \RR^n \times \RR\to \RR \times \cT Q \times \RR$ be a complete solution of the action-dependent Hamilton--Jacobi problem for $(\RR\times \T^*Q\times \RR, \dd t, \eta, H)$. Around each point $x\in \RR\times Q \times \RR^n$, there is a neighbourhood $U$ such that $\restr{\Phi}{U}\colon U \to V = \Phi(U)$ is a diffeomorphism. The functions $f_i = \pi_i \circ \restr{\Phi}{U}^{-1}\in \Cinfty(V),\, i\in \{1, \dotsc, n\}$ are constants of the motion. However, they are not in involution, that is, $\{f_i, f_j\}\neq 0$, for at least one pair $i\in \{1, \dotsc, n\}$, where $\{\cdot, \cdot\}$ is the Jacobi bracket defined by $(\restr{\dd t}{V}, \restr{\eta}{V})$. 
\end{theorem}

\begin{proof}
    Observe that
    \begin{equation}
        \Phi(U, \lambda) = \bigcap_{i=1}^n \, f_i^{-1}(\lambda_i)\,,
    \end{equation}
    The fact that $X_H$ is tangent to any of the submanifolds $\Phi(U, \lambda)$ implies that $X_H (f_i) = 0$ for every $i=1, \ldots, n$.
    Moreover, the Jacobi bracket of each pair of these functions is given by
    \begin{equation}
        \{f_i, f_j\} = \Lambda (\d f_i, \d f_j) - f_i \Rz (f_j) + f_j \Rz (f_i)\,,
    \end{equation}
    but
    \begin{equation}
        \Lambda (\d f_i, \d f_j) = \hat\Lambda(\d f_i)(f_j) = 0\, ,
    \end{equation}
    since $(\T \Ima \Phi_\lambda)^{\perp_\Lambda} = \hat\Lambda ((\T \Ima \Phi_\lambda)^\circ) \subset \T \Ima \Phi_\lambda$. Therefore, 
    \begin{equation}
        \{f_i, f_j\} = - f_i \Rz(f_j) + f_j \Rz(f_i)\,.
    \end{equation}
    By means of an analogous argument to the one used in the proof of \Cref{thm:complete_solution_involution_indep}, one concludes that $\{f_i, f_j\}$ cannot vanish for every $i, j= 1, \ldots, n$.
\end{proof}

From a complete solution of the Hamilton--Jacobi problem one can reconstruct the dynamics of the system. If $\sigma$ is an integral curve of the vector field $X_H^\gamma$, then $\Phi_\lambda \circ \sigma$ is an integral curve of $X_H$, thus recovering the dynamics of the original system.


\subsection{Example 1: freely falling particle with linear dissipation}

Consider a particle of time-dependent mass $m(t)$, with $m$ a positive-valued function of $t$, which is freely falling and subject to a dissipation linear in the velocity with proportionality constant $\gamma$. The Hamiltonian function $H:\RR\times \T^*\RR \times \RR\to \RR$ is given by
\begin{equation}
    H(t, q,p,z) = \frac{p^2}{2m(t)} + m(t)g q + \frac{\gamma}{m(t)} z\,,
\end{equation}
where the constant $g$ corresponds to the gravity. The Hamiltonian vector field of $H$ reads
\begin{equation}
\begin{aligned}
    X_H & = \parder{}{t} + \frac{p}{m(t)}\parder{}{q} - \left( m(t)g + \frac{\gamma}{m(t)} p \right)\parder{}{p}\\ 
    & \quad + \left(\frac{p^2}{2m(t)} - m(t)g q - \frac{\gamma}{m(t)} z\right)\parder{}{z}\,. 
\end{aligned}
\end{equation}
Its integral curves $c(r) = (t(r), q(r), p(r), z(r))$ satisfy the system of differential equations
\begin{equation}
\begin{array}{ll}
    \dot t = 1\,,\qquad & \dot q = \frac{p}{m(t)}\,,\\
    \dot p = -m(t)g - \frac{\gamma}{m(t)}p\,, & \dot z = \frac{p^2}{2m(t)} - m(t)g q - \frac{\gamma}{m(t)} z\,.
\end{array} 
\end{equation}
Combining the second and third equations yields
\begin{equation}
    \frac{\d}{\d t}(m(t)\dot q) = -m(t) g - \gamma\dot q\,.
\end{equation}
A conserved quantity can be use solve the action-dependent Hamilton--Jacobi equation. For the sake of simplicity, consider a conserved quantity $f$ which does not depend on the coordinates $q$ or $z$. In other words, $f$ is a solution to the partial differential equation 
\begin{equation}
    \parder{f}{t} - \left( m(t)g + \frac{\gamma}{m(t)} p \right)\parder{f}{p} = 0 \, .
\end{equation}
One can verify that
\begin{equation}
    f(t,q,p,z)=e^{\int _1^t\frac{\gamma }{m(s)}\d s} \left(p + g e^{-\int _1^t\frac{\gamma }{m(s)}\d s} \int _1^te^{\int _1^u\frac{\gamma}{m(s)}\d s}  m(u)\d u\right)
\end{equation}
is a solution to this partial differential equation, and hence it is a conserved quantity. Thus, it is possible to express the momentum $p$ as a function of $t$ and a real parameter $\lambda$, namely, 
\begin{equation}
    P(t,\lambda)=e^{-\int _1^t\frac{\gamma }{m(s)}\d s}\left(\lambda - ge^{-\int _1^t\frac{\gamma }{m(s)}\d s} \int _1^t e^{\int _1^u\frac{\gamma }{m(s)}\d s} m(u)\d u\right)\, .
\end{equation}
Hence, a complete solution $\Phi\colon \RR \times Q \times \RR \times \RR \to \RR \times \cT \RR \times \RR$ of the action-dependent Hamilton--Jacobi equation for $H$ is given by
\begin{equation}
\begin{aligned}
    \Phi(t, q, z, \lambda)
    & = \left(t, q,P(t,\lambda), z \right)\,. \\
    & = \left(t, q, e^{-\int _1^t\frac{\gamma }{m(s)}\d s}\left(\lambda - ge^{-\int _1^t\frac{\gamma }{m(s)}\d s} \int _1^t e^{\int _1^u\frac{\gamma }{m(s)}\d s} m(u)\d u\right), z \right)\,.
\end{aligned}
\end{equation}
In this case, equation \eqref{eq_coisotropic_coords} holds trivially, and therefore $\Ima \Phi_\lambda$ is coisotropic.


\subsection{Example 2: damped forced harmonic oscillator}

Consider the contact Hamiltonian system $(\RR \times \cT \RR \times \RR, \d t, \eta_{\RR}, H)$, where
\begin{equation}
    H(t,q,p,z) = \frac{p^2}{2m} + \frac{k}{2}q^2 - qF(t) + \frac{\gamma}{m} z \, .
\end{equation}
This system describes a harmonic oscillator with elastic constant $k$, friction coefficient $\gamma$ and subjected to an external time-dependent force $F(t)$ \cite{d.G.G+2022}.

The Hamiltonian vector field reads
\begin{equation}
\begin{aligned}
    X_H & = \parder{}{t} + \frac{p}{m}\parder{}{q} + \left( -kq + F(t) - \frac{p}{m}\gamma \right)\parder{}{p} \\
    & \quad + \left( \frac{p^2}{2m} - \frac{k}{2}q^2 + qF(t) - \frac{\gamma}{m} z \right)\parder{}{z}\,,
\end{aligned}
\end{equation}
Its integral curves $c\colon I\subseteq \RR \times \cT \RR \times \RR, \, c(r)=(t(r),q(r),p(r),z(r))$ satisfy
\begin{equation}
\begin{array}{ll}
    \dot t = 1\,,\qquad & \dot q = \frac{p}{m}\,,\\
    \dot p = -kq + F(t) - \frac{p}{m}\gamma\,, & \dot z = \frac{p^2}{2m} - \frac{k}{2}q^2 + qF(t) - \frac{\gamma}{m} z\,.
\end{array}
\end{equation}
Combining the second and the third equations above, yields the second-order differential equation
\begin{equation}
    m\ddot q + \gamma \dot q + kq = F(t) \,,
\end{equation}
which corresponds to a damped forced harmonic oscillator. One can verify that the function
\begin{equation}
\begin{aligned}
    g(t,q,p,z) & =  e^{\frac{\gamma  t}{2 m}} \left(\frac{\sinh \left(\frac{\kappa t}{2 m}\right) (2 k m q+\gamma  p)}{\kappa}+p \cosh \left(\frac{\kappa t}{2 m}\right)\right)\\
    & \quad -\int _1^t F(s) e^{\frac{\gamma  s}{2 m}} \left(\cosh \left(\frac{\kappa s}{2 m}\right)+\frac{\gamma  \sinh \left(\frac{\kappa s}{2 m}\right)}{\kappa}\right)\d s\, ,
\end{aligned}
\end{equation}
is a conserved  quantity, where $\kappa = \sqrt{\gamma^2 - 4km}$.
It is worth noting that, since $\sinh = x +\mathcal{O}(x^3)$ and $\cosh x=1+\mathcal{O}(x^2)$ near $x=0$, $\sinh(ix)=i\sin x$ and $\cosh (ix)=\cos x$, the equation above is well-defined and real-valued for any of $\kappa\in \mathbb{C}$.
Thus, it is possible to write $p$ in terms of $t,q,z$ and a real parameter $\lambda$ as
\begin{equation}
\begin{aligned}
    P(t,q, \lambda, z)
    & = \frac{e^{-\frac{\gamma  t}{2 m}}}{\gamma  \sinh \left(\frac{\kappa t}{2 m}\right)+\kappa \cosh \left(\frac{\kappa t}{2 m}\right)}\\
    & \quad \cdot \left(\kappa \int _1^te^{\frac{s \gamma }{2 m}} F(s) \left(\cosh \left(\frac{\kappa s}{2 m}\right)+\frac{\gamma  \sinh \left(\frac{\kappa s}{2 m}\right)}{\kappa}\right)\d s
    \right. \\ & \qquad \left.
    -2 k m q e^{\frac{\gamma  t}{2 m}} \sinh \left(\frac{\kappa t}{2 m}\right)+\kappa \lambda \right)\, .
\end{aligned}
\end{equation}
Consequently, a complete solution $\Phi\colon \RR \times Q \times \RR \times \RR \to \RR \times \cT \RR \times \RR$ of the action-dependent Hamilton--Jacobi equation for $H$ is given by
\begin{equation}
    \Phi\colon (t, q, \lambda, z) \mapsto \left(t, q, P\left(t, q,\lambda, z\right), z \right)\, .
\end{equation}
Obviously equation \eqref{eq_coisotropic_coords} is satisfied, hence $\Ima \Phi_\lambda$ is coisotropic. 

\chapter{Liouville--Arnol'd theorem for contact Hamiltonian systems}\label{ch:contact_integrability}

\insquote{Confused is of course the best state a mathematician can be in; the struggle out of that state is the primary drive for progress.}
{\emph{Dror Bar-Natan's Research Statement for 2005}}

The present chapter is devoted to the integrability of contact Hamiltonian systems. A Liouville--Arnol'd theorem for these systems is proven. More specifically, it is shown that, given a $(2n+1)$-dimensional completely integrable contact system, one can construct a foliation by $(n+1)$-dimensional Abelian Lie groups and induce action-angle coordinates in which the equations of motion are linearized. One important novelty with respect to previous attempts is that the foliation consists of $(n+1)$-dimensional coisotropic submanifolds given by the preimages of rays by the functions in involution. In order to prove the theorem, firstly a Liouville--Arnol'd theorem for homogeneous functions on exact symplectic manifolds is shown (which is a result of independent interests). Then, the symplectization is used to translate the problem from a co-oriented contact manifold into an exact symplectic manifold. 

The results of this chapter have been previously published in the preprint \cite{Colombo2023}.

\section[Liouville--Arnol'd theorem on exact symplectic manifolds]{Liouville--Arnol'd theorem for homogeneous functions on exact symplectic manifolds}\label{sec:theorem_exact_symp}

Let $(M, \theta)$ be an exact symplectic manifold with Liouville vector field $\Delta$ and symplectic form $\omega = - \dd \theta$ (see \Cref{subsec:exact_symplectic}). Hereinafter, it will be assumed that the exact symplectic manifolds have no \emph{singular points}, that is, those in which $\Delta$ (or, equivalently, $\theta$) vanish. Notice that those points form a closed subset which is contained on a submanifold of dimension at most half of the dimension of the original manifold. In the case that $M= \cT  Q$ is a cotangent bundle with $\theta_Q$ the canonical one-form and $\Delta_Q$ the canonical Liouville vector field, the zero section would be removed. Recall that, if $Q$ has local coordinates $(q^i)$ and $(q^i, p_i)$ are the induced bundle coordinates on $\cT Q$, then $\theta_Q = p_i \dd q^i$ and $\Delta_Q = p_i \partial/\partial p_i$.

\begin{definition}\label{def:integrable_systems_homogeneous}
    A \emph{homogeneous integrable system} is a triple $(M, \theta, F)$, where $(M, \theta)$ is an exact symplectic manifold and $F=(f_1, \ldots, f_n)\colon M\to \RR^{n}$ is a map such that the functions $f_1, \ldots, f_n$ are in independent, in involution and homogeneous of degree $1$ on $M$. The functions $f_1, \ldots, f_n$ are called \emph{integrals}.
\end{definition}

In other words, a homogeneous integrable system is a completely integrable system whose symplectic form is exact and such that the integrals are homogeneous of degree~$1$.

If there are singularities, such as critical points of the integrals, on a zero measure subset $S$ of $M$, it suffices to restrict to the dense open subset $M_{0} = M\setminus S$.

\begin{theorem}[Liouville--Arnol'd theorem for homogeneous functions on exact symplectic manifolds]
\label{theorem:LA_exact_symp}
    Let $(M, \theta, F)$ be an $2n$-dimensional homogeneous integrable system with $F=(f_1, \ldots, f_n)$. Given $\Lambda\in \RR^n$, assume that $M_\Lambda = F^{-1}(\Lambda)$ is connected (if not, replace $M_\Lambda$ by a connected component).
    Let $U$ be an open neighbourhood of the level set $M_\Lambda=F^{-1}(\Lambda)$ (with $\Lambda\in \RR^n$) such that:
    \begin{enumerate}
        \item $f_1, \dotsc, f_n$ have no critical points in $U$,
        \item the Hamiltonian vector fields $X_{f_1}, \dotsc, X_{f_n}$ are complete,
        \item the submersion $F\colon U \to \R^n$ is a trivial bundle over a domain $V\subseteq \R^n$.
    \end{enumerate}
    Then, $U$ is diffeomorphic to $W=\TT^k \times \R^{n-k}\times V$, where $\TT^k$ denotes the $k$-dimensional torus. Furthermore, there is a chart $(\hat{U}\subseteq U; y^i, A_i)$ of $M$
    such that:
    \begin{enumerate}
        \item $A_i = M_i^j f_j$, where $M_i^j$ are homogeneous functions of degree $0$ depending only on $f_1, \ldots, f_n$,
        \item $(y^i, A_i)$ are canonical coordinates for $\theta$, that is, $ \theta = A_i \dd y^i$,
        \item the Hamiltonian vector fields of the functions $f_i$ read 
        \begin{equation}
            X_{f_i} = N_i^j \frac{\partial}{\partial y^j}\, ,
        \end{equation}
        with $(N_i^j)$ the inverse matrix of $(M_i^j)$.
    \end{enumerate}
    In this chart, the integral curves of $X_{f_j}$ are given by
    \begin{equation}
    \begin{aligned}
        & \dot y^i = \Omega^i (A_1, \ldots, A_n)\, ,\\
    & \dot A_i = 0\,  ,
    \quad i= 1,\ldots, n\, ,
    \end{aligned}
    \end{equation}
    where $\Omega^i\coloneqq N^i_j$.
\end{theorem}

The coordinates $(y^i)$ and $(A_i)$, for $i=1,\dotsc n$, are called \emph{angle coordinates} and \emph{action coordinates}, respectively.



In order to prove \Cref{theorem:LA_exact_symp} some lemmas will be required.

\begin{lemma}\label{lemma:Arnold_tangent_fields}
    Let $M$ be an $n$-dimensional differentiable manifold and let $X_1, \ldots, X_n\in \X(M)$ be linearly independent vector fields. If these vector fields are pairwise commutative and complete, then $M$ is diffeomorphic to $\TT^k\times \R^{n-k}$ for some $k\leq n$, where $\TT^k$ denotes the $k$-dimensional torus.
\end{lemma}
See Lemmas 1.3 and 1.4 in reference \cite{B.F2004} for the proof.



\begin{lemma}\label{lemma:linear_combination_functions}
    Let $(M, \theta)$ be an $2n$-dimensional exact symplectic manifold. Suppose that the functions $f_i,\ i=1,\ldots, n,$ on $M$ are functionally independent, in involution and homogeneous of degree 1. Assume that their Hamiltonian vector fields $X_{f_i}$ are complete. Then, there exists $n$ functions $g_i = M^j_i f_j \in \Cinfty(M)$ which verify the same assumptions as $f_i$ and such that:
    \begin{enumerate}
        \item $M^j_i$ for $i, j\in {1, \ldots, n}$ are homogeneous functions of degree $0$ and they depend only on $f_1, \ldots, f_n$,
        \item $X_{g_1}, \ldots, X{g_k}$ are infinitesimal generators of $\Sp^1$-actions and their flows have period 1,
        \item $X_{g_{k+1}}, \ldots, X_{g_n}$ are infinitesimal generators of $\RR$-actions.
    \end{enumerate}
    Here $k$ is the dimension of the isotropy subgroup from the action generated by the Hamiltonian vector fields $X_{f_i}$. 
\end{lemma}

\begin{proof}
    The restriction of the vector fields $X_{f_1}, \ldots, X_{f_n}, \Delta$ to each level set $M_\Lambda=F^{-1}(\Lambda)$ of $F=(f_1, \ldots, f_n)$ generate the Abelian Lie algebra
    \begin{equation}
    \begin{aligned}
         \mathfrak{g} & = \operatorname{Lie} (\TT^k \times \RR^{n-1}) \oplus \operatorname{Lie}(\RR^\times) \\
         & \cong \operatorname{Lie} (\Sp^1) \oplus \cdots\oplus  \operatorname{Lie} (\Sp^1) \oplus \operatorname{Lie} (\RR) \oplus \cdots \oplus \operatorname{Lie} (\RR) \oplus \operatorname{Lie}(\RR^\times)\, , 
    \end{aligned}
    \end{equation}
    where there are $k$ copies of $\operatorname{Lie} (\Sp^1)$ and $n-k$ copies of $\operatorname{Lie} (\RR)$. Hence, one can choose new generators given by linear combinations of $\restr{X_{f_i}}{M_\Lambda}, \restr{\Delta}{M_\Lambda}$ which are adapted to the direct sum. More specifically, there exist $Y_1, \ldots, Y_n \in \X(M_\Lambda)$ with $Y_i = M^j_i \restr{X_{f_j}}{M_\Lambda}$, for some real parameters $M^j_i$, such that
    \begin{itemize}
        \item for $1\leq i \leq k$ the vector field $Y_i$ is the generator of the $i$-th copy of $\operatorname{Lie} (\Sp^1)$,
        \item for $k+1\leq i \leq n$ the vector field $Y_i$ generates the $(i-k)$-th copy of $\operatorname{Lie} (\RR)$,
        \item $[Y_i, \Delta]=0$ for all $i\in \{1, \ldots, n\}$. 
    \end{itemize}
    Without loss of generality, the generators of $\operatorname{Lie} (\Sp^1)$ can be assumed to have flows with period $1$. (If that is not the case it suffices to rescale $Y_1, \ldots, Y_k$.) 
    Since the level sets $M_{\Lambda}$ foliate $M$, these vector fields can be extended to the whole manifold, namely,
    \begin{equation}
        Y_i = M^j_i (f_1, \ldots, f_n) X_{f_j}\in \X(M)
    \end{equation}
    where $M^j_i$ are functions which depend only on $f_1, \ldots, f_n$. 

    Since the functions $f_i$ are homogeneous of degree $1$ their Hamiltonian vector fields are infinitesimal homogeneous symplectomorphisms. In other words, $\theta$ is $\mathfrak{g}$-invariant. Hence, $Y_1, \ldots, Y_n$ are infinitesimal homogeneous symplectomorphisms. By \Cref{proposition:homogeneous_Hamiltonian}, each vector field $Y_i=X_{g_i}$ is the Hamiltonian vector field of
    \begin{equation}
        g_i = \theta (Y_i) = M^j_i f_j\, ,
    \end{equation}
    which is a homogeneous function of degree $1$. Therefore, $M^j_i$ are homogeneous functions of degree $0$. The fact that 
    \begin{equation}
        X_{\{h_i, h_j\}_\theta} = - [X_{h_i}, X_{h_j}]\, ,
    \end{equation}
    for all $h_i, h_j\in \Cinfty(M)$, implies that the functions $g_1, \ldots, g_n$ are in involution. Finally, taking into account that the functions $M^j_i$ are fixed on each $M_\Lambda$ and $Y_i$ are tangent to $M_{\Lambda}$, the completeness of $Y_i$ follows from the completeness of $X_{f_i}$.  
\end{proof}

\begin{lemma}\label{lemma:global_section}
    Let $\pi \colon P\to M$ be a $G$-principal bundle over a connected and simply connected manifold. Suppose there exists a connection one-form $A$ such that the horizontal distribution $\mathrm{H}$ is integrable. Then $\pi \colon M\to N$ is a trivial bundle and there exists a global section $\chi \colon N\to M$ such that $\chi^\ast A = 0$.
\end{lemma}

\begin{proof}
    Let $m=\dim P$ and $n=\dim M$. Since $\mathrm{H}$ is integrable, around each point $p\in P$ there is a chart $(V; x^1, \ldots, x^m)$ such that the integral manifolds of $\mathrm{H}$ are given by the equations $x^{n+1}=c_{n+1}, \ldots, x^m = c_m$, where $c_{n+1}, \ldots, c_m$ are constants. 
    Without loss of generality, consider the leave with $c_a = 0$ for every $a$.
    Denoting by $(x^1, \ldots, x^n)$ the local coordinates in $U= \pi(V)$ around $x=\pi(p)$, one can define a local section $\chi\colon U \to P_u = \pi^{-1}(U)$ given by $\chi(x^1, \ldots, x^n) = (x^1, \ldots, x^n, 0, \ldots, 0)$. Therefore, the image of $\chi$ is the leave of the foliation that goes through $p$ in $P_U$, and a submanifold of $P$ whose tangent space at each point is the horizontal space at that point, namely, $\T_{\chi(x)} \Ima \chi = \mathrm{H}_{\chi(x)}$. 

    Let $N_p$ denote the leave that goes through $p$ from the foliation of $P$ by integrable manifolds of $\mathrm{H}$. Then, $N_p \cap \pi^{-1}(x)\cap V = \{\chi(x)\} = \{p\}$. Therefore, $N_p \cap \pi^{-1}(x)$ is totally disconnected. Observe that $\restr{\pi}{N_p}$ is a local diffeomorphism onto its image. The local inverses look like the section $\chi$ constructed above.
    
    Since $M$ is a connected manifold, it is path-connected. Given a fixed point $x=\pi(p)$, let $\gamma_y\colon [0,1] \to M$ be the smooth curve with $\gamma_y(0)=x$ and $\gamma_y(1)=y$ for each $y\in M$. There exists a unique horizontal lift $\tilde{\gamma}_y\colon [0,1]\to P$ such that $\tilde{\gamma}_y(0)=p$ and the velocity vectors of $\tilde{\gamma}_y$ are horizontal, namely, $\dot{\tilde{\gamma}}_y(t) \in \mathrm{H}_{\tilde{\gamma}_y(t)}$ for all $t\in [0,1]$ (see Theorem 5.8.2 in \cite{Hamilton2017}). The fact that $\restr{\pi}{N_p}(\tilde{\gamma}_y(1)) = y$ for an arbitrary point $y\in M$ implies that $\restr{\pi}{N_p}\colon N_p \to M$ is surjective.

    To show that $\restr{\pi}{N_p}\colon N_p \to M$ is injective, suppose that there exists a $q\neq p$ in $N_p \cap \pi^{-1}(x)$. Since $N_p$ is a connected submanifold of $P$, there exists a path $\gamma$ in $N_p$ connecting $p$ with $q$. The simple connectedness of $M$ makes $\restr{\pi}{N_p}(\gamma)$ homotopic to the point $x$. The homotopy lifting property implies that $\gamma$ is homotopic to $p$, contradicting the fact that $N_p \cap \pi^{-1}(x)$ is totally disconnected.

    Since $\restr{\pi}{N_p}$ is an injective and surjective local diffeomorphism, it is a global diffeomorphism from $N_p$ to $M$. As a consequence, $\chi = \restr{\pi}{N_p}^{-1}\colon M \to N_p \subset P$ is a global section of $\pi\colon P \to M$. This implies that $\pi\colon P \to M$ is a trivial bundle and $\chi^\ast A = 0$. 
\end{proof}

\begin{proof}[Proof of \Cref{theorem:LA_exact_symp}]
    By \Cref{lemma:Arnold_tangent_fields}, $M_\Lambda$ is diffeomorphic to $\TT^k \times \RR^{n-k}$. Without loss of generality, assume that $X_{f_1}, \ldots, X_{f_k}$ are infinitesimal generators of $\Sp^1$-actions whose flows have period 1, and that $X_{g_{k+1}}, \ldots, X_{g_n}$ are infinitesimal generators of $\RR$-actions (see \Cref{lemma:linear_combination_functions}). 
    
    Let $\Ldist = \ker \theta$ and $\overline{U} = \left\{x\in U \mid f_i (x)\neq 0 \ \forall\, i \text{ and } \theta (x) \neq 0\right\}$. By hypothesis, $F\colon U \to V$ is a trivial bundle. Hence, $U\cong V \times \TT^k \times \RR^{n-k}$ can be endowed with a Riemannian metric $g$ given by the product of flat metrics in $V\subseteq \RR^n,\, \TT^k$ and $\RR^{n-k}$. The metric $g$ is flat and invariant by the Lie group action of $\TT^k \times \RR^{n-k}$. Consider the distribution
    \begin{equation}
        \Ldist^\theta = \big(\Ldist \cap \gen{X_{f_i}}_{i=1}^n\big)^{\perp_g} \cap \Ldist\, ,
    \end{equation}
    where $\perp_g$ denotes the orthogonal complement with respect to the metric $g$. This distribution is:
    \begin{enumerate}
        \item invariant by the Lie group action of $\TT^k \times \RR^{n-k}$,
        \item contained in $\Ldist$,
        \item complementary to the vertical bundle.
    \end{enumerate} 
    Indeed,
    \begin{equation}
        \ker \T F = \bigcap_{i=1}^n \ker \dd f_i = \gen{X_{f_i}}_{i=1}^n\, ,
    \end{equation}
    and 
    \begin{equation}
        \Ldist^\theta_x \oplus \gen{X_{f_i}(x)}_{i=1}^n = \T_x M\, ,
    \end{equation}
    for every $x\in \overline{U}$. Moreover, $F\colon \overline{U} \to \overline{U}/(\TT^k \times \RR^{n-k})$ is a principal bundle and $\Ldist^\theta$ is a principal connection with connection one-form $\theta$. In addition, the fact that $\theta\wedge \dd \theta = 0$ implies that $\Ldist$ is integrable. The fact that the orthogonal complement of an integrable distribution with respect to a flat metric is integrable implies that $\Ldist^\theta$ is integrable. Let $\hat{U}\subseteq \overline{U}$ be an open subset of $\overline{U}$ such that $\hat{V} = F(\hat{U})$ is simply connected. By Lemma~\ref{lemma:global_section}, there exists a global section $\chi$ of $F\colon \hat{U} \to \hat{V} \cong \hat{U}/(\TT^k \times \RR^{n-k})$ such that $\chi^\ast \theta = 0$. 

    Let $\Phi\colon \TT^{k} \times \RR^{n-k} \times M \to M$ denote the Abelian Lie group action defined by the flows of $X_{f_i}$. For each point $x\in M_{\Lambda} = F^{-1}(\Lambda)$, the angle coordinates $(y^i(x))$ are determined by
    \begin{equation}
        \Phi\big(y^i(x), \chi(F(x))\big) = x\, .
    \end{equation} 
    Notice that $(y^i, f_i)$ are coordinates in $\hat{U}$ adapted to the foliation of $M$ in $M_\Lambda$. 
    In these coordinates,
    \begin{equation}
        \theta = A_i \dd y^i + B^i \dd f_j\, , \quad X_{f_i} = \frac{\partial}{\partial y^i}\, ,
    \end{equation}
    for some functions $A_i$ and $B^i$. Contracting $\theta$ with $X_{f_i}$ yields $A_i = f_i$. 

    Finally, notice that the image of the section $\chi$ is given by the intersection of level sets of angle coordinates $y^i$, namely, $\Ima \chi = \cap_{i=1}^n (y^i)^{-1}(\mu_i)$. Hence, 
    \begin{equation}
        0 = \chi^\ast \theta = B^i \dd f_i\, .
    \end{equation}
    Since $\mu_i$'s are arbitrary, the functions $B^i$ are identically zero on all the manifold $M$. In conclusion, $\theta = f_i \dd y^i$. 
\end{proof}

The process for constructing the action-angle coordinates can be retrieved from the proof of the theorem.

\begin{remark}[Construction of action-angle coordinates]\label{remark:construction_coords_exact}
    In order to construct action-angle coordinates in a neighbourhood $U$ of $M_{\Lambda}$, one has to carry out the following steps:
    \begin{enumerate}
        \item Fix a section $\chi$ of $F\colon U \to V$ such that $\chi^\ast \theta = 0$.
        \item Compute the flows $\phi_t^{X_{f_i}}$ of the Hamiltonian vector fields $X_{f_i}$.
        \item Let $\Phi\colon \R^{n}\times M \to M$ denote the action of $\R^{n}$ on $M$ defined by the flows, namely,
        \begin{equation}
            \Phi(t_1, \ldots, t_n; x) = \phi_{t_1}^{X_{f_1}}\circ \cdots \circ\phi_{t_n}^{X_{f_n}}(x)\, .
        \end{equation}
        \item It is well-known that the isotropy subgroup $G_{\chi(\Lambda)(\Lambda)}=\{g\in \R^{n}\mid \Phi(g,\chi(\Lambda))=\chi(\Lambda)\}$, forms a lattice (that is, a $\mathbb{Z}$-submodule of $\R^{n}$). Pick a $\mathbb{Z}$-basis $\{e_1,\ldots, e_m\}$, where $m$ is the rank of the isotropy subgroup.
        \item Complete it to a basis $\mathcal{B}=\{e_1, \ldots, e_m, e_{m+1}, \ldots, e_{n}\}$ of $\R^{n}$.
        \item Let $(M_i^j)$ denote the matrix of change from the basis $\{X_{f_i}(\chi(\Lambda))\}$ of $\T_{\chi(\Lambda)} M_\Lambda\simeq \RR^{n}$ to the basis $\{\partial/\partial y^i\}$. The action coordinates are the functions $A_i = M_i^j f_j$.
        \item The angle coordinates $(y^i)$ of a point $x\in M$ are the solutions of the equation
        \begin{equation}
             x = \Phi\big(y^i e_i;\chi \circ F(x)\big)\, .
        \end{equation}
    \end{enumerate}
 \end{remark}

\section{Liouville--Arnol'd theorem for contact Hamiltonian systems}\label{sec:main_theorem}

\begin{definition}\label{def:integrable_systems}
    A \emph{completely integrable contact system} is a triple $(M, \eta, F)$, where $(M, \eta)$ is a co-oriented contact manifold and 
    \begin{equation}
        F=(f_0, \ldots, f_n)\colon M\to \RR^{n+1}
    \end{equation}
    is a map such that the functions $f_0, \ldots, f_n$ are in involution and $\T F$ has rank at least $n$ on a dense open subset $M_{0}\subseteq M$. The functions $f_0, \ldots, f_n$ are called \emph{integrals}.
\end{definition}

See \Cref{rem:characterization_integrable_systems} in \Cref{sec:coisotropic} for alternative characterizations of completely integrable contact systems.

Henceforth, it will be assumed that $M_0=M$. Since $M_0$ is a dense open set, one can remove the points in $M\setminus M_0$ if it is necessary.

For each $\Lambda\in \R^{n+1}\setminus\{0\}$, let $\ray{\Lambda}$ denote the ray generated by $\Lambda$, namely, 
\begin{equation}
    \ray{\Lambda} = \{x \in \R^{n+1}\mid \exists\, r\in\R_+ \colon x=r \Lambda\}\, .
\end{equation} 
Consider the subsets $M_\Lambda = F ^{-1}(\Lambda)$
and $ M_{\ray{\Lambda}} = F^{-1}(\ray{\Lambda})$ of $M$.

\begin{theorem}[Liouville--Arnol'd theorem for contact Hamiltonian systems]\label{thm:main_theorem}
    Let $(M, \eta, F)$ be a completely integrable contact system, where $F=(f_0, \ldots, f_n)$.
    Assume that the Hamiltonian vector fields $X_{f_0}, \ldots, X_{f_n}$ are complete. 
    Given $\Lambda \in \R^{n+1}\setminus\{0\}$,
    let $B \subseteq \RR^{n+1}\setminus\{0\}$ be an open neighbourhood of $\Lambda$. 
    Let $\pi\colon U\to M_{\ray{\Lambda}}$ be a tubular neighbourhood of $M_{\ray{\Lambda}}$ such that $\restr{F}{U}\colon U \to B$ is a trivial bundle over a domain $V \subseteq B$.
    Then:
    
    \begin{enumerate}
        \item The submanifold $M_{\ray{\Lambda}}$ is coisotropic, invariant by the Hamiltonian flow of $f_\alpha$, and diffeomorphic to $\mathbb{T}^k\times \R^{n+1-k}$ for some $k\leq n$. In particular, if $M_\Lambda$ is compact, then $k=n$. 
        \item There exist coordinates $(y^0, \ldots, y^n, \tilde A_1, \ldots, \tilde A_n)$ on $U$ such that the Hamiltonian vector fields of the functions $f_\alpha$ read
        \begin{equation}
            X_{f_\alpha} = \overline{N}_\alpha^\beta X_{f_\beta}\, ,
        \end{equation}
        where $\overline{N}_\alpha^\beta$ are functions depending only on $\tilde{A}_1, \ldots, \tilde{A}_n$.
        Hence, the integral curves of $X_{f_\beta}$ are given by
        \begin{equation}\label{eq:Hamilton_action_angle_symp}
            \begin{array}{ll}
                \dot y^\alpha = \Omega^\alpha(\tilde A_1, \ldots, \tilde A_n)\, ,\quad & \alpha=0, \ldots, n\, ,\\
                \dot{\tilde{A}}_i = 0\, ,  &i=1, \ldots, n\, ,
            \end{array}
        \end{equation}
        where $\Omega^\alpha\coloneqq \overline{N}^\alpha_\beta$.
        \item There exists a nowhere-vanishing function $A_0\in \Cinfty(U)$ and a conformally equivalent contact form $\tilde \eta = \eta/A_0$ such that $(y^i, \tilde A_i, y^0)$ are Darboux coordinates for $(M, \tilde \eta)$, namely, $\tilde{\eta} = \dd y^0 - \tilde{A}_i \dd y^i$.
    \end{enumerate}
\end{theorem}

The coordinates $(y^\alpha)$ and $(\tilde{A}_i)$, for $\alpha=0,\dotsc, n$ and $i=1,\dotsc, n$, are called \emph{angle coordinates} and \emph{action coordinates}, respectively. In physical applications, one of the integrals is regarded as the Hamiltonian function (which represents the energy of the system), and equations \eqref{eq:Hamilton_action_angle_symp} are the equations of motion.

There are functions $A_0, \ldots, A_n\in \Cinfty(U)$ such that the action coordinates are given by $\tilde{A_i}=-A_i/A_0\, (i=1, \ldots, n)$. These functions are the projection of the action coordinates from the symplectization of the completely integrable contact system (see the proof below for more details). In terms of these functions and the angle coordinates $(y^i)$, the original contact form reads $\eta = A_i \dd y^i$.


The proof of \Cref{thm:main_theorem} will consist of the following steps:
\begin{enumerate}
    \item Symplectize the completely integrable contact system $(M, \eta, F)$, obtaining the homogeneous integrable system $(M^\Sigma, \theta, F^\Sigma)$.
    \item Ensure that sufficient conditions hold for $F$ so that the Liouville--Arnol'd theorem can be applied to $(M^\Sigma, \theta, F^\Sigma)$.
    \item Apply the Liouville--Arnol'd theorem for homogeneous functions on exact symplectic manifolds (\Cref{theorem:LA_exact_symp}).
    \item Construct action-angle coordinates for $(M, \eta, F)$ from the ones for $(M^\Sigma, \theta, F^\Sigma)$.
\end{enumerate}

It is worth remarking that, in order to obtain coordinates on $M$ from the ones on $M^\Sigma$, the coordinates on $M^\Sigma$ must be homogeneous functions of degree 1. This is not guaranteed if one applies the Liouville--Arnol'd theorem for non-compact invariant submanifolds by Fiorani, Giachetta and Sardanashvily \cite{F.G.S2003,F.G.S2003a}. \Cref{theorem:LA_exact_symp} has to be utilised instead.

The symplectization relates dissipated quantities with respect to $(M, \eta, H)$ and conserved quantities with respect to $(M^\Sigma, \theta, H^\Sigma)$. As a matter of fact, an immediate consequence of \Cref{thm:symp_functions} is the following.
\begin{proposition}\label{proposition:dissipated_conseved}
    Let $(M, \eta, H)$ be a contact Hamiltonian system. Suppose that $\Sigma\colon M^\Sigma \to M$ is a symplectization such that $\theta = \sigma (\Sigma^\ast \eta)$ is the symplectic potential on $M^\Sigma$. Then, a function $f$ on $M$ is an dissipated quantity with respect to $(M, \eta, H)$ if and only if $f^\Sigma$ is a conserved quantity with respect to $(M^\Sigma, \theta, H^\Sigma)$.
\end{proposition}

Additionally, there is a one-to-one correspondence between homogeneous integrable systems and completely integrable contact systems.

\begin{proposition}
    Let $(M, \eta)$ be a co-oriented contact manifold. Suppose that $\Sigma\colon M^\Sigma \to M$ is a symplectization such that $\theta = \sigma (\Sigma^\ast \eta)$ is the symplectic potential on $M^\Sigma$. Then, $(M^\Sigma, \theta, F^\Sigma)$, with $F^\Sigma=-\sigma(\Sigma^\ast F)$, is a homogeneous integrable system if and only if $(M, \eta, F)$ is a completely integrable contact system.
\end{proposition}

\begin{proof}[Proof of \Cref{thm:main_theorem}]
    Hereinafter, consider the symplectization
    \begin{equation}
        \Sigma\colon M^\Sigma =  M\times \R_+\to M\, ,
    \end{equation}
    where $\Sigma$ is the canonical projection on the first component of the product manifold, and $\R_+$ denotes the positive real half-line. Its conformal factor is $\sigma=s$, with $s$ the global coordinate of $\R_+$.
    Let $F^\Sigma=(f_0^\Sigma, \ldots, f_n^\Sigma)\colon M\times\R_+\to \R^{n+1}$, where  $f_\alpha^\Sigma = -\sigma \left(\Sigma^\ast f_\alpha\right)$. Let $\Lambda\in \R^{n+1}\setminus\{0\}$. Then,
    \begin{equation}
    \begin{aligned}
        (F^\Sigma)^{-1}(\Lambda) 
        & = \{(x,s)\in  M\times\R_+\mid - s F \circ \Sigma(x,s) = \Lambda\} \\
        & = \left\{(x,s)\in  M\times\R_+\mid  F (x) = -\frac{\Lambda}{s}\right\}\, ,
    \end{aligned}
    \end{equation}
    which implies that
    \begin{equation}
        \Sigma \big( (F^\Sigma)^{-1}(\Lambda) \big) 
        = \left\{x \in  M \mid \exists\, s\in \R_+\colon\   F (x) = -\frac{\Lambda}{s}\right\} 
        = M_{\ray{-\Lambda}}\, .
    \end{equation}
    Since $X_{f_\alpha^\Sigma}$ are tangent to $(F^\Sigma)^{-1}(\Lambda)$, the Hamiltonian vector fields $X_{f_\alpha}=\Sigma_\ast X_{f_\alpha^\Sigma}$ are tangent to $M_{\ray{-\Lambda}}$.
    
    Assume $\Lambda\neq 0$.
    Since $\dd f_\alpha^\Sigma = \sigma \left(\Sigma^\ast \dd f_\alpha\right) + \Sigma^\ast \left(f_\alpha\right) \dd \sigma$ and $\sigma$ is functionally independent of $\Sigma^\ast f_\alpha$, it suffices that $\dd f_\alpha$ have rank $n$ for $\dd f_\alpha^\Sigma$ to have rank $n+1$. The fact that $f_\alpha^\Sigma$ are functionally independent implies that $X_{f_\alpha^\Sigma}$, and then $X_{f_\alpha}$, are linearly independent. On the other hand, since $f_\alpha$ are in involution, their Hamiltonian vector fields commute. Hence, by \Cref{lemma:Arnold_tangent_fields}, $M_{\ray{\Lambda}}$ is diffeomorphic to $\TT^k\times \R^{n+1-k}$ for some $k\leq n+1$.

    The tangent space $\T M_{\ray{\Lambda}}$ is spanned by $X_{f_{\alpha}}$. Hence, its annihilator, $\T M_{\ray{\Lambda}}^\circ$, is spanned by the one-forms $\Omega_{\alpha\beta}=f_\alpha \dd f_\beta - f_\beta \dd f_\alpha$. Indeed, using the identity $ \{f,g\} = X_f(g) + g\Reeb(f)$, one can write
    \begin{equation}
        \contr{X_{f_\gamma}} \Omega_{\alpha \beta} = f_\alpha X_{f_\gamma} (f_\beta) - f_\beta X_{f_\gamma} f_\alpha
        = f_\alpha \{f_\gamma, f_\beta\} - f_\beta \{f_\gamma, f_\alpha\} = 0\, ,
    \end{equation}
    for any $\alpha, \beta, \gamma \in \{0, \ldots, n\}$. Since $\T F$ has rank at least $n$, there are at least $n$ independent forms $\Omega_{\alpha\beta}$. Thus, $\T M_{\ray{\Lambda}}^{\perp_\Lambda}$ is spanned by the vector fields $Y_{\alpha \beta}= \lsharp(\Omega_{\alpha \beta})$. Then, 
    \begin{equation}
    \begin{aligned}
        \Omega_{\alpha \beta} (Y_{\gamma \delta}) 
        & = f_\alpha f_\gamma \Lambda ( \dd f_\beta, \dd f_\delta )
        - f_\beta f_\gamma \Lambda ( \dd f_\alpha, \dd f_\delta )
        \\ & \quad 
        - f_\alpha f_\delta \Lambda ( \dd f_\beta, \dd f_\gamma )
        + f_\beta f_\delta \Lambda ( \dd f_\alpha, \dd f_\gamma ) \\
        & = f_\alpha f_\gamma \left\{ f_\beta,  f_\delta \right\}
        - f_\beta f_\gamma \left\{  f_\alpha,  f_\delta  \right\}
        \\ & \quad 
        - f_\alpha f_\delta \left\{ f_\beta,  f_\gamma  \right\}
        + f_\beta f_\delta \left\{  f_\alpha,  f_\gamma  \right\} = 0\, ,
    \end{aligned}
    \end{equation}
    which implies that $\T M_{\ray{\Lambda}}^{\perp_\Lambda} \subseteq \T M_{\ray{\Lambda}}$, that is, $M_{\ray{\Lambda}}$ is coisotropic. 

    Since $f_\alpha$'s are in involution with respect to the Jacobi bracket, $f_\alpha^\Sigma$ are in involution with respect to the Poisson bracket. The Hamiltonian vector fields $X_{f_\alpha^\Sigma}$ are complete since $X_{f_\alpha}$ are complete. 
    Observe that the map $F^\Sigma$ is given by $F^\Sigma(x, r) = -r F(x)$, for each $(x, r) \in M \times \RR_+$.
    Let $\tilde{U} = \Sigma^{-1}(U)$. Since the map $\restr{F}{U}\colon U \to B$ is a trivial bundle over a domain $V \subseteq B$, the map $\restr{F^\Sigma}{\tilde{U}}\colon \tilde{U} \to B$ is also a trivial bundle over $V$.
    Finally, as it has been proven above, $\dd f_\alpha^\Sigma$ has rank $n+1$.

    By the preceeding paragraph, all the hypotheses of \Cref{theorem:LA_exact_symp} hold for the functions $f_\alpha^\Sigma$ on the exact symplectic manifold $(M^\Sigma, \theta)$. Hence, there are coordinates $(y^\alpha_\Sigma, A_\alpha^\Sigma)$ on $\tilde U$ such that 
    \begin{equation}
        \theta = A_\alpha^\Sigma \dd y^\alpha_\Sigma\, ,\qquad
        A_\alpha^\Sigma = M^\beta_\alpha f_\beta^\Sigma\, , 
    \end{equation}
    where $M_\alpha^\beta$ are homogeneous functions of degree $0$ depending only on the functions $(f_0^\Sigma, \ldots, f_n^\Sigma)$ (see \Cref{lemma:linear_combination_functions}), and 
    \begin{equation}
        X_{f_\alpha^\Sigma} = N_{\alpha}^\beta \frac{\partial}{\partial y^\beta_\Sigma}\, , 
    \end{equation}
    where $(N^\alpha_\beta)$ denotes the inverse matrix of $(M^\alpha_\beta)$.

    By construction, $A_\alpha^\Sigma$ are homogeneous functions of degree 1, and therefore there exist functions $A_\alpha$ on $M$ such that $A_\alpha^\Sigma = -\sigma \left(\Sigma^\ast A_\alpha\right)$. 
    Similarly, since $y^\alpha_\Sigma$ are homogeneous of degree 0, there exist functions $y^\alpha$ on $M$ such that $y^\alpha_\Sigma = \Sigma^\ast y^\alpha$. 
    Since $\sigma \left(\Sigma^\ast \eta\right) = \theta$, the contact form is given by
    \begin{equation}
        \eta = A_\alpha \dd y^\alpha\, .
    \end{equation}
    The functions $A_\alpha$ are dissipated quantities since they are linear combinations of dissipated quantities. Moreover,
    \begin{equation}
        f_\alpha = \overline{M}^\beta_\alpha A_\beta\, , \quad X_{f_\alpha} = \overline{N}_\alpha^\beta \frac{\partial}{\partial y^\beta}\, ,
    \end{equation}
    where $\overline{M}_\alpha^\beta$ and $\overline{N}_\alpha^\beta$ are functions on $M$ such that $M_\alpha^\beta = \Sigma^\ast \overline{M}_\alpha^\beta$ and $N_\alpha^\beta = \Sigma^\ast \overline{N}_\alpha^\beta$. Hence, in coordinates $(y^\alpha, A_i)$ the contact Hamilton equations read
    \begin{equation}
    \begin{aligned}
        &\dot y^\alpha = \Omega^\alpha \, ,\\
        &\dot{\tilde{A}}_i = 0\, ,
    \end{aligned}
    \end{equation}
    where $\Omega^\alpha = \overline{N}^\alpha_\beta$ if $f_\beta\equiv h$ is the Hamiltonian function.

    Since $\Lambda\neq 0$, there is at least one nonvanishing $f_\alpha$. Hence, there is at least one nonvanishing $A_\alpha$. By relabeling the $A_\alpha$ if necessary, without loss of generality one can assume that $A_0\neq 0$. Defining $\tilde A_i = -A_i/A_0$ and 
    \begin{equation}
        \tilde \eta = \frac{1}{A_0} \eta = \dd y^0 - \tilde A_i \dd y^i\, ,
    \end{equation}
    one can observe that $(y^i, \tilde A_i, y^0)$ are Darboux coordinates for the co-oriented contact manifold $(M, \tilde \eta)$.
\end{proof}

\begin{remark}
    In order to compute the action-angle coordinates for $(M, \eta, F)$, one has to carry out the following steps:
    \begin{enumerate}
        \item Construct the symplectization $(M^\Sigma, \theta, F^\Sigma)$, where $M^\Sigma = M\times \RR_+, \, \theta = r \Sigma^\ast \eta$ and $F^\Sigma = -r \Sigma^\ast F$, with $\Sigma \colon M^\Sigma \to M$ the canonical projection of $\RR_+$.
        \item Compute the action-angle coordinates $(y^\alpha_\Sigma, A_\alpha^\Sigma)\, , \alpha \in \{0, \ldots, n\}$ for $(M^\Sigma, \theta, F^\Sigma)$ (see \Cref{remark:construction_coords_exact}).
        \item The angle coordinates for $(M, \eta, F)$ are functions $y^\alpha$ such that $y^\alpha_\Sigma = \Sigma^\ast y^\alpha$. Similarly, one has the functions $A_\alpha$ such that $A_\alpha^\Sigma = -r \Sigma^\ast A_\alpha$. 
        \item At least one of the functions $(A_\alpha)$ is non-vanishing. By relabeling the indices if necessary, assume that $A_0\neq 0$. The action coordinates are given by $\tilde{A}_i = - A_i/A_0, \, i\in \{1, \ldots, n\}$. 
    \end{enumerate}
\end{remark}

These steps have been sketched out in order to make clear that the proof is constructive, and for applying them on an illustrative example (see \Cref{sec:example}). Nevertheless, this method is by no means computationally efficient, since it requires computing the Hamiltonian flows of the integrals. In future works, methods for obtaining action-angle coordinates will be studied, such as their possible relation with solutions of the Hamilton--Jacobi equation.

\section{Integrable systems, Legendrian and coisotropic submanifolds}\label{sec:coisotropic}
The aim of this section is to use the properties of coisotropic submanifolds in order to produce alternative characterizations of completely integrable contact systems. This provides a richer information about the geometry of contact completely integrable systems. In addition, these results will be employed in the next section for comparing the results from this dissertation with the ones in previous literature.

\begin{proposition}\label{prop:coisotropic}
    Let $(M, \Lambda, E)$ be a Jacobi manifold. A submanifold $N \hookrightarrow M$ is coisotropic if and only if for each point $x\in N$ there is a neighbourhood $U$ of $x$ in $M$ such that all functions $f,g:M \to \RR$ that are constant on $N\cap U$ satisfy
    \begin{equation}
        \restr{\Lambda(\dd f, \dd g)}{N\cap U} = 0\, .
    \end{equation}
\end{proposition}

\begin{proof}
    Let $x\in N$ and let $U$ be a neighbourhood of $x$ in $M$ such that $N\cap U$ is a level set of functions $f_i\colon U\to \R,\ i=1,\ldots, k$. Then, $\T_x N = \ker \{\dd_x f_i\}_i$, in other words, $(\T_x N)^\circ = \langle \dd_x f_i \rangle$, and thus $(\T_x N)^{\perp_\Lambda} = \langle \lsharp \dd_x f_i \rangle$. Hence, $(\T_x N)^{\perp_\Lambda}\subseteq \T_x N$ if and only if
    \begin{equation}
        0 = \dd_x f_j (\lsharp \dd_x f_i) = \Lambda(\dd_x f_i, \dd_x f_j)\, ,
    \end{equation}
    for every $i, j\in \{1, \ldots, k\}$.
    Since $x$ is arbitrary, the result follows.
\end{proof}

Let $S\colon \R^{n+1}\setminus \{0\} \to \Sp^n$ be the canonical projection on the sphere, that is, $S\colon x \mapsto \tfrac{1}{\norm{x}} x$, where $\norm{\cdot}$ denotes the Euclidean norm. Equivalently, $S$ can be understood as the map that assigns to each point $x\in\R^{n+1}\setminus \{0\}$ the equivalence class of $x$ under homothety, that is, $x\sim y$ if there exists an $r\in \RR_+$ such that $rx=y$.

\begin{lemma}
    Let $N\hookrightarrow M$ be a coisotropic submanifold of a co-oriented contact manifold $(M, \eta)$. Then, there exists a local Hamiltonian vector field $X_h$ on $M$ which is tangent to $N$ and such that $h$ does not vanish on $N$.
\end{lemma}

\begin{proof}
    Let $\cdist = \ker \eta$ be the contact distribution.
    Let $x \in N$ and $v_x \in \T_x M$ such that $v_x \notin \T_x N$ and $v_x \notin \cdist_x$. Using a Darboux chart $(W; q^i, p_i, z)$ centered on $x$ one can write
    \begin{equation*}
        v_x = a^i \frac{\partial}{\partial q^i} + b_i \frac{\partial}{\partial p_i} + c \frac{\partial}{\partial z}\, .
    \end{equation*}
    The Hamiltonian vector field of $h = a^i p_i - b_i q^i - c\in \Cinfty(W)$ is given by
    \begin{equation}
        X_h = a^i \frac{\partial}{\partial q^i} + b_i \frac{\partial}{\partial p_i} + (b_i q^i + c) \frac{\partial}{\partial z}\, .
    \end{equation}
    Hence, $X_h$ extends $v_x$. Since $\T N \cap \cdist$ is closed, there exist a neighbourhood $U\subseteq W$ of $x$ such that $X_h$ is not tangent to $\cdist$ on $\T(N\cap U)\subseteq\T N \setminus (\T N \cap \cdist)$. Therefore, $h=-\eta(X_h) \neq 0$ and $X_h$ is tangent to $N\cap U$.
\end{proof}

\begin{definition}
    Let $(M, \eta)$ be a co-oriented contact manifold with Jacobi bracket $\{\cdot, \cdot\}$.
    A map $\hat F\colon M \to \Sp^n$ is called \emph{dissipative} at a regular value $\ray{\Lambda}$ if there exists a map $F=(f_\alpha)_\alpha\colon M\to \R^{n+1}$ such that $\hat F=S\circ F$ and the functions $f_\alpha$ are in involution, namely, $\{f_\alpha, f_\beta\}=0$ for every $\alpha, \beta \in \{0, \ldots, n\}$.
\end{definition}

\begin{theorem}\label{thm:characterizations_coisotropic}
    Let $(M, \eta)$ be a co-oriented contact manifold, let $\Lambda \in \R^{n+1}\setminus \{0\}$. Let $F=(f_0, \dotsc, f_n)\colon M \to \R^{n+1}$ be a map such that $\operatorname{rank} \T F\geq n$ at $M_{\ray{\Lambda}}=F^{-1}(\ray{\Lambda})$. Then, the following statements are equivalent:
    \begin{enumerate}
        \item \label{item:coisotropic} $M_{\ray{\Lambda}}$ is coisotropic.
        \item \label{item:brackets} $f_\alpha\jacBr{f_\beta,f_\gamma}_{\eta} + f_\gamma\jacBr{f_\alpha,f_\beta}_{\eta} + f_\beta\jacBr{f_\gamma,f_\alpha}_{\eta} = 0$ on $M_{\ray{\Lambda}}$, for each $\alpha, \beta, \gamma \in \{0, \dotsc, n\}$.
        \item \label{item:involution} There exist a contact form $\bar{\eta}$ conformally equivalent to $\eta$ such that $\jacBr{f_\alpha,f_\beta}_{\bar{\eta}} = 0$, for each $\alpha, \beta \in \{0, \dotsc, n\}$.
        \item \label{item:sphere} $\hat F=S\circ F$ is dissipative at $\ray{\Lambda}$.
    \end{enumerate}
    Here $\jacBr{\cdot, \cdot}_{\eta}$ and $\jacBr{\cdot, \cdot}_{\bar{\eta}}$ denote the Jacobi brackets defined by $\eta$ and $\bar{\eta}$, respectively.
\end{theorem}

\begin{proof}
  At least one function $f_\alpha$ is non-zero in a neighbourhood of any point of $M_{\ray{\Lambda}}$. Thus, without loss of generality, one can assume that $f_0$ does not vanish and define $g_i = f_i/f_0$ for $i \in \set{1,\ldots,n}$. Notice that working locally is enough and one can extend the result using a partition of unity.

    The first step is to show that \ref{item:coisotropic} and \ref{item:brackets} are equivalent.
    Since $M_{\ray{\Lambda}}$ is a level set of the $g_i$, one can apply \Cref{prop:coisotropic}, obtaining
    \begin{equation}
        0 = \Lambda(\dd g_i, \dd g_j) =  \Lambda \left(\dd \left(\frac{f_i}{f_0}\right), \dd \left(\frac{f_j}{f_0}\right)\right)\
    \end{equation}
    on $M_{\ray{\Lambda}}$.
    Hence, 
    \begin{equation}
    \begin{aligned}
        0 & = f_0^3\Lambda \left(\dd \left(\frac{f_i}{f_0}\right), \dd \left(\frac{f_j}{f_0}\right) \right)
        \\ &
        = f_0\Lambda(\dd f_i, \dd f_j) - f_i \Lambda(\dd f_0, \dd f_j) -  f_j \Lambda(\dd f_i, \dd f_0) 
        \\ &
        = f_0 \jacBr{f_i, f_j} -f_i  \jacBr{f_0, f_j} - f_j \jacBr{f_i, f_0}
    \end{aligned}
    \end{equation}
    on $M_{\ray{\Lambda}}$.
    Since the choice of $f_0$ is arbitrary, $M_{\ray{\Lambda}}$ is coisotropic if and only if
    \begin{equation*}
        f_\alpha\jacBr{f_\beta,f_\gamma} + f_\gamma\jacBr{f_\alpha,f_\beta} + f_\beta\jacBr{f_\gamma,f_\alpha} = 0
    \end{equation*}
    holds on $M_{\ray{\Lambda}}$.

    The next step is to prove that \ref{item:brackets} implies \ref{item:involution}. In order to show that there is a contact structure $\bar{\eta}$, the first step is to construct a Hamiltonian vector field $X_h$ so that it is tangent to $M_{\ray{\Lambda}}$ and, hence $X_h(g_i) = 0$. The fact that $X_h$ is the Reeb vector field of $\eta_0 = -\eta/h$ implies that
    \begin{equation*}
        \jacBr{g_i, g_j}_{\eta_0} = 0\, .
    \end{equation*}

    Define $ \bar{\eta} = f_0 \eta_0$. Then,
    \begin{equation*}
        \jacBr{f_i, f_j}_{\bar{\eta}} = f_0 \jacBr{f_i/f_0, f_j/f_0}_{\eta_0} =  f_0\jacBr{g_i, g_j}_{\eta_0} = 0\, .
    \end{equation*}

    For showing that \ref{item:involution} implies \ref{item:coisotropic} and \ref{item:brackets}, one has to take into account that being coisotropic is invariant under conformal changes of the contact form (see Remarks \ref{remark:Jacobi_equivalent} and \ref{remark:conformal_equivalence_contact}). Hence, $M_{\ray{\Lambda}}$ is coisotropic for $\eta$ if and only if it is coisotropic for $\bar \eta$. Using that, as it has been proven above, \ref{item:coisotropic} and \ref{item:brackets} are equivalent, this holds whenever
    \begin{equation*}
        f_\alpha\jacBr{f_\beta,f_\gamma}_{\bar\eta} + f_\gamma\jacBr{f_\alpha,f_\beta}_{\bar\eta} + f_\beta\jacBr{f_\gamma,f_\alpha}_{\bar\eta} = 0\, ,
    \end{equation*}
    which clearly holds since $\jacBr{f_\alpha,f_\beta}_{\bar\eta} = 0$.

    The last step is to prove the equivalence between \ref{item:involution} and \ref{item:sphere}. Let $\bar{\eta} = a \eta$. Without loss of generality, one can assume that $a > 0$ (changing the sign of the contact form would only change the sign of the brackets, but the functions would still be in involution).

    One has that $\hat{F} = S \circ  \restr{F'}{M\setminus (F')^{-1}(0)}$, where $F' = (f_0', \ldots, f_n')\colon M \to \RR^{n+1}$, if and only if $\ray{F(x)} = \ray{F'(x)}$ for all $x \in M$. Equivalently, $f_\alpha = a f'_\alpha$ for some positive function $a\in \Cinfty(M)$. Thus, $\hat{F}$ is dissipative at $\ray{\Lambda}$ if and only if
    \begin{equation}
        \jacBr{f'_\alpha, f'_\beta}_{\eta} = \jacBr{a f_\alpha, a f_\beta}_{\eta} = a \jacBr{f_\alpha, f_\beta}_{\bar{\eta}}\, ,
    \end{equation}
    on $M_{\ray{\Lambda}}$.
    Hence, $\hat{F}$ is dissipative if and only if there exist some contact form conformally equivalent to $\eta$ that makes the functions $f_\alpha$ be in involution. 
\end{proof}

\begin{remark}\label{rem:characterization_integrable_systems}
Let $(M, \eta)$ be a co-oriented contact manifold.
Using \Cref{thm:characterizations_coisotropic}, one can give the following characterizations of a completely integrable contact system.
    \begin{enumerate}
        \item A triple $(M,\eta, F)$, where $F=(f_\alpha): M \to \RR^{n+1}$ such that $\jacBr{f_\alpha,f_\beta} = 0$ and $\T F$ has rank at least $n$ (that is~a completely integrable system according to \Cref{def:integrable_systems}).

        \item A triple $(M,\eta, F)$, where $F=(f_\alpha): M \to \RR^{n+1}$ such that $f_\alpha\jacBr{f_\beta,f_\gamma} + f_\gamma\jacBr{f_\alpha,f_\beta} + f_\beta\jacBr{f_\gamma,f_\alpha} = 0$ at least $n$.

        \item A triple $(M, \eta, \hat{F})$, where $\hat{F}: M \to \Sp^n$ such that $\hat{F}$ is a submersion and has coisotropic fibers.
    \end{enumerate}

\end{remark}

\section{Other definitions of contact integrable systems}\label{sec:other_definitions}

Several definitions of contact integrable systems can be found in the literature. Nevertheless, to the best of the author's knowledge, there has been no previous investigation into the concept of integrability for contact Hamiltonian systems without imposing restrictive assumptions on the dynamics. As it is well-known, in symplectic and Poisson manifolds, a collection of functions $f_1, \ldots, f_k$ are in involution if and only if their Hamiltonian vector fields $X_{f_i}$ are tangent to the level set of those functions. However, this is not the case in co-oriented contact manifolds unless one assumes that every function is preserved by the flow of the Reeb vector field. This leads to dynamics that cannot describe dissipative phenomena. Instead of considering level sets of the functions in involution, the approach of this dissertation considers preimages of rays $M_{\ray{\Lambda}}$, so that the contact Hamiltonian vector fields $X_{f_\alpha}$ are tangent to $M_{\ray{\Lambda}}$ without needing to assume that $\Reeb (f_\alpha)=0$. Furthermore, in previous works concerning integrability of contact systems the submanifolds to which $X_{f_\alpha}$ are tangent are assumed to be compact. In several examples of contact Hamiltonian systems this is not the case.

\begin{itemize}
    \item Boyer~\cite{Boyer2011} introduced a concept of completely integrable system for the so-called good Hamiltonians, that is, the Hamiltonian function is preserved along the flow of the Reeb vector field. More specifically, the author says that a contact Hamiltonian system $(M, \eta, h)$ is completely integrable if there exists $n+1$ independent functions $h, f_1, \ldots, f_n$ in involution such that $X_h(h)=0$ and $X_h(f_i)=0,\, i=1, \ldots, n$. This definition implies that $\Reeb(h)=0$. This condition is overly restrictive from a dynamical systems perspective because it excludes systems with energy dissipation.
    
    \item Jovanovi\'{c}~\cite{Jovanovic2012} (see also~\cite{J.L2023}) considered noncommutative integrability for the flows of contact Hamiltonian vector fields. However, the functions considered are assumed to be invariant under the Reeb flow.
    
    \item Miranda~\cite{Miranda2014} (see also~\cite{Miranda2005}) obtained action-angle coordinates for the dynamics generated by the Reeb vector field on a co-oriented contact manifold. It is worth mentioning that the Hamiltonian vector field of a nonvanishing function $f$ with respect to $\eta$ is the Reeb vector field of a conformal contact form $\bar \eta = -\eta/f$. Nevertheless, the author assumes that the Reeb vector field is an infinitesimal generator of an $\Sp^1$-action, which restricts the dynamics one can consider to periodic orbits.
    
    \item Khesin and Tabachnikov~\cite{K.T2010} call a foliation co-Legendrian when it is transverse to $\cdist$ and $T\mathcal{F}\cap \cdist$ is integrable. Additionally, they define an integrable system as a particular case of a co-Legendrian foliation with some extra regularity conditions.
    In a previous paper, it was shown that if $N$ is an $(n+1)$-dimensional co-Legendrian submanifold, then it is also a coisotropic submanifold (see Proposition~5.9 from \cite{deLeon2023}).

    \item Banyaga and Molino~\cite{B.M1993} studied completely integrable contact forms of toric type. More specifically, they considered contact forms on $M$ for which the space of first integrals of the Reeb field determines a singular fibration $\pi\colon M \to W$ defined locally by the action of a torus of dimension $n+1$ of contact transformations, where $W$ is then the space of orbits of this action. A classification of compact connected toric contact manifolds was done by Lerman~\cite{Lerman2003}.
    
    \item Geiges, Hedicke and Sa\u{g}lam~\cite{G.H.S2024} recently introduced a notion of Bott integrability for Reeb flows on contact 3-manifolds. They call a Reeb flow Bott integrable if there exists a Morse--Bott function $f\colon M \to \R$ which is invariant under the Reeb flow. They proof that a closed, oriented 3-manifold admits a Bott-integrable Reeb flow if and only if it is a graph manifold.
\end{itemize}


\section{Example}\label{sec:example}


Consider the co-oriented contact manifold $(M, \eta)$, with $M=\RR^3\setminus\{0\}$ and $\eta= \dd z - p \dd q$ in canonical coordinates $(q, p, z)$. The functions 
\begin{equation}
    h = p\, , \quad f = z
\end{equation}
are in involution. Indeed, their Hamiltonian vector fields are given by
\begin{equation}
    X_h = \frac{\partial}{\partial q}\, , \quad  
    X_f = -p \frac{\partial}{\partial p} - z  \frac{\partial}{\partial z}\, .
\end{equation}
Thus, 
\begin{equation}
    \{h, f\} = X_h(f) + f \Reeb(h) = 0\, .
\end{equation}
Additionally, $h$ and $f$ are functionally independent, that is, $\dd h \wedge \dd f  = \dd p \wedge \dd z \neq 0$. Hence, $(M, \eta, F)$ is a completely integrable contact system, with $F = (h, f)$. Since $F\colon (q, p, z) \mapsto (p, z)$ is the canonical projection, $F\colon \RR^3 \to \RR^2$ is a trivial bundle and the hypotheses of \Cref{thm:main_theorem} are satisfied. 

Consider the symplectization $\Sigma \colon M^\Sigma = M \times \RR_+\to M$ with $\Sigma$ the canonical projection. Let $(q, p, z, r)$ denote the bundle coordinates of $M^\Sigma$. In these coordinates, the conformal factor reads $\sigma =r$. Therefore, $\theta = r \dd z- rp \dd q$ is the symplectic potential on $M^\Sigma$, and the symplectizations of $h$ and $f$ are $h^\Sigma = -rp$ and $f^\Sigma = -rz$. Their Hamiltonian vector fields are
\begin{equation}
    X_{h^\Sigma} = \frac{\partial}{\partial q}\, , \quad  
    X_{f^\Sigma} = -p \frac{\partial}{\partial p} - z  \frac{\partial}{\partial z} + r \frac{\partial}{\partial r}\, .
\end{equation}
Consider a section $\chi\colon \RR^2 \to M^\Sigma$ of $F^\Sigma=(h^\Sigma, f^\Sigma)$ such that $\chi^\ast \theta = 0$. Suppose that $\chi$ reads
\begin{equation}
    \chi(\Lambda_1, \Lambda_2) = \big( \chi_q(\Lambda_1, \Lambda_2),  \chi_p(\Lambda_1, \Lambda_2),  \chi_z(\Lambda_1, \Lambda_2),  \chi_r(\Lambda_1, \Lambda_2)\big)\, .
\end{equation}
The condition of $\chi$ being a section yields
\begin{equation}
   \Lambda_1 \chi_z = \Lambda_2 \chi_p\, , \quad \chi_r \chi_p = \Lambda_1\, ,
\end{equation}
Moreover, the condition $\chi^\ast \theta = 0$ implies that
\begin{equation}
    \dd \xi_q = \frac{\Lambda_2}{\Lambda_1} \dd \left(\log \left( \frac{\Lambda_2}{\Lambda_1} \chi_p\right)\right)\, .
\end{equation}
For instance, one can choose $\chi_r = \Lambda_2/\Lambda_1$ so that
\begin{equation}
    \chi(\Lambda_1, \Lambda_2) = \left(0, \frac{\Lambda_1}{\Lambda_2}, 1, \Lambda_2\right)\, ,
\end{equation}
in the points where $\Lambda_2\neq 0$. The Lie group action $\Phi\colon \RR^2 \times M^\Sigma \to M^\Sigma$ defined by the flows of $X_{h^\Sigma}$ and $X_{f^\Sigma}$ is given by
\begin{equation}
    \Phi(t, s; q, p, z, r) = \left(q+t, p e^{-s}, z e^{-s}, r e^{s}\right)\, ,
\end{equation}
whose isotropy subgroup is the trivial one. The angle coordinates $(y^0_\Sigma, y^1_\Sigma)$ of a point $x\in M^\Sigma$ are determined by
\begin{equation}\label{eq:flow_example}
    \Phi\left(y^0_\Sigma, y^1_\Sigma, \chi(F(x))\right) = x\, .
\end{equation}
If the canonical coordinates of $x$ are $(q, p, z, r)$, then
\begin{equation}
    \chi\circ F(x) = \left(0, \frac{p}{z}, 1, rz\right)\, ,
\end{equation}
and equation~\eqref{eq:flow_example} reads
\begin{equation}
    \left(y^0_\Sigma, \frac{p}{z} e^{-y^1_\Sigma}, e^{-y^1_\Sigma}, rz e^{y^1_\Sigma}\right) = (q, p, z, r)\, .
\end{equation}
As a consequence,
\begin{equation}
    y^0_\Sigma = q\, , \quad y^1_\Sigma = - \log z\, .
\end{equation}
Since the isotropy subgroup is trivial, the action coordinates coincide with the functions in involution, namely, 
\begin{equation}
    A_0^\Sigma = h^\Sigma = -rp\, , \quad A_1^\Sigma = f^\Sigma = -rz\, .
\end{equation}
Projecting to $M$ yields the functions
\begin{equation}
    y^0=q\, , \quad y^1 = - \log z\, , \quad A_0 = h = p\, , \quad A_1 = f = z\, .
\end{equation}
Recall that $\chi$ is defined for points where $\Lambda_2 \neq 0$, that is, the projection by $F^\Sigma$ of points with $z\neq 0$. The action coordinate is 
\begin{equation}
    \tilde{A} = - \frac{A_0}{A_1} = -\frac{p}{z}
\end{equation}
In the coordinates $(y^0, y^1, \tilde{A})$ the Hamiltonian vector fields read
\begin{equation}
    X_h = \frac{\partial}{\partial y^0}\, , \quad X_f = \frac{\partial}{\partial y^1}\, ,
\end{equation}
and there is a conformal contact form given by
\begin{equation}
    \tilde{\eta} = - \frac{1}{A_1} \eta = \dd y^1 - \tilde{A}  \dd y^0\, .
\end{equation}
In these coordinates, contact Hamilton equations are written as
\begin{equation}
    \dot y^0 = 1\, , \quad \dot y^1 = 0\, , \quad \dot{\tilde{A}} = 0\, .
\end{equation}
Integrating them yields the curves
\begin{equation}
    c(t) = \left(y^0(t), y^1(t), \tilde{A}(t) \right) = \left(y^0(0)+t, y^1(0), \tilde{A}(0) \right)\, .
\end{equation}
Similarly, 
\begin{equation}
    \chi(\Lambda_1, \Lambda_2) = \left( \frac{\Lambda_2}{\Lambda_1} , 1, \frac{\Lambda_2}{\Lambda_1}, \Lambda_1 \right)
\end{equation}
is a section of $F^\Sigma$ in the points where $\Lambda_1\neq 0$. Performing analogous computations as above one obtains the action-angle coordinates
\begin{equation}
    \hat{y}^0 = q - \frac{z}{p}\, , \quad \hat{y}^1 = - \log p\, , \quad \hat{A} = - \frac{z}{p}\, ,
\end{equation}
such that
\begin{equation}
    X_h = \frac{\partial}{\partial \hat{y}^0}\, , \quad X_f = \frac{\partial}{\partial \hat{y}^1}\, , \quad \hat{\eta} = - \frac{1}{p} \eta = \dd \hat{y}^0 - \hat{A} \dd \hat{y}^1\, . 
\end{equation}

\chapter{Stability of contact Hamiltonian systems}\label{ch:contact_stability}

The contents of the present chapter are part of a joint work with Prof.~Javier de Lucas and Bartosz M.~Zawora from the University of Warsaw, which is currently unpublished as of the submission of this dissertation.


Let $M$ be an $n$-dimensional manifold and $X\in \X(M)$ a vector field. In the following paragraphs, balls in $\RR^n$ will be identified with the neighbourhoods in $M$ to which they are homeomorphic. A ball of radius $r$ centered at $x$ will be denoted by $\BB_{r,x}$.

A point $x_e\in M$ is called an \emph{equilibrium point} of $X$ if $X(x_e)=0$. An equilibrium point $x_e$ is \emph{stable} if, for every $t_0\in \RR$ and any ball $\BB_{\epsilon,x_e}$, there exists a ball $\BB_{\delta(\epsilon),x_e}$, such that every integral curve $x(t)$ of $X$ with $x(t_0)\in \BB_{\delta(\epsilon),x_e}$ satisfies that $x(t)\in \BB_{\epsilon,x_e}$ for all time $t\geq t_0$.  

An equilibrium point $x_e$ is \emph{asymptotically stable} if it is stable and  there exists an open neighbourhood $\BB_{r,x_e}$ of $x_e$ such that every integral curve $x(t)$ of $X$ with some $t_0$ satisfying $x(t_0)\in \BB_{r,x_e}$ converges to $x_e$. 

The following classical result is essential in stability theory.

\begin{theorem}\label{thm:Lyapunov}
   Let $X$ be a vector field on $M$ such that $X(x_0)=0$ at some $x_0\in M$. If there exists a function $V :
     U \to \R$ defined on some open neighbourhood $U$ of $x_0$ such that
    \begin{enumerate}
         \item \label{Lyapunov_1} $V (x_0) = 0$ and $V (x) > 0$ for $x \in U \setminus \{x_0 \}$,  
        \item \label{Lyapunov_2}  $X(V) (x) \leq 0$ for $x \in U \setminus \{x_0 \}$,
    \end{enumerate}
    then $x_0$ is stable. If additionally $X(V) (x) < 0$ for $x \in U \setminus
    \{x_0 \}$, then $x_0$ is asymptotically stable.
\end{theorem}

Refer to \cite{Vidyasagar2002, Wiggins2003} for the proof. This theorem justifies the following definition.

\begin{definition}
    A function $V\colon U \to \R$ satisfying~\ref{Lyapunov_1} and \ref{Lyapunov_2} in \Cref{thm:Lyapunov} is called a \emph{Lyapunov function}.
    If $X(V) (x) < 0$ for $x \in U \setminus \{x_0 \}$, the function $V$ is called a \emph{strict Lyapunov function}.
\end{definition}

Lyapunov functions for contact Hamiltonian systems may be constructed from dissipated quantities.

\begin{proposition}\label{proposition:Lyapunov_dissipated}
  Let $(M, \eta, H)$ be a contact Hamiltonian system with Reeb vector field $\Reeb$ such that $X_H(x_0)=0$ at $x_0 \in M$. Suppose that $f_1, \ldots, f_k$ are dissipated quantities. If $\Reeb (H)(x_0)>0$ at an isolated point $x_0\in \bigcap_{i=1}^k f_i^{-1}(0)$, then $x_0$ is asymptotically stable.
\end{proposition}

\begin{proof} Since $H$ is smooth and $\Reeb (H)(x_0)>0$ by assumption, there exists an open neighbourhood $U$ of $x_0$ where $\Reeb (H)>0$. Since $x_0$ is an isolated point, one can assume that $\bigcap_{i=1}^k f_i^{-1}(0)\cap U =\{x_0\}$ by choosing a small enough $U$.
Define $V : x\in U \mapsto \sum^k_{i=1}f_i^2(x)\in \R$. Then, $V
  (x_0) = 0, V (x) > 0$ for $x\in U\setminus \{x_0\}$, and 
  \begin{equation}
    \dot{V} (x) = (X_H V) (x) = \sum_{i=1}^k(X_H f_i^2)(x) = - 2 (\Reeb (H) V)(x) < 0,
  \end{equation}
  for $x \in U \setminus \{x_0 \}$, because $x_0$ is isolated in $\bigcap_{i=1}^k f_i^{-1}(0)$. From \Cref{thm:Lyapunov}, the point $x_0$ is asymptotically stable.
\end{proof}

\begin{example}
    Consider the contact Hamiltonian system $(\RR^3, \eta, H)$, with
            \begin{equation}
                \eta = \dd z - p \dd q\, ,\quad  H = \frac{p^2}{2} + \frac{q^2}{2} + z\, .
            \end{equation}
           The Hamiltonian vector field of $H$ is
            \begin{equation}
                X_H = p \parder{}{q} - (q + p) \parder{}{p} + \left(\frac{p^2}{2} - \frac{q^2}{2} - z\right) \parder{}{z}\, ,
            \end{equation}
            which vanishes at $0$.
            The function
            \begin{equation}
                f = z - \frac{pq}{2}
            \end{equation}
            is a dissipated quantity. The zero level sets of $H$ and $f$ intersect solely at $0$, namely, $H^{-1}(0)\cap f^{-1}(0) = \{0\}$.
            Since $\Reeb (H) = 1$ everywhere (in particular, $\Reeb (H)(0)>0$), it follows that $0$ is an asymptotically stable equilibrium point of $X_H$.
\end{example}

The problem with applying \Cref{proposition:Lyapunov_dissipated} is to determine a method to ensure that the intersection of the level sets of several dissipated functions has an isolated point. A necessary condition for verifying this property is as follows.

\begin{proposition}
 Let $f_1, \ldots ,f_k \in \Cinfty(M)$ be such that $f_i(x_0)=0$ for $i=1,\ldots, k$ and $\dim M\geq k+1$.
  If $x_0$ is an isolated point of $\bigcap_{i=1}^kf_i^{-1}(0)$, then 
  \begin{equation}\label{eq:NecCon}
  \restr{\dd f_1}{x_0}\wedge \cdots \wedge\restr{\dd f_k}{x_0}=0.
  \end{equation}
\end{proposition}

\begin{proof}
  The proof is accomplished by reduction to contradiction. Assume that $\restr{\dd f_1}{x_0}\wedge \cdots \wedge \restr{\dd f_k}{x_0} \neq 0$.
Then,  on some neighbourhood $U$ of $x_0$, the map $\Phi\colon U\ni x \mapsto (f_1(x), \ldots, f_k(x))\in \R^k$ is regular and $\dim M>k$, and hence $\restr{\dd f_1}{U}\wedge \cdots \wedge \restr{\dd f_k}{U} \neq 0$. Thus, $\Phi^{-1}(0)= \cap^n_{i=1}f_i^{-1}(0)\cap U$ is a $k$-codimensional submanifold  and $x_0$ is not an isolated point of $\bigcap_{i=1}^kf_i^{-1}(0)$.
\end{proof}

Note that \eqref{eq:NecCon} is a necessary but not sufficient condition. Whenever $\Phi^{-1}(0)$ is a submanifold of dimension bigger than one, the point $x_0$ is not isolated. This means that all theorems in the literature give conditions ensuring that $\Phi^{-1}(0)$ is a submanifold of dimension bigger than zero, which will show that $x_0$ is not an isolated point of $\bigcap^k_{i=1}f^{-1}_i(0)$. This happens, for instance, when $\Phi$ has a constant rank around $x_0$ and one applies the constant rank theorem.

Due to the above, hereafter it will be assumed that $\restr{\dd f_1}{x_0}\wedge \ldots\wedge \restr{\dd f_k}{x_0}=0$. Moreover, the rank of $\Phi:x\in M\mapsto f_1(x),\ldots,f_k(x)\in \RR^k$ should not be locally constant around $x_0$ when $k<\dim M$.

A sufficient condition to ensure that $x_0$ is an isolated point in 
$ \bigcap_{i=1}^k f_i^{-1}(0) \,$ can be given via Lagrange multipliers.

\begin{proposition}\label{prop:isolated_Lagrange_multipliers}
Let $f_1, \ldots ,f_k \in \Cinfty(M)$ with $k< \dim M$ be such that $f_i(x_0)=0$ for some $x_0\in M$, all $i=1,\ldots, k$, and $\dim \langle \restr{\dd f_1}{x_0},\ldots, \restr{\dd f_{k}}{x_0} \rangle=k-1$. Without loss of generality, assume that $\restr{\dd f_1}{x_0},\ldots,\restr{\dd f_{k-1}}{x_0}$ are linearly independent. If  $g = f_k + \lambda_1 f_1 + \cdots + \lambda_{k-1} f_{k-1}$, where $\lambda_1,\ldots,\lambda_{k-1}$ are Lagrange multipliers, has a strict minimum or maximum at $x_0$, then $x_0$ is an isolated point of 
  $ \bigcap_{i=1}^k f_i^{-1}(0) \, .$
\end{proposition}
\begin{proof}
  By construction, $g(x_0)=0$. If $x_0$ is a constrained local strict minimum or maximum of $g$, then there exists a neighbourhood $U$ of $x_0$ in  $f^{-1}_1(0)\cap\ldots\cap f^{-1}_{k-1}(0)$ such that $g(x)\neq 0$ for all $x\in U\setminus\{x_0\}$. Consequently, $f_1(x), \ldots, f_k (x)$ cannot vanish simultaneously at any $x
  \in U\backslash\{x_0\}$. In conclusion,
  \begin{equation}
    \bigcap_{i=1}^k f_i^{-1}(0) \cap \widehat{U} = \{x_0\} \,
  \end{equation}
  for any open subset $\widehat{U}$ in $M$ such that $\widehat{U}\cap\bigcap_{i=1}^{k-1} f_i^{-1}(0)=U$.
\end{proof}

Combining \Cref{proposition:Lyapunov_dissipated,prop:isolated_Lagrange_multipliers} yields the following result.
\begin{theorem}\label{thm:dissipated_stability}
  Let $(M, \eta, H)$ be a contact Hamiltonian system and let $x_0\in M$ be an equilibrium point of $X_H$. Suppose that $f_1, \ldots ,f_k \in \Cinfty(M)$ are dissipated functions for $X_H$ such that $f_i(x_0)=0$ for  $i=1,\ldots, k$, and $\dim \langle \restr{\dd f_1}{x_0},\ldots, \restr{\dd f_{k}}{x_0} \rangle=k-1$. Without loss of generality, assume that $\restr{\dd f_1}{x_0},\ldots,\restr{\dd f_{k-1}}{x_0}$ are linearly independent (if they are not, it suffices to relabel the indices). If the function $g = f_k + \lambda_1 f_1 + \cdots + \lambda_{k-1} f_{k-1}$, where $\lambda_1,\ldots,\lambda_{k-1}$ are Lagrange multipliers, has a strict minimum or maximum at $x_0$, then $x_0$ is asymptotically stable.
\end{theorem}

\begin{remark}
    To determine the Lagrange multipliers, since $\restr{\dd f_i}{x_0}$ are linearly independent for $i=1,\ldots,k-1$, and $\dim\langle\restr{\dd f_1}{x_0},\ldots,\restr{\dd f_{k}}{x_0} \rangle=k-1,$ one has that
    \begin{equation}
      \restr{\dd f_k}{x_0} = -\sum_{i=1}^{k-1} \lambda_i \restr{\dd f_i}{x_0}\, ,
    \end{equation}
    for some fixed $\lambda_1,\ldots,\lambda_{k-1}$.
\end{remark}

The theory of stability for Hamiltonian dynamics on symplectic manifolds is well-known (see for instance Chapter~8 from \cite{A.M2008}).
Given a contact Hamiltonian system $(M, \eta, H)$, the zero level set of the Hamiltonian function is a symplectic manifold  if $\Reeb (H) \neq 0$ on $H^{-1}(0)$ (see Theorem~3.3 in \cite{B.L.M+2020}). More specifically, the symplectic structure on $H^{-1}(0)$ is the restriction of $\d\eta$ to $H^{-1}(0)$. However, $\restr{X_H}{H^{-1}(0)}$ is not a Hamiltonian vector field with respect to $\omega$. As a matter of fact, $\restr{X_H}{H^{-1}(0)}$ is, up to reparametrisation, a Liouville vector field, namely, 
\begin{equation}
  \restr{X_H}{H^{-1}(0)} = -\Reeb (H) \Delta\,  ,
\end{equation}
where $\Delta$ is the unique vector field on $H^{-1}(0)$ so that $\contr{\Delta}\omega=\restr{\eta}{H^{-1}(0)}$. This makes the study of the reduced system more complicated as the standard methods for Hamiltonian systems are not available.

Equilibrium points for contact Hamiltonian systems can be characterized through their associated Hamiltonian functions, which will be used later on. The following simple proposition also appears in \cite{Montaldi2023}.

\begin{proposition}
\label{Th::equilibrium_point}
    Let $(M, \eta, H)$ be a contact Hamiltonian system. Then, $x_e\in M$ is an equilibrium point of $(M, \eta, H)$ if and only if $H(x_e)=0$ and $\restr{\dd H}{\ker \eta_{x_e}}=0$.
\end{proposition}
\begin{proof}
    If $x_e$ is an equilibrium point of $X_H$, that is, $X_H(x_e)=0$, 
    then $H(x_e) = -\eta(X_H)(x_e) = 0$. Taking into account that
    \begin{equation}
      \flat_{\eta_{x_e}}\big(X_H(x_e)\big) = \dd H(x_e) - (\Reeb (H) + H) (x_e) \eta_{x_e}\, ,
    \end{equation}
    it follows that $\restr{\dd H}{\ker \eta_{x_e}}=0$. Conversely, if $H(x_e)=0$, then $X_H(x_e)\in \ker \eta_{x_e}$. Moreover, the assumption $\restr{\dd H}{\ker \eta_{x_e}}=0$ implies that $\contr{X_H}\dd \eta (x_e)= 0$. As $\T_{x_e} M = \ker \eta_{x_e} \oplus \ker \dd \eta_{x_e}$, it follows  that $X_H(x_e)=0$.
\end{proof}

Combining \Cref{thm:dissipated_stability} and \Cref{Th::equilibrium_point}, one has the following proposition.
\begin{proposition}
\label{Prop:AsymStab}
  Let $(M, \eta, H)$ be a contact Hamiltonian system and let $f\in\Cinfty(M)$ be a dissipated function. Consider an equilibrium point $x_0\in M$ such that
  \begin{enumerate}
    \item $
    f(x_0)=0$,\, $\restr{\dd f}{x_0}\neq 0$
 
    \item $\restr{\dd f}{x_0} + \lambda \restr{\dd H}{x_0}=0$,
    \item $x_0$ is a strict maximum or minimum of the function $\tilde f = f + \lambda H$.
  \end{enumerate}
  Then, $x_0$ is an asymptotically stable equilibrium point of $X_H$.
\end{proposition}

Note that $\Reeb (H) (x_0) \neq 0$ follows from the combination of conditions $\restr{\dd H}{\ker \eta_{x_0}}=0$ and $\restr{\dd H}{x_0}\neq 0$, which follows from condition ii), together with the fact that $\T_{x_0} M = \ker \eta_{x_0} \oplus \ker \d\eta_{x_0}$.

\part{Systems with impacts}\label{part:impacts}
\chapter{Hybrid systems}\label{ch:hybrid_systems}

Essentially, a hybrid system is a dynamical system 
whose evolution depends on a coupling between variables that take values in a continuum and variables that take values in a finite or countable set.
These systems can be employed to model mechanical systems with collisions. Furthermore, they have been used to model UAVs (unmanned aerial vehicles) systems \cite{L.S.K2013}, bipedal robots \cite{W.G.C+2018}. Additional examples include control systems with a Boolean variable, such as the temperature of a room that can be controlled by switching on or off a heater \cite{v.S2000}, or electric circuits with a power control \cite{G.S2012}. 

This chapter introduces the so-called hybrid systems, particularizing for hybrid (forced and unforced) Hamiltonian and Lagrangian systems. Moreover, as a novelty, hybrid contact Hamiltonian and Lagrangian systems are defined.

A \emph{(simple) hybrid system} is a $4$-tuple $\hybrid = (M, X, S, \Delta)$, formed by a manifold $M$, a vector field $X\in \X(M)$, a submanifold $S$ of codimension 1 or greater, and a smooth embedding $\Delta\colon S \to M$. The manifold $M$ is called the \emph{domain}, the submanifold $S$ is called the \emph{switching surface}, and the embedding $\Delta$ is called the \emph{impact map}. The dynamics generated by $\hybrid$ are the curves $c\colon I\subseteq \RR\to M$ such that
\begin{equation}
\begin{array}{ll}
    \dot{c}(t) = X(c(t))\, , \quad & \text{if } c(t)\notin S\, ,\\
    c^+(t) = \Delta(c^-(t))\, , \quad  & \text{if } c(t)\in S\, ,
\end{array}
\end{equation}
where $c^-$ and $c^+$ denote the states immediately before and after the times when the integral curves of $X$ intersect $S$, namely,
\begin{equation}
    c^-(t) = \lim_{\tau\to t^-} c(\tau)\, , \quad  c^+(t) = \lim_{\tau\to t^+} c(\tau)\, .
\end{equation}
A curve $c$ satisfying these equations will be called an \emph{integral curve of the hybrid dynamics}.

\section{Hybrid Hamiltonian systems}

A hybrid dynamical system $(M,X,S,\Delta)$ is said to be a \emph{hybrid Hamiltonian system} and denoted by $\hybrid_H$ if the domain $M\subseteq\cT Q$ is a zero-codimensional submanifold of the cotangent bundle $\cT Q$ of a manifold $Q$, the switching surface $S$ projects onto a codimension-one submanifold $\pi_Q(S)$ of $Q$, the impact map preserves the base point, namely, $\pi_Q \circ \Delta = \pi_Q$, and $X=X_H$ is the Hamiltonian vector field of the function $H\in \Cinfty(\cT Q)$ with respect to the canonical symplectic form $\omega_Q$. The Hamiltonian system $(\cT Q, \omega_Q, H)$ is called the \emph{underlying Hamiltonian system}.

Similarly, given a forced Hamiltonian system $(Q, H, \alpha)$ with forced Hamiltonian vector field $X\Halpha$ and a zero-codimensional submanifold $M\subseteq \cT Q$, the hybrid system $(M,X\Halpha,S,\Delta)$  is called a \emph{hybrid forced Hamiltonian system} and denoted by $\hybrid\Halpha$. 
As in the unforced case, it is assumed that $\pi_Q(S)$ is a codimension-one submanifold of $Q$ and $\pi_Q\circ \Delta = \pi_Q$.
The system $(Q, H, \alpha)$ is called the \emph{underlying forced Hamiltonian system}.

Physically, $\pi_Q(S)$ can be usually regarded as the wall where the impact occurs, while $S$ is the subspace of the phase space in which positions are in the wall and momenta are pointing to the wall. The impact map represents the change of momenta produced in the instant of the impact, for instance, an elastic collision with a wall. 

\begin{remark}
    A solution of a simple hybrid system may experience a \emph{Zeno state} if infinitely many impacts occur in a finite amount of time. To exclude these types of situations, the set of impact times is required to be closed and discrete \cite{W.G.C+2018}. To ensure that, along the rest of the dissertation, it will be assumed that $\overline{\Delta({S})}\cap{S}=\emptyset$ (where $\overline{\Delta({S})}$ denotes the topological closure).
\end{remark}

In many examples, the switching surface and the impact map are determined as follows.

\begin{remark}[Newtonian--Hamiltonian impact law]\label{remark:Newtonian_Hamiltonian_impact_law}
   Consider a mechanical Hamiltonian function $H\in \Cinfty(\cT Q)$, namely,
    \begin{equation}
        H (q, p) = g^{-1}_q(p,p) - V(q)\, ,
    \end{equation}
    where $g$ is a Riemannian metric and $V$ a function on $Q$. Let $h\in \Cinfty(Q)$ be a function such that $0$ is a regular value, that is, $h^{-1}(0)$ is a submanifold of $Q$, called the \emph{constraint function}. These functions define a hybrid Hamiltonian system $\hybrid_H = (M, X_H, S, \Delta)$ as follows:
    \begin{enumerate}
        \item the domain is $M = \left\{(q, p)\in \cT Q  \mid h (q) \geq 0 \right\}$
        \item $X_H$ is the Hamiltonian vector field of $H$,
        \item the switching surface is
        \begin{equation}
            S = \left\{(q, p)\in \cT Q \mid h(q) = 0\, , \ \langle\langle p, \dd h_q\rangle\rangle_q < 0 \right\}\, ,
        \end{equation}
        \item the impact map is $\Delta (q, p) = \big(q, P_q(p)\big)$, with $P_q \colon \cT_q Q \to \cT_q Q$ being the map given by
        \begin{equation}
            P_q (p) = p - (1+e) \frac{\langle\langle p, \dd h_q\rangle\rangle_q}{\norm{\dd h_q}_q^2} \dd h_q\, ,
        \end{equation}
    \end{enumerate}
    where $\norm{\cdot}_q$ and $\langle\langle \cdot, \cdot\rangle\rangle_q$ denote the norm and the inner product defined by $g$ on $\cT Q$, and $e\in [0,1]$ is a constant called the \emph{coefficient of restitution}. In particular, $e=1$ and $e=0$ correspond to purely elastic and purely inelastic impacts, respectively. The dynamics described by $\Delta$ is known as the \emph{Newton's restitution law} 
    (see \cite{Brogliato1996}). Of course, by replacing $X_H$ with $X\Halpha$, a hybrid forced Hamiltonian system may also be described by this law.

    Observe that 
    \begin{equation}
    \begin{aligned}
        \Delta(S)
        & =\Bigg\{\left(q,p-(1+e)\frac{\langle\langle p,\dd h_q\rangle\rangle_q}{||\dd h_q||_q^{2}} \dd h_{q}\right)\in \cT Q\mid
        \\ & \qquad 
        h(q)=0 \hbox{ and }\langle\langle p,\dd h_q\rangle\rangle_q< 0\Bigg\}\, ,
    \end{aligned}
    \end{equation}
    and 
    \begin{equation}
        \begin{aligned}
            \overline{\Delta(S)}
            & =\Bigg\{\left(q,p-(1+e)\frac{\langle\langle p,\dd h_q\rangle\rangle_q}{||\dd h_q||_q^{2}} \dd h_{q}\right)\in \cT Q\mid
            \\ & \qquad 
            h(q)=0 \hbox{ and }\langle\langle p,\dd h_q\rangle\rangle_q\leq  0\Bigg\}\, ,
        \end{aligned}
    \end{equation}
    which implies that
    \begin{equation}
        \overline{\Delta(S)} \cap S 
        = \emptyset,
    \end{equation}
    Therefore, no Zeno effect occurs unless the impact is purely inelastic $(e=0)$.
\end{remark}

\section{Hybrid Lagrangian systems}

A hybrid dynamical system $(M,X,S,\Delta)$ is called to be a \emph{hybrid Lagrangian system} and denoted by $\hybrid_L$ if $M\subseteq\T Q$ is a submanifold of the tangent bundle $\T Q$ of a manifold $Q$ and $X=\sode_L$ is the Euler--Lagrange vector field of a regular Lagrangian function $\L\in \Cinfty(\T Q)$.  The system $(Q, L)$ is called the \emph{underlying Lagrangian system}.

Analogously, given a forced Lagrangian system $(Q, L, \beta)$ with forced Euler--Lagrange vector field $\sode\Lbeta$ and a submanifold $M\subseteq \T Q$, the hybrid system $(M,\sode\Lbeta,S,\Delta)$ is called a \emph{hybrid forced Lagrangian system} and denoted by $\hybrid\Lbeta$. The system $(Q, L, \beta)$ is called the \emph{underlying forced Lagrangian system}.

\begin{remark}[Newtonian--Lagrangian impact law]\label{remark:Newtonian_Lagrangian_impact_law}
    Consider a mechanical Lagrangian function $L\in \Cinfty(\T Q)$, namely,
     \begin{equation}
         L (q, v) = g_q(v, v) - V(q)\, ,
     \end{equation}
     where $g$ is a Riemannian metric and $V$ a function on $Q$. Let $h\in \Cinfty(Q)$ be a \emph{constraint function} such that $0$ is a regular value. These functions define a hybrid Lagrangian system $\hybrid_L = (M, \sode_L, S, \Delta)$ as follows:
     \begin{enumerate}
         \item the domain is $M = \left\{(q, v)\in \T Q  \mid h (q) \geq 0 \right\}$
         \item $\sode_L$ is the Euler--Lagrange vector field of $L$,
         \item the switching surface is
         \begin{equation}
             S = \left\{(q, v)\in \T Q \mid h(q) = 0\, , \ \dd h_q (v) \leq 0 \right\}\, ,
         \end{equation}
         \item the impact map is $\Delta (q, v) = \big(q, P_q(v)\big)$, with $P_q \colon \T_q Q \to \T_q Q$ being the map given by
         \begin{equation}
             P_q (v) = p - (1+e) \frac{\dd h_q (v)}{\norm{\dd h_q}_q^2} g^{-1}_q(\dd h_q)\, ,
         \end{equation}
     \end{enumerate}
     where $\norm{\cdot}_q$ denotes the norm defined by $g$ on $\T Q$, and $e\in [0,1]$ is the \emph{coefficient of restitution}.
 \end{remark}

\section{Hybrid contact Hamiltonian systems}

A natural generalization of the systems introduced above is to consider hybrid systems whose continuous dynamics is generated by a contact Hamiltonian vector field.

\begin{definition}
    A hybrid system $(M, X, S, \Delta)$ is called a \emph{hybrid contact Hamiltonian system} and denoted by $\hybrid_{\eta,\,  H}$ if $(M, \eta)$ is a contact manifold and $X$ is the contact Hamiltonian vector field of the Hamiltonian function $H\in \Cinfty(M)$ with respect to $\eta$. The triple $(M, \eta, H)$ will be called the \emph{underlying contact Hamiltonian system}.
\end{definition}

\begin{definition}
    A \emph{hybrid contact Lagrangian system} $\mathscr{H}_L$ is a hybrid contact Hamiltonian system with underlying contact Hamiltonian system $(\T Q\times \R, \eta_L, E_L)$, where $\eta_L$ and $E_L$ denote the contact form and the energy associated with a regular Lagrangian function $L\in \Cinfty(\T Q)$, respectively. In other words, the domain is of the form $M=\T Q\times \R$, and the vector field $X$ is the Herglotz--Euler--Lagrange vector field $\Gamma_L$ associated with the Lagrangian $L$.
\end{definition}

The dynamics generated by hybrid contact Lagrangian system can also be derived from a variational principle, the so-called Nonsmooth Herglotz principle (see \Cref{ch:nonsmooth_Herglotz}).

\chapter{Reduction of hybrid forced systems}\label{ch:hybrid_reduction}

This chapter discusses the reduction by symmetries of hybrid forced systems. Firstly, the symplectic reduction scheme is adapted to hybrid forced Hamiltonian system. After that, the Routhian reduction for hybrid forced Lagrangian systems with cyclic coordinates is studied and illustrated with an example. These results have been published in the preprint \cite{Colombo2022}. Some preliminary results were also previously published in the conference paper \cite{EyreaIrazu2022}.

\section{Reduction of hybrid forced Hamiltonian systems}

Let $\hybrid\Halpha=(M, X\Halpha, S, \Delta)$ be a hybrid forced Hamiltonian system with underlying forced Hamiltonian system $(Q, H, \alpha)$. Consider a Lie group $G$ with Lie algebra $\mathfrak{g}$, and denote by $\mathfrak{g}^\ast$ the dual of $\mathfrak{g}$.

\begin{definition}
    A Lie group action $\psi\colon G\times Q \to Q$ is called a \emph{hybrid action for $\hybrid\Halpha$} if its cotangent lift $\psi^{\cT} \colon G\times \cT Q \to \cT Q$ satisfies the following conditions:
    \begin{enumerate}
        \item the Hamiltonian function $H$ is $\psi^{\cT}$-invariant, namely, $\psi_g^{\cT \ast} H = H$ for all $g\in G$,
        \item the restriction $\restr{\psi^{\cT}}{G\times S}$ is a Lie group action of $G$ on $S$,
        \item the impact map is equivariant with respect to this action, that is, 
        \begin{equation}
                \Delta \circ \restr{\psi_g^{\cT}}{S} = \psi_g^{\cT} \circ \Delta\, ,
        \end{equation}
        for any $g\in G$.
    \end{enumerate}
\end{definition}

Since $\psi^{\cT}$ is the cotangent lift of an action, it is a symplectic action. In fact, it is an exact symplectic action, that is, $\psi^{\cT\ast}_g \theta_Q = \theta_Q$ for all $g\in G$.

\begin{definition}\label{def:generalized_hybrid_momentum_map}
        Let $\psi\colon G\times Q \to Q$ be a hybrid action for $\hybrid\Halpha$. A momentum map $\mommap\colon \cT Q \to \mathfrak{g}^\ast$ for the cotangent lift action $\psi^{\cT}$ is called a \emph{generalized hybrid momentum map} if, for each connected component $C\subseteq S$ and for each regular value $\mu_-$ of $\mommap$, there is another regular value $\mu_+$ such that
        \begin{equation}
                \Delta \big(\restr{\mommap}{C}^{-1}(\mu_-)\big) \subset \mommap^{-1}(\mu_+)\, .
        \end{equation}
        In particular, if $\mu_- = \mu_+$ it is called a \emph{hybrid momentum map}. 

        In other words, $\mommap$ is a generalized hybrid momentum map if, for every point in the connected component $C$ of the switching surface $S$ such that the momentum before the impact takes a value of $\mu_-$, the momentum will take a value $\mu_+$ after the impact; and it is a hybrid momentum map if its value does not change with the impacts.

        A \emph{hybrid regular value} of $\mommap$ is a regular value of both $\mommap$ and $\restr{\mommap}{S}$.
\end{definition}

\begin{remark}
        In the case of the Newtonian impact law (see \Cref{remark:Newtonian_Hamiltonian_impact_law}), a momentum map $\mommap$ is a hybrid momentum map if and only if 
        \begin{equation}
            \mommap\left(q, p-(1+e)\frac{\langle\langle p,\dd h_q\rangle\rangle_q}{\norm{\dd h_q}_q^{2}} \dd h_{q} \right)=\mu_+,
        \end{equation}
        for every $(q, p)\in \cT Q$ such that $h(q)=0,\ \langle\langle p,\dd h_q\rangle\rangle_q\leq 0$ and $\mommap(q,p)=\mu_-$.
\end{remark}

For each $\mu\in \mathfrak g^*$, let $G_{\mu}$ denote the isotropy subgroup of $G$ in $\mu$ under the co-adjoint action, namely, $G_\mu = \left\{g\in G \mid \Ad_g^*\ \mu = \mu  \right\}$.

\begin{proposition}
        Let $\psi:G\times Q \to Q$ be a Lie group action of a connected Lie group $G$ on $Q$. 
        Let $\mommap\colon \cT Q \to \mathfrak{g}^\ast$ be an $\Ad$-equivariant momentum map for the cotangent lift action $\psi^{\cT}$. If $\Delta$ is equivariant with respect to $\psi^{\cT}$, and $\mu_-,\ \mu_+$ are regular values of $\mommap$ such that $\Delta \left(\restr{\mommap}{S}^{-1}(\mu_-)  \right)\subset \mommap^{-1}(\mu_+)$, then the isotropy subgroups in $\mu_-$ and $\mu_+$ coincide, that is, $G_{\mu_-}=G_{\mu_+}$.
\end{proposition}

\begin{proof}
        Let $g\in G_{\mu_-}$. The equivariance of $\mommap$ and $\Delta$ imply that, for each $g\in G_{\mu_-}$, 
        \begin{equation}
        \begin{aligned}
                \mommap \circ \Delta \left(\restr{\mommap}{S}^{-1}(\mu_-)  \right)
                &= \mommap \circ \Delta\circ \psi_g^{\cT} \left(\restr{\mommap}{S}^{-1}(\mu_-)  \right)
                \\&
                = \mommap \circ \psi_g^{\cT} \circ \Delta \left(\restr{\mommap}{S}^{-1}(\mu_-)  \right)
                \\&
                = \Ad_{g^{-1}}^* \circ\, \mommap \circ \Delta \left(\restr{\mommap}{S}^{-1}(\mu_-)  \right)\, .
        \end{aligned}
        \end{equation}
        Therefore, $g\in G_{\mu_+}$, making $G_{\mu_-}$ a Lie subgroup of $G_{\mu_+}$. 

        Now, observe that $G_\mu$ has the same dimension, for each $\mu\in \mathfrak g^*$. Therefore, the identity components of $G_{\mu_-}$ and $G_{\mu_+}$ coincide. The assumption that $G$ is connected implies that $G_{\mu_-}$ and $G_{\mu_+}$ are equal to their identity components. As a consequence, $G_{\mu_-}=G_{\mu_+}$.
\end{proof}

Recall that
\begin{equation}
        \mathfrak{g}_\alpha = \left\{\xi \in \mathfrak{g} \mid \alpha(\xi_{\cT Q}) = 0, \, \contr{\xi_{\cT Q}} \dd \alpha = 0\right\}
\end{equation} 
is the Lie subalgebra of $\mathfrak{g}$ which leaves $\alpha$ invariant (see \Cref{prop:Lie_subalgebra_forced_Hamiltonian}). Let $G_\alpha\subseteq G$ denote unique connected Lie subgroup of $G$ with Lie algebra $\mathfrak{g}_\alpha$. Without loss of generality, assume hereafter that $G_\alpha = G$ (if that is not the case, one has to replace $G$ by $G_\alpha$ as the Lie group considered).

If $\mommap$ is a generalized hybrid momentum map and $\mu_-$ and $\mu_+$ are hybrid regular values, then the following diagram commutes:
\begin{equation}
\begin{tikzcd}[column sep=1.3cm, row
sep=1.2cm]
\mommap^{-1}(\mu_-)\arrow[d,hook'] &  \mommap\mid_{S}^{-1}(\mu_-)\arrow[r,"\Delta\mid_{\mommap^{-1}(\mu_-)}"] \arrow[l,swap,hook']\arrow[d,hook'] & \mommap^{-1}(\mu_+)\arrow[d,hook']\\
\cT Q  & S \arrow[l,swap,hook'] \arrow[r,"\Delta"] &  \cT Q  
\end{tikzcd} \, ,
\end{equation} 
where $\mommap^{-1}(\mu)$ and $\restr{\mommap}{S}^{-1}(\mu)$ are embedded submanifolds of $T^{*}Q$ and $S$, respectively. The hook arrows $\hookrightarrow$ in the diagram denote the corresponding canonical inclusions.

Since the $G_{\mu_0}$-action restricts to a free and proper action on $S$, $(S)_{\mu_i}=\mommap\mid_{S}^{-1}(\mu_i)/G_{\mu_0}$ is 
a smooth manifold. Clearly, it is
a submanifold of\\ $\mommap^{-1}(\mu_i)/G_{\mu_0}$. Since $\Delta$ is equivariant, it induces an embedding $(\Delta)_{\mu_i}:(S)_{\mu_i} \rightarrow \mommap^{-1}(\mu_{i+1})/G_{\mu_0}$.


\begin{theorem}\label{theorem_reduction_autonomous}
        Let $\hybrid\Halpha=(M, X\Halpha, S, \Delta)$ be a hybrid forced Hamiltonian system with underlying forced Hamiltonian system $(Q, H, \alpha)$. Let $\psi\colon G\times Q \to Q$ be a hybrid action of a connected Lie group $G$ on $Q$. Assume that the cotangent lift action $\psi^{\cT}\colon G \times \cT Q \to \cT Q$ is free and proper.
        Suppose that $\alpha$ is $\psi^{\cT}$-invariant and $\mommap$ is an $\Ad$-equivariant generalized hybrid momentum map.
        Consider a sequence $\left\{\mu_i  \right\}$ of hybrid regular values of $\mommap$, such that $\Delta \left(\restr{\mommap}{S}^{-1}(\mu_i)  \right)\subset \mommap^{-1}(\mu_{i+1})$.
        Let $G_{\mu_i}=G_{\mu_0}$ be the isotropy subgroup in $\mu_i$
        under the co-adjoint action. Then,
        \begin{enumerate}
        \item $\mommap^{-1}(\mu_i)$ is a submanifold of $T^{*}Q$ and $X\Halpha$ is tangent to it.
        \item  The reduced space $M_{\mu_i}\coloneqq \mommap^{-1}(\mu_i)/G_{\mu_0}$ is a symplectic manifold, whose symplectic structure $\omega_{\mu_i}$ is uniquely determined by  $\pi^{*}_{\mu_i}\omega_{\mu_i}=\incl^{*}_{\mu_i}\omega_{Q},$ where $\pi_{\mu_i}:\mommap^{-1}(\mu_i)\rightarrow M_{\mu_i}$ and $\contr {\mu_i}:\mommap^{-1}(\mu_i)\hookrightarrow T^{*}Q$ denote the canonical projection and the  canonical inclusion, respectively. 
        \item 
        $(H,\alpha)$ induces a reduced forced Hamiltonian system $(H_{\mu_i}, \alpha_{\mu_i})$ on $M_{\mu_i}$, given by $H_{\mu_i}\circ\pi_{\mu_i}=H\circ \contr {\mu_i}$ and $\pi^{*}_{\mu_i}\alpha_{\mu_i}=\incl^{*}_{\mu_i}\alpha$. Moreover, the forced Hamiltonian vector field $X\Halpha$ projects onto $X_{H_{\mu_i},\, \alpha_{\mu_i}}$.
        \item $\mommap\mid_{S}^{-1}(\mu_i)\subset S$ reduces to a submanifold of the reduced space
        $(S)_{\mu_i}\subset \mommap^{-1}(\mu_i)/G_{\mu_0}.$ 
        \item 
        $\restr{\Delta}{\mommap^{-1}(\mu_i)}$ reduces to a map $(\Delta)_{\mu_i}:(S)_{\mu_i} \rightarrow \mommap^{-1}(\mu_{i+1})/G_{\mu_0}.$
        \end{enumerate}
        {Therefore, after the reduction procedure, one obtains a sequence of reduced hybrid forced Hamiltonian systems $\left\{\hybrid\Halpha^{\mu_i}  \right\}$, where 
        $\hybrid\Halpha^{\mu_i}=(\mommap^{-1}(\mu_i)/G_{\mu_0},$ $X_{H_{\mu_i},\, \alpha_{\mu_i}},(S)_{\mu_i},(\Delta)_{\mu_i})$.}

        \begin{equation} 
        \begin{footnotesize}
        \begin{tikzcd}
        \cdots \arrow[r] & \mommap^{-1}(\mu_i) \arrow[dd]        &  & \restr{\mommap}{S}^{-1}(\mu_i) \arrow[dd] \arrow[rr, "\restr{\Delta}{\mommap^{-1}(\mu_i)}"] \arrow[ll, hook] &  & \mommap^{-1}(\mu_{i+1}) \arrow[dd]        & \cdots \arrow[l, hook] \\
                        & {} \arrow[d]                      &  &                                                                                               &  &                                       &                        \\
        \cdots \arrow[r] & \frac{\mommap^{-1}(\mu_i)}{G_{\mu_0}} &  & \left(S\right)_{\mu_i} \arrow[rr, "\left(\Delta\right)_{\mu_i}"] \arrow[ll, hook]         &  & \frac{\mommap^{-1}(\mu_{i+1})}{G_{\mu_0}} & \cdots \arrow[l, hook]
        \end{tikzcd}
        \end{footnotesize}
        \end{equation}

\end{theorem}

\begin{proof}
        The first three assertions follow from \Cref{thm:symplectic_point_reduction} and the Hamiltonian counterpart of \Cref{theorem:reduction_forced}.

        Since the $G_{\mu_0}$-action restricts to a free and proper action on $S$, the quotient $(S)_{\mu_i}=\mommap\mid_{S}^{-1}(\mu_i)/G_{\mu_0}$ is 
        a smooth manifold. Moreover, it is
        a submanifold of $\mommap^{-1}(\mu_i)/G_{\mu_0}$. Since $\Delta$ is equivariant, it induces an embedding $(\Delta)_{\mu_i}:(S)_{\mu_i} \rightarrow \mommap^{-1}(\mu_{i+1})/G_{\mu_0}$.
\end{proof}

The value of the momentum map will, in general, be modified in the collisions with the switching surface. As a matter of fact, this will be the case unless the momentum map is a hybrid momentum map. Suppose that the collisions take place at the times $\{\tau_i\}$. Therefore, the reduced Hamiltonian $H_{\mu_i}$ and the reduced external force $\alpha_{\mu_i}$ will have to be defined in each time interval $I_i=[\tau_{i}, \tau_{i+1}]$, and they will depend on the value of the momentum $\mu_i$ after the collision at time $\tau_i$. The same will occur for the reduced impact map $\Delta_{\mu_i}$ and the reduced impact surface $S_{\mu_i}$. 
Consequently, the reduction procedure yields a sequence of reduced hybrid forced Hamiltonian systems $\{(M_{\mu_i}, X_{H_{\mu_i},\, \alpha_{\mu_i}}, S_{\mu_i}, \Delta_{\mu_i})\}$.

\section{Reduction of hybrid forced Lagrangian systems}
The reduction picture in the Lagrangian side can now be obtained from the Hamiltonian one by adapting the Routhian reduction (see \Cref{sec:Routh}). Let $\hybrid_{\Lbeta} = (M, \sode\Lbeta, S, \Delta)$ be a hybrid forced Lagrangian system with underlying forced Lagrangian system $(Q, L, \beta)$. In the same fashion as in the Hamiltonian side, a Lie group action $\psi\colon G\times Q\to Q$ will be called a \emph{hybrid action} if its tangent lift $\psi^\T \colon G \times \T Q \to \T Q$ verifies the following conditions:
\begin{itemize}
	\item the Lagrangian function $L$ is $\psi^{\T}$-invariant, that is, $L\circ \psi^{\T}=L$,
	\item The restriction $\restr{\psi^{\T}}{G\times S}$ is a Lie group  action of $G$ on $S$.
	\item the impact map $\Delta$ is equivariant with respect to the previous action, namely, $\Delta\circ \psi^{\T}_g\mid_{S}=\psi^{\T}_g\circ \Delta$.
\end{itemize}

Assume that the Lagrangian $L$ is hyper-regular (which is always the case for mechanical Lagrangians). Then, since $\psi^{\T}$ is a hybrid action under which $L$ is invariant, the Legendre transform $\FF L\colon \T Q\to \cT Q$ is a diffeomorphism such that:
\begin{itemize}
	\item it is equivariant with respect to $\psi^{\T}$ and $\psi^{\cT}$,
	\item it preserves the level sets of the momentum map, that is,
	\begin{equation}
                \FF L (\mommap_L)^{-1}({\mu_i}))=\mommap^{-1}({\mu_i}) \, ,
        \end{equation}
	\item it is a symplectomorphism, namely, $(\FF L)^*\omega_{Q}=\omega_L$.
\end{itemize}
It follows that the map $\FF L$ reduces to a symplectomorphism $(\FF L)_{\text{red}}$ between the reduced spaces. 

Consider the Lie subalgebra $\mathfrak{g}_{\beta}=\{\xi\in \mathfrak{g}:\beta(\xi_{Q}^{c})=0,\,\, \contr {\xi_{Q}^{c}}\dd \beta=0\}$ of $\mathfrak g$ (see \Cref{prop:Lie_subalgebra_forced_Lagrangian}), and let $G_{\beta}$ be the unique connected Lie subgroup it generates. 

\begin{proposition}
        If $(L, \beta)$ is the Lagrangian counterpart of $(H, \alpha)$ (that is, $E_L = H \circ \FF L$ and  $\beta = \FF L^* \alpha$), then $G_{\beta}= G_\alpha$.
\end{proposition}
\begin{proof}
        For any $g\in G_{\alpha}$, 
        \begin{equation}
        \begin{aligned}
        \left( \psi_g^{TQ}  \right)^* \beta
        &=  \left( \psi_g^{TQ}  \right)^* \circ \FF L^* \alpha
        = \left( \FF L \circ \psi_g^{TQ}  \right)^* \alpha
        = \left( \psi_g^{T^*Q} \circ \FF L  \right)^* \alpha\\
        & = \FF L^* \circ \left( \psi_g^{T^*Q}  \right)^* \alpha
        = \FF L^* \alpha 
        = \beta\, ,
        \end{aligned}
        \end{equation}
        where the equivariance of $\FF L$ has been used. Consequently, $g \in G_{\beta}$. Similarly one can show that, for any $g \in G_{\beta}$, $g\in G_\alpha$.
\end{proof}

Assuming $L$ to be $G$-regular and making use of a principal connection on the principal bundle $Q\to Q/G$, one could make further identifications and perform Routhian reduction (see \Cref{sec:Routh} and references therein). In what follows, only $\Sp^1$ or $\RR$-actions will be considered. For these cases, the symmetry can be interpreted as the presence of cyclic coordinate, and thus the classical Routhian reduction is sufficient.

\subsection{Reduction of hybrid forced Lagrangian systems with a cyclic coordinate}

If both $L$ and $\beta$ are $\Sp^1$-invariant, one recovers the classical notion of a cyclic coordinate (see \Cref{example:cyclic_coord_forced}). The case $G=\RR$ is analogous, and if $G$ is a product of $\Sp^1$ or $\R$ one can iterate the procedure. Since $G=\Sp^1$ is Abelian, the isotropy subgroup is $G_{\mu_i}=G$ for every $\mu_i \in \mathfrak g^*$. 

Assume that $(L,\beta)$ has a cyclic coordinate $\theta$ and denote by $x$ the rest of the coordinates in $Q$, namely, $L$ and $\beta$ are of the form
\begin{equation}
        L = L (x, v_\theta, v_x)\, , \quad \beta = \beta_x (x, v_\theta, v_x) \dd x\, ,
\end{equation}
where $(\theta, x, v_\theta, v_x)$ are the bundle coordinates in $\T Q$ induced by the coordinates $(\theta, x)$. The momentum map can be identified with the function $\mommap_L = \tparder{L}{v_\theta}$. Assuming that $\partial^2 L/ \partial v_\theta^2\neq 0$, one can express $v_\theta$ as a function of $x, v_x$, and the fixed regular value of the momentum map ${\mu_i}$, namely, $v_\theta = \phi(x, v_x, \mu_i)$.
The reduced space $\mommap_L^{-1}({\mu_i})/\Sp^1$ can be identified with $\T(Q/\Sp^1)$, and its bundle coordinates with $(x, v_x)$. Similarly, the reduced switching surface $S_{\mu_i}$ can be identified with a submanifold of $\T(Q/\Sp^1)$, and the reduced impact map can be identified with a map $(\Delta_L)_{\mu_i}\colon S_{\mu_i}\to \T(Q/\Sp^1)$. 
If the connection on the bundle $Q\rightarrow Q/\Sp^1=M$ is chosen to be the canonical flat connection, then the Routhian $R_{\beta}^{\mu_i}\in \Cinfty(\T(Q/\Sp^1))$ and the reduced external force $\beta^{\mu_i}\in \Omega^1(\T(Q/\Sp^1))$ can be written as
\begin{align} \label{eqq5}
  & R_{\beta}^{\mu_i}(x,v_x) = L\big(x,v_x, \phi(x, v_x, \mu_i)\big)-{\mu_i}\phi(x, v_x, \mu_i)\, ,\\
  & \beta^{\mu_i}(x,v_x) = \beta_x\big(x,v_x, \phi(x, v_x, \mu_i)\big) \dd x\, .
\end{align}

Since the value of the momentum map, generally, changes with the collisions the reconstruction procedure will be more challenging. In order to make use a reduced solution to reconstruct the original dynamics, the reduced hybrid data have to be computed after each collision. That is, once the reduced solution for the time interval between two collision events, say between $t=\tau_{i} $ and $t=\tau_{i+1},$ has been obtained, this solution has to be reconstructed to obtain the new momentum after the collision at $\tau_{i+1}$. After that, this new momentum has to be used in order to build a new reduced hybrid system whose solution should be obtained until the next collision event at $\tau_{i+2}$, and so forth. In order to reconstruct the dynamics of the original hybrid system from the reduced hybrid systems, essentially, one has to impose the value of the momentum at each interval $I_i$ between impacts.

\begin{example}[Rolling disk with dissipation hitting fixed walls]\label{example:disk_reduction}
        Consider a homogeneous circular disk of radius $R$ and mass $m$ moving in the vertical plane $xOy$ (see \cite[Example 8.2]{I.d.L+1997}, and also \cite[Example 3.7]{I.d.L+2001}). Let $(x, y)$ be the coordinates of the centre of the disk and $\varphi$ the angle between a point of the disk and the axis $Oy$. In other words, the configuration space is 
        $Q=\RR^2\times \Sp^1$, with coordinates $(x, y, \varphi)$. Let $(x, y, \varphi, v_x, v_y, v_\varphi)$ denote the induced bundle coordinates on $\T Q$.
        The dynamics of the system is determined by the forced Lagrangian system $(\RR^2\times \Sp^1, L, \beta)$, where
        \begin{equation}
                L=\frac{1}{2}m(v_x^2+v_y^2+k^2v_\varphi^2)\, , 
        \end{equation} 
        and
        \begin{equation}
                \beta = -2c \left(v_x xy-v_y  x^2\right) \dd x + 2c\left(v_y  xy-v_x y^2\right)\dd y\, .
        \end{equation}
        The free motion of the disk $c(t) = (x(t), y(t), \varphi(t))$ is given by the forced Euler--Lagrange equations:
        \begin{equation}
                m\ddot x=-2c(\dot{y}  x^2- \dot{x} xy)\, , \quad  
                m\ddot y=2c(\dot{x} y^2- \dot{y}  xy)\, , \quad
                \ddot\varphi=0\, .
        \end{equation}

        Consider the Lie group action of~$\Sp^1 \times \Sp^1$ on $Q$ given by 
        \begin{equation}
                (\alpha_1, \alpha_2; x, y, \varphi)\mapsto (\cos \alpha_1\ x- \sin \alpha_1\ y, \sin \alpha_1\ x + \cos \alpha_1\ y, \varphi + \alpha_2)
        \end{equation}
        Note that $L$ and $\beta$ are invariant under the lifted action on $\T Q$. The corresponding momentum map is
        \begin{equation}
                \mommap_L=(mx v_y- my v_x, mk^2 v_r)\, .
        \end{equation}

        Let $(r, \theta)$ be polar coordinates in $\RR^2$, namely,
        \begin{equation}
                x = r \cos \theta\, , \quad y = r \sin \theta\, .
        \end{equation}
        Denoting by $(r, \theta, \varphi, v_r, v_\theta, v_\varphi)$ the induced bundle coordinates in $\T (\RR^2 \times \Sp^1)$, the forced Lagrangian system reads
        \begin{equation}
                L = \frac{m}{2}(v_r^2+ r^2 v_\theta^2+k^2 v_\varphi^2)\, , \quad
                \beta = 2cr^3 v_{\theta}\dd r\, .
        \end{equation}
        In these coordinates, the forced Euler--Lagrange equations for a curve $c(t)=(r(t), \theta(t), \varphi(t))$ can be written as
        \begin{equation}
                m\frac{\dd (r^2\dot{\theta})}{\dd t}=0\, , \quad
                mk^2 \ddot{\varphi}=0\, .
        \end{equation}
        The momentum map now reads
        \begin{equation}
                \mommap_L = (mr^2v_\theta,mk^2v_\varphi)\, .
        \end{equation}
        By observing the forced Euler--Lagrange equations in polar coordinates, it is clear that $\mommap_L$ is preserved along the dynamics. 
        
        Suppose that $\mu = (\mu_1, \mu_2)$ is a regular value of $\mommap_L$. Since $\partial^2 L /\partial v_\theta^2 \neq 0$ and $\partial^2 L /\partial v_\varphi^2 \neq 0$, one can express $v_\theta$ and $v_\varphi$ as functions of $\mu$ and the rest of the coordinates.
        The quotient $\mommap_L^{-1}(\mu)/(\Sp^1 \times \Sp^1)$ can be identified with $\T \RR$, with coordinates $(r, v_r)$.
        The Routhian and the reduced external force take the form
        \begin{equation}
                R^{\mu} = \frac{m}{2}v_r^2-\frac{\mu_1^2}{2mr^2}-\frac{\mu_2^2}{2mk^2}\, , \quad \beta^{\mu}=2cr\frac{\mu_1} {m}\dd r\, .
        \end{equation}
        The reduced forced Euler--Lagrange equations for $(R^\mu, \beta^\mu$) seek for curves $r(t)$ in $\RR$ such that
        \begin{equation}
                \ddot r=\frac{\mu_1^2}{m^2r^3}-2cr\frac{\mu_1} {m^2}\, .
        \end{equation}

        Suppose that there are two rough walls at the axis $y=0$ and at $y=h$, where $h=\alpha R$ for some constant $\alpha>1$. Assume that the impact with a wall is such that the disk rolls without sliding and that the change of the velocity along the $y$-direction is characterized by an elastic constant $e$. When the disk hits one of the walls, the impact map is given by {(see {\cite[Example 8.2]{I.d.L+1997}}, and also {\cite[Example 3.7]{I.d.L+2001}})}
        \begin{equation}\label{eq:impact_map_disk_Cartesian}
               \Delta\colon \left(v_x^-, v_y^-, v_\varphi^-  \right)
                \mapsto \left(\frac{R^2 v_x^- +k^2 R v_\varphi^-}{k^2+R^2}, - ev_y^-, \frac{R v_x^- +k^2 v_\varphi^-}{k^2+R^2}  \right)
        \end{equation} 
        and the switching surface is $S = C_1 \cup C_2$, with 
        \begin{equation}
        \begin{aligned}
                & C_1 = \{(x,y,\varphi,v_x,v_y,v_\varphi)\mid y=R,\,  v_x =Rv_{\varphi} \hbox{ and } v_y<0\}\, ,\\
                & C_2 = \{(x,y,\vartheta,p_x,p_y,p_\vartheta)\mid y=h-R,\,  v_x =Rv_{\varphi} \hbox{ and } v_y>0\}\, .
        \end{aligned}
        \end{equation}
        Here the condition $v_x =Rv_{\varphi}$ comes from the nonholonomic constraint of the walls, whereas the conditions on the sign of $v_y$ ensure that the $y$-component of the velocity points towards corresponding the wall. Since the impact map changes the sign of $v_y$, the Zeno effect is avoided as long as $e\neq 0$.

        For the sake of simplicity, in what follows it will be assumed that $e=1$. It is worth noting that, despite the fact this corresponds to an elastic collision, the momentum map will not be preserved in the impact.


        In polar coordinates, using that
        \begin{equation}
               r^2=x^2+y^2\, ,\quad  \theta=\arctan \left(\frac{y}{x}\right)\, ,
        \end{equation}
        together with \eqref{eq:impact_map_disk_Cartesian} yields
        \begin{equation}
        \begin{aligned}
                v_\theta^+&=\frac{1}{1+(y/x)^2} \left(\frac{v_y^+ x-y v_x^+ }{x^2}\right)
                =\frac{1}{r^2} \left(-v_y^- x - y\frac{R^2 v_x^-+k^2 R v_\varphi^-}{R^2+k^2}\right)
                \\&
                =\frac{1}{r^2} \left(-v_y^- x - yv_x^- \right)
                =-v_\theta^-\, , \\
                v_r^+ &= \frac{1}{r} (x v_x^+ + y v_y^+)
                = \frac{1}{r} (x v_x^- - y v_y^-)
                \\& 
                = (2\cos^2 \theta -1) v_r^- -2r \sin \theta \cos \theta v_\theta^-
                \\& 
                = (2\cos^2 \theta -1) v_r^- + 2 \cos \theta (Rv_\varphi^- -v_r^- \cos \theta) 
                \\&
                = -v_r^- + 2 \cos \theta R v_\varphi^-  
                \, ,\\
                v_\varphi^+&=\frac{R v_x^- +k^2 v_\varphi^-}{k^2+R^2} 
                = \frac{R \left(\cos \theta\, v_r^-  -r\sin \theta\, v_\theta^- \right) +k^2 v_\varphi^-}{k^2+R^2} = v_\varphi^-\, .
                \label{eq:impact_map_disk_polar}
        \end{aligned}
        \end{equation}
        The connected componnents of the switching surface can be written as
        \begin{equation}
        \begin{aligned}
                C_1 & = \left\{(r,\theta,\varphi,v_r,v_\theta,v_\varphi)\mid r\sin \theta=R\, , \quad 
                v_r \cos \theta - r v_\theta \sin \theta=Rv_{\varphi}
                \right. \\ &\left.\quad
                \hbox{ and } 
                v_r \sin \theta + r v_\theta \cos \theta < 0
                \right\}\, , \\
                C_2 & = \left\{(r,\theta,\varphi,v_r,v_\theta,v_\varphi)\mid r\sin \theta=h-R\, , \quad 
                v_r \cos \theta - r v_\theta \sin \theta=Rv_{\varphi}
                \right. \\ &\left.\quad
                \hbox{ and } 
                v_r \sin \theta + r v_\theta \cos \theta > 0
                \right\}\, .
        \end{aligned}
        \end{equation}

        In polar coordinates, it is evident that the momentum map $\mommap_L = (mr^2 v_\theta, mk^2 v_\varphi)$ is a generalized hybrid momentum map but not a hybrid momentum map, that is, $\mommap_L(q_1, v_{q_1}^-)=\mommap_L(q_2, v_{q_2}^-)$ implies that $\mommap_L(q_1, v_{q_1}^+)=\mommap_L(q_2, v_{q_2}^+)$ but $\mommap_L(q_1, v_{q_1}^+)\neq \mommap_L(q_1, v_{q_1}^-)$.

        Let $\mu^-=(\mu_1^-,\mu_2^-)$ and $\mu^+(\mu_1^+,\mu_2^+)$ be the value of the momentum map before and after the impact, respectively. Equations~\eqref{eq:impact_map_disk_polar} lead to the following relation between these values of the momentum map:
        \begin{equation}
                \mu_1^+ = -\mu_1^-\, , \quad \mu_2^+ = \mu_2^-\, .
        \end{equation}
        Taking into account that $v_\theta^\pm=\mu_1^\pm/mr^2$ and $v_\varphi^\pm=\mu_2^\pm/mk^2$, the reduced connected components of the switching surface can be written as
        \begin{equation} \label{eq:switching_surface_disk_reduced}
        \begin{aligned}
        C_{1,\mu^{-}}&=\Big\{(r,v_r)\mid r\sin \gamma=R\, , \
        v_r \cos \gamma - \frac{\mu_1^-}{mr} \sin \gamma=R\frac{\mu_2^-}{mk^2}\, ,
        \\ &  \quad
        \hbox{ and }
        v_r \sin \gamma + \frac{\mu_1^-}{mr} \cos \gamma < 0
        \hbox{ for some } \gamma \in [0, 2\pi)\Big\}\, , \\
        C_{2,\mu^{-}}&=\Big\{(r,v_r)\mid r\sin \gamma=h-R\, , \
        v_r \cos \gamma -  \frac{\mu_1^-}{m} \sin \gamma=R\frac{\mu_2^-}{mk^2}\, ,
        \\ &  \quad
        \hbox{ and }
        v_r \sin \gamma + \frac{\mu_1^-}{mr} \cos \gamma > 0
        \hbox{ for some } \gamma \in [0, 2\pi)\Big\}\, ,
        \end{aligned}
        \end{equation}
        and the reduced impact map reads
        \begin{equation}
        \begin{aligned}
                \Delta_{\mu^-}\colon v_r^- & \mapsto  
                (2\cos^2 \gamma -1) v_r^- -2r \sin \theta \cos \gamma \frac{\mu_1^-}{mr^2} \\
                & =\quad  -v_r^- +2\cos \gamma R \frac{\mu_2^-}{mk^2}\, ,
        \end{aligned}
        \end{equation}
        where $\gamma$ is determined by the relation between $v_r^-, \, \mu_1^-$ and $\mu_2^-$. 
\end{example}

The theory developed in this chapter may also be extended to non-autonomous hybrid forced Lagrangian systems in a natural manner (see \cite{Colombo2022}).

\chapter{Integrability of hybrid systems}\label{ch:integrability_hybrid}

The aim of this chapter is to provide a notion of complete integrability for hybrid Hamiltonian systems and a way to construct action-angle variables for these systems. These results were previously published in the preprint \cite{Lopez-Gordon2023a}.

If a $2n$-dimensional Hamiltonian system $(\cT Q, \omega_Q, H)$ is completely integrable, then the Hamiltonian flows of the integrals $f_1, \ldots, f_n$ define an Abelian Lie group action of $\RR^n$ on $\cT Q$. In other words, the Hamiltonian vector fields $X_{f_1}, \ldots, X_{f_n}$ are the infinitesimal generators of the action. Hence, the momentum map can be identified with the map $F=(f_1, \ldots, f_n)\colon \cT Q\to \RR^n$. The combination of this with the notion of generalized hybrid momentum map (see \Cref{def:generalized_hybrid_momentum_map}) motivates the following definition.

\begin{definition}\label{def:generalized_hybrid_constants}
    Let $\hybrid=(M,X, S, \Delta)$ be a hybrid dynamical system. A function $f \colon M \to \RR$ is called a \emph{generalized hybrid constant of the motion} if 
    \begin{enumerate}
        \item $X (f) = 0$,
        \item For each connected component $C\subseteq S$ and each $a\in \Ima f$, there exists a $b\in \Ima f$ such that
        \begin{equation}\label{eq:condition_generalized_hybrid_constant}
            \Delta\left(\restr{f}{C}^{-1}(a)\right)  \subseteq f^{-1}(b)\, .
        \end{equation}
    \end{enumerate}
    In particular, $f$ is called a \emph{hybrid constant of the motion} if, in addition, $b=a$ for each  $a\in \Ima f$.

    In other words, a function $f$ is a generalized hybrid constant of the motion if the value of $f$ after an impact in $C$ is uniquely determined by its value before the impact; and $f$ is called a hybrid constant of the motion if the value of $f$ does not change in the impacts with $C$.
\end{definition}

\begin{definition}\label{def:completely_integrable_hybrid}
Let $Q$ be an $n$-dimensional manifold.
A \emph{completely integrable hybrid Hamiltonian system} is a $5$-tuple
$(M, X_H, S,  \Delta, F)$, formed by a hybrid Hamiltonian system $\hybrid_H=(M, X_H, S,  \Delta)$ with underlying Hamiltonian system $(\cT Q, \omega_Q, H)$, together with a map $F=(f_1, \ldots, f_n)\colon M \to \RR^n$ such that $\operatorname{rank} \T_x F = n$ almost everywhere and the functions $f_1, \ldots, f_n$ are generalized hybrid constant of the motion and in involution, that is $\{f_i, f_j\}=0$ for all $i, j\in \{1, \ldots, n\}$.
\end{definition}

By the Liouville--Arnol'd Theorem (\Cref{theorem:Liouville-Arnold}), on a neighborhood $U_\Lambda$ of each regular level set $M_\Lambda = F^{-1}(\Lambda)$ there exists action-angle coordinates $(\varphi^i, s_i)$ such that the dynamics are given by
\begin{equation}
    \dot{\varphi}^i(t) = \Omega^i(s_1, \ldots, s_n)\, ,\quad  \dot{s_i}(t) = 0\, ,
\end{equation}
for $(\varphi^i(t), s_i(t))\in U_\Lambda \setminus S$. On the other hand, condition~\eqref{eq:condition_generalized_hybrid_constant} implies that, for each level set $M_\Lambda$ and each connected component $C\subseteq S$, there exists a $\Lambda'\in \RR^n$ such that $\Delta (M_\Lambda\cap C) \subset M_{\Lambda'}=F^{-1}(\Lambda')$. In other words, the impact map takes invariant level sets of $F$ into level sets. Since each level set is uniquely determined by the action coordinates $s_i$ and vice versa, this implies that, for impacts occurring on a fixed connected component of the switching surface, the value of the action coordinates after the impact depends exclusively on their value before the impact, and they are independent of the value of the angle coordinates. This can be summarized as follows.

\begin{theorem}\label{theorem:Liouville_Arnold_hybrid}
    Consider a completely integrable hybrid Hamiltonian system $(M, S, X_H, \Delta)$, with underlying Hamiltonian system $(\cT Q, \omega_Q, H)$ and $F=(f_1, \ldots, f_n)$, where $n=\dim Q$. Let $M_{\Lambda}$ be a regular level set of $F$, namely, $M_\Lambda= F^{-1}(\Lambda)$ for $\Lambda\in \RR^n$ such that $\operatorname{rank} \T_x F = n$ for all $x\in M_{\Lambda}$. Then:
    \begin{enumerate}
        \item Each regular level set $M_\Lambda$ is a Lagrangian submanifold of $\cT Q$, and it is invariant with respect to the flows of $X_H, X_{f_1}, \ldots, X_{f_n}$.
        \item Any compact connected component of $M_\Lambda$ is diffeomorphic to an $n$-dimensional torus $\mathbb{T}^n$.
        \item For each regular level set $M_\Lambda$ and each connected component $C\subseteq S$, there exists a $\Lambda'\in \RR^n$ such that $\Delta(M_\Lambda \cap C) \subset M_{\Lambda'} = F^{-1}(\Lambda')$. 
        \item On a neighbourhood $U_\lambda$ of $M_\Lambda$ there are coordinates $(\varphi^i, s_i)$ such that
        \begin{enumerate}
            \item they are Darboux coordinates for $\omega$, namely, $\omega = \dd \varphi^i \wedge \dd s_i$,
            \item the action coordinates $s_i$ are functions depending only on the integrals $f_1, \ldots, f_n$,
            \item the continuous part hybrid dynamics are given by
             $$\dot \varphi^i = \Omega^i(s_1, \ldots, s_n),\qquad \dot s_i = 0\, .
             $$
             \item In these coordinates, for each connected component $C\subseteq S$, the impact map reads $\Delta \colon (\varphi^i_-, s_i^-)\in M_\Lambda \cap C\mapsto  (\varphi^i_+, s_i^+)\in M_{\Lambda'}$, where $s_1^+, \ldots, s_n^+$ are functions depending only on $s_1^-, \ldots, s_n^-$.
        \end{enumerate}
    \end{enumerate}
\end{theorem}
 
\begin{remark}
    This notion of integrability may be useful for studying the dynamics of some hybrid Hamiltonian systems. Nevertheless, unlike integrable systems in the usual sense, an integrable hybrid Hamiltonian system may be chaotic. For instance, a billiard (that is, a particle undergoing free motion within a plane, encountering collisions with a wall) is an integrable hybrid Hamiltonian system and there are chaotic billiards \cite{C.M2006}.
\end{remark}

\section{Examples}
\subsection{Rolling disk with a harmonic potential hitting fixed walls}\label{example:disk_integrability}

Consider a homogeneous circular disk of radius $R$ and mass $m$ moving in the plane. The configuration space is $Q = \RR^2 \times \Sp^1$, with canonical coordinates $(x, y, \theta)$. The coordinates $(x, y)$ represent then position of the center of the disk, while the coordinate $\theta$ represents the angle between a fixed reference point of the disk and the $y$-axis.
Suppose that the Hamiltonian function $H\colon \cT Q \to \RR$ of the system is
\begin{equation}\label{eq:Hamiltonian_disk_oscillator}
    H = \frac{1}{2m} (p_x^2 + p_y^2) + \frac{1}{2mk^2} p_\theta^2 + \frac{1}{2} \Omega^2 (x^2+y^2)\, ,
\end{equation}
where $(x, y, \theta, p_x, p_y, p_\theta)$ are the bundle coordinates in $\cT(\RR^2\times \Sp^1)$. 
Here $m$ is the mass of the disk, $k$ is a constant such that $mk^2$ is the moment of inertia of the disk, and $\Omega$ is a constant such that $\Omega/\sqrt{m}$ is the frequency of oscillation.
Consider that there are two rough walls situated at $y=0$ and at $y=h>R$. Assume that the impact with a wall is such that the disk rolls without sliding and that the change of the velocity along the $y$-direction is characterized by an elastic constant $e$. Then, the switching surface is $S=C_1 \cup C_2$, where
\begin{equation}\label{eq:impact_surface_disk}
\begin{aligned} 
    & C_1 = \{(x,y,\vartheta,p_x,p_y,p_\vartheta)\mid y=R,\,  p_x=R p_{\vartheta}/k^2 \hbox{ and } p_y<0\}\, ,\\
    & C_2 = \{(x,y,\vartheta,p_x,p_y,p_\vartheta)\mid y=h-R,\,  p_x=R p_{\vartheta}/k^2 \hbox{ and } p_y>0\}\, ,
\end{aligned}
\end{equation}
and the impact map $\Delta\colon S\to \cT Q$ is given by 
\begin{equation}\label{eq:impact_map_disk}
    \Delta \colon \left(p_x^{-}, p_y^{-}, p_{\theta}^{-}\right) \mapsto\left(\frac{R^2 p_x^{-}+ R p_{\theta}^{-}}{k^2+R^2},-e p_y^{-}, k^2\frac{R p_x^{-}+p_{\theta}^{-}}{k^2+R^2}\right)
\end{equation}
Here the condition $p_x=R p_{\vartheta}/k^2$ comes from the nonholonomic constraint of the walls, whereas the conditions on the sign of $p_y$ ensure that the $y$-component of the momenta points towards corresponding the wall.
The switching surface and the impact map are the Hamiltonian counterparts of those from \Cref{example:disk_reduction}.

For simplicity's sake, assume hereafter that $m=R=k=\Omega=1$. 
The functions
\begin{equation}\label{eq:action_coords_disk}
    f_1 = \frac{p_x^2 + x^2}{2}\, , \quad
    f_2 = \frac{p_y^2 + y^2}{2}\, , \quad
    f_3 = \frac{p_\theta^2}{2}\, ,
\end{equation}
are conserved quantities with respect to the Hamiltonian dynamics of $H$. Moreover, they are in involution and functionally independent almost everywhere. 

Let $F=(f_1, f_2, f_3)\colon \cT(\RR^2 \times \Sp)\to \RR^3$. It is clear that, for $\Lambda\neq 0$, the level sets $F^{-1}(\Lambda)$ are diffeomorphic to $\Sp \times \Sp \times \RR$.
In the intersection of their domains of definition, the functions
\begin{equation}\label{eq:angle_coords_disk}
    \phi^1 = \arctan\left(\frac{x}{p_x}\right)\, , \quad 
    \phi^2 = \arctan\left(\frac{y}{p_y}\right)\, , \quad 
    \phi^3 = \frac{\theta}{p_\theta}
\end{equation}
are coordinates on each level set $F^{-1}(\Lambda)$ for $\Lambda\neq 0$. Moreover, observe that $(\phi^i, f_i),\, i\in \{1, 2, 3\}$ are Darboux coordinates for the canonical symplectic form $\omega_Q$, namely, $\omega_Q = \dd \phi^i\wedge \dd f_i$.
In these coordinates, the Hamiltonian function reads
\begin{equation}
    H = f_1 + f_2 + f_3\, .
\end{equation}
Hence, its Hamiltonian vector field is simply
\begin{equation}
    X_H = \frac{\partial}{\partial\phi^1} + \frac{\partial}{\partial\phi^2} + \frac{\partial}{\partial\phi^3}\, .
\end{equation}


Equations~\eqref{eq:impact_surface_disk}, \eqref{eq:impact_map_disk} and \eqref{eq:action_coords_disk}, imply that
\begin{equation}
\begin{aligned}
    \Delta \left(f_1^{-1}(\mu_1^-)\cap C_1\right) 
    &= \left\{\left(x, R, \theta, p_x, -e p_y, \frac{k^2}{R} p_x\right)\mid
    \right.\\ &\left.\qquad
    p_x^2+x^2 = 2\mu_1^-,\, x, p_y \in \RR, \, \theta \in \Sp^1 \right\} 
    \\ &
    \subset f_1^{-1}(\mu_1^-)\, , \\
    \Delta \left(f_2^{-1}(\mu_2^-)\cap C_1\right) 
    &= \left\{\left(x, R, \theta, p_x, -e p_y, \frac{k^2}{R} p_x\right)\mid
    \right.\\ &\left.\qquad 
    (-ep_y)^2+R^2 = 2e^2\mu_2^- +(1-e^2)R^2 = 2 \mu_2^+, 
    \right.\\ &\left.\qquad
    x, p_y \in \RR, \, \theta \in \Sp^1 \right\} 
    \\ &
    \subset f_2^{-1}(\mu_2^+)\, , \\
    \Delta \left(f_3^{-1}(\mu_3^-)\cap C_1\right) & \subset f_3^{-1}(\mu_3^-)\, ,
\end{aligned}
\end{equation}
and analogously for $C_2$.
Thus, $f_1$ and $f_3$ are hybrid constants of the motion, while $f_2$ is a generalized hybrid constant on the motion. In particular, for a purely elastic impact ($e=1$), $\mu_2^+=\mu_2^-$ and $f_2$ is a hybrid constant on the motion

The next step is to compute how the impact map modifies the action-angle coordinates. 
Equations~\eqref{eq:impact_map_disk} and \eqref{eq:action_coords_disk}, imply that
\begin{equation}
\begin{aligned}
    & f_1 \circ \Delta = \frac{1}{2}\left(\frac{R^2 p_x+ R p_{\theta}}{k^2+R^2}\right)^2+\frac{x^2}{2}= \frac{p_x^2+x^2}{2}\, , \\
    & f_2 \circ \Delta =  \frac{e^2p_y^2+y^2}{2}\, , \\
    & f_3 \circ \Delta = \frac{1}{2}\left(k^2\frac{R p_x+ p_{\theta}}{k^2+R^2}\right)^2 = \frac{p_\theta^2}{2}\, ,
\end{aligned}
\end{equation}
taking into account that $p_\theta = k^2 p_x/R$ for points in the impact surface $S$.
Thus, 
\begin{equation}
\begin{aligned}
    f_1 \circ \Delta = f_1\, , \quad 
    f_3 \circ \Delta = f_3\, ,
\end{aligned}
\end{equation}
On the other hand,
\begin{equation}
    \restr{f_2}{S} = \frac{p_y^2+a^2}{2}\, ,
\end{equation}
where $a=R$ or $a=h-R$ for impacts on $C_1$ and $C_2$, respectively. Thus, along $S$, one can write $p_y^2 = 2 f_2 - a^2$. Hence,
\begin{equation}
    f_2 \circ \Delta = \frac{2e^2 f_2 -e^2 a^2 + a^2}{2} = e^2 f_2 + \frac{1-e^2}{2} a^2\, .
\end{equation}
The combination of equations~\eqref{eq:impact_map_disk} and \eqref{eq:angle_coords_disk} yields
\begin{equation}
\begin{aligned}
    & \phi^1 \circ \Delta  = \arctan\left(x\frac{k^2+R^2}{R^2 p_x+ R p_{\theta}}\right) = \arctan\left(\frac{x}{p_x}\right)= \phi^1 \, , \\
    & \phi^2 \circ \Delta = \arctan\left(\frac{y}{-ep_y}\right)
    = -\arctan\left(\frac{y}{ep_y}\right)\, , \\
    & \phi^3 \circ \Delta = \theta \frac{k^2+R^2}{k^2(R p_x+ p_{\theta})} 
    = \frac{\theta}{p_\theta} = \phi^3 \, , 
\end{aligned}
\end{equation}
using that $p_\theta = k^2 p_x/R$ for points in the impact surface $S$.
One can write 
\begin{equation}
    \frac{y}{p_y} = \tan \phi^2 \, ,
\end{equation}
and therefore
\begin{equation}
    \phi^2 \circ \Delta = -\arctan\left(\frac{\tan \phi^2 }{e}\right)\, .
\end{equation}

The last step is to write the impact surface in terms of the action-angle coordinates $(\phi^i, f_i)$. One can write $y= \sqrt{2 f_2} \sin \phi^2$.
The fact that $y=a$ (where $a=R$ or $a=h-R$) on $S$ implies that
\begin{equation}
    \restr{\left(2f_2 \sin^2 \phi^2\right)}{S} =  a^2\, .
\end{equation}
Similarly, one can write $p_x = \sqrt{2 f_1} \cos \phi^1$, and thus
\begin{equation}
    \restr{f_3}{S} = \frac{2k^4 f_1 \cos^2 \phi^1}{R^2}\, ,
\end{equation}
using that $p_\theta = k^2 p_x/R$ on $S$. Consequently, in the action-angle coordinates, the impact surface reads
\begin{equation}
\begin{aligned}
    S & = \left\{\left(\phi^i, f_i\right)\mid 2f_2 \sin^2 \phi^2 = R^2 \text{ and } f_3 = \frac{2k^4 f_1 \cos^2 \phi^1}{R^2} \right\} \\
    & \cup \left\{2f_2 \sin^2 \phi^2 = (h-R)^2 \text{ and } f_3 = \frac{2k^4 f_1 \cos^2 \phi^1}{R^2} \right\}\, .
\end{aligned}
\end{equation} 
The relations between the coordinates before, $(\phi^i_-, f_i^-)$, and after, $(\phi^i_+, f_i^+)$, are
\begin{equation}
\begin{array}{lll}
    \phi^1_+ = \phi^1_-\, , \quad 
    & \phi^2_+ = -\arctan\left(\frac{\tan \phi^2_-}{e}\right)\, , \quad
    & \phi^3_+ = \phi^3_-\, , \\ \\
    f_1^+ = f_1^- \, , \quad
    & f_2^+ = e^2 f_2 + \frac{1-e^2}{2} a^2\, , \quad
    & f_3^+ = f_3^-\, ,
\end{array}
\end{equation}
where $a=R$ or $a=h-R$ depending on the wall where the impact takes place.

\subsection{Pendulum hitting a surface} 
Consider a pendulum mounted on the floor.
The configuration space is $Q=\mathbb{S}$ with generalized coordinate $\theta$. Coordinates on $\cT \mathbb{S}$ are denoted by $(\theta,p)$. The Hamiltonian function of the system $H:\cT \mathbb{S}\to\mathbb{R}$ is given by 
$$H(\theta,p)=\frac{p^{2}}{2ml^2}+mgl(1-\cos\theta)\, .$$ 
Henceforth, assume that $m=g=l=1$.
The vector field describing the continuous-time dynamics is the Hamiltonian vector field $X_H$ of $H$ with respect to the canonical symplectic structure, namely,
\begin{equation}
    X_H = p \frac{\partial}{\partial \theta} - \sin \theta \frac{\partial}{\partial p}\, .
\end{equation}
The switching surface is given by 
$$C=\{(\theta,p)\in \cT \mathbb{S}|\cos\theta=0 \hbox{ and }p\geq 0\}\, .$$
The impact map $\Delta:S\to \cT \mathbb{S}$ is given by 
\begin{equation}\label{eq:impact_map_pendulum_canonical}
    \Delta(\theta,p)=(\theta,-ep)\, ,
\end{equation}
where $e\in [0,1]$  denotes the coefficient of restitution. In particular, for a perfectly elastic impact $e=1$, and for a perfectly plastic
impact $e=0$. In other words, the coordinates before, $(\theta^-, p^-)$, and after, $(\theta^+, p^+)$, the impact are related by
\begin{equation}\label{eq:impact_map_pendulum_canonical_2}
    \theta^+ = \theta^-\, , \quad p^+ = -e p^-\, .
\end{equation}

 Therefore the system \begin{equation*}
    \Sigma_{\mathscr{H}}:
    \left\{\begin{array}{ll}\dot{\theta}(t)=\frac{p}{ml^2},\dot{p}=-mgl\sin\theta, & \hbox{ if } \cos\theta(t)\neq 0,\, p(t)> 0,\\ \theta^{+}=\theta^{-},\,p^{+}=-ep^{-},&\hbox{ if } \cos\theta(t)=0,\, p(t)\geq 0, 
    \end{array}\right.
  \end{equation*}
is a hybrid Hamiltonian system.

The Hamiltonian function is a conserved quantity with respect to the continuous dynamics, that is, $X_H (H) = 0$. Moreover, it is a hybrid constant of the motion (that is, $H \circ \Delta = H$) if and only if the coefficient of restitution is $e=1$. 

It is convenient to write the level sets of $H$ in terms of a parameter $\kappa$, which can be interpreted as half of the energy. There are two types of invariant submanifolds, the so-called \emph{libration} and \emph{rotation} cases, corresponding to the level sets $H^{-1}(2\kappa)$ for $\kappa< 1$ and $\kappa>1$, respectively.

In the following, the analysis will be restricted to the libration case ($\kappa<1$).
In that case, the action coordinate is given by (see \cite{Brizard2013})
\begin{equation}\label{eq:action_coordinate_libration}
\begin{aligned}
    J_\ell(\theta, p) & = \frac{8}{\pi}\left[\mathrm{E}\left(\frac{H(\theta,p)}{2}\right)
    \right.\\ &\quad \left.
    -\left(1-\frac{H(\theta,p)}{2}\right) \mathrm{K}\left(\frac{H(\theta,p)}{2}\right)\right]\, ,
\end{aligned}
\end{equation}
where $\mathrm{K}$ and $\mathrm{E}$ denote the complete elliptic integrals of the first and second kinds, respectively. 
The canonical coordinates as a function of $\kappa$ and the angle coordinate $\zeta_\ell$ are
\begin{equation}\label{eq:angle_coordinate_libration}
\begin{aligned}
    & \theta\left(\kappa, \zeta_{\ell}\right)=2 \arcsin \left[\sqrt{\kappa} \sn\left(\frac{2 \mathrm{K}(\kappa)}{\pi} \zeta_{\ell} \mid \kappa\right)\right]\, , \\
    & p\left(\kappa, \zeta_{\ell}\right)=2 \sqrt{\kappa} \cn\left(\frac{2 \mathrm{K}(\kappa)}{\pi} \zeta_{\ell} \mid \kappa\right)\, ,
\end{aligned}
\end{equation}
where $\sn$ and $\cn$ denote the Jacobi elliptic functions. Refer to \cite{Lawden1989, O.L.B+2010} for more details about elliptic functions.

The manifold $\cT \mathbb{S} \simeq \mathbb{S} \times \mathbb{R}$ is foliated by leaves which are diffeomorphic to $\mathbb{S}$. The action coordinate $J_\ell$ determines the leave of the foliation, while the angle coordinate $\zeta_\ell$ is a coordinate on the leave. Hence, the condition of hybrid momentum map implies that if $\Delta (\zeta_\ell^-, J_\ell^-) =  (\zeta_\ell^+, J_\ell^+)$, then $\Delta (\tilde{\zeta}_\ell^-, J_\ell^-) =  (\tilde{\zeta}_\ell^+, J_\ell^+)$ for all $(\zeta_\ell^-, J_\ell^-), (\tilde{\zeta}_\ell^-, J_\ell^-)\in C$. In other words, the action coordinate in the instant after the impact, $J_\ell^+$, depends only on the action coordinate in the instant before the impact, $J_\ell^-$. This means that if the leaves of the foliation are invariant under the hybrid dynamics, that is, if the initial conditions are in one leave, after the impact the system will remain in the same leave.

By equations \eqref{eq:impact_map_pendulum_canonical} and \eqref{eq:action_coordinate_libration}, this condition is verified in the case of a completely elastic impact ($e=1$). As a matter of fact, in that case the impact map does not modify the action coordinate, namely, $J_\ell^+ = J_\ell ^-$. Similarly, $\Delta$ does not change $\kappa$. By equations~\eqref{eq:impact_map_pendulum_canonical_2} and \eqref{eq:angle_coordinate_libration}, the relation between the angle coordinates before, $\zeta_\ell^-$, and after, $\zeta_\ell^+$, the impact is given by
\begin{equation}\label{eq:impact_map_pendulum_libration}
\begin{aligned}
    & \sn\left(\frac{2 \mathrm{K}(\kappa)}{\pi} \zeta_{\ell}^+ \mid \kappa\right)
    =\sn\left(\frac{2 \mathrm{K}(\kappa)}{\pi} \zeta_{\ell}^- \mid \kappa\right)\, , \\
    & \cn\left(\frac{2 \mathrm{K}(\kappa)}{\pi} \zeta_{\ell}^+ \mid \kappa\right) 
    = - \cn\left(\frac{2 \mathrm{K}(\kappa)}{\pi} \zeta_{\ell}^- \mid \kappa\right)\, ,
\end{aligned}
\end{equation}
where it has been taken into account that $\arcsin\colon [-1,1] \to \RR$ is an injective function.

In the low energy limit ($\kappa\ll 1$), the Jacobi elliptic functions behave like trigonometric functions, namely, $\sn(x| \kappa) = \sin x + \mathcal{O}(\kappa)$ and $\cn(x| \kappa) = \cos x + \mathcal{O}(\kappa)$. Hence, equations~\eqref{eq:impact_map_pendulum_libration} can be approximated by
\begin{equation}\label{eq:impact_map_pendulum_low_energy}
\begin{aligned}
    & \sin\left(\frac{2 \mathrm{K}(\kappa)}{\pi} \zeta_{\ell}^+ \right)
    = \sin\left(\frac{2 \mathrm{K}(\kappa)}{\pi} \zeta_{\ell}^- \right) + \mathcal{O}(\kappa)\, , \\
    & \cos\left(\frac{2 \mathrm{K}(\kappa)}{\pi} \zeta_{\ell}^+ \right) 
    = - \cos\left(\frac{2 \mathrm{K}(\kappa)}{\pi} \zeta_{\ell}^- \right) + \mathcal{O}(\kappa)\, ,
\end{aligned}
\end{equation}
whose solution is $\zeta^+ = \pi - \zeta^- + \mathcal{O}(\kappa)\, (\operatorname{mod} 2\pi)$.

\chapter{Hamilton--Jacobi theory for hybrid systems}\label{ch:hybrid_HJ}

The aim of this chapter is to develop a Hamilton--Jacobi theory for nonholonomic and forced hybrid dynamical systems. See \Cref{sec:Hamilton-Jacobi} for a review on Hamilton--Jacobi theory.
A Hamilton--Jacobi for hybrid Hamiltonian systems has been studied by Clark \cite{Clark2020}, whose approach is here extended to forced and nonholonomic hybrid systems. Moreover, the notion of complete solution of the Hamilton--Jacobi problem for hybrid Hamiltonian (forced, or nonholonomic) systems is presented. The results are illustrated with several examples.

The results of the present chapter were previously published in the preprint \cite{Colombo2022b}.

\section{Hamilton--Jacobi theory for hybrid Hamiltonian systems}

\begin{definition}\label{def:solution_hybrid_HJ}
    Let $\hybrid_H= (M, X_H, S, \Delta)$ be a hybrid Hamiltonian system with underlying Hamiltonian system $(\cT Q, \omega_Q, H)$. Consider a set  $\mathfrak{U} = \left\{(U_k,\gamma_k)  \right\}$ whose elements are pairs $(U_k,\gamma_k)$, where $U_k\subseteq Q\setminus \pi_Q(S)$ are open subsets of $Q$ and $\gamma_k \in \Omega^1 (U_k)$ are closed one-forms. 
    The set $\mathfrak{U}$ is called a \emph{solution of the hybrid Hamilton--Jacobi problem} for $\hybrid_H$ if:
    \begin{enumerate}
        \item For each $k$, $\gamma_k$ is a solution of the Hamilton--Jacobi equation for $H$ on $U_k$, namely,
        \begin{equation}
            \dd (H \circ \gamma_k) = 0\, . 
            \label{HJ_eq_hybrid}
        \end{equation}
        \item If $x\in \partial U_k \cap \pi_Q(S)$, then there exists an index $l$ such that $x\in \partial U_l \cap \pi_Q(S)$ and $\gamma_l$ and $\gamma_k$ are $\Delta$-related, that is,
        \begin{equation}
            \lim_{y\to x} \gamma_l(y) = \Delta \left( \lim_{y\to x} \gamma_k(y) \right)\, .
            \label{Delta_related_condition}
        \end{equation}
        The indices $k$ and $l$ may be different or not. This will depend on the impact map of the problem.
    \end{enumerate}
\end{definition}

Unlike continuous systems, where a solution of the Hamilton--Jacobi equation is a single one-form $\gamma$, in the hybrid Hamilton--Jacobi equation a family of one-forms $\gamma_k$ is required. Each one-form $\gamma_k$ will determine the dynamics between the $k$-th and $(k+1)$-th impacts (regarding the initial conditions as the ``$0$-th impact''). For simplicity's sake, along this dissertation it will be assumed that Zeno effect does not occur so that the impacts are discrete and $k\in \NN$. Nevertheless, one could consider systems that experience Zeno by replacing the integer index $k$ with a real parameter. It is also worth remarking that the open subsets $U_k$ could have non-empty intersection. As a matter of fact, it is possible to have $U_k = Q \setminus \pi_Q(S)$ for all $k$.

\begin{remark}\label{remark:def_Clark}
    Clark's \cite[Definition VIII.1]{Clark2020} definition additionally requires the condition 
    \begin{equation}
        \lim_{y\to x} \mathcal{A}_l (y)  = \lim_{y\to x} \mathcal{A}_k (y),  
    \end{equation}
    where $\mathcal{A}_k\in \Cinfty(U_k)$ and $\mathcal{A}_l\in \Cinfty(U_l)$ are functions such that $\gamma_k = \dd \mathcal A_k$ and $\gamma_l = \dd \mathcal A_l$. This condition, albeit important for optimal control purposes, is irrelevant for the aims for the present chapter. Therefore, it will not be required.
\end{remark}

The geometric interpretation of the hybrid Hamilton--Jacobi problem is as follows. Given a collection of open subsets $U_k\subset Q \setminus \pi_Q(S)$ and closed one-forms $\gamma_k\in \Omega^1 (U_k)$, one can project the vector fields $\restr{X_H}{\cT U_k}\in \mathfrak{X}(\cT U_k)$ along $\gamma_k(U_k)$, obtaining the vector fields 
\begin{equation}
    X_H^{\gamma_k}\coloneqq \T \pi_Q \circ X_H \circ \gamma_k \in \mathfrak{X}(U_k) \, .
\end{equation}
Additionally, it is assumed that $\restr{X_H}{\cT U_k}$ and $X_H^{\gamma_k}$ are $\gamma_k$-related, that is, $\T\gamma_k \circ X_H^{\gamma_k} =  \restr{X_H}{\cT U_k}\circ \gamma$. In other words, the following diagram commutes:
\begin{equation}
\begin{tikzcd}
    \cT U_k \arrow[rr, "\restr{X_H}{\cT U_k}"] \arrow[d, "\pi_{Q\mid \cT U_k}"] &  & \T \cT U_k \arrow[d, "\T\pi_{Q\mid \cT U_k}"']    \\
    U_k \arrow[rr, "X_H^{\gamma_k}"'] \arrow[u, "\gamma_k", bend left, shift left]    &  & \T U_k \arrow[u, "\T\gamma_k"', bend right, shift right]
\end{tikzcd}
\end{equation}
Therefore, $\gamma_k$ is a solution of the Hamilton--Jacobi equation \eqref{eq:HJ_geometric} for $\restr{H}{U_k}$. Moreover, the one-forms $\gamma_l$ and $\gamma_k$ have to be $\Delta$-related on $\partial U_k \cap \partial U_l \cap \pi_Q(S)$, that is, they have to satisfy equation~\eqref{Delta_related_condition}.

\begin{theorem}[Hybrid Hamilton--Jacobi theorem]\label{theorem:HJ_hybrid}
    Consider a hybrid Hamiltonian system $\hybrid_H= (M, X_H, S, \Delta)$ with underlying Hamiltonian system $(\cT Q, \omega_Q, H)$.
    Let $\mathfrak{U}$ be a set formed by pairs $(U_k,\gamma_k)$, where $U_k\subseteq Q\setminus \pi_Q(S)$ are open subsets of $Q$ and $\gamma_k \in \Omega^1 (U_k)$ are closed one-forms. 
    Then, the following statements are equivalent:
    \begin{enumerate}
        \item The set $\mathfrak{U}$ is a solution of the hybrid Hamilton--Jacobi problem for $\hybrid_H$. 
        \item For every continuous and piecewise smooth curve $c\colon \RR\to Q$ such that 
    \begin{enumerate}
        \item $c\big((t_k, t_{k+1})\big)\subset U_k$, 
        \item $c$ intersects $\pi_Q(S)$ at $\{t_k\}_k$,
        \item $c$ satisfies the equations
    \begin{align}
    &\dot c(t) = T \pi_Q \circ X_H \circ \gamma_k \circ c(t), \qquad t_k<t<t_{k+1},\\
    & \gamma_{k+1} \circ c(t_{k+1}) = \Delta \circ \gamma_k \circ c(t_{k+1}),
    \end{align}
    \end{enumerate}
    the curve $\tilde{c}\colon \RR \to \cT Q$ given by $\tilde{c}(t) = \gamma_k \circ c(t)$ for $t\in[t_k, t_{k+1})$ is an integral curve of the hybrid dynamics.
    \end{enumerate}
\end{theorem}

This theorem, with the additional assumption from Remark~\ref{remark:def_Clark}, was proven in \cite{Clark2020} (Theorem VIII.3).

\begin{proof}
    As it was presented in \Cref{sec:Hamilton-Jacobi}, equation~\eqref{HJ_eq_hybrid} holds if and only if, for each integral curve $c\colon I\subseteq \RR\rightarrow U_k$ of $X_{H}^{\gamma_k}$, the curve $\gamma_k\circ c$ is an integral curve of $\restr{X_H}{\cT U_k}$. Since the dynamics $x(t)$ of $\hybrid_H$ are given by the integral curves of $X_H$ for $x(t)\notin S$, the curves $x(t)=\gamma_k \circ c(t)$ are the hybrid dynamics on $\cT U_k$.
    
    On the other hand, if $c$ is a continuous curve that intersects $\pi_Q(S)$ at $\{t_k\}_k$ and $c(t_{k+1}) \in \partial U_k \cap \partial U_{k+1} \cap \partial \big(\pi_Q(S)\big)$, then $\gamma_{k+1}$ and $\gamma_k$ are $\Delta$-related if and only if
    \begin{equation}
        \gamma_{k+1} \circ c(t_{k+1}) = \Delta \circ \gamma_k \circ c(t_{k+1}) \, .
    \end{equation}
\end{proof}

\begin{definition}\label{def:complete_sols}
    Let $Q$ be an $n$-dimensional manifold and consider a hybrid Hamiltonian system $\hybrid_H= (M, X_H, S, \Delta)$ with underlying Hamiltonian system $(\cT Q, \omega_Q, H)$.
    A \emph{complete solution of the hybrid Hamilton--Jacobi problem for $\hybrid_H$}  is a set $\mathfrak{U}_{\mathrm{com}} = \left\{(U_k,\Phi_k)  \right\}$ formed by pairs $(U_k,\Phi_k)$, where $U_k\subseteq Q\setminus \pi_Q(S)$ are open subsets and $\Phi_k \colon U_k \times \RR^n \to \cT U_k$ are local diffeomorphisms such that, for each $\lambda\in \RR^n$ and each index $k$,
    \begin{enumerate}
        \item the map $({\gamma_{k}})_{\lambda} = \Phi_k(\cdot, \lambda) \colon U_k \to \cT U_k$ is a solution of the Hamilton--Jacobi equation for $\restr{H}{U_k}$,
        \item there exists a $\mu \in \R^n$ and an index $l$ for which
        \begin{equation}
            \Ima \restr{\big(\Delta \circ ({\gamma_{k}})_{\lambda}\big)}{\pi_Q(S)}\subset \Ima\,  ({\gamma_{l}})_{\mu}\, .
        \end{equation}
    \end{enumerate}

\end{definition}

Before the first impact takes place, the initial value of the parameters $\lambda\in \RR^n$ will be fixed by the initial conditions of the dynamics, and the subsequent values of these parameters will be determined by the relation $\operatorname{Im} (\Delta \circ ({\gamma_{k}})_{\lambda})\subseteq \operatorname{Im} ({\gamma_{l}})_{\mu}$.

\begin{remark}\label{remark:Liouville_integrability_hybrid}
    Let $\mathfrak{U}_{\mathrm{com}} = \left\{(U_k,\Phi_k)  \right\}$ be a complete solution of the hybrid Hamilton--Jacobi problem for $\hybrid_H= (M, X_H, S, \Delta)$. By definition, for each $k$, $\Phi_k\colon U_k\times \R^n\to \cT U_k$ is a complete solution of the hybrid Hamilton--Jacobi problem for $\restr{H}{\cT U_k}$. Therefore, each of the open subsets $\cT U_k\subset \cT Q$ has a foliation on Lagrangian submanifolds invariant under the flow of $\restr{X_H}{\cT U_k}$. These submanifolds are given by $\Ima\Phi_k(\cdot, \lambda)$ for each $\lambda\in \R^n$.
    The condition $\Ima (\Delta \circ ({\gamma_{k}})_{\lambda})\subseteq \Ima ({\gamma_{l}})_{\mu}$ implies that the impact map $\Delta$ maps the border of each Lagrangian submanifold of $\cT U_k$ into a subset of a Lagrangian submanifold of $\cT U_l$. If $U_k$ is an open cover of $Q\setminus \pi_Q(S)$, then the Lagrangian submanifolds $\operatorname{Im} ({\gamma_{l}})_{\mu}$ will foliate the phase space $\cT Q$. The parameters $\mu$ and $\lambda$ are related by the condition $\operatorname{Im} (\Delta \circ ({\gamma_{k}})_{\lambda})\subseteq \operatorname{Im} ({\gamma_{l}})_{\mu}$.
    Hence, obtaining a complete solution of the hybrid Hamilton--Jacobi problem can be useful for computing the dynamics of certain hybrid Hamiltonian systems. The motion between one impact and the next one will be quasi-periodic (as a consequence of Liouville--Arnold theorem). Nevertheless, an integrable hybrid Hamiltonian system may be chaotic. For instance, a billiard (that is, a particle undergoing free motion within a plane, encountering collisions with a wall) is an integrable hybrid Hamiltonian system and there are chaotic billiards \cite{C.M2006}.
\end{remark}

It is also worth mentioning that in \cite{Clark2020} (completely) integrable hybrid Hamiltonian systems are defined in terms of Lagrangian submanifolds. 
In that reference (see also \cite{C.O2021}), a submanifold $\mathcal{L}\subset \cT Q$ is called a hybrid Lagrangian submanifold if $\mathcal{L}\setminus \partial \mathcal{L}$ is a Lagrangian submanifold of $\cT Q$ and $\pi_Q(\partial \mathcal{L})$ is contained in $\pi_Q(\partial S)$ (see Definition V.III.6 in \cite{Clark2020}). 
More concretely (cf.~Definitions V.III.5 and V.III.6 in \cite{Clark2020}), a hybrid Hamiltonian system $\hybrid_H= (\cT Q, X_H, S, \Delta)$ is called completely integrable if there exists a foliation $\{\mathcal{L}_\alpha\}$ of $\cT Q$ such that, for each leaf $\mathcal{L}_\alpha$,
\begin{enumerate}
   \item \label{Clark_integrable_1_2} $\mathcal{L}_\alpha$ is a hybrid Lagrangian submanifold of $\cT Q$,
    \item \label{Clark_integrable_3} $\mathcal{L}_\alpha\subseteq H^{-1}(E)$ for some $E\in \RR$,
    \item \label{Clark_integrable_4} $\Delta(\partial \mathcal{L}_\alpha\cap S)\subseteq \mathcal{L}_\alpha$.
\end{enumerate}

Essentially, condition \ref{Clark_integrable_4} means that $\Delta$ maps points from the border of each Lagrangian leave into points inside the leave.

Let $I\subseteq \NN$ be a set of indices, $\{U_k\}_{k\in I}$ an open cover of $Q\setminus\pi_Q(S)$, and $\left\{\Phi_k\colon U_k \times \R^n \to \cT U_k \right\}$ be a complete solution of the hybrid Hamilton--Jacobi problem for $\hybrid_H$. Let $\alpha_k\in \RR^n$ such that $\Ima \big(\Delta \circ \Phi_k(\cdot, \alpha_k)\big) \subseteq \Ima \big( \Phi_l(\cdot, \alpha_l)\big)$ for each $k,l\in I$. \Cref{def:solution_hybrid_HJ,def:complete_sols} imply that $\mathcal{L}_\alpha = \cup_{k\in I} \Ima \Phi_k(\cdot, \alpha_k)$ satisfies conditions \ref{Clark_integrable_1_2}. 
However, condition \ref{Clark_integrable_3} only holds if
\begin{equation}\label{eq:Clark_integrable}
    H\circ \Phi_k(\cdot, \alpha_k)=H\circ \Phi_l(\cdot, \alpha_l)=E\, ,
\end{equation}
for each $k\in I$. Additionally, condition \ref{Clark_integrable_4} holds if $\alpha_k = \alpha_l$.
In summary, a hybrid Hamiltonian system with a complete solution of its Hamilton--Jacobi problem is completely integrable (in the sense of Clark) if and only if the open subsets $U_k$ cover $Q\setminus\pi_Q(S)$, $\alpha_k = \alpha_l$ and equation~\eqref{eq:Clark_integrable} holds.

\begin{remark}\label{remark_reconstruction}
    Solutions of the Hamilton--Jacobi problem can be used to reconstruct the dynamics of a hybrid Hamiltonian system $\hybrid_H= (\cT Q, X_H, S, \Delta)$ as follows:
    \begin{enumerate}
        \item\label{reconstruction_1} Solve the Hamilton--Jacobi equation \eqref{HJ_eq_hybrid} for each $(\gamma_k)_{\lambda_k}$.
        \item\label{reconstruction_2} Label the open subsets $U_k$ so that the initial condition is in $U_0$ and between $k$-th and $(k+1)$-th impacts the system is in $U_k$. Let $t_k\in \R$ denote the instants of the impacts, that is, $\lim_{t\to t_k^-} c_k(t) \in \pi_Q(S)$. 
        \item\label{reconstruction_3} Use that $(\gamma_{k+1})_{\lambda_{k+1}}$ and $(\gamma_k)_{\lambda_k}$ are $\Delta$-related to compute $\lambda_{k+1}$ in terms of $\lambda_k$, where $\lambda_0$ is determined by the initial conditions.
        \item\label{reconstruction_4} Compute the integral curves $c_k\colon (t_k, t_{k+1}) \to U_k$ of $X_H^{\gamma_k}$. 
        \item\label{reconstruction_5} The hybrid dynamics are given by $\gamma_k\circ c_k(t)$ for each interval $(t_k, t_{k+1})$.
    \end{enumerate}
\end{remark}

\begin{example}[Circular billiard]
    Consider the hybrid Hamiltonian system $\hybrid_H= (M , {S}, X_{H}, \Delta)$, with underlying Hamiltonian system $(\cT \RR^2, \omega_{\RR^2}, H)$, where 
    \begin{equation}
    \begin{aligned}
        & H = \frac{1}{2} \left(p_x^2+p_y^2\right)\, ,\\
        & M = \left\{(x,y, p_x, p_y) \in \cT \RR^2 \mid x^2+y^2\leq 1  \right\}\, , \\
        & S = \left\{(x,y, p_x, p_y) \in \cT \RR^2 \mid x^2+y^2=1 \hbox{ and } x p_x + y p_y >0 \right\} \, \\
    \end{aligned}
    \end{equation}
    and the impact map $\Delta\colon (x, y, p_x^-, p_y^-)\mapsto (x, y, p_x^+, p_y^+)$ is given by
    \begin{equation}
        p_x^+ = p_x^- -2x\left(xp_x^- + y p_y^-\right)\, \quad
        p_y^+ = p_y^- -2y\left(xp_x^- + y p_y^-\right)\, .
    \label{impact_map_billiard}
    \end{equation}
    They can be determined from the Newtonian law (see \Cref{remark:Newtonian_Hamiltonian_impact_law}) with the constraint function $h = 1-x^2-y^2$.
    Suppose that an impact occurs at $(x^\ast, y^\ast)\in \pi_{\RR^2}(S)$. 
    Let $\gamma^-$ and $\gamma^+$ denote the solutions of the Hamilton--Jacobi problem before and after the impact, respectively, with $\gamma^\pm = \gamma_x^\pm \dd x + \gamma_y ^\pm \dd y$.  The Hamilton--Jacobi equation \eqref{HJ_eq_hybrid} is written 
    \begin{equation}
    \begin{aligned}
        0 &= \frac{1}{2}  \dd \left(\left(\gamma_x^\pm\right)^2 + \left(\gamma_y^\pm\right)^2  \right)\\
        &=\left(\gamma_x^\pm \frac{\partial \gamma_x^\pm} {\partial x} + \gamma_y^\pm \frac{\partial \gamma_y^\pm} {\partial  x}  \right) \dd x
        +\left(\gamma_x^\pm \frac{\partial \gamma_x^\pm} {\partial y} + \gamma_y^\pm \frac{\partial \gamma_y^\pm} {\partial  y}  \right) \dd y\, .
    \end{aligned}
    \end{equation}
    Due to the form of this partial differential equation, it is natural to assume that $\gamma$ is separable, that is, $\gamma_x^\pm$ and $\gamma_y^\pm$ depend only on $x$ and on $y$, respectively.
    Then,
    \begin{equation}
    \begin{aligned}
        &\gamma_x^\pm \frac{\partial \gamma_x^\pm} {\partial x} = 0\, ,\quad \gamma_y^\pm \frac{\partial \gamma_y^\pm} {\partial  y} = 0\, ,
    \end{aligned}
    \end{equation}
    and thus $\gamma_x^\pm = \lambda_1^\pm$ and $\gamma_y^\pm = \lambda_2^\pm$, where $\lambda_1^-, \lambda_1^+, \lambda_2^-, \lambda_2^+$ are constants. The parameters $\lambda_1^-$ and $\lambda_2^-$ determining the solution $\gamma^-$ before the impact are given by the initial values of the momenta, namely, $\lambda_1^-=p_x(0)$ and $\lambda_2-= p_y(0)$. The parameters of the solution $\gamma^+$ after the impact are determined via the impact map \eqref{impact_map_billiard}:
    \begin{equation}
        \lambda_1^+ = \lambda_1^- -2x^\ast \left(x^\ast \lambda_1^- + y^\ast  \lambda_2^-\right)\, , \quad
        \lambda_2^+ = \lambda_2^- -2y^\ast \left(x^\ast \lambda_1^- + y^\ast  \lambda_2^-\right)\, .
    \end{equation}



    A complete solution of the hybrid Hamilton--Jacobi problem for $\hybrid_H$ is given by $\left\{(\gamma^-)_{\lambda^-},(\gamma^+)_{\lambda^+}\right\}$, where $(\gamma^\pm)_{\lambda^\pm}=\lambda_1^\pm \dd x + \lambda_2^\pm \dd y$. Then, $f_1^\pm(x,y,p_x,p_y)=\pi_1 \circ (\Phi^{\pm})^{-1}(x,y,p_x,p_y)=p_x$ and $f_2^\pm(x,y,p_x,p_y)=\pi_2 \circ (\Phi^{\pm})^{-1}(x,y,p_x,p_y)=p_y$ are generalized hybrid constants of the motion. The one-forms $\gamma^-$ and $\gamma^+$ determine Lagrangian submanifolds of $\cT \RR^2$, namely, namely,
    \begin{equation}
        \operatorname{Im} \gamma^\pm
        = \left\{ (x, y, p_x, p_y)\in \cT \RR^2\times \mid p_x = \lambda_1^\pm,\, p_y = \lambda_2^\pm\right\}\, .
    \end{equation}
\end{example}

\begin{example}[Rolling disk hitting fixed walls]\label{example_disk}
    Consider the hybrid Hamiltonian system $\hybrid_H=(\cT(\R^2 \times \Sp^1), X_H, \Delta, S)$ with underlying Hamiltonian system $(\cT(\R^2 \times \Sp^1), \omega_{\R^2 \times \mathbb{S}^1}, H)$, where
    \begin{equation}
        H = \frac{1}{2m} \left(p_x^2 +  p_y^2\right) +\frac{1}{2mk^2} p_\vartheta^2\, .
    \end{equation}
    The switching surface $S=C_1\cup C_2$ and the impact map $\Delta$ are given by equations~\eqref{eq:impact_surface_disk} and \eqref{eq:impact_map_disk}, respectively.

    Assume that there exists a separable solution of the hybrid Hamilton--Jacobi problem, that is, a collection of one-forms $\gamma_i=\gamma_{x,i}(x) \dd x + \gamma_{y,i}(y) \dd y + \gamma_{\vartheta,i}(\vartheta) \dd \vartheta$. Equation~\eqref{HJ_eq_hybrid} implies that
    $\gamma_i=a_i \dd x + b_i \dd y + c_i \dd y$, where $a_i, b_i, c_i$ are constants. The relation between these constants is given by equation~\eqref{Delta_related_condition}, namely,
    \begin{align}
        &a_{i+1} = \frac{R^2 a_i + R c_i}{k^2+R^2},\,b_{i+1} = -e b_i,\hbox{ and }c_{i+1} = k^2\frac{R a_i + c_i}{k^2+R^2}.
    \end{align}
    Consequently, $\{(\gamma_k)_{(a_k,b_k,c_k)}\}$ is a complete solution of the hybrid Hamilton--Jacobi problem for $\hybrid_H$. The initial values $(a_0, b_0, c_0)$ correspond with the initial values $(p_x(0), p_y(0), p_{\vartheta}(0))$ of the momenta at time zero.
    Each one-form $\gamma_i$ determines a Lagrangian submanifold of $\cT (\RR^2\times \mathbb{S}^1)$, namely,
    \begin{equation}
        \operatorname{Im} \gamma_i
        = \left\{ (x, y, \vartheta, p_x, p_y, p_\vartheta)\in \cT (\RR^2\times \mathbb{S}^1) \mid p_x = a_i,\, p_y = b_i,\, p_\vartheta = c_i\right\}\, .
    \end{equation}
\end{example}

It is worth remarking that in the examples above the separability condition has been imposed since it is a natural \textit{ansatz} for Hamiltonians of the form $H = \sum_{i=1}^n A_i p_i^2$ for some constants $A_i$. However, this is by no means a requirement of the theory.

\section{ Hamilton--Jacobi theory for hybrid forced Hamiltonian systems}

\begin{definition}
    Let $\hybrid_H= (M, X\Halpha, S, \Delta)$ be a hybrid forced Hamiltonian system with underlying forced Hamiltonian system $(Q, H, \alpha)$. Consider a set  $\mathfrak{U} = \left\{(U_k,\gamma_k)  \right\}$ whose elements are pairs $(U_k,\gamma_k)$, where $U_k\subseteq Q\setminus \pi_Q(S)$ are open subsets of $Q$ and $\gamma_k \in \Omega^1 (U_k)$ are closed one-forms. The set $\mathfrak{U}$ is called a \emph{solution of the hybrid Hamilton--Jacobi problem} for $\hybrid\Halpha$ if:
    \begin{enumerate}
        \item For each $k$, $\gamma_k$ is a solution of the forced Hamilton--Jacobi equation \eqref{eq:HJ_forced} on $U_k$, namely,
        \begin{equation}
            \dd (H \circ \gamma_k) = -\gamma_k^\ast \alpha \, . 
            \label{HJ_eq_hybrid_forc}
        \end{equation}
        \item If $x\in \partial U_k \cap \pi_Q(S)$, then there exists an $l$ such that $x\in \partial U_l \cap \pi_Q(S)$ and $\gamma_l$ and $\gamma_k$ are $\Delta$-related, that is,
        \begin{equation}
            \lim_{y\to x} \gamma_l(y) = \Delta \left( \lim_{y\to x} \gamma_k(y) \right).
        \label{Delta_related_condition_forc}
        \end{equation}
    \end{enumerate}
\end{definition}

\begin{theorem}[Hybrid forced Hamilton--Jacobi theorem]\label{HJ_theorem_hybrid_forced}
    Let $\hybrid\Halpha= (M, {S}, X\Halpha, \Delta)$ be a hybrid forced Hamiltonian system, $U_k\subset Q\setminus \pi_Q(S)$ be open subsets of $Q$ and $\mathfrak{U} = \left\{(U_k, \gamma_k)  \right\}$ be a family of closed one-forms $\gamma_k\in \Omega^1(U_k)$.  Then, the following statements are equivalent:
    \begin{enumerate}
        \item The set $\mathfrak{U}$ is a solution of the hybrid Hamilton--Jacobi equation for $\hybrid\Halpha$. 
        \item For every continuous and piecewise smooth curve $c\colon \R\to Q$ such that 
        \begin{enumerate}
            \item $c\big((t_k, t_{k+1})\big)\subset U_k$, 
            \item $c$ intersects $\pi_Q(S)$ at $\{t_k\}_k$,
            \item $c$ satisfies the equations
        \begin{equation}
        \begin{aligned}
            &\dot c(t) = T \pi_Q \circ  X_{H,F} \circ \gamma_k \circ c(t)\, , \qquad t_k<t<t_{k+1}\, ,\\
            & \gamma_{k+1} \circ c(t_{k+1}) = \Delta \circ \gamma_k \circ c(t_{k+1})\, ,
        \end{aligned}
        \end{equation}
        \end{enumerate}
        the curve $\tilde{c}\colon \RR \to \cT Q$ given by $\tilde{c}(t) = \gamma_k \circ c(t)$ for $t\in[t_k, t_{k+1})$ is an integral curve of the hybrid dynamics.
    \end{enumerate}
\end{theorem}


\textit{Mutatis mutandis}, the proof is identical to the one of \Cref{theorem:HJ_hybrid}.

Complete solutions are defined as in the unforced case (see \Cref{def:complete_sols}). The dynamics of $\hybrid\Halpha$ can be reconstructed by solving the forced Hamilton--Jacobi equation \eqref{HJ_eq_hybrid_forc} for each $(\gamma_k)_{\lambda_k}$ and following steps \ref{reconstruction_2}-\ref{reconstruction_5} from Remark \ref{remark_reconstruction}. (Observe that $\T\pi_Q(X\Halpha)=\T\pi_Q(X_H)$, and thus step \ref{reconstruction_4} does not change.)

    Given a complete solution $\left\{(U_k, \Phi_k)\right\}$, each $\cT U_k$ can be foliated into Lagrangian submanifolds invariant under the flow of $\restr{X_{H,F}}{\cT U_k}$. Moreover, the impact map $\Delta$ maps the borders of the Lagrangian submanifolds of $\cT U_k$ into subsets of the Lagrangian submanifolds of $\cT U_l$. In addition, the functions $f_{k,i}= \pi_i~\circ~\Phi_k^{-1}\colon \cT U_k\to \R$ are constants of the motion for $\left(\restr{H}{U_k}, \restr{\alpha}{U_k}\right)$ but, in general, they are not hybrid constants of the motion. 
    {Therefore, obtaining a complete solution of the hybrid Hamilton--Jacobi problem may simplify solving the equations of motion for $\hybrid\Halpha$. For instance, one may choose coordinates $(x_i, y_i)$ in $U_k$, where $x_i = f_{k,i}$, so that half of the $2n$ equations of motion on $U_k$ become simply $\restr{X_{H,F}}{U_k} (x_i) = 0$.}

\begin{example}[Rolling disk with dissipation hitting fixed walls]\label{example_disk_forced}
    Consider the hybrid system from \Cref{example_disk}. Suppose that now the disk is additionally subject to the external force
    \begin{equation}
        \alpha = B p_y \dd y\, ,
    \end{equation}
    where $B$ is some constant. 
    Then, a one-form $\gamma_k = \gamma_{k,x} \dd x + \gamma_{k,y} \dd y + \gamma_{k,\vartheta} \dd \vartheta$ satisfies equation~\eqref{HJ_eq_hybrid_forc} if and only if
    \begin{align}
        &\frac{1}{m} \left[\gamma_{k,x} \frac{\partial \gamma_{k,x}}{\partial x}
        + \gamma_{k,y} \frac{\partial \gamma_{k,y}}{\partial x}
        + \frac{1}{k^2} \gamma_{k,\vartheta} \frac{\partial \gamma_{k,\vartheta}}{\partial x}
        \right] = 0\, ,\\ 
        &\frac{1}{m} \left[\gamma_{k,x} \frac{\partial \gamma_{k,x}}{\partial y}
        + \gamma_{k,y} \frac{\partial \gamma_{k,y}}{\partial y}
        + \frac{1}{k^2} \gamma_{k,\vartheta} \frac{\partial \gamma_{k,\vartheta}}{\partial y}
        \right] = B \gamma_{k,y}\, ,\\
        &\frac{1}{m} \left[\gamma_{k,x} \frac{\partial \gamma_{k,x}}{\partial \vartheta}
        + \gamma_{k,y} \frac{\partial \gamma_{k,y}}{\partial \vartheta}
        + \frac{1}{k^2} \gamma_{k,\vartheta} \frac{\partial \gamma_{k,\vartheta}}{\partial \vartheta}
        \right] = 0\, ,\\ 
    \end{align}
    Suppose that $\gamma_k$ is separable, namely, $\gamma_{k,x}=\gamma_{k,x}(x), \gamma_{k,y}=\gamma_{k,y}(y), \gamma_{k,\vartheta}=\gamma_{k,\vartheta}(\vartheta)$. Then,
    \begin{align}
        &\gamma_{k,x} \frac{\dd \gamma_{k,x}}{\dd x} = 0
        \, \gamma_{k,y} \frac{\dd \gamma_{k,y}}{\dd y} = Bm \gamma_{k,y},\hbox{ and }\frac{1}{k^2} \gamma_{k,\vartheta} \frac{\dd \gamma_{k,\vartheta}}{\dd \vartheta} = 0\, ,
    \end{align}
    which implies that
    \begin{equation}
        \gamma_k = a_k\ \dd x + (Bm y + b_k)\ \dd y + c_k\ \dd \vartheta\, .
    \end{equation}
    The relation between $(a_k, b_k, c_k)$ and $(a_l, b_l, c_l)$ is given by equation~\eqref{Delta_related_condition_forc}, namely,
    \begin{align}
        & a_l = \frac{R^2 a_k   + R c_k}{k^2+R^2}\, ,\\
        & Bmy + b_l = -e(Bmy+b_k)\, ,\\
        & c_l = \frac{R a_k   + c_k}{k^2+R^2}\, .
    \end{align}
    Thus,
    \begin{equation}
        b_l 
        = \left\{ 
        \begin{array}{ll}
            -eb_k   & \text{for the wall at } y=0,  \\
            -(e+1)Bmh-eb_k   & \text{for the wall at } y=h\, .
        \end{array}
        \right.
    \end{equation}
    Consequently, $\{(\gamma_k)_{(a_k,b_k,c_k)}\}$ is a complete solution of the hybrid Hamilton--Jacobi problem for $\hybrid\Halpha=(\cT(\R^2 \times \Sp^1), X\Halpha, \Delta, S)$.
\end{example}

\section{Hamilton--Jacobi theory nonholonomic hybrid systems}
\subsection{Hamilton--Jacobi theory for nonholonomic continuous systems}
Let $(Q,L)$ be a mechanical Lagrangian system, namely,
\begin{equation}
    L(q,v) = \frac{1}{2} g_q(v,v) - V(q)\, ,
\end{equation}
where $g$ is a Riemannian metric on $Q$.
Let $(\cT Q, \omega_Q, H)$, where 
\begin{equation}
    H(q,p) = \frac{1}{2} g_q^{-1}(p,p) + V(q)\, ,
\end{equation}
be the Hamiltonian counterpart of $(Q, L)$, with $g_q^{-1}\colon \cT_q Q \times \cT_q Q\to \RR$ denoting the inner product defined by $g$ on $\cT_q Q$.

Suppose that the system is subject to the (linear) nonholonomic constraints given by the distribution
\begin{equation}
    D = \left\{ v \in \T Q \mid \mu^a(v) = 0,\ a=1,\ldots, k \right \}\, ,
\end{equation}
where $\mu^a=\mu^a_i(q) \dd q^i$ are constraint one-forms.
Denote by $C = \FF L (D)$ the associated codistribution. The triple $(Q, H, C)$ will be called a \emph{nonholonomic Hamiltonian system}.
The \emph{nonholonomic vector field} $X_H^{\mathrm{nh}}$ of $H$ is given by
\begin{equation}
    X_H^{\mathrm{nh}} = \sharp_{\omega_Q}\left(\dd H - \lambda_a \, \mu^a\right)\, ,
\end{equation}
with the constraint 
\begin{equation}
    \T\pi_Q\left(X_H^{\mathrm{nh}}\right) \in \Gamma(D)\,.
\end{equation}
Here, $\flat_{\omega_Q}\colon Y\in \X(\cT Q)\mapsto \contr{Y} \omega_Q\in \Omega^1(\cT Q)$ denotes the isomorphism defined by the canonical symplectic form $\omega_Q$, and $\lambda_a$ are Lagrange multipliers. Defining the vector fields $Z_a = \sharp_{\omega_Q}\mu^a$, one can write $X_H^{\mathrm{nh}} = X_H - \lambda_a Z_a$.
Locally, $Z_a = \mu^a_i \tparder{}{p_i}$, and thus the projections on $Q$ of the nonholonomic and the Hamiltonian vector fields of $H$ coincide, namely, $\T\pi_Q\left(X_H^{\mathrm{nh}}\right)= \T\pi_Q\left(X_H\right)$.


\begin{theorem}[Ohsawa and Bloch, 2009]\label{HJ_theorem_nh}
    Assume that $D$ is a \textit{completely nonholonomic distribution}, that is,
    \begin{equation}
        \T Q = \left\langle \left\{D, [D, D], [D, [D, D]], \ldots \right\} \right\rangle\, .
    \end{equation}
    Let $\gamma$ be a one-form on $Q$ such that $\Ima \gamma \subset C$ and $\dd \gamma(v,w)=0$ for any $v,w\in D$. Then, the following statements are equivalent:
    \begin{enumerate}[label=\roman*)]
        \item For every integral curve $c$ of $T\pi_Q \circ X_H \circ \gamma$, the curve $\gamma \circ c$ is an integral curve of $X_H^{\mathrm{nh}}$.
        \item The one-form $\gamma$ satisfies the nonholonomic Hamilton--Jacobi equation:
        \begin{equation}\label{eq:HJ_nh}
            H \circ \gamma = E\, ,
        \end{equation}
        where $E$ is a constant.
    \end{enumerate}
\end{theorem}
See \cite{O.B2009} for the proof.

\begin{remark}
    In the general case, where $D$ is not necessarily completely nonholonomic,the Hamilton--Jacobi equation is
    \begin{equation}
       \dd\left( H \circ \gamma\right) \in D^\circ\, ,
    \end{equation}
    since it is not possible to apply Chow's theorem (see \cite{I.d.M2008}). For an approach based on Lie algebroids, refer to \cite{d.M.M2010}.
\end{remark}

\begin{remark}
    In \cite{d.M.V2014}, the authors considered the Hamilton--Jacobi equation for a Hamiltonian system on an almost-Poisson manifold $(E, \Lambda)$ fibered over a base manifold $Q$. Moreover, they defined complete solutions for the Hamilton--Jacobi equation, and showed that from a complete solution one can define $m=\rank E$ independent functions in involution with respect to the almost-Poisson bracket (see Definition~4.1 and Proposition~2 in \cite{d.M.V2014}, respectively). In particular, in Subsection 5.2 of that paper, they consider the nonholonomic almost-Poisson bracket. Therefore, a complete solution of the Hamilton--Jacobi equation for a nonholonomic system provides $m=\rank C$ independent functions in involution with respect to the nonholonomic bracket. However, since almost-Poisson brackets do not necessarily verify the Jacobi identity, these functions in involution are not associated with commuting vector fields as in the Poisson case. Nevertheless, an isomorphism between the nonholonomic bracket and the so-called Eden bracket has been recently shown (see~\cite{d.L.L+2024, d.L.L+2023a}), namely,
    \begin{equation}
        \{f, g\}_{\mathrm{nh}} = \{f \circ \gamma , g \circ \gamma\} \circ \incl_C\, ,
    \end{equation}
    for $f, g\in \Cinfty(C)$, where $\gamma\colon \cT Q \to C$ is an orthogonal projector with respect to the metric appearing in the mechanical Hamiltonian and $\incl_C\colon C \hookrightarrow \cT Q$ denotes the canonical inclusion. Therefore, roughly speaking, nonholonomic dynamics can be understood as free Hamiltonian dynamics for a modified Hamiltonian function restricting to the appropriate subspaces. Consequently, complete solutions of the Hamilton--Jacobi problem for a nonholonomic system seem to be related to complete integrability in the sense of Liouville. This relation will be explored in future works.
\end{remark}

\subsection{Hamilton--Jacobi theory for nonholonomic hybrid systems}

\begin{definition}
    Let $(Q, H, C)$ be a nonholonomic Hamiltonian system. A hybrid system $(\cT Q, X_{H}^{\mathrm{nh}}, S, \Delta)$ is called a \emph{nonholonomic hybrid system with underlying nonholonomic Hamiltonian system $(Q, H, C)$} and denoted by $\hybrid_{\mathrm{nh}}$. As in the case of hybrid Hamiltonian systems, it will be assumed that the switching surface $S$ projects onto a codimension-one submanifold $\pi_Q(S)$ of $Q$, and that the impact map preserves the base point, namely, $\pi_Q \circ \Delta = \pi_Q$.
 
    Consider a set $\mathfrak{U} = \left\{(U_k,\gamma_k)  \right\}$ whose elements are pairs $(U_k,\gamma_k)$, where $U_k\subseteq Q\setminus \pi_Q(S)$ are open subsets of $Q$ and $\gamma_k \in \Omega^1 (U_k)$ are one-forms such that $\Ima\gamma_k\subset C$ and $\dd \gamma_k(v,w)=0$ for each $v, w \in D$.  
    The set $\mathfrak{U}$ will be called a \emph{solution of the hybrid Hamilton--Jacobi problem for $\hybrid_{\mathrm{nh}}$} if:
    \begin{enumerate}
        \item For each $k$, $\gamma_k$ is a solution of the nonholonomic Hamilton--Jacobi equation \eqref{eq:HJ_nh} on $U_k$, namely,
        \begin{equation}
            H \circ \gamma_k = E_k\, . 
            \label{HJ_eq_hybrid_nh}
        \end{equation}
        \item If $x\in \partial U_k \cap \pi_{Q}(S)$, then there exists an $l$ such that $x\in \partial U_l \cap \pi_{Q}(S)$ and $\gamma_l$ and $\gamma_k$ are $\Delta$-related, that is,
        \begin{equation}
            \lim_{y\to x} \gamma_l(y) = \Delta \left( \lim_{y\to x} \gamma_k(y) \right)\, .
        \label{Delta_related_condition_nh}
        \end{equation}
    \end{enumerate}
\end{definition}

\begin{theorem}[Hybrid nonholonomic Hamilton--Jacobi theorem]\label{HJ_theorem_hybrid_nh}
    Consider a hybrid nonholonomic system $\hybrid_{\mathrm{nh}}= (\cT Q, X_{H}^{\mathrm{nh}}, S, \Delta)$ with underlying nonholonomic Hamiltonian system $(Q, H, C)$. 
    Let $\mathfrak{U}$ be a set formed by pairs $(U_k,\gamma_k)$, where $U_k\subseteq Q\setminus \pi_Q(S)$ are open subsets of $Q$ and $\gamma_k \in \Omega^1 (U_k)$ are one-forms such that $\Ima\gamma_k\subset C$ and $\dd \gamma_k(v,w)=0$ for each $v, w \in D$.  
    Then, the following statements are equivalent:
    \begin{enumerate}
        \item The set $\mathfrak{U}$ is a solution of the hybrid Hamilton--Jacobi equation for $\hybrid_{\mathrm{nh}}$. 
        \item For every continuous and piecewise curve $c\colon \R\to Q$ such that 
        \begin{enumerate}
            \item $c\big((t_k, t_{k+1})\big)\subset U_k$, 
            \item $c$ intersects $\pi_Q(S)$ at $\{t_k\}_k$,
            \item $c$ satisfies the equations
        \begin{align}
        &\dot c(t) = \T \pi_Q \circ  X_H^{\mathrm{nh}} \circ \gamma_k \circ c(t), \qquad t_k<t<t_{k+1},\\
        & \gamma_{k+1} \circ c(t_{k+1}) = \Delta \circ \gamma_k \circ c(t_{k+1}),
        \end{align}
        \end{enumerate}
        then the curve $\tilde{c}\colon \RR \to C$ given by $\tilde{c}(t) = \gamma_k \circ c(t)$ for $t\in[t_k, t_{k+1})$ is an integral curve of the hybrid dynamics.
    \end{enumerate}
\end{theorem}

\begin{proof}
    By \Cref{HJ_theorem_nh}, equation~\eqref{HJ_eq_hybrid_nh} holds if and only if, for each integral curve $c\colon \RR\rightarrow U_k$ of $X_{H}^{\gamma_k}$, the curve $\gamma_k\circ c\colon \RR\rightarrow \cT U_k$ is an integral curve of $X^{\mathrm{nh}}_{H\mid \cT U_k}$. Since the dynamics $x(t)$ of $\hybrid_{\mathrm{nh}}$ are given by the integral curves of $X_H^{\mathrm{nh}}$ for $x(t)\notin S$, the curves $x(t)=\gamma_k \circ c(t)$ are the hybrid dynamics on $\cT U_k\cap C$.
    
    On the other hand, if $c$ is a continuous curve that intersects $\pi_Q(S)$ at $\{t_k\}_k$ and $c(t_{k+1}) \in \partial U_k \cap \partial U_l \cap \partial \big(\pi_{Q}(S)\big)$, then $\gamma_{k+1}$ and $\gamma_k$ are $\Delta$-related if and only if
    \begin{equation}
        \gamma_{k+1} \circ c(t_{k+1}) = \Delta \circ \gamma_k \circ c(t_{k+1}) \, .
    \end{equation}
\end{proof}


\begin{definition}\label{def:complete_sols_nh}
    Let $m=\operatorname{rank} D=\operatorname{rank} C$. 
    Consider a family of solutions $({\gamma_{k}})_{\lambda}$ depending on $m$ additional parameters $\lambda\subset \RR^{m}$, and suppose that $\left\{\Phi_k\colon U_k \times \R^{m} \to \cT U_k\cap C \right\}$, where $\{\Phi_{k}(q, \lambda)=({\gamma_{k}})_{\lambda}(q)\}$, is a family of local diffeomorphisms. The family $({\gamma_{k}})_{\lambda}(q)$ is called a \emph{complete solution of the hybrid Hamilton--Jacobi problem} for $\hybrid_{\mathrm{nh}}$ if, for each $\lambda \in \R^{m}$, there exists a $\mu \in \R^{m}$ such that $\Ima (\Delta \circ ({\gamma_{k}})_{\lambda})\subseteq \Ima  ({\gamma_{l}})_{\mu}$.
\end{definition}

Solutions of the Hamilton--Jacobi problem can be used to reconstruct the dynamics of a nonholonomic hybrid Hamiltonian system (see \Cref{remark_reconstruction}).

\subsection{Examples}
\subsubsection{The nonholonomic particle.}
    Consider the nonholonomic Lagrangian system $(\R^3, L, D)$, with Lagrangian function
    \begin{equation}
        L = \frac{1}{2} \left(v_x ^2 + v_y^2 + v_z^2\right)\, ,
    \end{equation}
    and constraint distribution
    \begin{equation}
        D = \left\langle\left\{ \frac{\partial}{\partial x} + y \frac{\partial}{\partial z}, \frac{\partial}{\partial y} \right\}\right\rangle\, . 
    \end{equation}
    Its Hamiltonian counterpart is the nonholonomic Hamiltonian system $(\R^3, H, C)$, where
    \begin{equation}
        H = \frac{1}{2} \left(p_x ^2 + p_y^2 + p_z^2\right)\, ,
    \end{equation}
    and
    \begin{equation}
        C = \left\langle\left\{ \dd x + y \dd z, \dd y\right\}\right\rangle\, .
    \end{equation}
    Let $\gamma = \gamma_x \dd x + \gamma_y \dd y + \gamma_ z \dd z$ be a solution of the nonholonomic Hamilton--Jacobi equation \eqref{eq:HJ_nh}. The condition $\Ima \gamma \subset C$ means that $\gamma$ is a linear combination of $\dd x + y \dd z$ and $\dd y$, which implies that $\gamma_z=y \gamma_x$. The Hamilton--Jacobi equation can then be written as
    \begin{equation}
        E = \frac{1}{2} \left(\gamma_x^2 + \gamma_y ^2 + \gamma_z ^2\right)
         = \frac{1}{2} \left(\gamma_x^2 + \gamma_y ^2 + y^2 \gamma_x^2\right),
    \end{equation}
    which implies that
    \begin{equation}
        \gamma_y = \pm \sqrt{2E-(1+y^2) \gamma_x^2},
    \end{equation}
    Therefore,
    \begin{equation}
        \gamma = \gamma_x \dd x \pm \sqrt{2E-(1+y^2) \gamma_x^2} \dd y + y \gamma_x \dd z.
    \end{equation}
    Finally, one has to impose the condition $\dd \gamma(v,w)$ for any $v, w \in D$. Since $\dd \gamma$ is bilinear and alternating, it suffices to impose that 
    \begin{equation}
        \dd \gamma \left(\frac{\partial}{\partial y}, \frac{\partial}{\partial x} + y \frac{\partial}{\partial z}\right) = 0\, .
    \end{equation}
    Suppose that $\gamma_x$ only depends on $y$. Then,
    \begin{equation}
        \dd \gamma \left(\frac{\partial}{\partial y}, \cdot \right)
        = \frac{\dd \gamma_x}{\dd y} \dd x + \left( \gamma_x + y \frac{\dd \gamma_x}{\dd y}\right) \dd z\, ,
    \end{equation}
    and
    \begin{equation}
         \dd \gamma \left(\frac{\partial}{\partial y}, \frac{\partial}{\partial x} + y \frac{\partial}{\partial z}\right) 
         = (1+y^2) \frac{\dd \gamma_x}{\dd y} + y \gamma_x\, .
    \end{equation}
    Therefore,
    \begin{equation}
        \gamma_x = \frac{\lambda}{\sqrt{1+y^2}}\, ,
    \end{equation}
    for some $\lambda\in \RR$, and then
    \begin{equation}
        \gamma =  \frac{\lambda}{\sqrt{1+y^2}} \dd x 
        \pm \sqrt{2E- \lambda^2} \dd y
        + \frac{\lambda y}{\sqrt{1+y^2}} \dd z\, .
    \end{equation}

    Now suppose that there are two walls at the planes $y=0$ and $y=a$ and that the particle impacts with them with an elastic constant $e$, that is,
    \begin{equation}
    \Delta (p_x, p_y, p_z) = (p_x, -ep_y, p_z)\, .
    \end{equation}
    This impact map can be derived from the Newtonian impact law (see \Cref{remark:Newtonian_Hamiltonian_impact_law}) for a constraint function $h=y$.
    Let $\gamma_i$ denote the solution of the Hamilton--Jacobi equation between the $i$-th and $(i+1)$-th impacts, namely,
    \begin{equation}
        \gamma_i =  \frac{\lambda_i}{\sqrt{1+y^2}} \dd x 
        \pm \sqrt{2E_i- \lambda_i^2} \dd y
        + \frac{\lambda_i y}{\sqrt{1+y^2}} \dd z\, .
    \end{equation}
    Then, the condition $\gamma_{i+1}=\Delta \circ \gamma_i$ implies that
    \begin{equation}
    \begin{aligned}
        & \frac{\lambda_{i+1}}{\sqrt{1+y^2}} = \frac{\lambda_i}{\sqrt{1+y^2}}\, ,\\
        & \sqrt{2E_{i+1}- \lambda_{i+1}^2} = e\sqrt{2E_i- \lambda_i^2}\, ,\\
        & \frac{\lambda_{i+1} y}{\sqrt{1+y^2}} = \frac{\lambda_i y}{\sqrt{1+y^2}}\, ,
    \end{aligned}
    \end{equation}
    that is,
    \begin{equation}
    \begin{aligned}
        & \lambda_{i+1} = \lambda_i\, ,\\
        & E_{i+1} = e^2 E_i + \frac{1+e^2}{2} \lambda_{i+1}^2\, .
    \end{aligned}
    \end{equation}
    Clearly, fixing $\lambda_i$ and $E_i$ also fixes $\lambda_{i+1}$ and $E_{i+1}$, that is,
    \begin{equation}
        \Ima (\Delta \circ ({\gamma_{i}})_{(\lambda_i, E_i)})\subset \Ima  ({\gamma_{i+1}})_{(\lambda_{i+1}, E_{i+1})}.
    \end{equation}
    Moreover, $(x, y, z, \lambda_i, E_i)\mapsto \gamma_i$ is a local diffeomorphism, making $(\gamma_i)_{(\lambda_i, E_i)}$ a complete solution.

    The Hamiltonian vector field of $H$ is
    \begin{equation}
        X_H = p_x \frac{\partial}{\partial x} + p_y \frac{\partial}{\partial y}
        + p_z \frac{\partial}{\partial z}\, .
    \end{equation}
    Therefore, its projection is 
    \begin{equation}
    \begin{aligned}
        X_H^{\gamma_i} 
        & = \gamma_{i,x} \frac{\partial}{\partial x} + \gamma_{i,y} \frac{\partial}{\partial y} + \gamma_{i,z} \frac{\partial}{\partial z}\\
        & = \frac{\lambda_{i}}{\sqrt{1+y^2}} \frac{\partial}{\partial x} \pm \sqrt{2E_{i}- \lambda_{i}^2} \frac{\partial}{\partial y} + \frac{\lambda_{i} y}{\sqrt{1+y^2}} \frac{\partial}{\partial z}\, ,
    \end{aligned}
    \end{equation}
    whose integral curves $\sigma_i(t) = (x_i(t), y_i(t), z_i(t))$ are given by
    \begin{equation}
    \begin{aligned}
        \dot x_i(t) = \frac{\lambda_{i}}{\sqrt{1+y(t)^2}}\, ,\quad \dot y_i(t) = \pm \sqrt{2E_{i}- \lambda_{i}^2}\, , \quad 
        \dot z_i(t) = \frac{\lambda_{i} y(t)}{\sqrt{1+y(t)^2}}\, .
    \end{aligned}
    \end{equation}
    Hence,
    \begin{equation}
    \begin{aligned}
        & x_i(t) 
        = \frac{\lambda_i}{A_i}\left[ \operatorname{arcsinh}\left(A_i t+y_{i,0}\right) -\operatorname{arcsinh}\left(y_{i,0}\right)\right] + x_{i,0}\, ,\\
        & y_i(t) = A_i t + y_{i,0}\, ,\\
        & z_i(t) = \frac{\lambda_i}{A_i} \left[\sqrt{(A_i t+y_{i,0})^2+1}- \sqrt{y_{i,0}^2+1}\right]+z_{i,0}\, ,
    \end{aligned}
    \end{equation}
    if $A_i\neq 0$, and
     \begin{equation}
    \begin{aligned}
        & x_i(t) = \frac{\lambda_{i}}{\sqrt{1+y_{i,0}^2}} + x_{i,0}\, ,\\
        & y_i(t) = y_{i,0}\, ,\\
        & z_i(t) = \frac{\lambda_{i} y_{i,0}}{\sqrt{1+y_{i,0}^2}}+z_{i,0}\, ,
    \end{aligned}
    \end{equation}
    if $A_i=0$,
    where $A_i=\pm \sqrt{2E_{i}- \lambda_{i}^2}, \, x_{i+1,0}=x_{i}(t_{i}),\, y_{i+1,0}=y_{i}(t_{i})$ and $z_{i+1,0}=z_{i}(t_{i})$, for $i=0, \ldots, n$, with $t_0=0$ and $t_i$ the instant of the $i$-th impact. Fixing $\lambda_i,\ E_i$ and the initial conditions $(x_{0,0},y_{0,0},z_{0,0})$, one can determine any trajectory of the hybrid system. These trajectories coincide with the ones obtained by computing the integral curves of the nonholonomic Euler--Lagrange vector field $\sode_{(L, D)}$ (see \cite[Example 4.3.4]{AnahorySimoes2021}).

\subsubsection{The generalized rigid body}

Consider a mechanical system with a Lie group as configuration space, namely $Q=G$. Let $\mathfrak g$ denote the Lie algebra of $G$ and $\mathfrak g^\ast$ its dual. Its Lagrangian is the left-invariant function $L\colon \T G \simeq G\times \mathfrak{g} \to \R$ given by $L(g, v_g) = \ell(g^{-1} v_g)$, where $\ell\colon \mathfrak g \to \R$ is the reduced Lagrangian, defined by
\begin{equation}
    \ell(\xi) = \frac{1}{2} I_{ij} \xi^i \xi^j\, ,
\end{equation}
for $\xi = (\xi^1, \ldots, \xi^n)\in \mathfrak g$, where $I_{ij}$ are the components of the (positive-definite and symmetric) inertia tensor $\mathbb I \colon \mathfrak g \to \mathfrak g^\ast$ (see, for instance, \cite{B.M.Z2005, F.B.Z2014, Z.B2000}).

The Legendre transform of $L$ is given by $\FF L \colon \left(g, \xi^i\right) \mapsto \left(g, I_{ij} \xi^j\right)$, so the Hamiltonian function $H\colon G \times \mathfrak g^\ast \to \R$ is
\begin{equation}
    H = \frac{1}{2} I^{ij} \eta_i\, \eta_j\, ,
\end{equation}
where $I^{ij}$ are the components of the inverse of $\mathbb I$, and $\eta = (\eta_1, \ldots, \eta_n)\in \mathfrak g^\ast$.

The constrained generalized rigid body is subject to the left-invariant nonholonomic constraint
\begin{equation}
   D_\mu = \left\{(g,\xi) \in G\times \mathfrak g \mid  \langle \mu, \xi \rangle = \mu_i\, \xi^i = 0\right\}\, ,
\end{equation}
where $\mu=(\mu_1, \ldots, \mu_n)$ is a fixed element of $\mathfrak g^\ast$ and $\langle \cdot, \cdot \rangle$ denotes the natural pairing between a Lie algebra and its dual. Then,
\begin{equation}
   C_\mu = \FF L(D_\mu) = \left\{(g,\eta) \in G\times \mathfrak g^\ast \mid  
   \eta_i I^{ij} \mu_j =0 \right\}\, .
\end{equation}
Since $\mathbb{I}$ is positive-definite, $\FF L=\flat_{\mathbb{I}}$ is a diffeomorphism and $C_\mu$ is well-defined globally.
A solution of the nonholonomic Hamilton--Jacobi problem is a one-form $\gamma\colon G \to G\times \mathfrak g^\ast,\ g\mapsto (g, \gamma_1(g), \ldots, \gamma_n (g))$ satisfying
\begin{equation}
\begin{aligned}
    & H\circ \gamma = \frac{1}{2} I^{ij} \gamma_i \gamma_j = E\, , \\
    & I^{ij} \gamma_i \mu_j = 0\, , \\
    & \dd \gamma_{\mid D\times D} = 0\, .
\end{aligned}
\end{equation}

Hereinafter, consider the lie group $G=\mathrm{SO(3)}$. Let $\{e_1, e_2, e_3\}$  be the canonical basis of $\mathfrak{so}(3)\simeq \R^3$, whose Lie brackets are
\begin{equation}
    [e_1, e_2] = e_3\, , \quad [e_1, e_3] = -e_2\, , \quad [e_2, e_3] = e_1\, ,
\end{equation}
and let $\{e^1, e^2, e^3\}$ be its dual basis.
For simplicity's sake, assume that
\begin{equation}
    \mathbb{I} = I e^1\otimes e^1 + I e^2\otimes e^2 + I e^3\otimes e^3\, ,
\end{equation}
and thus
\begin{equation}
    H(g,\eta) = \frac{1}{2I^2} \left(\eta_1^2 +  \eta_2^2 +  \eta_3^2 \right)\, .
\end{equation}
The nonholonomic distribution is given by
\begin{equation}
\begin{aligned}
    \mathcal D_\mu = \left\{(g, \xi) \in \mathrm{SO(3)} \times \mathfrak{so}(3)\mid {\mu}_i \xi^i = 0 \right\}
    = \gen{\left\{{\mu}_2 e_1 -{\mu}_1 e_2,\ {\mu}_3 e_1 - {\mu}_1 e_3 \right\}}\, .
\end{aligned}
\end{equation}
A solution of the Hamilton--Jacobi problem satisfies
\begin{equation} \label{HJ_eqs_SO3}
    \begin{aligned}
        & \gamma_1^2 +  \gamma_2^2 + \gamma_3^2 = 2E I^2, \\
        & \gamma_1 \mu_1 + \gamma_2 \mu_2 + \gamma_3 \mu_3  = 0, \\
        & \dd \gamma \left({\mu}_2 e_1 -{\mu}_1 e_2,\ {\mu}_3 e_1 - {\mu}_1 e_3 \right) = 0.
\end{aligned}
\end{equation}
Using the first two of the equations~\eqref{HJ_eqs_SO3}, one can write
\begin{equation}
\begin{aligned}
    & \gamma_2 = \frac{\pm\mu_{3}\sqrt{ 2 E I^2 \left(\mu_{2} ^2+\mu_{3} ^2\right)-\gamma_{1} ^2 \left(\mu_{1} ^2+\mu_{2} ^2+\mu_{3} ^2\right)}-\mu_{1}  \mu_{2}  \gamma_{1}}{\mu_{2} ^2+\mu_{3} ^2}\, ,\\
    & \gamma_3 = \frac{\pm \mu_{2} \sqrt{ 2 E I^2\left(\mu_{2} ^2+\mu_{3} ^2\right)-\gamma_{1} ^2 \left(\mu_{1} ^2+\mu_{2} ^2+\mu_{3} ^2\right)}-\mu_{1} \mu_{3}   \gamma_{1}}{\mu_{2} ^2+\mu_{3} ^2}\, .
\end{aligned}
\end{equation}
Utilising the \textit{ansatz} $\gamma_1=\lambda_1$ for some $\lambda_1\in \R$, it is possible to write
\begin{equation}
    \gamma = \lambda_1 e^1 + \frac{\mu_{3} \lambda_2-\mu_{1}  \mu_{2}    \lambda_1 }{\mu_{2} ^2+\mu_{3} ^2} e^2 
    + \frac{\mu_{2}  \lambda_2 -\mu_{1} \mu_{3}  \lambda_1}{\mu_{2} ^2+\mu_{3} ^2} e^3\, ,
\end{equation}
where $\lambda_2 = \pm \sqrt{ 2 E I^2 \left(\mu_{2} ^2+\mu_{3} ^2\right)-\lambda_{1} ^2 \left(\mu_{1} ^2+\mu_{2} ^2+\mu_{3} ^2\right)}$. Moreover, using that 
\begin{equation}
    \dd \gamma (X, Y) = \liedv{X} \contr{Y} \gamma + \contr{[X,Y]} \gamma - \liedv{X} \contr{Y} \gamma
\end{equation}
for $X= {\mu}_2 e_1 -{\mu}_1 e_2$ and $Y = {\mu}_3 e_1 - {\mu}_1 e_3$, one can verify that the third of the equations~\eqref{HJ_eqs_SO3} is satisfied.

The Euler angles $(\alpha, \beta, \varphi)$ can be used as a coordinate system for $\mathrm{SO}(3)$, that is, the action of an element $g\in\mathrm{SO}(3)$ with coordinates $(\alpha, \beta, \varphi)$ on a point $(x,y, z)\in \R^3$ by rotations is given by the matrix multiplication
\begin{equation}
    \begin{pmatrix}
        \mathrm{c}_\beta \mathrm{c}_\varphi    
        & \mathrm{s}_\alpha \mathrm{s}_\beta  \mathrm{c}_\varphi  -  \mathrm{c}_\alpha  \mathrm{s}_\varphi
        & \mathrm{c}_\alpha \mathrm{s}_\beta \mathrm{c}_\varphi +  \mathrm{s}_\alpha \mathrm{s}_\varphi
        \\
         \mathrm{c}_\beta  \mathrm{s}_\varphi 
        & \mathrm{s}_\alpha \mathrm{s}_\beta  \mathrm{s}_\varphi +  \mathrm{c}_\alpha \mathrm{c}_\varphi
        & \mathrm{c}_\alpha \mathrm{s}_\beta \mathrm{s}_\varphi - \mathrm{s}_\alpha  \mathrm{c}_\varphi
        \\ 
        -\mathrm{s}_\beta 
        & \mathrm{s}_\alpha \mathrm{c}_\beta 
        & \mathrm{c}_\alpha \mathrm{c}_\beta
    \end{pmatrix}
    \begin{pmatrix}
        x \\ y \\ z
    \end{pmatrix}\, ,
\end{equation}
where $\mathrm{c}_x = \cos x$ and $\mathrm{s}_x=\sin x$.
Define the switching surface as the codimension-1 submanifold $S$ of $\mathrm{SO(3)}\times \mathfrak{so}(3)^\ast$ given by
\begin{equation}
    S = \left\{ \left(\alpha, \beta, \varphi, \eta_1, \eta_2, \eta_3\right)\in \mathrm{SO(3)}\times \mathfrak{so}(3)^\ast\mid \alpha=0 \right\}\, ,
\end{equation}
and the impact map $\Delta\colon S \to \mathrm{SO(3)}\times \mathfrak{so}(3)^\ast$ by
\begin{equation}
    \Delta \colon 
    \left(0, \beta, \varphi, \eta_1, \eta_2, \eta_3\right) \mapsto
    \left(0, \beta, \varphi, \varepsilon \eta_1, \eta_2,  \eta_3 \right)\, ,
\end{equation}
for some constant $\varepsilon$. Note that this impact map does not preserve the nonholonomic distribution. In other words, it is possible that after the impact the momenta no longer satisfy the nonholonomic constraints. This type of impact maps could be employed to model dynamical systems whose constraints are modified. For instance, a ball that switches from rolling without sliding to moving freely. Nevertheless, the present example is just an illustrative academic example that pretends to show the theory developed.

Let $\gamma^-$ and $\gamma^+$ denote the solutions to the Hamilton--Jacobi equation before and after the impact, respectively, where
\begin{equation}
    \gamma^\pm = \lambda_1^\pm e^1 + \frac{\mu_{3} \lambda_2^\pm-\mu_{1}  \mu_{2}    \lambda_1^\pm }{\mu_{2} ^2+\mu_{3} ^2} e^2 
    + \frac{\mu_{2}  \lambda_2^\pm -\mu_{1} \mu_{3}  \lambda_1^\pm}{\mu_{2} ^2+\mu_{3} ^2} e^3\, ,
\end{equation}
Then, $\gamma^+$ and $\gamma^-$ are $\Delta$-related if and only if
\begin{align}
    & \lambda_1^+ = \varepsilon \lambda_1^-\, ,\\
    & \mu_{3}\lambda_2^+ - \mu_{1}  \mu_{2}   \lambda_1^+ = \mu_{3}\lambda_2^- - \mu_{1}  \mu_{2}   \lambda_1^-\, ,\\
    & \mu_{2}  \lambda_2^+ -\mu_{1} \mu_{3}  \lambda_1^+ = \mu_{2}  \lambda_2^- -\mu_{1} \mu_{3}  \lambda_1^-\, ,
\end{align}
that is,
\begin{align}
    & \lambda_1^+ = \varepsilon \lambda_1^-\, ,\\
    & \lambda_2^+ = \lambda_2^- + (\varepsilon-1) \frac{\mu_1\mu_2}{\mu_3}\lambda_1^-\, , \\
   & \lambda_2^+ = \lambda_2^- + (\varepsilon-1) 
   \frac{\mu_1\mu_3}{\mu_2}\lambda_1^- \, ,
\end{align}
which has solutions if $\mu_3=\pm \mu_2$ or if $\varepsilon=1$. Therefore, $\hybrid_{\mathrm{nh}} = (\cT Q, C_\mu, X_{H}^{\mathrm{nh}}, S, \Delta)$ is an integrable nonholonomic hybrid system for $\mu=(\mu_1, \mu_2, \pm \mu_2)\in \mathfrak{so}(3)^\ast$ or for $\Delta$ the canonical inclusion.

In the case that $\mu=(\mu_1, \mu_2, \mu_2)\in \mathfrak{so}(3)^\ast$, the nonholonomic distribution is 
\begin{equation}
    D_\mu 
    = \left\langle \left\{a_2 e_1 -a_1 e_2,\ a_2 e_1 - a_1 e_3 \right\} \right\rangle\,.
\end{equation}
The solutions to the Hamilton--Jacobi equation are
\begin{equation}
    \gamma^\pm = \lambda_1^\pm e^1 + \frac{\lambda_2^\pm-\mu_{1}    \lambda_1^\pm }{2\mu_{2}} e^2 
    + \frac{\lambda_2^\pm -\mu_{1} \lambda_1^\pm}{2\mu_{2}} e^3\, .
\end{equation}

\chapter{Contact systems with impulsive constraints}\label{ch:contact_impulsive}

A geometrical formulation for mechanical systems with one-sided constraints was developed by Lacomba and Tulczyjew \cite{L.T1990}.
Ibort, de León, Lacomba, Marrero, Martín de Diego and Pitanga studied the geometrical aspects of Lagrangian systems subject to impulsive and one-sided constraints in a series of papers \cite{I.d.L+1997,I.d.L+1998,I.d.L+2001}. This was extended to the Hamiltonian formalism by Cortés and Vinogradov \cite{C.V2006}. Additionally, Cortés, de León, Martín de Diego and Martínez studied Lagrangian systems subject to generalized nonholonomic constraints \cite{C.d.M+2001,C.d.M+2001a}. The aim of this chapter is to go one step further by considering contact Lagrangian systems subject to impulsive forces and constraints, as well as instantaneous nonholonomic constraints which are not uniform along the configuration space.  A Carnot-type theorem for contact Lagrangian systems subject to impulsive forces and constraints is proven. This result characterizes the changes of energy due to contact-type dissipation and impulsive forces. In particular, the results by Ibort \textit{et al.} and Cortés \textit{et al.} are recovered from this formalism in the limit where there is no dissipation (that is, the Lagrangian is action-independent and the Herglotz--Euler--Lagrange equations yield the classical Euler--Lagrange equations).

The results from the present chapter were previously published in the article  \cite{Colombo2022a}.

\section{Contact Lagrangian systems subject to constraints}

See \Cref{sec:contact_Lagrangian} for an introduction to contact Lagrangian systems. 

Let $(Q, L)$ be a contact Lagrangian system. If the Lagrangian function is given by
\begin{equation}\label{natural}
  L (q, v, z) = \frac{1}{2}g_{q} (v,v) - V(q,z)\, ,
\end{equation}
where $g = g_{ij} \dd q^i \otimes \dd q^j$ is a (pseudo)Riemannian metric on $Q$, then the Herglotz--Euler--Lagrange equations~\eqref{eq:Herglotz_variational} yield
\begin{equation}\label{eqmetric}
  \ddot q^i(t) + \Gamma^i_{\ jk}\circ c(t) \dot q^j(t) \dot q^k(t) + g^{ij} \circ c(t)\frac{\partial V} {\partial q^j}\circ \sigma(t) + \dot q^i(t) \frac{\partial V} {\partial z}\circ \sigma(t) = 0\, ,
\end{equation}
for a curve $c(t)=(q^i(t))$ on $Q$, where $\sigma(t) = \big(c(t), \dot c(t), \zaction(c)(t))$ and $\Gamma^i_{\ jk}$ are the Christoffel symbols of the Levi--Civita connection $\nabla$ determined by $g$ and $(g^{ij})$ is the inverse matrix of $g_{ij}$. In other words, a curve $c\colon I \subset \RR \to Q$ is a solution of the Herglotz--Euler--Lagrange vector field $\sode_L$ if and only if
\begin{equation}
  \nabla_{\dot c (t)} \dot c(t) = -\grad V(c (t)) - \frac{\partial V} {\partial z} \big(c(t), \zaction(c)(t)\big)\ \dot c(t)\, ,
\end{equation}
where $\grad$ denotes the gradient with respect to $g$. Note that in the absence of potential, equations \eqref{eqmetric} are just the geodesic equations.

Suppose that the contact Lagrangian system $(Q, L)$ is restricted to certain (linear) constraints on the velocities modelled by a regular distribution $D$ on $Q$ of corank $k$. Then, $D$ may be locally described in terms of independent linear constraint functions $\left\{\Phi^{a}\right\}_{a=1, \ldots, k}$ by $D=\left\{(q, v) \in \T Q \mid \Phi^{a}(q, v)=0\right\}$, where $\Phi^{a}(q, v)=\Phi_{i}^{a}(q) v^{i}$. 
 More generally, one could consider constraints $\Phi_a$ also depending on $z$. Nevertheless, the author is not aware of any physical examples of constraints which depend on the variable $z$.

The dynamics $c\colon I\subseteq \RR \to Q$ of a contact Lagrangian system subject to nonholonomic constraints are given by the \emph{constrained Herglotz--Euler--Lagrange equations}:
\begin{equation} \label{Herglotz_eq_constrained}
\begin{aligned}
  &\frac{\dd }{\dd t} \parder{L}{v^i} \big(\sigma(t)\big)
  -  \parder{L}{q^i}  \big(\sigma(t)\big)
  -  \parder{L}{v^i}  \big(\sigma(t)\big) \parder{L}{z}  \big(\sigma(t)\big)
  \\ & \qquad \qquad
    =\lambda_{a}\big(c(t)\big) \Phi_{i}^{a}\big(c(t)\big) \, , \\
   \quad 
  & \Phi^{a}\big(c(t), \dot{c}(t)\big)=0\,,\quad a=1,\ldots,k\,,
\end{aligned}
\end{equation}
where $\sigma(t) = \big(c(t), \dot{c}(t), \zaction(c)(t) \big)$ and $\lambda_a$ are Lagrange multipliers. These equations can be obtained variationally from the Herglotz principle with constraints (see \cite{d.J.L2021}). For contact Hamiltonian systems with nonholonomic constraints refer to \cite{d.J2023}.

Note that if the Lagrangian is of the form \eqref{natural} then the constrained Herglotz equations are
\begin{equation}
\begin{aligned}
  \ddot q^i(t) &+ \Gamma^i_{\ jk}\big(c(t)\big) \dot q^j(t) \dot q^k(t) + g^{ij} \big(c(t)\big)\frac{\partial V} {\partial q^j}\big(c(t),\zaction(c)(t)\big) 
  \\ &
  + \dot q^i(t) \frac{\partial V} {\partial z}\big(c(t),\zaction(c)(t)\big) 
  = g^{ij}\big(c(t)\big) \lambda_a \big(c(t)\big)\Phi_j^a \big(c(t), \dot{c}(t)\big)\, .
\end{aligned}
\end{equation}
In other words, a curve $c\colon I\subseteq \RR \to Q$ satisfies the Herglotz variational principle if and only if
\begin{equation}
\begin{aligned}
  \nabla_{\dot c (t)} \dot c(t) &= -\grad V \big(c(t),\zaction(c)(t)\big) 
  \\ & \quad
  - \frac{\partial V} {\partial z} \big(c(t),\zaction(c)(t)\big)\ \dot c(t)
  + \lambda (c(t))\, ,\\
   \dot c(t) &\in D_{c(t)}\, ,
\end{aligned}
\end{equation}
where $\lambda$ is a section of $D^{\perp_g}$, and $D^{\perp_g}$ denotes the orthogonal complement of $D$ with respect to the metric $g$.


\section{Contact Lagrangian systems subject to impulsive forces and constraints}

In this section, the notion of impulsive constraints \cite{Brogliato1996,C.V2006,I.d.L+1997,I.d.L+1998,I.d.L+2001,L.T1990,Rosenberg1977,N.F2004} is extended to contact Lagrangian systems.

Consider a system of $n$ particles in $\RR^3$ such that the $j$-th particle has mass $m_j$. Let $(q^{3j-2}, q^{3j-1}, q^{3j})$ be the canonical coordinates describing the position of the $j$-th particle. Suppose that $F_j = (F^{3j-2}, F^{3j-1}, F^{3j})$ is the net force acting on the particle $j$. The equations of motion of the $j$-th particle in the interval $[t_0, t_1]\subset \RR$ are then given by
\begin{equation}
   \dot q^{k}(t_1) = \frac{1}{m_j} \int_{t_0}^{t_1} F^k(\tau)\ \dd \tau + \dot q^k (t_0), 
 \end{equation} 
 where $3j-2\leq k \leq 3j$. This equation is a generalization of the classical Newton's second law, since it allows one to consider the case of finite jump discontinuities \cite{Rosenberg1977}. This is the case of impulsive forces, which produce a non-zero impulse at some time instant. More precisely, if $F$ is impulsive then
 \begin{equation}
    \lim_{t\to t_0^+} \int_{t_0}^t F(t)\ \dd \tau = P \neq 0,
  \end{equation} 
for some instant $t_0$. This implies that the impulsive force has an infinite magnitude at $t_0$, but $P$ is well-defined and bounded. This can be expressed as
\begin{equation}
  \lim_{t\to t_0^+} F(t) = P\ \delta (t_0),
\end{equation}
where $\delta$ denotes the Dirac delta (see \cite{Dirac1947, Schwartz1966}). The impulsive forces may be caused by constraints, the so-called \emph{impulsive constraints}. Nonholonomic constraints of the form $\psi=0$ for $\psi = \psi_i(q) v^i$ have an associated constraint force given by $F_k = \mu\ \psi_k$, where $\mu$ is a Lagrange multiplier. Thus, the constraint is impulsive if
\begin{equation}
  \lim_{t\to t_0^+} \int_{t_0}^t \mu \ \psi_k\  \dd \tau = P_k \neq 0.
\end{equation}


The impulsive force may be due to a discontinuity at $t_0$ of $\psi_k$, of $\mu$ or of both. Henceforth, it will be assumed that the constraints $\psi$ are smooth and therefore the impulsive force is caused by a discontinuity of the Lagrange multiplier.

Consider a contact Lagrangian system subject to a set of impulsive linear nonholonomic constraints determined by the constraint functions $\left\{\psi^a  \right\}_{a=1}^r$. Equation~\eqref{Herglotz_eq_constrained} implies that
\begin{equation}\label{constrained_Herglotz_form}
  \frac{\dd p_i} {\dd t} \circ \sigma(t) = \frac{\partial L}{\partial q^{i}} \circ \sigma(t) + p_i \circ \sigma(t) \frac{\partial L}{\partial z} \circ \sigma(t) + \lambda_{a}\circ c(t) \psi_{i}^{a} \circ c(t) \, ,
\end{equation}
where $p_i = \partial L/ \partial v^i$ and $\sigma(t) = \big(c(t), \dot{c}(t), \zaction(c)(t)\big)$.
Then,
\begin{equation}
\begin{aligned}
   \lim_{t\to t_0^+} \int_{t_0}^t \frac{\dd p_i} {\dd\tau}\circ \sigma(\tau)\, 
  \delta q^i \, \dd \tau
   & = \lim_{t\to t_0^+} \int_{t_0}^t  \left[\left(\frac{\partial L}{\partial q^{i}} \circ \sigma(\tau) + p_i\circ \sigma(\tau) \frac{\partial L}{\partial z}\circ \sigma(\tau)
   \right. \right. \\ & \left. \left. \quad
    + \lambda_{a} \circ c(\tau) \psi_{i}^{a}\circ c(\tau)\right) \delta q^i\right] \dd \tau\, ,
\end{aligned}
\end{equation}
where $\delta q^i(t)$ are variations satisfying the constraints, that is, 
\begin{equation}\label{eq:constraint_variations}
  \psi_i^a\circ c(t) \, \delta q^i(t) = 0\, ,
\end{equation}
for all $q\in Q$. Equation~\eqref{eq:constraint_variations} implies that
\begin{equation}
  \psi_i^a\circ c(t_0) \, \delta q^i(t) + \mathcal{O}(t-t_0) = 0\, .
\end{equation}
Hence, the variations at $c(t)$ can be approximated by virtual displacements at $c(t_0)$, namely,
\begin{equation}
  \delta q^i (t) = \delta q^i(t_0) + \mathcal{O}(t-t_0)\, .
\end{equation}
Since $\partial L/\partial q^i$, $p_i$ and $\partial L/\partial z$ are bounded, the first two terms of the integral on the right-hand side vanish. The third term vanishes as well by equation~\eqref{eq:constraint_variations}. This implies that
\begin{equation}
  \left[p_i(t_0^+) - p_i (t_0)  \right] \delta q^i(t_0) = 0\, ,
\end{equation}
for variations $\delta_i$ satisfying the constraints \eqref{eq:constraint_variations}
Hence, the change of momentum $\Delta p_i\coloneqq p_i(t_0^+) - p_i (t_0)$ is given by
\begin{equation}
  \Delta p_i = \bar \mu_a \psi_i^a\, ,
\end{equation}
where $\bar \mu_a$ are Lagrange multipliers. If the Lagrangian is of the form~\eqref{natural}, then
\begin{equation}
  \Delta v^i \coloneqq v^i (t_0^+) - v^i (t_0^-) 
  = g^{ij} \bar \mu_a \psi_j^a\, .
\end{equation}

\subsection{Holonomic one-sided constraints}

Consider a contact Lagrangian system $(Q,L)$ subject to a holonomic one-sided constraint $\Psi(q)\geq 0$ (for instance, the collision with a fixed wall). This inequality determines a closed subset of $Q$, whose boundary $N$ is $1$-codimensional submanifold of $Q$. Suppose that the Lagrangian function $L$ is given by \eqref{natural}. Then, the equations of motion are
\begin{equation}
\begin{array}{ll}
     \nabla_{\dot \sigma (t)} \dot \sigma(t) = -\grad V(\sigma (t)) - \dfrac{\partial V} {\partial z} (\sigma(t))\ \dot \sigma(t)\, ,
    & \text{if } \Psi(\sigma(t))>0\, , \\ \\
    \Delta \dot \sigma(t) = \dot \sigma(t^+) - \dot \sigma(t^-) 
    \in \T_{\sigma(t)}^{\perp_g} N \, ,
     & \text{if } \Psi(\sigma(t))=0\, ,
\end{array}
\end{equation}
where $\T_{\sigma(t)}^{\perp_g} N$ is the orthogonal complement of $ \T_{\sigma(t)} N$ with respect to the metric $g$. 
In other words, if $\sigma(t)=(q^i(t))$ and $\psi(\sigma(t))=0$, then
\begin{equation}
  \dot q^i (t^+) - \dot q^i (t^-) = \bar \mu g^{ij} \frac{\partial \Psi } {\partial q^j}\, ,
\end{equation}
that is,
\begin{equation}
  \Delta \dot \sigma(t) = \bar \mu \grad \Psi\, ,
\end{equation}
where $\bar{\mu}$ is a Lagrange multiplier.

Consider the orthogonal projectors  $\qtilde: \T Q \to \T^{\perp_g} N$ and $\ptilde: \T Q \to \T N$.  
Locally,
\begin{equation}
  \qtilde (X) =  \frac{ g(\grad \Psi, X)}{g (\grad \Psi, \grad \Psi)} \grad \Psi \label{projector_Q_local}
\end{equation}
for any vector field $X$ on $Q$.

Suppose that the normal components of the velocities before and after the impact are related by $\dot \sigma(t^+)^{\perp_g} = -\alpha \dot \sigma(t^-)^{\perp_g}$, where $\alpha$ is the restitution coefficient. In other words, $\dd \Psi(\dot \sigma(t^+)) = - \alpha  \dd \Psi(\dot \sigma(t^-))$, or, equivalently, $\qtilde (\dot \sigma(t^+)) = - \alpha \qtilde (\dot \sigma(t^-))$. 
On the other hand,
\begin{equation}
  \ptilde (\dot \sigma(t^+)) =  \ptilde (\dot \sigma(t^-)),
\end{equation}
and hence
\begin{equation}
  \dot \sigma(t^+) = \left(\ptilde - \alpha \qtilde  \right) (\dot \sigma(t^-)). \label{change_velocity_projector}
\end{equation}


\begin{theorem}[Ibort \textit{et al.}, 2001]
  Under the assumptions above, let $T(v)$ denote the kinetic energy for the velocity $v$, namely $T(v) = \frac{1}{2} g(v, v)$. Then,
  \begin{equation}
    T_+(t) - T_-(t) = - \frac{1-\alpha}{1+\alpha}T_l(t)\, ,
  \end{equation}
  where $T_-(t)=T(\dot \sigma(t^-)),\ T_+(t)=T(\dot \sigma(t^+))$ and $T_l(t)=T(\dot \sigma(t^+)-\dot \sigma(t^-))$.
\end{theorem}
This result is known as the \emph{Carnot's theorem}. See \cite{I.d.L+2001} for the proof.

Suppose that $\Psi(\sigma(t_1))=0$ and $\Psi(\sigma(t))>0$ for $t\neq t_1$. Let $\sigma(t)= (q^i(t), \dot q^i(t), z(t))$.
Since the potential has no discontinuities, the instantaneous change of the energy is given by
\begin{equation}
  E_L \circ \sigma(t^+) - E_L \circ \sigma(t^-)
  = - \frac{1-\alpha}{1+\alpha}T_l(t)\, , \label{jump_energy}
\end{equation}
Therefore, 
\begin{equation}
  E_L \circ \sigma(t^+)
  = V \circ \sigma(t) - \frac{1-\alpha}{1+\alpha}T_l(t)\, .
\end{equation}
Equations~\eqref{eq:dissipation_Hamiltonian_time} and \eqref{jump_energy} imply the following.

\begin{proposition}
  Under the assumptions above, suppose that $\Psi(\sigma(t_1))=0$ and $\Psi(\sigma(t))>0$ for $t\neq t_1$. Then, the evolution of the energy $\sigma$ is given by
  \begin{equation}
    E_L \circ \sigma(t) =   e^{\left(\int_{t_0}^t -\Reeb_L(E_L)\,  \circ\, \sigma(\tau) \dd \tau \right)}  E_L  \circ \sigma(0)\, ,
  \end{equation}
  for $t_0<t<t_1$, and
  \begin{equation}
    E_L \circ \sigma(t) = e^{\left(\int_{t_1}^t -\Reeb_L(E_L)\,  \circ\, \sigma(\tau) \dd \tau \right)}\left[ V  \circ \sigma(t_1^-)-\dfrac{1-\alpha}{1+\alpha}T_l(t_1) \right]\, , 
  \end{equation}
  for $t>t_1$.
\end{proposition}

\subsection{Application: Rolling cylinder on a spring plane with an external force}
\label{cylinder_Carnot}




Consider a cylinder constrained to be above on a plane, in a gravitational field. Assume that the system is externally influenced by a force that depends linearly on the velocity between the cylinder and the plane (see \cite{Hagerty2001}). Subsequently, the cylinder may spin and slide when in contact with the plane. For simplicity's sake, it will be assumed that the mass distribution of the cylinder is uniform along its height. A restoring force is acting on the plane. The flexible plane is modelled as a large mass $M$, attached to a spring with spring constant $k$. 
The \emph{stance phase} of the cylinder is the dynamics of the cylinder in contact with the plane, and the \emph{aerial phase} is the dynamics of the cylinder not in contact with the plane.

The configuration space for the free motion of the cylinder is $Q=\RR^2\times\Sp^1\times\RR$. Let $m$, $I$ and $r$ be the mass, the rotational inertia about the center of mass, and the radius of the cylinder, respectively. Denote by $\gamma$ the distance from the center of mass to the center of the cylinder. Let $(x,y)\in\RR^2$ denote the horizontal and vertical position of the cylinder's center of mass, $\phi$ the angle through which the cylinder rotates about the center of mass and $h$ the vertical displacement of the rough plane from its equilibrium.
The dynamics of the system is determined by the action-dependent Lagrangian function
\begin{equation}
  L = \frac{1}{2}m(v_{x}^2+v_{y}^2)+\frac{1}{2}Mv_{h}^2+\frac{1}{2}Iv_{\phi}^2-\frac{1}{2}kh^2-mgy-Mgh+\beta z\, ,
\end{equation}
where $\beta$ is a positive constant.

During the stance phase (rolling motion) the system is constrained as 
\begin{equation}\label{constraints}
    x=\gamma\sin\phi+r\phi+x_0\,,\quad y=h+r+\gamma\cos\phi\, ,
\end{equation}
together with the holonomic one-side physical constraint that the cylinder cannot pass through the springly plane, that is,
\begin{equation}
  \Psi(q)=y-h-\gamma\cos\phi+r\sin\psi\geq 0\, ,
\end{equation}
for all $\psi\in \Sp^1$ where $(x,y,\phi,h)$ are the coordinates of $q\in Q$. No force is required to impose the constraint unless $\psi(q)=0$. This equality is maintained for $\psi=\pi$, while the cylinder is constrained to roll. 

The rolling constraints \eqref{constraints} can be characterized by the distribution 
\begin{equation}
  \overline{D}=\{v\in \T_q Q|\omega^1(v)=0,\,\omega^2(v)=0\}\, ,
\end{equation}
where 
$\omega^1 = \dd f_1$ and $\omega^2 = \dd f_2$, with 
\begin{equation}
  f_1 = x-\gamma\sin\phi-r\phi-x_0,\, \quad f_2 = y-h-r-\gamma\cos\phi\, ,
\end{equation}
which are the functions characterizing the holonomic constraints \eqref{constraints}. Since $\omega^1$ and $\omega^2$ are exact, $\overline{D}$ is an integrable distribution. 

In addition, the plane imposes positive forces on the cylinder which enforce the constraint $\Psi(q)=0$. The one-sided nature of the constraint is realized by two positive normal forces between the cylinder and the platform, $\mu_1(q,v)$ and $\mu_2(q,v)$, which will be specified below. The constraint distribution
\begin{equation}
  D=\{v\in \T_q Q|\omega^1(v)=0,\,\omega^2(v)=0,\, \mu_1(q,v)>0,\,\mu_2(q,v)>0\}\, .
\end{equation}
includes the rolling constraints as well as the restrictions due to the normal forces.

The aerial phase occurs when the cylinder is above the platform, that is, $\Psi(q)>0$. During this phase the trajectories $(x(t), y (t), \phi(t))$ of the cylinder are given by
\begin{equation}
  m\ddot{x}-\beta \dot x=0\, ,\quad m\ddot{y}- \beta \dot y=-mg\, ,\quad I\ddot{\phi}-\beta \dot \phi=0\, ,
\end{equation}
and no forces are required to satisfy the one-side holonomic constraint. In this phase, the cylinder shows circular motion about the center of mass while falling in a gravitational field.

The stance phase occurs when the cylinder is in contact with the plane and there are positive normal forces between the plane and the cylinder, that is, $\Psi(q)=0$ and $v\in D_q$. 
For the stance phase motion, the constrained Herglotz equations \eqref{Herglotz_eq_constrained} read
\begin{equation}
\begin{aligned}
  & m\ddot{x}-m\beta\dot{x}=\mu_1\, ,\\
  & m\ddot{y}+mg-m\beta\dot{y}=\mu_2\, ,\\
  & I\ddot{\phi}-I\beta\dot{\phi}=-(r+\gamma\cos\phi)\mu_1-\mu_2\gamma\sin\phi\, ,\\
  & m\ddot{h}+Mg+kh-M\beta\dot{h}=-\mu_2\, ,
\end{aligned}
\end{equation}
together with the constraints $x=\gamma\sin\phi+r\phi+x_0$, $y=h+r+\gamma\cos\phi$, $y-\gamma\cos\phi+r\sin\psi\geq h$. By differentiating these last three constraints,  after some computations, and solving for the Lagrange multipliers (that is, the normal forces between the plane and the cylinder) $\mu_1$ and $\mu_2$ as functions of $h,\phi,\dot{h},\dot{\phi}$, one can obtain 
\begin{equation}
\begin{aligned}
    \mu_1=&-\Upsilon(\phi)\left(2\beta\dot{\phi}(\gamma\cos\phi+r)
    \right.\\&\left.
    +\gamma\sin\phi\dot{\phi}^2\right)+\frac{\Upsilon(\phi)mM\gamma^2\sin\phi}{\Gamma(\phi)}\left(-2\beta\sin\phi\dot{\phi}+\frac{kh}{\gamma M}+\cos\phi\dot{\phi}^2\right)\, , \\
    \mu_2=&-\frac{mMI}{\Gamma(\phi)}\left(-2\beta\gamma\sin\phi\dot{\phi}-\frac{kh}{M}-\gamma\cos\phi\dot{\phi}^2+I^{-1}(r+\gamma\cos\phi)\mu_1\right),
\end{aligned} 
\end{equation}
with 
\begin{equation}
  \Upsilon(\phi)=\frac{mI\Gamma(\phi)}{m^2M\gamma\sin\phi(r+\gamma\cos\phi)+\Gamma(\phi)(I+m(r+\gamma\cos\phi)^2)}
\end{equation}
and $\Gamma(\phi)=I(M+m)-mM\gamma^2\sin^2\phi$.

Utilising equation~\eqref{projector_Q_local}, the projector $\qtilde: \T Q \to \T^{\perp_g} N$ can be written as
\begin{equation}
  \qtilde (X) = 
  \frac{X^y-X^h+\gamma \sin \phi X^\phi}{\frac{1}{m}-\frac{1}{M}+\frac{1}{I} (\gamma \sin \phi)^2} \left(\frac{1}{m} \frac{\partial  } {\partial y} -\frac{1}{M} \frac{\partial  } {\partial h} + \frac{1}{I} \gamma \sin \phi \frac{\partial  } {\partial \phi}  \right),  
\end{equation}
for $X=X^x \tparder{}{x} + X^y \tparder{}{y} + X^h\tparder{}{h} + X^\phi  \tparder{}{\phi}$. Combining this with equation~\eqref{change_velocity_projector} yields
\begin{equation}
\begin{aligned}
  &\dot x^+ = \dot x^-\, , \\
  &\dot y^+ = \dot y^- - (1+\alpha)  \frac{1}{m} 
  \frac{\dot y^- - \dot h^- +\gamma \sin \phi_1 \dot \phi^-}{\frac{1}{m}-\frac{1}{M}+\frac{1}{I} (\gamma \sin \phi)^2}\, , \\
  &\dot h^+ = \dot h^- + (1+\alpha)  \frac{1}{M} 
  \frac{\dot y^- - \dot h^- +\gamma \sin \phi_1 \dot \phi^-}{\frac{1}{m}-\frac{1}{M}+\frac{1}{I} (\gamma \sin \phi)^2}\, , \\
  &\dot \phi^+ = \dot \phi^- - (1+\alpha)  \frac{\gamma \sin \phi}{I} 
  \frac{\dot y^- - \dot h^- +\gamma \sin \phi_1 \dot \phi^-}{\frac{1}{m}-\frac{1}{M}+\frac{1}{I} (\gamma \sin \phi)^2}.
\end{aligned}
\end{equation}
Here $\dot x^\pm = \dot x(t_1^\pm),\ \dot y^\pm = \dot y(t_1^\pm),\  \dot h^\pm = \dot h(t_1^\pm),\ \dot \phi^\pm = \dot 
\phi(t_1^\pm)$ and $\phi=\phi(t_1)$, where $t_1$ is the time instant at which the impact occurs. Let $c(t)$ be a trajectory of the system and $\sigma(t) = \big(c(t), \dot{c}(t), \zaction(c)(t)\big)$.
The evolution of the energy is given by
\begin{equation}
  E_L \circ \sigma(t) = e^{\beta t} E \circ \sigma(0)\, , 
\end{equation}
for $t<t_1$, and
\begin{equation}
  E_L\circ \sigma(t) = e^{\beta t} 
  \left[V \circ \sigma(t_1) - \frac{1-\alpha}{1+\alpha} T_l(t_1)  \right]\, ,
\end{equation}
for $t>t_1$, where
\begin{equation}
  V \circ \sigma(t_1) = \frac{1}{2}k h(t_1)^2+mgy(t_1)+Mgh(t_1)-\beta z(t_1)\, ,
\end{equation}
and
\begin{equation}
\begin{aligned}
  T_l(t_1)
  &=\frac{1}{2} (1+\alpha)^2  
  \left[  \frac{\dot y^- - \dot h^- +\gamma \sin \phi_1 \dot \phi^-}{\frac{1}{m}-\frac{1}{M}+\frac{1}{I} (\gamma \sin \phi)^2}\right]^2
  \\ &\quad\cdot
  \left(\frac{1}{m}+\frac{1}{M}+\frac{1}{I} (\gamma \sin \phi)^2  \right)\, .
\end{aligned}
\end{equation}

\section{Contact Lagrangian systems with instantaneous nonholonomic constraints}
The concept of distribution can be generalized by not requiring its rank to be constant (see \Cref{subsec:generalized_distributions}).
Let $D^\circ$ be a generalized differentiable codistribution on $Q$.
The codistribution induces a decomposition of $Q$ into regular and singular points, namely, $Q= R\cup S$. Let $R_c$ be a connected component of $R$. Let $D_c^\circ$ denote the restriction of the codistribution $D^\circ$ to base points on $R_c$. By definition, $D_{c}^\circ$ is a regular codistribution. The annihilator of $D_c^\circ$ is denoted by $D_c$, namely,
\begin{equation}
  (D_c)_q = \left\{v\in \T_q Q \mid \alpha(v) = 0\ \forall\, \alpha \in (D_c)_q\right\}\, ,
\end{equation}
for each $q\in R_c$. Similarly, $D$ is defined as the annihilator of $D^\circ$.

Given a contact Lagrangian system $(Q, L)$ constrained to $D$, one can apply the theory for nonholonomic contact Lagrangian systems developed by de León, Jiménez and Lainz in \cite{d.J.L2021} in order to solve the problem along each connected component of $R$. However, if the motion reaches a singular point, the rank of $D$ can vary suddenly, and the equations of motion can no longer be derived from the Herglotz principle with constraints. As a matter of fact, an impulsive force can emerge because of the change of rank of $D$. 
The rank of $D^\circ$, that is, the corank of $D$, is equal to the number of constraints.

Consider a trajectory of the system $c(t)$ which reaches a singular point at $t_0$, that is, $c(t_0)\in S$, such that $c((t_0- \varepsilon, t_0))\subset R$ and $c((t_0, t_0+\varepsilon))\subset R$ for sufficiently small $\varepsilon>0$. Let $\rho_\pm=\rho(c(t_0\pm\varepsilon))$ and $\rho_0=\rho(c(t_0))$. Recall that regular points are local maximums of $\rho$, and hence there are three possible cases that can occur at $t_0$:
\begin{enumerate}
  \item $\rho_-=\rho_0<\rho_+$,
  \item $\rho_->\rho_0=\rho_+$,
  \item $\rho_->\rho_0$ and $\rho_+>\rho_0$.
\end{enumerate}
In the first and third cases, the trajectory must satisfy, immediately after the point $c(t_0)$, a greater number of constraints which were not present before. This leads to an impulsive force which imposes the new constraints on the motion. Hereinafter, it will be assumed that the singular points are in one of this two cases, namely, $\rho_0<\rho_+$.

Given the codistribution $D^\circ$, one can introduce an associated distribution $D^\ell$ on $\T Q\times \RR$, whose annihilator is given by 
\begin{equation}
  D^{\ell^\circ } = \pi^\ast D^\circ,
\end{equation}
where $\pi\colon \T Q \times \RR \to Q$ is the canonical projection.

\begin{theorem}
Let $\sode_{L,D}$ be a vector field on $\T Q\times \RR$ such that
\begin{equation}
\begin{aligned}
  &\flat_{\eta_L} (\sode_{L,D})- \dd E_L + \left(E_L + \Reeb_L(E_L)  \right) \eta_L \in D^{\ell^\circ}\, , \\
 &\Ima \sode_{L,D}|_{D\times \RR} \subset {T} \left( D\times \RR \right)\, ,
\end{aligned}
\label{eqs:sode_constrained}
\end{equation}
where $\flat_{\eta_L}\colon Y\in \X(\T Q\times \RR)\mapsto (\contr{Y} \eta_L) \eta_L + \contr{Y} \dd \eta_L\in \Omega^1(\T Q\times \RR)$ denotes the isomorphism defined by the contact form $\eta_L$.
Then:
\begin{enumerate}
\item $\sode_{L,D}$ is a \textsc{sode},
\item the integral curves of $\sode_{L,D}$ are solutions of the contrained Herglotz equations \eqref{Herglotz_eq_constrained}.
\end{enumerate}
\end{theorem}

\begin{proof}
  Since $D^\circ$ is a differentiable codistribution, for each $q\in Q$ there exists a local neighbourhood such that
  $\restr{D}{U}=\gen{ \left\{\psi^1, \ldots, \psi^m  \right\}}$ for some one-forms $\psi^i \in \Omega^1(U)$, where $m$ is the local maximum of $\rho$ at $U$. The rest of the proof is identical to the one of Theorem 6 from reference~\cite{d.J.L2021}.
\end{proof}

Let $\mathcal S$ be the generalized distribution on $\T Q\times \RR$ defined by $\sharp_{\eta_L} (D^{\ell^\circ})$. Let $Y_a$ be the local vector fields on $\T Q\times \RR$ given by
\begin{equation}
  \flat_{\eta_L} (Y_a) = \tilde \psi^a\, ,
\end{equation}
where $\tilde \psi^a = \pi^\ast \psi^a = \psi^a_i \dd q^i$ are one-forms on $\T Q\times \RR$, with $\pi\colon \T Q \times \RR \to Q$ the canonical projection.
Clearly, $\mathcal S_q$ is generated by $\left\{\restr{Y_a}{q}  \right\}$.
These vector fields locally read
\begin{equation}
  Y_a = -W^{ij} \psi^a_j \frac{\partial  } {\partial \dot q^i}.
\end{equation}
Consider the condition
\begin{equation}
  \mathcal{S} \cap \T \left(D\times \RR  \right) = \left\{0  \right\}\, . \label{condition_uniqueness}
\end{equation}

Define $Z_a = \sharp_{\eta_L} \tilde{\psi}^a$ for each $a \in \{1, \ldots, m\}$.
Then, $X=X^b Z_b$ is a vector field tangent to $\mathcal S$. The vector field $X$ is a section of $\T\left(D\times \RR\right)$ if and only if
\begin{equation}
  0 = \dd \bar \psi^a(X) = \psi^a_i (X^b Z_b)^i = -\psi^a_i W^{jk} \psi^b_k X^b,
\end{equation}
where $\bar \psi^b=\psi^b_i(q)\dot q^i\ (b=1,\ldots, r)$ are functions on $Q\times \RR$.
Consider the matrix
\begin{equation}
  \left(\mathcal{C}_{ab}\right) = -\left( W^{ij} \psi^a_i \psi^b_j  \right). 
\end{equation}

Observe that $\dim \mathcal S_{(v_q, z)}=\rho(q)$ and $\dim T_{(v_q,z)}(D \times \RR)=2n+1-\rho (q)$ for each $(v_q, z)\in D\times \RR$. 
If condition \eqref{condition_uniqueness} holds, this implies that
\begin{equation}
  \mathcal S \oplus \T (D \times \RR) = \T_{D \times \RR} (\T Q \times \RR),
\end{equation}
where $\T_{D \times \RR} (\T Q \times \RR)$ consists of the tangent vectors of $\T Q\times \RR$ at points of $D \times \RR$.
Therefore, it is natural to introduce the following projectors:
\begin{equation}
\begin{aligned}
  &\hat{\mathcal{P}}: \T_{D \times \RR}(\T Q \times \RR) \rightarrow \T(D \times \RR)\, ,\\
  &\hat{\mathcal{Q}}: \T_{D \times \RR}(\T Q \times \RR) \rightarrow \mathcal{S}\, .
\end{aligned}
\end{equation}
Let $X= \hat{\mathcal{P}}(\sode_L|_{D \times \RR})$. By construction, $\Ima(X)\in \T(D\times \RR)$. On the other hand, at the points in $D \times \RR$, one can write
\begin{equation}
\begin{aligned}
  \flat_{\eta_L}(X)-\dd E_{L}+\left(E_{L}+\mathcal{R}_{L}\left(E_{L}\right)\right) \eta_{L} 
  &=\flat_{\eta_L}\left(\sode_{L}-\hat{\mathcal{Q}}\left(\sode_{L}\right)\right)-\dd E_{L}
  \\ & \quad 
  +\left(E_{L}+\mathcal{R}_{L}\left(E_{L}\right)\right) \eta_{L} \\
  &=-\flat_{\eta_L}\left(\hat{\mathcal{Q}}\left(\sode_{L}\right)\right) \in D^{\ell^{\circ}},
\end{aligned}
\end{equation} 
Hence, $X$ satisfies equations \eqref{eqs:sode_constrained}, which means that $X= \sode_{L,D}$.

The next step is to compute an explicit expression of $\sode_{L, D}$. Let $Y$ be a vector field on $\T Q \times \RR$. Then, choosing a local basis $\left\{\beta_{i}\right\}$ of $\T(D \times\RR)$ one may write the restriction of $Y$ to $D \times \RR$ as 
\begin{equation}
  \restr{Y}{D \times \RR}=Y^{i} \beta_{i}+\lambda^{a} Z_{a}\, ,
\end{equation}
Therefore, 
\begin{equation}
  \dd \bar \psi^b (Y) = \lambda^a \mathcal{C}_{ba}\, ,
\end{equation}
and the coefficients $\lambda^a$ are given by
\begin{equation}
  \lambda^a = \mathcal{C}^{ba} \dd \bar \psi^b(Y)\, ,
\end{equation}
where $(\mathcal C^{ab})$ denotes the inverse matrix of $(\mathcal C_{ab})$. Thus,
\begin{equation}
\begin{aligned}
  &\hat{\mathcal{Q}}\left(\restr{Y}{D \times \RR}\right)=\mathcal C^{b a} ~\dd  \bar{\psi}^{b}(Y) Z_{a}\, , \\  
  &\hat{\mathcal{P}}\left(\restr{Y}{D \times \RR}\right)=\restr{Y}{D \times \RR}-\mathcal C^{b a} ~\dd  \bar{\psi}^{b}(Y) Z_{a}.
\end{aligned}
\end{equation}
This yields the following result.
\begin{proposition}
  If $\sode_L$ is the Herglotz--Euler--Lagrange vector field of $L$, then 
  \begin{equation}
    \sode_{L, D}=\restr{\sode_{L}}{D \times \RR}-\mathcal C^{b a} ~\dd  \bar{\psi}^{b}\left(\sode_{L}\right) Z_{a}.
  \end{equation}
\end{proposition}


Consider now a contact Lagrangian system $(Q, L)$ with mechanical Lagrangian function $L(q, \dot q, z)=\frac{1}{2}g(\dot q, \dot q)-V(q,z)$. 
Given a trajectory $c(t)$, let $D_{c(t_0)}^{\circ -}$ and $D_{c(t_0)}^{\circ +}$ be the vector subspaces of $\cT_{c(t_0)}Q$ given by
\begin{equation}
\begin{aligned}
  D_{c(t_0)}{\circ -} 
  &\coloneqq \left\{\alpha \in \cT_{c(t_0)}Q \mid
  \exists \tilde \alpha: (t_0-\varepsilon, t_0)\to \cT Q 
  \right.\\ & \quad \left. \text{such that }
  \tilde \alpha(t)\in D_{c(t)} 
  \text{ and } \lim_{t\to t_0^-} \tilde \alpha(t) = \alpha
     \right\}\, ,
\end{aligned}
\end{equation}
and 
\begin{equation}
\begin{aligned}
  D_{c(t_0)}^{\circ +}
   & \coloneqq \left\{\alpha \in \cT_{c(t_0)}Q \mid
    \exists \tilde \alpha: (t_0, t_0+\varepsilon)\to \cT Q 
    \right.\\ & \quad \left. \text{such that }
    \tilde \alpha(t)\in D_{c(t)} 
    \text{ and } \lim_{t\to t_0^+} \tilde \alpha(t) = \alpha
       \right\}\, .  
\end{aligned}
\end{equation}
Then,
\begin{equation}
  \left(D_{c(t_0)}^{\circ \pm}  \right)^{\perp_g}
  = \lim_{t\to t_0^\pm} \left(D_{c(t)^\circ}  \right)^{\perp_g},
\end{equation}
where the superscript ${\perp_g}$ denotes the orthogonal complement with respect to the bilinear form induced by the metric $g$ on $\cT_{c(t_0)}Q$, and the limits $(D^{\perp_g})^\pm$ are defined as in the case of $D^\pm$. 

Equation~\eqref{constrained_Herglotz_form} implies that
\begin{equation}
   \left(\frac{\dd p_i} {\dd t} \circ \sigma(t) - \frac{\partial L} {\partial q^i} \circ \sigma(t) - \frac{\partial L} {\partial \dot q^i} \circ \sigma(t) \frac{\partial L} {\partial z}\circ \sigma(t)  \right)\ 
   \dd q^i
   \in D_{c(t)}^\circ\, ,
\end{equation}
where $\sigma(t) = \big(c(t), \dot{c}(t), \zaction(c)(t)\big)$. Thus, 
\begin{equation}
 \lim_{t\to t_0^+} \int_{t_0}^t  \left(\frac{\dd p_i} {\dd t} - \frac{\partial L} {\partial q^i} - \frac{\partial L} {\partial \dot q^i} \frac{\partial L} {\partial z}  \right)\ 
   \dd q^i\ \dd t
   = \left[p_i(t_0^+)-p_i(t_{0})  \right] \dd q^i
   \in 
    D_{q(t)}^{\circ +}\, .
\end{equation}
Additionally, the `post-impact' momentum $p(t_0^+)$ must satisfy the new constraints imposed by $D_{q(t_0)}^{\circ +}$. Therefore, the change of momentum is determined by the following equations:
\begin{subequations}
\begin{flalign}
    & \left[p_i(t_0^+)-p_i(t_{0})  \right] \dd q^i
   \in D_{q(t)}^{\circ +},
   \label{change_momentum_distribution}
   \\ 
   & p_i(t_0^+)\ \dd q^i \in \left(D_{q(t)}^{\circ +}  \right)^{\perp_g}.
\end{flalign}
\end{subequations}
A momentum jump occurs if the `pre-impact' momentum does not satisfy the constraints imposed by $D_{q(t)}^{\circ +}$, that is,
\begin{equation}
   p_i(t_0^-)\ \dd q^i \notin \left(D_{q(t)}^{\circ +}  \right)^{\perp_g}.
\end{equation}

Let $m=\max\{\rho_-, \rho_+\}$. Then, there exists a neighbourhood $U$ of $q(t_0)$ and one-forms $\psi^1, \ldots, \psi^m$ such that $D_q = \gen{ \left\{\psi^i(q)  \right\}}_{i=1}^m$ for any $q\in Q$.
Assume that $\psi^1, \ldots, \psi^{\rho_+}$ are linearly independent at the regular posterior points. Obviously, these one-forms are linearly dependent at $q(t_0)$. In order to simplify the notation, hereinafter let $\psi^a=\psi^a(q(t))$. The constraints read
 \begin{equation}
   \psi^a_i\dot q^i(t_0^+)
   =\psi^a_{i} g^{ij} p_j(t_0^+) = 0\, ,
   \label{constraint_local}
 \end{equation}
 for $a=1, \ldots, \rho_+$, where $c(t) = (q^i(t))$ and $p_i(t) = \tparder{L}{v^i} \circ \sigma(t)$. The metric $g$ induces the decomposition $\cT_q Q= D_q^\circ \oplus D_q^{\circ {\perp_g}}$, with the projectors
 \begin{equation}
 \begin{aligned}
   &\mathcal P_q : \cT_q Q \to D_q^{\circ {\perp_g}}\, , \\
   & \mathcal Q_q : \cT_q Q \to D_q^{\circ}.
 \end{aligned}
 \end{equation}
 Consider the matrix $(\mathcal C^{ab}) = (\psi^a_i g^{ij} \psi^b_j)$, in other words, $\mathcal C= \psi g^{-1} \psi^T$, and let $(\mathcal C_{ab})$ denote its inverse matrix. The projector $\mathcal P_q$ is given by
 \begin{equation}
   \mathcal P_q(\alpha) 
   = \alpha -\mathcal C_{ab} g^{ij} \psi^a_i \alpha_j \psi^b\, ,
 \end{equation}
 for $\alpha=\alpha_i \dd q^i \in \cT_q Q$.
Equation~\eqref{constraint_local} implies that
\begin{equation}
  \mathcal P_{q(t)}\left(p_i(t)\dd \restr{q^i}{q(t)}\right)=p_i(t)\dd \restr{q^i}{q(t)} \, .
\end{equation}
Therefore,
\begin{equation}
\begin{aligned}
  p_i(t_0^+) \dd \restr{q^i}{q(t_0^+)}
  & = \lim_{t\to t_0^+} p_i(t) \dd \restr{q^i}{q(t)}\\
  & = \lim_{t\to t_0^+} \mathcal P_{q(t)}\left(p_i(t_0^+)\dd \restr{q^i}{q(t)}\right)
  \in \left( D_{q(t_0)}^{\circ +}  \right)^{\perp_g}\, .
\end{aligned}
\end{equation}
On the other hand, equation~\eqref{change_momentum_distribution} implies that
\begin{equation}
   \lim_{t\to t_0^+} \mathcal P_{q(t)}\left( \left(p_i(t_0+) - p_i(t_0)  \right)\dd \restr{q^i}{q(t)}\right) = 0\, ,
\end{equation}
and hence the change of momentum is given by
\begin{equation}
  p_i(t_0^+) \dd \restr{q^i}{q(t_0)}
  =  \lim_{t\to t_0^+} \mathcal P_{q(t)} \left(  p_i(t_0^-) \dd \restr{q^i}{q(t_0)} \right)\, .
  \label{eq_change_momentum_P_matrix}
\end{equation}
Locally, this can be expressed as
\begin{equation}
  p_i(t_0^+) = p_i(t_0^-) - \lim_{t\to t_0^+} \sum_{a,b,j,k}  
  \left.\left(\mathcal C_{ab} \psi^a_j g^{jk} \psi^b_i\right)\right|_{q(t)} p_k(t_0^-)\, ,
\end{equation}
or, in matrix form,
\begin{equation}
  p(t_0^+)= \left[\operatorname{Id}- \lim_{t\to t_0^+} \left. \left(\psi^T \mathcal C^{-1} \psi g^{-1}  \right)\right|_{q(t)}  \right] p(t_0^-)\, .
  \label{eq_change_momentum_C_matrix}
\end{equation}

\begin{example}[Rolling cylinder on a spring plane with an external force]
  Consider the system from \Cref{cylinder_Carnot}. The constraints \eqref{constraints} define the generalized codistribution $D^\circ$ given by
  \begin{equation}
    D_{(x, y, \phi, h)}^\circ = \gen{\left\{\dd x-(r+\gamma\cos\phi)\dd \phi, \dd y-\dd h+(\gamma\sin\phi)\dd \phi\right\}}\, ,
  \end{equation}
  for the base points in which $y-h-\gamma \cos \phi=0$, and $D_{(x, y, \phi, h)}^\circ = \{0\}$ elsewhere.
  The instantaneous change of momentum that occurs when the cylinder transitions from the aerial phase to the stance phase can be computed via equation~\eqref{eq_change_momentum_P_matrix}. The constraints can be expressed in matrix form as
  \begin{equation}
      \phi = \left(
      \begin{array}{cccc}
        1 & 0 & -\gamma  \cos (\phi )-r & 0 \\
        0 & 1 & -\gamma  \sin (\phi ) & -1 \\
      \end{array}
      \right)\, .
  \end{equation}
  Therefore, the matrix $\mathcal C$ is given by
  \begin{equation}
      \mathcal C =\phi g^{-1} \phi^T 
      = \left(
      \begin{array}{cc}
      \frac{(-\gamma  \cos (\phi )-r)^2}{I}+\frac{1}{m} & -\frac{\gamma  \sin (\phi ) (-\gamma  \cos (\phi )-r)}{I} \\
      -\frac{\gamma  \sin (\phi ) (-\gamma  \cos (\phi )-r)}{I} & \frac{\gamma ^2 \sin ^2(\phi )}{I}+\frac{1}{m}+\frac{1}{M} \\
      \end{array}
      \right)\, ,
  \end{equation}
  and the projector $\mathcal P$ is
  \begin{equation}
    \mathcal P
      = \operatorname{Id}_4-\phi^T \mathcal C^{-1} \phi g^{-1}\, .
  \end{equation}
  Hence, 
  \scriptsize
  \begin{align}
      p_x^+&=\frac{2 m (\gamma  \cos (\phi )+r) [\gamma  \sin (\phi ) (m p_h^--M p_y^-)+(m+M) (\gamma  p_x^- \cos (\phi )+p_x^- r+p_\phi^-)]}{2 (m+M) \left(I+m r^2\right)+\gamma ^2 m (m+2 M)+\gamma  m (\gamma  m \cos (2 \phi )+4 r (m+M) \cos (\phi ))}\, , \\
      p_y^+&=\frac{m}{(m+M) \left(I+m r^2+\gamma  m \cos (\phi ) (\gamma  \cos (\phi )+2 r)\right)+\gamma ^2 m M \sin ^2(\phi )}\\
      &\times\left[(p_h^-+p_y^-) \left(I+m r^2+\gamma  m \cos (\phi ) (\gamma  \cos (\phi )+2 r)\right)
      \right.\\&\left.
      -\gamma  M \sin (\phi ) (\gamma  p_x^- \cos (\phi )+p_x^- r+p_\phi^-)+\gamma ^2 M p_y^- \sin ^2(\phi )\right]\, , \\
      p_\phi^+&=\frac{2 I (\gamma  \sin (\phi ) (m p_h^--M p_y^-)+(m+M) (\gamma  p_x^- \cos (\phi )+p_x^- r+p_\phi^-))}{2 (m+M) \left(I+m r^2\right)+\gamma ^2 m (m+2 M)+\gamma  m (\gamma  m \cos (2 \phi )+4 r (m+M) \cos (\phi ))}\, , \\
      p_h^+&=\frac{M}{2 (m+M) \left(I+m r^2\right)+\gamma ^2 m (m+2 M)+\gamma  m (\gamma  m \cos (2 \phi )+4 r (m+M) \cos (\phi ))}\\
      &\times \left[2 (p_h^-+p_y^-) \left(I+m r^2\right)+\gamma  m (4 r (p_h^-+p_y^-) \cos (\phi )
      \right.\\&\left.
      +2 \sin (\phi ) (\gamma  p_x^- \cos (\phi )+p_x^- r+p_\phi^-)+\gamma  p_y^- \cos (2 \phi ))+\gamma ^2 m (2 p_h^-+p_y^-)\right]\, .
  \end{align}
  \normalsize
  Making use of the relations
  \begin{equation}
      p_x^\pm =  m \dot x^\pm,\quad
      p_y^\pm =  m \dot y^\pm,\quad
      p_\phi^\pm =  I\dot \phi^\pm,\quad
      p_h^\pm =  M \dot h^\pm
  \end{equation}
  defined by the Legendre transform, one can obtain the instantaneous change of velocity.
\end{example}

\begin{example}[The rolling sphere with dissipation]
  Consider a homogeneous sphere rolling on a plane. The configuration space is $Q=\RR^{2} \times \mathrm{S O}(3)$ (see 
  \cite{C.d.M+2001a} for theaction-independent counterpart of this example). Let $(x, y)$ denote the position of the centre of the sphere and let $(\varphi, \theta, \psi)$ denote the Eulerian angles.

  Assume that the plane is smooth if $x<0$ and absolutely rough if $x>0$. On the smooth part, the motion of the ball is free, whereas when it reaches the rough half-plane, the sphere rolls without slipping. Suppose that the motion of the sphere, both on the smooth and rough half-planes, has a dissipation linear in the velocities. The contact Lagrangian of the system is
  \begin{equation}
    L = \frac{1}{2}\left[v_{x}^{2}+v_{y}^{2}+k^{2}\left(\omega_{x}^{2}+\omega_{y}^{2}+\omega_{z}^{2}\right)\right] - \beta Z\, ,
  \end{equation}
  where $\omega_{x}, \omega_{y}$ and $\omega_{z}$ are the angular velocities with respect to the inertial frame, given by
  \begin{equation}
  \begin{aligned}
    &\omega_{x}=v_{\theta} \cos \psi+v_{\varphi} \sin \theta \sin \psi\, , \\
    &\omega_{y}=v_{\theta} \sin \psi-v_{\varphi} \sin \theta \cos \psi\, ,\\
    &\omega_{z}=v_{\varphi} \cos \theta+v_{\psi}\, ,
  \end{aligned}
  \end{equation}
  $Z$ is the action, and $\beta$ is a positive constant.

  The condition of rolling without sliding is given by
  \begin{equation}
  \begin{array}{l}
    \phi^{1}=v_{x}-r \omega_{y}=0, \\
    \phi^{2}=v_{y}+r \omega_{x}=0.
  \end{array}
  \end{equation}
 Let $q^1,\, q^2$ and $q^3$ be quasi-coordinates such that their associated velocities are the angular velocities, namely, $v_ q^1 = \omega_x,\ v_ q^2= \omega_2$ and $v_ q^3= \omega_3$. The generalized distribution $D$
  characterizing the constraints has annihilator $D^\circ$ given by
  \begin{equation}
    D_{(x, y, \phi, \theta, \psi)}^\circ=\left\{\begin{array}{ll}
    \{0\}\,, & \text { if } x \leqslant 0\, , \\
    \gen{\left\{\dd x-r ~\dd  q^{2}, ~\dd  y+r ~\dd  q^{1}\right\}}\,, & \text { if } x>0\, .
  \end{array}\right.
  \end{equation}
  The set of regular points of the distribution has two connected components, namely,
  \begin{equation}
  \begin{aligned}
    &R_1 = \left\{(x, y, \varphi, \theta, \psi )\in Q\mid x<0  \right\}\, , \\  
    &R_2 = \left\{(x, y, \varphi, \theta, \psi )\in Q\mid x>0  \right\},
  \end{aligned}
  \end{equation}
  while the line $\left\{x=0  \right\}$ belongs to the singular set of $D$.
  On $R_1$ the equations of motion are
  \begin{equation}
  \begin{aligned}
    &  \ddot x + \beta \dot x  = 0\, ,\\
    &  \ddot y + \beta \dot y  = 0\, , \\
    & k^2 \dot \omega_a + \beta \omega_a = 0, \quad a=x,y,z.
  \end{aligned}
  \label{eqs_sphere_R1}
  \end{equation}
  On $R_2$ the equations of motion are
  \begin{equation}
  \begin{aligned}
    &  \ddot x + \beta \dot x  = \lambda_1\, , \\
    &  \ddot y + \beta \dot y  = \lambda_2\, , \\
    & k^2 \dot \omega_x + k^2 \beta \omega_x = r\lambda_2\, , \\
    & k^2 \dot \omega_y + k^2 \beta \omega_y = -r\lambda_1\, , \\
    & k^2 \dot \omega_z + k^2\beta \omega_z = 0\, , \\
    & \dot x - r\omega_y = 0\, , \\
    & \dot y+ r \omega_x = 0.
  \end{aligned}
  \label{eqs_sphere_R2}
  \end{equation}

  Assume that the sphere starts its motion at some point in $R_1$ with positive velocity in the $x$-direction, namely, $x(0)=x_0,\, y(0)=y_0,\, \omega_a(0)=(\omega_a)_0\, (a=x,y,z)$ such that $x_0<0$ and $\dot x_0>0$. Integrating equations~\eqref{eqs_sphere_R1} yields
  \begin{equation}
  \begin{aligned}
    &  x(t)=\frac{\dot x_0}{\beta } \left(1-e^{-\beta  t}\right)+ x_0\, , \\
    &  y(t)=\frac{\dot y_0}{\beta }\left(1-e^{-\beta  t}\right)  +y_0\, , \\
    & \omega_a(t)=e^{-\frac{\beta  t}{k^2}}(\omega_a)_0, \quad a=x,y,z,
  \end{aligned}
  \label{eqs_sphere_R1_integrated}
  \end{equation}
  for $x(t)<0$. At time $\bar{t}=-x_0/\dot x_0$ the sphere reaches the rough {surface} of the plane, where the codistribution $D^\circ$ is no longer zero, so the sphere is forced to roll without sliding.

  The instantaneous change of momentum can be computed by means of equation~\eqref{eq_change_momentum_P_matrix}. 
  The constraints can be expressed in matrix form as
  \begin{equation}
      \phi = (\phi_i^a) =  \begin{pmatrix}
          1 & 0 & 0 & -r & 0\\
          0 & 1 & r & 0 & 0
      \end{pmatrix}, 
  \end{equation}
  so the matrix $\mathcal C$ is given by
  \begin{align}
      \mathcal C &=\phi g^{-1} \phi^T 
      = \begin{pmatrix}
          1+\frac{r^2}{k^2} & \\
        &1+\frac{r^2}{k^2},
      \end{pmatrix}
      =\frac{k^2+r^2}{k^2} \mathrm{Id}_2
  \end{align}
  and the projector $\mathcal P$ is
  \begin{align}
      \mathcal P
      &= \operatorname{Id}_5-\phi^T \mathcal C^{-1} \phi g^{-1}  
      =\left(
      \begin{array}{ccccc}
      \frac{r^2}{k^2+r^2} & 0 & 0 & \frac{r}{k^2+r^2} & 0 \\
      0 & \frac{r^2}{k^2+r^2} & -\frac{r}{k^2+r^2} & 0 & 0 \\
      0 & -\frac{k^2 r}{k^2+r^2} & \frac{k^2}{k^2+r^2} & 0 & 0 \\
      \frac{k^2 r}{k^2+r^2} & 0 & 0 & \frac{k^2}{k^2+r^2} & 0 \\
      0 & 0 & 0 & 0 & 1 \\
      \end{array}
      \right).
  \end{align}
  Hence, 
  \begin{align}
      &\left(p_{x}\right)_{+}=\frac{r^{2}\left(p_{x}\right)_{-}+r\left(p_{2}\right)_{-}}{r^{2}+k^{2}}\, , \\
      &\left(p_{y}\right)_{+}=\frac{r^{2}\left(p_{y}\right)_{-}-r\left(p_{1}\right)_{-}}{r^{2}+k^{2}}\, , \\
      &\left(p_{1}\right)_{+}=\frac{-r k^{2}\left(p_{y}\right)_{-}+k^{2}\left(p_{1}\right)_{-}}{r^{2}+k^{2}}\, , \\
      &\left(p_{2}\right)_{+}=\frac{r k^{2}\left(p_{x}\right)_{-}+k^{2}\left(p_{2}\right)_{-}}{r^{2}+k^{2}}\, , \\
      &\left(p_{3}\right)_{+}=\left(p_{3}\right)_{-}.
  \end{align}
  Introducing the quasi-coordinates $q^1, q^2$ and $q^3$ such that $\dot q^1= \omega_x,\, \dot q^2= \omega_y$ and $\dot q^3= \omega_z$, the momenta read
  \begin{equation}
  \begin{array}{lll}
      p_{x}=\frac{\partial L}{\partial \dot x}=\dot{x}, 
      &p_{y}=\frac{\partial L}{\partial \dot y}=\dot{y}, \\ 
      p_{1}=\frac{\partial L}{\partial \dot q^1}=k^{2} \omega_{x}, 
      &p_{2}=\frac{\partial L}{\partial \dot q^2}=k^{2} \omega_{y}, 
      &p_{3}=\frac{\partial L}{\partial \dot q^3}=k^{2} \omega_{z},
  \end{array}
  \end{equation}
  and thus the instantaneous change of velocity is given by
  \begin{equation}
  \begin{aligned}
      &\dot{x}_{+}=\frac{r^{2} \dot{x}_{-}+r k^{2}\left(\omega_{y}\right)_{-}}{r^{2}+k^{2}}\, , \\
      &\dot{y}_{+}=\frac{r^{2} \dot{y}_{-}-r k^{2}\left(\omega_{x}\right)_{-}}{r^{2}+k^{2}}\, , \\
      &\left(\omega_{x}\right)_{+}=\frac{-r \dot{y}_{-}+k^{2}\left(\omega_{x}\right)_{-}}{r^{2}+k^{2}}\, , \\
      &\left(\omega_{y}\right)_{+}=\frac{r \dot{x}_{-}+k^{2}\left(\omega_{y}\right)_{-}}{r^{2}+k^{2}}\, , \\
      & (\omega_z)_+=(\omega_z)_{-}\, .
      \label{change_velocities_sphere}
  \end{aligned}
  \end{equation}
  The penultimate step is to integrate equations~\eqref{eqs_sphere_R2} with the initial conditions $x(\bar{t})=x_1,\, y(\bar{t})=y_1,\, \dot x(\bar{t})=\dot x_+,\, \dot y(\bar{t})= \dot y_+,\, \dot z(\bar{t})=\dot z_+,\, \omega_{a}(\bar{t})=(\omega_a)_+\, (a=x,y,z)$.
  The equations of motion in $R_2$ obtained are
  \begin{align}
      x(t)&=\frac{e^{-\beta t}}{\beta ^2 \left(k^2+r^2\right)}
      \left[e^{\beta  t} \left(\lambda_1  \left(k^2+r^2\right) (\beta  t-\beta  \bar{t}-1)
      \right.\right.\\&\qquad \left.\left.
      +\beta  k^2 (r (\omega_y)_-+\beta  x_1)
      +\beta  r^2 (\dot x_-+\beta  x_1)\right)
      \right.\\&\qquad \left.
      +e^{\beta  \bar{t}} \left(k^2 (\lambda_1 -\beta  r (\omega_y)_-)+r^2 (\lambda_1 -\beta  \dot x_-)\right)\right]\, , \\
      y(t)&=\frac{e^{-\beta  t} }{\beta ^2 \left(k^2+r^2\right)}
      \left[e^{\beta  t} \left(\lambda_2  \left(k^2+r^2\right) (\beta  t-\beta  \bar{t}-1)
      \right.\right.\\&\qquad \left.\left.
      +\beta  k^2 (\beta  y_1-r (\omega_x)_-)
      +\beta  r^2 (\dot y_-+\beta  y_1)\right)
      \right.\\&\qquad \left.
      +e^{\beta  \bar{t}} \left(k^2 (\lambda_2 +\beta  r (\omega_x)_-)+r^2 (\lambda_2 -\beta  \dot y_-)\right)\right]\, , \\
      \omega_x(t)&=\frac{1}{\beta }{\lambda_2  r-\frac{e^{\frac{\beta  (\bar{t}-t)}{k^2}} \left(k^2 (\lambda_2  r-\beta  (\omega_x)_-)+\lambda_2  r^3+\beta  r \dot y_-\right)}{k^2+r^2}}\, , \\
      \omega_y(t)&= \frac{1}{\beta }{\frac{e^{\frac{\beta  (\bar{t}-t)}{k^2}} \left(k^2 (\lambda_1  r+\beta  (\omega_y)_-)+\lambda_1  r^3+\beta  r \dot x_-\right)}{k^2+r^2}-\lambda_1  r}\, , \\
      \omega_z(t)&=(\omega_z)_- e^{\frac{\beta  (\bar{t}-t)}{k^2}}\, .
  \end{align}
  Finally, the Lagrange multipliers $\lambda_1$ and $\lambda_2$ are obtained by imposing
  $\dot x - r\omega_y = 0$ and  $\dot y+ r \omega_x = 0$, yielding
  \begin{align}
      \lambda_1&=\frac{\beta  r \left(k^2 (\omega_y)_-+r \dot x_-\right) \left(e^{\beta  \bar{t}}-e^{\frac{\beta  (\bar{t}-t)}{k^2}+\beta  t}\right)}{\left(k^2+r^2\right) \left(r^2 e^{\frac{\beta  (\bar{t}-t)}{k^2}+\beta  t}-r^2 e^{\beta  t}-e^{\beta  t}+e^{\beta  \bar{t}}\right)}\, , \\
      \lambda_2&=\frac{\beta  r \left(r \dot y_--k^2 (\omega_x)_-\right) \left(e^{\beta  \bar{t}}-e^{\frac{\beta  (\bar{t}-t)}{k^2}+\beta  t}\right)}{\left(k^2+r^2\right) \left(r^2 e^{\frac{\beta  (\bar{t}-t)}{k^2}+\beta  t}-r^2 e^{\beta  t}-e^{\beta  t}+e^{\beta  \bar{t}}\right)}\, .
  \end{align}
\end{example}

\chapter{Nonsmooth Herglotz principle}\label{ch:nonsmooth_Herglotz}

\newcommand{\curve}{c}
\newcommand{\Leg}{\FF L}

\insquote{It [science] has as its highest principle and most coveted aim the solution of the problem to condense all natural phenomena which have been observed and are still to be observed into one simple principle, that allows the computation of past and more especially of future processes from present ones. ...Amid the more or less general laws which mark the achievements of physical science during the course of the last centuries, the principle of least action is perhaps that which, as regards form and content, may claim to come nearest to that ideal final aim of theoretical research.}{Max Planck, as quoted by Morris Kline, \emph{Mathematics and the Physical World} (1959)}

The possible trajectories considered both in Hamilton’s and Herglotz's principles are usually smooth curves. Nevertheless, many mechanical systems of interest have non-smooth trajectories. As a matter of fact, the trajectories described by a system with impacts are not smooth. Mechanical systems with impacts are usually modeled as hybrid systems (see \Cref{ch:hybrid_systems,ch:hybrid_reduction,ch:hybrid_HJ}). The disadvantage of this approach is that the impact map --the map characterizing the change of velocity in the instant of the impact-- has to be defined \textit{ad hoc} or obtained in some phenomenological fashion, for instance, by the Newtonian impact law (see \Cref{remark:Newtonian_Hamiltonian_impact_law}). 
Another approach consists on considering an impulsive constraint acting on the instant of the impact (see \Cref{ch:contact_impulsive})
Alternatively, in order to characterize the dynamics of a mechanical system with impacts, one can consider a variational principle for which the possible curves are not smooth at certain points. An extension of Hamilton’s Principle to a nonsmooth setting was developed by Fetecau, Marsden, Ortiz and West \cite{F.M.O+2003}. Inspired by their approach, in this chapter a non-smooth Herglotz principle for contact Lagrangian systems with impacts is developed. The main advantage of this approach is that one can characterize the dynamics of the system with dissipation and impacts by means of just the variational principle, without the need of considering additional forces, maps or constraints. 

The results of the present chapter were previously published in the conference paper~\cite{Lopez-Gordon2023}.

\section{Herglotz principle for nonsmooth action-dependent Lagrangians}\label{sec_herglotz}

\subsection{Herglotz principle for nonsmooth Lagrangians} 

Since trajectories with impacts are not smooth curves, the space of curves will no longer be a smooth manifold. Therefore, Herglotz variational principle cannot be generalized to a nonsmooth setting in a straightforward manner. In order to overcome this issue, the problem can be extended to the nonautonomous framework. This allows to define a path space $\mathcal M$ which is indeed a smooth manifold. In the nonsmooth Herglotz principle, variations on this submanifold will be considered. The position and time variables are regarded as functions of a parameter $\tau$. In this way, the impact can be fixed in $\tau$ space while remaining variable in both configuration and time spaces.

Let $(Q, L)$ be a regular contact Lagrangian system. Consider a submanifold with boundary $C \subseteq Q$, which represents the subset of admissible configurations. Let $\partial C$ denote the boundary of $C$.
The \emph{extended path space} is defined as
\begin{equation}
  \mathcal{M} = \mathcal{T} \times \mathcal{Q}\left([0,1], \tau_{a}, \partial C, Q\right)\, ,
\end{equation}
where
\begin{equation}
\begin{aligned} 
  \mathcal{T} = \left\{c_{t} \in \Cinfty([0,1], \mathbb{R}) \mid c_{t}^{\prime}>0 \text { in }[0,1]\right\}\, ,
\end{aligned}
\end{equation}
and
\begin{equation}
\begin{aligned}
  \mathcal{Q}\big([0,1], \tau_{a}, & \partial C, Q\big) =
  \left\{c_{q}\colon [0,1] \rightarrow Q \mid c_{q} \text { is a } \Czero,\right.\\
  &\left. \text { piecewise } \Ctwo \text { curve}\, , 
  \right.\\
  &\left.c_{q}(\tau) \text { has only one singularity at } \tau_{a}, c_{q}\left(\tau_{a}\right) \in \partial C\right\}\, .
\end{aligned}
\end{equation}
One can show that these spaces are smooth manifolds \cite{F.M.O+2003}. 
Here $\displaystyle{c_t^\prime(0) = \lim_{t\to 0^+} c_t^\prime(t)}$ and $\displaystyle{c_t^\prime(1) = \lim_{t\to 1^-} c_t^\prime(t)}$ are understood, and similarly for $c_q^\prime$ and higher derivatives.

A \emph{path} $c \in \mathcal{M}$ is a pair $c=\left(c_{t}, c_{q}\right)$. Given a path, the \emph{associated curve} $q\colon \left[c_{t}(0), c_{t}(1)\right] \rightarrow Q$ is given by $q(t) = c_q \circ c_t^{-1} (t)$. Notice that, since $c_t^\prime>0$, each $c_t\in \mathcal{T}$ is injective and thus invertible (if it is not surjective, it suffices to restrict the codomain), and therefore $q$ is well-defined.

The instant of the impact $\tau_a\in (0,1)$ is fixed in the $\tau$ space, but can vary in the $t$ space according to $t_a=c_t(\tau_a)$. The tangent space at $c_q\in \mathcal Q$ is given by
\begin{align}
  \T_{c_q} \mathcal{Q}=&\left\{v\colon [0,1] \rightarrow \T Q \mid v \text { is a } \Czero \text { piecewise } \Ctwo \operatorname{map},\right.\\ & \left.\qquad\qquad\qquad v\left(\tau_{a}\right) \in \T_{c_q\left(\tau_{a}\right)} \partial C\right\}\, .
\end{align}

Let $\hat \Omega \left(q_1, q_2, [0,1]  \right)\subset \mathcal{M}$ denote the subset of curves such that $c_q(0)=q_1$ and $c_q(1)=q_2$.
Consider the operator
\begin{equation}
  \hat{\zaction}\colon  \hat \Omega \left(q_1, q_2, [0,1]  \right) \to \mathcal{T}
\end{equation}
that assigns to each $c=(c_q, c_t)\in \mathcal{M}$ the solution of the following Cauchy problem:
\begin{equation}
\begin{aligned}        
  & \frac{\dd  \hat{\mathcal {Z}}(c)} {\dd  \tau} (\tau)
  = L \left( c_q(\tau), \frac{c_q'(\tau)}{c_t'(\tau)}, \hat{\mathcal {Z}}(c) (\tau)   \right) c_t'(\tau) \, ,\\
  &\hat{\mathcal {Z}}(c) (0) = \hat{z}_0\, ,
\end{aligned}
\end{equation}
and denote by $\hat{\mathcal {A}}$ the functional
\begin{equation}
\begin{aligned} 
  \hat{\mathcal {A}}:   \hat \Omega \left(q_1, q_2, [0,1]  \right) &\to \RR\\
  c & \mapsto \hat{\mathcal {Z}}(c) (1)\, .
\end{aligned}
\end{equation}

\begin{theorem}[Nonsmooth Herglotz variational principle]
  \label{theorem:nonsmooth_lagrangian}
  Let $(Q, L)$ be a regular contact Lagrangian system with Lagrangian energy $E_L$.
  Given a curve in $c=(c_q, c_t)\in \hat \Omega \left(q_1, q_2, [0,1]  \right)$ consider the curve $\chi\colon [0,1] \to \T Q\times \RR$ given by 
  \begin{equation}
    \chi(\tau) = \left(c_q(\tau), \frac{c_q'(\tau)}{c_t'(\tau)}, \hat{\zaction}(c)(\tau)\right) \, .
  \end{equation}
  Then, $c$ is a critical point of $\hat{\action}$ if and only if
  \begin{equation}
    \frac{\partial L} {\partial q^i} (\chi(\tau)) 
    -\frac{\dd  } {\dd t} \frac{\partial L} {\partial v^i } (\chi(\tau))  
    +\frac{\partial L} {\partial v^i } (\chi(\tau))  \frac{\partial L} {\partial  z} (\chi(\tau))   = 0\, , 
  \end{equation}
  for $\tau \in [0, \tau_a) \cup (\tau_a,1]$, and, for any $w = w^i\tparder{}{q^i} \in \T_{c_q(\tau_a)}\partial C$,
  \begin{equation}
  \begin{aligned}
    & \frac{\partial L} {\partial v^i} (\chi(\tau_a^-)) w^i
    = \frac{\partial L} {\partial v^i} (\chi(\tau_a^+)) w^i\, , \\
    & E_L (\chi(\tau_a^-))   = E_L (\chi(\tau_a^+))\, ,
  \end{aligned} \label{conditions_impact}
  \end{equation}
where $\displaystyle{\chi(\tau_a^\pm) = \lim_{\tau \to \tau_a^\pm} \chi(\tau)}$.
\end{theorem}

It is worth mentioning that the adjective ``nonsmooth'' refers to the fact that the curves in $\widehat \Omega(q_1, q_2, [0,1])$ are not differentiable, even though the Lagrangian function $L$ is differentiable.

\begin{proof}
Let $c=(c_q,c_t)\in  \hat \Omega \left(q_1, q_2, [0,1]  \right)$ be a curve. 
Consider a smoothly parametrized family of curves $c^\lambda=(c_q^\lambda, c_t^\lambda)\in \hat \Omega \left(q_1, q_2, [0,1]  \right)$ such that $c^0=c$ and 
\begin{equation}
  u(\tau) = \left.\frac{\dd c_q^\lambda(\tau)} {\dd \lambda} \right|_{\lambda=0}
  \, ,\qquad
  \theta(\tau) = \left.\frac{\dd c_t^\lambda(\tau)} {\dd \lambda} \right|_{\lambda=0}\, .
\end{equation}
In coordinates, denote $u(\tau)= \big(u^i(\tau)\big)$ for $i\in \{1, \ldots, n\}$.
Let $\varphi=\T_{c} \hat{\zaction}(u,\theta)$, so that $\T_c \hat{\action}(u, \theta)=\varphi(1)$. Observe that $\varphi(0)=0$, since $\hat{\zaction}(c^\lambda)(0)= \hat{z}_0$ for every $\lambda$. 
Thus,
\begin{equation}
\begin{aligned} 
  \varphi^{\!\prime}(\tau)
  & = \left. \frac{\dd  } {\dd \tau} \frac{\dd  } {\dd \lambda}   \hat{\zaction}(c^\lambda(\tau))\right|_{\lambda=0}
  = \left.  \frac{\dd  } {\dd \lambda} \frac{\dd  } {\dd \tau}  \hat{\zaction}(c^\lambda(\tau))\right|_{\lambda=0}\\
  &= \left.  \frac{\dd  } {\dd \lambda}  \left[
    L \left( c_q^\lambda(\tau), \frac{c_q^{\lambda \prime}(\tau)}{c_t^{\lambda \prime}(\tau)}, \hat{\mathcal {Z}}(c_q^\lambda, c_t^\lambda) (\tau)   \right) c_t^{\lambda \prime}(\tau) 
    \right] \right|_{\lambda=0}\\
  & = \left[ \frac{\partial L} {\partial q^i} (\chi(\tau)) u^i(\tau) 
      + \frac{\partial L} {\partial v^i } (\chi(\tau)) \frac{1}{c_t'(\tau)} u^{i \prime} (\tau)  \right.\\ &\left. \quad
      -\frac{\partial L} {\partial v^i } (\chi(\tau)) \frac{c_q'(\tau)}{c_t'(\tau)^2} \theta(\tau)
      + \frac{\partial L} {\partial z} (\chi(\tau)) \varphi(\tau)
   \right] c_t'(\tau)\\
&  +  L \left( c_q^\lambda(\tau), \frac{c_q^{\lambda \prime}(\tau)}{c_q^{\lambda \prime}(\tau)}, \hat{\mathcal {Z}}(c_q^\lambda, c_t^\lambda) (\tau)   \right) \theta'(\tau).
\end{aligned}
\end{equation}

An integrating factor for this ordinary differential equation is 
\begin{equation}
  \mu (\tau) = \exp \left( - \int_0^\tau \frac{\partial L} {\partial z} (\chi(s))\ c_t'(s) \dd s  \right)\, .
\end{equation}
Therefore,
\begin{equation}
\begin{aligned} 
    \varphi(\tau) \mu(\tau) 
    &= \int_0^\tau \mu(s) c_t'(s)
    \left[ \frac{\partial L} {\partial q^i} (\chi(s)) u^i(s) 
      \right.\\&\left.+ \frac{\partial L} {\partial v^i } (\chi(s)) \frac{1}{c_t'(s)} u^{i \prime} (s)-\frac{\partial L} {\partial v^i } (\chi(s)) \frac{c_q'(s)}{c_t'(s)^2} \theta'(s) \right] \ \dd s \\&+ \int_0^\tau \mu(s) L (\chi(s)) \theta'(s)\ \dd s\\
    &= \int_0^\tau \mu(s) c_t'(s)  
    \left[ \frac{\partial L} {\partial q^i} (\chi(s))u^i(s)
    \right.\\ &\left.
+ \frac{\partial L} {\partial v^i } (\chi(s)) \frac{u^{i \prime}(s)}{c_t'(s)}
       \right] \ \dd s   \\  
  & \quad + \int_0^\tau \mu(s) \theta'(s)
  \left[ L (\chi(s)) 
  -\frac{\partial L} {\partial v^i } (\chi(s)) \frac{c_q'(s)}{c_t'(s)}
    \right]\dd s\, .
\end{aligned}
\end{equation}
Integrating by parts and taking into account that $u(0)=u(1)=0$ and $\theta(0)=\theta(1)=0$ yields
\allowdisplaybreaks
\begin{align} 
  \varphi(1) \mu (1)
    & = \int_0^{\tau_a} \mu(s) c_t'(s)  u^i(s)
    \left[ \frac{\partial L} {\partial q^i} (\chi(s)) 
    - \frac{\dd  } {\dd t} \frac{\partial L} {\partial v^i } (\chi(s))  \right.\\
    & \quad\left.+\frac{\partial L} {\partial v^i } (\chi(s))  \frac{\partial L} {\partial  z} (\chi(s)) \right] \, \dd s   \\  
  & \quad - \int_0^{\tau_a} \mu(s) \theta(s) 
  \frac{\dd  } {\dd s} \left[ L (\chi(s)) 
  -\frac{\partial L} {\partial v^i } (\chi(s)) \frac{c_q'(s)}{c_t'(s)}
    \right] \dd s\\
    & \quad + \int_0^{\tau_a} \mu(s) \theta(s) 
  \left[ L (\chi(s)) 
  \right.\\ &\left.\qquad
  -\frac{\partial L} {\partial v^i } (\chi(s)) \frac{c_q'(s)}{c_t'(s)}
    \right] \frac{\partial L} {\partial z} (\chi(s))\ c_t'(s)  \, \dd s\\
  & \quad +\int_{\tau_a}^1 \mu(s) c_t'(s)  u^i(s)
  \left[ \frac{\partial L} {\partial q^i} (\chi(s)) 
  -\frac{\dd  } {\dd t} \frac{\partial L} {\partial v^i } (\chi(s))  \right.\\
  &\qquad\left.+\frac{\partial L} {\partial v^i } (\chi(s))  \frac{\partial L} {\partial  z} (\chi(s)) \right] \, \dd s   \\  
  & \quad - \int_{\tau_a}^1 \mu(s) \theta(s) 
  \frac{\dd  } {\dd s} \left[ L (\chi(s)) 
  -\frac{\partial L} {\partial v^i } (\chi(s)) \frac{c_q'(s)}{c_t'(s)}
    \right] \, \dd s\\
    & \quad + \int_0^{\tau_a} \mu(s) \theta(s) 
  \left[ L (\chi(s)) 
  \right.\\ &\left.\qquad
  -\frac{\partial L} {\partial v^i } (\chi(s)) \frac{c_q'(s)}{c_t'(s)}
    \right] \frac{\partial L} {\partial z} (\chi(s))\, c_t'(s)\,   \dd s\\
  &\quad 
  +\left. \frac{\partial L} {\partial v^i} (\chi(s)) u^i(s) \mu(s) \right|_{\tau_a^+}^{\tau_a^-}
  \\&+ \left.\mu(s) \theta(s)  \left[ L (\chi(s)) 
  -\frac{\partial L} {\partial v^i } (\chi(s)) \frac{c_q'(s)}{c_t'(s)}
    \right] \right|_{\tau_a^+}^{\tau_a^-}\, .
\end{align}
Taking into account that $\mu(\tau)$ and $c_t'$ are nonzero, the fundamental lemma of calculus of variations implies that $\varphi(1)$ vanishes for every $u$ and every $\theta$ (that is, $c$ is a critical point of $\hat{\action}$) if and only if
\begin{subequations}
\begin{flalign}
  &\frac{\partial L} {\partial q^i} (\chi(\tau)) 
    -\frac{\dd  } {\dd t} \frac{\partial L} {\partial v^i } (\chi(\tau))  
    +\frac{\partial L} {\partial v^i } (\chi(\tau))  \frac{\partial L} {\partial  z} (\chi(\tau))   = 0\, ,\\
    &\frac{\dd  } {\dd \tau} \left[ L (\chi(\tau)) 
  -\frac{\partial L} {\partial v^i } (\chi(\tau)) \frac{c_q'(\tau)}{c_t'(\tau)}
    \right]\\
\qquad &= \left[ L (\chi(\tau)) 
  -\frac{\partial L} {\partial v^i } (\chi(\tau)) \frac{c_q'(\tau)}{c_t'(\tau)}
    \right] \frac{\partial L} {\partial z} (\chi(\tau))\ c_t'(\tau)\, ,
\end{flalign}
\end{subequations} for $\tau \in [0, \tau_a) \cup(\tau_a,1]$, and
\begin{subequations}
\begin{flalign}
   \frac{\partial L} {\partial v^i} (\chi(\tau_a^-))  w^i
  &= \frac{\partial L} {\partial v^i} (\chi(\tau_a^+)) w^i\, , \\
   L (\chi(\tau_a^-)) 
  &-\frac{\partial L} {\partial v^i } (\chi(\tau_a^-)) \frac{c_q'(\tau_a^-)}{c_t'(\tau_a^-)}
 \\& = L (\chi(\tau_a^+)) 
  -\frac{\partial L} {\partial v^i } (\chi(\tau_a^+)) \frac{c_q'(\tau_a^+)}{c_t'(\tau_a^+)}\, ,
\end{flalign}
\end{subequations}
for any $w = w^i \tparder{}{q^i} \in \T_{c_q(\tau_a)}\partial C$. 
In other words,
  \begin{flalign}
    &\frac{\partial L} {\partial q^i} (\chi(\tau)) 
      -\frac{\dd  } {\dd t} \frac{\partial L} {\partial v^i } (\chi(\tau))  
      +\frac{\partial L} {\partial v^i } (\chi(\tau))  \frac{\partial L} {\partial  z} (\chi(\tau))   = 0\, , \label{eq:Herglotz_nonsmooth_a}
      \\
      &\frac{\dd  } {\dd t} E_L (\chi(\tau)) 
      =  E_L (\chi(\tau)) \frac{\partial L} {\partial z} (\chi(\tau)) \label{eq:Herglotz_nonsmooth_b}\, ,
  \end{flalign}
  for $\tau \in [0, \tau_a) \cup(\tau_a,1]$, and
  \begin{subequations}
  \begin{flalign}
    & \frac{\partial L} {\partial v^i} (\chi(\tau_a^-))  w^i = \frac{\partial L} {\partial v^i} (\chi(\tau_a^+)) w^i\, , \\
    & E_L (\chi(\tau_a^-)) = E_L (\chi(\tau_a^+)) \, ,
  \end{flalign}
  \end{subequations}
  for any $w = w^i \tparder{}{q^i} \in \T_{c_q(\tau_a)}\partial C$. Finally, observe that equation~\eqref{eq:Herglotz_nonsmooth_a} implies equation~\eqref{eq:Herglotz_nonsmooth_b}, making the latter redundant. Indeed, one can write
  \begin{equation}
  \begin{aligned} 
    \frac{\dd  } {\dd t} E_L (\chi(\tau))
    &= \frac{\dd  } {\dd t}
    \left( \frac{\partial L} {\partial v^i} (\chi(\tau))  \frac{(c_q^i)'(\tau)}{c_t'(\tau)} - L(\chi(\tau))  \right) \\ 
    & =  \left( \frac{\dd  } {\dd t}\frac{\partial L} {\partial v^i} (\chi(\tau))  - \frac{\partial L} {\partial q^i} (\chi(\tau)) \right) \frac{(c_q^i)'(\tau)}{c_t'(\tau)}\\
    & \quad - \frac{\partial L} {\partial z}(\chi(\tau)) \frac{\dd}{\dd t} \hat{\zaction}(c_q, c_t) (\tau) \\
    & =  \left( \frac{\dd  } {\dd t}\frac{\partial L} {\partial v^i} (\chi(\tau))  - \frac{\partial L} {\partial q^i} (\chi(\tau)) \right) \frac{(c_q^i)'(\tau)}{c_t'(\tau)}\\
    & \quad - \frac{\partial L} {\partial z}(\chi(\tau)) L (\chi(\tau)) \, ,
  \end{aligned}
  \end{equation}
  which in combination with equation~\eqref{eq:Herglotz_nonsmooth_a} yields
  \begin{equation}
  \begin{aligned}
    \frac{\dd  } {\dd t} E_L (\chi(\tau))
    & = \frac{\partial L} {\partial v^i } (\chi(\tau))  \frac{\partial L} {\partial  z} (\chi(\tau)) \frac{(c_q^i)'(\tau)}{c_t'(\tau)}\\
    & \quad - \frac{\partial L} {\partial z}(\chi(\tau)) L (\chi(\tau)) \\
    & = E_L (\chi(\tau)) \frac{\partial L} {\partial z}(\chi(\tau))\, .
  \end{aligned}
  \end{equation}
\end{proof}




\section{Example: Billiard with dissipation}\label{sec_billiard}

\begin{figure}[t]
  \centering
  \includegraphics[width=.6\linewidth]{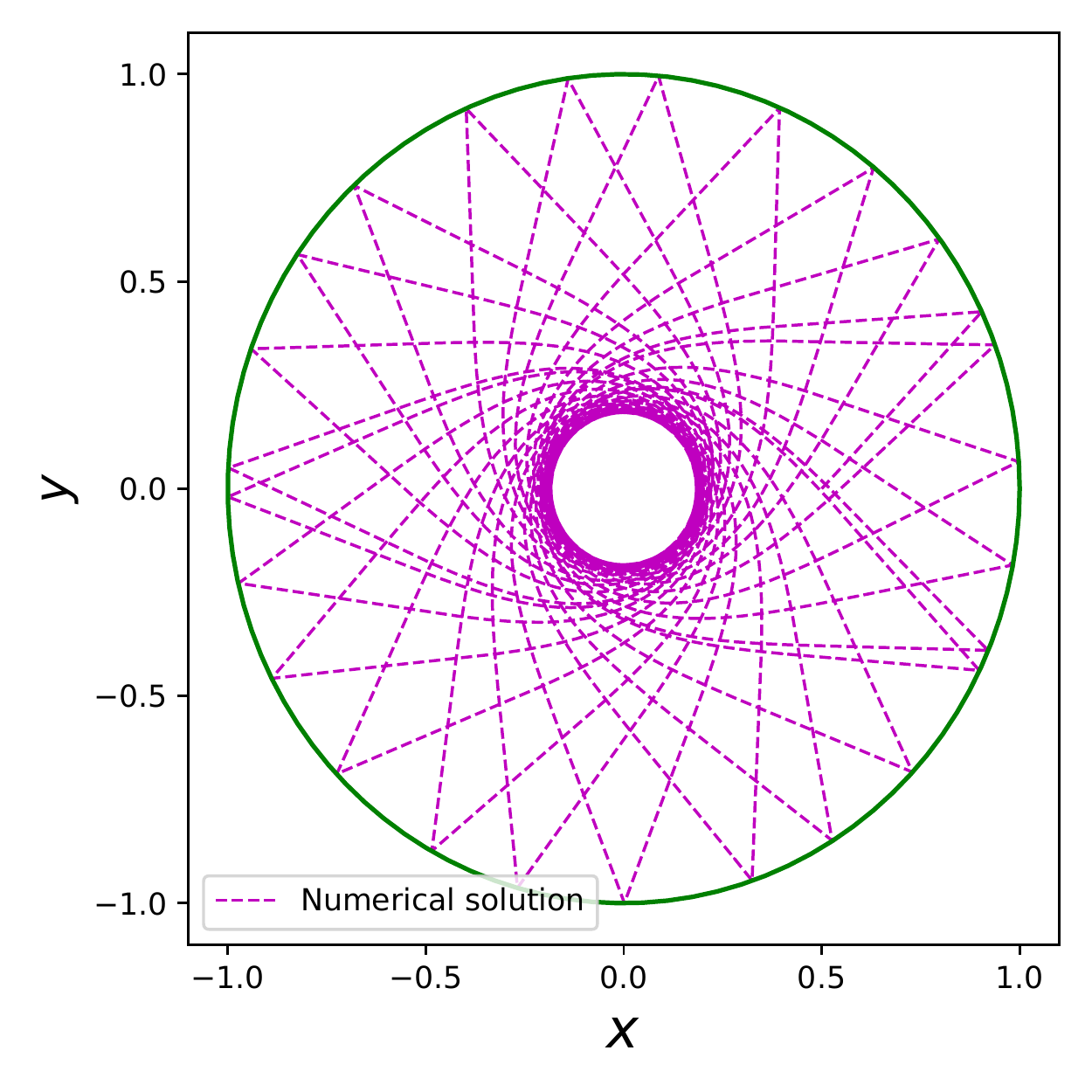}
  \caption{Numerical simulation for the trajectory of a particle in the billiard, with $\kappa = 10^{-4}$.}
  \label{fig:simulation_nonsmooth_circle}
\end{figure}

Consider the contact Lagrangian system $(\RR^2, L)$
\begin{equation}
  L (x, y, v_x, v_y, z)
  = \frac{1}{2} \left( v_x^2 + v_y^2  \right)
  - \kappa z\, ,
\end{equation}
for some constant $\kappa \in \RR$, where $(x, y, v_x, v_y, z)$ are the canonical bundle coordinates in $\TT \RR^2 \times \R$.
Suppose that the system is confined to the unit ball centered at the origin, namely,
\begin{equation}
  \BB = \{(x,y) \in \RR^2 \mid x^2+y^2\leq1\}\, .
\end{equation}
Its border is the unit circle
\begin{equation}
  \partial\BB = \Sp^1 = \{(x,y) \in \RR^2 \mid x^2+y^2 = 1\}\, .
\end{equation}
The tangent space to $\Sp^1$ is spanned by the generator of rotations on the plane, namely, 
\begin{equation}
  \T_{(x,y)} \Sp^1 = \left\langle y \parder{}{x} - x \parder{}{y} \right \rangle
\end{equation}

A curve $c\colon \RR \to \RR^2, \, c(t)=(x(t), y(t))$ is a critical point of the action if and only if it satisfies the Herglotz equations, namely,
\begin{equation}
\begin{array}{lll} 
  \ddot x(t) = - \kappa \dot x(t)\, ,
  &\ddot y(t) = - \kappa \dot y(t)\, .
\end{array}
\end{equation}
Their solutions for the initial conditions $x(0)=x_0,\ y(0)=y_0,\ \dot{x}(0)=v_{x,0},\ \dot{y}(0)=v_{y,0}$ 
are
\begin{equation}
\begin{aligned} 
    &x(t) = x_0 + \frac{v_{x,0}}{\kappa} \left( 1- e^{-t\kappa}  \right)\, ,\\
    &y(t) = y_0 + \frac{v_{y,0}}{\kappa} \left( 1- e^{-t\kappa}  \right)\, .
\end{aligned}
\end{equation}

Conditions \eqref{conditions_impact} can then be written as
\begin{equation}
\begin{aligned} 
  & \dot{x}^- -\frac{x}{y} \dot{y}^- = \dot{x}^+ -\frac{x}{y} \dot{y}^+\,  , \\
  & \frac{1}{2}\left( (\dot{x}^-)^2 + (\dot{y}^-)^2 \right) + \kappa \hat{\zaction}(c)(\tau_a^-)
  =  \frac{1}{2}\left( (\dot{x}^+)^2 + (\dot{y}^+)^2 \right) + \kappa \hat{\zaction}(c)(\tau_a^+)\, ,
\end{aligned}
\end{equation}
where $\dot{x}^\pm = \dot{x}(t_a^\pm)$ and $\dot{y}^\pm = \dot{y}(t_a^\pm)$. 
Assuming that $z_{\mid t_a^+}=z_{\mid t_a^-}$, one obtains
\begin{align}
    & \dot{x}^+ = \frac{-\dot{x}^- x^2+\dot{x}^- y^2-2 \dot{y}^- x y}{x^2+y^2}\, ,\\
    & \dot{y}^+ = \frac{-2 \dot{x}^- x y+\dot{y}^- x^2-\dot{y}^- y^2}{x^2+y^2}\, .
\end{align}
Consider the polar coordinates $(r, \theta)$ in $\RR^2$, namely,
\begin{equation}
  r = \sqrt{x^2+y^2}\, , \quad \theta = \arctan \left(\frac{y}{x} \right)\, , 
\end{equation}
and let $(r, \theta, v_r, v_\theta, z)$ be the induced bundle coordinates in $\TT \RR^2 \times \RR$. In these coordinates, the Lagrangian function reads
\begin{equation}
    L(r, \theta, v_r, v_\theta,z ) = \frac{1}{2}\left(v_r^2 + r^2 v_\theta^2 \right) - \kappa z \, .
\end{equation}
One can check that the function $\ell = r^2v_\theta$ is a dissipated quantity, namely,
\begin{equation}
    \frac{\dd}{\dd t}  \ell \circ c (t) = \frac{\partial L}{\partial z} \ell \circ c(t) = -\kappa \ell\circ c(t)\, ,
\end{equation}
along a solution $c$ of Herglotz equations. Therefore, $\ell \circ c (t) = \ell_0 e^{-\kappa t}$, where $\ell_0=\ell\circ c(0)$. 
The equations of motions outside the impact surface can thus be expressed as
\begin{equation}
\begin{aligned}
    &\ddot r = -\kappa \dot{r} + \frac{\ell_0}{r}e^{-\kappa t},\\
    &\dot{\theta} = \frac{\ell_0}{r^2}e^{-\kappa t}.   
\end{aligned}
\label{eqs_motion_polar}
\end{equation}
On the other hand, deriving $r(t)^2=x(t)^2+y(t)^2$ yields
\begin{equation}
  2r(t)\dot{r}(t)=2(x(t)\dot{x}(t) + y(t)\dot{y}(t))\, .
\end{equation}
Hence,
\begin{equation}
\begin{aligned}
    r\dot{r}^+ &= x\dot{x}^+ + y\dot{y}^+\\
    &= \frac{x\left(-\dot{x}^- x^2+\dot{x}^- y^2-2 \dot{y}^- x y\right)}{x^2+y^2} \\
    & + \frac{y\left(-2 \dot{x}^- x y+\dot{y}^- x^2-\dot{y}^- y^2\right)}{x^2+y^2} \\
    & = -\left( x\dot{x}^- +y \dot{y}^-\right) = -r\dot{r}^-,
\end{aligned}
 \label{eqs_impact_r}
\end{equation}
which implies that $\dot{r}^+ = -\dot{r}^-$. Similarly, deriving $\theta=\arctan(y/x)$ yields
\begin{equation}
\begin{aligned}
    \dot{\theta}^+ &= \frac{1}{1+(y/x)^2} \left(\frac{x\dot{y}^+  - y \dot{x}^+}{x^2}\right)\\
    &= \frac{x\left(-2 \dot{x}^- x y+\dot{y}^- x^2-\dot{y}^- y^2\right)}{(x^2+y^2)^2} \\
    &+ \frac{y\left(-\dot{x}^- x^2+\dot{x}^- y^2-2 \dot{y}^- x y\right)}{(x^2+y^2)^2} \\
    &= \frac{(-y\dot{x}^- + x \dot{y}^-)(x^2+y^2)}{(x^2+y^2)^2} = \dot{\theta}^-.
\end{aligned}
 \label{eqs_impact_theta}
\end{equation}
Making use of equations~\eqref{eqs_motion_polar}, \eqref{eqs_impact_r} and \eqref{eqs_impact_theta}, a python numerical simulation has been performed for $\kappa = 10^{-4}$ and initial values $x(0)=0.5,\ y(0)=0,\ \dot{x}(0)=1,\ \dot{y}(0)=1$ (see \Cref{fig:simulation_nonsmooth_circle}).

\chapter{Conclusions and further work}\label{ch:further_work}

\insquote{Le savant n’est pas l’homme qui fournit les vraies réponses; c’est luis qui pose les vraies questions.}{Claude Lévi-Strauss, \emph{Le Cru et le Cuit} (1964)}

In this dissertation, different geometrical frameworks for non\hyphen{}conservative mechanical systems have been studied, with a special emphasis on the symmetries and integrability aspects of these systems. \Cref{part:forces,part:contact} were devoted to forced systems and to contact systems, respectively. Both these formalisms allow modelling certain mechanical systems which experience a dissipation of energy in a continuous manner along the evolution. On the other hand, \Cref{part:impacts} explores different geometrical frameworks for systems with impacts: hybrid systems, impulsive constraints and a variational principle considering nonsmooth curves. Most of these systems experience instantaneous changes of energy in the moments of the impacts. In addition, they may experience a continuous dissipation of energy along the time.

\section*{Summary of contributions}

The main novelties presented in this dissertation are the following:

\begin{itemize}
    \item A Liouville--Arnol'd theorem for contact Hamiltonian systems has been proven in \Cref{ch:contact_integrability}. In order to prove this theorem, a Liouville--Arnol'd theorem for homogeneous functions on exact symplectic manifold has also been proven. Unlike previous approaches towards contact integrability, the one on the present dissertation does not require the Hamiltonian function to be preserved by the Reeb flow.
    \item The symmetries on forced systems (\Cref{ch:forced_symmetries}) and time-dependent contact systems (\Cref{ch:contact_symmetries}) have been classified, showing the relations between them and with conserved or dissipated quantities.
    \item Reduction procedures for forced (\Cref{ch:forced_symmetries}) and hybrid (\Cref{ch:hybrid_reduction}) systems have been developed.
    \item Hamilton--Jacobi equations for forced (\Cref{ch:forced_HJ}), discrete forced (\Cref{ch:forced_discrete_HJ}), time-dependent contact (\Cref{ch:contact_HJ}) and hybrid systems (\Cref{ch:hybrid_HJ}) have been deduced. 
    \item The use of dissipated quantities for studying the stability of a contact Hamiltonian system via the Lyapunov method has been sketched out in \Cref{ch:contact_stability}.
    \item A notion of integrability for hybrid Hamiltonian systems has been proposed (\Cref{ch:integrability_hybrid}).
    \item In order to obtain a variational principle for dissipative systems with impacts, the Herglotz principle has been generalized for nonsmooth curves (\Cref{ch:nonsmooth_Herglotz}).
\end{itemize}

\section*{Future research}

There are several lines of research that can be derived from this thesis. Some of them are the following:

\begin{itemize}
    \item To delve on the study of completely integrable contact systems (in the sense of \Cref{def:integrable_systems}). Although the proof of the Liouville--Arnol'd theorem provides a method for computing action-angle coordinates, it is by no means a computationally efficient algorithm. It seems that the solutions of the Hamilton--Jacobi equation may be utilized for the computation of action-angle coordinates. Additionally, 
    it is pending to consider completely integrable contact systems with critical points, that is, non-regular values of $F=(f_\alpha)$ (see \cite{P.N2011}). In the same fashion as \Cref{def:integrable_systems}, one could define partial integrable and superintegrable contact systems. Furthermore, other structures employed in the study of classical integrable systems could be generalized to completely integrable contact systems. These include bi-Hamiltonian structures \cite{M.C.F+1997,C.M.P1993,F.M.P2000,Fernandes1994}, momentum polytopes \cite{Atiyah1982,Delzant1988,G.S1982,P.N2011} and Haantjes tensors \cite{T.T2022, R.T.T2022}. Additionally, the notion of completely integrable contact system could serve as a starting point to develop a Kolmogorov--Arnol'd--Moser (\textsc{kam}) theory for contact Hamiltonian systems \cite{Arnold1963a,Kolmogorov1954,Moser1962,Dumas2014,Chierchia2009}. 
    It is worth mentioning that there is a plethora of open problems regarding integrable systems, even in the finite-dimensional symplectic setting \cite{B.M.M+2018}.
    \item To consider (co)contact manifolds which are not necessarily co-orientable. In order to study a dynamical system on such manifolds, one can always assume that a contact form exists at least locally. Nevertheless, there are some applications for which considering a contact distribution instead of a contact form is more convenient. For instance, one could consider the reduction of a contact manifold by a group of symmetries which preserves the contact distribution but not the contact form \cite{Willett2002,G.G2023a}. Furthermore, a more general definition of cocontact manifold, in terms of distributions instead of one-forms, will be considered in future works. This new notion will provide a clearer geometrical interpretation of cocontact structures.
    \item A complete solution of the Hamilton--Jacobi equation for a $2n$-dimensional forced Hamiltonian system provides $n$ independent conserved quantities $f_i$, which are in involution with respect to the canonical Poisson bracket. If the Hamiltonian vector fields of $f_i$ are complete, then each of the level sets $\cap_{i=1}^n f_i^{-1}(\lambda_i)\, (\lambda_i\in \RR)$ will be diffeomorphic to a product of tori and real lines. It is thus natural to explore if, in a neighbourhood of each of those level sets, there exist some sort of ``action-angle'' coordinates in which the forced Hamiltonian dynamics become trivial.
    \item To extend the results of \Cref{part:impacts} to hybrid systems that experience Zeno effect, that is, infinitely many impacts occur in a finite amount of time \cite{A.Z.G+2006,D.F2018,L.A2008}. Such systems include the bouncing ball \cite{O.T2011,A.Z.G+2006}.
\end{itemize}

\chapter{Conclusiones y trabajo futuro}\label{ch:trabajo_futuro}


En esta disertación se han estudiado diferentes marcos geométricos para sistemas mecánicos no conservativos, con un particular énfasis en las simetrías e integrabilidad de dichos sistemas. Las Partes~\textsc{\ref{part:forces}} y \textsc{\ref{part:contact}} versan sobre los sistemas forzados y sobres los sistemas de contacto, respectivamente. Ambos formalismos permiten modelar ciertos sistemas mecánicos que experimentan una disipación de la energía de manera continua a lo largo de su evolución. Por otro lado, la Parte~\textsc{\ref{part:impacts}} explora diferentes marcos geométricos para sistemas con impactos: sistemas híbridos, ligaduras impulsivas y un principio variacional que considera curvas no diferenciables. La mayor parte de estos sistemas experimentan cambios instantáneos en la energía en el momento de los impactos. Asimismo, pueden experimentar una disipación de la energía continua a lo largo del tiempo.

\section*{Resumen de contribuciones}

Las principales novedades presentadas en esta disertación son las siguientes:

\begin{itemize}
    \item Se ha demostrado un teorema de Liouville--Arnol'd para sistemas hamiltonianos de contacto en el Capítulo~\ref{ch:contact_integrability}. Con el fin de probar este teorema, también se ha demostrado un teorema de Liouville--Arnol'd para funciones homogéneas en variedades simplécticas exactas. A diferencia de otras formas de abordar la integrabilidad en sistemas de contacto que se habían considerado previamente, la que se expone en esta disertación no requiere que la función hamiltoniana se preserve a lo largo del flujo de Reeb.
    \item Se han clasificado las simetrías en sistemas forzados (Capítulo~\ref{ch:forced_symmetries}) y en sistemas de contacto dependientes del tiempo (Capítulo~\ref{ch:contact_symmetries}), mostrando las relaciones entre ellas y con las cantidades conservadas o disipadas.
    \item Se han desarrollado procedimientos para la reducción de sistemas forzados (Capítulo~\ref{ch:forced_symmetries}) e híbridos (Capítulo~\ref{ch:hybrid_reduction}).
    \item Se han deducido ecuaciones de Hamilton--Jacobi para sistemas forzados (Capítulo~\ref{ch:forced_HJ}), sistemas discretos forzados (Capítulo~\ref{ch:forced_discrete_HJ}), sistemas de contacto dependientes del tiempo (Capítulo~\ref{ch:contact_HJ}) y sistemas híbridos (Capítulo~\ref{ch:hybrid_HJ}).
    \item Se ha presentado un primer esbozo de cómo emplear las cantidades disipadas para estudiar la estabilidad de un sistema hamiltoniano de contacto mediante la teoría de Lyapunov (Capítulo~\ref{ch:contact_stability}).
    \item Se ha propuesto una noción de integrabilidad para sistemas híbridos hamiltonianos (Capítulo~\ref{ch:integrability_hybrid}).
    \item Con el fin de obtener un principio variacional para sistemas disipativos con impactos, el principio de Herglotz se ha generalizo para curvas no diferenciables (Capítulo~\ref{ch:nonsmooth_Herglotz}).
\end{itemize}

\section*{Investigación futura}

Hay diversas líneas de investigación que pueden seguirse a partir de esta tesis. Algunas de ellas son las que siguen:

\begin{itemize}
    \item Profundizar en el estudio de los sistemas de contacto completamente integrables (en el sentido de la Definición~\ref{def:integrable_systems}). Aunque la demostración del Teorema de Liouville--Arnol'd proporciona un método para calcular las coordenadas de acción-ángulo, no es para nada un algoritmo computacionalmente eficiente. Aparentemente, las soluciones de la ecuación de Hamilton--Jacobi podrían emplearse para calcular las coordenadas de acción-ángulo. Asimismo, queda pendiente estudiar sistemas de contacto integrables con singularidades, esto es, valores no regulares de $F=(f_\alpha)$ (véase \cite{P.N2011}). De manera análoga a la Definición~\ref{def:integrable_systems} es posible definir sistemas de contacto parcialmente integrables o superintegrables. Además, otras estructuras empleadas para estudiar sistemas integrables clásicos podrían generalizarse para sistemas de contacto completamente integrables. Entre estas están las estructuras bihamiltonianas \cite{M.C.F+1997,C.M.P1993,F.M.P2000,Fernandes1994}, los politopos de momento \cite{Atiyah1982,Delzant1988,G.S1982,P.N2011} y los tensores de Haantjes \cite{T.T2022, R.T.T2022}. Es más, la noción de sistema de contacto completamente integrable podría servir como punto de partida para desarrollar una teoría de Kolmogorov--Arnol'd--Moser (\textsc{kam}) para sistemas hamiltonianos de contacto \cite{Arnold1963a,Kolmogorov1954,Moser1962,Dumas2014,Chierchia2009}. Cabe mencionar que hay una plétora de problemas abiertos concernien a los sistemas integrables, incluso en el caso simpléctico y de dimensión finita \cite{B.M.M+2018}.
    \item Considerar variedades de (co)contacto no necesariamente co-orientables. Con el fin de estudiar un sistema dinámico en tales variedades, siempre es posible asumir que existe, al menos localmente, una forma de contacto. No obstante, hay aplicaciones para las que resulta más conveniente considerar una distribución de contacto en lugar de una forma de contacto, por ejemplo, la reducción de una variedad de contacto por un grupo de simetrías que preserva la distribución pero no la forma \cite{Willett2002,G.G2023a}. Además, en futuros trabajos se estudiará una definición más general de variedad de cocontacto, en términos de distribuciones en lugar de formas, lo que propocionará una interpretación geométrica más clara de las estructuras de cocontacto.
    \item Una solución completa de la ecuación de Hamilton--Jacobi para un sistema hamiltoniano forzado con $2n$ dimensiones proporciona $n$ cantidades conservadas independentes $f_i$, las cuales están en involución con respecto al corchete de Poisson canónico. Si los campos hamiltonianos de $f_i$ son completos, cada uno de los conjuntos de nivel $\cap_{i=1}^n f_i^{-1}(\lambda_i)\, (\lambda_i\in \RR)$ es difeomorfo a un producto de toros y rectas reales. Por ello, resulta natural explorar si, en un entorno de dichos conjuntos de nivel, existen una suerte de coordenadas de <<acción-ángulo>> en las que las ecuaciones de Hamilton forzadas se vuelvan triviales.
    \item Extender los resultados de la Parte~\textsc{\ref{part:impacts}} para sistemas híbridos que experimenten efecto Zenón, esto es, que infinitos impactos puedan ocurrir en un intervalo de tiempo finito \cite{A.Z.G+2006,D.F2018,L.A2008}. Estos sistemas incluyen la bola que rebota \cite{O.T2011,A.Z.G+2006}.
\end{itemize}

\cleardoublepage
\defbibheading{bibintoc}[\bibname]{%
  \phantomsection
  \manualmark
  \markboth{\spacedlowsmallcaps{#1}}{\spacedlowsmallcaps{#1}}%
  \addtocontents{toc}{\protect\vspace{\beforebibskip}}%
  \addcontentsline{toc}{chapter}{\tocEntry{#1}}%
  \chapter*{#1}%
}
\printbibliography[heading=bibintoc]


\cleardoublepage\pagestyle{empty}

\hfill

\vfill

\pdfbookmark[0]{Colophon}{colophon}
\section*{Colophon}
This document was typeset using the typographical look-and-feel \texttt{classicthesis} developed by Andr\'e Miede and Ivo Pletikosić.
The style was inspired by Robert Bringhurst's seminal book on typography ``\emph{The Elements of Typographic Style}''.
\texttt{classicthesis} is available for both \LaTeX\ and \mLyX:
\begin{center}
\url{https://bitbucket.org/amiede/classicthesis/}
\end{center}

\bigskip

\noindent\finalVersionString


%
%

\thispagestyle{empty}
\cleardoublepage
\newpagecolor{RoyalBlue}
\begin{titlepage}
    \begin{addmargin}[-1cm]{-3cm}
    {\color{silvergrey}

    \begin{center}
        \large

        \hfill


        \begingroup
            \color{mintgreen}
            \LARGE\textdubf{THE GEOMETRY OF DISSIPATION} \\ \bigskip
        \endgroup

        \Large{Asier López-Gordón}

    \end{center}

    \vfill

    \noindent Dissipative phenomena manifest in multiple mechanical systems. In this dissertation, different geometric frameworks for modelling non-conservative dynamics are considered. The objective is to generalize several results from conservative systems to dissipative systems, specially those concerning the symmetries and integrability of these systems. More specifically, three classes of geometric frameworks modelling dissipative systems are considered: systems with external forces, contact systems and systems with impacts. The first two allow modelling a continuous dissipation of energy over time, while the latter also permits considering abrupt changes of energy in the instants of the impacts. 

    \vfill

    \begin{flushleft}
        \large 
        Thesis realized at the Institute of Mathematical Sciences (ICMAT)\\
        Supervised by Manuel de León\\
        Doctoral Programme in Mathematics\\
        Autonomous University of Madrid\\
        \vspace{4cm}

        \Large{Madrid, 2024}
    \end{flushleft}

    }
  \end{addmargin}
\end{titlepage}
















\end{document}